\documentclass[timesnewroman,12pt, twoside]{report}
\usepackage[titletoc]{appendix}
\usepackage[a4paper]{geometry}
\usepackage{authblk, lineno, color, lscape, url, sectsty}
\usepackage[autostyle]{csquotes}
\usepackage[dvipsnames]{xcolor}
\usepackage[labelfont=bf]{caption}
\usepackage{emptypage}
\usepackage{array, float, multirow, graphicx, setspace, cite, url, titlesec, tocloft}
\usepackage{amsmath, amsfonts, mathtools, notoccite}
\usepackage{amssymb}
\usepackage{chngcntr, rotating, pdflscape, enumitem, algpseudocode, algorithm, subcaption}
\usepackage[square,sort,comma,numbers,sort&compress]{natbib}
\usepackage[colorlinks=True,citecolor=blue,linkcolor=blue,anchorcolor=blue,filecolor=blue,urlcolor=blue]{hyperref}
\allowdisplaybreaks
\usepackage{tensor}
\usepackage{braket}
\usepackage{booktabs}
\usepackage{lipsum}
\usepackage[Sonny]{fncychap}
\usepackage{adjustbox}
\usepackage{makecell}
\usepackage{slashed}
\usepackage[normalem]{ulem}
\usepackage{wasysym}
\def\blankpage{%
	\clearpage%
	\thispagestyle{empty}%
	\addtocounter{page}{-1}%
	\null%
	\clearpage}
\begin{document}
\pagenumbering{roman}
\thispagestyle{empty}
\thispagestyle{empty}
\graphicspath{{Figures/PNG/}{Figures/}}
\begin{center}
{\bf {\Large Probing Beyond the Standard Model Scenarios in Long-baseline and  Astrophysical Neutrino Experiments} }
\end{center}
\vspace{0.3 cm}
\begin{center}
    {\it \large By}
\end{center}
\begin{center}
    {\bf {\Large Sudipta Das } \\ PHYS07201804013}
\end{center}
\begin{center}
\bf {{\large Institute of Physics, Bhubaneswar, India }}
\end{center}
\vskip 2.0 cm
\begin{center}
\large{
{ A thesis submitted to the }  \\
 {Board of Studies in Physical Sciences }\\

In partial fulfillment of requirements \\
For the Degree of } \\
{\bf  DOCTOR OF PHILOSOPHY} \\
\emph{of} \\
{\bf HOMI BHABHA NATIONAL INSTITUTE}
\vskip 1.0 cm
\begin{figure}[H]
	\begin{center}
    \includegraphics[width=4.0cm, height= 4.0cm]{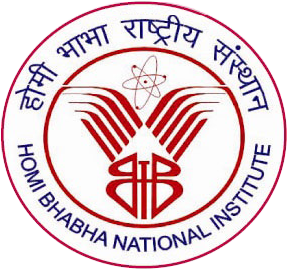}
	\end{center}
\end{figure}
\vskip 0.2 cm
{\bf \today}
\end{center}
\thispagestyle{empty}
\blankpage 

\oddsidemargin 0.0in \evensidemargin -0.0in
\begin{center}
	{\large\textbf{DEDICATIONS}}
\end{center}
\vspace*{3.0in}
\hspace*{2.0in}
\begin{center}
        {\Large \emph{Dedicated To My Parents}\\}
\hrule
\end{center}
    
\newpage  
\begin{center}
\large \bf
ACKNOWLEDGMENTS
\end{center}
\vspace{1.5cm}
This thesis dissertation is the culmination of hard work and effort I put forward over the years. However, this achievement was only possible with the support of many individuals. I want to take this opportunity to thank them all. 

\par First of all, I would like to express my sincere gratitude to my Ph.D. supervisor, Prof. Sanjib Kumar Agarwalla. Without his constant support and guidance throughout these years, my Ph.D. journey would not have concluded on a positive note. His continuous encouragement and motivation during difficult times helped me overcome numerous difficulties. He has been an outstanding mentor, and it has been a great pleasure learning from him, both academically and non-academically.

\par I am thankful to my collaborators: Prof. Mauricio Bustamante, Dr. Mehedi Masud, Dr. Anil Kumar, Dr. Alessio Giarnetti, and Prof. Davide Meloni. Their consistent support, help, and valuable suggestions have immensely contributed to my growth as a researcher. I am also grateful to Prof. Amol Dighe and Dr. Ujjal Kumar Dey for the useful discussions we had during my Ph.D. years. I would also like to thank Prof. Manimala Mitra for her support. I extend my thanks to my group collaborators Ashish da, Masoom, Pragyan, and Sadashiv for the fruitful discussions we had during our projects together.

\par  I acknowledge my thesis committee members, Prof. Suresh Patra, Dr. Kirtiman Ghosh, Prof. Aruna Kumar Nayak, and Prof. Sanjay Swain, for their useful comments and suggestions. I thank the Department of Atomic Energy (DAE), Govt. of India, for
their ﬁnancial support. I also acknowledge the support from the IOP administration, library, and computer center. 

\par I thank my group members Ritam, Anuj, Krishnamoorthi, Sharmistha, Gopal, and Thiru for the discussions and fun we had over the years. I also thank my batchmates Siddharth, Abhishek, Arpan, Arnob, Pritam, Chitrak, Mausam, Harish, Sameer, Aisha, Sandhya, and my seniors Saiyad Da, Sudarshan Da for their support and the enjoyable moments we shared during the initial years of my Ph.D. Special thanks to my friend Ankit for all the irrelevant discussions we had and for making campus life more enjoyable. I also thank Rahul, Ayaz, Vandana, Surojit, Prawi, and Abinash for their support. I acknowledge my juniors Moonsun, Rameswar, Subhadip, Debashish, Dipak, Suman, Amartya, Aswin, and Debabrata for their support and help.

\par I am deeply grateful to my loving parents and sisters for their presence in my life. Their unwavering support and encouragement have been a major driving force behind my accomplishments. Last but not least, I express my heartfelt thanks to my best friend and wife, Susmita, for being a constant companion throughout the final stages of my Ph.D. journey.
\vspace{1.2cm}
\begin{flushright}
	\textbf{(Sudipta Das)}
\end{flushright}

\newpage
\hypersetup{linkcolor=blue}
\tableofcontents
\numberwithin{equation}{chapter}
\numberwithin{figure}{chapter}
\numberwithin{table}{chapter}
\newpage 
\addcontentsline{toc}{chapter}{List of Figures}
\listoffigures
\blankpage 
\addcontentsline{toc}{chapter}{List of Tables}
\listoftables
\blankpage 
\setcounter{page}{1}
\pagenumbering{arabic}
\def\blue{\textcolor{blue}}
\chapter{Introduction}
\label{C1} 
\graphicspath{{Figures/Chapter-1figs/PDF/}{Figures/Chapter-1figs/}}

From ancient times, humans' urge to understand the Nature has been a major driving force behind the progress of mankind, which shapes modern world today. One of the most fascinating goals of the scientific community today is to explain the underlying symmetries of the Nature using a unified theory. The Standard Model of particle physics represents a significant step toward this goal, unifying the strong, weak, and electromagnetic forces. This theory has been remarkably successful in explaining many aspects of the Nature, from the basic building blocks of ordinary matter to the processes that power gigantic stars in the universe.

However, there are reasons to believe that the Standard Model is incomplete and represents a low-energy extension of a more comprehensive theory. One such reason is the non-zero mass of one of its constituent particles, the {\it neutrino}. In the framework of the Standard Model, neutrinos are categorized in the lepton sector and are considered massless. Nevertheless, experimental observations have conclusively established that neutrinos have a non-zero mass, indicating the need for a theory beyond the Standard Model (BSM). The evidence for massive neutrinos came from the phenomenon of {\it neutrino oscillation}, where neutrinos change their flavor as they travel through a medium. This thesis primarily explores the effects of BSM theories on neutrino oscillation and their possible impact on neutrino observation experiments. In this chapter, we begin with a basic introduction to neutrinos.

\section{History of neutrinos}
The story of neutrino began nearly a century ago when Prof. Wolfgang Pauli proposed a mysterious particle in order to solve the beta particle energy conundrum. During the 1920s, several experiments found the energy spectrum for beta decay to be continuous. However, it was expected that the electron emerging from the decay would have a constant energy, like the alpha and gamma decay. A continuous energy spectrum violates the conservation of energy. The mystery remained unsolved for many years until 1930 when physicist Wolfgang Pauli postulated an invisible particle emitted along with the electron in beta decay. Although the idea of the particle neutrino came from Pauli, it was included in the theory of beta decay by Enrico Fermi, who showed how neutrino emitted from the beta decay would lead to the energy conservation of the process, also explaining the continuous nature of the beta dacay~\cite{1934ZPhy...88..161F}. He also coined the term ``{\it neutrino}", which means ``the little neutral one" in Italian.

Pauli predicted that the particle would remain invisible, and it remained so for more than two decades until it was first observed by physicists Clyde Cowan and Frederick Reines in a reactor neutrino experiment in 1956. They used a nuclear reactor, which generated neutrinos with high density from fission reaction to detect the so-called ghost particle with a 10 ton detector. Frederick Reines received the Nobel Prize for the discovery in 1995. After its discovery, the quest for understanding the properties of neutrino began. The fundamental question was, how many types of neutrinos are there? Electron-type neutrino was already discovered in the reactor neutrino experiment. In 1962, a second type of neutrino, muon neutrino, was observed by a team of scientists at Columbia University. They used a powerful accelerator at Brookhaven National Laboratory as their source of neutrinos~\cite{Danby:1962nd}. The existence of a third type of neutrino was first predicted in 1975, when its charged counterpart, tau lepton, was discovered at the SLAC National Accelerator Laboratory. Later, in 2000, tau neutrino was discovered from the DONUT~(Direct Observation of a Nu Tau) experiment~\cite{DONUT:2000fbd}. 
Along with the discovery of different flavors of neutrinos, over these years, neutrinos from different natural sources like the sun, Earth's atmosphere, astrophysical bodies, the interior of the Earth, and artificial sources like accelerators, nuclear reactors were observed, which we are going to discuss in detail in the next section. 

In 1998, the atmospheric neutrino experiment Super-Kamiokande observed a significantly smaller number of atmospheric neutrinos going upwards after passing through the Earth than the predicted one~\cite{Super-Kamiokande:1998kpq}. Around the same time, several experiments observing solar neutrinos found similar deficits in their data~\cite{Super-Kamiokande:1998qwk,Cleveland:1998nv,GALLEX:1998kcz}. These two discrepancies later led to the discovery of neutrino oscillation, establishing that the neutrino changes its flavor along the path. A detailed formalism of neutrino oscillation is given in Chapter~\ref{C2}. Next, we briefly discuss the different sources of neutrinos.
\section{Neutrino sources}
\label{sec:neutrino_sources}

Neutrino is the second most abundant fundamental particle after photon. It is present in the universe since the time of the Big Bang and is continuously produced inside numerous celestial bodies in the vast cosmos. There are also artificial neutrino sources. The energy of neutrinos covers an extensive range, from sub-eV relic neutrinos produced during the Big Bang to as large as $> 10^{18}$ eV ultra-high energy astrophysical neutrinos. In the following, we list the natural and artificial sources of neutrinos.

\paragraph{Relic neutrinos:} In the early universe, about less than $10^{-6}$ seconds after the Big Bang, all the fundamental particles created a thermal bath. As the universe cooled down with its expansion, the interaction rate of the particles inside the thermal bath started decreasing and became insufficient for some particles to stay in equilibrium. Eventually, they decoupled from the thermal bath. When the age of the universe was around one second, neutrinos decoupled or froze out from the other fundamental particles, electrons, positrons, and photons, in thermal equilibrium. Those neutrinos are freely streaming through the universe at the present day, known as {\it relic} neutrinos. The present model suggests that the relic neutrino has a density of 330 per cubic centimeter. However, they have a very small energy~($\sim 10^{-5}$ eV). Due to the tiny energy, relic neutrino has not yet been observed at any detector. However, there is indirect evidence of their presence.

\paragraph{Solar neutrinos:} The Sun is a natural source of neutrinos that emits billions of neutrinos that pass through us every second. Solar neutrinos are produced inside the Sun through the nuclear fusion process, which is the source of its energy. Two fusion chains are involved in the neutrino production, namely, the $pp$ and the CNO~(Carbon-Nitrogen-Oxygen) chains. $pp$ chain generates 98.4\% of the total neutrinos produced inside the Sun. In the $pp$ chain, the electron-type neutrino is produced mainly {\it via} five processes. First, $pp$ neutrinos, when two protons undergo fusion to produce a deuterium nucleus, a positron, and an electron neutrino. Similarly, there are $pep$ neutrinos when two protons and an electron produce a deuterium and a neutrino. $hep$ neutrinos are generated from the fusion reaction of $\vphantom{He}^3 {\rm He}$ nuclei and a proton. Also, from the beta decay of  $\vphantom{Be}^7 {\rm Be}$, neutrinos are produced. The last one is the $\vphantom{B}^8 {\rm B}$ neutrinos from decay of $\vphantom{B}^8 {\rm B}$. 99.77\% of solar neutrinos produced from the $pp$ chain are the $pp$ neutrinos. From the CNO cycle, which contributes only 1.6\% of the total solar neutrino production, neutrinos are produced from the decay of $\vphantom{N}^{13} {\rm N}$
$\vphantom{O}^{15} {\rm O}$.
Solar neutrinos are excellent messengers from the Sun, carrying crucial information about their core. The energy of these neutrinos generally varies from sub-MeV to a few MeV, depending on the fusion process through which they are generated.

\paragraph{Atmospheric neutrinos:} Earth's atmosphere is one of the prominent natural sources of neutrinos. Cosmic rays, majorly consisting of protons with small fractions of alpha particles, heavy nuclei, and electrons, interact with the nuclei inside the atmosphere while passing through it. The resultant interaction produces charged pions and kaons, which decay to produce neutrinos {\it via} following the process
\begin{align}
\label{eq:atm_neutrino_prod}
&\pi^{\pm}/K^{\pm}\rightarrow \,{\mu^\pm} + \nu_{\mu}(\bar{\nu}_\mu)\,,\\\nonumber
&\mu^\pm\rightarrow e^+ + \bar{\nu}_\mu(\nu_\mu)+\nu_e(\bar{\nu}_e)\,. 
\end{align}
In the process, one electron neutrino and two muon neutrinos are produced, with the resulting flavor ratio, $f_e:f_\mu:f_\tau=1:2:0$. Note that there are other semileptonic decay modes of Kaon which contribute to small but relvent fracion of atmospheric neutrinos, shown as follows,
\begin{align}
\label{eq:atm_neutrino_prod_semileptonic}
&K^{+}/K^{0}\rightarrow \,\pi^0/\pi^- + \mu^+ + \nu_{\mu}\,,\\\nonumber
&K^{+}/K^{0}\rightarrow \,\pi^0/\pi^- + e^+ + \nu_{e}\,. 
\end{align}
The energy of the atmospheric neutrino varies from a few hundred MeV to a TeV range, depending on the energy of the cosmic particles. Since atmospheric neutrinos are produced throughout the atmosphere, their fluxes are expected to be isotropic around the Earth. Atmospheric neutrinos may reach the surface of the Earth directly from the atmosphere, known as downward-going neutrinos, or after passing through the Earth, they are called upward-going neutrinos.

\paragraph{Astrophysical neutrinos:} Astrophysical neutrinos are produced inside various cosmic accelerators with extremely high energy. Protons and nuclei are accelerated to a very high-energy~($\sim 10^{12}$ GeV) inside the accelerator and undergo $pp$ or $p\gamma$ collision with the ambient matter and radiation to produce pions. The pions decay to produce HE and UHE neutrinos.
The energy of the high-energy~(HE) astrophysical neutrinos generally varies between several hundred TeV to PeV range. However, the energy can go up to the EeV range, known as ultra-high energy~(UHE) neutrinos.
Possible candidates for the source of HE and UHE astrophysical neutrinos are blazers, starburst galaxies, pulsars, tidal disruption events, superluminous supernovae, and numerous others~\cite{Anchordoqui:2013dnh}.  Astrophysical neutrinos carry crucial information about various cosmic activities, and being weakly interacting particles, that information remains intact en route to the Earth.

\paragraph{Geoneutrinos:} Neutrinos are also produced inside the Earth. The interior of the Earth contains radioactive materials, mostly uranium, thorium, and potassium~($\vphantom{K}^{40} {\rm K}$). Radioactive decay of these materials leads to the production of electron antineutrino. The geoneutrino flux through the Earth depends on the distribution of radioactive material inside the Earth. The energy of geoneutrinos is in the order of the MeV.

\paragraph{Accelerator neutrinos:} Apart from the natural source of neutrinos, neutrinos are artificially produced in the various particle accelerator facilities. The production process of the accelerator neutrinos is similar to that of atmospheric neutrinos. The proton beam accelerated to a very high energy is bombarded to a stationary target. The resultant $pp$ collisions produce pions and kaons, which are channelized to the decay pipe using magnetic horns. The pions and kaons decay to produce $\nu_\mu(\bar{\nu}_\mu)$ and $\mu^+(\mu^-)$. Charge lepton is absorbed in the beam dump while neutrinos escape. Accelerator neutrinos generally have energy of a few hundred MeV up to tens of GeV. Neutrino energy and the nature of the flux depend on the number of protons bombarded on the target, known as protons on target~(P.O.T.), and the angle between the detector where the neutrino is observed, and the beam axis, known as the off-axis angle.

\paragraph{Reactor neutrinos:}  Nuclear reactors, which are a prominent source of energy, produce neutrinos from the nuclear fission of radioactive materials like uranium and thorium inside the reactor core. Nuclear fission generates a flux of electron antineutrino from the beta decay, which escapes from the reactor. Reactor neutrinos have energy in the MeV range.

\section{Neutrinos in the Standard Model}

The Standard Model~(SM) of particle physics is the basic theory in particle physics, classifying all the fundamental particles according to their mass, spin, charge, and interactions. Although there are reasons to believe that the theory is a low energy limit of a more fundamental theory, it explains the properties of the particles and their interactions in the Nature quite successfully. Additionally, the theory has been tested by a number of outstanding experiments with impressive precision.

In the SM, three neutrinos, $\nu_e,\nu_\mu$, and $\nu_\tau$, are categorized in the lepton sector, paired with their charged counterpart, electron, muon, and tau. They are chargeless and have mass exactly equal to zero. The massless nature of neutrinos suggests that the neutrino should have a definite parity. This was found experimentally from Wu's experiment~\cite{PhysRev.105.1413}, which observed asymmetry in the distribution of electrons emitted from the beta decay of $\vphantom{Co}^{60} {\rm Co}$, which confirms that the parity is violated in the beta decay. A similar observation was made from the experiment of Telegdi~\cite{PhysRev.105.1681.2}, where the decay of polarized muons resulted in an asymmetric distribution of electrons. Both of these experiments involved neutrinos, revealing that the neutrinos maximally violate parity and there are only left-handed neutrinos in Nature.

The evidence of the left-handed nature of neutrinos was crucial for formulating $V-A$ theory of weak interaction within the SM framework. In this theory, SU(2) doublets composed of the left-handed neutrinos and their charged counterpart, interact {\it via} weak interaction, which has the following form in SM Lagrangian,
\begin{align}
\mathcal{L} = -i\bar{L}_\alpha (ig \slashed{W}^a\tau^a-ig'Y_{L}\slashed{B})L_\alpha\,,
\end{align}
where, $L_\alpha$ is the SU(2) doublets, 
$W_{\mu}^a$ and $B_\mu$ are the $SU(2)_L$ and $U(1)_Y$ gauge bosons and $g$ and $g'$ are the corresponding coupling, $\tau^a=\frac{1}{a}\sigma^a$~($\sigma\rightarrow$ Pauli matrices), and $Y_L$ is hyperchagre of $U(1)_{Y}$ field. In the physical basis, the above interaction term generates charge-current and neutral-current interaction of neutrinos, as given below,
\begin{align}
\mathcal{L}_{\rm CC} &= \frac{g}{2\sqrt{2}}\left[\bar{l}\gamma^\rho(1-\gamma_5)\nu\right] W_\rho^-+h.c.\\
\mathcal{L}_{\rm NC} &= \frac{g}{2\cos\theta_{W}}\left[\bar{\nu}\gamma_\rho(1-\gamma_5)\nu\right] Z^\rho\,.
\end{align}
Here, $l$ are the charged leptons, $\theta_W$ denotes the weak mixing angle: $\tan\theta_{W}=\frac{g'}{g}$, $W^{-}$ and $Z$ are the physical gauge bosons, mediator for the charge-current and neutral-current interactions, respectively. Note that in the above two equations, the left-handedness of the fermions is ensured by the chiral projection operator, $P_L = \frac{1-\gamma_5}{2}$.

The masslessness of neutrinos in the SM framework results from the fact that neutrinos are always left-handed in nature. In the SM,  fermions acquire mass through the Yukawa interaction after the electroweak symmetry breaking. 
For the first-generation fermions, mass terms in the physical basis can be written as,
\begin{align}
\mathcal{L}_{\rm mass}= -m_e\bar{e}_L e_R-m_d\bar{d}_Ld_R- m_u\bar{u}_Lu_R+ h.c.\,,
\end{align}
where $m_e = \frac{y}{\sqrt{2}}v$, $v$ is the vacuum expectation value ($vev$) of Higgs boson after the symmetry breaking and $y$ denotes Yukawa coupling constant. Mass of the first generation quarks $m_u$ and $m_d$ are the first diagonal element of $\frac{v}{\sqrt{2}}M_u$ and $\frac{v}{\sqrt{2}}M_d$, respectively, where $M_u$ and $M_d$ are the quark mass matrices in the diagonal basis. These mass terms are called Dirac mass terms, which involve right-handed fermions. However, for neutrinos, there is no right-handed component in the SM framework and as a result, electroweak symmetry breaking fails to provide mass terms for neutrinos. Apart from the Dirac mass, having neutrino mass term in SM is still possible with left-handed neutrinos, known as Majorana mass term. The Majorana mass term uses the opposite helicity state of the antiparticle. However, this type of mass term violates the lepton number symmetry of the SM by two units. As a result,
	SM does not allow either Dirac or Majorana mass terms for the neutrinos.

Although the Standard Model of particle physics appears to be the ultimate theory that explains most of the phenomena occurring in the universe, the absence of neutrino mass term in SM is one of the prime reasons to believe that the theory is not complete. Evidence of non-zero neutrino mass from several experiments made people to think of the physics Beyond the Standard Model~(BSM).

\section{Massive neutrinos --- SM to BSM}

Among the many limitations of the Standard Model, the non-zero neutrino mass is one of the prominent missing pieces. It opens a window for BSM physics, with the major motivation to explain the non-zero neutrino mass. Before we discuss possible BSM scenarios that accommodate massive neutrinos and their consequences, let us discuss evidence of non-zero neutrino mass. 

The hint for massive neutrinos came from the experimental observation of the atmospheric and solar neutrino flux. The production process of the atmospheric neutrino is described in section~\ref{sec:neutrino_sources}. Eq.~\ref{eq:atm_neutrino_prod} suggest that the
expected ratio of muon type and electron type neutrinos, R$\left(\equiv\frac{N(\nu_\mu+\bar{\nu}_\mu)}{N(\nu_e+\bar{\nu}_e)}\right)$ should be equal to 2. However, in the 1990s, several experiments like Super-Kamiokande, Soudan2, and IMB observed the ratio to be significantly less than two.
This led to discrepancies between the expected and observed flux of the upward-going atmospheric neutrinos, known as {\it atmospheric neutrino anomaly}. 
The discrepancy was later explained through the flavor oscillation of muon-type neutrinos, which require non-zero neutrino mass. Similarly, in the solar neutrino sector, several experiments, like Homestake, Gallex, and SAGE, measuring neutrino flux from the Sun, also found a simpler deficit in the observed neutrino flux, which also signals the neutrino oscillation. At present, several experiments have confirmed neutrino oscillation, establishing non-zero neutrino mass.

However, as discussed in the previous section, the SM does not accommodate neutrino mass. So, there is a need to go beyond the SM framework, which will contain neutrino mass term as well as preserving basic structure of the SM. There are two possible types of mass terms that can generate neutrino mass. One is the Dirac mass term, which has the following structure in the Lagrangian
\begin{align}
\mathcal{L}_{\rm Dirac} = m_\nu\bar{\nu}_L\nu_R+ h.c.
\end{align}
For the Dirac mass term to exist, one needs to include right-handed neutrinos in the model, which are not there in the SM. The Majorana mass term has the form,
\begin{align}
\mathcal{L}_{\rm Majorana} = \frac{m_\nu}{2}\bar{\nu}_L\nu^c_L+h.c.\,,
\end{align}
where $\nu^c_L$ is the charge conjugate of the left-handed neutrino state. A Majorana mass term must be generated below the electroweak symmetry breaking scale since above this scale, left-handed neutrino transform under the $U(1)^{\rm Y}$ symmetry. Also, a Majorana mass term violates the lepton number. The simplest way to assign neutrino mass is to include a term that couples neutrino to the Higgs field, similar to their charge counterparts:
\begin{align}
{\rm Y}^\nu_{\alpha\beta}\bar{L}^\alpha.\tilde{H}\nu^\beta_R+h.c.\,,
\end{align}
where $L^{\alpha}$ is the SU(2) doublet, $\nu_{R}$ is the right-handed neutrinos of flavor $\beta$, and $\tilde{H}=i\sigma_2 H^\ast$. After the Higgs field acquires {\it vev}, this term would generate the Dirac mass of neutrinos. However, this term would require an extremely small coupling between the neutrinos and the Higgs field in order to explain the tininess of the neutrino mass.  Also, no symmetry prevents $\nu_R$ to have Majorana mass, which is missing here. For these two reasons, this scheme seems unnatural.

Various BSM models have been proposed in the literature that accommodate the neutrino mass. The seesaw scheme is the most popular one.
These models can generate mass for the left-handed fermions below the electroweak symmetry-breaking scale and account for the tininess of the neutrino mass without requiring abnormally small Higgs coupling.
There are three classes of seesaw models; type-i seesaw models are the simplest ones. In this model, right-handed neutrino are introduced, and the large Majorana mass term of the right-handed neutrino leads to a generation of small left-handed neutrino mass. For a simplified scenario, considering one left-handed and one right-handed neutrino, the relevant terms for the neutrino mass generation are
\begin{align}
{\rm Y}^\nu \bar{L}. \tilde{H}\nu_R+\frac{M}{2}\bar{\nu}_R \nu^c_R+h.c.\,,
\end{align} 
where the second term is the Majorana mass term of the right-handed neutrino. After the electroweak symmetry breaking, neutrino mass terms have the form,
\begin{align}
\label{eq:seesaw_1_mass}
m\bar{\nu}_L\nu_R+\frac{M}{2}\bar{\nu}_R \nu^c_R+h.c.
\end{align}
In the above expression, $m_\nu = \frac{{\rm Y}^\nu v}{\sqrt{2}}$, where $v$ is the {\it vev} of the SM Higgs, and $M>>m$. This term generates two Majorana mass terms after the diagonalization of the mass matrix\footnote{eq.~\ref{eq:seesaw_1_mass} can be written in matrix form,
	\begin{align*}
	\begin{pmatrix}
	\bar{\nu}_L & \bar{\nu}^c_R\\
	\end{pmatrix}\cdot
	\begin{pmatrix}
	0 & m\\
	m & M\\
	\end{pmatrix}\cdot\begin{pmatrix}
	\nu^c_L\\
	\nu_R\\
	\end{pmatrix}+h.c.\,\,.
	\end{align*}}
, $M$ and $\frac{m^2}{M}$. $M$ is the mass of the right-handed neutrino, and the $\ \frac {m^2}{M}$ accounts for left-handed neutrino mass, which is suppressed by $M$. As a result, even if $m$ is not small, the suppression factor would lead to the smallness of the neutrino mass.

In the type-II seesaw model, SM is extended with one additional Higgs triplet, $\Delta$, which has the following matrix representation
\begin{align}
\Delta = \begin{pmatrix}
\delta^+/\sqrt{2} & \delta^{++}\\
\delta^0 & -\delta^+/\sqrt{2}\\
\end{pmatrix}\,.
\end{align}
The relevant term in Lagrangian for neutrino mass is
\begin{align}
\mathcal{L}_\nu = -{\rm Y}^\nu\nu^\dagger_L(\delta^0)^\ast\nu^c_L+h.c.\,,
\end{align}
where ${\rm Y}_\nu$ is the coupling constant. After the symmetry breaking, $\delta^0$ acquire the {\it vev} $v_\Delta = \frac{\mu v^2}{\sqrt{2}M^2_{\Delta}}$, where coupling constant and $M_{\Delta}$ is mass of the field $\Delta$. So, the neutrino mass~(Majorana type) in this case is given by
\begin{align}
m_\nu = \frac{{\rm Y}^\nu\mu v^2}{\sqrt{2}M^2_{\Delta}}\,.
\end{align} 

The type-III seesaw model introduces a fermion SU(2) triplet N:
\begin{align}
N = \begin{pmatrix}
N_0/\sqrt{2} & N_+\\N_- & -N_0/\sqrt{2}
\end{pmatrix}\,,
\end{align}
where $N_+$, $N_-$ are Dirac fermion, whereas $N_0$ is a Majorna fermion. The triplet N has zero hypercharge. The part of Lagrangian corresponding to $N$ is
\begin{align}
\mathcal{L}_N = {\rm Tr} i \bar{N}\slashed{D} N + Y^N(\bar{L}\cdot i\sigma_2\cdot N\cdot H)+{\rm Tr}\frac{M_N}{2}\bar{N}N+h.c.\,
\end{align}
Neutrino mass term after the electroweak symmetry breaking is 
\begin{align}
\mathcal{L}_\nu = Y^N v \bar{\nu}_L N_0 +\frac{M_N}{2}N_0\bar{N}_0\,,
\end{align}
where $v$ is the {\it vev} of the Higgs field, and light neutrino mass is given by $m_\nu = Y^N v$.

Apart from seesaw models, there are a number of neutrino mass models in the literature that account for the non-zero neutrino mass. In this thesis, we do not focus on the mass-generation mechanism of the light neutrinos. We explore the possible impact of various Beyond the Standard model scenarios in the neutrino oscillation phenomena and the potential of the neutrino oscillation experiments to explore these BSM scenarios. 

\section{Layout of the thesis}
The thesis is organized as follows. In chapter~\ref{C2}, we discuss the theory of neutrino oscillation along with the current status of the neutrino oscillation parameters, and their measurements from the various experiments. In this chapter, we describe in detail the basic neutrino oscillation formalism in two-flavor and three-flavor cases for the neutrino propagating in a vacuum, as well as in ordinary matter. Also, we give brief details about various past, ongoing, and future neutrino oscillation experiments.
In chapter~\ref{C3}, we discuss in detail how various BSM scenarios can have a subleading impact on the neutrino flavor transition. The BSM scenarios discussed in this chapter are neutrino non-standard interactions, the presence of sterile neutrinos, non-unitarity of the neutrino mixing matrix, CPT and Lorentz violation, and neutrino decay. 
In chapter~\ref{C4}, we derive the expression of the effective/modified mixing angles for the neutrinos propagating in matter in the presence of neutral-current non-standard interactions and their implications in the neutrino oscillation probability. We study in detail the evolution of those modified mass and mixing parameters that help to explain various important features of the oscillation probability in the appearance and disappearance channel.
In chapter~\ref{C5}, we study the impact of possible non-unitary neutrino mixing~(NUNM) in neutrino flavor oscillation in the context of the next-generation neutrino oscillation experiments DUNE and T2HKK. Using these two setups, we estimate the limit on various NUNM parameters.
In chapter~\ref{C6}, we explore the effect of long-range interaction induced from the gauged $L_e-L_\mu$, $L_e-L_\tau$ and $L_\mu-L_\tau$ symmetries. We use flavor composition estimates from IceCube and future projections from IceCube-Gen2 and other upcoming neutrino telescopes to place estimated/projected constraints on the long-range interaction parameters for the three symmetries. Finally, we summarize the thesis in chapter~\ref{C7} with concluding remarks.


\newpage
\newcommand{\mue}{\nu_\mu\rightarrow\nu_e}
\newcommand{\mumu}{\nu_\mu\rightarrow\nu_\mu}
\newcommand{\pme}{P_{\nu_{\mu}\rightarrow\nu_{e}}}
\newcommand{\pmeb}{\bar{P}_{\mu e}}
\newcommand{\pmm}{P_{\nu_{\mu}\rightarrow\nu_{\mu}}}
\newcommand{\pmmb}{\bar{P}_{\mu \mu}}
\newcommand{\ldm}{\ensuremath{\Delta m_{31}^2}}
\newcommand{\sdm}{\ensuremath{\Delta m_{21}^2}}
\newcommand{\eem}{\ensuremath{\varepsilon_{e\mu}}}
\newcommand{\eet}{\ensuremath{\varepsilon_{e\tau}}}
\newcommand{\emt}{\ensuremath{\varepsilon_{\mu\tau}}}
\newcommand{\eee}{\ensuremath{\varepsilon_{ee}}}
\newcommand{\emm}{\ensuremath{\varepsilon_{\mu\mu}}}
\newcommand{\ett}{\ensuremath{\varepsilon_{\tau\tau}}}
\newcommand{\eff}{\ensuremath{\gamma-\beta}}
\newcommand{\txm}{\ensuremath{\theta_{12}^{m}}}
\newcommand{\tym}{\ensuremath{\theta_{13}^{m}}}
\newcommand{\tzm}{\ensuremath{\theta_{23}^{m}}}
\newcommand{\ldmm}{\ensuremath{\Delta m^2_{31,m}}}
\newcommand{\sdmm}{\ensuremath{\Delta m^2_{21,m}}}
\newcommand{\ie}{\textit{i.e.}}
\newcommand{\A}{\hat{A}}

\newcommand{\tx}{\ensuremath{\theta_{12}}}
\newcommand{\ty}{\ensuremath{\theta_{13}}}
\newcommand{\tz}{\ensuremath{\theta_{23}}}
\newcommand{\dcp}{\delta_{\mathrm{CP}}}

\newcommand{\capdef}{}
\newcommand{\mycaption}[2][\capdef]{\renewcommand{\capdef}{#2}
	\caption[#1]{{\footnotesize #2}}}
\makeatletter
\renewcommand{\fnum@table}{\textbf{\tablename~\thetable}}
\renewcommand{\fnum@figure}{\textbf{\figurename~\thefigure}}
\makeatother


\chapter{Neutrino oscillation}
\label{C2} 
The theory of neutrino oscillation was put forward by Bruno Pontecorvo long before it was experimentally established from solar and atmospheric neutrino data. The idea of this phenomenon primarily depends on two factors: first, mass and flavor eigenstates of neutrinos are not identical, and there is non-trivial mixing between these two eigenstates. Second, the mass-eigenstates of neutrinos are non-degenerate. As a result, different mass-eigenstates interfere with each other while propagating in a medium and can have a non-zero probability of changing their flavor when detected at the source. In this chapter, we explore the theory of neutrino oscillation and its current status. Also, we elaborate on several neutrino experiments and their role in the measurement of neutrino oscillation parameters. 

This chapter is organized as follows: section \ref{sec:oscillation_formalism} demonstrates the basic formalism of neutrino oscillation in vacuum. We discuss neutrino oscillation in the two-neutrino and the three-neutrino framework. In section~\ref{sec:osc_mat}, we describe the neutrino oscillation in the presence of the standard neutrino-matter interaction both in the two and three-flavor scenarios. Details of the various neutrino oscillation experiments and their role in measuring oscillation parameters are discussed in section~\ref{sec:osc_experiments}. The current standing of the six oscillation parameters is presented in section~\ref{sec:osc_param_status} along with the discussion about the measurement of the six oscillation parameters. In section~\ref{sec:summary_osc}, we summarize the chapter.

\section{Neutrino oscillation formalism}
\label{sec:oscillation_formalism}

The theory of neutrino oscillation is mainly based on its two properties. First, the basis in which neutrino produced or detected {\it via} weak interaction differs from the basis in which neutrino propagates. The former is the weak interaction basis or the flavor basis, while the latter is called propagation basis or mass basis. Neutrino in the flavor eigenstate is a linear combination of all the mass eigenstates as shown below,
\begin{equation*}
\ket{\nu_\alpha} = \sum_{i=1,..n} U^\ast_{\alpha i}\ket{\nu_i}\,,
\end{equation*}
where $U_{\alpha i}$ is the element of the matrix that connects the flavor basis of the neutrinos to its mass basis. The flavor eigenstate of the neutrino is an admixture of the mass eigenstates, and vice versa. The matrix $U$ is called the mixing matrix. In Nature, there are three flavors of neutrinos detected so far, namely, electron neutrino~$(\nu_e)$, muon neutrino~$(\nu_\mu)$, and tau neutrino~$(\nu_\tau)$, and the three mass eigenstates are $\nu_1$, $\nu_2$, and $\nu_3$. The second essential property for the neutrino oscillation is the non-degenerate mass of neutrinos in the physical/mass basis. While propagating through the medium, different mass eigenstates interfere with each other, and consequently, their fraction in the flavor basis evolves with time. For a neutrino produced with flavor $\alpha$, after travelling a distance $x$,
evolution of the neutrino state is denoted as
\begin{align}
\label{eq:nu_evol}
\ket{\nu_\alpha (t)} = U^\ast_{\alpha j}e^{-iE_j t}\ket{\nu_j}\,,
\end{align}
such that $\ket{\nu_\alpha(t=0)}=\ket{\nu_\alpha}$. $E_j$ is the energy of the $j$-th eigenstate. For highly relativistic neutrino with momentum\footnote{It is the assumption that the each flavor of neutrino has a definite momentum $p$, {\it i.e.} all the mass eigenstate $\nu_i$~$(i=1,2,..n)$ has equal momentum. This assumption is motivated by the fact that all the mass eigenstates are propagating in the same direction from the same source.} $p\simeq E>>m_j$, energy $E_j$ can be approximated as
\begin{align}
\label{eq:E_approx}
E_j = \sqrt{p^2+m^2_j}\simeq p+\frac{m^2_j}{2 p}\simeq p+\frac{m^2_j}{2E}\,.
\end{align}
The amplitude of flavor transition of the neutrino of flavor $\alpha$ to $\beta$ is given by,
\begin{align}
\mathcal{A}(t)=\braket{\nu_\beta | \nu_\alpha (t)} = \sum_j U^\ast_{\alpha j} U_{\beta j} e^{-i E_j t}\,,
\end{align}
where we have used $\bra{\nu_\beta} 
= \sum_k U_{\beta k}\bra{\nu_k}$, 
and orthogonal property of the mass basis, {\it i.e.}, $\braket{\nu_i | \nu_j} =\delta_{ij}$. So, the oscillation probability from $\nu_\alpha$ to $\nu_\beta$ is the square of the transition amplitude given as follows,
\begin{align}
\label{eq:osc_prob1}
P(\nu_\alpha\to\nu_\beta) = |\mathcal{A}(t)|^2 = \sum_{j,k} U^\ast_{\alpha j} U_{\beta j} U_{\alpha k} U^\ast_{\beta k} e^{-i(E_i-E_j)t}\,.
\end{align}
In the ultrarelativistic limit, neutrinos propagate almost with the speed of light and one can use $t=L$, where $L$ is the propagation length. After simplification, we get a generalized expression of the oscillation probability in $n$ flavor scenario,
\begin{align}
\label{eq:osc_prob2}
P(\nu_\alpha\to\nu_\beta, L, E) = \delta_{\alpha\beta} - &4\sum_{j>k} \mathrm{Re} \left[U^\ast_{\alpha j} U_{\beta j} U_{\alpha k} U^\ast_{\beta k}\right]\sin^2\left(\frac{\Delta m^2_{jk}L}{4E}\right) \\\nonumber
&+2\sum_{j>k}\mathrm{Im}\left[U^\ast_{\alpha j} U_{\beta j} U_{\alpha k} U^\ast_{\beta k}\right]\sin\left(\frac{\Delta m^2_{jk}L}{2E}\right)\,.
\end{align}
In the above equation, we have used the unitarity condition of the mixing matrix $U$ $\sum_k U^\ast_{\alpha k}U_{\beta k} = \delta_{\alpha\beta}$. Also, we have used $E_j -E_k = \frac{\Delta m^2_{jk}}{2E}$, where $\Delta m^2_{jk} = m^2_j - m^2_k$~(see eq.~\ref{eq:E_approx}).
When the $\alpha\neq\beta$, {\it i.e.} neutrino of flavor $\alpha$ oscillates to flavor $\beta$, we call it appearance or transition probability, while in the case $\alpha=\beta$, probability of the neutrino to retain its flavor, called disappearance or survival probability. In case of survival probability, eq.~\ref{eq:osc_prob2}, simplifies to
\begin{align}
\label{eq:surv_prob1}
P(\nu_\alpha\to\nu_\alpha, L, E) = 1 - &4\sum_{j,k} |U_{\alpha j}|^2 |U_{\alpha k}|^2 \sin^2\left(\frac{\Delta m^2_{jk}L}{4E}\right)\,. 
\end{align}

Note that eqns.~\ref{eq:osc_prob1} and~\ref{eq:osc_prob2} are general expressions of the oscillation probabilities valid for any number of neutrinos. The following sections discuss the neutrino oscillation probabilities in the two-flavor and three-flavor scenarios.

\subsection{Neutrino oscillation in two neutrino flavor}
\label{sec:osc_two_flavors}

In the two flavor scenario with ($\nu_\alpha,\nu_\beta$) as flavor eigenstates and ($\nu_1,\nu_2$) as mass eigenstates, the mixing matrix can be parameterized in terms of a mixing angle $\theta$,
\begin{align}
\label{eq:two_flav_mix}
\begin{pmatrix}
\nu_\alpha \\ 
\nu_\beta
\end{pmatrix}
=\underbrace{\begin{pmatrix}
	\cos\theta & \sin\theta \\
	-\sin\theta & \cos\theta
	\end{pmatrix}}_{U}
\cdot
\begin{pmatrix}
\nu_1\\
\nu_2
\end{pmatrix}\,.
\end{align}
Now, using eq.~\ref{eq:osc_prob1}, one can derive the expression for the transition probability as,
\begin{align}
\label{eq:osc_prob_two_flav}
P(\nu_\alpha\to\nu_\beta) = \sin^2 2\theta\sin^2\left(\frac{\Delta m^2_{21}L}{4E}\right)\,,
\end{align}
where $\Delta m^2_{21}=m^2_2-m^2_1$. Expressing baseline length $L$ in km and neutrino energy $E$ in GeV in the above equation, transition probability takes the form,
\begin{align}
\label{eq:trans_prob}
P(\nu_\alpha\to\nu_\beta) = \sin^2 2\theta\sin^2\left(1.27\times\frac{\Delta m^2_{21}[\text{eV}^2]\,L[\mathrm{km}]}{E[\mathrm{GeV}]}\right)\,,
\end{align}

Similarly, the expression of the survival probability is given by
\begin{align}
\label{eq:surv_prob}
P(\nu_\alpha\to\nu_\alpha) = 1- \sin^2 2\theta\sin^2\left(1.27\times\frac{\Delta m^2_{21}[\text{eV}^2]\,L[\mathrm{km}]}{E[\mathrm{GeV}]}\right)\,.
\end{align}

\begin{figure}[h!]
	\centering
	\includegraphics[scale=0.95]{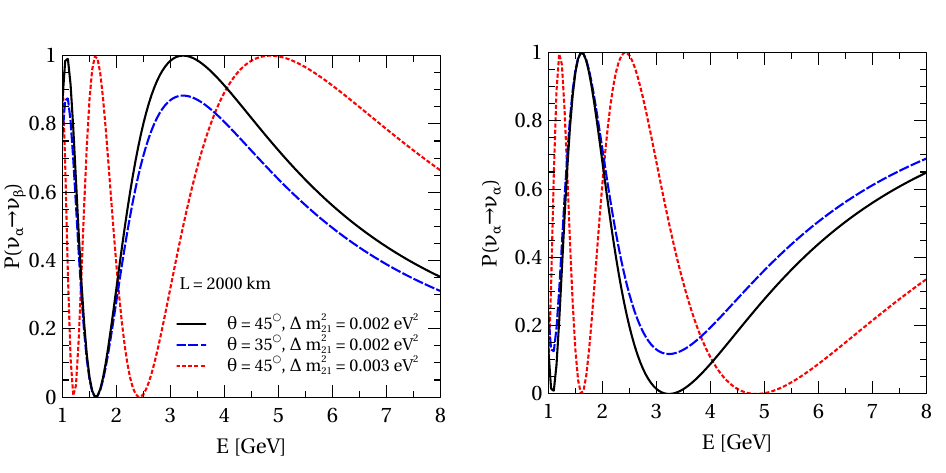}
	\mycaption{$\nu_\alpha\to\nu_\beta$ appearance~(left panel) and $\nu_\alpha\to\nu_\alpha$ disappearance~(right panel) probabilities as a function of neutrino energy $E$. We consider a baseline of 2000 km. The probabilities are computed for three sets of mixing angle and mass-squared difference values; $\theta = 45^{\circ}, \Delta m^{2}_{21} = 2.0\times 10^{-3}$ eV$^2$~(solid black line), $\theta = 35^{\circ}, \Delta m^{2}_{21} = 2.0\times 10^{-3}$ eV$^2$~(blue dashed line), $\theta = 45^{\circ}, \Delta m^{2}_{21} = 3.0\times 10^{-3}$ eV$^2$~(red dotted line).}
	\label{fig:prob_two_flav}
\end{figure}

Using eqns.~\ref{eq:trans_prob} and~\ref{eq:surv_prob}, we plot the neutrino oscillation probabilities as a function of energy for a baseline of $1000$ km. In this figure, we show the impact of the mixing angle and mass-squared difference in the oscillation probability. Solid black lines in both panels correspond to the value of $\theta = 45^\circ$ and mass-squared difference $\Delta m^2_{21} = 2.0\times 10^{-3}$ eV$^2$. We observe appearance probability is maximum, $P(\nu_\alpha\to\nu_\beta)\simeq 1$ at $E\simeq 1.6$ GeV, known as {\it oscillation maximum}. In the right panel, the disappearance probability is minimum around the same value of energy. The amplitude of the appearance/disappearance probability at the oscillation maximum is determined by the mixing angle $\theta$. In the blue dashed curves, where the mixing angle is $10^\circ$ less than that of the black solid curve, we observe a reduction~(enhance) in the appearance ~(disappearance) probability at the oscillation maximum. The position~(neutrino energy and baseline) of the oscillation maximum is determined by the value of the mass-squared difference. At the oscillation maximum, $\frac{\Delta m^2_{21}L}{4E}=(2n+1)\frac{\pi}{2}$~$(n=1,2,..)$. 
In the figure, we observe a shift in the energy at the oscillation maximum when the value of the mass-squared difference is increased, as shown by the red dotted line and the black solid line. To summarise, in the two-flavor scenario, neutrino oscillation is driven by two parameters: the mixing angle, which determines the oscillation amplitude, and the mass-squared difference, which determines the position of the oscillation maxima or oscillation in general. 
\subsection{Neutrino oscillation in three neutrino flavor}
\label{sec:osc_three_flavors}

In the case of three neutrino states, the flavor states are the three active neutrinos detected so far, $(\nu_e, \nu_\mu, \nu_\tau)$, and corresponding mass states are $(\nu_2,\nu_2, \nu_3)$. The mixing matrix, in this case, is a $3\times 3$ matrix. With unitary assumption, nine independent parameters of the matrix can be divided into three mixing angles and six complex phases. However, one can reduce five more phases by rephasing, resulting in four independent parameters. Similar to the Cabibbo–Kobayashi–Maskawa (CKM) parameterization in the quark sector, the mixing matrix in the three-neutrino scenario is expressed as the product of three rotation matrices as follows,
\begin{align}
\label{eq:PMNS_matrix}
U &= 
\begin{pmatrix}
1 & 0 & 0\\
0 & c_{23} & s_{23}\\
0 & -s_{23} & c_{23}\\
\end{pmatrix}
\cdot
\begin{pmatrix}
c_{13} & 0 & s_{13}e^{-i\delta_{\mathrm{CP}}}\\
0 & 1 & 0\\
-s_{13}e^{i\delta_{\rm CP}} & 0 & c_{13}\\
\end{pmatrix}
\cdot
\begin{pmatrix}
c_{12} & s_{12} & 0\\
-s_{12} & c_{12} & 0\\
0 & 0 & 1\\
\end{pmatrix}\\\nonumber
&=
\begin{pmatrix}
c_{12}c_{13} & s_{12}c_{13} & s_{13}e^{-i\delta_{\rm CP}}\\
-s_{12}c_{23}-c_{12}s_{23}s_{13}e^{i\delta_{\rm CP}} & c_{12}c_{23}-s_{12}s_{23}s_{13}e^{i\delta_{\rm CP}}
& s_{23}c_{13}\\
s_{12}s_{23}-c_{12}c_{23}s_{13}e^{i\delta_{\rm CP}}
& -c_{12}s_{23}-s_{12}c_{23}s_{13}e^{i\delta_{\rm CP}} & c_{23}c_{13}\,,
\end{pmatrix}
\end{align}
where, $c_{ij}=\cos\theta_{ij}$, $s_{ij}=\sin\theta_{ij}$. So the mixing matrix in this case is parameterized in terms of three mixing angles, $\theta_{23}$, $\theta_{13}$, and $\theta_{12}$; and one CP violating phase $\delta_{\rm CP}$, known as Pontecorvo–Maki–Nakagawa–Sakata~(PMNS) matrix. One can express the three mixing angles in terms of the mixing matrix elements, as shown below
\begin{align}
\frac{|U_{e2}|^2}{|U_{e1}|^2}=\tan^2\theta_{12},\hskip 0.5cm \frac{|U_{\mu 3}|^2}{|U_{\tau 3}|^2}=\tan^2\theta_{23},\hskip 0.5cm|U^2_{e3}|=\sin^2\theta_{13},\,.
\end{align}
Apart from the four mixing parameters, mass-squared differences between the three mass eigenstates, namely, $\Delta m^2_{31}=m^2_3-m^2_1$ and $\Delta m^2_{21} = m^2_2-m^2_1$, affect the neutrino oscillation.

Now, using eqns.~\ref{eq:osc_prob2} and~\ref{eq:surv_prob1}, one can estimate the oscillation probability in various oscillation channel~$\nu_\alpha\to\nu_\beta$, where $U_{\alpha i}$ are the elements of the matrix in eq~\ref{eq:PMNS_matrix}.
The expression of the $\nu_\alpha\to\nu_\beta$ appearance probability is given by~\cite{Agarwalla:2013tza}
\begin{align}
\label{eq:app_three_flav}
P(\nu_\alpha\to\nu_\beta) &= 4|U_{\alpha 2}|^2|U_{\beta 2}|^2\sin^2\frac{\Delta_{21}}{2}+4|U_{\alpha 3}|^2|U_{\beta 3}|^2\sin^2\frac{\Delta_{31}}{2}\\\nonumber
&+2 \,\mathrm{Re}(U^\ast_{\alpha 3} U_{\beta 3} U_{\alpha 2} U^\ast_{\beta 2})\left(4\,\sin^2\frac{\Delta_{21}}{2}\sin^2\frac{\Delta_{31}}{2}+\sin2\Delta_{21}\sin2\Delta_{31}\right)\\\nonumber
&+4\,\mathrm{J}_{(\alpha,\beta)}\left(\sin^2\frac{\Delta_{21}}{2}\sin2\Delta_{31}-\sin^2\frac{\Delta_{31}}{2}\sin2\Delta_{21}\right)\,,
\end{align}
where $\Delta_{ij} = \frac{\Delta m^2_{ij}L}{2E}\simeq 2.54\times\frac{\Delta m^2_{ij}L[\mathrm{km}]}{E[\mathrm{GeV}]}$. The quantity $J_{(\alpha,\beta)}$ is the Jarlskog invariant~\cite{PhysRevLett.55.1039}, defined as
\begin{align}
J_{(\alpha,\beta)} =  \mathrm{Im}(U^\ast_{\alpha 1} U_{\beta 1} U_{\alpha 2} U^\ast_{\beta 2})= \mathrm{Im}(U^\ast_{\alpha 2} U_{\beta 2} U_{\alpha 2} U^\ast_{\beta 3})= \mathrm{Im}(U^\ast_{\alpha 3} U_{\beta 3} U_{\alpha 1} U^\ast_{\beta 1})\,,
\end{align}
and $J_{(\alpha,\beta)} = - J_{(\beta,\alpha)}$.
The expression for the $\nu_\alpha\to\nu_\alpha$ disappearance probability is  
\begin{align}
\label{eq:disapp_three_flav}
P(\nu_\alpha\to\nu_\alpha) &= 1-4\,|U_{\alpha 2}|^2(1-|U_{\alpha 2}|^2)\sin^2\frac{\Delta_{21}}{2}-4\,|U_{\alpha 3}|^2(1-|U_{\alpha 3}|^2)\sin^2\frac{\Delta_{31}}{2}\\\nonumber
&+2\,|U_{\alpha 2}|^2 |U_{\alpha 3}|^2\left(4\,\sin^2\frac{\Delta_{21}}{2}\sin^2\frac{\Delta_{31}}{2}+\sin\Delta_{21}\sin\Delta_{31}\right)\,.
\end{align}
\begin{figure}[h!]
	\centering
	\includegraphics[scale=0.95]{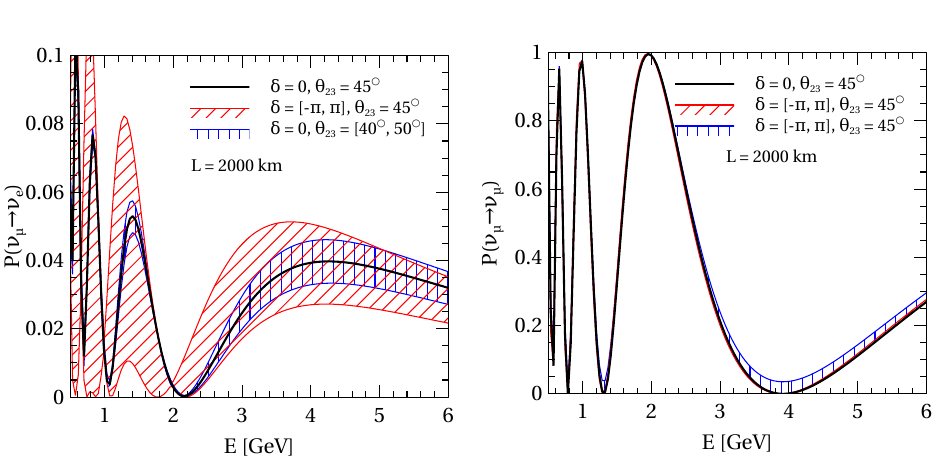}
	\mycaption{$\nu_\mu\to\nu_e$ appearance~(left panel) and $\nu_\mu\to\nu_\mu$ disappearance~(right panel) probabilities as a function of neutrino energy $E$ for a baseline $L=2000$~km. The black solid lines depict the oscillation probabilities computed with $\delta_{\rm CP} = 0$ and $\theta_{23}=45^\circ$. In the red bands, value of the $\delta_{\mathrm{CP}}$ is varied in the range $-180^\circ$ to $180^\circ$, while keeping $\theta_{23}$ constant at $45^\circ$. Blue bands show the oscillation probabilities with $\theta_{23}$ values in the range $40^\circ$ to $50^\circ$ and $\delta_{\rm CP} = 0^\circ$.
	The benchmark values of the other oscillation parameters are $\theta_{13}= 8^\circ$, $\theta_{12}=33^\circ$, $\Delta m^2_{21}=7.5\times 10^{-5}$ eV$^2$, $\Delta m^2_{31}=2.5\times 10^{-3}$ eV$^2$.}
	\label{fig:prob_three_flav}
\end{figure}

In fig.~\ref{fig:prob_three_flav}, we show the $\nu_\mu\to\nu_e$ appearance~(left panel) and $\nu_\mu\to\nu_\mu$ disappearance~(right panel) probabilities calculated using eqns.~\ref{eq:app_three_flav} and~\ref{eq:disapp_three_flav} as a funcion of energy for a benchmark baseline $2000$ km. These two oscillation channels are the major channels that are probed in the long-baseline neutrino experiments. Unlike the two-flavor scenario, the appearance probability now depends on the six oscillation parameters. In the figure, we explore the impact of two of the most uncertain oscillation parameters, namely, $\dcp$ and $\tz$.
The black solid line in both the panels corresponds to the oscillation probability computed with $\delta_{\rm CP} = 0$ and $\theta_{23}=45^\circ$. The benchmark value of the other oscillation parameters are $\theta_{13}= 8^\circ$, $\theta_{12}=33^\circ$, $\Delta m^2_{21}=7.5\times 10^{-5}$ eV$^2$, $\Delta m^2_{31}=2.5\times 10^{-3}$ eV$^2$. Red bands in both panels show the probabilities by varying $\dcp$ in the range $[-180^\circ,180^\circ]$, while keeping other oscillation parameters fixed at their benchmark values. In the blue bands, $\tz$ is varied in the range $[40^\circ, 50^\circ]$.

The observed features of the $\mue$ appearance probability shown in the left panel can be explained by a simplified approximated analytical expression of $P(\nu_\mu\to\nu_e)$ derived by expanding in terms of the small parameters $\alpha = \frac{\Delta m^2_{21}}{\Delta m^2_{31}}\simeq 0.03$ and $\sin\theta_{13} \simeq 0.1$. The expression after the expansion has the form
\begin{align}
\label{eq:app_prob_approx}
P(\nu_\mu\to\nu_e) &\simeq \sin^22\theta_{13}s^2_{23}\sin^2\Delta_{31}\\\nonumber
&-(\alpha\Delta_{31})\sin2\theta_{13}
\sin2\theta_{12}\sin2\theta_{23}\left[\sin(2\Delta_{31}+\delta_{\rm CP})-\sin\delta_{\rm CP}\right]\\\nonumber
&+(\alpha\Delta_{31})^2\sin^22\theta_{12}c^2_{23}\,.
\end{align}
The major contribution to flavor transition comes from the first term in the right-hand side~(RHS) of the above equation, as it is only $\sin^2\theta_{13}$ dependence. The second term with ($\alpha\sin\theta_{13}$) dependence has a subleading contribution to the oscillation probability. The least contribution comes from the last term, which has a $\alpha^2$ dependence. The neutrino energy and baseline at oscillation maxima are determined by the first term when $\Delta_{31} = (2n+1)\frac{\pi}{2}$~$(n=1,2,3,..)$. 
 We observe for varying $\dcp$ case~(red band) that the position of the oscillation maximum does not change significantly when $\delta_{\rm CP}$ is varied, but the amplitude of the oscillation can change significantly depending on the value of the CP phase. As we can see, $\delta_{\rm CP}$ dependence in the $\nu_\mu\to\nu_e$ appearance probability comes from the second term in eq.~\ref{eq:app_prob_approx}, which has a subleading contribution. The amplitude of the oscillation changes depending on the value of $\delta_{\rm CP}$. For the varying $\tz$ case (blue band), we also observe the noticeable impact of $\tz$ in the $\mue$ appearance probability, which mainly comes from the first term with a small contribution from the second term in eq.~\ref{eq:app_prob_approx}. From the overlap of the two bands, one would expect a degeneracy between two oscillation parameters $\dcp$ and $\tz$ in the $\mue$ channel, which would potentially impact their measurement through neutrino oscillation experiments.

In the right panel, we plot the $\mumu$ disappearance probability as a function of energy. 
The approximate analytical expression of the $\mumu$ disappearance probability using the series expansion is given by
\begin{align}
\label{eq:disapp_approx}
P(\mumu) &\simeq 1-\sin^22\tz\sin^2\Delta_{31}+(\alpha\Delta_{31}) c^2_{12}\sin^22\tz\sin2\Delta_{31}\\\nonumber
&+\mathcal{O}(\alpha^2)+\mathcal{O}(\alpha s_{13}) + \mathcal{O} (s^2_{13})\,.
\end{align}
It is clear from the above equation that the leading contribution to the $\mumu$ survival probability comes from the second term in the RHS of eq.~\ref{eq:disapp_approx}. The third term has a subleading effect in oscillation, while the other terms are highly suppressed. Similar to the appearance channel, neutrino energy and baseline at oscillation maxima are determined by the condition $\Delta_{31} = (2n+1)\frac{\pi}{2}$. The figure shows that $\dcp$ does not has a negligible impact on the $\mumu$ disappearance probability, as both the leading term and the subleading term do not have $\dcp$ dependence. However, $\tz$ has an observable impact on the survival probability at higher energy as the leading, and the subleading term contains $\sin^22\tz$. 

In this subsection, we discuss the oscillation probability of the neutrinos in the three-flavor scenario. For antineutrinos, expression of the oscillation probabilities is similar except the sign of the CP phase $\dcp$ is flipped, \ie~$P(\nu_\alpha\to\nu_\beta;\,\delta_{\rm CP})\rightarrow P(\nu_\alpha\to\nu_\beta;\,-\delta_{\rm CP})$.
\section{Neutrino oscillation in matter}
\label{sec:osc_mat}
In the previous sections, we discuss the formalism of the flavor transition of the neutrinos propagating in a vacuum. However, in a more realistic scenario, neutrinos propagate through the medium containing matter particles. In that case, forward coherent elastic scattering between neutrinos and the matter particle influence the flavor oscillation.

In the Standard Model~(SM) framework, neutrino interacts with the matter particle {\it via} charge-current~(CC) and neutral-current~(NC) interaction. In the case of ordinary matter with electrons and quarks, CC interactions involve only electrons from the ambient matter, as there are no muons and taus. Both electrons and quarks contribute to the NC interaction.
The effective Hamiltonian representing CC and NC interactions are
\begin{align}
\label{eq:CC_Ham}
\mathcal{H}^{\rm CC}_{eff} &= \frac{G_{ F}}{\sqrt{2}}\left[\bar{e}(p_1)\gamma^\rho(1-\gamma^5)\nu_e(p_2)\right]\left[\bar{\nu}_e(p_3)\gamma_\rho(1-\gamma^5)e(p_4)\right]\\\nonumber
& = \frac{G_{ F}}{\sqrt{2}}\left[\bar{\nu}_e(p_3)\gamma_\rho(1-\gamma^5)\nu_e(p_2)\right]\left[\bar{e}(p_1)\gamma^\rho(1-\gamma^5)e(p_4)\right]\,,
\end{align}
\begin{align}
\label{eq:NC_Ham}
\mathcal{H}^{\rm NC}_{eff} = \frac{G_{F}}{\sqrt{2}}\left[\bar{\nu}_e(p_1)\gamma^\rho(1-\gamma^5)\nu_e(p_2)\right]\left[\bar{f}(p_3)\gamma_\rho(g^f_V-g^f_A\gamma^5)f(p_4)\right]\,,
\end{align}
where $G_F$ is the Fermi coupling constant, and $g^f_V$, $g^f_A$;~$f \in (e, u, d)$, are the vector and axial vector coupling for neutrinos and charged leptons, respectively. Note that in eq.~\ref{eq:CC_Ham}, we have used Fierz Transformation to separate the neutrino and electrons contribution. An effective description of the interaction potential can be extracted by considering the expectation values of the charge fermions' degrees of freedom in the background matter,

\begin{align}
&\mathcal{H}^{\rm CC}_{eff} = -\sqrt{2}G_F\left[\bar{\nu}_{eL}(p)\gamma_\rho \nu_{eL}(p)\right]\braket{\bar{e}\gamma^\rho(1-\gamma^5)e}\,,\\
&\mathcal{H}^{\rm NC}_{eff} = -\sqrt{2}G_F\left[\bar{\nu}_{eL}(p)\gamma_\rho \nu_{eL}(p)\right]\braket{\bar{f}\gamma_\rho(g^f_V-g^f_A\gamma^5)f}\,,
\end{align}
where we use $p_2=p_3=p$ for forward scattering.
For the CC interaction,
\begin{align}
\braket{\bar{e}\gamma^0 e} &= N_e\label{eq:average_density}\\
\braket{\bar{e}\gamma^i e} &= \braket{{\bf v}_e}\\
\braket{\bar{e}\gamma^0 \gamma^5 e} &= \braket{\frac{\sigma_e {\bf p}_e}{E_e}}\\
\braket{\bar{e}\gamma^i\gamma^5 e} &= \braket{\sigma_e}\,,
\end{align}
where $N_e$, ${\bf v}_e$, $\sigma_e$, ${\bf p}_e$ are the density, average velocity, polarization, and momentum of the electrons in the background medium, respectively. For a non-relativistic unpolarized medium, only first term~(eq.~\ref{eq:average_density}) survives. After simplification, the expression of the effetive interaction Hamiltonian for the CC interaction is written as
\begin{equation}
H^{CC}_{eff} = V_{\rm CC}\times \left[\bar{\nu}_{eL}\gamma^\rho(1-\gamma^5)\nu_{eL}\right]\,.
\end{equation}
$V_{\rm CC}$ denotes the the neutrino-matter charge-current potential with expression,
\begin{align}
\label{eq:Vcc_1}
V_{\rm CC} = \sqrt{2}G_F N_e\,.
\end{align}
Similarly, neutral-current potential can be written as
\begin{align}
V^f_{\rm NC} &= \sqrt{2}G_{F} \left[g^e_V N_e+g^p_V N_p+g^n_V N_n\right]\\
&=-\frac{G_F}{\sqrt{2}}N_n\,,
\end{align}
where $g^e_V = -\frac{1}{2}+\sin^2\theta_{W}$, $g^p_V = \frac{1}{2}-\sin^2\theta_{W}$, and $g^n_V = -\frac{1}{2}$. We have considered the neutral matter background in the above equation. 

The propagation Hamiltonian for the neutrino traveling inside the matter, undergoing CC and NC interaction with the matter particles, can be written in two parts, as shown below,
\begin{align}
H = H_{\rm vac}+H_{\rm int}\,.
\end{align}  
The first term is the vacuum propagation Hamiltonian, while the second term represents the neutrino-matter interaction Hamiltonian. In the mass basis, vacuum Hamiltonian $H^m_{\rm vac}$ is a diagonal matrix with the energy of the neutrino mass eigenstates at the diagonal elements,  
\begin{align}
H^m_{\rm vac} = {\rm Diag.}\left(E_1, E_2,\,\ldots E_n\right) = E\times {\bf I} + {\rm Diag.} \left(\frac{m^2_1}{2E}, \frac{m^2_2}{2E}, \ldots \frac{m^2_n}{2E}\right)\,,
\end{align}
where $n$ is the number of neutrino states. Note that we have used eq.~\ref{eq:E_approx} in the last step of the above equation. In the flavor basis, vacuum Hamiltonian transforms as $H^f_{\rm vac} = U\cdot H^m_{\rm vac}\cdot U^\dagger$, where $U$ is the mixing matrix. Interaction Hamiltonian is generally written in the flavor basis. In the presence of CC and NC interaction, Interaction Hamiltonian can be written as 
\begin{align}
H_{\rm int} = {\rm Diag.}\left(V_{\rm CC}+V_{\rm NC},V_{\rm NC}, V_{\rm NC}, \ldots\right)\,.
\end{align}
Since only the electron type of neutrino undergoes this interaction, the CC interaction potential is present only in the first diagonal element of the $H_{\rm int}$. On the other hand, NC interaction potential is flavor-independent and present in all the diagonal elements. In the neutrino flavor transition amplitude, $V_{\rm NC}$ acts as a common phase, which is removed while estimating the oscillation probabilities, and only CC interaction potential influence the flavor transition.

In the presence of the neutrino-matter interaction, the vacuum mixing matrix $U$ no longer diagonalizes the neutrino propagation Hamiltonian. Consequently, the Hamiltonian on the mass basis is no longer diagonal.
In this case, estimating neutrino oscillation probability is not as straightforward as in the vacuum case. To estimate the neutrino flavor transition probability in the presence of matter effect, one needs to define a new basis $\ket{\nu_i^m}$ in which neutrino propagation Hamiltonian is diagonal, such that $H^m = {\rm Diag.}(E^m_1, E_2^m, \ldots) = (\tilde{U})^\dagger H \tilde{U}$ and $\ket{\nu_\alpha} = \tilde{U}\ket{\nu^m_i}$. The new basis is generally called {\it modified mass basis}, while the diagonalizing matrix $\tilde{U}$ is termed as {\it modified mixing matrix}. Following eq.~\ref{eq:E_approx}, eigenvalues of the Hamiltonian can be parameterized as $E^m_{i} =\frac{\tilde{m}_i^2}{2E}$, where $m^2_i$ are the {\it modified mass-squares}. Since the Hamiltonian is diagonal in the modified mass basis, one can now estimate the neutrino oscillation probability using the same formalism for the vacuum case, except replacing the elements of th vacuum mixing matrix $U$ with the corresponding elements of modified mixing matrix $\tilde{U}$ and energy of the mass eigenstates with the modified energy $E^m_i$. Using eqns.~\ref{eq:osc_prob2} and~\ref{eq:surv_prob1}, the expression for the appearance and disappearance probability for the neutrino propagating matter are
\begin{align}
\label{eq:osc_prob_mat}
P(\nu_\alpha\to\nu_\beta, L, E) = \delta_{\alpha\beta} - &4\sum_{j>k} \mathrm{Re} \left[\tilde{U}^\ast_{\alpha j} \tilde{U}_{\beta j} \tilde{U}_{\alpha k} \tilde{U}^\ast_{\beta k}\right]\sin^2\left(\frac{\Delta \tilde{m}^2_{jk}L}{4E}\right) \\\nonumber
&+2\sum_{j>k}\mathrm{Im}\left[\tilde{U}^\ast_{\alpha j} \tilde{U}_{\beta j} \tilde{U}_{\alpha k} \tilde{U}^\ast_{\beta k}\right]\sin\left(\frac{\Delta \tilde{m}^2_{jk}L}{2E}\right)\,.
\end{align}
\begin{align}
\label{eq:surv_prob_mat}
P(\nu_\alpha\to\nu_\alpha, L, E) = 1 - &4\sum_{j,k} |\tilde{U}_{\alpha j}|^2 |\tilde{U}_{\alpha k}|^2 \sin^2\left(\frac{\Delta \tilde{m}^2_{jk}L}{4E}\right)\,,
\end{align}
where $\Delta \tilde{m}^2_{ij} = \tilde{m}^2_i-\tilde{m}^2_j$ and $\tilde{U}_{\alpha i}$ are the elements of the modified mixing matrix.
 that above equations are valid for a medium with constant matter denisty. For a medium with with varying matter denisty, these expression would be modified.
In what follows, we discuss the neutrino oscillation in the two-flavor and three-flavor scenarios.
\subsection{Two-flavor case}
\label{subsec:two_flav_mat}
In the two-flavor scenario, neutrino propagation Hamiltonian in the flavor basis is written as
\begin{align}
H &=  U
\begin{pmatrix}
0 & 0\\
0 & \Delta m_{21}^2
\end{pmatrix}
U
+\begin{pmatrix}
V_{\rm CC} & 0\\
0 & 0
\end{pmatrix}\\\nonumber
& = \frac{\sdm}{2E}
\begin{pmatrix}
1- \cos2\theta+ A_{\rm CC} & \sin2\theta\\
\sin2\theta & 1+\cos2\theta
\end{pmatrix}\,,
\end{align}
where mixing matrix $U$ is defined in eq.~\ref{eq:two_flav_mix} and $A_{\rm CC} = \frac{2EV_{\rm CC}}{\sdm}$. 

Diagonlizing the Hamiltonian, expression of the energy eigenvalues or the modified mass-squares are
\begin{align}
\label{Em_two_flav}
E^m_{1} &= \frac{\tilde{m}^2_1}{2E} = \frac{V_{\rm CC}}{2}+\frac{\Delta m^2_{21}}{4E}\left(1-\sqrt{\sin^22\theta+\left(\cos2\theta-A_{\rm CC}\right)}\right)\\\nonumber
E^m_{2} &= \frac{\tilde{m}^2_2}{2E} = \frac{V_{\rm CC}}{2}+\frac{\Delta m^2_{21}}{4E}\left(1+\sqrt{\sin^22\theta+\left(\cos2\theta-A_{\rm CC}\right)}\right)\,.
\end{align}
The modified mixing matrix $U^m$ can be parameterized in terms of the modified mixing angle $\theta^m$, which has the form
\begin{align}
U^m = \begin{pmatrix}
\cos\theta^m & \sin\theta^m\\
-\sin\theta^m & \cos\theta^m
\end{pmatrix}\,,
\end{align}
where
\begin{align}
\label{eq:mod_mix_angle}
\sin2\theta^m = \frac{\sin2\theta}{\sqrt{\sin^22\theta+\left(\cos2\theta-A_{\rm CC}\right)^2}}\,.
\end{align}
It is interesting to note that when the condition $\cos2\theta = A_{\rm CC}$ is satisfied, RHS of the above equation becomes 1, and $\theta^m$ attains the maximal value, \ie~$45^\circ$ maximizing the oscillation amplitude. This is known as {\it resonance}, and from the resonance condition, one can derive neutrino energy at resonance
\begin{align}
E_{\rm res} = \frac{\sdm\cos2\theta}{2V_{\rm CC}}\,.
\end{align}

Now, using eq.~\ref{eq:osc_prob_two_flav}, the expression of the oscillation probability in the matter can  be written as
\begin{align}
\label{eq:trans_prob_mat}
P(\nu_\alpha\to\nu_\beta) &= \sin^2 2\theta^m\sin^2\left(1.27\times\frac{\Delta \tilde{m}^2_{21}L[\mathrm{km}]}{E[\mathrm{GeV}]}\right)
\end{align}

\begin{align}
\label{eq:surv_prob_mat2}
P(\nu_\alpha\to\nu_\alpha) = 1- \sin^2 2\theta^m\sin^2\left(1.27\times\frac{\Delta \tilde{m}^2_{21}L[\mathrm{km}]}{E[\mathrm{GeV}]}\right)\,.
\end{align}
\begin{figure}[h!]
	\centering
	\includegraphics[scale=0.95]{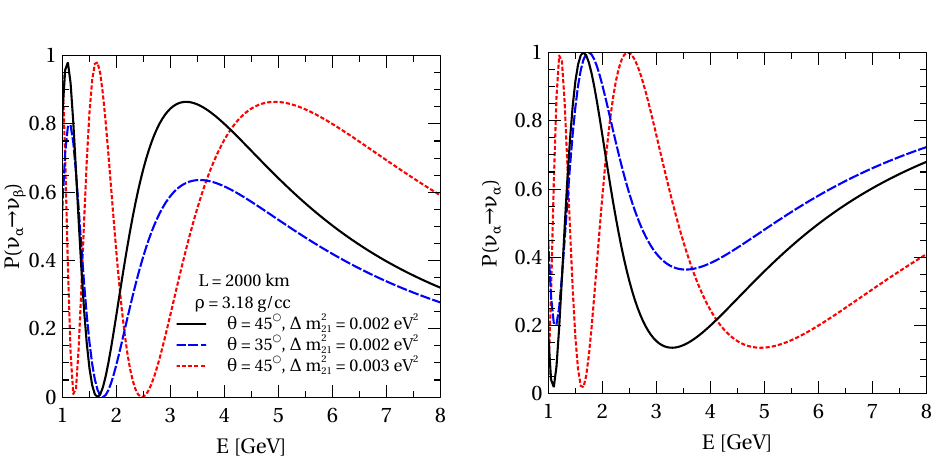}
	\mycaption{$\nu_\alpha\to\nu_\beta$ appearance~(left panel) and $\nu_\alpha\to\nu_\alpha$ disappearance~(right panel) probabilities in matter as a function of neutrino energy $E$. We consider a baseline of 2000 km in matter with constant density $\rho = 3.18$ g/cc. The probabilities are computed for three sets of mixing angle and mass-squared difference values; $\theta = 45^{\circ}, \Delta m^{2}_{21} = 2.0\times 10^{-3}$ eV$^2$~(solid black line), $\theta = 35^{\circ}, \Delta m^{2}_{21} = 2.0\times 10^{-3}$ eV$^2$~(blue dashed line), $\theta = 45^{\circ}, \Delta m^{2}_{21} = 3.0\times 10^{-3}$ eV$^2$~(red dotted line).}
	\label{fig:prob_two_flav_mat}
\end{figure}

To show the impact of matter effect, in fig.~\ref{fig:prob_two_flav_mat}, we reproduce fig.~\ref{fig:prob_two_flav} but in the presence of matter with a constant density of 3.18 g/cc. The value of the density is considered according to the preliminary reference Earth model~(PREM) profile~\cite{DZIEWONSKI1981297} of the Earth. As evident from eqns.~\ref{Em_two_flav} and \ref{eq:mod_mix_angle} that modified oscillation parameters $\theta^m$ and $\Delta \tilde{m}^2_{21}$ in matter changes with the strength of the matter potential and neutrino energy, amplitude of the oscillation probability and position of the oscillation maxima will be modified in presence of matter. In the left panel of fig.~\ref{fig:prob_two_flav_mat}, where we show the appearance probability, we observe a reduction in the oscillation amplitude for all the three sets of oscillation parameters as compared to the vacuum case shown in fig.~\ref{fig:prob_two_flav}. This happens because, in the presence of the new potential, the value of the modified mixing angle deviates from the maximal value of $45^\circ$. Also, there is a marginal shift in the position of the oscillation maxima due to the modification in $\sdm$ in matter. In the right panel, we show the survival probability in the presence of matter. Similar to appearance probability, we observe a reduction in the oscillation probability near the oscillation maxima. However, at the oscillation minima, survival probability is always equal to 1 as the last term in eq~\ref{eq:surv_prob_mat} is zero irrespective of the $\theta^m$ value.

This subsection discusses the oscillation probability for neutrino propagating in the matter. For antineutrinos, matter potential would have an additional negative sign as the particle number is negative in eq.~\ref{eq:Vcc_1}. One can repeat the above-discussed formalism with $V_{\rm CC}\to -V_{\rm CC}$ to estimate the oscillation probability for antineutrino in the matter. Also, note that in the case of antineutrino, resonance can not be attained when $\sdm$ is positive.

\subsection{Three-flavor case}
\label{subsec:three_flav_mat}

In the three-flavor scenario, neutrino propagation Hamiltonian in the flavor basis is written as
\begin{align}
H = U
\begin{pmatrix}
0 & 0 & 0\\
0 & \frac{\sdm}{2E} & 0\\
0 & 0 & \frac{\ldm}{2E}
\end{pmatrix}
U^\dagger
+
\begin{pmatrix}
V_{\rm CC} & 0 & 0\\
0 & 0 & 0\\
0 & 0 & 0
\end{pmatrix}\,.
\end{align}
The diagonalization of this Hamiltonian analytically is complex and involves several parameters. Various approximations can be used to diagonalize the Hamiltonian in order to estimate the oscillation probability in the three-flavor scenario. In chapter~\ref{C4}, we will discuss a formalism of an approximate analytical diagonalization of neutrino propagation Hamiltonian in the presence of standard CC interaction, as well as some new interaction beyond the Standard Model. Using the diagonalized Hamiltonian and the diagonalizing matrix or the effective mixing matrix, one can estimate the oscillation probability in the three-flavor case can be estimated using 
eq.~\ref{eq:app_three_flav} and~\ref{eq:disapp_three_flav}, and by replacing element of vacuum mixing matrix and mass-square differences by their modified counterparts in matter.
\begin{figure}[h!]
	\centering
	\includegraphics[scale=0.95]{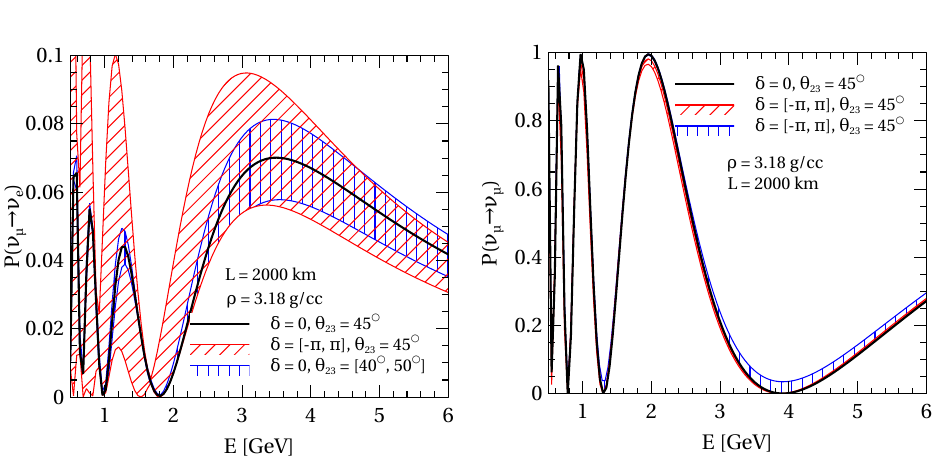}
	\mycaption{$\nu_\mu\to\nu_e$ appearance~(left panel) and $\nu_\mu\to\nu_\mu$ disappearance~(right panel) probabilities as a function of neutrino energy for a baseline $L=2000$~km in matter with a constant density $\rho = 3.18$ g/cc. The black solid lines depict the oscillation probabilities computed with $\delta_{\rm CP} = 0$ and $\theta_{23}=45^\circ$. In the red bands, value of the $\delta_{\mathrm{CP}}$ is varied in the range $-180^\circ$ to $180^\circ$, while keeping $\theta_{23}$ constant at $45^\circ$. Blue bands show the oscillation probabilities with $\theta_{23}$ values in the range $40^\circ$ to $50^\circ$ and $\delta_{\rm CP} = 0^\circ$.
		The benchmark values of the other oscillation parameters are $\theta_{13}= 8^\circ$, $\theta_{12}=33^\circ$, $\Delta m^2_{21}=7.5\times 10^{-5}$ eV$^2$, $\Delta m^2_{31}=2.5\times 10^{-3}$ eV$^2$.}
	\label{fig:prob_three_flav_mat}
\end{figure}

In fig.~\ref{fig:prob_three_flav_mat}, we reproduce fig.~\ref{fig:prob_three_flav} in the presence of matter effect considering a baseline 2000 km with constant matter density $\rho = 3.18$ g/cc. In the case of $\mue$ appearance channel, we observe an overall increase in transition probability due to the matter interaction. It happens because the value of the modified mixing angle $\theta^m_{13}$, which is present in the leading term in $\mue$ appearance probability~(see eq.~\ref{eq:app_prob_approx}) increases significantly in the presence of matter effect. Also, the impact of $\dcp$ broadens as the matter brings in fake $\dcp$ effect~\cite{Gonzalez-Garcia:2001snt}. Broadening in bands corresponds to the varying $\tz$ related to an increase in the value of $\tym$ in matter as $\tz$ is coupled with the $\ty$ in the leading term of the $\mue$ appearance probability. In the right panel, we do not observe any significant change in survival probability compared to the vacuum case. It is mainly because the leading term does not contain $\tym$, which is the major reason for enhancement in the appearance probability.

In the antineutrino case, matter potential $V_{\rm CC}$ as well as the CP phase $\dcp$ gains an extra negative sign, \ie, $P(\nu_\alpha\to\nu_\beta, V_{\rm CC}, \dcp)\rightarrow P(\nu_\alpha\to\nu_\beta, -V_{\rm CC}, -\dcp)$.
\section{Neutrino oscillation experiments}
\label{sec:osc_experiments}
After the discovery of the neutrino oscillation phenomena, numerous experiments came into existence with their major focus on establishing neutrino oscillation in the three-neutrino paradigm. Many of these experiments have already completed their run, and some are currently operating. Also, many outstanding experiments are in the pipeline, aiming to resolve existing issues in neutrino oscillation. These experiments operate in a diverse range of neutrino energy and use neutrino flux from different sources. Also, detection techniques vary in different experiments. In this section, we discuss various neutrino experiments that are crucial for our understanding of neutrino oscillation.

\subsection{Solar neutrino experiments}
\label{solar_neutrino_exp}

The neutrinos from the Sun undergo flavor transition on the way to the Earth. Given the large Earth-Sun distance, solar neutrinos presents an excellent opportunity to probe the neutrino oscillation. Generally, gallium and chlorine detectors are used to observe solar neutrinos. Also, big water Cherenkov detectors like Super-Kamiokande and SNO have the ability to observe higher energy components of the neutrino flux. Here, we discuss some of the solar neutrino experiments.

\paragraph{\bf Homestake:} 
This experiment~\cite{Cleveland:1998nv}, started its operation in 1965, was the first neutrino experiment to observe solar neutrinos. It was a chlorine detector of 615 ton mass comprising of perchlorate-ethylene, where the incoming $\nu_e$ interacts with $Cl^{37}$ to produce $Ar^{37}$ and an electron. Unstable $Ar^{37}$ decays back to $Cl^{37}$ by electron capture, emitting 2.82 keV X-ray spectra that are observed. After several years of running, this experiment produced the result for the average solar neutrino capture rate of $2.56\pm 0.25$ SNU~\footnote{SNU is the unit to measure the event rate, 1 SNU = $10^{-36}$ interactions per target atom per second.}. However, the Standard Solar Models~(SSM) predicts a capture rate of $8.1\pm1.2$ SNU, leading to an observed deficit of around 70\%. 

\paragraph{SAGE and Gallex:}
The SAGE (Soviet-American Gallium Experiment)~\cite{SAGE:1999nng} and GALLEX
(Gallium Experiment)~\cite{GALLEX:1992gcp} were pioneering experiments designed to detect solar neutrinos.
These two experiments use Gallium detectors to detect the low-energy component
of the solar neutrino flux using a technique similar to the Homestake experiment. In this case, radioactive $Ge^{71}$  is produced as a daughter particle, whose decay gives the signature of solar neutrino. SAGE observed a neutrino capture rate of $70.8\pm5.0$ SNU~\cite{SAGE:2009eeu}, in contrast to the SSM prediction of $129\pm9$ SNU. The observed rate of Gallex was $77.5\pm8$ SNU~\cite{GALLEX:1998kcz}, almost 40\% less than the SSM prediction validating the solar neutrino deficit. 

\paragraph{Super-Kamiokande:}
Unlike the previous two experiments that use radiochemical techniques to observe neutrinos, Super-Kamiokande~\cite{Super-Kamiokande:2002weg} is a 50 kiloton~(kt) water Cherenkov detector with a fiducial mass of 22.5 kt located at Kamioka mine in Japan. 
Solar neutrinos passing through the detector volume interact with the ambient electrons and nuclei to produce charge particles that move faster than light, emitting Cherenkov radiation. Photomultiplier tubes~(PMTs) placed on the wall of the detector record this radiation to observe the neutrino events. Super-Kamiokande contains 11,146 20-inch PMTs to detect the signal and 1,885 8-inch PMTs for outer veto.
Since the energy required to produce Cherenkov radiation by accelerating charge particles is large, it has a comparatively larger energy threshold of 5 MeV. However, it has a good directional resolution, which confirms the observed neutrino events are indeed from the Sun. Super-Kamiokande observed a capture rate of $0.45\pm0.02$ SNU~\cite{Super-Kamiokande:2005wtt}, almost a factor of half from the SSM prediction.

\subsection{Reactor neutrino experiments}
\label{subsec:reactor_neutrino_exp}
Reactor neutrinos are electron antineutrinos emerging from the  $\beta$-decay during nuclear fission inside the reactors.
The idea of observing neutrinos from the fission reaction inside the reactor was introduced during the 1950s~\cite{PhysRev.90.493}, in which large liquid scintillator detectors were proposed for the detection technique. The first evidence of reactor neutrinos, in fact, neutrinos overall, came from the experiments performed by  Fredrick Reines and Clyde Cowan at the Hanford and Savannah River nuclear reactor~\cite{PhysRev.92.830,PhysRevLett.45.1307}. After that, several outstanding reactor neutrino experiments like Palo Verde~\cite{PhysRevLett.84.3764}, KamLAND~\cite{KamLAND:2002uet}, CHOOZ~\cite{CHOOZ:1999hei}, Daya Bay~\cite{DayaBay:2012fng}, Double CHOOZ~\cite{DoubleChooz:2011ymz}, RENO~\cite{RENO:2012mkc} has improved our understanding of both reactor neutrino flux as well as neutrino oscillation. All these experiments probe the disappearance of $\bar{\nu}_e$ flux. Pioneering data from Daya Bay, Double CHOOZ has established a non-zero value of the smallest mixing angle $\theta_{13}$ with excellent precision. Here, we give a brief description of some of the reactor neutrino experiments.

\paragraph{\bf KamLAND:} 
KamLAND~(Kamioka Liquid Scintillator Antineutrino Detector) is a multipurpose scintillator detector with 1 kt mass placed at the Kamioka observatory in Japan. There were 55 nuclear reactor cores surrounding the detector with an average flux-weighted baseline of 80 km. The detector measures the neutrino capture rate as well as their energy. This experiment accomplished its primary goal of observing the oscillation of reactor $\bar{\nu}_e$. KamLAND results~\cite{KamLAND:2008dgz} highly favors the two neutrino oscillation hypotheses, rejecting the no-oscillation hypothesis and other theories proposed to explain solar neutrino deficit. These results are consistent with solar neutrino experiments. A combined analysis of KamLAND and SNO data gave the measurement of $\tan^2\theta_{12}=0.47^{+0.06}_{-0.05}$ and $\sdm = 7.59^{+0.21}_{0.21}\times 10^{-5}$ eV$^2$ with highest precision.

\paragraph{Daya Bay, Double CHOOZ, RENO:}
After two-flavor oscillation hypothesis was well-validated in the solar and atmospheric neutrino sectors\footnote{In the solar neutrino sector, experiments measured the value of $\theta_{12}$ and $\sdm$ known as solar neutrino parameters, whereas the atmospheric neutrino data confirmed neutrino oscillation with $\Delta m^2_{32}$ and $\theta_{23}$.}, measuring the non-zero value of the $\theta_{13}$ became essential in order to establish neutrino oscillation in a three-flavor framework. Among eight proposed experiments to measure $\theta_{13}$, three were finally constructed, namely, Daya Bay, Double CHOOZ, and RENO. Daya Bay is a China-based experiment consisting of 8 detector modules filled with 20 ton of liquid scintillators. Eight modules are clustered in three locations nearby, surrounded by six powerful nuclear reactor cores situated within a 1.9 km range. Double CHOOZ is the successor of the CHOOZ experiments based in Chooz, France. It had two detector modules filled with a gadolinium-doped liquid scintillator, placed at 400 m and 1050 m from the Chooz nuclear powerplant. RENO~(Reactor Experiment for Neutrino Oscillation) is also a gadolinium-doped liquid scintillator experiment with two detector modules of mass of 16.5 kt, placed at 294 m and 1383 m  from Hanbit Nuclear power Plant in South Korea. The results from Double CHOOZ~\cite{DoubleChooz:2011ymz} gave a hint toward non-zero $\theta_{13}$, along with accelerator neutrino experiments T2K~\cite{T2K:2011ypd}, and MINOS~\cite{MINOS:2011amj} around the same time. In 2012, Daya Bay measured a non-zero value of $\theta_{13}$ with a precision of 5.2$\sigma$~\cite{DayaBay:2012fng} later, which was confirmed later in the RENO experiment~\cite{RENO:2012mkc}.

\paragraph{JUNO:} JUNO~(Jiangmen Underground Neutrino Observatory) is a next-generation medium-baseline reactor neutrino experiments located in Kaiping, a small town in South China, 53 km away from the Yangjiang and Taishan nuclear power plants. The Central detector is a sphere of low background liquid scintillator held in the acrylic vessel. It is surrounded by the water Cherenkov detector, which acts as a muon veto system. A major goal of this experiment is to measure neutrino mass ordering, which will be discussed in the next section. Medium baseline~($\sim$ 53 km) of the experiment with unprecedented energy resolution gives this experiment a unique advantage over others. Due to the medium baseline, it is sensitive to both the mass splittings. Using the interplay between the two mass-splittings, it is expected to determine neutrino mass ordering, given that the experiment has an outstanding energy resolution~($\sim$ 3\% at 1 MeV neutrino energy). Also, due to its medium baseline, the matter effect does influence oscillation significantly. Apart from the determining neutrino mass ordering, JUNO is expected to have a leading contribution in the precision measurement of the oscillation parameters, $\sin^2\theta_{12}$, $\sdm$, and $\Delta m^2_{31}$. After six years of its run, JUNO will measure these three parameters with a precision of $\sim 6$\% or better~\cite{ParticleDataGroup:2018ovx}.

\subsection{Atmospheric neutrino experiments}
\label{subsec:atmospheric_neutrino_Exp}

As discussed in the last chapter, neutrinos are produced in the atmosphere by interacting cosmic particles with the atmospheric nuclei.
Those neutrinos are observed at the detector placed on the surface of the Earth. For upward-going atmospheric neutrinos, the baseline can be as large as the diameter of the Earth. Such a large distance would cause neutrinos to undergo flavor transition when they are observed at the detector. 
Here, we give a brief summary of the ongoing and upcoming atmospheric neutrino experiments.

\paragraph{Super-Kamiokande:}
Apart from detecting solar neutrinos, Super-Kamiokande also measures the atmospheric neutrino flux. In fact, Super-Kamiokande is the first neutrino experiment to observe atmospheric neutrino anomaly and measure neutrino oscillation in the atmospheric neutrino sector.
Details of the detector setup and working principle are already discussed in the context of solar neutrino experiments. 

\paragraph{IceCube DeepCore:}
The IceCube is a neutrino observatory located at the geographic South Pole of the Earth. It uses underground ice between 1.5 km and 2.5 km to observe neutrinos of astrophysical and atmospheric origin. High-energy neutrinos passing through the ice produce secondary charge particles after deep-inelastic scattering~(DIS) with the nuclei. Like the water Cherenkov detector, high-energy secondary particles emit Cherenkov radiation, which is recorded by the Digital Optical Modules~(DOM) placed inside the detector arranged in vertical strings. Although the IceCube primarily observes high-energy astrophysical neutrinos, the bottom-center part of the detector is suitable for observing atmospheric neutrinos. This extension of the IceCube detector is known as IceCube DeepCore and majorly focuses on studying atmospheric neutrino oscillation. IceCube Deepcore contains closely spaced strings with a horizontal spacing of 42-72 m and less vertical spacing between the DOMs~($\sim 7$ m) attached to the strings~\cite{IceCube:2011ucd}. The large number of DOMS in DeepCore reduces its energy threshold~($\sim 10$ GeV) for neutrino detection, making it suitable for detecting atmospheric neutrino events.
DeepCore is currently collecting data and contributing crucially to the measurement of atmospheric oscillation parameters $\theta_{23}$ and $\Delta m^2_{32}$~\cite{IceCubeCollaboration:2023wtb}.

\paragraph{KM3NeT/ORCA:} ORCA~(Oscillation Research with Cosmics in the Abyss)~\cite{Katz:2013svu} is one of the two major components of the KM3NeT~(Cubic Kilometre Neutrino Telescope)~\cite{KM3Net:2016zxf} experiment, located at seabed off the shore of Toulon, France. It is the low-energy component aimed at exploring atmospheric neutrino oscillation in the neutrino energy range [1-100]~GeV. It is also a water Cherenkov detector that will use seawater to produce Cherenkov radiation. The radiation is detected at the DOMs containing 31 PMTs. A set of 18 DOMs are attached to the strings that are placed inside the sea, known as Detection Unit~(DU). The full ORCA detector setup is expected to have 115 DUs, which will be installed in phases, with an expected fiducial mass of 7 Mtons. After the first phase of installation, 6 DUs have been installed, which are currently under operation. Latest results on the measurement of $\sin^2\theta_{23}$ and $\Delta m^2_{31}$ with 355 days of data is reported in ref.~\cite{psf2023008035}.

\paragraph{INO-ICAL:} India-based Neutrino Observatory (INO)~\cite{INO:2006vde} is a proposed atmospheric neutrino experiment at the Theni district of Tamilnadu, India. It will use a 50 kt Iron Calorimeter~(ICAL) detector divided into three modules placed inside at least 1 km of rock coverage. The active component of the detector contains Resistive Plate Chambers~(RPCs) of 2 m$\times$ 2 m dimension placed between the iron layers. The atmospheric neutrinos have energy $\sim$ 1-25 GeV. Additionally, the detector will be magnetized to 1.5 Tesla in the plane of iron layers. $\mu^+(\mu^-)$ produced in the detector after the CC interaction of atmospheric $\nu_\mu(\bar{\nu}_\mu)$, passes through many layers of the iron plate, producing hits in the RPCs. Each hit generated at the RPCs will be recorded by the data acquisition system~(DAQ) to reconstruct the directionality of the events as well as energy deposition in each iron layer. The magnetic field is applied to discriminate the tracks of $\mu^+$ and $\mu^-$, providing a unique advantage to the detector. At present, a prototype of the detector, called mini-ICAL, has been fully operational since 2018 at Madurai, India~\cite{Khindri:2022elz}. In ref.~\cite{ICAL:2015stm}, collaboration discussed the physics potential of INO-ICAL.
 
 \subsection{Long-baseline experiments}
 \label{subsec:long_baseline_exp}
 
As the term long-baseline suggests, long-baseline~(LBL) neutrino experiments are characterized by their large baseline as compared to the other experiments that use artificial neutrino sources.
The typical length of the baseline of these experiments can vary from one hundred to several thousand kilometers. These experiments use accelerator neutrinos with energy in the sub-GeV to multi-GeV range as its source. 
The outgoing neutrino flux from the accelerator travels through the earth's matter to reach the far detector placed at several hundred or thousand kilometers, where the neutrino flux is measured to study neutrino oscillation. In accelerator neutrino experiments, control over the primary proton energy, the geometry of the target, and the horn current allows us to adjust the beam spectrum with great flexibility. Also, the possibility of having a near detector close to the accelerator, which will measure the unoscillated neutrino flux with good precision, helps to reduce the systematic uncertainties of the far detector. Here, we discuss the details of some LBL neutrino experiments.
\paragraph{T2K:}
T2K~(Tokai to Kamioka)~\cite{T2K:2011qtm} is an ongoing LBL experiment based in Japan. It started taking data in 2009. The experiment uses neutrino flux from the J-PARC accelerator facility at Tokai on the east coast of Japan, which uses a proton beam of 770 kW average power. The neutrino flux is measured at the Super-Kamiokande detector located at Kamioka mine, 295 km from the J-PARC, with $2.5^{\circ}$ off-axis angle\footnote{Off-axis angle is the angle between the direction of with decaying neutrino and the decaying pions. The off-axis axis beam peaks at a certain energy-producing narrow band beam.}. The neutrino beam peaks at around $\sim 0.6$ GeV, which is the energy at first oscillation maxima corresponding to the baseline for T2K. Apart from the main detector, Super-Kamiokande. T2K also consists of two near-detector setups, ND280 and INGRID, that measure the unoscillated neutrino flux and reduce the systematic uncertainties related to the beam. T2K primarily probes $\mue$, $\mumu$, $\bar{\nu}_\mu\to\bar{\nu}_e$, and $\bar{\nu}_\mu\to\bar{\nu}_\mu$ oscillation channel. A major goal of T2HK is to establish CP violation in the lepton sector and improve the precision of oscillation parameters. The latest results of T2K reported in ref.~\cite{T2K:2023mcm}, which uses neutrino events, correspond to $1.97\times 10^{21}$ protons on target~(POT) in the neutrino mode and $1.62\times 10^{21}$ POT in the antineutrino mode.

\paragraph{NO$\nu$A:} NO$\nu$A~(NuMI Off-axis $\nu_e$ Appearance)~\cite{NOvA:2004blv, NOvA:2019cyt} is the USA based LBL experiment presently collecting data. Neutrino flux from the NuMI (Neutrinos as Main Injector)~\cite{Adamson:2015dkw} beam facility at Fermilab operating at 700 kW beam power is used as the source.
The far detector is placed at a distance of 810 km from the source with an off-axis angle $0.8^\circ$. The average energy of the neutrino beam is around 2 GeV. The detector is a tracking calorimeter of 14 kt fiducial mass, consisting of rigid polyvinyl chloride (RPV) cells filled with liquid scintillator material. The experiment also contains a near detector setup placed 1 km near the source with the same off-axis angle at the far detector. Major oscillation channels that are explored by this experiment are $\mue$ and $\bar{\nu}_\mu\to\bar{\nu}_e$. Similar to the T2K experiment, a significant goal of NO$\nu$A includes probing CP violation and neutrino mass ordering. Also, it plays a crucial role in the precision measurement of the neutrino oscillation parameters~\cite{NOvA:2021nfi}.

\paragraph{DUNE:} DUNE~(Deep Underground Neutrino Experiment)~\cite{DUNE:2020lwj} is a next-generation LBL experiment currently under construction. The proposed neutrino source will be an on-axis wide-band beam with multi-GeV energy generated at Fermilab with an average proton beam power of 1.2 MW and $1.1\times10^{21}$ POT per year. The far detector is a Liquid Argon Time Projection Chember~(LArTPC) with a fiducial mass of 40 kt, which will be placed at Homestake mine in South Dakota, 1300 km from the source. The experiment will have equal runtime in the neutrino and antineutrino modes. DUNE will majorly explore neutrino oscillation in the $\mue$, $\mumu$, $\bar{\nu}_\mu\to\bar{\nu}_e$, and $\bar{\nu}_\mu\to\bar{\nu}_\mu$ channels. Apart from these four channels, DUNE will also have the potential to probe $\nu_\mu\to\nu_\tau$ and $\bar{\nu}_\mu\to\bar{\nu}_\tau$ channel, as the incoming neutrino beam will have enough energy to produce {\it tau}. DUNE is expected to have three modules of near detector setup~\cite{DUNE:2021tad}. The first will be an LArTPC detector placed at 574 m from the source. The second is a high-pressure Gaseous Argon time projection chember~(GArTPC) with 1 kt mass, and the third is System for On-Axis Neutrino Detection~(SAND). Apart from SAND, the other two detectors can be moved to different off-axis of the beam to have better information about the source. A major goal of the DUNE experiment is to determine the neutrino mass ordering and measure the oscillation parameters with unprecedented precision. The physics potential of the DUNE experiment to probe neutrino oscillation is discussed in ref.~\cite{DUNE:2020jqi}.

\paragraph{T2HK:} T2HK~(Tokai to Hyper-Kamiokande)~\cite{Hyper-KamiokandeWorkingGroup:2014czz,Hyper-Kamiokande:2018ofw} is the successor T2K which is expected to start collecting data in the upcoming decade. Similar to T2K, it will use a neutrino beam from the J-PARC facility in Japan but correspond to the proton beam power of 1.3 MW and $27\times 10^{21}$ POT in its 10 years of proposed runtime. T2HK will have the same baseline as T2K, {\,  i.e.}~295 km. The far detector will be the upgrade of Super-Kamiokande, known as Hyper-Kamiokande~(Hyper-K). It is a huge water Cherenkov detector with 187 kt of fiducial mass. Similar to T2K, the far detector will be placed at $2.5^{\circ}$ off-axis angle, receiving a narrow band beam peaked at 0.6 GeV. Unlike DUNE, T2HK is expected to have unequal runtime in the neutrino and antineutrino modes, with 2.5 years of runtime in the neutrino mode and 7.5 years of runtime in the antineutrino mode. T2HK will also probe $\mue$, $\mumu$, $\bar{\nu}_\mu\to\bar{\nu}_e$, and $\bar{\nu}_\mu\to\bar{\nu}_\mu$ oscillation channels. T2HK will have two near detector setups; one is ND280, placed at 280 m from the source with the same off-axis angle as the Hyper-K~\cite{Hyper-Kamiokande:2018ofw}. This detector will measure the unoscillated neutrino flux. The other detector is the Intermediate Water Cherenkov Detector~(IWCD) with a 1 kt mass placed at 1km from the source. IWCD can be moved vertically, which will allow to measure neutrino flux at different off-axis angles. T2HK is expected to have outstanding potential to probe neutrino oscillation~\cite{Hyper-KamiokandeProto-:2015xww}.

\subsection{Astrophysical neutrino experiments}
\label{subsec:astro_expt}
The terrestrial neutrino sources, including the artificial ones, generally have energy in the order of MeV to GeV, except for atmospheric neutrinos, a small fraction of which can have several TeV energy environments. Astrophysical neutrinos are the only source of HE and UHE neutrinos.
The detection of astrophysical neutrinos is crucial for our understanding of the very little-known universe. Many experiments are working in this direction, and several others have been proposed. Here, we discuss a few of them:

\paragraph{IceCube:} IceCube experiment is playing the leading role in observing astrophysical neutrinos and their physics potential. As discussed in the context of IceCube DeepCore, the experiment is located in the geometric South Pole of the Earth. It is a cubic kilometer of ice instrumented with 5160 DOMs arranged in 60 strings vertically placed inside the ice. Each DOM contains a 10-inch photomultiplier. High-energy astrophysical neutrinos interact with ambient nuclei in the ice to produce secondary charged particles that generate Cherekov radiation. DOMs collect photons from the radiation and convert them to electrical signals, which are sent to processors at the surface. Signals are analyzed further to reconstruct the flavor, directionality, and energy of the event. Generally, $\mu^{\pm}$ produced from the CC interaction of $\nu_\mu$ produces a track-like signal, which is known as {\it track-like} events. A small fraction of $\nu_\tau$ events can also generate track-like events as the final state {\it tau} can decay into muons. For $\nu_e$ and $\nu_\tau$, final state leptons ($e$ and $\tau$) and hadrons showers around the interaction vertex, known as {\it cascade like} events. Also, CC interaction involving $\nu_\tau$ produces two cascades, one from the shower of final state hadrons and the other from the decay of final state $\tau$. These events are known as {\it double cascade} events. Analysis of the neutrino events can give information about the flavor composition of the astrophysical neutrinos on Earth, which includes the neutrino oscillation effect in the path.
Presently, IceCube collaboration has released 7.5 years of data containing High-Energy Starting Events\footnote{High-Energy Starting Events~(HESE) are neutrino events with reconstructed neutrino energy $E> 60$~TeV, and the event vertex contained within the detector volume.}~(HESE)~\cite{IceCube:2020wum}.
Also, the direction and the energy of the events can reveal crucial information about its possible source. Recently, IceCube collaboration has confirmed the first candidate of the high-energy astrophysical neutrino sources, the blazar TXS
0506+056~\cite{IceCube:2018dnn,IceCube:2018cha} and the Seyfert galaxy NGC 1068~\cite{IceCube:2022der}.
\paragraph{IceCube-Gen2:} IceCube-Gen2 is the proposed successor of IceCube, which is expected to start taking data in the next decade. It will have almost eight times the volume of the IceCube experiment, with 120 new strings installed with the present IceCube setup. With such a large volume of the detector equipped with a highly advanced photo and radio sensor, it will have the potential to observe around 1 million neutrino events per year, spreading across $10^{12}$ order of energy. 
\paragraph{Baikal-GVD:} Baikal-GVD~\cite{Baikal-GVD:2020xgh} is a next-generation neutrino telescope of 1.5 cubic kilometers proposed volume located at the southern part of the lake Baikal, Siberia. The experiment has been under construction since 2011. Seven clusters with a total of 2016 optical modules~(OMs) have been installed. Each cluster is an independent unit of the detector, containing 8 strings with 36 OMs attached to each string. The primary detection method of this experiment is similar to that of IceCube. The experiment has already detected at least one cascade-like neutrino event of astrophysical origin with reconstructed energy 91 TeV~\cite{Baikal-GVD:2020irv}.
\paragraph{KM3NeT/ARCA:} ARCA~(Astroparticle Research with Cosmic in the Abyss)~\cite{KM3Net:2016zxf} is the component of the KM3NeT experiment observing astrophysical neutrinos. It is located in the Mediterranean Sea,  100 km offshore from Sicily, Italy, with a depth of 3500 m  depth. It will comprise of two building blocks. Each block consists of 115 strings, with each string containing 18 OMs. Like the IceCube experiment, PMTs inside each OM will detect the Cherenkov light generated by relativistic charged particles produced from the neutrino-nuclei interactions. AT present ARCA detector is functional with 28 detection units. After the completion of phase-2 construction, it will have 230 detection units with an effective volume of 1 km$^3$. The potential of the detector to explore neutrino astrophysics is discussed in ref~\cite{Aiello:2024jbp}.
\paragraph{P-ONE:} P-ONE~(Pacific Ocean Neutrino experiment)~\cite{P-ONE:2020ljt} is a proposed multi-cubic kilometer neutrino telescope to be installed in the Cascadia basin off Vancouver Island. It will use Ocean Networks Canada infrastructure for communication. It has planned to deploy the first segment containing ten strings, with each containing twenty optical modules and two calibration modules in 2023-2024. After the completion, the experiment will include around 70 strings arranged in a cylindrical structure with a volume corresponding to 1 km height and 1 km radius.

\section{Current status of the oscillation parameters}
\label{sec:osc_param_status}

The discovery of the non-zero value of $\theta_{13}$ establishes neutrino oscillation in the three-flavor framework. This takes us to the precision era of neutrino physics, where one of the major goals of the neutrino oscillation experiments is to measure the six oscillation parameters in the three neutrino paradigm~($\theta_{12}$, $\theta_{13}$, $\theta_{23}$, $\dcp$, $\ldm$, and $\sdm$) at the CKM level precision. With the existing data from various neutrino oscillation experiments, we have already measured a few oscillation parameters with outstanding precision. There are a few parameters, namely $\dcp$ and $\tz$, that are still loosely constrained.
Upcoming data from the presently running and future neutrino experiments will improve the precision of these parameters further. In this section, we discuss the present status and the future projection of the oscillation parameters in the three-neutrino paradigm.

\begin{table}
	\centering
	\begin{tabular}{|c|c|c|}
		\hline
		\text{Parameter}&\text{Main contribution from}&\text{Other contributions from}
		\\
		\hline
		$\theta_{12}$ & SOL & KamLAND
		\\
		$\theta_{23}$ & LBL+ATM & -
		\\
		$\theta_{13}$& REAC & (LBL+ATM) and (SOL+KamLAND)
		\\
		$\delta$ & LBL & ATM
		\\
		$\Delta m_{21}^2$& KamLAND & SOL
		\\
		$|\Delta m_{31}^2|$& LBL+ATM+REAC& -
		\\
		MO & (LBL+REAC) and ATM & COSMO and $0\nu\beta\beta$
		\\\hline
	\end{tabular}
	\mycaption{The list of the experiments showing major contributions in the measurement of the six oscillation parameters. This table is taken from the ref.~\cite{deSalas:2020pgw}.}
	\label{tab:exps}
\end{table}

\begin{figure}[h!]
	\centering
	\includegraphics[scale = 0.95]{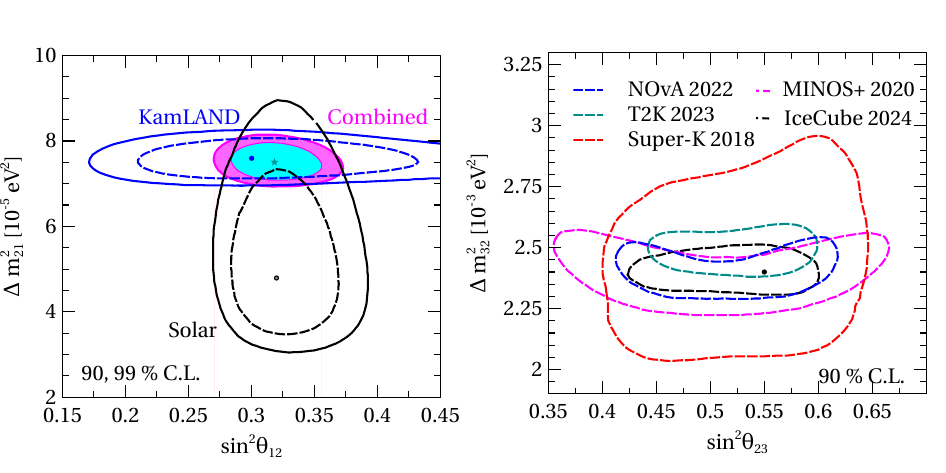}
	\mycaption{Allowed parameter region in the $\sin^2\tx$- $\sdm$~(left panel) and $\Delta m^2_{32}-\sin^2\theta_{23}$ plane~(right panel) at the 90\%~(dashed lines)  confidence level. Solid lines in left panel show the results at 99\% confidence level. In the left panel, the black line shows the allowed region from the analysis of the solar neutrino data~\cite{Cleveland:1998nv,Kaether:2010ag,SAGE:2009eeu,Borexino:2013zhu,Super-Kamiokande:2010tar}, while the blue line corresponds to the KamLAND data~\cite{KamLAND:2010fvi}. The magenta colored regions are used for the combined analysis of Solar and KamLAND. The left panel is redrawn using the information using the information from the ref.~\cite{deSalas:2020pgw}. In the right panel, each colored curve shows the allowed region for the atmospheric oscillation parameters using the data from atmospheric experiment NO$\nu$A~\cite{NOvA:2021nfi}, T2K~\cite{T2K:2023smv}, Super-Kamikande~\cite{pablo_fernandez_2021_5779075}, MINOS+~\cite{MINOS:2020llm}, IceCube Deepcore~\cite{IceCube:2024xjj}.
	Each curve is redrawn using the information from ref.~\cite{IceCube:2024xjj}.}
	\label{fig:sol_KamLand}
\end{figure}

In table~\ref{tab:exps}~\cite{deSalas:2020pgw}, we list the experiments from the different neutrino sectors that have contributed to the measurement of the oscillation parameters. The parameters $\tx$ and $\sdm$, also called the solar oscillation parameters, are mainly constrained by the solar neutrino experiments and KamLAND data. The left panel of fig.~\ref{fig:sol_KamLand} shows allowed regions in the ($\sin^2\tx$, $\sdm$) plane using the solar neutrino experiments and KamLAND individually, and using their combination. The figure shows that KamLAND has better precision in the $\sdm$ measurement while solar neutrino data is more sensitive to $\sin^2\theta_{12}$. Since the KamLAND experiment has a fixed average baseline~($\sim$180 km), it can precisely determine the value of $\sdm$ as compared to the solar neutrino experiments. At the same time, solar neutrino data are more sensitive to the oscillation amplitude, which allows them to measure $\sin^2\theta_{12}$ with good precision. In addition to the present data, the prospective data from the JUNO experiment will improve the accuracy on these two parameters further up to the sub-percent level~\cite{JUNO:2022mxj}. 

The data from atmospheric and LBL neutrino experiments have major contributions in measuring $\theta_{23}$ and $\Delta m^2_{31}$, known as the atmospheric neutrino oscillation parameters. Atmospheric and LBL neutrino experiments operate with multi-GeV neutrino energy and have a baseline that can be as large as several thousand kilometers, which are sensitive to $\ldm$ and $\tz$. In the right panel of fig.~\ref{fig:sol_KamLand}, we show the present constraints in the plane of these two parameters. The figure shows recent data from LBL experiment T2K~\cite{T2K:2023smv} and atmospheric neutrino experiment IceCube DeepCore~\cite{IceCube:2024xjj} have placed the most stringent constraint of these two parameters. Future data from IceCube-Upgrade and next-generation LBL experiments DUNE and T2HK will improve the precision of these two parameters.
 
\begin{figure}
	\includegraphics[scale = 0.95]{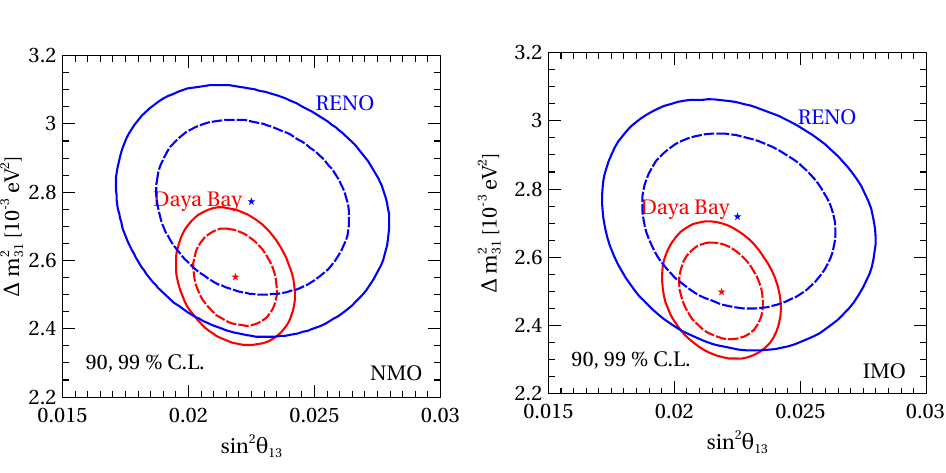}
	\mycaption{Allowed parameter region in the $\sin^2\ty$- $\ldm$ plane at the 90\%~(dashed lines) and 99\%~(solid lines) confidence level using data from the reactor neutrino experiments Daya Bay~\cite{DayaBay:2018yms}~(red line) and RENO~\cite{RENO:2020dxd}~(blue line). The left~(right) panel shows the allowed parameter space for normal~(inverted) neutrino mass ordering. This figure is redrawn using the information from ref.~\cite{deSalas:2020pgw}.}
	\label{fig:reactor_th13}
\end{figure}

$\theta_{13}$ was the last mixing angle whose non-zero value was discovered, which enabled us to connect the solar and atmospheric neutrino oscillation. Reactor neutrino experiments, with medium to small baseline and almost negligible matter effect, are most sensitive to $\theta_{13}$, also called reactor mixing angle. In fig.~\ref{fig:reactor_th13}~\cite{deSalas:2020pgw}, allowed regions in the plane of $\ty$ and $\ldm$ are shown from the analysis of Daya Bay~\cite{DayaBay:2018yms} and RENO~\cite{RENO:2020dxd} data with their complete run. The left and right panels are for the normal and inverted mass ordering. Since these experiments are not sensitive to neutrino mass ordering, allowed regions are almost identical for the two cases.

Among six oscillation parameters, the CP violation phase $\dcp$ is the least precisely measured parameter. LBL experiments play a major role in measuring this parameter, along with some contributions from the atmospheric neutrino experiments. With the present data, one of the two CP conserving values, $\dcp = 0$, is excluded at $3\sigma$ confidence level for both the mass ordering. In contrast, the other CP conserving value $\dcp=\pi$ is still allowed for normal mass ordering~{NMO}.
Although the LBL experiments T2K and NO$\nu$A have a leading contribution in the $\dcp$ measurement, there is a tension between their results~\cite{Himmel:2020,Dunne:2020,T2K:2021xwb,NOvA:2021nfi}. For the NMO case, T2K analysis prefers a value of $\dcp$ around $3\pi/2$. However, from the NO$\nu$A analysis, although there is no any particular preference for $\dcp$, but it disfavors the best-fit value of T2K for the same value of $\theta_{23}$. The possible source of this discrepancy can be statistical fluctuation or systematic error. Also, it can hint toward some new physics scenarios.
With the preliminary joint analysis of the two experiments, tension still persists.
The tension is expected to be resolved with more data from the two experiments in the near future. 

From the discussion above, it is clear that several neutrino oscillation experiments can measure one oscillation parameter. A global analysis of the oscillation data, which analyzes data from the experiments from the different neutrino sectors, may significantly improve the precision of oscillation parameters. There are several groups~\cite{NuFIT,Esteban:2020cvm,deSalas:2020pgw,Capozzi:2021fjo} in the neutrino physics community working in this direction. 
\begin{table}
	\centering
	\begin{tabular}{|c|ccc|}
	\hline\hline
	Osc. Parameters & best-fit$\pm 1\sigma$ & $3\sigma$ range &  $1\sigma$ precision~(\%) \\[1mm]
	\hline
	$\Delta m^2_{21} [10^{-5}$eV$^2$]  &  $7.42^{+0.21}_{-0.20}$  &  6.82 -- 8.04  &  2.7  \\[4mm]

	$|\Delta m^2_{3l}| [10^{-3}$eV$^2$] (NMO)  &  $2.51^{+0.027}_{-0.027}$  &  2.43 -- 2.593  &  1.08  \\[1mm]
	$|\Delta m^2_{3l}| [10^{-3}$eV$^2$] (IMO)  &  $-2.490^{+0.026}_{-0.028}$  &  -2.574 -- -2.410  &  1.08  \\[4mm]
	$\sin^2\theta_{12}$         &  $0.304^{+0.012}_{-0.012}$  &  0.269 -- 0.343  &  4.0 
	\\[4mm]
	$\sin^2\theta_{23} $       (NMO)  &  $0.450^{+0.019}_{-0.016}$  &  0.408 -- 0.603  &  7.2  \\[1mm]
	$\sin^2\theta_{23} $       (IMO)  &  $0.570^{+0.016}_{-0.022}$  &  0.410 -- 0.613  & 5.90   \\[4mm]
	%
	$\sin^2\theta_{13} $       (NMO)  &  $0.0225^{+0.00062}_{-0.00062}$  &  0.02060 -- 0.02435  &  2.77  \\[1mm]
	$\sin^2\theta_{13}$       (IMO)  &  $0.02241^{+0.00074}_{-0.00062}$  &  0.02055 -- 0.02457  &  2.99  \\[4mm]
	%
	$\delta(\circ)$                        (NMO)  &  $230^{+36}_{-25}$  &  144 -- 350  &  14.9  \\[1mm]
	$\delta(^\circ)$                        (IMO)  &  $278^{+22}_{-30}$  &  194 -- 345  &  9.05  \\[1mm]
	%
	\hline\hline
	\end{tabular}
\mycaption{Best-fit value of the oscillation parameters along with theri $3\sigma$ allowed range and $1\sigma$ precision using the global analysis of the oscillation data~\cite{Esteban:2020cvm,NuFIT}. $1\sigma$ precision is calculated using the procedure discussed in ref.~\cite{Capozzi:2021fjo}.
	For the NMO case, $\Delta m^2_{3l} = \Delta m^2_{31}$, and for the IMO case $\Delta m^2_{3l} = \Delta m^2_{32}$.}
\label{tab:global-fit-data}
\end{table}
In table~\ref{tab:global-fit-data}, we show the present best-fit values and allowed $3\sigma$ range of the six oscillation parameters from the global oscillation analysis done by the NuFit group~\cite{NuFIT,Esteban:2020cvm}. In the last column, we estimate the relative $1\sigma$ precision, which is calculated as one-sixth of $3\sigma$ uncertainty divided by the best-fit value of the oscillation parameter. As we can see, the parameters $\theta_{12}$, $\theta_{13}$, $\sdm$ and the absolute value of $\ldm$ have been measured quite precisely with $1\sigma$ relative precision less than 5\%. The atmospheric mixing angle $\theta_{23}$ has a comparatively large uncertainty, and as mentioned earlier, $\dcp$ is loosely bounded with the present data. 

Without the precise measurement of the uncertain parameters mentioned in the previous paragraph, neutrino oscillation in the three-flavor framework is incomplete. The task of the next-generation neutrino oscillation experiment would be to determine these parameters with utmost accuracy. Here, we discuss a few of the such major issues that exist in the three-neutrino paradigm:

\paragraph{Octant of the $\theta_{23}$:} 
The atmospheric mixing angle determines the mixing between the flavor states $\nu_\mu$, $\nu_\tau$, and mass states $\nu_2$, $\nu_3$. When $\theta_{23}=45^\circ$, mixing is maximal. Although the maximal mixing scenario is not completely ruled out, there are hints toward the non-maximal value of the $\theta_{23}$ from the global fit analyses~\cite{NuFIT,Esteban:2020cvm,deSalas:2020pgw,Capozzi:2021fjo}.
As shown in table~\ref{tab:global-fit-data},
$3\sigma$ range allowed value of $\theta_{23}$ varies between $\sim 0.4$ to $0.6$, \ie~ $\theta_{23}\in [40^\circ, 50^\circ]$, which allows both the octant of $\tz$, \ie~$\tz<45^\circ$~(upper octant) and $\tz>45^\circ$~(lower octant). Determining the octant of $\tz$ is important as it may have crucial implications in various neutrino mass models~\cite{Mohapatra:2006gs,Albright:2006cw,Albright:2010ap,King:2013eh}. However, measuring the octant of $\tz$ is difficult because the primary oscillation channel in the atmospheric neutrino experiments, $\mumu$, is proportional to $\sin^22\theta_{23}$~(see eq.~\ref{eq:disapp_approx}) in the leading term.
As a result, this channel is almost insensitive to deviation from maximal mixing. Nonetheless, data from LBL experiments which probe $\mue$~($\bar{\nu}_\mu\to\bar{\nu}_e$) oscillation~(see eq.~\ref{eq:app_prob_approx}), can break the octant degeneracy, as leading term in these channels is proportional to $\sin^2\theta_{23}$~\cite{Antusch:2004yx,Hagiwara:2006nn,Chatterjee:2013qus,Agarwalla:2013hma,Agarwalla:2021bzs}. Future atmospheric and LBL experiments will play the leading role in measuring the octant of $\theta_{23}$.

\paragraph{Value of $\dcp$:} As discussed earlier, the value of $\dcp$ is the most poorly measured parameter. With the present data, the $3\sigma$ range for $\dcp$ allows almost 57\%~(41\%) of all possible values of this parameter with NMO~(IMO). $\dcp$ is primarily measured by the combined analysis of $\mue$ and $\bar{\nu}_\mu\to\bar{\nu}_e$ in experiments, as appearance probability in the neutrino and antineutrino mode has a shift in the opposite direction. The LBL experiments that probe $\nu_e$ appearance both in the neutrino and antineutrino mode can measure the value of $\dcp$. However, the LBL experiments face a significant amount of matter effect, which brings in a  {\it fake} CP violation~\cite{Gonzalez-Garcia:2001snt}, making it difficult for the $\dcp$ measurement. Additionally, the large uncertainty in the measurement of $\tz$ and also the undetermined neutrino mass ordering bring in new degeneracies between those parameters in the analysis, which reduces the sensitivity of the LBL experiments towards determining the value of $\dcp$. With the increasing volume of the data from the ongoing T2K and NO$\nu$A, limits on $\dcp$ are reducing constantly. Also, the upcoming LBL experiment DUNE and T2HK will play a crucial role in the measurement of $\dcp$.

\paragraph{Neutrino mass ordering:} The absolute value of the two mass squared splittings $\sdm$ and $\ldm$ have been measured with considerably good accuracy. However, the information we have regarding these two parameters is not complete. Although the sign of $\sdm$ is known to be positive, \,  i.e., $m_2>m_1$ from the solar neutrino experiments, the sign of the other mass splitting $\ldm$ is not known yet. As a result, there are two possibilities for neutrino mass ordering, namely, normal mass ordering~(NMO), $m_1<m_2<m_3$ and inverted mass ordering~(IMO), $m_3<m_1<m_2$. Neutrino oscillation in a vacuum is insensitive to the neutrino mass ordering. However, the matter effect can potentially distinguish between the two mass orderings. Also, the interference effect from the two mass-squared differences in the oscillation experiments can reveal the true neutrino mass-ordering. The ongoing atmospheric neutrino experiments like IceCube Deepcore and KM3Net/ORCA exploiting their large earth matter effect playing a significant role in this direction~\cite{psf2023008007,KM3NeT:2021ozk}. Moreover, LBL experiments, which face a considerable matter effect, can be sensitive to mass ordering. At present, both T2K and NO$\nu$A slightly prefer normal mass ordering over the inverted one~\cite{Cao:2023hwq,T2K:2023smv,NOvA:2021nfi}. Despite the effort from these outstanding experiments, till now, there is no conclusive preference over a particular mass-ordering~\cite{Gariazzo:2022ahe}. The upcoming JUNO experiment will have excellent sensitivity to mass ordering. With its sub-percent level of energy resolution, JUNO is expected to determine the neutrino mass-ordering above 3$\sigma$ C.L. with six years of data taking~\cite{Heinz:2024dyt}. In addition, the upcoming LBL experiment will also have an outstanding potential to determine the neutrino mass ordering~\cite{Hyper-Kamiokande:2016srs,DUNE:2020jqi,Panda:2023rxa}.


\section{Summary}
\label{sec:summary_osc}

In this chapter, we discuss the basic formalism of the neutrino oscillation. We derive the expressions of the oscillation probabilities for the neutrinos propagating in vacuum, both in the two-flavor and three-flavor cases. We plot oscillation probabilities as a function of energy to show the role of various oscillation parameters in flavor transition. In the two-flavor scenario, the mixing angle determines the amplitude of the oscillation probability, whereas the mass-square difference influences the position of the oscillation maxima. In the three-flavor case, there are six oscillation parameters that influence the neutrino oscillation in various channels. The relevance of the oscillation parameters in the three-flavor scenario depends on orders of baseline and neutrino energies. 
We also elaborate on how the standard CC interaction of neutrinos with ambient matter can modify the neutrino oscillation and the concept of resonance due to the matter effect. We give a brief description of various neutrino experiments and their role in the field of neutrino physics. A global analysis of the data from past and ongoing neutrino oscillation experiments has measured the six oscillation parameters in a three-neutrino framework with good precision. We give a brief review of the present status of the oscillation parameters along with best-fit values from the global-fit analysis. Finally, we discuss the unsolved issues in the three neutrino frameworks, which are the primary aims of the next-generation neutrino oscillation experiments.

The current status of the oscillation parameters opens up opportunities to study various Beyond the Standard Model~(BSM) physics scenarios using the neutrino oscillation experiments. Though these new physics effects should not be dominant over neutrino oscillation, given the small uncertainties in oscillation parameters, they can appear at the sub-dominant level. In the next chapter, we are going to discuss such scenarios in detail.

\blankpage
\chapter{BSM studies in neutrino oscillation}
\label{C3} 
\section{Introduction}
The evidence for the massive neutrinos from the neutrino oscillation experiments is a groundbreaking signal for BSM physics. Although this fact was established more than two decades ago, the present scenario lacks a complete, well-established theoretical framework that can accommodate the tiny nature of the neutrino mass. As discussed in chapter~\ref{C1}, various BSM models have been proposed to describe neutrino mass. Neutrino oscillation can be a great tool to test these models. Many of these mass models involve extra neutrino states, mostly heavy right-handed neutrinos, that can have essential implications in the neutrino flavor transitions. Additionally, there are models in the BSM sector that induce new neutrino-matter interactions. This new interaction potential can modify the neutrino propagation through the matter and, consequently, the neutrino flavor transition. Also, violation of any fundamental laws of nature may leave some imprints in the neutrino oscillation that may have some observable effect in the experiments. In a way, neutrino oscillation phenomena offer excellent opportunities in probing BSM physics as well as testing fundamental physics.

As mentioned in the last chapter, the current status of the oscillation parameters allows enough room for BSM studies. In fact, it provides a perfect opportunity for BSM studies. We have a concrete three-flavor neutrino oscillation framework with few well-measured oscillation parameters and few others having a significant uncertainty associated with it. The possibility of having a subleading new physics effect is allowed by the uncertain oscillation parameters, which increases the potential for new physics studies. Moreover, there are few anomalies that exist in the neutrino data, signaling the possibility of the BSM effect in neutrino oscillation. There are efforts to explain the tension between T2K and NO$\nu$A data using BSM physics~\cite{Choubey:2018cfz,Chatterjee:2020kkm,deGouvea:2022kma,Konwar:2024nwc}. The well-known short-baseline~(SBL) anomaly, Gallium anomaly, and reactor antineutrino anomalies may also hint towards a BSM physics with the presence of eV-scale sterile neutrinos~\cite{Gariazzo:2015rra,Giunti:2019aiy,Diaz:2019fwt,Boser:2019rta}. In the SBL anomaly, various short-baseline experiments like LSND, MiniBooNE found an excess of $\bar{\nu}_e$ events at lower neutrino energy regions, which possibly due to $\bar{\nu}_\mu\to\bar{\nu}_e$ oscillation at shorter baseline than the expected from the standard three-neutrino oscillaton framework~\cite{LSND:2001aii,MiniBooNE:2008yuf,PhysRevD.64.112007,MiniBooNE:2020pnu,MiniBooNE:2013uba,MiniBooNE:2018esg}.
Similarly, Gallium neutrino experiments, like SAGE and Gallex, also observed a deficit in the $\nu_{e}$ events~\cite{Giunti:2022btk,SAGE:2009eeu}.
A number of reactor neutrino experiments observed a discrepancy in the measured neutrino flux from the predicted one~\cite{Mention:2011rk,Giunti:2006bj,Giunti:2010zu,Giunti:2012tn}. Although the recent reactor neutrino flux model refinement~\cite{Giunti:2021kab,Elliott:2023cvh} has almost resolved the reactor neutrino anomaly, the discrepancies in the SBL and Gallium experiments data are still an open problem~\cite{Barinov:2022wfh,MicroBooNE:2022sdp,Denton:2021czb,Arguelles:2021meu}. Moreover, recent results from the BEST~\cite{Gavrin_2017} experiments have favored the Gallium anomaly~\cite{Barinov:2022wfh}. 

The next-generation neutrino oscillation experiments, like,  DUNE, Hyper-K, JUNO, IceCube-Gen2, KM3NeT, 
will observe neutrino events with an unprecedented level of precision. Although the primary goal of these experiments would be to resolve the three major unknowns in the oscillation parameters, namely, $\tz$ octant, leptonic CP violation, and neutrino mass ordering, they will have unprecedented potential to explore various BSM scenarios.
There are proposed experiments like Fermilab SBN programme~\cite{MicroBooNE:2015bmn}, JSNS2~\cite{JSNS2:2013jdh} IsoDAR~\cite{Alonso:2022mup}, PROSPECT-II~\cite{PROSPECT:2021jey} which are dedicated to probing the above-mentioned neutrino anomalies.

New physics effects in the neutrino oscillation experiments can appear through three processes, namely, neutrino production, neutrino propagation, and detection. 
If the BSM physics modifies the interaction rate from the neutrino production, that will essentially modify the outgoing neutrino flux from the source. Similarly, the BSM physics effect can influence the neutrino cross-section and, hence, modify the observed event rate at the detector. Both the process will affect the outcome of the neutrino oscillation experiments. Also, neutrino detectors and accelerators can be great probes for BSM searches, such as heavy sterile neutrinos and neutrino dark matter interaction. The other avenue for the BSM study in neutrino oscillation is the new physics effects in neutrino propagation. There are various new physics scenarios that can affect the neutrino propagation. A BSM neutrino-matter interaction will modify the neutrino flavor conversion. Also, the new physics scenarios like neutrino decay, the presence of extra neutrino states, and violation of fundamental laws, like Lorentz and CPT violation, could potentially affect the neutrino propagation and neutrino oscillation probability. In this chapter, we discuss a few of such scenarios in detail, mainly focussing on how they would modify the neutrino oscillation.

This chapter is organized as follows. In section~\ref{sec:NSI}, we discuss in detail about neutrino non-standard interaction~(NSI). We show how charge-current NSI and neutral-current NSI in matter affect the neutrino oscillation. Neutrino oscillation phenomenology in the presence of eV-scale and keV-scale sterile neutrinos is discussed in section~\ref{sec:sterile_neutrinos}. The impact of non-unitarity of the standard three-flavor neutrino mixing matrix in the neutrino flavor transition is described in section~\ref{sec:NUNM_formalism}. In section~\ref{lorentz_and_CPT_violation}, we show how violation of CPT and Lorentz symmetry would have impact on the neutrino flavor conversion. Finally, in section~\ref{sec:neutrino_decay}, we discuss the phenomenology of neutrino decay.



\section{Neutrino non-standard interaction}
\label{sec:NSI}
As discussed in the previous section, SM framework allows interactions of neutrinos with ordinary matter particles {\it via} charge-current and neutral current interactions. The coupling and mass of the mediator of the interactions have been well-measured in various experiments. However, given the outstanding precision on the these parameters as well as the magnificent potential of the currently running and upcoming neutrino experiments in measuring neutrino oscillation parameters, it is now inevitable to search for the BSM interactions of neutrinos. At the theoretical frontier, various models~\cite{PhysRevD.40.1569, Davidson:1993qk,Gavela:2008ra,Dorsner:2016wpm,Babu:2019mfe} has been proposed that induces BSM interactions with the ordinary matter particles. One such interaction is non-standard interaction (NSI) of neutrinos, which was introduced by Wolfenstein in 1978~\cite{PhysRevD.17.2369}. Although the details of NSI involves various BSM models, in effective field theory framwork, NSI is represented as effective six-dimensional four-fermion operators, which can be of charge-current (CC) or neutral current type, as shown below
\begin{align}
&\mathcal{L}_{\mathrm{NC-NSI}} = -2\sqrt{2} G_{F} \sum_{\alpha, \beta, f, C} \varepsilon^{fC}_{\alpha\beta} ({\bar{\nu}_{\alpha}}\gamma^{\mu}P_{L}\nu_{\beta}) (\bar{f}\gamma_{\mu}P_Cf), 
\label{eq:Lnsi} \\
&\mathcal{L}_{\mathrm{CC-NSI}} = -2\sqrt{2} G_{F} \sum_{\alpha, \beta, f^{\prime}, f, C} \varepsilon^{ff^{\prime}C}_{\alpha\beta} ({\bar{\nu}_{\alpha}}\gamma^{\mu}P_{L}l_{\beta}) (\bar{f}^{\prime}\gamma_{\mu}P_Cf), 
\label{eq:Lnsi_CC}\,,
\end{align}
where $\nu_\alpha\,(\alpha= e,\mu,\tau)$ are the three active neutrino flavor, $l_{\alpha}$ is their charged counterpart, and $f$ denotes the chaged fermions in the ordinary matter. $P_C\,(C=L,R)$ are the chiral projections operators. In the above equations, the strength of new interactions is represented by the parameter $\varepsilon^{f,f',C}_{\alpha\beta}$ ($\varepsilon^{f,C}_{\alpha\beta}$) for CC-NSI (NC-NSI). The hermiticity of these interactions imposes the following conditions:
\begin{equation}
\varepsilon_{\alpha\beta}^{fC} \;=\; (\varepsilon_{\beta\alpha}^{fC})^* \;, \;\;\;\;\;\;
\varepsilon_{\alpha\beta}^{ff^{\prime}C} \;=\; (\varepsilon_{\beta\alpha}^{ff^{\prime}C})^* \;.
\end{equation}

\subsection{CC-NSI}

Eq.~\ref{eq:Lnsi_CC} shows the effective expression for CC-NSI. Neutrino produced at the sources or observed at the detector can affected by the CC-NSI,
\begin{align}
\ket{\nu^s_\alpha} &= \ket{\nu_\alpha} + \sum_{\beta=e,\mu,\tau} \varepsilon^s_{\alpha\beta} \ket{\nu_\beta} 
\label{eq:CC_NSI_source}\\
\bra{\nu^d_\beta} &= \bra{\nu^d_{\beta}} + \sum_{\alpha = e,\mu,\tau} \varepsilon_{\alpha\beta}\bra{\nu_{\alpha}}\,.
\end{align}
$\nu^s_\alpha$ and $\nu^d_{\alpha}$ are the neutrino flavor eigenstate at the source and the detector, respectively, and $\varepsilon^s_{\alpha\beta}\, (\varepsilon^d_{\alpha\beta})$ are the element of the NSI matrices at the source (detector). From eq.~\ref{eq:Lnsi_CC}, it is clear that CC-NSI parameters at the source~($\varepsilon^s_{\alpha}$) and the detector~($\varepsilon^d_{\alpha}$) which depends on the process of production or detection, are not same when $f\neq f'$. As a result, flavor eigenstates at the source and the detector are not orthonormal.

In the presence of CC-NSI, neutrino flavor transition probability is given by~\cite{Ohlsson:2008gx,Meloni:2009cg}
\begin{align}
P(\nu^s_\alpha\rightarrow\nu^d_\beta,L) = \left|\sum_{\gamma,\delta,i}(1+\varepsilon^d)_{\gamma\beta}(1+\varepsilon^s)_{\alpha\delta}U_{\delta i}U_{\gamma i}e^{i\frac{m^2_i L}{2E}}\right|\,.
\label{eq:Prob_cc_nsi}
\end{align} 

From the above equations, one can observe that in the presence of CC-NSI, there is non-zero neutrino flavor transition probability even when $L\rightarrow 0$, depending on the strength of the NSI parameters, that means neutrino can change its flavor at the source even before it starts oscillating during propagation. In literature, it is known as zero-distance effect~\cite{Giarnetti:2020bmf}. After expanding eq.~\ref{eq:Prob_cc_nsi} in the limit $L\rightarrow0$, we get~\cite{Ohlsson:2012kf}
\begin{align}
P(\nu^s_\alpha\rightarrow\nu^d_\beta,0) = \sum_{i,j} J^i_{\alpha\beta}J^{j\ast}_{\alpha\beta}\,,
\end{align}
where,
\begin{align}
J^i_{\alpha\beta} = U_{\alpha i}U^\ast_{\beta i}+\sum_{\gamma}\varepsilon^s_{\alpha \gamma} U_{\gamma i}U_{\beta i}+\sum_{\gamma}\varepsilon^d_{\gamma \beta} U_{\alpha i}U_{\gamma i}+\sum_{\gamma,\delta}\varepsilon^s_{\alpha \gamma}\varepsilon^d_{\delta \beta}U_{\gamma i}U_{\delta i}\,.
\end{align}
In the absence of production and detection NSI parameters, {\it i.e.} $\varepsilon^s_{\alpha\beta}=\varepsilon^d_{\alpha\beta}=0$, one can get usual expression of the standard oscillation probability in eq.~\ref{eq:Prob_cc_nsi}, and zero-distance effect vanishes.

\subsection{NC-NSI}

Eq.~\ref{eq:Lnsi} shows the general expression for the NC-NSI.
Similar to the standard matter potential, NC-NSI generates an effective potential modifying the neutrino propagation Hamiltonian, as shown below,
\begin{equation}
H_{\rm NSI}=\left[U\begin{pmatrix}
0&0&0\\
0&\frac{\Delta m^2_{21}}{2E}&0\\
0&0&\frac{\Delta m^2_{31}}{2E}\\
\end{pmatrix}U^{\dagger}+V_{\text{CC}}\begin{pmatrix}
1+\varepsilon_{ee}&\varepsilon_{e\mu}&\varepsilon_{e\tau}\\
\varepsilon^{\ast}_{e\mu}&\varepsilon_{\mu\mu}&\varepsilon_{\mu\tau}\\
\varepsilon^{\ast}_{e\tau}&\varepsilon^{\ast}_{\mu\tau}&\varepsilon_{\tau\tau}\\
\end{pmatrix}\right] \;,
\label{eq:H_nsi}
\end{equation}
where $V_{\text{CC}}$ is the standard charge-current potential in matter and the effective NSI parameters $\varepsilon_{\alpha\beta}$ are parameterized as 
\begin{equation}
\varepsilon_{\alpha\beta} = \sum_{f}\varepsilon^{f}_{\alpha\beta}\frac{N_f}{N_e}\,,
\end{equation}
where, $\varepsilon_{\alpha\beta}^{f}= \varepsilon_{\alpha\beta}^{f,L}+\varepsilon_{\alpha\beta}^{f,R}$, and $N_{f}\, (f=e,u,d)$ are the number density of fermion $f$ in matter. Note that standard CC interaction is included in eq.~\ref{eq:H_nsi} in the (1,1) element of the interaction matrix (second term). 

NC-NSI will necessarily
impact the neutrino evolution inside the matter, consequently, the neutrino flavor transition probability. Due to the addition of nine new parameters (including the phases of the off-diagonal terms) in the neutrino Hamiltonian, estimating neutrino oscillation probability analytically is very complex~\cite{Kopp:2007ne}. One way to calculate the oscillation probability in the presence of NC-NSI parameters is to diagonalize the $H_f$ in eq.~\ref{eq:H_nsi}, which is hermitian in nature
$\tilde{U}H_f \tilde{U}^\dagger=\text{Diag.}\left(\frac{\tilde{m}_1^2}{2E},\frac{\tilde{m}_2^2}{2E},\frac{\tilde{m}_3^2}{2E}\right)$. Neutrino flavor transition probablity can be calculated in terms of the digonalizing matrix $\tilde{U}$ and the eigenvalues $\frac{\tilde{m}_i^2}{2E}$,
\begin{align}
P(\nu_\alpha\rightarrow\nu_\beta,L) = \left|\sum_{i}\tilde{U}_{\alpha i}\tilde{U}^\ast_{\beta i} e^{-i\frac{\tilde{m}_i^2 L}{2E}}\right|^2\,.
\end{align}

Note that the expression of the oscillation probabilities in the presence of NC-NSI as shown in the above equation, is similar to the expression of the oscillation probability in vacuum (see eq.~\ref{eq:osc_prob2}), except elements of the mixing matrix in vacuum are now replaced by the element of the modified mixing matrix ($\tilde{U}$) in matter with NC-NSI, and mass squares terms are replaced by the eigenvalues of $H_{f}$, which we call modified mass-squares.
A detailed procedure to diagonalize the Hamiltonian analytically in the presence of NC-NSI parameters has been discussed in chapter~\ref{C4}.

\section{eV-scale sterile neutrinos}
\label{sec:sterile_neutrinos}

The existence of three active neutrinos in Nature has been well-established from LEP and several other experiments. Also, the neutrino mass and mixing parameters in the three-neutrino framework are being measured with excellent precision using the global oscillation data\cite{Esteban:2020cvm,deSalas:2020pgw}. However, as discussed previously, few anomalies in the experimental data hints toward the existence of more than three neutrinos. 
The data from short-baseline~(SBL) experiments like LSND~\cite{LSND:2001aii}, MiniBooNE~\cite{MiniBooNE:2018esg}, reactor neutrino experiments~\cite{Mention:2011rk}, and gallium experiments~\cite{GALLEX:1998kcz,Abdurashitov:2005tb} can not be explained in the standard three neutrino framework, and hint toward the existence of extra neutrino states~\cite{Abazajian:2012ys,Palazzo:2013me,Capozzi:2016vac,Giunti:2019aiy}. These experiments predict the presence of an extra neutrino state of eV-scale mass, which successfully solve these anomalies. Also, X-ray astrophysical observations from XMM-Newton and Chandra space telescope~\cite{Boyarsky:2014jta,Bulbul:2014sua} suggests presence of KeV-scale sterile neutrino~\cite{Abazajian:2014gza,Ng:2015gfa,Schneider:2016uqi}, which successfully explains unidentified emission line in the X-ray spectra of the galaxy clusters. Other neutrino experiments, such as Daya Bay, MINOS/MINOS+, T2K, SuperKamiokande, and IceCube, have explored the possibility of the presence of sterile neutrino in Nature and have placed tight constraints on the relevant parameters. However, a definite ``smoking gun" signature is yet to be discovered.

Theoretically, the number of sterile neutrinos can be arbitrary, with mass varying from sub-eV to TeV range. The ``sterile" term comes from the idea that they do not interact with SM particles with any fundamental interaction except for gravitation. However, it takes part in mixing with the active neutrinos, which makes it possible to probe sterile neutrinos in the neutrino oscillation experiments. In the literature, neutrino oscillation in the presence of one eV-scale sterile neutrinos has been studied widely in the literature~\cite{Klop:2014ima,Berryman:2015nua,Gandhi:2015xza,Agarwalla:2016mrc,Choubey:2017cba,Agarwalla:2018nlx,Miranda:2020syh,Fiza:2021gvq}, which we call the 3+1 model in neutrino oscillation.

As discussed earlier, anomalous measurements from LSND, MiniBooNE, and reactor neutrino experiments hint toward the presence of an eV-scale sterile neutrino state. In such a scenario, the neutrino mixing matrix will be a $4\times 4$ matrix as follows
\begin{equation}
\label{eq:U_sterile}
U =   \tilde R_{34}  R_{24} \tilde R_{14} R_{23} \tilde R_{13} R_{12}\,, 
\end{equation} 

where $R_{ij}$ ($\tilde R_{ij}$) represent a real (complex) $4\times4$ rotation matrix
with mixing angle $\theta_{ij}$. For example, for the mixing angle $\theta_{14}$, matrix $R_{14}$ and $\tilde{R}_{14}$ are defined as 

\begin{eqnarray}
\label{eq:R_ij_2dim}
R^{4\times4}_{14} =
\begin{pmatrix}
c_{14} & 0 &0 & s_{ij}  \\
0 & 1 & 0 & 0\\
0 & 0 & 1 & 0\\
- s_{ij}  & 0 & 0 & c_{ij}
\end{pmatrix} ,
\,\,\,\,\,\,\,   
\tilde R^{4\times4}_{14} =
\begin{pmatrix}
c_{ij} &  0 & 0 &\tilde s_{ij}\\
0 & 1 & 0 & 0\\
0 & 0 & 1 & 0\\
- \tilde s_{ij}^*  & 0 & 0 & c_{ij}
\end{pmatrix}
\,,    
\end{eqnarray}
where $c_{ij} = \cos\theta_{ij}$, $s_{ij}=\sin\theta_{ij}$, and $\tilde{s}_{ij}=\sin\theta_{ij}e^{-i\delta_{ij}}$. Apart from the four standard mixing parameters, four additional parameters are introduced due to sterile neutrino, three mixing angles $\theta_{14}$, $\theta_{24}$, $\theta_{34}$ and one complex phase after rephasing. One can get back the standard unitary matrix in the three-neutrino scenario by setting new mixing angles $\theta_{14}$, $\theta_{24}$, and $\theta_{34}$ to zero. Under small mixing angle assumption of $\theta_{14}$, $\theta_{24}$, and $\theta_{13}$, one gets a simplified form of new mixing matrix elements, such as $|U_{e3}|^2=s^2_{13}$, $|U_{e4}|^2=s^2_{14}$,
$|U_{\mu 4}|^2=s^2_{24}$,
$|U_{\tau4}|^2=s^2_{34}$.

In the short-baseline limit, appearance probabilities are approximated as 
\begin{align}
P(\nu_\alpha\rightarrow\nu_\beta) \approx \sin^22\theta_{\alpha\beta }\sin^2\frac{\Delta m^2_{41} L}{4E}\,,
\end{align}
where $\sin^22\theta_{\alpha\beta} = |U_{\alpha 4}|^2|U_{\beta 4}|^2\approx s^2_{24}s^2_{14}$. Similarly, for the $\nu_\mu\rightarrow\nu_\mu$ disappearance channel, relevant mixing matrix elements are $U_{e4}$ and $U_{\mu 4}$.

\begin{align}
P(\nu_\alpha\rightarrow\nu_\alpha) \approx 1-\sin^22\theta_{\alpha\alpha}\sin^2\frac{\Delta m^2_{41} L}{4E}\,,
\end{align}
where $\sin^22\theta_{\alpha\alpha}=4|U_{
	\alpha 4}|^2(1-|U_{\alpha 4}|^2)$. So we observe that $\bar{\nu}_e$ disappearance in the reactor neutrino experiments majorly depends on only the element $|U_{e4}|$.

\section{Non-unitary neutrino mixing}
\label{sec:NUNM_formalism}
Non-unitary neutrino mixing (NUNM) is another new physics scenario that naturally emerges from the presence of heavy sterile neutrino states. For instance, the smallness of neutrino mass is successfully explained by the seesaw models, effectively introducing right-handed heavy neutrinos. In the low-scale type-I seesaw models, the mass of the heavy neutrino can be
as large as of the order TeV, well above the electroweak scale. Such a large mass of the heavy right-handed neutrinos prohibits them from taking part in neutrino oscillation.
Despite the decoupling 
from active neutrino oscillations, imprint of the heavy neutrinos can be realized through investigating the non-unitarity of the active neutrino mixing. 
In the literature, non-unitary of the lepton mixing matrix has been extensively studied~\cite{Escrihuela:2015wra,Miranda:2016wdr,Ge:2016xya,Blennow:2016jkn,Escrihuela:2016ube,Fong:2017gke,DeGouvea:2019kea,Miranda:2019ynh,Martinez-Soler:2019noy,Ellis:2020hus,Hu:2020oba,Forero:2021azc,Coloma:2021uhq,Chatterjee:2021xyu}, exploring its impact on the measurement of the standard oscillation parameters, as well as potential of the neutrino experiments to constrain the relevant NUNM parameters.

In presence of the extra neutrino states ($n>3$), it is natural to assume that $n\times n$ mixing matrix $U$ preserves unitary property i.e. $U^{n\dagger} U^{n} = U^{n} U^{n\dagger}=I$. One can parameterize $U^n$ as a product of the complex rotation matrices $R_{ij} (\theta_{ij},\delta_{ij})$ following Okubo's notation~\cite{Okubo:1962zzc}
\begin{align}
\widetilde{U}^{n\times n} & = R_{n-1\,n}\cdot R_{n-2\,n}\ldots R_{3\,n}\cdot R_{2\,n}\cdot R_{1\,n}\ldots R_{3\,n-1}\cdot R_{2\,n-1}\cdot R_{1\,n-1}\ldots R_{23}\cdot R_{13} \cdot R_{12}\,.
\label{eq:Un_1}
\end{align}

Furthermore, the $n\times n$ matrix $U^n$ can be decomposed into four submatrices based on their mixing patterns:
\begin{align}
U^n & = \left(
\begin{array}{cc}
N & S \\
V & T
\end{array}
\right) \equiv \left(
\begin{array}{c|c}
N^{3 \times 3} & S^{3 \times (n-3)} \\
\hline
V^{(n-3) \times 3} & T^{(n-3) \times (n-3)}
\end{array}
\right)\,.
\label{eq:Un_2}
\end{align}
Here, $N$ is the $3\times 3$ active neutrino mixing matrix, $S$ and $V$ depict the mixing between active and heavy neutrinos, and $T$ characterizes the mixing among the heavy neutrinos. The unitarity of $U^n$ implies that the submatrix $N$ is not unitarity anymore. The non-unitary property of $N$ can be probed in neutrino experiments operating at the multi-GeV energy scale. To parameterize $N$, one can express $U^n$ as the product of two matrices, $U^n=U^{N-3}\times U^{SM}$, where $U^{SM}$ corresponds to the last three matrices in eq.~\ref{eq:Un_1} depicting standard neutrino mixing, namely $R_{23}\cdot R_{13} \cdot R_{12}$, and $U^{NU}$ encompasses the remaining matrices. Since the $R_{ij},R_{kl}$ commutes when $i\neq k,l$ and $j\neq k,l$, $U_{NU}$ can be written as,
\begin{align}
\label{eq:U_nu}
U^{N-3} =& \underbrace{R_{n-1\,n}R_{n-2\,n}...R_{4\,n}R_{n-2\,n-1}...R_{4\,n-1}...R_{4\,5}}_{I}\times\\\nonumber
& \underbrace{R_{3\,n}R_{2\,n}R_{1\,n}R_{3\,n-1}R_{2\,n-1}R_{1\,n-1}...R_{3\,4}R_{2\,4}R_{1\,4}}_{II}\,.
\end{align}
In the above equation, we divide $U^{NU}$ into two parts, one containing rotation matrices in $(i,j;\,\,i\,\,\text{and}\,\,j >4)$ plane, and other contains same in the plane $(i,j;\,\,i\,\,\text{or}\,\, j<=3)$, respectively. It is evident that the first part~($I$) has no impact on the active neutrino mixing matrix $N$. Consequently, $N$ is influenced solely by the second part~($II$) of eq.~\ref{eq:U_nu}, which exhibits a lower triangular structure in the top $3\times 3$ submatrix. For instance, in the case of $n=4$, the structure of the second part is given by:
\begin{align}
R_{34}\cdot R_{24} \cdot R_{14} = 
\left(\begin{array}{ccc|c}	
c_{14}                            & 0                        & 0              & \tilde{s}_{14}                         \\
-\tilde{s}_{14}\tilde{s}_{24}              & c_{24}              & 0              & c_{14}\tilde{s}_{24}             \\
-c_{24}s_{14}\tilde{s}_{34} & -\tilde{s}_{24}\tilde{s}_{34} & c_{34}   & c_{14}c_{24}\tilde{s}_{34} \\ \hline
-c_{24}\tilde{s}_{14}c_{34} & -\tilde{s}_{24}\tilde{s}_{34} & -\tilde{s}_{34} & c_{14}c_{24}c_{34}
\end{array}
\right)\,,
\label{eq:U_4}
\end{align}
where $c_{ij}=\cos_{ij}$ and $\tilde{s}_{ij}= \sin\theta_{ij}e^{-i\delta_{ij}}$. One can parameterize the non-unitary active neutrino mixing matrix as 
\begin{align}
N=\eta\times U\,,
\label{eq:Un_3}
\end{align}
where, $U$ is the standard unitary $3\times 3$ matrix, and $\eta$ is a lower triangular $3\times 3$ matrix that introduces the non-unitarity. We parameterize $\eta$ as
\begin{align}
\eta=(I+\alpha)\times U\,,\hspace{1cm} \alpha = \begin{pmatrix}
\alpha_{11}&0&0\\
\alpha_{21}&\alpha_{22}&0\\
\alpha_{31}&\alpha_{32}&\alpha_{33}
\end{pmatrix}\,,
\end{align}
where $I$ is the identity matrix, $\alpha_{ij}$ are denoted as Non-Unitary Neutrino Mixing (NUNM) parameters. From eq.~\ref{eq:U_4}, it is clear that diagonal parameters of $\alpha$ ($\alpha_{ii}\,,i=1,2,3$) are real, whereas off-diagonal parameters are complex in nature. Consequently, the NUNM matrix introduces nine additional parameters, supplementing the four standard mixing parameters. 

In the NUNM scenario, the neutrino propagation Hamiltonian is expressed as
\begin{align}
H_{\rm NU} & = \frac{1}{2E}
\left(\begin{array}{ccc}
0 & 0               & 0              \\
0 & \Delta m^2_{21} & 0              \\
0 & 0               & \Delta m^2_{31} 
\end{array}\right)\;
+\;\; N^\dagger\cdot 
\left(\begin{array}{cccc}
V_{CC} + V_{NC} & 0       & 0 \\
0           &  V_{NC} & 0 \\
0           & 0       & V_{NC}
\end{array}\right)\cdot N\,,
\label{eq:H_nunm}
\end{align}

where $V_{CC}$ and $V_{NC}$ represent the standard charge-current and neutral-current interaction potentials, respectively. Consequently, the neutrino flavor transition probability in the presence of NUNM is given by,
\begin{align}
P(\nu_\alpha\rightarrow\nu_\beta) = \left|(Ne^{-i\,H\,L}N^\dagger)_{\alpha\beta}\right|^2\,.
\end{align}
The neutrino flavor transition probability in the NUNM scenario now consist of nine additional parameters, introducing potential new degeneracies with the standard oscillation parameters.
In chapter~\ref{C5}, we derive the expressions of the neutrino oscillation probabilities in relevant channels within the context of the long-baseline experiment using perturbative expansion. From the above expression, one can notice that in the NUNM case, there is a non-zero probability of neutrino flavor transition even when the baseline $L$ is very small or approximately zero, commonly referred to as the {\it zero-distance effect}. 

\begin{table}
	\centering
	\begin{tabular}{|c|c|c|c|c|c|c|}
		\hline\hline
		Parameter&$\alpha_{11}$&$\alpha_{22}$&$\alpha_{33}$&$|\alpha_{21}|$&$|\alpha_{31}|$&$|\alpha_{32}|$ \\
		\hline
		$90\%$ C.L. bound & $< 0.031$&$<0.005$& $< 0.110$ &$< 0.013$&$<0.033$ &$<0.009$\\
		\hline\hline
	\end{tabular}
	\mycaption{$90\%$ C.L. limits on NUNM parameters from analysis of the oscillation data from short-baseline experiments, NOMAD and NuTeV, as well as the long-baseline experiments from MINOS/MINOS+, T2K, and NO$\nu$A. Limits are taken from the ref.~\cite{Forero:2021azc}.}
	\label{table:bound_nunm}
\end{table}

It is clear from the above discussion that neutrino oscillation experiments have the potential to probe non-unitarity of the neutrino mixing matrix.
In table~\ref{table:bound_nunm}, we present the current constraints on Non-Unitary Neutrino Mixing (NUNM) parameters, derived in ref.~\cite{Forero:2021azc} using the latest data of both short-baseline and long-baseline experiments, at the $90\%$ confidence level. 
It is noteworthy to mention that the NUNM parameters have also been constrained by the analysis of current flavor and electroweak precision observables, as discussed in ref.~\cite{Fernandez-Martinez:2016lgt,Blennow:2016jkn,Blennow:2023mqx}. Remarkably, constraints from lepton flavor-violating (LFV) processes on the NUNM parameters appear considerably stronger than those arising from neutrino oscillation experiments~\cite{Blennow:2016jkn,Blennow:2023mqx}.

\section{CPT and Lorentz violation}
\label{lorentz_and_CPT_violation}

CPT and Lorentz symmetry are the fundamental symmetries of the Nature. Violation of these symmetries will have a profound consequence in the modern physics. CPT conservation in nature have three key elements: Lorentz invariance, locality, and hermiticity of the Hamiltonian. Therefore, any deviation from these symmetries can potentially result in violation of CPT symmetry. Over the last few decades, these two symmetries has been put under scrutiny in several sectors of particle physics, with neutrinos being the prominent one.

CPT conservation implies, particle and antiparticle would necessarily have same mass, and equal lifetime in case of unstable particles. So, CPT violation can be tested by measuring the mass differences between the particles and antiparticles. One such measure ment on mass of of neutral kaons put a stringent bound of the CPT invariance~\cite{PhysRevLett.74.4376}:
\begin{align}
\frac{|m(K^0)-m(\bar{K}^0)|}{m_K}<0.6\times 10^{-8}\,.
\end{align}
However, one should intrepet this limit carefully. Since CPT violation do not have a complete theory, considering kaon mass as the scale in the denominator would be arbitrary. Also, neutral kaon is not an elementary particle, so many features in QCD involves in this test, which would imply testing QCD properties than a fundamental symmetry of fermions. However, neutrino being an elementary particle, holds a potential to probe CPT violation. But, as mentioned before, in the absence of a complete theory on CPT violation, interpretation of this result would not be meaningfull. Neutrino oscillation experiments can probe CPT violation from the measurement of mass-squared differences and mixing angles. Comparing the mass-splitting and mixing angles with neutrino and antineutrino data separately can give a measurement of CPT violation. In the literature, various works has been done along this direction~\cite{Barenboim:2009ts,Barenboim:2017ewj,Kaur:2020ggv,Ternes_2021,CAPOLUPO2019298}. In ref.~\cite{Barenboim:2017ewj}, authors provide latest limit on the CPT violation in the neutrino sector through mass and mixing angles:
\begin{align}
|\Delta m^2_{21}-\Delta \bar{m}^2_{21}|&< 4.7\times 10^{-5} \text{eV}^2\\
|\Delta m^2_{31}-\Delta \bar{m}^2_{31}|&< 3.7\times 10^{-4} \text{eV}^2\\
|\sin^2\theta_{12}-\sin^2\bar{\theta}_{12}|&<0.12\\
|\sin^2\theta_{13}-\sin^2\bar{\theta}_{13}|&<0.03\\
|\sin^2\theta_{23}-\sin^2\bar{\theta}_{23}|&<0.12\,.
\end{align}
Note that these bounds are derived with the assumption of normal mass ordering, in case of inverted mass-ordering, bounds may change.

Lorentz invariance is the other fundamnetal symmetry which has been tested by pioneering experiments. The standard model of particle physics as well as the general theory of relativity preserves Lorentz symmetry. However, there exist models in string theory and loop quantum gravity that allows Lorentz violation 
at the Planck scale. At the low energy, such Lorentz invariance violation~(LIV) is suppressed by the Planck scale mass ($1/M_P$). In the Standard Model Extension~(SME) framework, LIV can be realized at the low energy neutrino experiments introducing new interactons in the minimal SME framework involving neutrinos. In this framework, relevant terms in the Lagrangian dansity showing Gauge invariant LIV interaction of left-handed neutrinos are
\begin{align}
\mathcal{L}_{\rm LIV} & = -\frac{1}{2}\left[a^{\mu}_{\alpha\beta}\,\overline{\psi}_\alpha\,\gamma_{\mu}\,P_L\,\psi_\beta - i c^{\mu\nu}_{\alpha\beta}\,\overline{\psi}_\alpha\,\gamma_{\mu}\,\partial_\nu P_L\,\psi_\beta \right] + h.c.\,, 
\label{eq:LIV-1}
\end{align}
where $P_L$ is the projection operator. The coefficiants $a^\mu_{\alpha\beta}$ and $c^{\mu\nu}_{\alpha\beta}$ are the CPT-violating and CPT-conserving parameters, respectively. The effective Hamiltonian for the neutrinos propagating in vacuum can be written  as
\begin{align}
H_{ij} = E\delta_{ij} +\frac{\delta m^2_{ij}}{2E}+\frac{1}{E}(a^\mu_Lp_\mu+c^{\mu\nu}_Lp_\mu p_\nu)_{ij}\,,
\end{align}
where $p_\mu$ and $E$ are the four momenta and the energy of the neutrino states. So, neutrino propagating in vacumm will have impact due to LIV in Nature. Including the neutrino interaction with matter particles, effective Hamiltonian for the neutrino propagating inside matter in the flavor basis can be written as
\begin{align}
H_{\rm LIV} = \; \frac{1}{2E} 
U\left(\begin{array}{ccc}
0 & 0 & 0                 \\
0 & \Delta m^{2}_{21} & 0 \\
0 & 0 & \Delta m^{2}_{31} \\
\end{array}\right)U^{\dagger}
+&\left(\begin{array}{ccc}
a_{ee} & a_{e\mu} & a_{e\tau} \\
a^*_{e\mu} & a_{\mu\mu} & a_{\mu\tau} \\
a^*_{e\tau} & a^*_{\mu\tau} & a_{\tau\tau}
\end{array} \right) \nonumber \\
-&\frac{4}{3} E
\left(
\begin{array}{ccc}
c_{ee} & c_{e\mu} & c_{e\tau} \\
c^*_{e\mu} & c_{\mu\mu} & c_{\mu\tau} \\
c^*_{e\tau} & c^*_{\mu\tau} & c_{\tau\tau}
\end{array}
\right) 
+ 
\left(\begin{array}{ccc}
V_{\text{CC}} & 0 & 0 \\ 
0 & 0 & 0 \\ 
0 & 0 & 0
\end{array}\right).
\label{eq:LIV-3}
\end{align}
The second term in the above equation correspond to the CPT-violating LIV parameters, whereas the third term contains CPT-conserving LIV parameters. The last term is the standard charge-current interaction term. Note that the structure of the Hamiltonian in the presence of the LIV parameters have similar structure as that of the NC-NSI parameters in eq.~\ref{eq:H_nsi}, except new physics terms~(second and third terms) here are not associated with the charge-current interaction potential. So, at the probability level, presence of LIV can be in principle map with the oscillation in the presence of NC-NSI parameters. A comparative study between the LIV and NC-NSI parameters has been done in ref.~\cite{Sahoo:2022nbu,Majhi:2022fed}.

At present, neutrino experiments like IceCube, Super-K, and T2K has put stringent bound on the LIV different LIV parameters. In Table~\ref{tab:existing_bounds_liv}, existing bound on different LIV parameters are listed. Note that IceCube has significantly stringent constraint of CPT-conserving LIV parameters, than Super-K due to its larger neutrino energy.

\begin{table}[h!]
	\centering
	\begin{center}
		\begin{adjustbox}{width=1\textwidth}
			\begin{tabular}{|c| c| c| c|}
				\hline \hline
				\multicolumn{4}{|c|}{CPT-violating LIV parameters} \\ \hline
				Experiments & $a_{e\mu} ~[  10^{-23} \,\rm GeV ]$ &  $a_{e\tau} ~[ 10^{-23} \,\rm  GeV ]$ & $a_{\mu\tau} ~[ 10^{-23} \,\rm GeV ]$ \\ \hline
				\multirow{2}{*}{95\% C.L. limits} & $\mathrm{Re}(a_{e\mu}) < 1.8$~\cite{Super-Kamiokande:2014exs}  & $\mathrm{Re}(a_{e\tau}) < 4.1$~\cite{Super-Kamiokande:2014exs} & $\mathrm{Re}(a_{\mu\tau}) < 0.65$~\cite{Super-Kamiokande:2014exs} \\
				
				& $\mathrm{Im}(a_{e\mu}) < 1.8$~\cite{Super-Kamiokande:2014exs}  & $\mathrm{Im}(a_{e\tau}) < 2.8$~\cite{Super-Kamiokande:2014exs} & $\mathrm{Im}(a_{\mu\tau}) < 0.51$~\cite{Super-Kamiokande:2014exs} \\
				\hline
				\multirow{2}{*}{99\% C.L. limits} &\multirow{2}{*}{--} & \multirow{2}{*}{--} & $|\mathrm{Re}(a_{\mu\tau})| < 0.29$\cite{IceCube:2017qyp} \\ 
				&\multirow{2}{*}{--} & & $|\mathrm{Im}(a_{\mu\tau})| < 0.29$~\cite{IceCube:2017qyp} \\ 
				\hline 
				\hline
				\multicolumn{4}{|c|}{CPT-conserving LIV parameters} \\ \hline 
				
				Experiments & $c_{e\mu}\,[10^{-27}]$ &  $c_{e\tau}\,[10^{-27}]$ & $c_{\mu\tau}\,[10^{-27}]$ \\ \hline
				
				\multirow{2}{*}{Super-K (95\% C.L.)} & $\mathrm{Re}(c_{e\mu}) < 8.0$~\cite{Super-Kamiokande:2014exs}  & $\mathrm{Re}(c_{e\tau}) < 930$~\cite{Super-Kamiokande:2014exs} & $\mathrm{Re}(c_{\mu\tau}) < 4.4$~\cite{Super-Kamiokande:2014exs} \\
				
				& $\mathrm{Im}(c_{e\mu}) < 8.0$~\cite{Super-Kamiokande:2014exs}  & $\mathrm{Im}(c_{e\tau}) < 1000$~\cite{Super-Kamiokande:2014exs} & $\mathrm{Im}(c_{\mu\tau}) < 4.2$~\cite{Super-Kamiokande:2014exs} \\
				
				\hline
				
				\multirow{2}{*}{IceCube (99\% C.L.)} &\multirow{2}{*}{--} & \multirow{2}{*}{--} & $|\mathrm{Re}(c_{\mu\tau})| < 0.39$~\cite{IceCube:2017qyp} \\ 
				&\multirow{2}{*}{--} & & $|\mathrm{Im}(c_{\mu\tau})| < 0.39$~\cite{IceCube:2017qyp}
				\\\hline\hline
				
			\end{tabular}
		\end{adjustbox}
	\end{center}
	\mycaption{Existing constraints on the off-diagonal CPT-violating and CPT-conserving LIV parameters from Super-K~\cite{Super-Kamiokande:2014exs} and IceCube~\cite{IceCube:2017qyp}. This table is taken taken from ref.~\cite{Agarwalla:2023wft}.}
	\label{tab:existing_bounds_liv}
\end{table}

\section{Neutrino decay}
\label{sec:neutrino_decay}

Descovery of non-zero and non-degenerate mass of neutrinos has point toward finite lifetime of neutrinos. However, within the SM framework, weak-interaction-mediated lifetime of neutrino are extremely high, exceeding the lifetime of the universe by several orders. There is a possibility of faster decay, in the presence of new interaction mediated by light particles~\cite{Acker:1992eh,PhysRevD.25.774,NUSSINOV1987171,Kim:1990km,Biller:1998nc}. The present limits on the decay parameters $\tau_i/m_i\,\,(i=1,2,3)$ of various neutrino mass eigenstates are $\tau_3/m_3<2.9\times 10^{-10}\,\,\text{s/eV}$ from atmospheric and long-baseline data~\cite{Gonzalez-Garcia:2008mgl}, $\tau_2/ m_2< 7.2\times 10^{-4}\,\text{s/eV}$ from solar neutrino data~\cite{Berryman:2014qha,Picoreti:2015ika}. Data from the supernova 1987A have also given stringent limits on neutrino lifetime~\cite{FRIEMAN1988115,PhysRevLett.58.1490,PhysRevLett.58.1494}. These limits are mostly applied to $\nu_2$ and $\nu_1$ as only $\bar{\nu}_e$ was observed.

There are various models in literature that allows neutrino decay. Most prominent one is Majoron model, which allows a coupling between  neutrinos and a  massless scalar called Majoron $(\phi)$~\cite{CHIKASHIGE1981265,GELMINI1981411,PhysRevD.25.774,PhysRevD.45.R1}, as shown below:
\begin{align}
\mathcal{L}_{\text{int}} = \sum_{i\neq j} (g_s)_{ij}\bar{\nu}_j\nu_i \phi+i (g_p)_{ij} \bar{\nu}_j\gamma_5\nu_i\phi + h.c.
\label{eq:L_dec}
\end{align}
The coupling can be scaler ($g_s$) type or pseudo-scalar ($g_p$) type.
Above terms introduce neutrino decay {\it via} $\nu_i\rightarrow\nu_j+\phi$. From the above equation, it is clear that both helicity-conserving ($\nu_i\rightarrow\nu_j$ or $\bar{\nu}_i\rightarrow\bar{\nu}_j$) and helicity violating $(\nu_i\rightarrow\bar{\nu}_j)$ decay may occur. In case of Dirac nuetrinos, decay products can be active (helicity conserving case) as well as a sterile neutrinos (helicty-violating case). Active neutrinos decaying to a sterile neutrinos is called {\it invisible decay}, and if the decay product contains another active neutrinos of smaller mass, it is known as {\it visible decay}. The expression for the partial decay width from $\nu_i$ to $\nu_j$ state can be written as~\cite{Kim:1990km}
\begin{align}
\Gamma_{ij} = \frac{m_i m_j}{16 \pi E_i}\{(g_s)^2_{ij}[f(x_{ij})+k(x_{ij})]+(g_p)^2_{ij}[h(x_{ij})+k(x_{ij})]\}\,,
\end{align}
where $x_{ij} = m_i/m_j$ and the functions $f(x_{ij}), k(x_{ij})$, and $h(x_{ij})$ are defined as follows
\begin{align}
f(x) &= \frac{x}{2}+2+\frac{2}{x}\log x-\frac{2}{x^2}-\frac{1}{2x^3}\\
h(x) &= \frac{x}{2}-2+\frac{2}{x}\log x+\frac{2}{x^2}-\frac{1}{2x^3}\\
k(x) &= \frac{x}{2}-\frac{2}{x}\log x -\frac{1}{2x^3}\,.
\end{align}

For the helicity conserving scenario ~($\nu_i\rightarrow\nu_j+\phi$ or $\bar{\nu}_i\rightarrow\bar{\nu}_j+\phi$), $f(x)$ and $h(x)$ contribute, while $k(x)$ shows the contribution from helicity violating scenario ($\nu_i\rightarrow\bar{\nu_j}+\phi)$.

Neutrino decaying while its propagation will have impact on the neutrino flavor transition. Neutrino propagation Hamiltonian in this case can be written as 
\begin{equation}
H_{\rm dec}=\left[U^r\begin{pmatrix}
\frac{m^2_1}{2E}&0&0\\
0&\frac{m^2_2}{2E}-i\Gamma_2&0\\
0&0&\frac{m^2_3}{2E}-i\Gamma_3\\
\end{pmatrix}(U^r)^{\dagger}+V_{\text{CC}}\begin{pmatrix}
1&0&0\\
0&0&0\\
0&0&0\\
\end{pmatrix}\right] \;,
\label{eq:H_dec}
\end{equation}
where $\Gamma_i=\sum_j \Gamma_{ij}$ is the full decay width of the {\it i}th mass eigenstates in the lab frame. The superscript $r$ with the mixing matrix $U$ denotes the helicity.  In case of invisible decay, neutrino flavor transition probability in vacuum is written as~\cite{Coloma:2017zpg}
\begin{eqnarray}
P( \nu_\alpha^r \to \nu_\beta^r )  =  \bigg | U^{r}_{\alpha 1}(U^r_{\beta 1})^* +& U^r_{\alpha 2}(U^{r}_{\beta 2})^*e^{-i\Delta m^2_{21}L/(2E)} e^{-\Gamma_2 L/2} + \\
&U^r_{\alpha 3}(U^{r}_{\beta 3})^* e^{- i\Delta m^2_{31}L/(2E)} e^{-\Gamma_3 L/2} \bigg |^2  \nonumber \,.
\end{eqnarray}
In case of invisible decay, oscillation probability will be depleted since the fraction of $\nu_i$ is reduced with decay, as indicated by the exponential decay factor associated with the mass-squared terms. The amount of depletetion depends on the decay rate and the propagation distance. In case of visible decay, active neutrinos decay to another active neutrino states which are observable at the detector. As a result, there will be regeneration of the active neutrino after the decay. Amplitude for this regeneration due to visible decay is written as
\begin{equation}
\mathcal A_{\nu_\alpha^r \to \nu_\beta^s} (E_\alpha, E_\beta) = \sum_{i>j} (U^r_{\alpha i})^* U^s_{\beta j}
e^{-i E_\beta (L - L')} e^{-i E_\alpha L'} e^{-\Gamma^{rs}_{ij} L'/2}\sqrt{\Gamma^{rs}_{3j}}\sqrt{W^{rs}_{ij}}\, .
\label{eq:Avac}
\end{equation}
Here, $L'$ is the distance where decay takes place, $E_\beta$ is energy of daughter neutrino after the decay and the normalized energy distribution term $W^{rs}_{ij}=\frac{1}{\Gamma^{rs}_{ij}} \frac{d\Gamma^{rs}_{ij} (E_\alpha, E_\beta)}{dE_{\beta}}$.
Flavor transition probability in case of visible decay can be written as
\begin{eqnarray}
\frac{d P_{\nu_\alpha^r \to \nu_\beta^s}}{dE_\beta} & = & 
P_{\nu_\alpha^r \to \nu_\beta^s}^{\rm inv}(E_\alpha)\, \delta(E_\alpha - E_\beta) \delta_{rs}
+  \Delta P^{\rm vis}_{ \nu_\alpha^r \to \nu_\beta^s } (E_\alpha, E_\beta) \, .
\label{eq:P1}
\end{eqnarray}
The last term $\Delta P^{vis}_{\nu_\alpha^r \to \nu_\beta^s}$ that brings in the effects from the visible decay is defined as
\begin{align}
\Delta P^{\rm vis}_{\nu_\alpha^r \to \nu_\beta^s } (E_\alpha, E_\beta) & = \int_0^L \mid \mathcal A_{\nu_\alpha^r \to \nu_\beta^s} (E_\alpha, E_\beta)\mid^2 dL'
\end{align}
Neutrino oscillation probability in the presence of visible decay depends on a number of parameters apart from the standard oscillation parameters, such as, energy of the neutrinos after the deacay, position of the decay.

\section{Summary}

This chapter discusses the impact of various BSM scenarios on neutrino oscillation probability. We briefly describe the theoretical motivation and experimental anomalies that hint toward possible BSM physics in the neutrino sector. BSM physics can influence the neutrino flavor transition through the new neutrino-matter interaction that will impact neutrino evolution through the matter. We give a detailed description of such non-standard interaction. Also, the existence of additional neutrino states in Nature would modify the mixing between its flavor and mass eigenstates, which will have an essential effect on neutrino oscillation. The presence of an eV-scale sterile neutrino would lead to the addition of six more oscillation parameters, which can be probed through the oscillation experiments. Also, the presence of heavy sterile neutrinos can be investigated by studying the non-unitarity of the active neutrino matrix, which we discuss in this chapter. New physics scenarios like neutrino decay that are allowed by various BSM models can also be probed in various neutrino oscillation experiments. We also discuss how the violation of fundamental symmetries like Lorentz and CPT would affect neutrino propagation and flavor transition.


\newpage 

\chapter{Evolution of standard neutrino oscillation parameters in the presence of NC-NSI}
\label{C4} 
\section{Introduction}

Understanding the impact of various oscillation parameters in the neutrino oscillation probabilities becomes intricate after introducing standard neutrino-matter interactions. Moreover, if non-standard interactions (NSI) are present in the Nature, the complexity of this task increase further. Assuming a constant line-averaged matter density and a fixed baseline, several works exist in literature that have formulated approximate analytical formulas to estimate the probabilities of neutrino oscillations in the presence of standard interactions (SI)~\cite{Petcov:1986qg,Kim:1986vg,Arafune:1996bt,Arafune:1997hd,Ohlsson:1999xb,Freund:2001pn,Cervera:2000kp,Akhmedov:2004ny,Asano:2011nj,Agarwalla:2013tza,Minakata:2015gra,Denton:2016wmg} and SI+NSI~\cite{GonzalezGarcia:2001mp,Ota:2001pw,Yasuda:2007jp,Kopp:2007ne,Ribeiro:2007ud,Blennow:2008eb,Kikuchi:2008vq,Meloni:2009ia,Agarwalla:2015cta}.

In order to enhance our understanding of neutrino oscillation probabilities for a given baseline ($L$) and/or neutrino energy ($E$) under the influence of standard interactions (SI) or both SI and NSI, it is crucial to establish a clear understanding of how different mixing angles and mass-squared differences are altered within matter as a function of energy for a given baseline. The straightforward approximate analytical expressions to illustrate the changes in mass-mixing parameters within matter will allows us to explore various essential aspects observed in neutrino oscillations with a broader and more transparent way. This simplified and more intuitive approach to comprehending neutrino oscillation phenomena in the presence of neutrino-matter interactions is likely to open pathways for unraveling the intricate correlations or degeneracies that might exist among different oscillation and NSI parameters. 
In this chapter, we focus on comprehensive exploration of the modifications of the effective oscillation parameters in the presence of standard and non-standard matter effect in neutrino oscillation. We diagonalize the effective neutrino Hamiltonian in the presence of SI as well as NSI, and 
derive simple analytical expressions of those effective parameters. We discuss in detail various crucial feature of these modified parameters that will lead to the significant impact on the neutrino oscillations in presence of the matter effect for various benchmark choices of baselines and energies.
We demonstrate how the simple approximate
analytical expressions for the modified oscillation
parameters in matter help us to estimate 
the baselines and energies for which we 
have the maximal matter effect in 
$\nu_{\mu} \to \nu_{e}$ oscillation channel
in the presence of various NSI parameters. Also, we discuss the impact of the NSI parameters from the (2,3) block of the NSI matrix in the $\mumu$ disappearnce channel.

This chapter is organized as follows. In section~\ref{sec:diag_Heff}, we present our formalism for diagonalizing the effective neutrino Hamiltonian in the presence of both SI and SI+NSI. We also provide the expressions for the modified oscillation parameters. Section~\ref{sec:evol_angles} explores in detail how mixing angles altered as a function of energy, focusing on specific benchmark choices of baselines. Similar discussions for the modified mass-squared differences are done in section~\ref{sec:evol_mass}. We demonstrate how the various NSI parameters would modify the energy and baseline at $\theta_{13}$-resonance in section~\ref{sec:resonance_appearance_channel}.
In section~\ref{sec:nsi_appearance_channel}, we examine the impact of the off-diagonal NSI parameters in the $\mue$ appearance channel. A detailed discussion on the impact of the NSI parameters in the (2,3) block of the NSI matrix on the $\mumu$ disappearance channel is done in section~\ref{sec:nsi_disappearance_channel}. Finally, in Section~\ref{sec:summary}, we summarize our findings.

\section{Diagonailzation of effective neutrino Hamiltonian in presence of NSI}
\label{sec:diag_Heff}

The effective neutrino propagation Hamiltonian 
in matter in presence of all the lepton-flavor-conserving 
and lepton-flavor-violating neutral-current non-standard interaction (NC-NSI) can be written as
\begin{equation}
\label{eq:Heff}
H_f = \frac{1}{2E}\left[U\begin{pmatrix}
0&0&0\\
0&\Delta m^2_{21}&0\\
0&0&\Delta m^2_{31}\\
\end{pmatrix}U^{\dagger}+2EV_{\rm CC}\begin{pmatrix}
1+\varepsilon_{ee}&\varepsilon_{e\mu}&\varepsilon_{e\tau}\\
\varepsilon^{\ast}_{e\mu}&\varepsilon_{\mu\mu}&\varepsilon_{\mu\tau}\\
\varepsilon^{\ast}_{e\tau}&\varepsilon^{\ast}_{\mu\tau}&\varepsilon_{\tau\tau}\\
\end{pmatrix}\right],
\end{equation}
where,
$\Delta m^2_{21}$ and 
$\Delta m^2_{31}$ are 
the solar and atmospheric mass-squared differences, 
respectively. $U$ is the 
PMNS matrix in vacuum~\cite{Pontecorvo:1957qd,Maki:1962mu,Pontecorvo:1967fh},
which can be parametrized as the product of three $3\times3$  rotation matrices $R(\theta_{ij},\delta)$
in the following fashion
\begin{equation}
U =  R_{23}(\theta_{23},0)\;R_{13}(\theta_{13},\delta_{\mathrm{CP}})\;R_{12}(\theta_{12},0) \;.
\label{eq:U-parametrization}
\end{equation}
In eq~\ref{eq:Heff}, the last term contains the standard charge-current and non-standard neutral-current neutrino-matter interaction. $V_{\rm CC}$ is the standard 
$W$-exchange interaction potential in matter  defined in eq.~\ref{eq:Vcc_1}. The parameters $\varepsilon_{ij}$ depicts the strength of NC-NSI parameters.

We parameterize the effective Hamiltonian $H_f$ by subtracting the common physical phase $I (\equiv \eee V_{\rm CC})$ 
from the right-hand side (R.H.S.) of eq.~\ref{eq:Heff}. This would help us to separate the contribution of standard CC interaction and the NSI parameter $\eee$ in the oscillation probability without altering the results. The effective Hamiltonian now has to form
\begin{equation}
\label{eq:Heff1}
H_{f}=\Delta_{31}\left[U\begin{pmatrix}
0&0&0\\
0&\alpha&0\\
0&0&1\\
\end{pmatrix}U^{\dagger}+\hat{A}\begin{pmatrix}
1&\varepsilon_{e\mu}&\varepsilon_{e\tau}\\
\varepsilon^{\ast}_{e\mu}&\beta&\varepsilon_{\mu\tau}\\
\varepsilon^{\ast}_{e\tau}&\varepsilon^{\ast}_{\mu\tau}&\gamma\\
\end{pmatrix}\right] \;,
\end{equation}
where, $\Delta_{31} \equiv \Delta m^2_{31}/2E$,
$\alpha \equiv \Delta m^2_{21}/\Delta m^2_{31}$,
$\hat{A} \equiv 2EV_{\rm CC}/\Delta m^2_{31}$.
We define the effective diagonal NC-NSI parameters as 
$\beta \equiv \emm - \eee$ 
and $\gamma \equiv \ett - \eee$.

The elements of the $H_f$ can be written as 
\begin{align}
(H_{f})_{11} & \,=\, \Delta_{31} \, [\alpha s^2_{12}c^2_{13}+s^2_{13}+\hat{A}]\\
(H_{f})_{12} & \,=\, \frac{\Delta_{31}}{2} \, [\sin{2\theta_{13}}s_{23}(1-\alpha s^2_{12})+\alpha\sin{2\theta_{12}}c_{13}c_{23}+2\varepsilon_{e\mu}\hat{A}] \\
(H_{f})_{13} & \,=\, \frac{\Delta_{31}}{2} \, [\sin{2\theta_{13}}c_{23}(1-\alpha s^2_{12})-\alpha\sin{2\theta_{12}}c_{13}s_{23}+2\varepsilon_{e\tau}\hat{A}] \\
(H_{f})_{22} & \,=\, \frac{\Delta_{31}}{2} \, [\alpha c^2_{12}+c^2_{13}+\alpha s^2_{12}s^2_{13}+\cos{2\theta_{23}}(\alpha c^2_{12}-\alpha s^2_{12}s^2_{13}-c^2_{13}) \nonumber \\
	           & -\alpha\sin{2\theta_{12}}s_{13}\sin{2\theta_{23}} +2\beta\hat{A}]\\
(H_{f})_{23} & \,=\, \frac{\Delta_{31}}{2}[\sin{2\theta_{23}}(c^2_{13}-\alpha c^2_{12}+\alpha s^2_{12}s^2_{13})-\alpha\sin{2\theta_{12}}s_{13}\cos{2\theta_{23}}+2\varepsilon_{\mu\tau}\hat{A}] \\
(H_{f})_{33} & \,=\, \frac{\Delta_{31}}{2}[\alpha c^2_{12}+c^2_{13}+\alpha s^2_{12}s^2_{13}+\cos{2\theta_{23}}(c^2_{13}-\alpha c^2_{12}+\alpha s^2_{12}s^2_{13}) \nonumber \\
	           & +\alpha\sin{2\theta_{12}}s_{13}\sin{2\theta_{23}}+2\gamma\hat{A}]\,,
\end{align}
where $c_{ij}\rightarrow\cos\theta_{ij}$, $s_{ij}\rightarrow\sin\theta_{ij}$. 
We retain the terms of all orders in $\sin\theta_{13}$ and $\alpha$
which are essential in light of the large value of $\theta_{13}$. Also, for simplification, we take $\dcp=0$ in our calculation; later in this section, we show the final expression of the modified oscillation parameters for non-zero values of $\dcp$.

\subsection{Diagonalization of $H_f$}
\label{subsec:diag_Hf}
The calculation of neutrino oscillation probability in the presence of matter effect involves diagonalization of the effective Hamiltonian in matter. To do this, we use an approximate analytical method of successive rotation, where we apply three subsequent rotations in three planes: $R_{23} (\theta^m_{23})$ in (2,3) plane, 
$R_{13} (\theta^m_{13})$ in (1,3) plane, and $R_{12} (\theta^m_{12})$ in (1,2) plane. $\tzm$, $\tym$, and $\txm$ are the rotation angle in each plane.  
After each rotation, we make off-diagonal elements of the corresponding plane to zero. The product of these 
rotation matrices construct a $3 \times 3$ unitary matrix
\begin{equation}
\tilde{U} \equiv R_{23}\left(\theta_{23}^m\right)R_{13}\left(\theta_{13}^m\right)R_{12}\left(\theta_{12}^m\right) \,,
\label{eq:new-U}
\end{equation}
such that it approximately diagonalizes $H_f$~\footnote{After the final rotation, (1,3) and (2,3) elements of the rotated effective Hamiltonian remain non-zero. However, these elements are very small ($O(\Delta m^2_{31}\times\alpha^2)$ or $O(\Delta m^2_{31}\times \alpha s^2_{13}$) compared to the diagonal elements ($O(\Delta m^2_{31})$). So, we neglect these two off-diagonal elements and consider that the effective Hamiltonian is approximately diagonal.}. Since the modified mixing matrix $\tilde{U}$ has the same parameterization as the PMNS matrix, we term the rotation angles $\theta^m_{ij}$ as modified mixing angles.

\subsection{The expressions of the modified mass-mixing parameters with no $\delta_{\rm CP}$}
\label{subsec:mixing_param_dcp_0}
Each subsequent rotation during diagonalization process gives the expression for the modified mixing angle corresponding to the plane of rotation. The expressions of the three mixing angles are as follows
\begin{equation}
\tan 2\theta^m_{23}\simeq\frac{(c^2_{13}-\alpha c^2_{12}+\alpha s^2_{12}s^2_{13})\sin2\theta_{23}-\alpha s_{13}\sin2\theta_{12}\cos2\theta_{23}+2\varepsilon_{\mu\tau}\hat{A}}{(c^2_{13}-\alpha c^2_{12}+\alpha s^2_{12}s^2_{13})\cos2\theta_{23}+\alpha s_{13}\sin 2\theta_{12}\sin2\theta_{23}+(\gamma-\beta)\hat{A}} \;,
\label{eq:th23m}
\end{equation}
\begin{equation}
\tan2\theta^m_{13}\simeq\frac{\sin2\theta_{13}(1-\alpha {s_{12}}^2)\cos\Delta\theta_{23}-\alpha\sin 2\theta_{12}c_{13}\sin\Delta \theta_{23}+2(\varepsilon_{e\mu}s^m_{23}+\varepsilon_{e\tau}c^m_{23})\hat{A}}{(\lambda_3-\hat{A}-\alpha s^2_{12}c^2_{13}-s^2_{13})} \;,
\label{eq:th13m}
\end{equation}
\begin{equation}
{\tan2\theta^m_{12}}\simeq
\frac{{c^m_{13}}[\alpha\sin2\theta_{12}c_{13}\cos\Delta\theta_{23}+\sin2\theta_{13}(1-\alpha s_{12}^2)\sin\Delta\theta_{23}+2(\varepsilon_{e\mu}c^m_{23}-\varepsilon_{e\tau}s^m_{23})\hat{A}]}{(\lambda_2-\lambda_1)} \;,
\label{eq:th12m}
\end{equation}
where, $\Delta\theta_{23} \equiv \theta_{23}-\theta^m_{23}$ is the deviation of the modified mixing angle $\theta_{23}$ from its vacuum value.
In the above equations, $\lambda_{1}$, $\lambda_{2}$, and $\lambda_3$ take the following forms:
\begin{align}
\label{eq:lambda_3}
\lambda_3=&\nonumber \frac{1}{2}\Big[c^2_{13}+\alpha c^2_{12}+\alpha s^2_{12}s^2_{13}+(\beta+\gamma)\hat{A} \\
&+\frac{(\gamma-\beta)\hat{A}+\alpha\sin2\theta_{12}s_{13}\sin2\theta_{23}+(c^2_{13}-\alpha c_{12}^2+\alpha s_{12}^2s_{13}^2)\cos2\theta_{23}}{\cos2\theta^m_{23}}\Big] \;,
\end{align}
\begin{align}
\label{eq:lambda_2}
\lambda_2=&\nonumber\frac{1}{2}\Big[\alpha c^2_{12}+c^2_{13}+\alpha s^2_{12}s^2_{13}+(\beta+\gamma)\hat{A} \\
&-\frac{(\gamma-\beta)\hat{A}+\alpha\sin2\theta_{12}s_{13}\sin2\theta_{23}+(c^2_{13}-\alpha c_{12}^2+\alpha s_{12}^2s_{13}^2)\cos2\theta_{23}}{\cos2\theta^m_{23}}\Big] \;,
\end{align}
\begin{align}
\label{eq:lambda_1}
\lambda_1=\frac{1}{2}\Big[\lambda_3+\hat{A}+s^2_{13}+\alpha s^2_{12}c^2_{13}
-\frac{\lambda_{3}-\hat{A}-s^2_{13}-\alpha s^2_{12}c^2_{13}}{\cos2\theta^m_{13}}\Big] \;.
\end{align}

Eigenvalues of the neutrino Hamiltonian after the full diagonalization can be parameterized as a function of modified mass-squares ($m^2_{i,m},\,\,i=1,2,3$). The expression of the eigenvalues is as follows
\begin{align}
\frac{m^2_{3,m}}{2E} \simeq &\frac{\Delta_{31}}{2}\Big[\lambda_{3}+\hat{A}+s^2_{13}+\alpha s^2_{12}c^2_{13}+\frac{\lambda_{3}-\hat{A}-s^2_{13}-\alpha s^2_{12}c^2_{13}}{\cos2\theta^m_{13}}\Big] \label{eq:m3} \,, \\
\frac{m^2_{2,m}}{2E} \simeq &\frac{\Delta_{31}}{2}\Big[\lambda_1+\lambda_2-\frac{\lambda_1-\lambda_2}{\cos2\theta^m_{12}}\Big] \label{eq:m2} \,, \\
\frac{m^2_{1,m}}{2E} \simeq &\frac{\Delta_{31}}{2}\Big[\lambda_1+\lambda_2+\frac{\lambda_1-\lambda_2}{\cos2\theta^m_{12}}\Big] \,.
\label{eq:m1}
\end{align}
One can obtain expressions for the modified mass-squared
differences as $\ldmm \equiv m^2_{3,m} - m^2_{1,m}$ 
and $\sdmm \equiv m^2_{2,m} - m^2_{1,m}$. 

We reiterate that in the above calculation, we assume $\dcp=0$ for simplification. Also, the above expressions of the modified mixing angles are valid for neutrinos with NMO. For the antineutrino case, $\A$ will get a negative sign, and for the IMO case, the sign of $\alpha$ as well as $\A$ will be flipped.

		\begin{figure}[h!]
			\centering
			\includegraphics[scale=0.85]{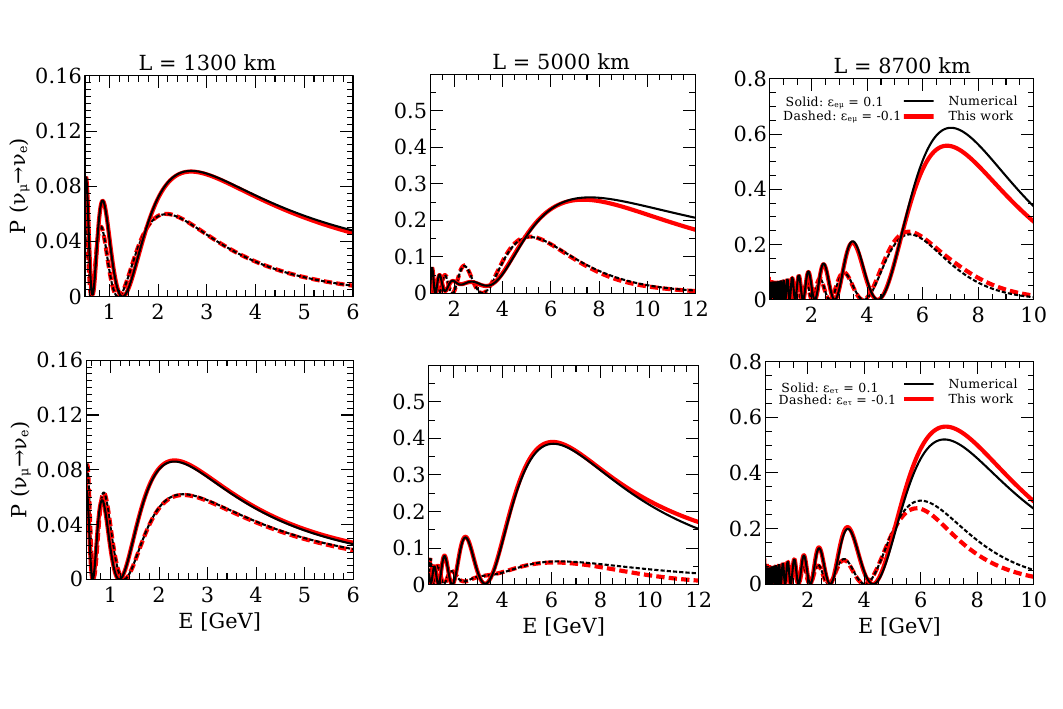}
			\mycaption{Comparison of $\nu_{\mu} \to \nu_{e}$ appearance probability estimated using our approximate analytical expressions of the modified mixing parameters (red curves), and the same derived from numerical calculations performed with GLoBES (black curves) in the presence of NC-NSI parameters $\varepsilon_{e\mu}$ (upper panels) and $\varepsilon_{e\tau}$ (lower panels) one-at-a-time.
            The three distinct columns correspond to the three different baselines: 1300 km (left panels), 5000 km (middle panels), and 8700 km (right panels). Solid curves are generated using a NSI parameter value of 0.1, while dashed curves correspond to a value of -0.1 for the NSI parameters. The best-fit values of the standard oscillation parameters are:  $\theta_{23}= 45^\circ$, $\theta_{13}=8.61^\circ$, $\theta_{12}= 33.8^\circ$, $\dcp=0^\circ$, $\Delta m^2_{31}=2.52\times10^{-3}$ eV$^2$ (NMO), and $\Delta m^2_{21}=7.39\times 10^{-5}$ eV$^2$.}
			\label{fig:comp_app}
		\end{figure}

		\begin{figure}[h!]
			\centering
			\includegraphics[scale=0.85]{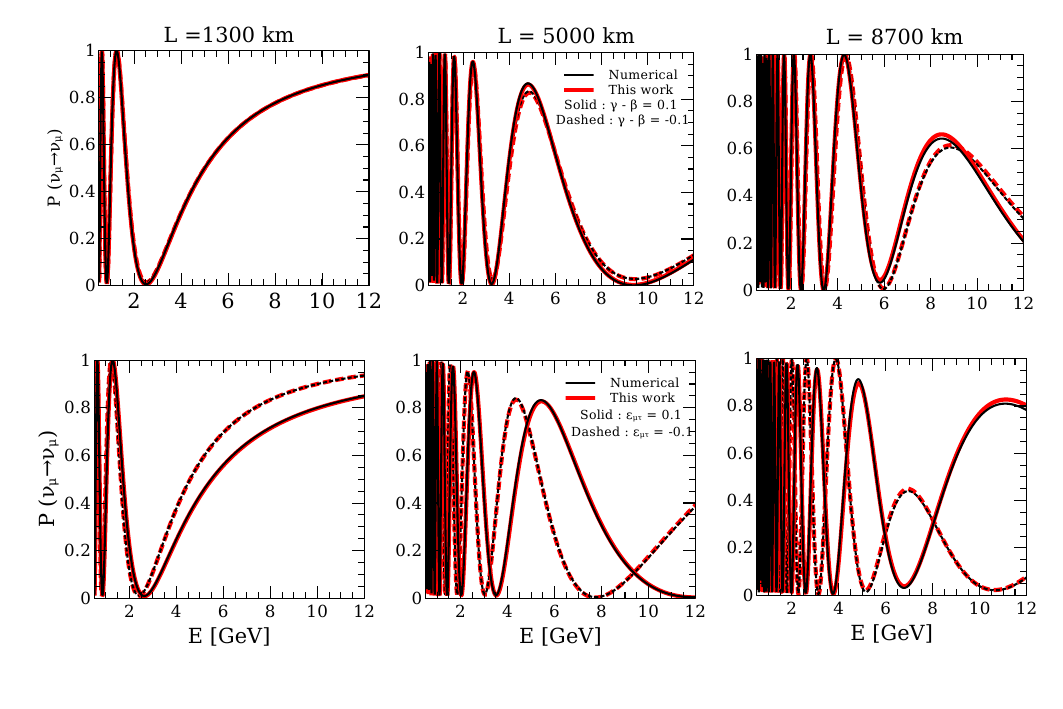}
			\mycaption{Comparison of $\mumu$ disappearance probability estimated using our approximate analytical expressions of the modified mixing parameters (red curves), and the same derived from numerical calculations performed with GLoBES (black curves) in the presence of effective parameter $\gamma-\beta$ (upper panels) and $\emt$ (lower panels) one-at-a-time.
            The three distinct columns correspond to the three different baselines: 1300 km (left panels), 5000 km (middle panels), and 8700 km (right panels). Solid curves are generated using a NSI parameter value of 0.1, while dashed curves correspond to a value of -0.1 for the NSI parameters. The best-fit values of the standard oscillation parameters are:  $\theta_{23}= 45^\circ$, $\theta_{13}=8.61^\circ$, $\theta_{12}= 33.8^\circ$, $\dcp=0^\circ$, $\Delta m^2_{31}=2.52\times10^{-3}$ eV$^2$ (NMO), and $\Delta m^2_{21}=7.39\times 10^{-5}$ eV$^2$.}
			\label{fig:comp_disapp}
		\end{figure}

Using the expressions of the modified oscillation parameter, one can calculate the oscillation probabilities in the presence of NC-NSI parameters in matter by replacing the standard vacuum oscillation parameters in the probability expressions with the corresponding modified oscillation parameters in matter (eqs.~[\ref{eq:th23m}-\ref{eq:m1}]). To check the validity of our expressions, we calculate neutrino oscillation probability using the expressions we derived in this section and compare with the exact probability calculated numerically using the GLoBES software~\cite{Huber:2004ka, Huber:2007ji}, as shown in Fig.~\ref{fig:comp_app} for $\mue$ appearance channel and Fig.~\ref{fig:comp_disapp} for $\mumu$ disappearance channel. For the appearance channel, we show the comparison in the presence of two off-diagonal NSI parameters ($\eem$, $\eet$), which affect this channel most. For the same reason, we show the comparison in the presence of the effective NSI parameter $\gamma-\beta$ and $\emt$ in the $\nu_\mu$ disappearance channel. Also, to see the validity of our expression in the different matter effect regions, we show our results for three distinct baselines, as mentioned in the three columns of the two figures. We observe that in all cases, the approximate oscillation probability calculated using the expression of the modified mixing parameters matches significantly well with the exact oscillation probability calculated numerically.

\subsection{The modified mass-mixing parameters with non-zero $\dcp$}
\label{subsec:mixing_param_non-zero_dcp}
As mentioned earlier, in subsection~\ref{subsec:mixing_param_dcp_0}, we have shown the expression of the modified mixing parameters in the CP conserving scenario, $\ie$, $\dcp=0$. In this subsection, we do the same, but we take non-zero value of $\dcp$ and consider $\tz=45^\circ$ for simplification. To diagonalize the effective Hamiltonian in the non-zero $\dcp$ case, we apply three complex rotations: $R_{23} (\theta^m_{23}, 0)$, $R_{13} (\theta^m_{13}, \delta_{\mathrm{CP}}^{m})$, and  $R_{12} (\theta^m_{12}, 0)$ successively, where  $R_{ij} (\theta^m_{ij}, \phi^{m})$ is the complex rotation matrix in the $(i,j)$ plane with angle $\theta^m_{ij}$ and phase $\phi^{m}$. The expression for the modified mass and mixing parameters in non-zero $\dcp$ case are 
\begin{align}
	\tan 2\theta^m_{23}\,\,=\,\,&\frac{c^2_{13}-\alpha c^2_{13}+2\emt\A}{(\gamma-\beta)\A},\label{eq:th23_dcp}\\
	\nonumber\\
	\tan2\theta^m_{13}\,\,=\,\, &\frac{1}{\sqrt{2}(\lambda_3-\A-s^2_{13}-\alpha s^2_{12}c^2_{13})}\big[\{\sin2\theta_{13}(1-\alpha s^2_{12})\cos\delta_{\mathrm{CP}}(c^m_{23}+s^m_{23})\nonumber\\
	&-\alpha\sin2\theta_{12}c_{13}
	(c^m_{23}-s^m_{23})+2\sqrt{2}(\eem s^m_{23}+\eet c^m_{23})\A\}^2\nonumber\\
	&+\{\sin2\theta_{13}(1-\alpha s^2_{12})\sin\delta_{\mathrm{CP}}(c^m_{23}+s^m_{23})\}^2\big]^{1/2},
	\label{eq:th13_dcp}\\
	\nonumber\\
	\tan2\theta^m_{12}\,\,=\,\, &\frac{c^m_{13}}{\sqrt{2}(\lambda_2-\lambda_1)}\big[\{\sin2\theta_{13}(1-\alpha s^2_{12})\cos\delta_{\mathrm{CP}}(c^m_{23}+s^m_{23})+\alpha\sin2\theta_{12}c_{13}(c^m_{23}-s^m_{23})\nonumber\\
	&+2\sqrt{2}(\eem c^m_{23}-\eet s^m_{23})\A\}^2+\{\sin2\theta_{13}(1-\alpha s^2_{12})\sin\delta_{\mathrm{CP}}(c^m_{23}+s^m_{23})\}^2\big]^{1/2},
	\label{eq:th12_dcp}\\
	\nonumber\\
	&\tan \delta_{\mathrm{CP}}^m \,\,=\,\, \tan \delta_{\mathrm{CP}}.\label{eq:dcp}
	\end{align}
	In the above equations, the $\lambda$'s take the form,
	\begin{align}
	\lambda_3 \,\,=\,\, & \frac{1}{2}\left[c^2_{13}+\alpha c^2_{12}+(\beta+\gamma)\A +\sqrt{(c^2_{13}-\alpha c^2_{12}+2\emt\A)^2+(\gamma-\beta)^2\A^2}\right],\label{eq:lmda3_dcp}\\
	\nonumber\\
	\lambda_2 \,\,=\,\, & \frac{1}{2}\left[c^2_{13}+\alpha c^2_{12}+(\beta+\gamma)\A -\sqrt{(c^2_{13}-\alpha c^2_{12}+2\emt\A)^2+(\gamma-\beta)^2\A^2} \right],\label{eq:lmda2_dcp}\\
	\nonumber\\
	\lambda_1 \,\,=\,\, & \frac{1}{2}\left[\lambda_3+\A+s^2_{13}+\alpha s^2_{12}c^2_{13}-\frac{\lambda_3-\A-s^2_{13}-\alpha s^2_{12}c^2_{13}}{\cos2\tym}\right]\label{eq:lmda1_dcp}.
	\end{align}

 The expression for the modified mass-squares is the same as eqs.~[\ref{eq:m3}-\ref{eq:m1}], except now $\cos 2\tym$ and $\cos 2\txm$ will depend on the value of $\dcp$.
 
 \begin{figure}[h!]
 	\centering
 	\includegraphics[scale=1.0]{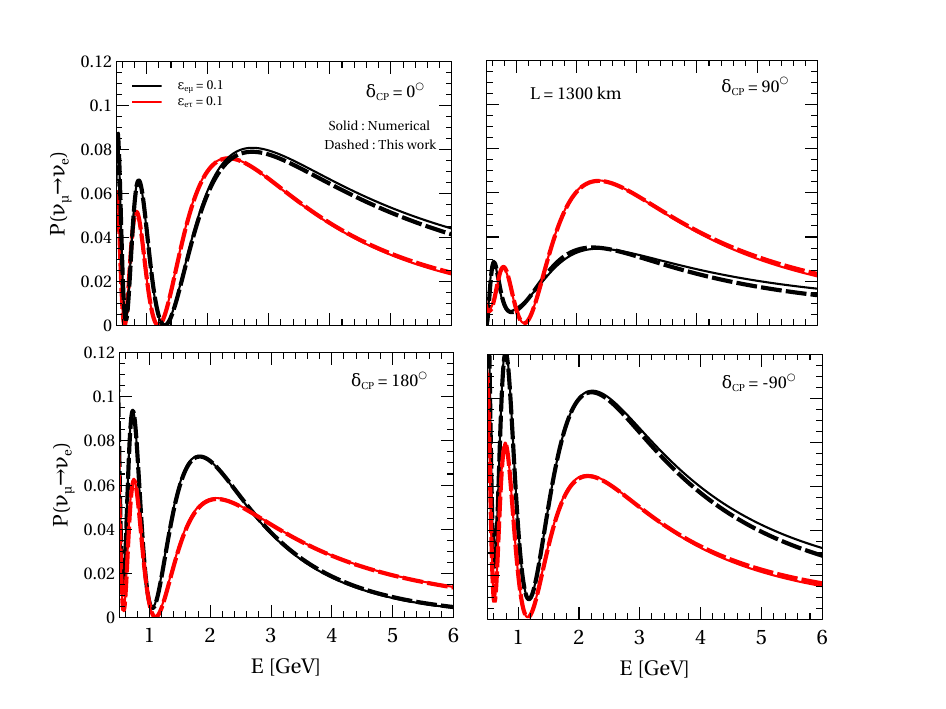}
 	\vspace*{-1.0cm}
 	\mycaption{Comparison of $\nu_{\mu} \to \nu_{e}$ appearance probability estimated from our approximate analytical expressions (dashed curves), and the numerical calculations obtained from GLoBES (solid curves) in the presence of NC-NSI parameters $\varepsilon_{e\mu}$ (black curves) and $\varepsilon_{e\tau}$ (red curves) in the presence of non-zero $\delta_{\mathrm{CP}}$. We consider our benchmark values of $\delta_{\mathrm{CP}}$: $0^{\circ}$ (upper left panel), $90^{\circ}$ (upper right panel), $180^{\circ}$ (lower left panel), and $-90^{\circ}$ (lower right panel). We assume $L=1300$ km, $\theta_{23}=45^{\circ}$, and NMO. The values of the other benchmark three-flavor oscillation parameters are taken from Table~\ref{table:vac_nsi}.}
 	\label{fig:Pmue-non-zero-dcp} 
 \end{figure}

 From eqs.~[\ref{eq:th23_dcp}-\ref{eq:lmda1_dcp}], we observe that, modified mixing angle $\tzm$ is independent of the value of the $\dcp$. Also, CP phase $\dcp$ itself does not altered in the matter in our parameterization. In fig.~\ref{fig:Pmue-non-zero-dcp}, we show a comparison between $\nu_{\mu}\rightarrow\nu_{e}$ appearance probabilities calculated using our approximate analytical expressions given in eqs.~\ref{eq:th23_dcp} to \ref{eq:dcp} (see dashed curves) and full numerical probabilities obtained from the publicly available software GLoBES (see solid curves). Here, we assume $L\,\,=\,\,1300$ km, $\theta_{23} = 45^{\circ}$, normal mass ordering (NMO). The values of the other benchmark three-flavor oscillation parameters are taken from Table~\ref{table:vac_nsi}.


\section{Evolution of mixing angles}
\label{sec:evol_angles}
In this subsection, we explore in detail how the three modified mixing angles varies with energy for some benchmark value of the baselines. For simplification, we use the CP conserving case or use $\dcp=0^\circ$, though we already mentioned in the previous section how our observation would be modified for a non-zero value of $\dcp$. Also, the study performed in this section is for neutrinos and NMO. One can do a similar analysis in the antineutrino and IMO cases by flipping the sign of $\A$ and ($\alpha$, $\A$), respectively. We use the benchmark value of the oscillation parameters given in table~\ref{table:vac_nsi} for the analysis. For the value of the NSI parameters, we choose a benchmark value of 0.2 for all the parameters. Although this value is larger than the present constraints on most NC-NSI parameters, we use this value for illustration purposes and to show its impact on the evolution of the oscillation parameters.

\begin{table}
\centering
\begin{tabular}{|c|c|c|c|c|c|}
\hline\hline
$\theta_{23}$&$\theta_{13}$&$\theta_{12}$&$\delta_{\mathrm{CP}}$&$\Delta m^2_{21} [\mathrm{eV^2}]$&$\Delta m^2_{31} [\mathrm{eV^2}]$ \\
\hline
 $45^{\circ}$, 	&$8.61^{\circ}$&$33.8^{\circ}$&$ 0^{\circ}$&$7.39\times 10^{-5}$&$2.52\times 10^{-3}$ \\
\hline\hline
\end{tabular}
\mycaption{The values of the oscillation parameters used in our analysis. 
The values of the other parameters are consistent with the present best-fit
values as obtained in various global fit 
studies~\cite{Marrone:2021,NuFIT,Esteban:2020cvm,deSalas:2020pgw}.
We assume normal mass ordering (NMO) throughout the chapter.}
\label{table:vac_nsi}
\end{table}

\subsection{Evolution of $\theta^m_{23}$}
\label{subsubsec:th23}
The expression describing the 
evolution of the effective mixing angle $\theta^m_{23}$
is shown in eq.~\ref{eq:th23m}. To have deeper analytical insights,
we further simplify this expression by neglecting the small terms which are proportional to $\alpha s_{13} \sim 10^{-3}$ in 
eq.~\ref{eq:th23m}. Now, the expression showing 
the evolution of $\theta_{23}$ in matter in the 
presence of NSI takes the form
\begin{align}
\label{eq:th23_1}
\tan 2\theta^m_{23} \simeq \frac{(c^2_{13}-\alpha c^2_{12})\sin2\theta_{23}+2\varepsilon_{\mu\tau}\hat{A}}{(c^2_{13}-\alpha c^2_{12})\cos2\theta_{23}+(\eff) \hat{A}},
\end{align}
where $\gamma-\beta\equiv(\ett-\emm)$. From the above expression, we observe two key features as listed below
\begin{itemize}

\item
Only NSI parameters from the (2,3) block ($\emt$ 
and an effective NSI parameter $\eff \equiv \ett-\emm$) 
of the NSI Hamiltonian affects to the evolution 
of $\theta^m_{23}$.

\item
$\tzm$ does not explicitly depend on the standard charge current interaction $\A$ (it is always associted with the NSI parameters). So, in the limiting case of all NSI parameters equal to zero (which in this case removes the standard 
matter effect $\hat{A}$ also), one would get back the vacuum mixing angle (\ie, $\tzm = \theta_{23}$) irrespective of energy, baseline, and the octant of $\theta_{23}$. In other words, it implies that $\tzm$ does not run in the presence of standard matter effect. Note that in the exact expression of $\tzm$ in eq.~\ref{eq:th23m}, due to the presence of the tiny terms proportional to 
$\alpha s_{13}$, $\tzm$ slightly deviates from its vacuum value even in the presence of SI.

\end{itemize}

\begin{figure}[htb!]
\centering
\includegraphics[scale=0.6]{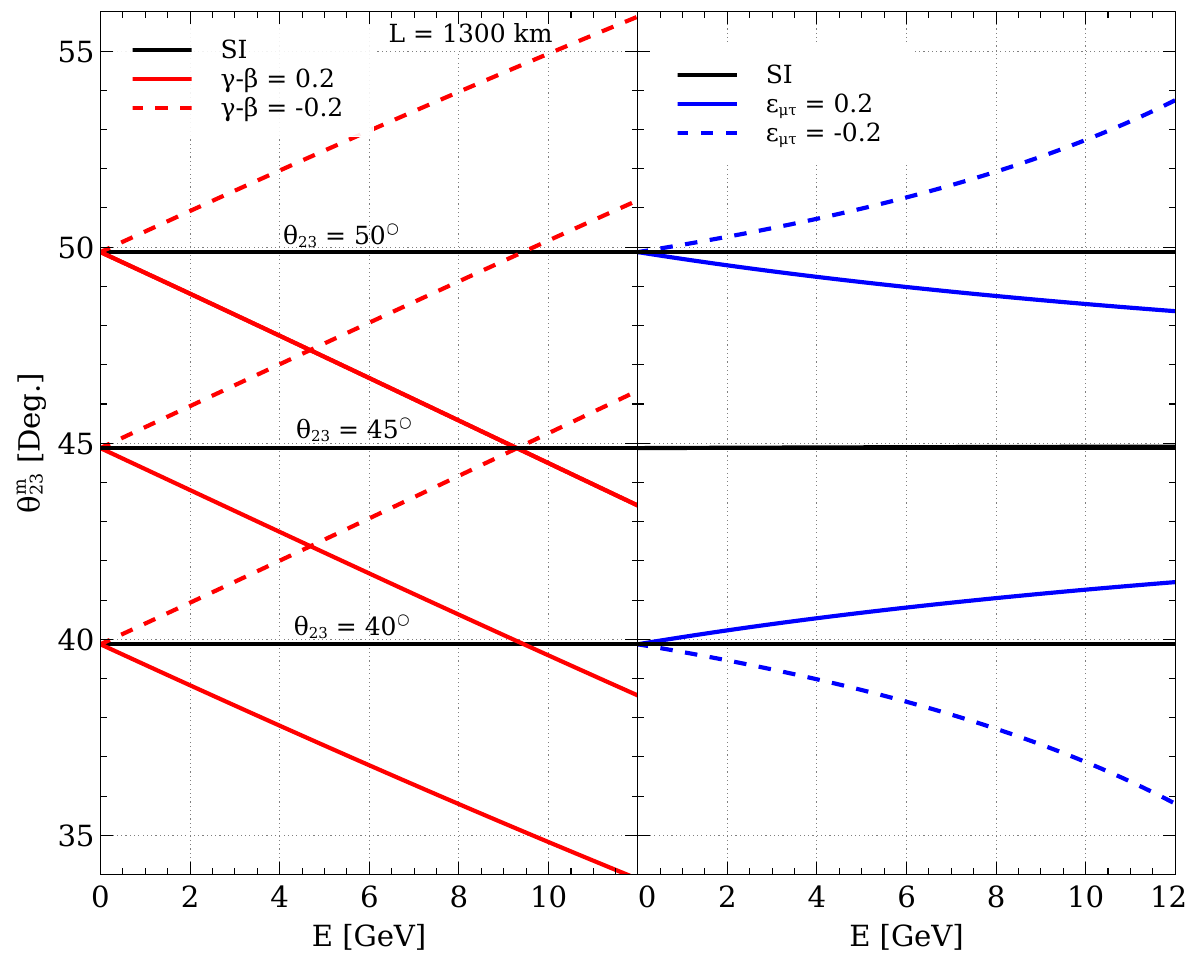}
\mycaption{
Evolution of $\theta^m_{23}$ in matter (eq.\ \ref{eq:th23m}) 
as a function of neutrino energy in of SI case  as well as in  the presence of NSI.
Solid black curve in each panel represents the SI case while 
the other curves correspond to the SI+NSI cases 
with positive (solid lines) and negative (dashed lines) 
values of NSI parameters. In the left column, we show 
the modification in the presence of NSI parameter 
$(\gamma-\beta)$, while the right column depicts 
the effect of $\emt$. We consider $L$ = 1300 km 
and assume NMO. We present results for three 
different values of $\theta_{23}$ in vacuum:
$40^{\circ}$ (lower octant), $45^{\circ}$ (maximal value), $50^{\circ}$ (upper octant). 
The values of the other oscillation parameters in vacuum are taken 
from Table~\ref{table:vac_nsi}.}
\label{fig:th23_1}
\end{figure}

In figure~\ref{fig:th23_1}, we show the variation of  $\tzm$ with energy in the presence of NSI parameters $\gamma-\beta$ and $\emt$ one-at-a-time, considering a baseline of 1300 km. In the SI case, as mentioned earlier, $\tzm$ does not show any variation with energy as the dependence of the matter parameter $\A$ vanishes in the absence of any NSI parameter.
In the presence of $\gamma-\beta$, we see a monotonic decrease~(increase) of $\tzm$ with neutrino energy when the strength of the NSI parameter is positive (negative), which can be easily understood from the approximate expression given in eq.~\ref{eq:th23_1} for all three benchmark choices of $\theta_{23}$. 
In the presence of $\emt$, we see almost no variation  in the $\tzm$ for $\theta_{23} = 45^\circ$. It is because in eq.~\ref{eq:th23_1}, denominator vanishes when $\theta_{23}$ is maximal, and $\tzm$ becomes $45^\circ$ irrespective of the strength of $\emt$. Interestingly, we see the opposite behavior of $\tzm$ with energy in the different octant of $\theta_{23}$ --- it decreases in the upper octant for positive values of $\emt$ while increasing in the lower octant. This happens because the term in the denominator of eq.~\ref{eq:th23_1} flipped sign as the octant of $\tz$ changes, reversing the impact of $\emt$. 

\begin{figure}[h!]
\centering
\includegraphics[height=6cm,width=14cm]{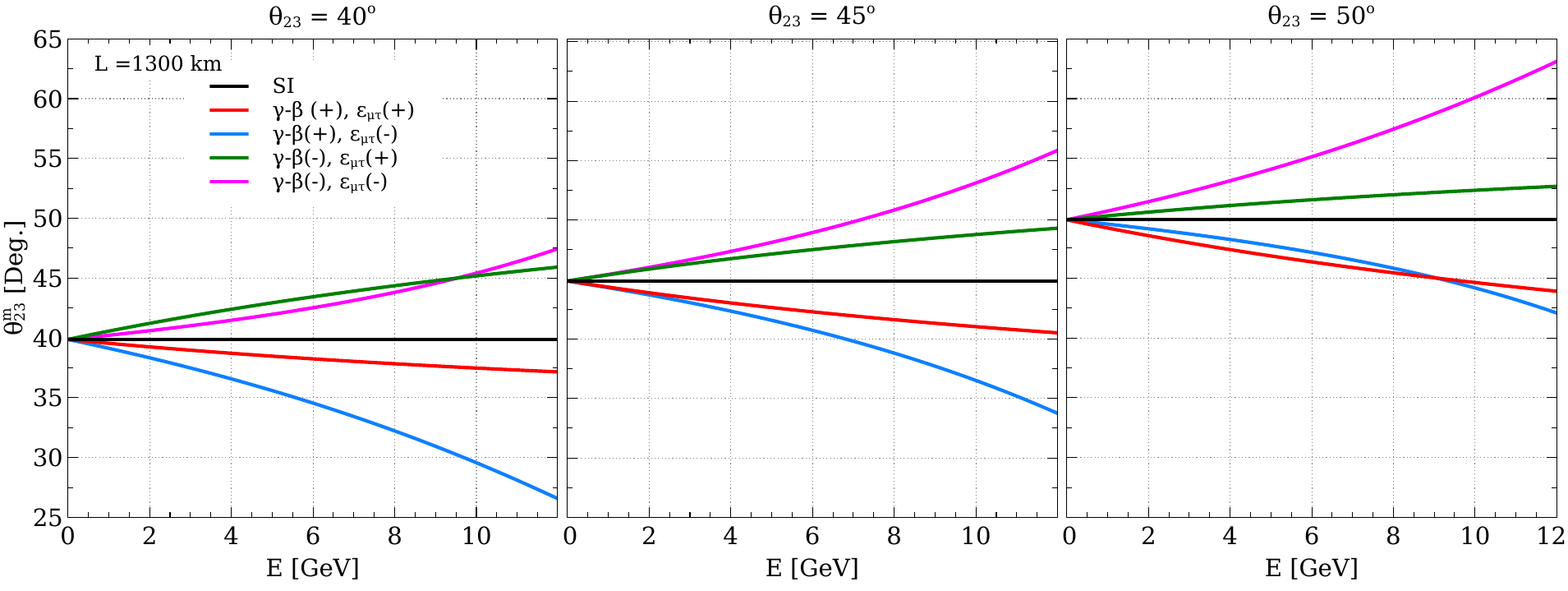}
\mycaption{
Evolution of $\theta^m_{23}$ (given in eq.~\ref{eq:th23m}) 
with neutrino energy in matter with SI and NSI considering 
both $(\gamma-\beta)$ and $\varepsilon_{\mu\tau}$ 
non-zero at-a-time. Black curve in each column represents 
the SI case while the other curves show the cases with 
four possible combinations of the sign of $(\gamma-\beta)$ 
and $\varepsilon_{\mu\tau}$ with magnitude 0.2. 
The left, middle, and right column correspond 
to the evolution considering three values of 
$\theta_{23}$ in vacuum, $40^{\circ}$, $45^{\circ}$, 
and $50^{\circ}$, respectively. We consider $L$ = 1300 km 
and assume NMO. Values of the oscillation parameters 
in vacuum used in this plot are taken from 
Table~\ref{table:vac_nsi}.}
\label{fig:th23_2}
\end{figure}

In figure~\ref{fig:th23_2}, we show the evolution of $\tzm$ when both $\gamma-\beta$ are non-zero at-a-time, considering different combination of their sign. We note from Fig.\ \ref{fig:th23_2} 
that in the presence of $(\eff)$ with a negative (positive) sign, 
$\tzm$ monotonically increases (decreases) with energy 
irrespective of the sign of $\emt$ and the octant of $\theta_{23}$. 
We also observe that for lower (higher) octant, the decrease (increase) is the steepest when $(\eff)$ is positive (negative) with negative value of $\emt$. For maximal mixing, the modification of $\tzm$ appears symmetric around the SI case since the term with $\cos2\theta_{23}$ in the denominator of eq.~\ref{eq:th23m} vanishes. 

In case of IMO 
with neutrino ($\nu$, IMO), the effect of each 
NSI parameters in $\tzm$ evolution is reversed 
(\ie, if $\theta^m_{23}$ increases with energy 
in presence of a particular NSI parameter 
with normal ordering of mass, in case of 
inverted mass ordering $\tzm$ will decrease with energy).
This happens since the term $\hat{A}$ associated 
with each NSI parameter changes its sign in case of IMO. 
Also, in case of antineutrino propagation with 
inverted mass ordering ($\bar{\nu}$, IMO), 
the change in $\tzm$ is almost the same as that 
of neutrino propagation with NMO ($\nu$, NMO).
This is because of the fact that in both cases, 
sign of $\hat{A}$ is the same.  

\subsection{Evolution of $\theta^m_{13}$}
\label{subsubsec:th13}

Eq.~\ref{eq:th13m} show the expression of the effective parameter $\tym$. From the equation, it is clear that all the five NSI parameters influence the evolution of $\tym$. In addition, unlike $\tzm$, the standard matter effect explicitly contributes to its running. We check that the value of the $\tz$ has marginal dependence on the evolution of $\tym$ when it is in the range [$40^\circ$, $50^\circ$].
So, for our analysis, we simplify the expression in eq.~\ref{eq:th13m} by considering $\tz=45^\circ$. The expression
of $\tym$ now have the form
\begin{equation}
\label{eq:th13_1}
\tan2\theta^m_{13} \simeq \frac{\sin2\theta_{13}(1-\alpha {s_{12}}^2)(s^m_{23}+c^m_{23})-\alpha\sin 2\theta_{12}c_{13}(c^m_{23}-s^m_{23})+2\sqrt{2}(\varepsilon_{e\mu}s^m_{23}+\varepsilon_{e\tau}c^m_{23})\hat{A}}{\sqrt{2}(\lambda_3-\hat{A}-\alpha s^2_{12}c^2_{13}-s^2_{13})},
\end{equation}
where,
\begin{equation}
\label{eq:lmda3_4}
\lambda_3 = \frac{1}{2}\bigg[{c_{13}}^2+\alpha {c_{12}}^2+\alpha s^2_{12}s^2_{13}+(\beta+\gamma)\hat{A}+\frac{(\gamma-\beta)\hat{A}+\alpha\sin2\theta_{12}s_{13}}{\cos2\theta^m_{23}}\bigg].
\end{equation} 

\begin{figure}[htb!]
\centering
\includegraphics[scale=0.80]{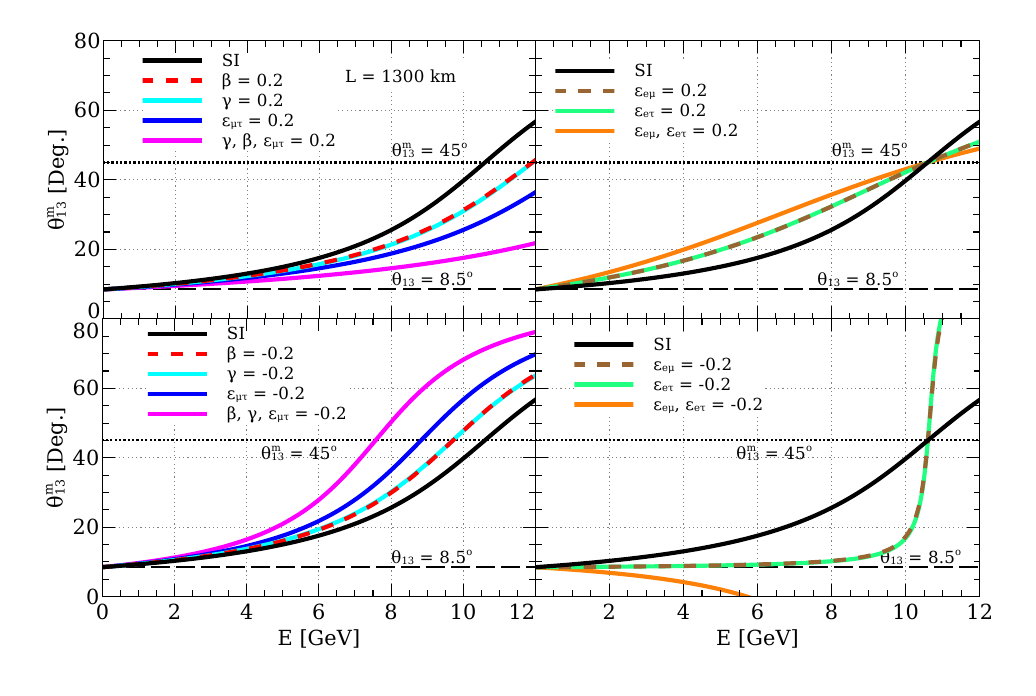}
\mycaption{Evolution of 
$\tym$ (given in eq.\ \ref{eq:th13_1}) with energy in presence of SI and NSI in matter. The solid black curve in each panel shows the SI case while the other curves correspond to the variation in the presence of SI+NSI. In the top (bottom) row, the NSI have been considered with a benchmark value of 0.2 (-0.2). The left column depicts the presence of NSI parameters in (2,3) block while the right column shows the effect of $\eem$ and $\eet$. We have used $L$ = 1300 km and assumed NMO in the plot. Values of oscillation parameters in vacuum are given in Table~\ref{table:vac_nsi} with 
$\theta_{23} =  45^{\circ}$.}
\label{fig:th13_1} 
\end{figure}
In figure~\ref{fig:th13_1}, we show the variation of $\tym$ as a function of energy in SI case as well as in the presence of NSI parameters one-at-a-time and various combinations of them. In the SI case, we observe a monotonic increase of $\tym$ with energy, which can go more than $45^\circ$ at around 10~GeV of neutrino energy for DUNE-like baselines. The presence of NSI parameters from the (2,3) block of the NSI matrix ($\gamma$, $\beta$, and $\emt$) one-at-a-time with positive (negative) strength suppress (enhance) the evolution as these parameters appear in the denominator in eq.~\ref{eq:th13_1}. The presence of all three parameters at at-a-time enhances/suppresses the evolution further. 
In the presence of NSI parameters $\eem$ and $\eet$ with positive (negative) strength one-at-a-time, we see an increasing deviation from the SI case at the lower value of energy. Interestingly, as the value of $\tym$ reaches $45^\circ$, the deviation from the SI case starts to reduce, and at $\tym=45^\circ$, it exactly overlaps with the SI line, irrespective of the sign of NSI parameters. This observation is valid for any strength of $\eem$ or $\eet$. One can understand this from the expression of $\tym$ in eq.~\ref{eq:th13_1}. In the denominator, with the increase in energy, the value of matter parameter $\A$ increases, reducing the overall value of the denominator. At a particular value of energy, the denominator becomes zero, \ie~$\tym=45^\circ$. We term this as a $\tym$-resonance. We discuss this in detail in subsection~\ref{sec:resonance_appearance_channel}.

In the case of IMO, the behavior of $\tym$ in SI as well as in SI+NSI case is significantly different from the NMO case for neutrino. It can be understood from the ($\lambda_3-\hat{A}$) term in the denominator of eq.~\ref{eq:th13_1}. Since $\hat{A}$ changes its sign, the denominator increases with energy; consequently, the value of the $\tym$ decreases from its vacuum value. However, in case of antineutrino ($\bar{\nu}$) propagation and inverted mass ordering ($\bar{\nu}$, IMO), since $\hat{A}$ does not change its sign, variation of $\tym$ is almost similar to neutrino ($\nu$, NMO) case.


\subsection{Evolution of $\theta^m_{12}$}
\label{subsubsec:th12}

Eq.~\ref{eq:th12m} shows the expression for the evolution of $\txm$. Here also, we simplify the equation by assuming $\tz=45^\circ$. The expression for $\txm$ now has the form
\begin{equation}
\label{eq:th12_1}
\tan2\theta^m_{12} \simeq \frac{{c^m_{13}}\big[\alpha\sin2\theta_{12}c_{13}(c^m_{23}+s^m_{23})+\sin2\theta_{13}(1-\alpha s_{12}^2)(c^m_{23}-s^m_{23})+2\sqrt{2}(\varepsilon_{e\mu}c^m_{23}-\varepsilon_{e\tau}s^m_{23})\hat{A}\big]}{\sqrt{2}(\lambda_2-\lambda_1)},
\end{equation}
where,
\begin{align}
&  \lambda_2 = \frac{1}{2}\bigg[\alpha c^2_{12}+c^2_{13}+\alpha s^2_{12}s^2_{13}+(\beta+\gamma)\hat{A}-\frac{(\gamma-\beta)\hat{A}+\alpha\sin2\theta_{12}s_{13}}{\cos2\theta^m_{23}}\bigg], \\
&\lambda_1=\frac{1}{2}\Big[\lambda_3+\hat{A}+s^2_{13}+\alpha s^2_{12}c^2_{13}
-\frac{\lambda_{3}-\hat{A}-s^2_{13}-\alpha s^2_{12}c^2_{13}}{\cos2\theta^m_{13}}\Big].
\end{align}
Similar to $\tym$, all the five NSI parameters and standard interaction affect the evolution of $\txm$.
\begin{figure}[htb!]
\centering
\includegraphics[scale=0.80]{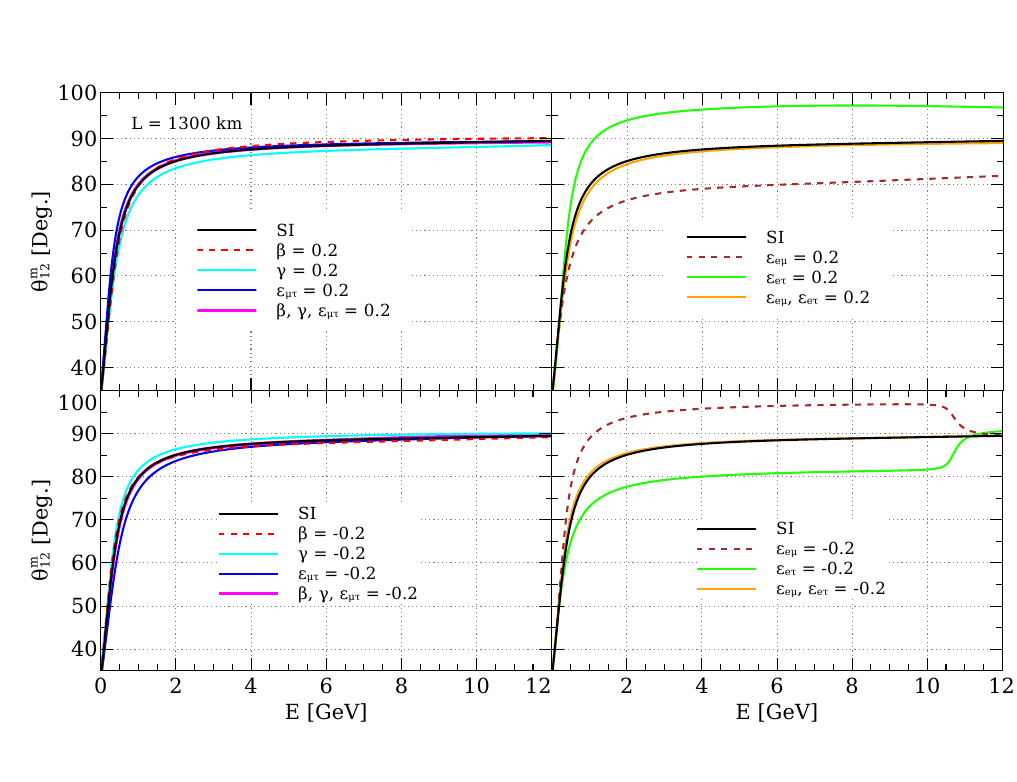} 
\mycaption{Evolution of $\txm$ (given in eq.~\ref{eq:th12_1}) 
with energy in presence of SI and NSI in matter. 
The solid black curve in each panel shows the SI case 
while the other curves correspond to the variation in the 
presence of SI+NSI. In the top (bottom) row, the positive 
(negative) values of the NSI are considered with a 
benchmark value of 0.2 (-0.2). The left column depicts 
the presence of NSI parameters in (2,3) block while 
the right shows the effect of $\eem$ and $\eet$. 
We consider $L$ = 1300 km and assume NMO
to prepare this plot. Values of oscillation parameters 
in vacuum used in this plot are taken from 
Table.~\ref{table:vac_nsi} with $\theta_{23} = 45^{\circ}$.}
\label{fig:th12_1}
\end{figure}
In figure~\ref{fig:th12_1}, we show the evolution of $\txm$ as a function of neutrino energy. For the SI case, at low energy ($E \le 1.5-2$ GeV), we see a very steep increase in the value of $\txm$. As the energy increases, it starts to saturate at around $90^\circ$. We explain this feature using the expression in eq.~\ref{eq:th12_1}, at small energies, 
$\lambda_{1}$ being close to $\lambda_{2}$, $\txm$ shows a very steep increase and then quickly saturates and approaches to $90^{\circ}$ approximately. Saturation occurs primarily for two reasons. First, with increase in energy, $\lambda_{1}$ moves away from $\lambda_{2}$, resulting in a large denominator in the R.H.S. of 
eq.~\ref{eq:th12_1}. Secondly, $\tym$ rises with energy (see Fig.~\ref{fig:th13_1} and the relevant discussions in subsec.~\ref{subsubsec:th13}) and so the overall factor $c_{13}^{m}$ in 
eq.~\ref{eq:th12_1} decreases. In the presence of the NSI parameters from (2,3) block of the NSI matrix, we do not observe any significant deviation from the SI. This is because the contribution from these NSI parameters is canceled out at the denominator at the leading order. In the presence of the off-diagonal NSI parameters $\eem$ and $\eet$, the value at which $\txm$ saturates differs from that of the SI case, which can be explained by their presence in the numerator. When $\eem$ ($\eet$) is positive, the saturation value is less (greater) than the SI case~($\sim 90^\circ$). When both the parameters are present at the same time, their contribution is canceled out, and the value of $\txm$ saturates at the same value as in the SI case. Interestingly, at energies around 10 GeV, we observe a sudden decrease (increase) of $\txm$ 
in the presence $\eem$ ($\eet$) with negative strength. It happens due to the presence of the term $\cos\tym$ in the numerator of the R.H.S. of eq.~\ref{eq:th12_1}, which reduces rapidly to a very small value around that energy (see figure~\ref{fig:th13_1} and related discussion in subsection~\ref{subsubsec:th13}).
		
Unlike $\tym$, the $\txm$ shows similar behavior in SI as well as in SI+NSI cases for neutrino propagation with IMO ($\nu$, IMO). Also, it shows completely different behavior in case of antineutrino propagation with inverted mass ordering ($\bar{\nu}$, IMO). It can be understood from the fact that in case of IMO, sign of first and third terms in the numerator of eq.~\ref{eq:th12_1} gets flipped, and in the denominator, the sign of $\lambda_1$ gets changed. Since the effect from other remaining terms are very small, both numerator and denominator change their sign, and as a result, $\txm$ remains 
the same as in the case of ($\nu$, NMO). In case of ($\bar{\nu}$, IMO), only first term in the numerator changes its sign, $\lambda_1$ in the denominator remains the same as in case of ($\nu$, NMO). As a result, we see a completely different behavior of $\txm$. 
		
\section{Evolution of mass-squared differences}
\label{sec:evol_mass}
Eqns.~[\ref{eq:m3}-\ref{eq:m1}] show the expression of the diagonalized Hamiltonian, which can be written in terms of mass-squares. The expressions of the modified mass-squared differences can be derived from these three equations: $\ldmm \equiv m^2_{3,m} - m^2_{1,m}$ 
and $\sdmm \equiv m^2_{2,m} - m^2_{1,m}$.
\begin{figure}[htb!]
\centering
\includegraphics[scale=0.80]{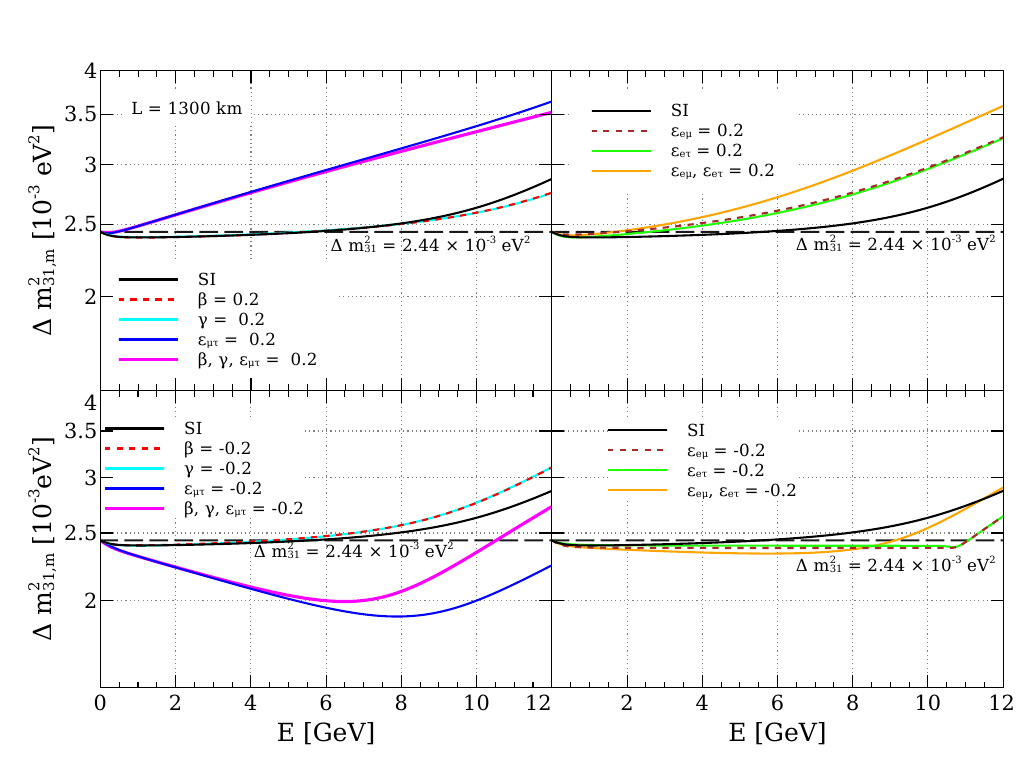}
\mycaption{Variation of $\ldmm$ ($\equiv m^2_{3,m}-m^2_{1,m}$) 
as obtained from eqns.~\ref{eq:m3}-\ref{eq:m1} is shown with energy 
in SI case and SI+NSI cases. Top (bottom) row corresponds 
to the positive (negative) NSI with strength 0.2. The solid black curve 
in each panel shows the SI case while the other curves show the modification in presence of NSI. The left column depicts the presence of 
various NSI parameters in (2,3) block while the right shows the effect 
of $\eem$ and $\eet$. We consider $L = 1300$ km and the values 
of the oscillation parameters used in this plot are taken from 
Table~\ref{table:vac_nsi}. We assume $\theta_{23} = 45^{\circ}$ and NMO.}
\label{fig:delm31}
\end{figure}

In figure~\ref{fig:delm31}, we show the evolution of $\ldmm$ as a function of energy in the SI case as well as in the presence of the NSI parameters. In the SI case, we observe that the magnitude of $\ldmm$ does not change at lower energy~($\le$ GeV). At high energy, we observe an increasing trend and can go up to $3.0\times 10^{-3}$ eV$^2$ at a neutrino energy around 12 GeV. In the presence of $\emt$, we observe a steady increase in the value of $\ldmm$ with energy, while the presence of $\beta$ or $\gamma$ shows a small and similar change in $\ldmm$. The presence of $\eem$ or $\eet$ shows identical effects and makes $\ldmm$ rise with a steeper rate. Both $\eem$ and $\eet$, when present together, generate an additive effect and further elevate the steepness of $\ldmm$. Flipping the sign of NSI parameters shows a reverse effect in the running of $\ldmm$.
\begin{figure}[htb!]
\centering
\includegraphics[scale=0.80]{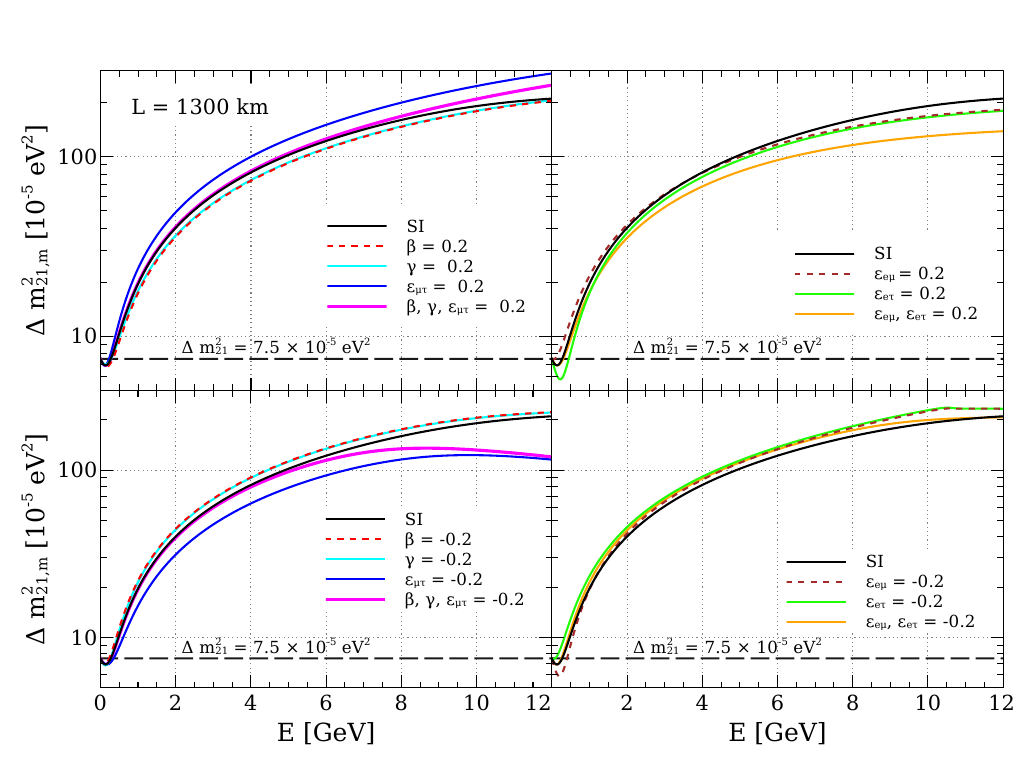}
\mycaption{Variation of $\sdmm$ ($\equiv m^2_{2,m}-m^2_{1,m}$) 
as obtained from Eqs.~\ref{eq:m3}-\ref{eq:m1} is shown with energy 
in SI case and SI+NSI cases. Top (bottom) row corresponds to the 
positive (negative) NSI with strength 0.2. The solid black curve 
in each panel shows the SI case while the other curves show 
the modification in presence of NSI. The left column depicts the presence 
of various NSI parameters in (2,3) block while the right shows the 
effect of $\eem$ and $\eet$. We consider $L = 1300$ km and 
the values of the oscillation parameters used in this plot are taken 
from Table~\ref{table:vac_nsi}. We assume $\theta_{23} = 45^{\circ}$ and NMO.}
\label{fig:delm21}
\end{figure}

Figure~\ref{fig:delm21} shows the evolution of $\sdmm$ as a function of energy in the SI case and in the presence of the NSI parameters. Unlike $\ldmm$, here we observe a huge increase in the value of $\sdmm$ with energy --- goes up to the order of $10^{-3}$ eV$^2$ at an energy around 10 GeV. In other words, the variation of $\sdmm$ can make itself comparable in magnitude with that of $\ldmm$. This is valid in the SI case and also in the presence of the NSI parameters. In the presence of the NSI parameters, we do not observe any significant variation in the feature of the evolution of $\sdmm$. Interestingly, depending on the sign of the NSI parameter, the magnitude of $\sdmm$ in the presence of NSI is slightly higher or lower than in the presence of SI. In the presence of negative $\emt$ (when present singly or together with negative $\beta$ and $\gamma$), we see a deviation from the SI case at higher energy, which can be understood from the variation of $\lambda_1$ with energy. 
With negative $\eem$ or $\eet$, however, at $E\ge$ 10.5 GeV $\sdmm$ becomes almost constant. It happens due to a sudden increase in the value of $\tym$ around that energy 
which results in saturation of the value of $\lambda_1$.

In the case of IMO, the evolution of $\sdmm$ is almost the same as ($\nu$, NMO) case for both neutrino and antineutrino propagation. However, IMO leads to a significant change in the evolution of $\ldmm$ for both neutrino and antineutrino propagation, which is obvious because the vacuum value $\ldm$ changes its sign. 

\section{$\theta_{13}$-resonance in matter and in the presence of NSI}
\label{sec:resonance_appearance_channel}
As discussed in subsection~\ref{subsubsec:th13}, in the presence of matter, the value of $\tym$ reaches the maximal value at some value of neutrino energy, which will have important implications in the neutrino oscillation probability. We term this as $\theta_{13}$-resonance. In this subsection, we estimate the neutrino energy at which $\ty$-resonance occurs in the presence of NSI.

$\theta_{13}$-resonance occurs when the denominator of the eq.~\ref{eq:th13_1} vanishes. Using this condition, one can write
\begin{equation}
\label{eq:resonance}
\hat{A} = \lambda_{3}-\alpha s^{2}_{12} c^{2}_{13} - s^{2}_{13} \,.
\end{equation}
Since the expression for $\lambda_3$ only contains NSI parameters from the (2,3) block, only these three parameters would affect the resonance and modify the resonance energy compared to the SI case. We know that for the SI case, under the one mass scale dominance
(OMSD) approximation ($\Delta m^2_{31}L/4E>>\Delta m^2_{21}L/4E$), 
the resonance occurs at an energy $E_{\text{res}}$ 
such that~\cite{Akhmedov:2004ny},
\begin{equation}
\label{eq:Eres_SM}
\left[E_{\text{res}}^{\text{SI}}\right]_\text{OMSD} = \frac{\Delta m^2_{31}\cos2\theta_{13}}{2V_{CC}} \,.
\end{equation}
Now, we discuss how this expression would change in the presence of NSI. Again, we consider $\theta_{23} = 45^{\circ}$. First, we simplify the expression of $\lambda_3$, replacing $\cos2\theta^m_{23}$ from eq.~\ref{eq:th23m}. 
we obtain,
\begin{align}
\lambda_3&\simeq\frac{1}{2}\bigg[c^2_{13}+\alpha c^2_{12}+\alpha s^2_{12}s^2_{13}+(\beta+\gamma)\hat{A} \nonumber \\
&+\sqrt{\{\alpha s_{13}\sin2\theta_{12}+(\gamma-\beta)\hat{A}\}^2+\{c^2_{13}-\alpha c^2_{12}+\alpha s^2_{12}s^2_{13}+2\varepsilon_{\mu\tau}\hat{A}\}^2}\bigg] \,.
\end{align}
Here, we neglect the small terms 
proportional to $\alpha s^2_{13}, (\eff)^{2} \hat{A}^{2}$ 
and the cross-term proportional to $\alpha \hat{A} (\eff) s_{13}$. 
We get the following simpler expression for $\lambda_{3}$, 
\begin{align}  
\label{eq:lmda3_3}    
\lambda_{3} \simeq c^2_{13}+\frac{1}{2}(\beta+\gamma+2\varepsilon_{\mu\tau})\hat{A} \,.
\end{align}
Now, equating Eqs.~\ref{eq:resonance} and \ref{eq:lmda3_3},
we obtain a simple expression for 
the resonance energy in the presence of NSI;
\begin{equation}
\label{eq:res}
E_{\text{res}}^{\text{NSI}} \simeq \frac{\Delta m^2_{31}\cos2\theta_{13}}{2V_{CC}} \bigg[\frac{1-(\alpha s^2_{12} c_{13}^{2} /\cos 2\theta_{13})}{1- \frac{1}{2}(\beta+\gamma+2\varepsilon_{\mu\tau})}\bigg] 
= \left[E_{\text{res}}^{\text{SI}}\right]_\text{OMSD} \bigg[\frac{1-(\alpha s^2_{12} c_{13}^{2} /\cos 2\theta_{13})}{1- \frac{1}{2}(\beta+\gamma+2\varepsilon_{\mu\tau})}\bigg] .
\end{equation}
In the absence of any NSI parameter, we get back the resonance energy expression in the SI case shown in eq.~\ref{eq:Eres_SM}.
The term $\frac{1}{2}(\beta+\gamma+2\varepsilon_{\mu\tau})$ 
is the correction induced by the presence of NSI, 
while $\alpha s^2_{12} c_{13}^{2} /\cos 2\theta_{13}$ 
is the modification induced by relaxing the OMSD approximation. 
As discussed earlier, energy at $\theta_{13}$-resonance only depends on NSI parameters from the (2,3) block of the NSI matrix and is independent of $\eem$ and $\eet$. We made a similar observation in figure~\ref{fig:th13_1}, where we observe that the line corresponds to the $\tym$ running in the presence of $\eem$ or $\eet$ intersects with the corresponding SI line when the value of $\tym$ is $45^\circ$, signaling $\ty$-resonance is independent of these two NSI parameters. 
Also, we observe resonance energy depends on the matter density.
\begin{figure}[htb!]
\centering
\includegraphics[width=0.75\textwidth]{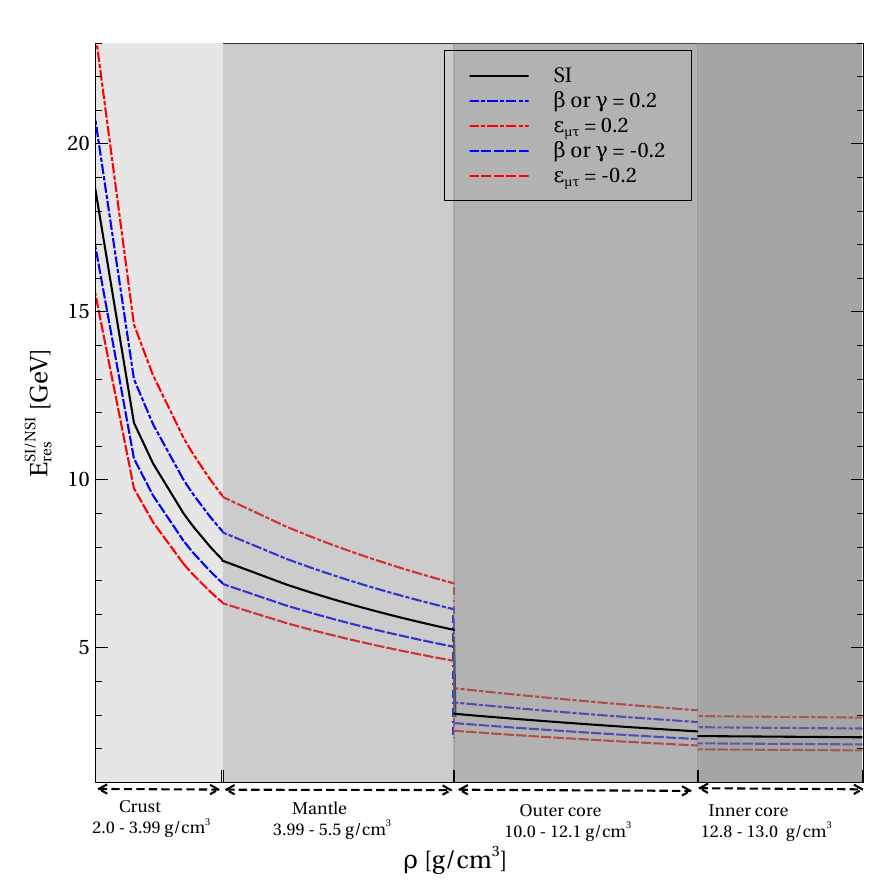}
\mycaption{The energy at $\theta_{13}$-resonance 
(see eq.~\ref{eq:res}) as a function of the Earth matter density $\rho$. The solid black line shows the SI case. The 
dot-dashed (dashed) lines correspond positive (negative)
values of the NSI parameters considering one NSI parameter at-a-time, as shown in the legend. The four grey shaded regions with varying intensity show the four different layers (crust, mantle, outer core, and inner core) and their corresponding densities inside the Earth~\cite{Dziewonski:1981xy}.
 The values of the benchmark oscillation parameters 
used in this plot are taken from Table~\ref{table:vac_nsi}. 
We assume $\theta_{23} = 45^{\circ}$ and NMO.}
\label{fig:res}
\end{figure}

In figure~\ref{fig:res}, we plot the energy at $\ty$-resonance as a function of the density profile of the Earth matter following Preliminary Reference Earth Model~(PREM). Neutrino energy required to attend $\theta_{13}$-resonance decreases as neutrinos travel deeper inside the Earth. The step-like behavior at the mantle-core boundary and outer-inner core boundary is due to the jump in the line-averaged matter density at those boundaries in the PREM profile. The presence of NSI in (2,3) block with positive (negative) strength reduces (increases) the resonance energy as compared to that of the SI case for a given matter density.
\section{Impact of NSI in $\nu_{\mu}\to\nu_e$ oscillation channel}
\label{sec:nsi_appearance_channel}
One of the prominent oscillation channel in long-baseline (LBL) neutrino experiments is the $\mue$ appearance channel. This channel plays a vital role in the outcome of the currently operating LBL experiments like T2K, NO$\nu$A. Also, this channel will play a dominant role in the upcoming experiments like DUNE and Hyper-K LBL setup, in the value of 
the CP phase, neutrino mass ordering, octant of $\theta_{23}$. In this subsection, we explore how standard and non-standard matter effect in the presence of NSI would affect the $\mue$ transition probability.
\begin{figure}[htb!]
\centering
\includegraphics[scale=0.95]{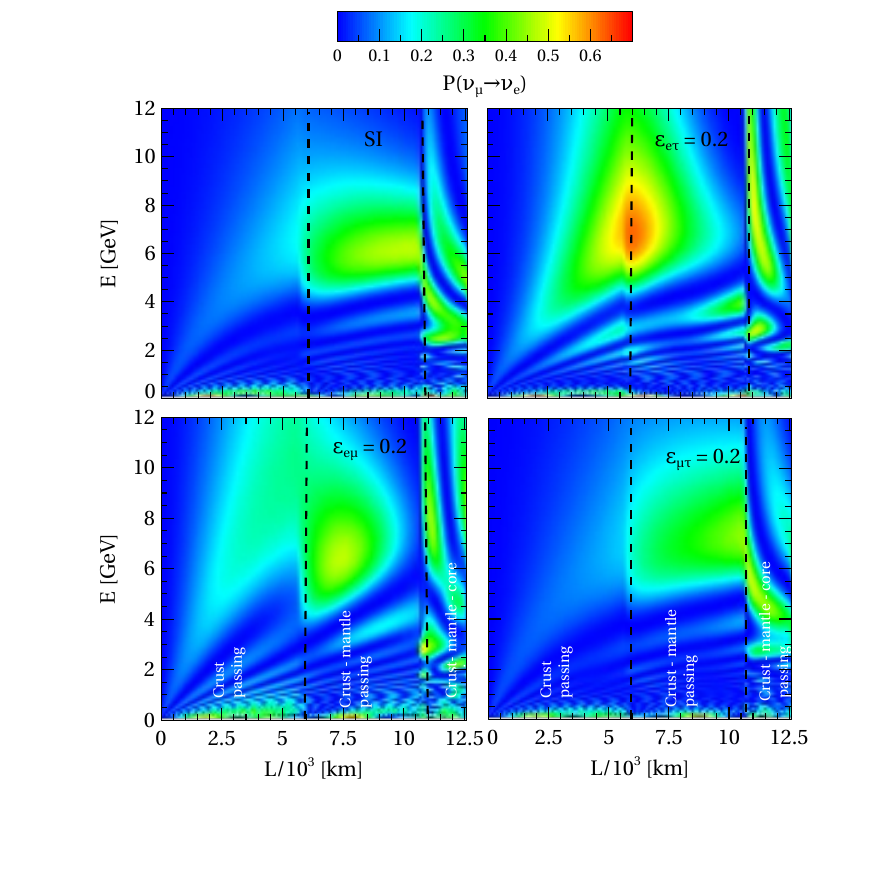}
\vspace*{-1.0cm}
\mycaption{Oscillograms of $\nu_{\mu} \rightarrow \nu_e$ 
transition probability as a function of baseline $L$ and 
energy $E$. Top left panel corresponds to SI case and 
other three panels correspond to the cases in presence 
of non-zero positive NSI parameters (taken one-at-a-time 
with a strength of 0.2 as shown in the legends). 
The values of the oscillation parameters used in this plot 
are taken from Table~\ref{table:vac_nsi} with 
$\theta_{23} = 45^{\circ}$ and NMO.}
\label{fig:osc_res_max}
\end{figure}

To discuss this, in figure~\ref{fig:osc_res_max}, we plot $\mue$ appearance probability as a function of neutrino energy $E$ and the baseline $L$ in the SI case as well as in the SI+NSI cases considering the off-diagonal NSI parameters one-at-a-time with strength 0.2. 
To calculate the oscillation probabilities, we first calculate the oscillation amplitude for each layer of the Earth, considering the well-known four-layered profile of the Earth. Density variation in each layer is taken from the PREM profile of the Earth. We multiply the oscillation amplitude through which neutrinos travel and calculate the oscillation probability.
We check that Fig.~\ref{fig:osc_res_max} shows 
very good agreement in both SI and SI+NSI cases with the 
exact three-flavor oscillation probabilities, which are calculated 
numerically using the GLoBES software~\cite{Huber:2004ka, Huber:2007ji}.
In the SI case, appearance probability is large inside the region with baselines passing through the crust+mantle of the Earth and having energy in~$\sim[4,9]$~GeV, where there is a large matter effect. We also observe oscillation maxima in the core passing baselines. In the presence of $\emt$, we do not observe any notable variation from the SI case, except the regions of oscillation maxima broaden. However, in presence of $\varepsilon_{e\mu}$ or $\varepsilon_{e\tau}$ this region shifts towards lower baselines. 
\begin{figure}[htb!]
\centering
\includegraphics[scale=0.9]{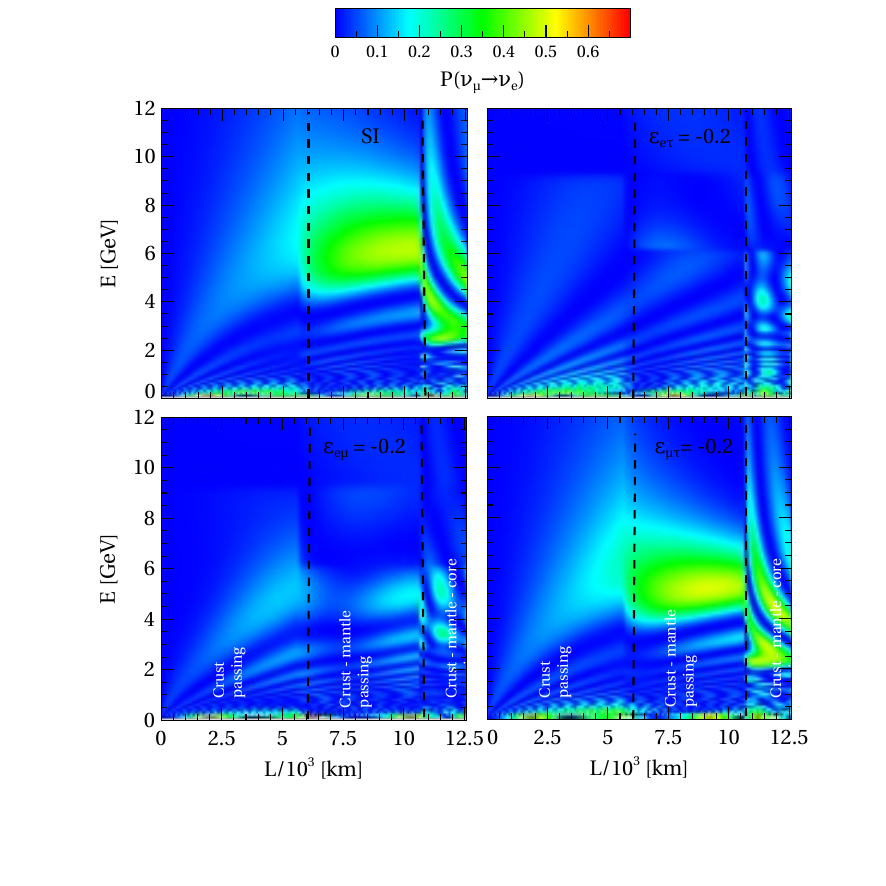}
\vspace*{-1.0cm}
\mycaption{Oscillogram of $\nu_{\mu} \rightarrow \nu_e$ 
transition probability as a function of baseline $L$ and 
energy $E$. Top left panel corresponds to SI case and 
other three panels correspond to the cases in presence 
of non-zero negative NSI parameters (taken one-at-a-time 
with a strength of 0.2 as shown in the legends). 
The values of the oscillation parameters used in this plot 
are taken from Table~\ref{table:vac_nsi} with 
$\theta_{23} = 45^{\circ}$ and NMO. }
\label{fig:osc_res_max_neg}
\end{figure}
To see the impact of the NSI parameters with negative strength, in figure~\ref{fig:osc_res_max_neg}, we plot the oscillogram with the negative strength of the off-diagonal NSI parameters. In the case of $\emt$ with strength -0.2, we observe a marginal change in the oscillogram with the regions of the oscillation maximum shrinking by a small amount. However, in the case of $\emt$ and $\eet$, we observe a drastic change in the oscillation pattern. The regions of the oscillation maximum completely vanished for $\varepsilon_{e\beta}=-0.2$~($\beta = \mu,\tau$), having almost no flavor oscillation around that region.

We explain this interesting observation analytically using the expressions of the modified oscillation parameters we derived in this section. First we write the expression $\mue$ appearance probability in presence of matter interactions, by replacing the vacuum oscillation parameters with the corresponding modified parameters. From figure~\ref{fig:th12_1}, we observe that at relevant energies, $\txm$ saturates to $90^\circ$. So, using the approximation $\txm\rightarrow 90^\circ$, $\mue$ appearance probability simplifies to
\begin{equation}
\label{eq:pme2}
P(\nu_{\mu}\rightarrow \nu_{e}) = \underbrace{\sin^2\theta_{23}^{m}}_{T_1} \,\, \underbrace{\sin^{2}2\theta^m_{13}}_{T_2} \,\, \underbrace{\sin^{2}\bigg[\frac{1.27\times\Delta m^{2}_{32,m} L}{E}\bigg]}_{T_3} \,.
\end{equation}
We have check that, dominant contribution in the $\mue$ oscillation probability in the presence of SI and SI+NSI comes from this term. Now using eq.~\ref{eq:pme2}, we explain our observation in figures~\ref{fig:osc_res_max} and~\ref{fig:osc_res_max_neg}. $\nu_e$ appearance probability would be maximum when the combination of $T_1$, $T_2$, and $T_3$ is maximum. 
\begin{figure}[htb!]
\centering
\includegraphics[scale=0.9]{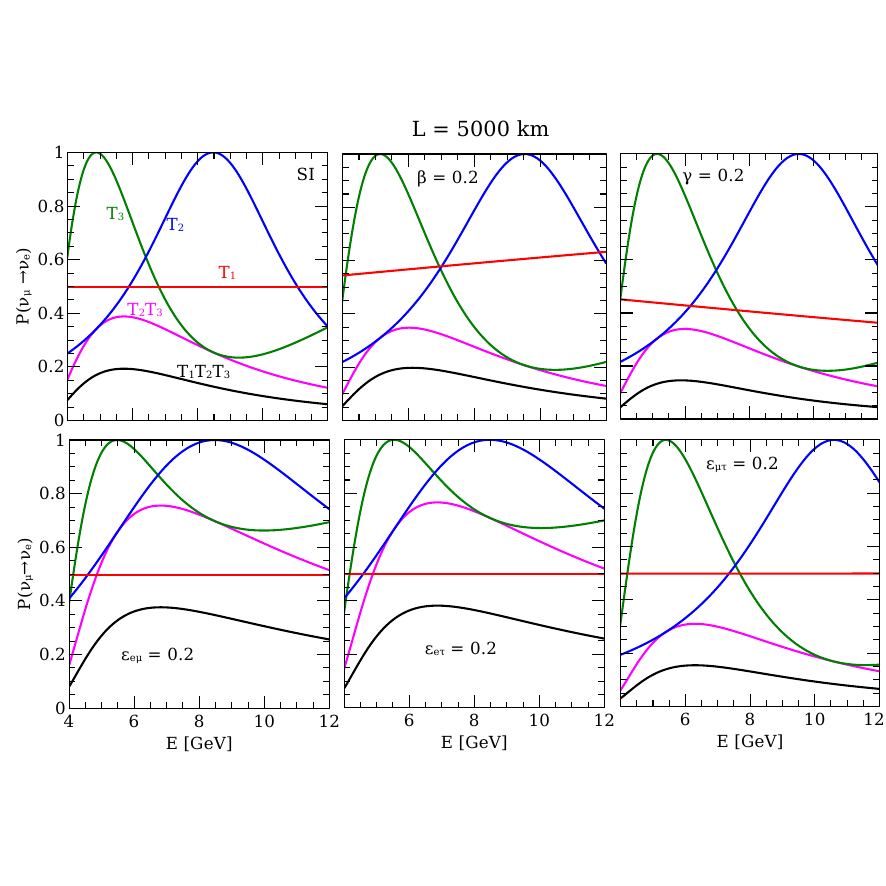}
\vspace*{-1.5cm}
\mycaption{Variation of $\nu_{\mu}\rightarrow\nu_{e}$ 
transition probability (Eq.~\ref{eq:pme2}) with energy 
under the approximation $\txm\rightarrow 90^{\circ}$ 
for a baseline of $L$ = 5000 km. $T_1$ (red curve), 
$T_2$ (blue curve), and $T_3$ (green curve) are the 
three terms defined in eq.~\ref{eq:pme2}. Various panels 
represent the SI case and SI+NSI cases as shown 
in the labels. To prepare this plot, the values of the
three-flavor oscillation parameters are taken from
Table~\ref{table:vac_nsi}. We assume $\theta_{23} = 45^{\circ}$ 
and NMO.}
\label{fig:Pmue_approx}
\end{figure}
In figure~\ref{fig:Pmue_approx}, we plot the term $T_1$, $T_2$, and $T_3$ separately, as well as their product for a illustrative baseline of 5000 km. Among the three terms, we observe $T_1$ shows the monotonic but least variation with energy in the presence of NSI parameters. So, the oscillation maximum is determine by the compbination $T_2\times T_3$ which can be seen in this figure --- maxima of $T_2\times T_3$ and $T_1\times T_2\times T_3$ are same. In other words, the maximum of 
$P^{m}_{\nu_{\mu}\rightarrow\nu_{e}}$ occurs at an energy 
determined by $T_2$ and $T_3$, not $T_1$. 
This feature is also valid for any other baselines.
So, $P^{m}_{\nu_{\mu}\rightarrow\nu_{e}}$ is maximum when 
both the following two conditions are satisfied simultaneously.
\begin{itemize}

\item 
$T_2 \equiv \sin^2\theta^m_{13} = 1$ \textit{i.e.},  $\tym = 45^{\circ}$ 
($\theta_{13}$-resonance condition). This condition is achieved in SI case
when $E = E_{\text{res}} =  \frac{\Delta m^2_{31}\cos2\theta_{13}}{2V_{CC}}$ 
(see eq.\ \ref{eq:Eres_SM}) with the OMSD approximation.

\item 
$T_3\equiv \sin^{2}\bigg[\frac{1.27\times\Delta m^{2}_{32,m} L}{E}\bigg] = 1$ 
for some energy $E=E^m_{\text{max}}$, such that
\begin{equation}
\label{eq:Emax}
E^m_{\text{max}} = \frac{1.27\times\Delta m^{2}_{32,m} L}{(2n+1)\pi/2} \quad \text{ with } n = 0, 1, 2...
\end{equation}

\end{itemize}
Thus, the condition $E_{\text{res}} = E^m_{\text{max}}$ gives the maximum matter effect, consequently the oscillation maximum~\cite{Banuls:2001zn,Gandhi:2004md,Gandhi:2004bj}. 

Now, we calculate the above condition for the SI case. First, we simplify the expression for $\Delta m^{2}_{32,m}$, applying OMSD condition and considering $\tz=45^\circ$. We get
\begin{equation}
\label{eq:delm32m_SI}
\Delta m^2_{32,m} = \Delta m^2_{31}\sqrt{(\lambda_3-\hat{A}-s^2_{13})+\sin^22\theta_{13}} \,.
\end{equation}
Now, using eq.~\ref{eq:delm32m_SI} in the expression 
of $E^m_{\text{max}}$ in eq.~\ref{eq:Emax}, the condition 
for the oscillation maximum $E_{\text{res}} = E^m_{\text{max}}$ 
is simplified further. Finally, we obtain a simple and compact
relation between the baseline ($L$) and the corresponding
matter density ($\rho$) where
maximum $\nu_{\mu}\rightarrow\nu_{e}$ transition 
probability in matter will occur~\cite{Gandhi:2004md},
\begin{equation}
\big(\rho \times L \big)_{\text{SI}} = \frac{(2n+1)\times\pi\times5.18\times10^3}{\tan2\theta_{13}} \hspace{0.1cm} \text{km g/cm$^3$} \,.
\label{eq:rho_l}
\end{equation} 

The above expression in eq.~\ref{eq:rho_l} is already defined in ref.~\cite{Gandhi:2004md}. Note that under the OMSD approximation, the resonance 
energy condition in eq.~\ref{eq:resonance} takes a very
simple form: $(\lambda_3 - \hat{A} -s^2_{13}) = 0$ and 
we make use of this expression in eq.~\ref{eq:delm32m_SI}
to obtain eq.~\ref{eq:rho_l}, which exactly matches with the
expression derived by the authors in ref.~\cite{Gandhi:2004md}.
In this subsection, we derive the expression is modified in the presence of the NSI parameters.
First, we use eqs.~\ref{eq:lambda_3}$-$\ref{eq:lambda_1} 
and Eqs.~\ref{eq:m3}$-$\ref{eq:m1} to derive the following two expressions 
for $m^{2}_{3,m}$ and $m^{2}_{2,m}$ under the OMSD approximation
and assuming $\theta_{23} = 45^{\circ}$.
\begin{align}
m^{2}_{3,m} &= \frac{\ldm}{2}\big[ 
\lambda_{3} + \hat{A} + s^{2}_{13} + T
\big] \,, \nonumber \\
m^{2}_{2,m} &= \frac{\ldm}{2}\big[ 
\lambda_{3} + \hat{A} + s^{2}_{13} - T
\big] \,, 
\end{align}
where,
\begin{equation}
T = \frac{1}{\sqrt{2}}\sqrt{2\big[\lambda_{3} - \hat{A} - s^{2}_{13}\big]^{2} 
+ \big[\sin 2\theta_{13}(c^{m}_{23}+s^{m}_{23}) + 2\sqrt{2}(\eem s^{m}_{23} + \eet c^{m}_{23})\hat{A} \big]^{2}} \,.
\end{equation}
Now, using the resonance energy condition (eq.~\ref{eq:resonance}), we get
\begin{align}
\Delta m^{2}_{32,m} = m^{2}_{3,m} - m^{2}_{2,m} 
= \ldm \big[ \frac{1}{\sqrt{2}}
\sin 2\theta_{13}(c^{m}_{23}+s^{m}_{23}) + 2(\eem s^{m}_{23} + \eet c^{m}_{23})\hat{A}
\big] \,.
\end{align}
Replacing $\Delta m^{2}_{32,m}$ in eq.~\ref{eq:Emax}, 
we finally have the following condition for the 
maximal $\nu_{\mu} \rightarrow \nu_{e}$ transition probability
in the presence of NC-NSI.
\begin{align}
\label{eq:rho_l_nsi}
&\big(\rho \times L \big)_{\text{NSI}}
\nonumber\\ 
& \simeq \frac{(2n+1)\pi \times 5.18 \times 10^{3}}{\tan 2\theta_{13}\big[\big\{1-\frac{1}{2}(\beta+\gamma+2\varepsilon_{\mu\tau})\big\}(\frac{c^{m}_{23}+s^{m}_{23}}{\sqrt{2}}) + \big\{2(\eem s^{m}_{23}+\eet c^{m}_{23})/\tan 2\theta_{13}\big\}\big]} \hspace{0.1cm} \text{km g/cm$^3$}\\
&=   \big(\rho\times L\big)_{\text{SI}} \bigg[ \frac{1}{\big\{1-\frac{1}{2}(\beta+\gamma+2\varepsilon_{\mu\tau})\big\}(\frac{c^{m}_{23}+s^{m}_{23}}{\sqrt{2}}) + \big\{2(\eem s^{m}_{23}+\eet c^{m}_{23})/\tan 2\theta_{13}\big\}}\bigg]\text{km g/cm$^3$}.  
\end{align} 
In the absence of any NSI parameters, we get back the same expression for SI case as in eq.~\ref{eq:rho_l}. The scond term in the denominator is the corrections from the NSI parameters in the (2,3) block of the NSI matrix, and the third term is the correction from $\eem$ and $\eet$. Since, value of the $\theta^m_{23}$ does not vary significantly from the $45^\circ$ (see figure~\ref{fig:th23_1}), we simplify the above equation further assuming $\tzm=45^\circ$ to get
\begin{align}    
&\big(\rho \times L \big)^{max}_{\text{NSI}} 
\simeq  \big(\rho\times L \big)_{\text{SI}}  \bigg[ \frac{1}{1-(\beta+\gamma+2\varepsilon_{\mu\tau})/2 + \sqrt{2}(\eem +\eet )/\tan 2\theta_{13}}\bigg]\text{km g/cm$^3$} \,.
\label{eq:rho_l_nsi_maximal}        
\end{align}  
\begin{figure}[htb!]
\centering
\includegraphics[width=00.75\textwidth]{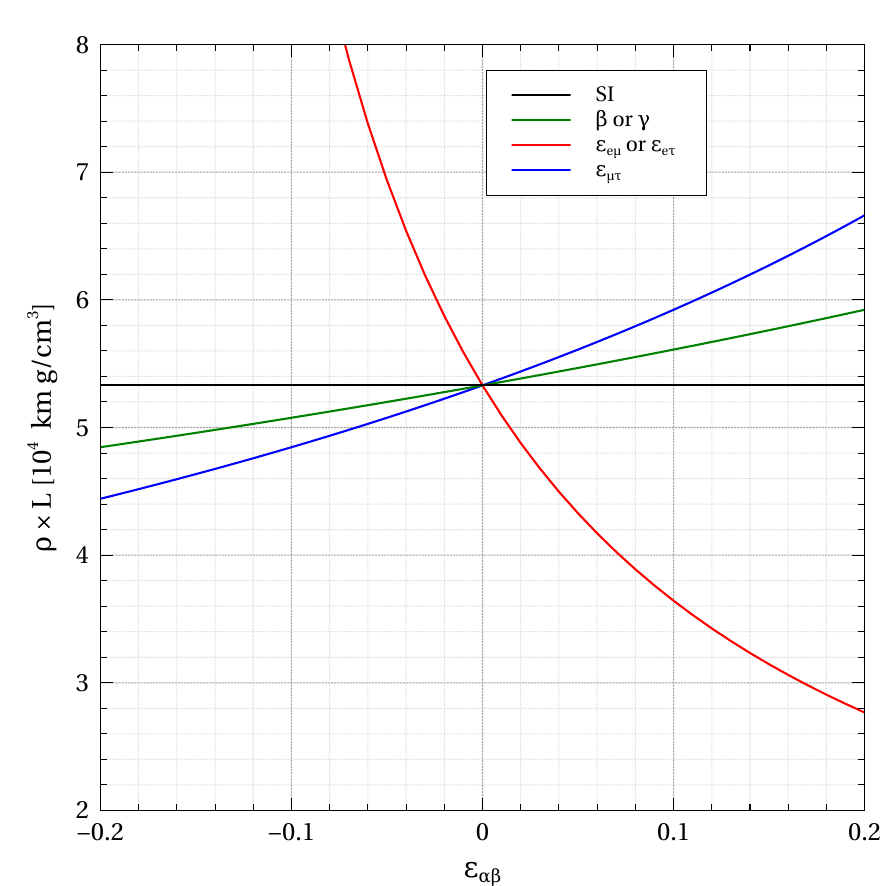}
\mycaption{\label{fig:rhol_cond}The product of the baseline ($L$) and the corresponding matter density ($\rho$) for which $\nu_{\mu}\rightarrow\nu_{e}$ transition probability 
attains the maximum value (see eq.~\ref{eq:rho_l_nsi_maximal}) as a function of various NSI parameters having strength in the range [-0.2, 0.2]. The solid black curve shows the SI case while the other colored curves correspond to the cases considering one NSI parameter at-a-time, as shown in the legend. The three-flavor oscillation parameters in vacuum 
are taken from Table~\ref{table:vac_nsi} with a choice 
of $\theta_{23} = 45^{\circ}$ and NMO.}
\label{fig:osc_max}
\end{figure}
In figure~\ref{fig:rhol_cond}, we plot the L.H.S. of eq.~\ref{eq:rho_l_nsi_maximal}~($\rho\times L$) by varying each each NSI parameters one-at-a-time in the range [-0.2, 0.2]. In the absence of any NSI (SI case), $\rho\times L$ for maximum $\mue$ oscillation probability is constant as shown by the solid black line. In the presence of the NSI parameter $\beta$ or $\gamma$, we see an increase in the value of $\rho\times L$ for maximum oscillation as the strength of the NSI parameter goes from  -0.2 to 0.2, which is clear from eq.~\ref{eq:rho_l_nsi_maximal}. $\emt$ shows similar behavior, except the slope is less, as compared to the $\beta$ or $\gamma$, due to the presence of a factor of 2 with $\emt$ in the denominator. However, in the presence of $\eem$ or $\eet$, we observe a completely opposite behavior; $\rho\times L$ shows a steep decrease as $\eem$ or $\eet$ varies from -0.2 to 0.2. The steep decreasing behavior can be explained by the factor $1/\tan 2\theta_{13}$ with $\eem$ or $\eet$ in the denominator with the minus sign in eq.~\ref{eq:rho_l_nsi_maximal}. So, it is clear that when $\eem$ or $\eet$ are positive with a strength around 0.2, the required value of $\rho\times L$ would be much smaller than that of the SI case. We observe the same feature in figure.~\ref{fig:osc_res_max}, where the region of maximum $\nu_{\mu}\rightarrow\nu_{e}$ transition in the presence of positive $\eem$ or $\eet$, shift towards the shorter baseline as compared to the SI case. In the case of negative $\eem$ or $\eet$, the required value of $\rho\times L$ for maximum $\nu_\mu\to\nu_e$ transition is so large that it is not attainable inside the Earth. 

It is evident from eq.~\ref{eq:rho_l_nsi_maximal} that since the role of $\eem$ and $\eet$ are on the same footing, the presence of $\eem$ induces an effect identical to that of $\eet$ with the same magnitude. However, the oscillograms (Figs.~\ref{fig:osc_res_max} and \ref{fig:osc_res_max_neg}) 
for $\eem$ and $\eet$ show different behavior in the presence of $\eem$ and $\eet$. This is because of the fact that 
$\txm$ saturates to a value higher or lower than $90^{\circ}$ in the presence 
of $\eem$ or $\eet$ (see figure~\ref{fig:th12_1}). Since $\txm$ is not exactly 
$90^{\circ}$, we have non-zero contributions from some other terms in
$\nu_{\mu} \to \nu_e$ oscillation probability expression, which affects the 
oscillograms in the presence of $\eem$ and $\eet$ in a different fashion.
In the presence of negative $\eem$ and $\eet$, we observe from figure.~\ref{fig:osc_res_max_neg} that we no longer achieve the maximum transition in $\nu_{\mu} \to \nu_e$ oscillation channel. It is because of the fact that in this case, the baseline length required for the maximum $\nu_{\mu} \rightarrow \nu_e$ appearance probability turns out to be longer than the Earth's diameter (see eq.~\ref{eq:rho_l_nsi_maximal}). Therefore, it is not possible to attain the maximum $\nu_{\mu} \rightarrow \nu_e$ transition inside the Earth for negative values of $\eem$ and $\eet$ as evident from figure.~\ref{fig:osc_res_max_neg}.
We observe from fig.~\ref{fig:osc_res_max} and fig.~\ref{fig:osc_res_max_neg} that in the presence of non-zero $\emt$, there are slight changes in $L$ and $E$ as compared to SI case for which we obtain maximum possible $\nu_{\mu} \rightarrow \nu_e$ transition.

\section{Impact of NSI in $\nu_{\mu}\to\nu_\mu$ disappearance channel channel}
\label{sec:nsi_disappearance_channel}
In the previous subsection, we discuss the impact of NSI parameters in the $\mue$ appearance channel. In LBL experiments, there is another crucial channel that plays an important role in determining the values of atmospheric oscillation parameters with high precision and is going to play an important role in neutrino mass ordering. In this subsection, we explore the impact of NSI parameters in the $\mumu$ disappearance channel. In this channel, NSI parameters from the (2,3) block of the NSI  matrix have the most significant influences. So, here we focus on the impact of these three NSI parameters ($\beta$, $\gamma$, and $\emt$). To understand the dominant features, we simplify the expression for $\mumu$ disappearance probability in vacuum assuming $\Delta_{21}\simeq 0$ and $\theta_{13}\simeq 0$, which gives 
\begin{align}
P(\nu_{\mu}\rightarrow\nu_{\mu})_{\rm vac}=1-\sin^22\theta_{23}\sin^2\bigg[\frac{\Delta m^2_{31}L}{4E}\bigg].
\label{eq:surv_1}
\end{align}
Now, we replace the vacuum oscillation parameters with the corresponding modified 
parameters in the presence of SI and NC-NSI, assuming the line-averaged constant
Earth matter density. Thus, eq~\ref{eq:surv_1} takes the form:
\begin{align}
P(\nu_{\mu}\rightarrow\nu_{\mu})= 1-\sin^22\theta^m_{23}\sin^2\bigg[\frac{\Delta m^2_{31,m}L}{4E}\bigg].
\label{eq:surv_mat}
\end{align}
Now, using OMSD approximation and $\theta_{13}\simeq 0$ in eq\ \ref{eq:th23m}, also implementing $\theta_{23}=45^\circ$, we simplify the expression of $\sin^22\tzm$ in above equation,
\begin{align}
\sin^22\theta^m_{23}=\frac{(1+2\varepsilon_{\mu\tau}\hat{A})^2}{[(\gamma-\beta)\hat{A}]^2+[1+2\varepsilon_{\mu\tau}\hat{A}]^2}\simeq\bigg[1-\frac{(\gamma-\beta)^2\hat{A}^2}{(1+2\emt\hat{A})^2}\bigg]. 
\label{eq:th23m_omsd}
\end{align}
To calculate the expressions for $\Delta m^2_{31,m}( = m^2_{3,m}-m^2_{1,m})$ we use  eq\ \ref{eq:m3} and eq\ \ref{eq:m1} and implement all the approximations. After simplification, we obtain,
\begin{align}
\Delta m^2_{31,m} = \ldm [\lambda_3-\lambda_2]
\simeq \ldm\bigg[1+2\emt\hat{A}+\frac{1}{2}\frac{(\eff)^2\hat{A}^2}{(1+2\emt\hat{A})}\bigg],
\label{eq:m31m}
\end{align}
where, we use the approximation $\theta^m_{12}\rightarrow\pi/2$ in the expression of $m^2_{1,m}$ in eq\ \ref{eq:m1}.
Now, implementing Eqs.\ \ref{eq:th23m_omsd} and \ref{eq:m31m} in eq~\ref{eq:surv_mat}, the simplified expression for $\nu_{\mu}\rightarrow\nu_{\mu}$ disappearance probability in presence of NSI parameters from (2,3) sector can be written as,
\begin{align}
P({\nu_{\mu}\rightarrow\nu_{\mu}})  = &1-\left[1-\frac{(\gamma-\beta)^2\hat{A}^2}{(1+2\varepsilon_{\mu\tau}\hat{A})^2}\right]\times\sin^2\left[\left\{1+2\varepsilon_{\mu\tau}\hat{A}+\frac{1}{2}\frac{(\gamma-\beta)^2\hat{A}^2}{(1+2\varepsilon_{\mu\tau}\hat{A})}\right\}\frac{\Delta m^2_{31}L}{4E}\right] \nonumber \\
=&\cos^2\left[\left\{1+2\varepsilon_{\mu\tau}\hat{A}+\frac{1}{2}\frac{(\gamma-\beta)^2\hat{A}^2}{(1+2\varepsilon_{\mu\tau}\hat{A})}\right\}\frac{\Delta m^2_{31}L}{4E}\right]\nonumber\\
&\hspace{5cm}+\frac{(\gamma-\beta)^2\hat{A}^2}{(1+2\varepsilon_{\mu\tau})^2}\times\sin^2\left[(1+2\varepsilon_{\mu\tau}\hat{A})\frac{\Delta m^2_{31}L}{4E}\right].
\label{eq:surv_3}
\end{align}
If we only consider the off-diagonal NSI parameter $\emt$, 
the expression boils down to the simplified expression already 
derived in~\cite{Mocioiu:2014gua}. From the approximate expression 
in eq.~\ref{eq:surv_3}, some broad features about the impact of NSI 
on the $\nu_{\mu}\rightarrow\nu_{\mu}$ survival channel can be observed.
We see that the parameter $(\eff)$ always appears in second order in 
eq.~\ref{eq:surv_3}, while other NSI parameter $\emt$ has a linear 
dependence. For the same reason, the sign of $(\eff)$, unlike the sign 
of $\emt$, does not affect the disappearance probability. 
Since the strength of NSI parameters is not very large, 
it is expected that the impact of $(\eff)$ will be always 
small compared to $\emt$.
\begin{figure}[h!]
\vspace{-1cm}
\centering
\includegraphics[scale =0.9]{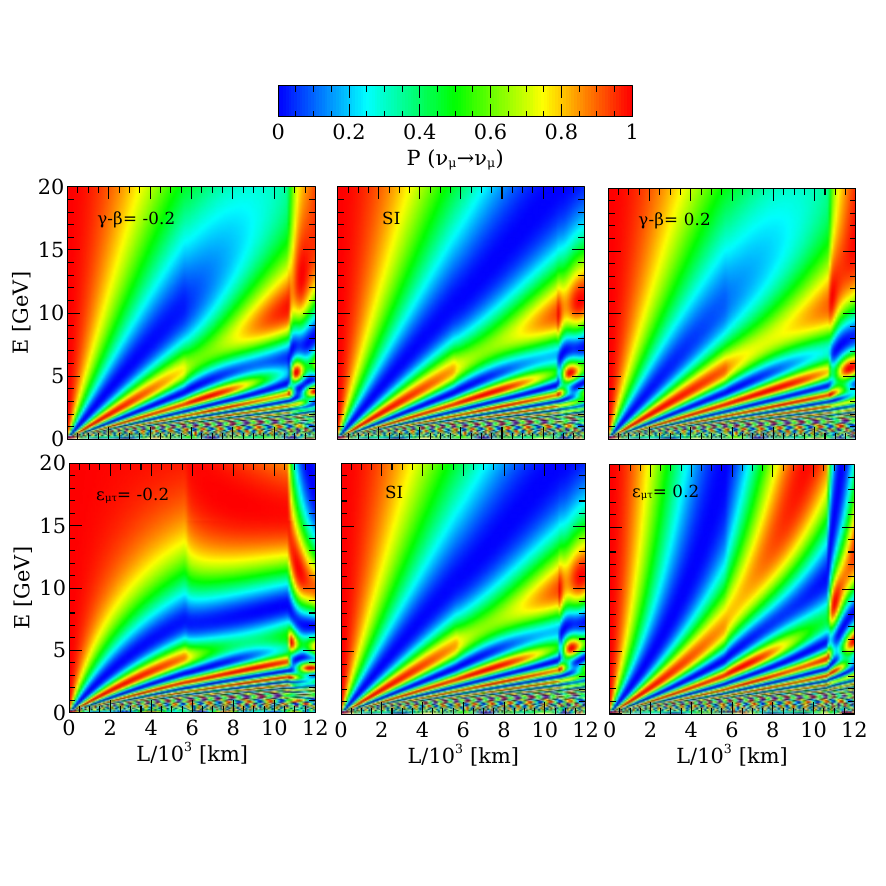}
\vspace*{-1.5cm}
\mycaption{Oscillogram of $\nu_{\mu}\rightarrow\nu_{\mu}$ 
survival probability as a function of baseline $L$ and energy $E$. 
Top and bottom rows correspond to the oscillogram in presence 
of NSI parameter $(\eff)$ and $\varepsilon_{\mu\tau}$, respectively. 
The middle column in both the rows shows the SI case. 	
The first (third) column depicts the presence of negative (positive) 
NSI with a strength of 0.2. Values of the oscillation parameters 
are taken from Table~\ref{table:vac_nsi}. We assume 
$\theta_{23}=45^{\circ}$ and NMO.}
\label{fig:surv_1}
\end{figure}

As mentioned earlier, the above expressions are derived under the OMSD approximation and $\ty\to 0$. To check whether these observations remain intact in the true three-flavor scenario with non-zero $\theta_{13}$, in figure~\ref{fig:surv_1}, we plot oscillogram for the $\nu_{\mu}\rightarrow\nu_{\mu}$ channel in the baseline (x-axis) and energy (y-axis) plane. To calculate the survival probability, we first consider the full three-flavor vacuum expression of  $\nu_{\mu}\rightarrow\nu_{\mu}$ probability without any approximation~\cite{Agarwalla:2013tza} and replace the vacuum parameters with their modified expressions in matter with NSI which we derive in this work. The top panels shows the effects in the presence of the NSI parameters ($\gamma-\beta$) compared to the SI case. We observe some differences in the oscillation pattern from the SI case, at large neutrino energy and a large baseline. However, as expected from the analytical expression in eq.~\ref{eq:surv_3}, we observe almost no change in the oscillation pattern when sign of the parameter $(\gamma-\beta)$ is flipped, except there are small changes at high energy, which appear due to non-zero $\theta_{13}$, that brings the matter effect into the picture and finite value of $\Delta_{21}$ causes some differences. In the presence of $\emt$, we observe a significant change in the oscillation pattern compared to the SI case. The oscillation valley --- denoted by the diagonal blue shaded region (where $\nu_{\mu}\rightarrow\nu_{\tau}$ transition probability in vacuum is maximum and because of that $\nu_{\mu}\rightarrow\nu_{\mu}$ survival probability is minimum), bends towards higher or lower energy for a given baseline depending of the sign of $\emt$. As we see discussed earlier in eq.~\ref{eq:surv_3}, we indeed observe that the sign of $\emt$ has a drastic role in the $\mumu$ disappearance probability, especially near the oscillation maximum. Now, the bending of the oscillation valley can be explained by the eq.~\ref{eq:surv_3}. In that expression, the value of 
$P^m_{\nu_{\mu} \rightarrow \nu_{\mu}}$ is mainly determined by the first term in R.H.S. since the second term is suppressed by the NSI parameters appearing quadratically. At the first term in R.H.S. of eq.~\ref{eq:surv_3}, minimum occurs at higher (lower) energy compared to the SI case for a given baseline $L$ when $\emt$ is present with positive (negative) strength\footnote{At oscillation dip, the argument of the cosine term in eq.~\ref{eq:surv_3} should be approximately equal to $(2n+1)\pi/2$ where $n$ = 0,1,2... This roughly implies that $(1+2\emt\hat{A})\frac{\Delta m^2_{31}L}{4E}\simeq\pi/2$.}. Depending on the sign of $\emt$, the regions representing the oscillation dip tend to bend upward or downward with the increase in baseline length compared to the SI case.

\section{Summary}
\label{sec:summary}

In this work, we present a formalism for a simplified approach to understanding various fascinating features of neutrino oscillation probabilities in the presence of all possible NC-NSI parameters. In order to do this, we derive simple, approximate analytical expressions of the effective/modified oscillation parameters in the presence of the NSI parameters. Features on the evolution of the effective oscillation parameters are reflected in the oscillation probabilities.

We diagonalize the effective neutrino Hamiltonian in the presence of all possible NC-NSI parameters, using a method of successive rotation in the plane (2,3), (1,3), and (1,2), respectively. This diagonalization process gives a simple expression of the modified mixing angles. Eigenvalues of the diagonalized Hamiltonian denotes the modified mass-squares.
We discuss the evolution of each modified oscillation parameter as a function of energy in the the presence of the NSI parameters with strength of 0.2.
We observe standard interactions does not affect $\tzm$; only the NSI parameters from the (2,3) block of the NSI matrix influence its evolution. 
For the maximal value of $\theta_{23}$ in vacuum, the change in $\tzm$ is negligible in the presence of $\emt$. If $\theta_{23}$ belongs to the upper octant, then $\tzm$ increases (decreases) for negative (positive) choices of $\emt$. We notice a completely opposite behavior if $\theta_{23}$ lies in the lower octant~(see figure~\ref{eq:th23_1}).
In the case of $\tym$, we observe even in the SI case for neutrinos, $\tym$ increases with energy and can go up to a very large value~($> 45^\circ$) at relevant energies. In the presence of the NSI parameters from the (2,3) block of the NSI matrix with positive~(negative) strength, the $\tym$ evolution is suppressed~(enhanced). In the presence of the $\eem$ and $\eet$ with positive (negative) strength, we see a steep increase~(decrease) in $\tym$ initially as compared to the SI case. But with incrase in energy deviation from the SI case starts to decrease till the value of $\tym$ reaches the maximal value $45^\circ$, known as $\theta_{13}$-resonance.
$\txm$ shows at steep increase in it value at low energy~($\lesssim 2$ GeV) and saturate to a value~$90^\circ$ at high energy in the SI case as well as in the presence of NSI parameters from the (2,3) block. The presence of $\eem$ ($\eet$), the value at which $\txm$ saturates decrease~(increase).
Such saturation of $\txm$ at around $90^\circ$ explains various interesting features in the matter-driven neutrino flavor transition.

We also study the evolution of modified mass-squared differences, $\ldmm$ and $\sdmm$.  In the SI case, $\ldmm$ increases slightly with energy. Even in the presence of NSI parameters, we do not observe and significant increase or decrease in $\ldmm$. Unlike $\ldmm$, $\sdmm$ shows a sharp increase with energy and increases up to two order~($10^{-3}$ eV$^2$) from the vacuum value in the SI case. The presence of NSI parameters does not cause any notable change in the evolution.

We discuss the impact of NSI parameters on the $\theta_{13}$-resonance. We derive a simple analytical expression showing a modification in the neutrino energy at $\theta_{13}$-resonance in the presence of NSI. We observe $\eem$ and $\eet$ do not have any influence on resonance energy. We conclude this a reason for the intersection between the line showing the evolution of $\tym$ with energy in the presence of $\eem$ and/or $\eet$ with that of the SI case (see figure~\ref{fig:th13_1}). 

We also explore the role of the off-diagonal NSI parameters in the $\mue$ appearance probability. We plot the oscillogram in the appearance channel in the SI case and in the presence of NSI parameters with both positive and negative strength. We observe that in the presence of positive $\eem$ or $\eet$, the region in the ($E,L$) plane shifts significantly toward smaller baselines. When the strength of $\eem$ and $\eet$ are negative, this region vanishes. We explain this behavior analytically by deriving a simple, compact expression for the line-average matter density and the baseline that will give the maximum appearance probability in the matter. We indeed observe in the presence of $\eem$ or $\eet$, the baseline required to have the oscillation maximum is significantly small compared to the SI case. When the sign of the $\eem$ or $\eet$ is flipped, the required value of $\rho\times L$ is so large that it can not be attained within the Earth.

Finally, we study in detail how the NSI parameters from the (2,3) block of the NSI matrix would affect the $\mumu$ disappearance channel. We derive the expression for the disappearance probability using the OMSD approximation and $\theta_{13}\to 0$. 
We found that the effective diagonal parameter~($\gamma-\beta$) appears in the second order, while the off-diagonal parameter $\emt$ shows up in the first order in the expression, thus having a dominant impact. To check the validity of these observations, we plot an oscillogram for the disappearance probability without any approximation~(\ref{fig:surv_1}). We conclude that our observations remain intact even at the non-zero $\theta_{13}$ limit and without OMSD approximation.
\blankpage 


\newcommand{\ame}{\ensuremath{\alpha_{21}}}
\newcommand{\ate}{\ensuremath{\alpha_{31}}}
\newcommand{\atm}{\ensuremath{\alpha_{32}}}
\newcommand{\aee}{\ensuremath{\alpha_{11}}}
\newcommand{\amm}{\ensuremath{\alpha_{22}}}
\newcommand{\att}{\ensuremath{\alpha_{33}}}


\chapter{Constraints on non-unitary neutrino mixing in next generation long-baseline experiments}
\label{C5} 

\section{Introduction}
In the present scenario, the status of the precision measurement of the neutrino oscillation parameters has been very promising. Global fit analyses~\cite{NuFIT,Esteban:2020cvm,deSalas:2020pgw,Capozzi:2021fjo} of the excellent data from various neutrino experiments have already been able to determine the values of the neutrino oscillation parameters with reasonable accuracy.
At present, the relative 1$\sigma$ precision on the mixing angles $\theta_{23}$, $\theta_{13}$, and $\theta_{12}$ lies in the range of ${\cal O}(3-7)\%$. 
For the mass-squared differences, the achieved relative 1$\sigma$ precision is around ${\cal O}(1-3)\%$. In this precision era of neutrino physics, it is inevitable that various possible BSM scenarios will be tested that challenge the standard oscillation framework. One such case is testing the unitarity of the neutrino mixing. In the standard three-neutrino framework, the mixing matrix is assumed to be unitary and is parameterized by four parameters: three mixing angles and one CP phase under PMNS parametrization. However, various neutrino mass models exist in the literature that demand non-unitary neutrino mixing (NUNM). The most general models are low-scale variants of the type-I seesaw mechanism that naturally explains the tininess of the neutrino mass. These models introduce heavy right-handed neutrino in its particle content in order to generate neutrino mass, which naturally induces non-unitarity of the neutrino mixing. However, in this work, we are not going to delve into the origin of the neutrino mass. Here, we explore if there is NUNM in Nature and to what extent neutrino oscillation experiments will be able to prove it. 
Previously, Considerable efforts have been made to study the possible impact of NUNM on the measurement of the leptonic CP phase $\delta_{\mathrm{CP}}$, determination of the neutrino mass-hierarchy~\cite{Li:2015oal,Parke:2015goa,Ge:2016xya,Miranda:2016wdr,Dutta:2016czj,Pas:2016qbg}, and in
estimating the performance of current and future long-baseline experiments to constrain the various NUNM parameters~\cite{Antusch:2006vwa,Miranda:2019ynh,DeGouvea:2019kea,Escrihuela:2015wra,Fernandez-Martinez:2016lgt,Forero:2021azc,Ellis:2020hus,Coloma:2021uhq,Hu:2020oba,Blennow:2016jkn,Abe:2017jit,Dutta:2019hmb}.

In our work, we focus on two upcoming LBL neutrino experiments, DUNE and Hyper-K, with one detector at Kamioka, Japan, which we call the Japanese detector (JD), and another detector in Korea, which we call the Korean detector (KD). We first explore the impact of NUNM on the neutrino oscillation probability. Then, we simulate these experiments to explore the sensitivity of these experiments on NUNM and present expected bounds on the NUNM parameters from these individual setups and their combinations.

The chapter is organized as follows. In section~\ref{sec:formalism}, we discuss the formalism and the parameterization of non-unitary neutrino mixing used in this work. In section~\ref{sec:osc_prob_NUNM}, we derive simple approximate analytical expressions of the oscillation probability in $\nu_{\mu}\rightarrow\nu_{e}$ and $\nu_{\mu}\rightarrow\nu_{\mu}$ oscillation channels in the presence of standard matter interaction.  
The impact of NUNM at the event level is discussed in section~\ref{sec:event}. Section~\ref{sec:simulation_details} provides the details of the numerical simulations performed in our analysis and some results on the expected signal events. The main results of our study are presented in section~\ref{sec:results}, where we illustrate the correlation of the standard oscillation parameters $\theta_{23}$ and $\delta_{\mathrm{CP}}$ with various NUNM parameters and give the expected constraints achievable by DUNE and T2HKK separately as well as their combination. In section~\ref{sec:ND}, we show the possible improvements in the bounds due to the presence of near detectors. 
Section~\ref{sec:nu_tau} shows the improved constraints on the NUNM parameters when we add $\nu_{\tau}$ events sample in the DUNE simulation. In section~\ref{sec:NU_marg}, we discuss in detail the impact on the bounds discussed in the main text of the simultaneous marginalization over all NUNM parameters.
Our concluding remarks are discussed in section~\ref{sec:conclusion}.

\section{NUNM Formalism}
\label{sec:formalism}
The non-unitarity neutrino mixing can be introduced to the oscillation of active neutrinos in various ways~\cite{Antusch:2006vwa, Xing:2012kh,Flieger:2019eor,Bielas:2017lok,Ellis:2020ehi,Hu:2020oba}. The most convenient one is the lower triangular formulation of the mixing matrix. In this formulation, the $3\times 3$ NUNM matrix is written as 
\begin{eqnarray}\label{eq:matriceN}
N=(I+\alpha)\,U_{PMNS}\,.
    \end{eqnarray}
In the above equation, $U_{PMNS}$ is the standard unitary neutrino mixing matrix. $\alpha$ is a lower triangular matrix that introduces non-unitarity to the neutrino mixing. We parameterize $\alpha$ as
\begin{eqnarray}
\alpha=
\begin{pmatrix}
\alpha_{11} & 0 & 0 \\
|\alpha_{21}|e^{i \phi_{21}} & \alpha_{22} & 0 \\
|\alpha_{31}|e^{i \phi_{31}} & |\alpha_{32}|e^{i \phi_{32}} & \alpha_{33} 
\end{pmatrix}\, \, .
\label{eq:NU-triangle-mat}
\end{eqnarray}
The diagonal parameters $\alpha_{ii}$ are real as discussed in section~\ref{sec:NUNM_formalism}, while off-diagonal parameters $\alpha_{ij}\,\,(i\neq j)$ are complex quantity with associated phases $\phi_{ij}$. As we discussed earlier, NUNM matrix $N$ is not unitary anymore. $i.e.$ $N^\dagger N\neq 1$. In the unitary limit, all the NUNM parameter vanishes to give standard unitary mixing $U_{PMNS}$ in eq.~\ref{eq:matriceN}.

The neutrino propagation Hamiltonian in the mass basis with NUNM is expressed as 
the mass basis is: 
\begin{eqnarray} \label{evol}
H= \frac{1}{2 E_\nu} \left[\left( \begin{array}{ccc}
                   0   & 0          & 0   \\
                   0   &  \Delta m^2_{21}  & 0  \\
                   0   & 0           &  \Delta m^2_{31}  
                   \end{array} \right)  +  N^{\dagger}
                  \left( \begin{array}{ccc}
            a_e + a_n      & 0 & 0 \\
            0  & a_n  & 0 \\
            0 & 0 & a_n
                   \end{array} 
                   \right) N\right]
\, .
\label{eq:matter}
\end{eqnarray}
The second term shows the neutrino interaction with matter in the mass basis. As usual, the matter potential parameters are given by
$a_e= 2 \sqrt{2} E_\nu G_F N_e$ and $a_n= -\sqrt{2} E_\nu G_F N_n$ where, $N_e$ and $N_n$ are the electron and the neutron number densities, respectively. Note that in this framework, the neutral current (NC) matter potential is necessary since the non-unitarity of the matrix $N$ does not allow the subtraction of an identity matrix proportional to $a_n$. 

So the neutrino flavor transition probability at a given baseline in the presence of NUNM can be written as 
\begin{eqnarray}
P_{\alpha\beta}=|(N e^{-i H L}N^\dagger)_{\beta\alpha}|^2.
\label{eq:prob}
\end{eqnarray}
On immediate consequence of the NUNM is that even if the baseline is negligibly small, there can be non-zero flavor transition probability, which is popularly known as zero-distance effect. This will have a crucial involvement in the near detector measurements, which we discuss in detail in the later sections.

\section{Neutrino flavor transition with NUNM}
\label{sec:osc_prob_NUNM}
In eq.~\ref{eq:prob}, we show a general expression for the neutrino oscillation probability. In this section, we derive the approximate analytical expressions of oscillation probability in the various oscillation channels in order to explore the impact of various NUNM parameters in $\mue$ appearance and $\mumu$ disappearance channel.

 We use perturbation theory in the small expansion parameters ($r,\,s,$ and $a$) defined as follows:
\begin{equation}
\label{eq:expansion_params}
\sin \theta_{13} = \frac{r}{\sqrt{2}}\,, \qquad \sin \theta_{12} = \frac{1}{\sqrt{3}}(1+s)\,,\qquad \sin \theta_{23} =  \frac{1}{\sqrt{2}}(1+a)\,,
\end{equation}
where $r,\, s$, and $a$ represent the deviation from the tri-bimaximal mixing values of the neutrino mixing parameters, which are $\sin \theta_{13}=0$, $\sin \theta_{23} = 1/\sqrt{2}$, and $\sin \theta_{12} = 1/\sqrt{3}$~\cite{King:2007pr,Pakvasa:2007zj}. At the present best-fit value of the oscillation parameters from the global oscillation analysis~\cite{NuFIT,Esteban:2020cvm,deSalas:2020pgw,Capozzi:2021fjo}, all three expansion parameters are of the order ${\cal O}(0.1)$. For our calculation, we define $\Delta_{31} = \Delta m_{31}^2 L/ 4E_\nu$,  $\Delta_e = a_e L/ 4E_\nu$ and $\Delta_n = a_n L/ 4E_\nu$. At the sub-GeV energy range, matter interaction terms $\Delta_e$ and $\Delta_n$ are of the order of 0.1. So, apart from $r,\, s$, and $a$, we use $\Delta_e$ and $\Delta_n$ as the expansion parameters in our calculation. Apart from these parameters, we use OMSD approximation($\Delta_{31}\,>>\,\Delta_{21}$, where $\Delta_{21}=\Delta m^2_{21}L/4E_{\nu}$) in while deriving the expression  of the oscillation probabilities.

Using the formalism mentioned above, we derive the approximate expression of the $\mue$ appearance channel as follows
\begin{align}
P(\nu_\mu \to \nu_e)&=\left(\frac{r^2}{\Delta_{31}}\right)\, \sin \Delta_{31} \left[(\Delta_{31}+2 \Delta_{e}) \sin \Delta_{31}-2 \Delta_{31} \Delta_{e} \cos\Delta_{31}\right]\nonumber + \\
   & \left(\frac{2 |\alpha_{31}|  r}{\Delta_{31}}\right) \Delta_n \sin\Delta_{31} \left[\cos (\delta_{\mathrm{CP}}-\phi_{31}) \sin\Delta_{31}-\Delta_{31}
   \cos (\delta_{\mathrm{CP}}+\Delta_{31}-\phi_{31})\right] \nonumber + \\
& \left(\frac{|\alpha_{21}| r}{\Delta_{31}} \right)  \,\left\{
\sin\Delta_{31}\left[2 \Delta_{31}   (\Delta_e+\Delta_n) \cos (\delta_{\mathrm{CP}}+\Delta_{31}  -\phi_{21})-\Delta_n \sin (\delta_{\mathrm{CP}}-\Delta_{31}  -\phi_{21})\right]+\right. \nonumber 
\\
& \qquad\qquad\left.\sin (\delta_{\mathrm{CP}}+\Delta_{31}  -\phi_{21}) \left[(-2 \Delta_{31} -2 \Delta_e+\Delta_n) \sin\Delta_{31}+2 \Delta_{31}  \Delta_e  \cos\Delta_{31}\right]\right\}  + \nonumber\\
& \left(\frac{|\alpha_{21}||\alpha_{31}|}{\Delta_{31} }\right) \Delta_n
\left[-2 \Delta_{31}  \sin (\phi_{21}-\phi_{31})+\cos (2 \Delta_{31} -\phi_{21}+\phi_{31})-\cos (\phi_{21}-\phi_{31})\right] \nonumber + \\ &
   \left(\frac{|\alpha_{21}|^2}{\Delta_{31}}\right)\left[\Delta_{31}-\Delta_n (1-\cos 2  \Delta_{31})\right]\label{eq:pme_anylit_main}\,.
\end{align}
From the above expression, we note the following crucial feature of the NUNM parameters in the $\mue$ appearance channel
\begin{enumerate}
    \item The appearance probability is primarily affected by two off-diagonal NUNM parameters, namely, $\ame$ and  $\ate$ at the leading order. The phases ($\phi_{j1},\,\,j=2,3$) associated with these two NUNM parameters always coupled to the standard CP phase $\dcp$, which may affect the $\dcp$ measurement in the experiments by bringing in new degeneracies.
    
    \item As mentioned earlier, there can be zero-distance effect in the presence of NUNM. In case of $\mue$ appearance channel, we derive the zero-distance probability by making $L\rightarrow 0$, which gives
    \begin{eqnarray}
P(\nu_\mu\to\nu_e; L = 0) &\sim& |\alpha_{21}|^2 \label{zerodistancemue}\,.
\end{eqnarray}

So, in the presence of non-zero $\ame$, there will be non-zero $\mue$ transition probability even when the baseline is negligible. We exploit this feature to constrain $|\ame|$ in the later sections.

   \item In eq.~\ref{eq:pme_anylit_main}, the terms with NUNM parameter $\ate$ always associated with neutral current matter term $\Delta_n$. This suggests that the impact of $\ate$ in the $\mue$ appearance channel depends on the NC matter effects neutrino experiences while propagating. 
   \end{enumerate}
   
In the vacuum, the expression the $\mue$ appearance probability leads to 
\begin{eqnarray}
P(\nu_\mu\to\nu_e)_{\rm vac} &=&
|\alpha_{21}|^2+r^2 \sin ^2\Delta_{31}-2 |\alpha_{21}| r 
\sin\Delta_{31} \sin (\delta_{\mathrm{CP}}+\Delta_{31}-\phi_{21})
\label{pmuevacuum} \,.
\end{eqnarray}
As we expected, impact of $\ate$ vanishes in the vacuum limit.

 \begin{table}
 	\centering
 	\begin{tabular}{|c|c|c|c|c|c|}
 		\hline\hline
 		$\theta_{23}$&$\theta_{13}$&$\theta_{12}$&$\delta_{\mathrm{CP}}$&$\Delta m^2_{21} [\mathrm{eV^2}]$&$\Delta m^2_{31} [\mathrm{eV^2}]$ \\
 		\hline
 		$45^{\circ}$ & $8.61^{\circ}$&$33.6^{\circ}$& $-90^{\circ}$ &$ 7.39\times 10^{-5}$&$2.52\times 10^{-3}$ \\
 		\hline\hline
 	\end{tabular}
 	\mycaption{The benchmark values of the oscillation parameters used in our analysis. These values are consistent with the present best-fit
 		values as obtained in various global fit 
 		studies~\cite{NuFIT,Esteban:2020cvm,deSalas:2020pgw,Capozzi:2021fjo}.
 		We assume normal mass ordering (NMO) of neutrino throughout this work.}
 	\label{table:vac}
 \end{table}

\begin{figure}[h!]
\centering
\includegraphics[width=1.05\textwidth]{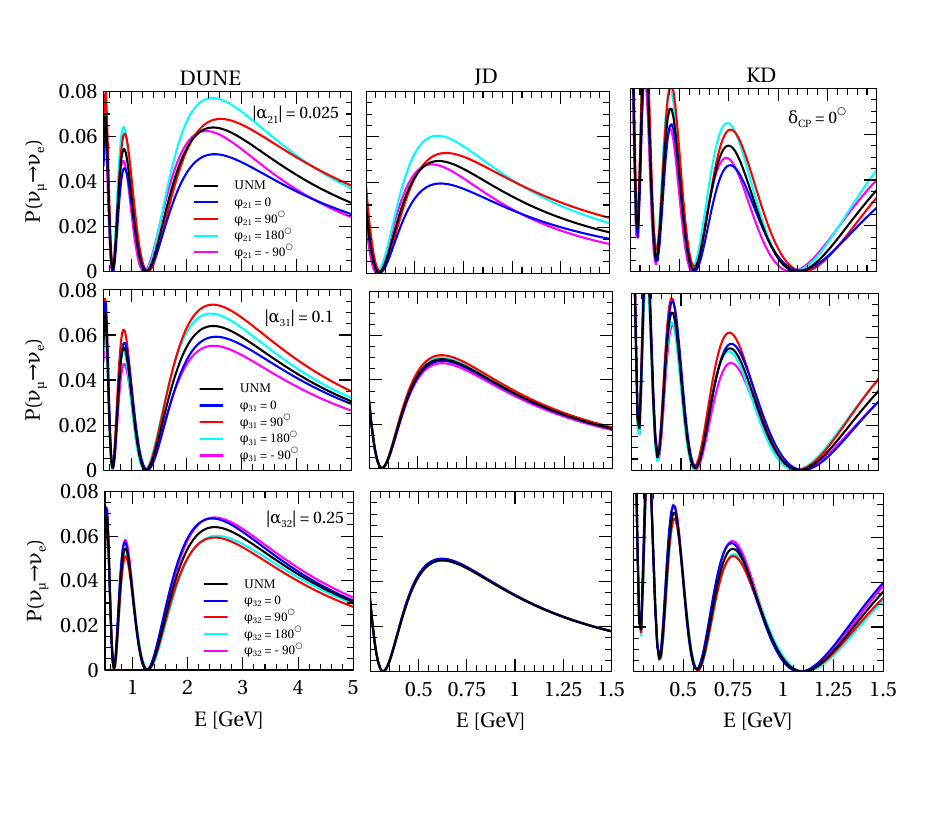}
\vspace{-0.8cm}
\mycaption{$\nu_{\mu}\rightarrow\nu_{e}$ appearance probability as a function of energy in the presence of off-diagonal NUNM parameters. Left, middle, and right columns correspond to the baselines of 1300 km (DUNE), 295 km (JD), and 1100 km (KD), respectively. The black curve in each panel shows the $\mue$ appearance probability in the unitary neutrino mixing (UNM) scenario. The four colored curves correspond to four benchmark values of the phases associated with off-diagonal NUNM parameters: $0^{\circ}$, $90^{\circ}$, $180^{\circ}$, and $-90^{\circ}$. We consider $\delta_{\mathrm{CP}} = 0^{\circ}$ and $\sin^2\theta_{23}=0.5$. The values of the other oscillation parameters are taken from table~\ref{table:vac}. }
\label{fig:prob_mue}
\end{figure}
In figure~\ref{fig:prob_mue}, we plot the $\mue$ oscillation probability as a function of neutrino energy in the NUNM scenario. For comparison, we also show the same in the standard unitary neutrino mixing~(UNM) scenario. In this plot, we calculate the oscillation probability using General long-baseline Experiment Simulator (GLoBES) software~\cite{Huber:2004ka,Huber:2007ji} along with the plug-in MonteCUBES~\cite{Blennow:2009pk}. As mentioned in the figure, we show the appearance probability for the baseline length ($L$) to the LBL experiments DUNE~($L=1300$ km), JD~($L=295$ km), and KD~($L=1100$ km). In this plot, we show the impact of the off-diagonal NUNM parameter $\ame$, $\ate$, and $\atm$ one-at-a-time, as we expect a comparatively larger influence from the first two (see eq.~\ref{eq:pme_anylit_main}) and for completeness we show the contribution from $\atm$. Note that for both JD and KD, we consider the neutrino energy range of 0 to 1.5 GeV, having a peak around 0.6 GeV. From the above plot, it is clear that the parameters $\ame$ show the largest impact on the $\mue$ appearance probability for all three setups. It is expected from the approximate expression of the appearance probability in eq.~\ref{eq:pme_anylit_main}, where we observe that $|\ame|$ appears at the leading order without coupling to the matter terms. As a result irrespective of the baseline length, which indirectly shows the strength of the matter effect neutrino experience, i=$\ame$ has a larger impact. However, for $\ate$, DUNE, and KD, which have a larger baseline, get affected significantly and show a substantial deviation from the oscillation probability in the UNM scenario. T2HK, with a comparatively smaller baseline, shows a marginal impact due to $\ate$. As we explained earlier, the parameter $\ate$ is always coupled to the NC matter effect terms in the appearance probability expression in eq.~\ref{eq:pme_anylit_main}. Since for setups like DUNE and KD, neutrino faces larger matter effects, it shows a larger impact, and we observe very little changes in the appearance probability in the case of JD. In eq.~\ref{eq:pme_anylit_main}, we do not observe a dependence of $\atm$ up to the order of the NUNM parameter we consider for derivation. However, it may appear in higher-order terms, which
is why we observe some impact of $\atm$ in the appearance probability mainly for DUNE and KD, signaling that it also coupled to matter effect terms.

\begin{figure}[h!]
\centering
\includegraphics[width=\textwidth]{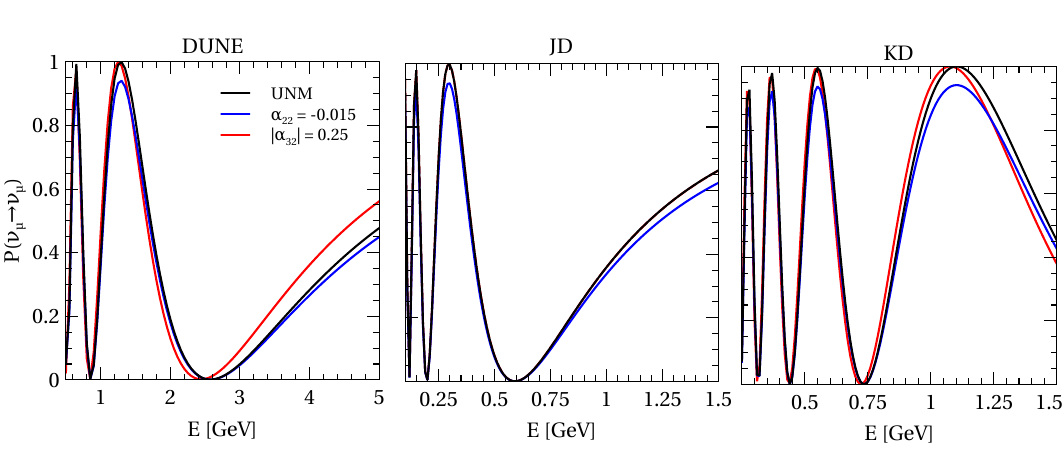}
\vspace{-0.5cm}
\mycaption{$\nu_{\mu}\rightarrow\nu_{\mu}$ disappearance probabilities as a function of neutrino energy in the presence of the NUNM parameters $\alpha_{22}$ and $|\alpha_{32}|$ assuming $\phi_{32}=0$ one at a time. Left, middle, and right columns correspond to the baselines of 1300 km (DUNE), 295 km (JD), and 1100 km (KD), respectively. We consider $\delta_{\mathrm{CP}} = 0^{\circ}$ and $\sin^2\theta_{23}=0.5$. The values of the other oscillation parameters are taken from table~\ref{table:vac}.}
\label{fig:prob_mumu}
\end{figure}

Following the similar formulation as $\mue$ channel, for the $\nu_\mu\rightarrow\nu_{\mu}$ disappearance channel, we get:
\begin{eqnarray}
 P(\nu_\mu\to \nu_\mu) &=&\cos^2\Delta_{31} (1+ 4 \alpha_{22})-2 |\alpha_{32}| \Delta_n \sin2\Delta_{31} \cos\phi_{32} + 4 a^2 \sin^2\Delta_{31} + \nonumber \\
 && 2 |\alpha_{21}|^2 \cos\Delta_{31} \left[2 \sin\Delta_{31}(\Delta_e+\Delta_n)+\cos\Delta_{31}\right] + 6 \alpha_{22}^2 \cos^2\Delta_{31} - \nonumber \\
 &&2 |\alpha_{21}| r \sin (2 \Delta_{31}) (\Delta_e+\Delta_n) \cos(\delta_{\mathrm{CP}}-\phi_{21}) + \nonumber\\
 &&\left(\frac{8 a}{\Delta_{31}}\right)(\alpha_{22}-\alpha_{33}) \Delta_n \sin \Delta_{31} (\sin \Delta_{31}-\Delta_{31} 
 \cos \Delta_{31}) \nonumber \\
 &&-2 |\alpha_{21}| |\alpha_{31}| \Delta_n  \sin2 \Delta_{31} \cos (\phi_{21}-\phi_{31})
 -8 \alpha_{22} |\alpha_{32}| \Delta_n\sin2\Delta_{31} \cos\phi_{32} \nonumber \\
 &&-2 |\alpha_{32}| \alpha_{33} \Delta_{n} \sin2\Delta_{31}\cos\phi_{32} \label{eq:pmm_anylit_main}\,.
\end{eqnarray}
From the above expression, we note the following:
\begin{enumerate}
    \item  All the NUNM parameters $\alpha_{ij}$ except $\alpha_{11}$ enter into the probability expression. However, three off-diagonal parameters appear at the second order.
    \item zero-distance expression of the $\nu_{\mu}\rightarrow\nu_{\mu}$ survival probability, given by:
\begin{eqnarray} \label{pmmzero}
P(\nu_\mu\to\nu_\mu; L = 0) &\sim& 1+ 2 |\alpha_{21}|^2+6 \alpha_{22}^2+4 \alpha_{22}\,.
\end{eqnarray}
\end{enumerate}
In the vacuum limit, the expression simplifies to
\begin{eqnarray}
P(\nu_\mu\to\nu_\mu)_{\rm vac}&=& \cos^2\Delta_{31} \left(1+2 |\alpha _{21}|^2 +4 \alpha_{22}+ 6\alpha_{22}^2 \right) + 4 a^2 \sin ^2\Delta_{31}\,, \label{pmmvacuum}
\end{eqnarray}
which agrees with the expression derived in ref.~\cite{Escrihuela:2015wra}.

In figure~\ref{fig:prob_mumu}, we show the exact $\nu_{\mu}\rightarrow\nu_{\mu}$ oscillation probabilities as a function of energy for the baseline lengths corresponding to DUNE (left panel), JD (middle panel), and KD (right panel) setups. Similar to the $\mue$ case, the black solid curves correspond to the UNM case, while the red and blue curves show the presence of $\amm$ and $\atm$, respectively with strength reported in the legend\footnote{Even though the analytical expression of $P_{\mu\mu}$ reported in Eq.~\ref{eq:pmm_anylit_main} shows the presence of other NUNM parameters,
we have numerically checked that they do not have any significant impact.}, one at a time. The impact of these two NUNM parameters can be understood from our approximated analytical expressions in Eq.~\ref{eq:pmm_anylit_main}.
When matter effects are negligible (for example, in the middle panel of figure~\ref{fig:prob_mumu}), we expect that the parameter $\amm$ 
dominates the deviation from UNM since it appears already in first order  in 
$P_{\mu\mu}$. This remains true when matter parameters are switched on; the relevant difference compared to the vacuum case relies on the fact that also $|\atm|$ enter at first order, although suppressed by $\Delta_{n}$. Thus, we expect that  
for DUNE and KD, one can see deviations from the UNM predictions,  as visible in figure~\ref{fig:prob_mumu}. Note that the impact of $\alpha_{32}$ is amplified by the larger benchmark value compared to the choice for $\alpha_{22}$.

\begin{figure}[h!]
\centering
\includegraphics[width=\textwidth]{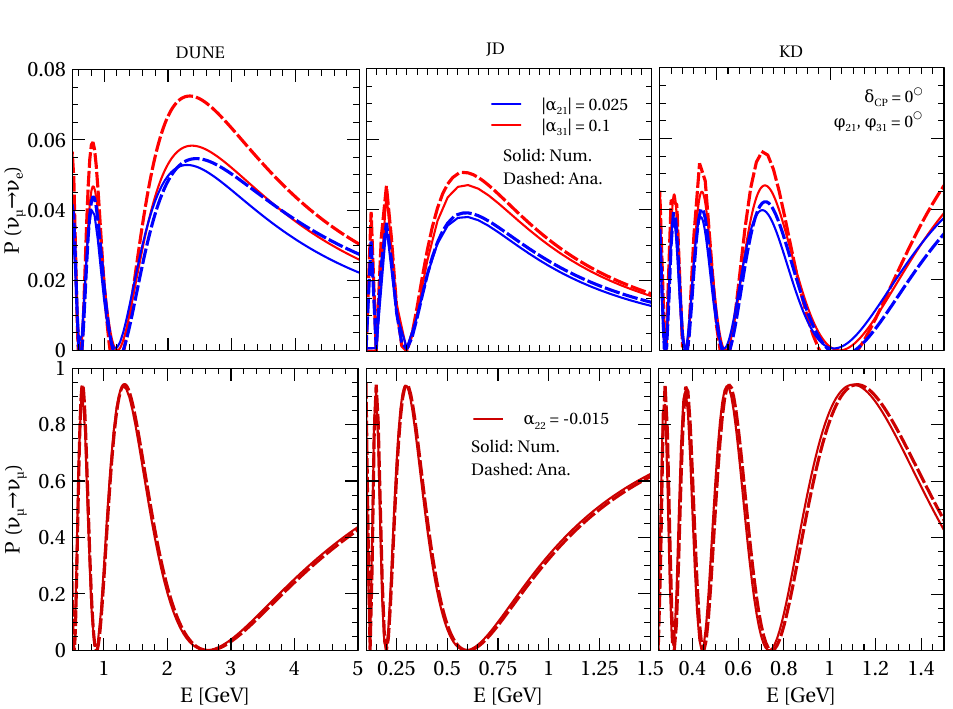}
\mycaption{Upper panels show the $\nu_{\mu}\rightarrow\nu_{e}$ appearance probability in presence of the NUNM parameters $|\alpha_{21}|\,=\,0.025$ and $|\alpha_{31}|\,=\,0.1$. Lower panels depict the $\nu_{\mu}\rightarrow\nu_{\mu}$
disappearance probability for $\alpha_{22}\,=\,0.015$. 	
Left, middle, and right columns correspond to the baselines of 1300 km (DUNE), 295 km (JD), and 1100 km (KD), respectively.
Solid curves in each panel show the oscillation probabilities obtained numerically, while the dashed curves are the probabilities obtained analytically using Eq.~\ref{eq:pme_anylit_main} and Eq.~\ref{eq:pmm_anylit_main}. We consider the values of  $\delta_{\mathrm{CP}}$ and the phases $\phi_{21}$, $\phi_{31}$ equal to zero. The values of the other standard oscillation parameters are taken from table~\ref{table:vac}. }
\label{fig:prob_mueCheck}
\end{figure}
To check the validity the approximate analytical expressions of the $\mue$ appearance and $\mumu$ disappearance probability derived in this section, in figure~\ref{fig:prob_mueCheck}, we plot the $\mue$ and $\mumu$ oscillation probability using the expressions derived in eq.~\ref{eq:pme_anylit_main} and eq.~\ref{eq:pmm_anylit_main} and compare with the same calculated numerically using GLoBES software. In each panel, solid curves are obtained numerically from the full (not expanded) Hamiltonian, while dashed curves correspond to the oscillation probabilities obtained analytically. The value of the CP phase $\delta_{\mathrm{CP}}$ and phases $\phi_{21}$, $\phi_{31}$ are fixed to zero, while the values of the other standard oscillation parameters are taken from table\ \ref{table:vac}. As we can see, the agreement between the full solid and the analytical dashed curves is generally very good for both probabilities. The expression for the probabilities for other oscillation channel is given in appendix~\ref{appendix:osc_prob_nunm}.\\
\section{Key features of the experiments}
\label{sec:experiments}

In the earlier sections, we discuss in detail how NUNM affect the $\mue$ appearance and $\mumu$ disappearance probability. The long-baseline experiments, which primarily probe $\nu_e$ appearance and $\nu_\mu$ disappearance channel, can be a excellent mode to quantify the NUNM in the neutrino oscillation. The fixed baseline and a focused neutrino beam will allow us to have different features of the experiments under control. The upcoming long-baseline experiments will increase the neutrino events statistics with the help of high precision detectors. Two such experiments which are currently under construction are DUNE and T2HK. For the latter is being taken into account the possibility of a second far detector placed approximately at the second oscillation maximum. These two complementary experiments are going to improve the measurements of the oscillation parameters and the searches of new physics effects in a significant way. Consequently, it can provide us excellent opportunity to probe NUNM. In section~\ref{sec:osc_experiments}, basic features of these two experiments have been discussed briefly. In th following, we give a detailed discussion on the important feature of these two setups, which will be essential for our simulation purpose.

\subsection{DUNE}

DUNE (Deep Underground Neutrino Experiment) is a next-generation long-baseline experiment which will play an important role in solving the existing puzzles in neutrino oscillation as well as in increasing the precision on the measured neutrino oscillation parameters. It will use an on-axis, high-intensity, wide-band neutrino beam traveling a distance of 1284 km from Fermilab to South Dakota. For this baseline, they consider an average matter density of $2.85$ g/cm$^{3}$.
 The far detector (FD) is a liquid argon time projection chamber (LArTPC) of 40 kt fiducial mass placed underground at the Homestake mine. 
 According to the recent Technical Design Report (TDR)~\cite{DUNE:2021cuw}, DUNE will use a proton beam having 1.2 MW power which will deliver $1.1\times10^{21}$ proton on target (P.O.T.) per year. DUNE considers a total run-time of 10 years with 5 years each in neutrino and antineutrino modes. However,
 following most of the existing studies related to DUNE, we present our results using a total run-time of 7 years with 3.5 years each in neutrino and antineutrino modes.
Beside the far detector, the possibility to have three modules of near detectors (ND) has been proposed~\cite{DUNE:2021tad}. The first one would be a liquid Argon Time Projection chamber (TPC) situated at a distance of 574 m from the source. It will measure the flux and cross-section of the neutrinos. The second one would be a multi purpose detector (MPD) equipped with a magnetic spectrometer with one ton High-Pressure Gaseous Argon TPC, which will be useful to study possible new physics signals. The third one would be 
the System for On-Axis Neutrino Detection (SAND). It will be made up of a former KLOE magnet and a calorimeter which will track the outgoing particles.

Like other long-baseline experiments, DUNE will probe mainly $\nu_e$ appearance channel, and $\nu_\mu$ disappearance channel. Backgrounds to the appearance channel are the $\nu_e$ beam contamination and the misidentified $\nu_\mu$, $\nu_\tau$, and NC events. The signal systematic normalization error have been chosen to be 2\%. For the disappearance channel, background to the signal are misidentified $\nu_\tau$ and NC events. The signal error is 5\%. Efficiency functions as well as smearing matrices have been provided by the DUNE collaboration in Ref.~\cite{DUNE:2021cuw};

\subsection{T2HK (JD) and T2HKK (JD+KD)}

T2HK (Tokai to Hyper-Kamiokande) is another promising next-generation long-baseline experiment which will play a very important role in $\delta_{\mathrm{CP}}$ measurement as well as in the study of various BSM physics~\cite{Kelly:2017kch,Agarwalla:2018nlx,Choubey:2017ppj}. In the T2HK setup~\cite{Abe:2011ts,Hyper-KamiokandeWorkingGroup:2014czz,Hyper-KamiokandeProto-:2015xww}, an intense neutrino beam from the J-PARC proton synchroton facility will be detected at the Hyper-Kamiokande (HK) detector situated at a distance of 295 km from the source. The power of the proton beam at the source is 
1.3 MW which will produce $27\times10^{21}$ P.O.T. in its total 10 years run-time.  The detector is a water Cherenkov (WC) detector with 187 kt of fiducial mass.
To have an equal contribution from the neutrino and the antineutrino signal events, the proposed running time of the experiment is 2.5 years (25\% exposure) in neutrino mode and 7.5 years (75\% exposure) in antineutrino mode.
The HK detector will be placed at $2.5^{\circ}$ off-axis angle from the neutrino beam-line to receive a narrow band beam with energy peaked at around the first oscillation maximum ($\sim 0.6$ GeV).

There is also a proposal~\cite{Seo:2019dpr} to have another identical detector which will be situated in Korea, 1100 km far from J-PARC. We assume that this detector will also be placed at an off-axis angle of $2.5^{\circ}$ from the neutrino beamline (similar to JD) and operate around the second oscillation maximum with a baseline of 1100 km and an average neutrino energy of 0.6 GeV. Combination of the T2HK setup along with the detector in Korea is known as T2HKK. However, in our work, we call the Hyper-Kamiokande as the Japanese detector (JD) and the detector in Korea as the Korean detector (KD). For both JD and KD baselines, we assume an average matter density of 2.8 g/cm$^3$.
 In the following, we will show our numerical results on the sensitivity to the various NUNM parameters separately for each detectors as well as for their combination.

In addition to these two detectors, two more near detectors, namely, the ND280 detector and Intermediate Water Cherenkov detector (IWCD), have been proposed. ND280 will be located at 280 m from the source with the same off-axis angle as the far detectors. It will be helpful to measure the flux of the unoscillated neutrino beam and study the neutrino cross-sections~\cite{T2K:2019bbb}. IWCD is a water Cherenkov detector of mass of 1 kt and possibly located at a distance of 1 km from the source~\cite{Drakopoulou:2017qdu, Wilson:2020trq}. One advantage of this detector is that it can be moved vertically to take data at different off-axis angles.  
In table~\ref{tab:exp_details}, we summarize the relevant information of all the experimental setups discussed in this section.

For the two T2HKK far detectors JD and KD, background in the appearance channels are the $\nu_e$ from the beam contamination and misidentified $\nu_\mu$ and NC events. Signal systematic uncertainties are 5\% normalization and 5\% calibration errors. In the disappearance channel, backgrounds to the signal are misidentified $\nu_e$ and NC events. Systematic uncertainties are 3.5\% normalization and 5\% calibration errors. Efficiencies and energy resolutions are taken from the ref.~\cite{Hyper-Kamiokande:2016srs}.

\begin{table}[htb!]
\resizebox{\columnwidth}{!}{%
    \centering
    \begin{tabular}{|c|c|c|}
    \hline \hline
       Characteristics  & DUNE & JD/KD  \\
       \hline \hline
       Baseline (km) & 1285 & 295 (1100)\\
       \hline
       $\rho_{\mathrm{avg}}$ (g/cm$^{3}$) & 2.848 & 2.7 (2.8)\\
       \hline
       Beam & LBNF~\cite{DUNE:2020lwj} & J-PARC~\cite{Hyper-Kamiokande:2018ofw}\\
       \hline
       Beam Type & wide-band, on-axis & narrow-band, 2.5$^{\circ}$ off-axis\\
       \hline
       Beam Power & 1.2 MW & 1.3 MW\\
       \hline
       Proton Energy & 120 GeV & 30 GeV\\
       \hline
       P.O.T./year & 1.1 $\times$ 10$^{21}$ & 2.7 $\times$ 10$^{22}$\\
       \hline
       Flux peaks at (GeV) & 2.5  & 0.6  \\
       \hline
       1$^{\mathrm{st}}$ ( 2$^{\mathrm{nd}}$) oscillation maxima & \multirow{2}{*}{2.6 (0.87) }& \multirow{2}{*}{0.6 (0.2) / 1.8 (0.6)}  \\
       for appearance channel (GeV) & &\\
       \hline
       Detector mass (kt) & 40, LArTPC & 187 each, water Cherenkov\\
       \hline
       Runtime ($\nu + \bar{\nu}$) yrs & 5 + 5 & 2.5 + 7.5\\
       \hline
       Exposure (kt$\cdot$MW$\cdot$yrs)  & 480 & 2431\\
       \hline
       Signal Norm. Error (App.) & 2\% & 5\%\\
       \hline
       Signal Norm. Error (Disapp.) & 5\% & 3.5\%\\
       \hline
       Binned-events & \multirow{2}{*}{~\cite{DUNE:2021cuw} }& \multirow{2}{*}{~\cite{Hyper-Kamiokande:2016srs}} \\
       matched with & &\\
       \hline \hline
    \end{tabular}}
\mycaption{Essential features of DUNE~\cite{DUNE:2021cuw} and JD/KD~\cite{Hyper-Kamiokande:2016srs}  experiments used in our simulation.}
 \label{tab:exp_details}
\end{table}

\section{Event level discussion}
\label{sec:event}
In this section, we estimate the expected signal events from DUNE and JD and KD with their full exposure in the the presence of NUNM. In order to perform our numerical simulations, we use the General long-baseline Experiment Simulator (GLoBES) package~\cite{Huber:2004ka,Huber:2007ji} along with the plug-in MonteCUBES~\cite{Blennow:2009pk}. In table~\ref{tab:exp_details}, we summarize the relevant information of the two experiments considered in our simulation.
\begin{table}

	\centering
	
	\begin{tabular}{|ccc|*{6}{c|}}
		
		\hline\hline
		
		\multicolumn{3}{|c|}{\multirow{2}{*}{}} & \multicolumn{3}{|c}{$\nu_e$ appearance} & \multicolumn{3}{|c|}{$\bar\nu_e$ appearance}\\ 
		
		\cline{4-9}
		
		\multicolumn{3}{|c|}{} & DUNE & JD & KD & DUNE & JD & KD \\
		
		\hline\hline
		
		\multicolumn{3}{|c|}{UNM} & 1259 & 1836 & 169 & 221 & 767 & 31 \\
		
		\cline{1-9}
		
		\hline\hline
		
		\multicolumn{3}{|c|}{NUNM} & DUNE & JD & KD & DUNE & JD & KD \\
		\hline\hline
		
		\multicolumn{3}{|c|}{$|\alpha_{21}|$ (= 0.025)} & 1328 & 1893 & 169 & 232 & 756 & 29 \\
		\hline
		\multicolumn{3}{|c|}{$|\alpha_{31}|$ (= 0.1)} & 1420 & 1893 & 187  & 264 & 754 & 30\\
		\hline
		\multicolumn{3}{|c|}{$|\alpha_{32}|$ (= 0.25)} & 1300 & 1855 & 172 & 213 & 756 & 30\\
		\hline
		\multicolumn{3}{|c|}{$\alpha_{11}$ (= -0.02)} & 1203 & 1761 & 162 & 214 & 738 & 29 \\
		\hline
		\multicolumn{3}{|c|}{$\alpha_{22}$  (= -0.015)} & 1223 & 1782 & 164 & 215 & 744 & 30 \\
		\hline
		\multicolumn{3}{|c|}{$\alpha_{33}$ (= -0.15)} & 1208 & 1817 & 168  & 227 & 779 & 30\\
		
		\hline
		\hline
		
	\end{tabular}
	
	\mycaption{Comparison of the total signal rate for the $\nu_{e}$ and $\bar{\nu_{e}}$ appearance channels in DUNE, JD, and KD setups in UNM case as well as in presence of various NUNM parameters. The relevant features of these facilities are given in table~\ref{tab:exp_details}. The values of the standard oscillation parameters used to calculate event rate are quoted in table~\ref{table:vac}. The phases associated with the off-diagonal NUNM parameters are considered
		to be zero.}
	
	\label{tab:total_events_app}
	
\end{table}

The number of expected signal events for both $\mue$ and $\mumu$ oscillation channels simulated here are summarized in tables~\ref{tab:total_events_app} and \ref{tab:total_events_dis}, respectively, where the cases of UNM and NUNM (for some benchmark values of the $\alpha_{ij}$ parameters) are reported.
The impact of NUNM parameters on the number of events is fully in agreement with our analytical discussions. First of all, as shown in Eq.~\ref{eq:pme_anylit_main}, the appearance channel is mainly influenced by $|\alpha_{21}|$, even in vacuum. This reflects in an  enhancement of the number of events by roughly 5\% in both JD and DUNE. 
On the other hand, the NUNM parameters $|\alpha_{31}|$ and $\alpha_{33}$ 
are also relevant but they are coupled to the matter potentials, so we expect them to be relevant primarily for DUNE, where matter effects are more important: in fact, $|\alpha_{31}|$  causes an increase in the number of events up to 10\%, while $\alpha_{33}$ provokes a small but visible reduction of the order of  4\%.
Finally, some impact on the number of signal events is also given by $\alpha_{11}$, even though it only appears at higher orders in our perturbative expansion and has not been displayed (but it present in the vacuum probabilities reported in ref.~\cite{Escrihuela:2015wra}). 
Note that the number of $\nu_e$ and $\bar{\nu}_e$ events in KD is only slightly influenced by the NUNM parameters due to fact that the experiment works close to the second oscillation maximum of the atmospheric oscillation ($\nu_{\mu}\rightarrow\nu_{\tau}$), where the $\nu_{\mu}\rightarrow\nu_e$ appearance probability approached one of its minima and the effects of new physics are suppressed.

\begin{table}
	
	\centering
	
	\begin{tabular}{|ccc|*{6}{c|}}
		
		\hline\hline
		
		\multicolumn{3}{|c|}{\multirow{2}{*}{}} & \multicolumn{3}{|c}{$\nu_\mu$ disappearance} & \multicolumn{3}{|c|}{$\bar\nu_\mu$ disappearance}\\ 
		
		\cline{4-9}
		
		\multicolumn{3}{|c|}{} & DUNE & JD & KD & DUNE & JD & KD \\
		
		\hline\hline
		
		\multicolumn{3}{|c|}{UNM} & 10359 & 9064 & 1266 & 6034 & 8625 & 1144 \\
		
		\cline{1-9}
		
		\hline\hline
		
		\multicolumn{3}{|c|}{NUNM} & DUNE & JD & KD & DUNE & JD & KD \\
		\hline\hline
		
		\multicolumn{3}{|c|}{$|\alpha_{21}|$ (= 0.025)} & 10371 &9074 & 1264 & 6045 & 8640 & 1149 \\
		\hline
		\multicolumn{3}{|c|}{$|\alpha_{31}|$ (= 0.1)} & 10351 & 9062 & 1261  & 6035 & 8627  & 1168\\
		\hline
		\multicolumn{3}{|c|}{$|\alpha_{32}|$ (= 0.25)} & 10978 & 9203 & 1255 & 6005 & 8467 & 1158\\
		\hline
		\multicolumn{3}{|c|}{$\alpha_{11}$ (= -0.02)} & 10359 & 9064 & 1266 & 6034 & 8625 & 1145 \\
		\hline
		\multicolumn{3}{|c|}{$\alpha_{22}$  (= -0.015)} & 9748 & 8531 & 1192 & 5681 & 8120 & 1077 \\
		\hline
		\multicolumn{3}{|c|}{$\alpha_{33}$ (= -0.15)} & 10406 & 9077 &  1268 & 6040 & 8619 & 1145 \\
		\hline
		\hline
		
	\end{tabular}
	
	\mycaption{Comparison of the total signal rate for the $\nu_{\mu}$ and $\bar{\nu_{\mu}}$ disappearance channels in DUNE, JD, and KD setups in UNM case as well as in presence of various NUNM parameters. The relevant features of these facilities are given in table~\ref{tab:exp_details}. The values of the standard oscillation parameters used to calculate event rate are quoted in table~\ref{table:vac}. The phases associated with the off-diagonal NUNM parameters are considered
			to be zero.}
\label{tab:total_events_dis}
\end{table}

For the disappearance channel, the parameter $\alpha_{22}$, which enters at the first perturbative order in eq.~\ref{eq:pmm_anylit_main}, produces  a reduction of about 6\% in the number of events for all three detectors. This can be roughly understood from the fact that the standard disappearance  probability is multiplied by $4\alpha_{22}=0.06$, which causes a reduction by a similar factor in the number of events. The other relevant NUNM parameter is $|\alpha_{32}|$ which, being coupled to matter potential in eq.~\ref{eq:pmm_anylit_main}, can cause a $\sim$ 6\% increase of events especially in DUNE. The other parameters at their benchmark values only have a negligible impact on the number of disappearance events.

\begin{figure}[h!]
\centering
\includegraphics[width=\textwidth]{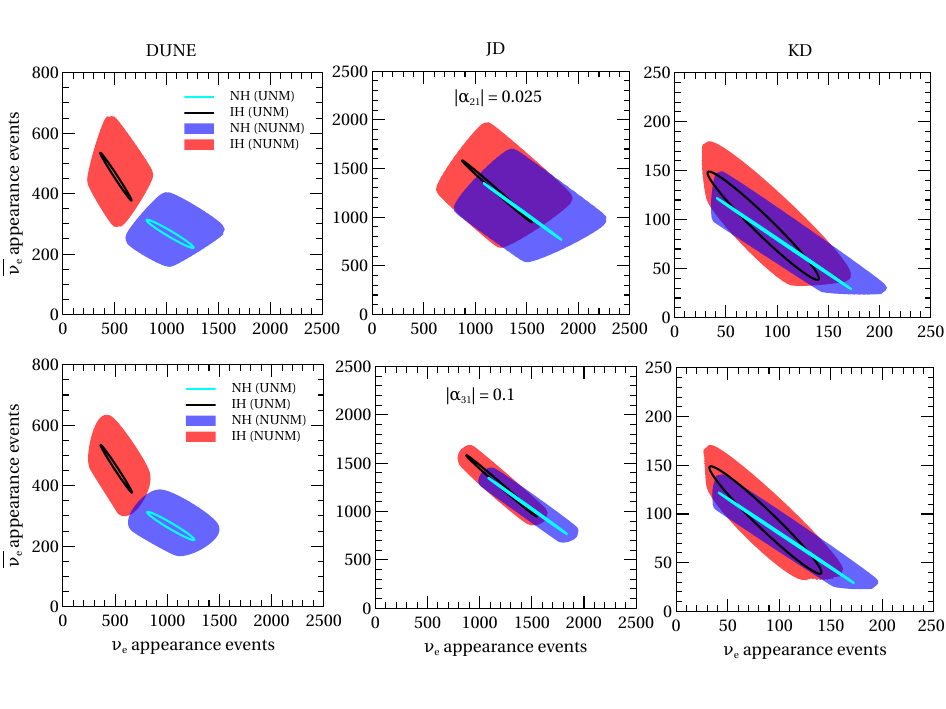}
\mycaption{Bi-event plots for the DUNE (left panel), JD (middle panel), and KD (right panel). The ellipses in each panel are obtained by varying the CP phase $\delta_{\mathrm{CP}}$ in the range $[-180^{\circ},\, 180^{\circ}]$ assuming $\alpha_{ij} = 0$. The colored blobs in the upper  panels show the bi-events in the presence of $|\alpha_{21}|$ with a strength of 0.025, while the lower panels depict the impacts of $|\alpha_{31}|$ with a strength of 0.1. To obtain the blobs with non-zero $\alpha_{ij}$,
	we vary the associated NUNM phases and the CP phase $\delta_{\mathrm{CP}}$ in their allowed range of $-\,180^{\circ}$ to $180^{\circ}$.
	 The values of the other standard three-flavor oscillation parameters are taken from table~\ref{table:vac}.}
\label{fig:bi-event}
\end{figure}

In figure~\ref{fig:bi-event}, we show the relation between the appearance events in neutrino and antineutrino modes at the three detectors discussed in this paper, namely DUNE, JD, and KD as shown by left, middle, and right panels.
The cyan (black) ellipses in each panel correspond to the standard interaction case (obtained varying the CP phase $\delta_{\mathrm{CP}}$ in the range $[-180^{\circ},\, 180^{\circ}]$), referring to the  normal (inverted) mass ordering.
 The red (blue) colored blobs in the upper  panels show the bi-events in the presence of the non-unitarity parameters $|\alpha_{21}|$ with strength 0.025 for the NMO (IH) case. 
The plots show that DUNE is in principle expected to be able to distinguish the mass hierarchy even in presence of NUNM. On the other hand, the two standard model curves and the two shadowed regions overlap in the case of JD (center panels) and KD (right panels). For these, the amplitude of the shadowed regions when $\phi_{31}$ is varied is much smaller than the DUNE one since, as it is clear from the analytical formulas, $|\alpha_{31}|$ is always coupled to matter effects, which are smaller for the two J-PARC based experiments.

\section{Simultaion details}
\label{sec:simulation_details}
We estimate the sensitivity of the above mentioned experimental setups to constrain various NUNM parameters
using Poissonian $\chi^2$ defined as follows
\begin{equation}
\chi^2 (\vec{\lambda}, \xi_{s}, \xi_{b}) = \min_{\vec\rho, \xi_{s}, \xi_{b}}~\Big[{{2\sum_{i=1}^{n}}\big(y_{i}-x_{i}-x_{i} \text{ln}\frac{y_{i}}{x_{i}}\big) + \xi_{s}^2 + \xi_{b}^2}~\Big]\,.
\end{equation} 
The above equation provides median sensitivity of the experiment using $n$ number of reconstructed energy bins. The quantity $y_i$ is defined as
\begin{equation}
y_{i} = N^{th}_{i}
(\vec\rho)~[1 + \pi^{s}\xi_{s}] + N^{b}_{i}(\vec\rho)~[1+\pi^b\xi_{b}],
\end{equation}
where $N^{th}_{i}$ is the number of expected event rate at the $i$-th energy bin with the theory we want to test. In our case, is is the expected event rate at the $i$-th bin in the presence of the NUNM in the theory with the set od oscillation parameters $\vec\lambda$ = 
$\{\theta_{12}, \theta_{13}, \theta_{23}, \Delta{m}^2_{21}, \Delta{m}^2_{31}, \delta_{\rm CP}, \alpha_{\alpha\beta}, \phi_{ij}\}$. $N^{b}_{i}$ is the background event rate at the $i$-th energy bins. 
$\xi_{s}$ and $\xi_{b}$ are the systematic pulls on the signal and background, respectively. In the analysis, e $\chi^2$ is marginalized over the set of parameters $\vec\rho$ and the systematic pulls ($\xi_{s}$ and $\xi_{b}$) in the theory. The variables $\pi^s$ and $\pi^b$ denote the normalization error on the signal and background event rate, whose corresponding values for DUNE and Hyper-K are listed in table~\ref{tab:exp_details}. $x_{i} = N^{obs}_{i} + N^{b}_{i}$ shows the prospective data from the experiments, where $N^{obs}_{i}$ and  $N^b_{i}$ are the number of charge-current (CC) signal events and the number of background events, respectively. 



We measure the sensitivity of the experiments on the NUNM parameters in terms of $\Delta \chi^2$ as defined below,
\begin{equation}
	\Delta \chi^2 = \underset{(\theta_{23},\,\delta_{\mathrm{CP}},\,\phi_{ij},\,\lambda_1,\, \lambda_2)}{\mathrm{min}} \,\bigg[\chi^2(\alpha_{ij}\neq0)-\chi^2(\alpha_{ij}=0)\bigg]\,\, ,
\end{equation}
where $\chi^{2} (\alpha_{ij}\neq0)$ and $\chi^2(\alpha_{ij}=0)$ are calculated by fitting the prospective data assuming NUNM ($\alpha_{ij}\neq0$) and UNM $(\alpha_{ij}=0)$. Note that $\chi^2(\alpha_{ij}=0) \approx 0$ because the statistical fluctuations are suppressed to obtain the median sensitivity of a given experiment in the frequentist approach~\cite{Blennow:2013oma}.
In our simulations, the true values of the standard oscillation parameters have been chosen as in table~\ref{table:vac}, while the true values of the NUNM parameters are set to zero. Since the presence of non-unitarity has basically the same effect on the number of events in the case of NMO and IMO, as shown in figure~\ref{fig:bi-event}, we consider only the NMO case. The computation of the $\Delta\chi^2$
is based on the pull method~\cite{Huber:2002mx,Fogli:2002pt,Gonzalez-Garcia:2004pka} implemented in the GLoBES software.
We study the NUNM parameters by fixing the mixing angles $\theta_{12}$, $\theta_{13}$, and two mass-squared differences $\Delta m^2_{21}$, and $\Delta m^2_{31}$ both in data and theory at their best fit values as given in table~\ref{table:vac}. 
We check that the marginalization over the atmospheric mass-squared difference $\Delta m^2_{31}$ does not have any significant effect on our analysis. On the other hand, the only notable effect of the marginalization over the reactor mixing angle $\theta_{13}$ (which has a very small experimental uncertainty of 3\%), is the worsening of the $\alpha_{11}$ bound at the level of 15\%. This is due to the fact that there is a correlation between these two parameters, which appear in a term proportional to $\alpha_{11}^2\sin^2 2\theta_{13}$ in the $\nu_\mu\to\nu_e$ transition probability as shown in ref.~\cite{Escrihuela:2015wra}. 
Finally, we marginalized $\theta_{23}$ in its current 3$\sigma$ allowed range~\cite{Capozzi:2021fjo}, which is approximately [$40^{\circ},\, 50^{\circ}$] and the CP phase $\delta_{\mathrm{CP}}$ in its entire possible range $[-180^{\circ},\, 180^{\circ}]$. We keep both these parameters with true values as in table~\ref{table:vac}. Moreover, we consider one NUNM parameter at a time, $\ie$,  when a parameter is taken into account the others are considered to be zero. 

\section{Results}
\label{sec:results}

\subsection{Correlations in test $(\theta_{23}-\alpha_{ij})$ and test $(\delta_{\mathrm{CP}}-\alpha_{ij})$ planes}
\label{subsec:correlation}
\begin{figure}[h!]
\centering
\includegraphics[width=\textwidth]{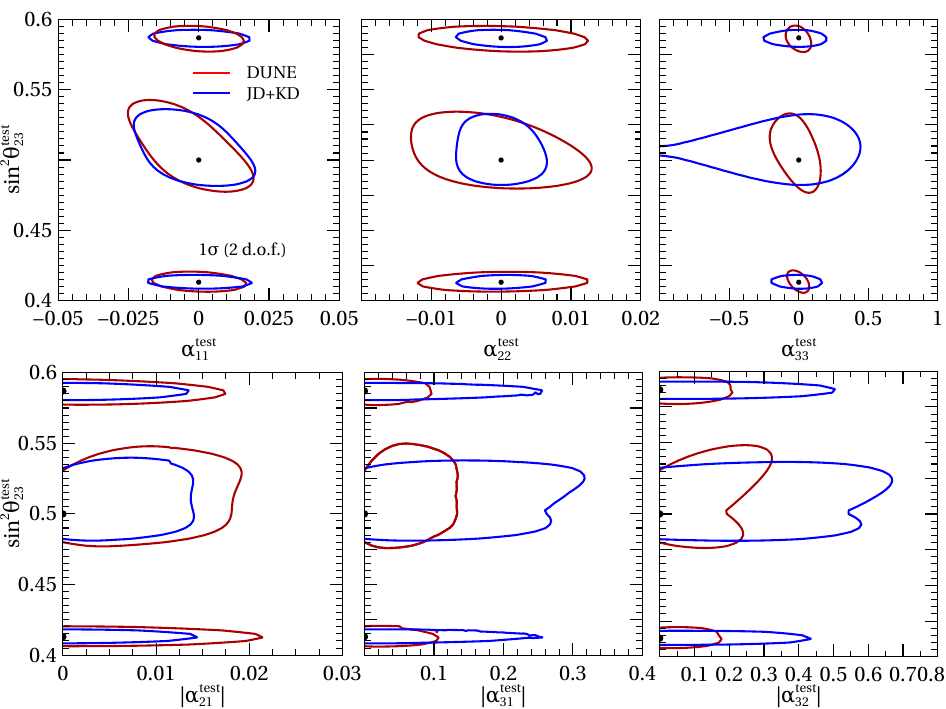}
\mycaption{1$\sigma$ (2 d.o.f.) C.L. contours in the $(\theta_{23}$ - $\alpha_{ij})$ planes for DUNE (red curves) and JD+KD (blue curves). The true values of $\theta_{23}$ considered here are $\theta_{23}=40^{\circ},\, 45^{\circ}$, and $50^{\circ}$ (shown by the black dots in each panel). True value of $\delta_{\mathrm{CP}}$ is considered to be $-\,90^{\circ}$. In the fit, $\delta_{\mathrm{CP}}$ and the phases associated with off-diagonal NUNM parameters (see lower panels) are marginalized over the range of [-$180^{\circ}$, $180^{\circ}$]. The other oscillation parameters are fixed at their best fit values (see table~\ref{table:vac}).}
\label{fig:th23_corr}
\end{figure}

In figure~\ref{fig:th23_corr}, we show the correlation between various NUNM parameters and the atmospheric mixing angle $\theta_{23}$. 
 Three different true values of $\theta_{23}$ (shown by the black dot in each panel) have been analyzed, namely, $\theta_{23}=40^{\circ}$ ($\sin^2\theta_{23} = 0.413$) in the lower octant, the maximal value $\theta_{23}=45^{\circ}$ ($\sin^2\theta_{23} = 0.5$), and $\theta_{23}=50^{\circ}$ ($\sin^2\theta_{23} = 0.586$) in the upper octant. 
Contours in each panel correspond to 1$\sigma$ (2 d.o.f.) C.L. obtained using the setups using the DUNE and JD+KD setups.
We observe that the correlation between $\sin^2\theta_{23}$ and $\alpha_{11}$ (top left panel) is almost the same for both DUNE and JD+KD setups. In the $(\theta_{23}-\alpha_{22})$ and $(\theta_{23}-\alpha_{21})$ planes, the allowed regions for JD+KD are smaller than the ones corresponding to DUNE, which suggest that the J-PARC based experiments will have a better sensitivity on these parameters (see the following subsection). For the other three NUNM parameters, DUNE shows much better sensitivity than JD+KD, due to the larger matter effects which couples to the NUNM parameters $\alpha_{33}$, $|\alpha_{31}|$, and $|\alpha_{32}|$. In particular, for maximal value of $\theta_{23}$, there are no lower limits on $\alpha_{33}$ from JD+KD. One interesting point to be noted is that, in the presence of non-unitary mixing, $\sin^2\theta_{23}$ can be constrained better for non-maximal true values of $\theta_{23}$.

\begin{figure}[tbp]
\centering
\includegraphics[width=\textwidth]{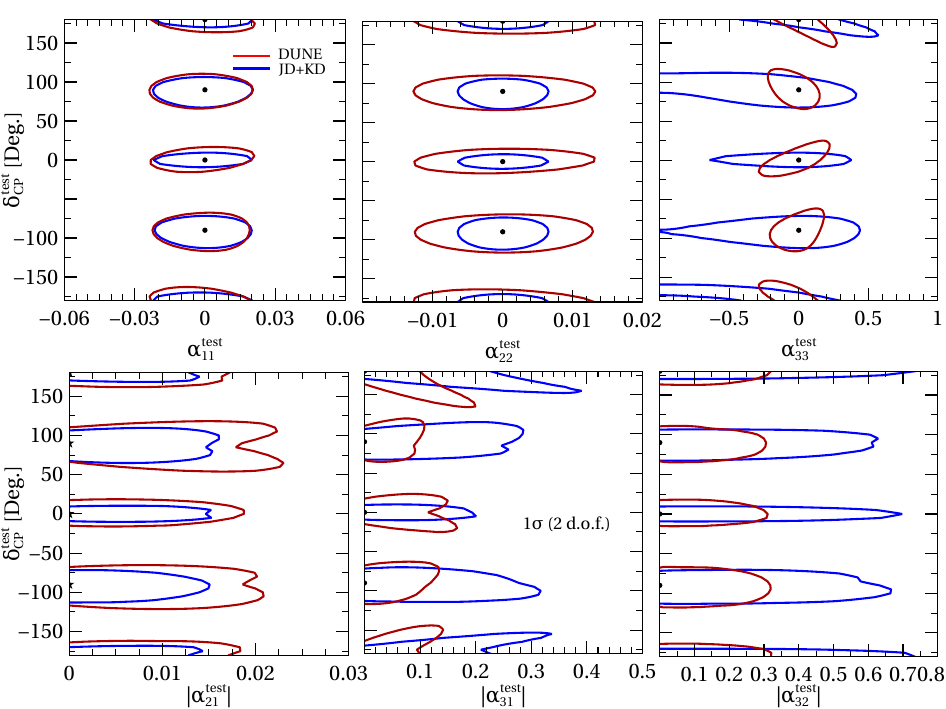}
\mycaption{
1$\sigma$ (2 d.o.f.) C.L. contours in the $(\delta_{\mathrm{CP}}$-$\alpha_{ij})$ planes for DUNE (red curves) and JD+KD (blue curves). The true values of $\delta_{\mathrm{CP}}$ considered here are $\delta_{\mathrm{CP}}=0^{\circ},\, 90^{\circ},\, 180^{\circ},\, -90^{\circ}$ (shown by the black dots in each panel). The true value of $\theta_{23}$ is $45^{\circ}$, while its test value is kept free in the range [$40^{\circ}$, $50^{\circ}$]. In the fit, the phases associated with off-diagonal NUNM parameters (see lower panels) are marginalized over the range of [-$180^{\circ}$, $180^{\circ}$]. The other oscillation parameters are fixed at the values as given in table~\ref{table:vac}.
}
\label{fig:dcp_corr}
\end{figure}

In figure~\ref{fig:dcp_corr}, we show the correlation of various NUNM parameters with the CP phase $\delta_{\mathrm{CP}}$.
In each panel, the black dots correspond to the true values of $\delta_{\mathrm{CP}}=0^{\circ},\, 90^{\circ},\, 180^{\circ}$, and $-\,90^{\circ}$ and the benchmark value chosen for the $\alpha_{ij}$ under consideration. The red contours refer to the sensitivity at 1$\sigma$ (2 d.o.f.) C.L. expected to be obtained by DUNE setup and the blue curve shows the same for JD+KD combination. The plots follow a similar behavior already seen in figure~\ref{fig:th23_corr}. In particular, the allowed regions for the two experiments are almost the same in the case of $\alpha_{11}$, smaller for JD+KD in the case of $\alpha_{22}$ and $|\alpha_{21}|$ and, finally, much smaller for DUNE in the case of the NUNM parameters $\alpha_{33}$, $|\alpha_{31}|$, and $|\alpha_{32}|$. Note that when true $\delta_{\mathrm{CP}}$ = $90^{\circ}$, $180^{\circ}$, and $-\,90^{\circ}$, the J-PARC based detectors will not be able to set any lower limit on $\alpha_{33}$.

\subsection{Constraints on non-unitary neutrino mixing parameters}
\label{subsec:num_results}
In this subsection, we present our numerical results showing the expected constraints on the six NUNM parameters ($\alpha_{ij}$) that DUNE, JD, KD, and JD+KD setups can place. 
We derive bounds on $\alpha_{ij}$ following the simulation method as discussed in section~\ref{sec:simulation_details}. As discussed earlier, while estimating the constraints, we marginalize over the most uncertain oscillation parameters ($\theta_{23}$, $\delta_{\mathrm{CP}}$) and the phases associated with the off-diagonal NUNM parameters ($\phi_{ij}$) in the fit. We also minimize over the systematic pulls on signal ($\lambda_1$) and background ($\lambda_2$).

\begin{figure}[h!]
\centering
\includegraphics[width=\textwidth]{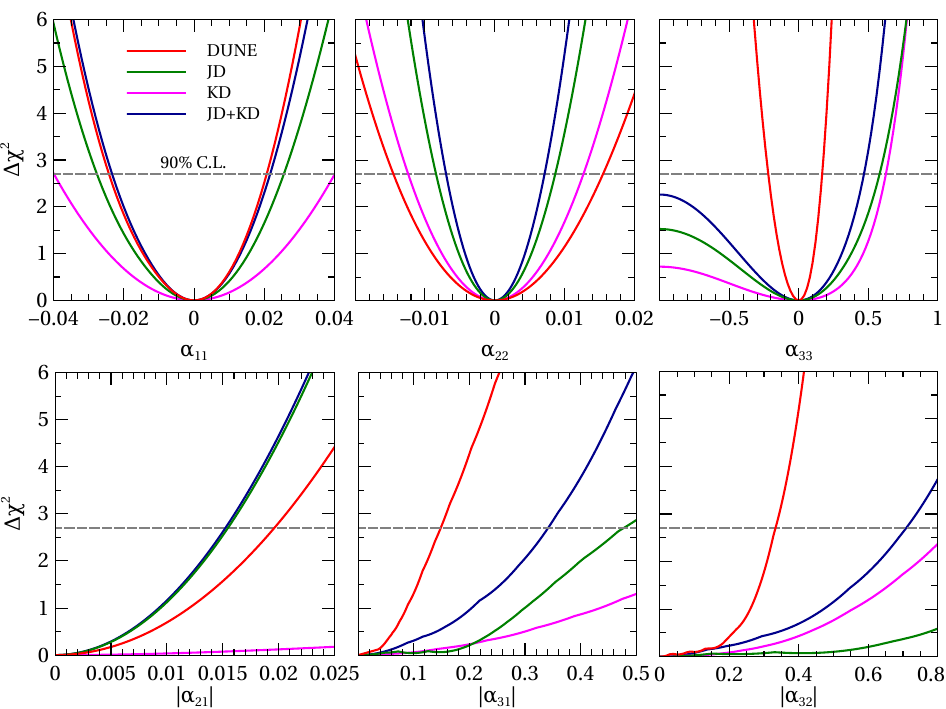}
\mycaption{ Expected limits on the NUNM parameters
	from DUNE (red curves), JD (green curves), KD (magenta curves), and JD+KD (blue curves).
	The upper (lower) panels are for the diagonal (off-diagonal) NUNM parameters one at a time.
	True values of $\theta_{23}$ and $\delta_{\mathrm{CP}}$ are $45^{\circ}$ and $-\,90^{\circ}$, respectively. 
	For the diagonal NUNM parameters, we marginalize over $\theta_{23}$ in the range [$40^{\circ},   50^{\circ}$] and $\delta_{\mathrm{CP}}$ in the range [$-180^{\circ}, 180^{\circ}$] in the fit. For the off-diagonal NUNM parameters, apart from $\theta_{23}$ and $\delta_{\mathrm{CP}}$, we also marginalize over the associated NUNM phases in the range of $-180^{\circ}$ to $180^{\circ}$.}
\label{fig:chi_sq}
\end{figure}
In figure~\ref{fig:chi_sq}, we plot the $\Delta \chi^2$ function for the six NUNM parameters analyzed in our paper considering only the far detectors in a given setup. Upper (Lower)  panels show the results for the diagonal (off-diagonal) NUNM parameters. 
We observe that for $\alpha_{11},$DUNE and JD+KD place similar constraints on $\aee$. The sensitivity to this parameter comes from two contributions: disappearance of intrinsic $\nu_e$ beam and the $\nu_e$ appearance.
 For both of these contributing channels, 
 DUNE has better systematics and JD+KD has more statistics. As a result, limits on $\alpha_{11}$ is found to be almost the same for the two setups. However, for $\amm$ (upper middle panel), JD+KD has significantly better sensitivity compared to DUNE setup. This is because $\amm$ is mainly constrained by the disappearance channels, which due to the large statistics, is primarily limited by the systematic uncertainties. Since the normalization error for this channel is 3.5\% (5\%) for JD+KD (DUNE), it is clear that JD+KD can put better limit than DUNE. We have checked that if we consider the same amount of systematic uncertainties for both the setups, DUNE shows slightly better sensitivity than T2HKK.
 
For $|\ame|$, DUNE and JD+KD have comparable sensitivities (see lower left panel). The slightly better limit on $|\alpha_{21}|$ achieved in the case of JD+KD as compared to DUNE, is due to the fact that JD+KD has larger statistics in the appearance channels. For the other three parameters $|\ate|$, $|\atm|$ and $\att$, which enter the $\nu_{\mu}\rightarrow\nu_{e}$ appearance channel through matter parameters $\Delta_{e}$ and $\Delta_n$ (see eq.~\ref{eq:pme_anylit_main}), DUNE outperforms JD+KD setups because of its large matter effects.

\begin{table}[thb!]
\begin{center}
 \begin{adjustbox}{width=\textwidth}
 \begin{tabular}{|c|c|c|c|c|c|c|}
\hline\hline
&DUNE&JD&KD&JD+KD&JD+KD+DUNE&T2K+NO$\nu$A\\ 
\hline
$\alpha_{11}$ & [-0.020, 0.020] &[-0.025, 0.025]&[-0.040, 0.040]&   [-0.022, 0.022] & [-0.017, 0.017]& [-0.06, 0.06]\\
\hline
$\alpha_{22}$& [-0.014, 0.014]& [-0.0087, 0.0087]&[-0.013, 0.013]& [-0.007, 0.007] & [-0.006, 0.006] & [-0.02, 0.02]\\
\hline
$\alpha_{33}$& [-0.2, 0.17] & $<$ 0.6 & $<$ 0.63 & $<$ 0.476  & [-0.17, 0.17]& $<$ 0.64\\
\hline
$|\alpha_{21}|$ &  $<$ 0.022 & $<$ 0.015 & $<$ 0.10  & $<$ 0.016 & $<$ 0.012 & $<$ 0.06\\ 
\hline
$|\alpha_{31}|$& $<$ 0.15  & $<$ 0.48 & $<$ 0.70 & $<$ 0.34 &$<$ 0.11 & $<$ 2.20\\
\hline
$|\alpha_{32}|$& $<$ 0.33 & $<$ 1.2 & $<$ 0.85 & $<$ 0.71& $<$ 0.27 & $<$ 1.4\\
\hline\hline
\end{tabular}
 \end{adjustbox}
\mycaption{Bounds on the NUNM parameters at 90\% C.L. (1 d.o.f.) using DUNE (second column), JD (third column), KD (fourth column), and JD+KD (fifth column). Sixth column shows the results for the combination of DUNE and JD+KD. Last column depicts results using the full exposure of T2K and NO$\nu$A.}
\label{Tab:constraints}
\end{center}
\end{table}

We summarize our results in table~\ref{Tab:constraints}, where we give the bounds on the six NUNM parameters at 90\% C.L. for the various long-baseline  experimental setups discussed in this paper. As clear from our previous discussion,  the expected constraints on NUNM parameters from DUNE is better than the other two experiments JD and KD (and their combination) except for the parameters $\amm$ and $|\ame|$, where JD has better sensitivity than DUNE.
Finally, in the sixth column of the table, we give the final constraints on the NUNM parameters by combining the expected results from  DUNE and JD+KD setups. As we have anticipated, the bounds experience a general improvements by $\sim\,20$\%, with the precise magnitude depending on the parameter under consideration.

For a comparison with the ongoing long-baseline experiments, we also add the expected constraints from the combination of the T2K and NO$\nu$A setup in the last column. For T2K, we consider a total exposure of 84.4 kt-MW-yrs with 22.5 kt detector mass, 750 kW proton beam power, 5 years run-time (2.5 years each for neutrino and antineutrino mode). 
 For NO$\nu$A, the considered exposure is 58.8 kt-MW-yrs with 14 kt detector mass, 700 kW proton beam power, and 6 years for total run-time (3 years each for neutrino and antineutrino modes).
Due to the limited statistics, we observe that the expected constraints from T2K+NO$\nu$A setup is worse than the DUNE or JD+KD setup.

\begin{table}
	\centering
	\begin{tabular}{|c|c|c|}
		\hline\hline
		Parameter & DUNE (3.5 yrs+3.5 yrs)& DUNE (5 yrs+5 yrs)\\ 
		\hline
		$\alpha_{11}$& [-0.020, 0.020] & [-0.020, 0.018] \\
		\hline
		$\alpha_{22}$ & [-0.014, 0.014] & [-0.013, 0.013]\\
		\hline
		$\alpha_{33}$ & [-0.2, 0.17] & [-0.19, 0.15]\\
		\hline
		$|\alpha_{21}|$ & $<$ 0.022 & $<$ 0.016\\ 
		\hline
		$|\alpha_{31}|$ & $<$ 0.15 & $<$ 0.12\\
		\hline
		$|\alpha_{32}|$ & $<$ 0.33 & $<$ 0.31\\
		\hline\hline
	\end{tabular}
	\mycaption{
		90\% C.L. (1 d.o.f.) limits on  the NUNM parameters considering two different exposures of DUNE: total run-time of 7 years (see second column) and 10 years (see third column) equally divided in neutrino and antineutrino modes.}
	\label{tab:DUNE_runtime}
\end{table}

Note that if some information coming from the near detector (for an example, measurement of the initial neutrino flux) are used to analyze the far detector data then the constraints on $\alpha_{11}$ and $\alpha_{22}$ may be modified (see section \ref{sec:ND} for a detailed discussion). 
	However, in this section, we adopt a different strategy, where we simulate the far detector data alone to set limits on the NUNM parameters. In principle, this approach is valid if the initial flux can be predicted by some theoretical calculation or measured at an experiment which is insensitive to neutrino oscillation phenomena. In fact, the MINOS/MINOS+ experiment adopted this strategy where the oscillation data at the far detector was analyzed using information from the MINERvA flux predictions~\cite{MINERvA:2016iqn}. In future, if somehow we can apply this approach for DUNE and T2HKK, then in principle, one can estimate the limits on the NUNM parameters using only the far detector data.

	At the same time, we understand that the assumptions that we take in our paper for the systematic uncertainties at the far detector may be too optimistic if we do not use the near detector to measure the initial flux.
		To have a better understanding on this issue, we perform some study and find that limits on $\alpha_{11}$ and $\alpha_{22}$ are mainly governed by the systematic uncertainties. In other words, the expected constraints on these two NUNM parameters are proportional to the systematic uncertainties that we consider in our simulation. Therefore, it may be possible to predict what would be the limits on $\alpha_{11}$ and $\alpha_{22}$ for a given assumptions on systematic uncertainties.

We compare our results summarized in table~\ref{Tab:constraints} with the bounds reported in ref.~\cite{Escrihuela:2016ube}\footnote{In order to get their results, the authors of ref.~\cite{Escrihuela:2016ube} left free the standard oscillation parameters $\theta_{23}$, $\delta_{\mathrm{CP}}$, and $\Delta m^2_{31}$ and the NUNM parameters $\aee$, $\ame$, $\amm$. Conversely, in our work we marginalize over $\delta_{\mathrm{CP}}$ and $\theta_{23}$ only, but we have checked that the marginalization over $\Delta m^2_{31}$ does not have any significant impact.}
We observe that the bound we achieve from the DUNE+JD+KD (or DUNE+T2HKK) setup for the diagonal $\aee$ is $\sim$\,80\% better than the bound quoted in ref.~\cite{Forero:2021azc}. In the $\amm$ case, the two results are comparable, with a slightly better limit when the global neutrino data analysis is performed. 
For the remaining diagonal parameter $\att$, NC data from MINOS/MINOS+ give a 60\% stronger bound \cite{Forero:2021azc}  compared to the one expected from the DUNE+JD+KD setup. As for $\ame$, the authors of the ref.~\cite{Forero:2021azc} make use of the triangular inequality as well as the data from the short-baseline experiments; this allows to constrain the mentioned parameter very tightly. However, due to the large statistics and good systematics of DUNE and JD+KD setups, we can achieve an almost similar bound without using any external hypothesis on the relations between the $\alpha_{ij}$. On the other hand,
constraints on $\ate$ and $\atm$ in ref.~\cite{Escrihuela:2016ube} are substantially better than the ones we obtain from DUNE+JD+KD setup. Also, in these cases, the triangular inequalities which link them to the diagonal NUNM parameters  play an important role, together with the short-baseline experiments limits on the $\nu_\tau$ appearance. However, it is important to stress that all our results are obtained in a complete model independent fashion, relying only on the expected data from DUNE and T2HKK. We check that for our best setup, namely DUNE+T2HKK, the only parameter whose limit gets improved when we consider these inequalities is $|\alpha_{32}|$ because of the stringent bound on $\alpha_{22}$. Since $|\alpha_{21}|$ is already tightly constrained, we do not see any improvement in its limit using these inequalities. As far as the bound on $|\alpha_{31}|$ is concerned, since the limit  on the diagonal parameter $\alpha_{33}$ is very poor, we also do not see any improvement.

Recently, the DUNE collaboration~\cite{DUNE:2021cuw}  exploited the possibility of increasing the exposure of the experiment from 336 kt-MW-yrs to 480 kt-MW-yrs (corresponding to an increase of the  data taking time  from 7 years to 10 years with 5 years in neutrino mode and 5 years in antineutrino mode). In Table~\ref{tab:DUNE_runtime}, we compare our previous constraints from the DUNE experiment, table~\ref{Tab:constraints}, with those obtained in the $(5 + 5)$ years configuration. We observe that the constraints on  all six NUNM parameters improve by small amount except for $|\ame|$, which shows a significant improvement. This happens because the higher run-time increases statistics of the $\nu_{\mu}\rightarrow\nu_e$ appearance channel, which is the one driving the $\alpha_{21}$ sensitivity. On the other hand, the $\nu_\mu\rightarrow\nu_{\mu}$ disappearance channel is almost already saturated by systematics after 3.5 years + 3.5 years of running. This leads to only small improvements on the other NUNM parameters sensitivities.

\section{Benefits of near detectors}
\label{sec:ND}
Near detectors (ND) are a fundamental component for long-baseline neutrino experiments. Indeed, a detector placed very close to the beam source (from hundreds of meters to a few kilometers) is able to monitor the neutrino beam, measuring the flavor composition, and the total number of neutrinos emitted from the source. 
Near detectors are not expected to improve any of the standard oscillation parameter measurements, since at such short distances, oscillations do not develop for neutrinos with energies in the GeV range.
 However, in some new physics scenarios, in which, oscillation probabilities contain  zero-distance terms, near detectors can be used to constrain non-standard parameters in a very straightforward way. This is the case of the non-unitarity framework under discussion where, as already mentioned in section~\ref{sec:osc_prob_NUNM}, at vanishing distances we have zero-distance terms in case $\nu_{\mu}\rightarrow\nu_{e}$ appearance channel: $
P_{\mu e}^{L = 0} \sim |\alpha_{21}|^2$, and $\nu_{\mu}\rightarrow\nu_{\mu}$ disappearance channel:
$P_{\mu\mu}^{L = 0} \sim 1+ 2 |\alpha_{21}|^2+6 \alpha_{22}^2+4 \alpha_{22}\,.$
Thus, we can expect that T2HKK and DUNE near detectors would be able to constrain two parameters $|\alpha_{21}|$ and $\alpha_{22}$ from $\nu_{\mu}\rightarrow\nu_e$ appearance and $\nu_{\mu}\rightarrow\nu_{\mu}$ disappearance channel, respectively, but also $\alpha_{11}$ considering the $\nu_e$ beam contamination (see eq.~\ref{eq:zero-distance} in appendix~\ref{appendix:osc_prob_nunm}). So, in this section, we analytically infer the order of magnitude of bounds implied by ND measurements.
Let us  consider the total number of events of a given channel as~\cite{Giarnetti:2020bmf}:
\begin{equation}
    N= N_0 P_{\alpha\beta}(\alpha_{ij})\,,
\end{equation}
where the normalization factor $N_0$ includes all the detector properties. For an oscillation channel $\nu_\alpha\to\nu_\beta$, $N_0$ can be defined as:
\begin{eqnarray}
\label{eq:asirate2}
N_0 &=&\int_{E_\nu} dE_\nu \,\sigma_\beta(E_\nu)\,\frac{d\phi_\alpha}{dE_\nu}(E_\nu) \,\varepsilon_\beta(E_\nu)\,,
\end{eqnarray} 
where, $\sigma_\beta$ denotes the production cross-section of  the $\beta$ lepton, $\varepsilon_\beta$ represents the detector efficiency, and $\phi_\alpha$ stands for the initial neutrino flux of flavor $\alpha$. If we want to put bounds on new physics parameters, we can use a simple $\chi^2$ test with a gaussian $\chi^2$ defined as 
\begin{equation}
    \chi^2=\frac{(N_{obs}-N_{fit})^2}{\sigma^2}\,,
\end{equation} 
where, $\sigma$ represents the uncertainty on the number of events; in this case, neglecting the backgrounds, we get:
\begin{equation}
    \chi^2=\frac{N_0^2}{\sigma^2}\left[\delta_{\alpha\beta}- P_{\alpha\beta}(\alpha_{ij}^{fit})\right]^2\,.
\label{chi}
\end{equation}
For the $\nu_{\mu}\rightarrow\nu_{\mu}$ disappearance channel, the leading term of the probability is $P_{\mu\mu}^{L=0}=1+4\alpha_{22}$. Therefore, the $\chi^2$ assumes the form:
\begin{equation}
    \chi^2=\frac{16 N_0^2 \alpha_{22}^2}{\sigma^2}.
\end{equation}
At a chosen confidence level, represented by a cut at $\chi^2_{0}$, it is possible to exclude the region satisfying
\begin{eqnarray}
|\alpha_{22}|>\frac{\sqrt{\chi^2_0}\sigma}{4 N_0}.
\label{chi_dis}
\end{eqnarray}

\begin{table}
	
	\centering
	
	\begin{tabular}{|cc|*{4}{c|}}
		
		\hline\hline
		
		\multicolumn{2}{|c|}{\multirow{2}{*}{}} & \multicolumn{2}{|c}{$\alpha_{11}$} & \multicolumn{2}{|c|}{$\alpha_{22}$}\\
		
		\cline{3-6}
		
		\multicolumn{2}{|c|}{Expt.} & w/o norm. & w/ norm. & w/o norm. & w/ norm. \\
		
		\hline\hline
		
		\multicolumn{2}{|c|}{DUNE} & [-0.02, 0.02] & [-0.043, 0.034] &  $[-0.014, 0.014]$& [-0.036, 0.048]   \\
		
		
		\hline\hline
		
		\multicolumn{2}{|c|}{JD+KD} &  $[-0.022, 0.022]$ & [-0.048, 0.040]  &  $[-0.007, 0.007]$& [-0.038, 0.050] \\
		\hline\hline
		
		\multicolumn{2}{|c|}{DUNE+JD+KD} & $[-0.017, 0.017]$ & [-0.036, 0.026]  &  $[-0.006, 0.006]$& [-0.026, 0.039]   \\
		\hline

		\hline
		\hline
		
	\end{tabular}
	\mycaption{90\% C.L. (1 d.o.f.) limits on the NUNM parameters $\aee$ and $\amm$ for the two setups, DUNE, JD+KD, and combination of them. Second  and fourth column correspond to the constraints assuming only far detector. Third and fifth columns correspond to the constraints using the FD and ND correlation (or with normalization factor in the oscillation probability). }
	\label{Table:cons_w_norm}
\end{table}

\begin{figure}[h!]
\centering
\includegraphics[width=\textwidth]{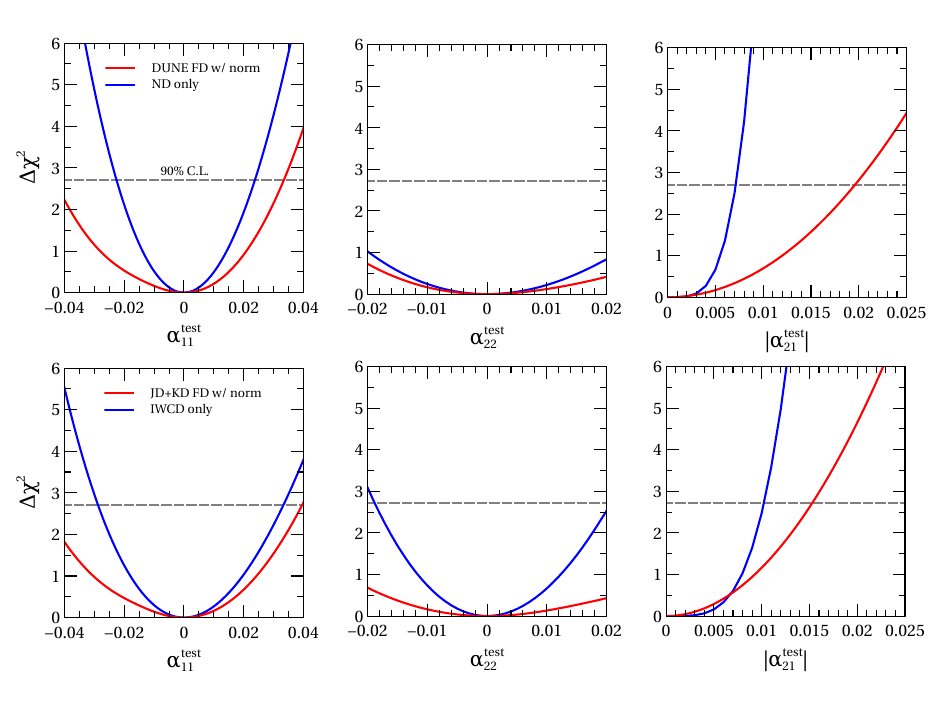}
\mycaption{Upper panels show the improvements in the sensitivities to $\alpha_{11}$, $\alpha_{22}$, and $|\alpha_{21}|$ due to the presence of 67 tons LArTPC near detector placed at distance of 574 meters from the neutrino source for DUNE. Lower panels portray the same for JD+KD having a 1 kt water Cherenkov  near detector placed at a distance of 1 km from J-PARC. The blue curves show the performance with only near detectors. The red curves represent the combined sensitivities due to both near and far detectors. True values of the standard oscillation parameters are taken from table~\ref{table:vac}.
We obtain our results marginalizing over $\delta_{\mathrm{CP}}$ in the range $[-180^{\circ},\,180^{\circ}]$ and $\theta_{23}$ in the range $[40^{\circ},\,50^{\circ}]$ in the fit. We also marginalize over the associated phase $\phi_{21}$ in the range $[-180^{\circ},\,180^{\circ}]$ for the off-diagonal NUNM parameter $|\alpha_{21}|$.}
\label{fig:ND-DUNE}
\end{figure}

Since the disappearance channel is expected to produce a huge number of events at the near detector, one can consider the uncertainty to be dominated by systematic errors $\sigma_{sys}$. Thus, it is possible to approximate $\sigma\,\sim\,N_0 \sigma_{sys}$, where $N_0$ represents the number of events in absence of zero-distance effects, being the disappearance probability in that case equal to 1. This allows to simplify  eq.~\ref{chi_dis} as follows:
\begin{eqnarray}
|\alpha_{22}|>\frac{\sqrt{\chi^2_0}\sigma_{sys}}{4}\,,
\end{eqnarray}
which tells us that, neglecting backgrounds effects, the near detector limits would be of the order of the chosen systematic uncertainty. A similar approach can be used for the $\nu_e\to\nu_e$ oscillation channel, which arises from the $\nu_e$ beam  contamination, obtaining an inequality for $|\alpha_{11}|$ of the similar form:
\begin{eqnarray}
|\alpha_{11}|>\frac{\sqrt{\chi^2_0}\sigma_{sys}^{\nu_e}}{4}\,,
\label{a11_ND}
\end{eqnarray}
where $\sigma_{sys}^{\nu_e}$ refers to the systematic uncertainty on the $\nu_e\to\nu_e$ transition.\\
For the appearance channel, the zero-distance probability reads $P_{\mu e}^{L=0}=|\alpha_{21}|^2$; the $\chi^2$ function can therefore be written as:
\begin{equation}
    \chi^2=\frac{N_0^2 |\alpha_{21}|^4}{\sigma^2}\,,
\end{equation}
and the excluded region is expected to be determined by the following relation:
\begin{eqnarray}
|\alpha_{21}|<\sqrt[4]{\frac{\chi_0^2 \sigma^2}{N_0^2}}\,.
\end{eqnarray}
Since the number of events at the near detector is in principle very small (being only caused by new physics), the uncertainty is dominated by statistics. Thus, given a certain number of observed events, $\sigma\,\sim\,\sqrt{N_{obs}}$ and the excluded values of $|\alpha_{21}|$ reduced to:
\begin{eqnarray}
 |\alpha_{21}|<\sqrt[4]{\frac{\chi_0^2 N_{obs}}{N_0^2}}\,,
 \label{mue_ND}
\end{eqnarray}
suggesting that the bounds are very sensitive to the number of events and to the  running time of the experiment. 

Both DUNE and T2HKK will have near detectors~\cite{Miranda:2018yym,DUNE:2021tad,Coloma:2021uhq} which may play a crucial role to probe various new physics scenarios including the possibility of non-unitarity of the PMNS matrix which is the main thrust of this work. In our analysis, for DUNE, we consider a 67 tons LArTPC near detector placed at a baseline of 574 meters from Fermilab~\cite{DUNE:2021tad}. For JD+KD, we consider a 1 kt water Cherenkov near detector located at a baseline of 1 km from J-PARC which is known as IWCD~\cite{Drakopoulou:2017qdu, Wilson:2020trq}.
In order to simulate their responses, we scale the far detector fluxes for ND baselines and take into account their fiducial masses. We follow a very conservative approach as far as the systematic uncertainties at the near detectors are concerned. We multiply the FD systematic uncertainties by a factor of three and consider them as inputs for the ND.
 In DUNE near detector, we expect $\mathcal{O}(10^7)$ $\nu_\mu$ and $\bar{\nu}_{\mu}$ events, which provide bounds on $\alpha_{22}$. DUNE can place stringent constraints on $\alpha_{11}$ and $\alpha_{21}$ using $\mathcal{O}(10^6)$ $\nu_e$ and $\bar{\nu}_e$ events at ND, which stem from both intrinsic $\nu_e$ ($\bar{\nu}_e$) beam contamination and via $\nu_\mu\to\nu_e$ ($\bar{\nu}_\mu\rightarrow\bar{\nu}_e$) appearance caused due to zero-distance effect. 
  For the NDs, we consider their appropriate baselines, fiducial masses, and systematic uncertainties which we assume to be larger than the systematic uncertainties considered for the FDs. 
 
 \begin{table}
 	\centering
 	\begin{adjustbox}{width=\textwidth}
 		\begin{tabular}{|c|c|c|c|c|}
 			\hline\hline
 			Parameter & DUNE (ND) & DUNE FD w/ norm. &  IWCD & JD+KD w/ norm.\\ 
 			\hline
 			$\alpha_{11}$& [-0.020, 0.024] & [-0.043, 0.034] & [-0.029, 0.033] & [-0.048, 0.040] \\
 			\hline
 			$\alpha_{22}$ & [-0.033, 0.037] & [-0.036, 0.048] & [-0.019, 0.020] & [-0.038, 0.050] \\
 			\hline
 			$|\alpha_{21}|$ & $<$ 0.007 & $<$ 0.022 & $<$ 0.01 & $<$ 0.015 \\
 			\hline\hline
 		\end{tabular}
 	\end{adjustbox}
 	\mycaption{90\% C.L. (1 d.o.f.)	bounds on the NUNM parameters $\aee$, $\amm$, and $|\ame|$ obtained with and without near detectors in DUNE and JD+KD. Note that IWCD is the near detector for JD+KD setup.}
 	\label{tab:NearDetector}
 \end{table}
 
 Before discussing the limits that the near detectors would be able to set using their own data, we want to study the effect of the ND flux measurements on the far detector constraints. Indeed, if the initial neutrino flux is measured at the near detector and then extrapolated to the far detector, the probability which could be inferred at the far detector is the effective probability defined as
	\begin{equation}
		P^{\text{eff}}_{\alpha\beta} = \frac{P_{\alpha\beta}}{P^{L=0}_{\alpha\alpha}}\,.
		\label{eq:eff_prob}
	\end{equation}
The $P^{L=0}_{\alpha\alpha}$ term that appears in the denominator is the survival probability initial neutrino flavor at the source or the zero-distance term which act as a normalization factor. If we normalize the $\nu_\mu\to\nu_\mu$ survival probability at the far detector using the zero-distance term in eq.~\ref{pmmzero}, it is observed that the contribution from $\alpha_{22}$ gets canceled at the leading order. As a result, sensitivity to the parameter $\alpha_{22}$ is worsened for a given setup.
The same happens for $\alpha_{11}$, whose contribution in the effective $\nu_e\to\nu_e$ disappearance probability is canceled at the leading order (see eq.~\ref{eq:e-disapp} and~\ref{eq:zero-distance}). Since the sensitivity to this parameter arises partially due to the intrinsic $\nu_e$ that we have in the beam to begin with, the near detector normalization causes a deterioration of $\alpha_{11}$ limits.
In table~\ref{Table:cons_w_norm}, we show how the constraints on $\aee$ and $\amm$ would be modified when taking into account the FD and ND correlation for the three setups namely, DUNE, JD+KD, and DUNE+JD+KD. We observe that the bound on $\aee$ is increased by a factor of almost two when we consider the correlation between the FD and ND. For $\amm$, the bound is deteriorated at least three times compared to the FD case only. We have checked that no other NUNM parameter is affected significantly if we consider the FD and ND correlation. Indeed, the non-diagonal parameters and $\alpha_{33}$ can be constrained using appearance channels (for which we do not have full cancellations in the effective probabilities) or using the interplay with matter effects, which are not developed at the near site. 
 
The bounds obtained using the above-mentioned near detectors are shown in figure~\ref{fig:ND-DUNE} for DUNE and T2HKK together with the results we got using the FD data with effective probabilities. For $\alpha_{11}$, NDs of the two setups can put bounds better than the one set by the FDs considering the ND normalization due to the very high statistics and the strong $\alpha_{11}$ dependence of the zero-distance probability. The improvement is roughly a factor of two for DUNE and 60\% for T2HKK (see table~\ref{tab:NearDetector}). Note that the obtained limits are in agreement with the predictions deduced from eq.~\ref{a11_ND}, once we insert a normalization uncertainty of 6\% (15\%) for DUNE (JD+KD).  

For the second diagonal parameter $\alpha_{22}$ we also observe a similar situation, in which the ND alone can put more stringent bounds than the far detector when the normalization is considered, despite of the increased systematics. The improvement can be quantified as roughly 25\% in DUNE and a factor of two in T2HKK. Once again, the analytical predictions from eq.~\ref{a11_ND} are sufficiently recovered by the numerical simulations, considering that the near detectors normalization systematics are 15\% for DUNE and 10.5\% for T2HKK. Note that the bounds from the far detector data alone would be considerably better than the near detector ones. 

Finally, for the NUNM parameter $|\alpha_{21}|$ the near detectors bounds are considerably better than the far detector ones, due to the zero-distance effect outlined in eq.~\ref{zerodistancemue}. 
In particular, the limits are $\sim$ 3 times smaller than the one set by the far detector in the DUNE facility and $\sim$ 70\% smaller in the case of T2HKK. Considering a number of observed events of  ${\cal O}(10)$, and taking into account that we expect $N_0\sim10^6$ per year \cite{DUNE:2021tad}, our analytic estimate  for $|\alpha_{21}|$ is comparable with the numerical results.

\section{Improvements due to $\nu_\tau$ events at DUNE}
\label{sec:nu_tau}
The $\nu_\tau$ production at accelerator experiments is very challenging since the charged current interactions of such particles with nuclei have an energy threshold of 3.1 GeV. Thus, many proposed long-baseline experiments are not able to detect such neutrinos\footnote{In this respect, the OPERA setup provided neutrinos with an average  energy close to 13 GeV~\cite{OPERA:2010pne}.}.
However, the DUNE neutrino spectra will have peak at around 2.5 GeV (differently from the T2HKK where $E_\nu \simeq 0.6$ GeV) and the most energetic neutrinos of the beam will have enough energy to produce $\tau$ leptons. Recently, different studies~\cite{Machado:2020yxl,Ghoshal:2019pab,DeGouvea:2019kea,Martinez-Soler:2021sir} take into account the possibility of including the $\nu_\tau$ ($\bar{\nu}_\tau$) sample in the DUNE analysis. 

The recognition of such events could be in principle possible due to the imaging capabilities of LArTPC detectors. Because of a relatively small number of neutrinos with an energy above the production threshold and of the short lifetime of the $\tau$ leptons which could make the recognition of the $\tau$ interaction and the decay vertices a difficult task, the $\nu_\tau$ appearance channel is not really useful to constrain the standard oscillation parameters. However, when new physics affects the oscillation probabilities, it has been shown in Refs.~\cite{Ghoshal:2019pab,DeGouvea:2019kea,Denton:2021mso,Denton:2021rsa} that the NUNM parameters $\alpha_{33}$ and $|\alpha_{32}|$ constraints can be improved even by the small number of $\nu_\tau$ events. 

\begin{figure}[h!]
\centering
\includegraphics[width=\textwidth]{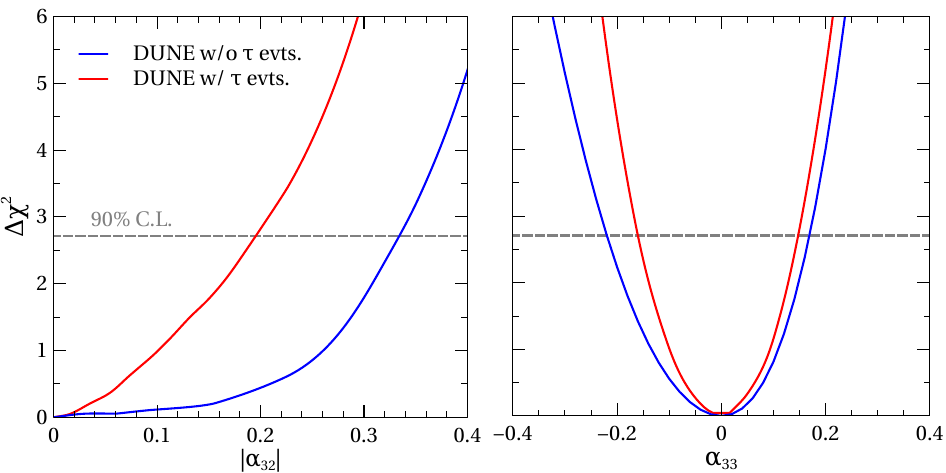}
\mycaption{Comparison between the DUNE sensitivities on $|\alpha_{32}|$ (left panel) and $\alpha_{33}$ (right panel) when $\nu_\tau$ appearance channel is included in the analysis (red lines) and the case where no $\tau$ events are analyzed (blue lines). True values of the standard oscillation parameters are taken from table~\ref{table:vac}. All results have been obtained marginalizing over $\delta_{\mathrm{CP}}$ in the range $[-180^{\circ}, 180^{\circ}]$ and $\theta_{23}$ in the range $[40^{\circ}, 50^{\circ}]$. For $|\alpha_{32}|$ (left panel), we also marginalize over $\phi_{32}$ in the range $[-180^{\circ}, 180^{\circ}]$. }
\label{fig:tau}
\end{figure}

\begin{table}
	\centering
	\begin{tabular}{|c|c|c|}
		\hline\hline
		Parameter & w/o $\nu_{\tau}$ appearance & w/ $\nu_\tau$ appearance\\ 
		\hline
		$\alpha_{33}$ & [-0.2, 0.17] & [-0.16, 0.15] \\
		\hline
		$|\alpha_{32}|$ & $<$ 0.33 & $<$ 0.19 \\
		\hline\hline
	\end{tabular}
	\mycaption{90\% C.L. limits on the NUNM parameters $\alpha_{33}$ and $|\alpha_{32}|$ from the DUNE setup.
		Second (third) column shows the results without (with) $\tau$ in the analysis.}	
		\label{tab:DUNE_taus}
\end{table}
In case of non-unitary neutrino mixing, the $\nu_\mu\to\nu_\tau$ oscillation probability mainly depends on the three parameters $|\alpha_{32}|$, $\alpha_{33}$ and $\alpha_{22}$ (see eq.~\ref{eq:P_mutau}). While we expect that the sensitivity to $\alpha_{22}$ will not be improved by the small number of $\nu_\tau$ events, the other two poorly constrained NUNM parameters $|\alpha_{32}|$ and $\alpha_{33}$ could take advantage of $\nu_\mu\to\nu_\tau$ oscillation channel. We include $\tau$ events in our analysis in the following fashion.
\begin{itemize}
    \item For the hadronic decays of $\tau$ events having a branching ratio of 65\%, a 30\% signal efficiency has been assumed and 10\% of the NC events are considered as background~\cite{DeGouvea:2019kea}.
   
    \item For $\tau$ decaying to electron with a branching ratio of 17.4\%, we assume a signal efficiency of 30\% considering $\nu_e$ events as possible background. We consider the signal to background ratio of 2.45 in our analysis~\cite{Ghoshal:2019pab}.
    
\end{itemize}
The muonic decays have not been taken into account since the discrimination of the number of background events would be too large compared to the signal events~\cite{DeGouvea:2019kea,Ghoshal:2019pab}. The total number of $\nu_\tau$ ($\bar{\nu}_\tau$) events in DUNE is expected to be roughly 72 (37) per year.
The normalization error for the signal is taken to be 20\%. The results of our analysis for $\alpha_{33}$ and $|\alpha_{32}|$ are shown in Fig.~\ref{fig:tau} and table~\ref{tab:DUNE_taus}. The allowed range of $\alpha_{33}$, which appears only at the second order in the probability, is reduced of $\sim$ 13\% by the inclusion of the new oscillation channel, and the new limits are set into the range [-0.16, 0.15] (see table~\ref{tab:DUNE_taus}). On the other hand, the sensitivity to $|\alpha_{32}|$, which impacts linearly the $\nu_\tau$ appearance probability, is significantly improved: in this case, the new upper bound is roughly 60\% smaller than the one set by the standard oscillation channels, namely $|\alpha_{32}|<0.19$. 

\section{Impact of marginalization of the other NUNM parameters}
\label{sec:NU_marg}
 \begin{figure*}[h!]
 	\centering
 	\includegraphics[width=\textwidth]{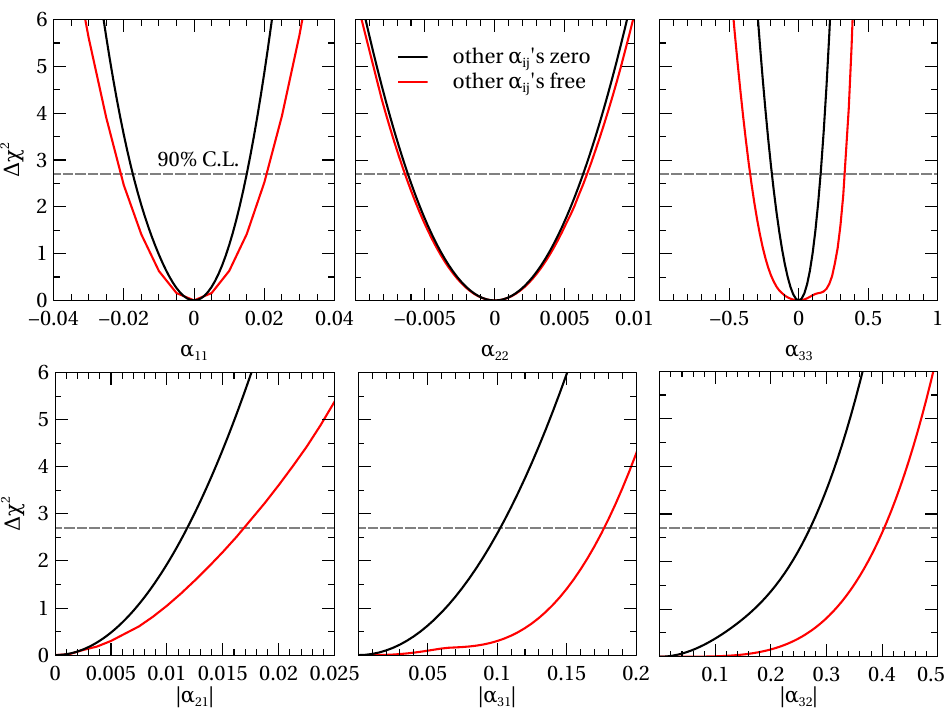}
 	\mycaption{The combined sensitivity of DUNE and JD+KD setups to the NUNM parameters. True values of the standard oscillation parameters are given in table~\ref{table:vac}. We marginalize over $\theta_{23}$ and $\delta_{\mathrm{CP}}$ in the fit (see text for details). Black curves show the sensitivity when only one NUNM parameter is considered at a time, while the others are taken to be zero. Red curves in all the panels correspond to the case when all other NUNM parameters are kept free in the fit.  \label{fig:NU-param-marg}} 
 \end{figure*}
 
In this section, we derive the constraints on the NUNM parameters obtained, differently from the procedure adopted in the main text, marginalizing over the other NUNM parameters $\alpha_{ij}$. Our numerical results are shown in figure~\ref{fig:NU-param-marg}, where we show the impact of the marginalization over various NUNM parameters on the sensitivity of the DUNE+JD+KD setup. 
 In each panel, black curves show the sensitivity for a particular NUNM parameter, when all other NUNM parameters are fixed in the fit. The red curves show the situation when all un-displayed NUNM parameters are marginalized in the fit with no priors.
The 90\% C.L. (1 d.o.f.) constraints on various NUNM parameters are summarized in table~\ref{tab:NUNM_marg}.
 \begin{table}
 	\centering
 	\begin{tabular}{|c|c|c|}
 		\hline\hline
 		Parameter & Other $\alpha_{ij}$'s zero & Other $\alpha_{ij}$'s free \\ 
 		\hline
 		$\alpha_{11}$& [-0.017, 0.017] & [-0.02, 0.02] \\
 		\hline
 		$\alpha_{22}$ & [-0.006, 0.006] &  [-0.006, 0.006] \\
 		\hline
 		$\alpha_{33}$ & [-0.20, 0.17] & [-0.35, 0.33] \\
 		\hline
 		$|\alpha_{21}|$ & $<$ 0.012  & $<$ 0.017\\ 
 		\hline
 		$|\alpha_{31}|$ & $<$ 0.11 & $<$ 0.18\\
 		\hline
 		$|\alpha_{32}|$ & $<$ 0.27 & $<$ 0.40\\
 		\hline\hline
 	\end{tabular}
 	\mycaption{90\% C.L.(1 d.o.f.) limits on various NUNM parameters. Second column shows the constraints considering only one NUNM parameter at a time, while other NUNM parameters are assumed to be zero in the fit. Third column depicts the bounds on a given NUNM parameter when all other NUNM parameters and the phases associated with the off-diagonal parameters are kept free in the fit. True values of the standard oscillation parameters are taken from table~\ref{table:vac}. We marginalize over $\theta_{23}$ and $\delta_{\mathrm{CP}}$ in the fit (see text for details). \label{tab:NUNM_marg}} 
 \end{table}
We observe that the marginalization over all other NUNM parameters worsen the bounds on all of them but $\alpha_{22}$. This is due to the fact that, as it can be seen in eq. \ref{eq:pmm_anylit_main}, $\alpha_{22}$ appears at the leading order, with no correlations to the other $\alpha_{ij}$.
The other diagonal parameter $\alpha_{11}$ shows only a marginal deterioration at the level of 15\% since, as before, its correlation with the other NUNM parameters are mild in $\nu_e$ appearance probability as shown in ref.~\cite{Escrihuela:2015wra}.
We see from table~\ref{tab:NUNM_marg} that there is a considerable deterioration in the sensitivities for $\alpha_{33}$, $|\alpha_{21}|$, $|\alpha_{31}|$, and $|\alpha_{32}|$ when we marginalize over the un-displayed NUNM parameters in the fit.  
This happens because all these four parameters are strongly correlated among them and with other two NUNM parameters $\alpha_{11}$ and $\alpha_{22}$ (see eqns.~\ref{eq:pme_anylit_main} and \ref{eq:pmm_anylit_main}).
Note that $|\alpha_{21}|$ does not have any correlation with the other NUNM parameters in $\nu_\mu\rightarrow\nu_e$ appearance channel but it is strongly correlated with $\alpha_{22}$, $|\alpha_{31}|$, and $|\alpha_{32}|$ in $\nu_\mu\rightarrow\nu_\mu$ disappearance channel which reduces its sensitivity by roughly 40\%, when we keep the other NUNM parameter free in the fit.
Similar correlations among the NUNM parameters are also responsible for worsening the bounds on $|\alpha_{31}|$ by 60\%, $|\alpha_{32}|$ by 45\%, and by a factor of two for $\alpha_{33}$ when we marginalize over the other NUNM parameters in the fit.

\section{Summary}
\label{sec:conclusion}

The data from the ungoing neutrino oscillation experiments has led us to a magnificent precision on neutrino mixing angles and rapidly increasing knowledge on $\delta_{\mathrm{CP}}$. 
In the current scenario,
it seems quite natural to ask if there is any violation of the unitary property of $3\times 3$ PMNS mixing matrix, which may be related to the existence of new mass eigenstates of neutrinos.
In this context, there exists one well-known parameterization in the literature, which takes into account the non-unitary neutrino mixing (NUNM) by introducing a lower triangular matrix with three real and three complex parameters, $\alpha_{ij}$.

To study the impact of these NUNM parameters on various oscillation channels, we derive simple 
approximate analytical expressions for the oscillation probabilities in matter in the atmospheric regime ($\Delta_{31}>>\Delta_{21}$).
Our perturbative expansions are valid up to second order in the small deviations from the mixing angles by their tri-bimaximal values, second order in the NUNM parameters, and first order in the matter potential. We have shown that the $\nu_{\mu}\rightarrow\nu_e$ appearance probability mainly depends on $|\alpha_{21}|$, but when matter potential is large, the impact of $|\alpha_{31}|$ and $|\alpha_{32}|$ can also be significant. On the other hand, the $\nu_{\mu}\rightarrow\nu_{\mu}$ disappearance probability mainly relies on $\alpha_{22}$, but sub-leading dependencies on $|\alpha_{21}|$ in vacuum and $\alpha_{33}$, $|\alpha_{31}|$, and $|\alpha_{32}|$ in matter are also present. The only parameter that does not appear in our formulae is $\alpha_{11}$, which is only relevant at the higher-orders in our perturbative expansions  as shown in Ref.~\cite{Escrihuela:2015wra} for the vacuum case.

In this work, we analyze in detail, the impact of possible NUNM in the context of long-baseline experiments DUNE and T2HK having one detector in Japan (JD) and a second detector in Korea (KD), and the combination of these two detectors, known as T2HKK or JD+KD. First, we show how the  various NUNM parameters ($\alpha_{ij}$) are correlated with the oscillation parameters $\theta_{23}$ and $\delta_{\mathrm{CP}}$ for these setups.
Then, we estimate in detail the sensitivities of these experiments to place direct, model-independent, competitive constraints on the six NUNM parameters  at 90\% confidence level. 
 The wide-band neutrino beam in DUNE encompassing both first and second oscillation maxima allows us to measure the NUNM parameters at several $L/E$ values in the presence of significant matter effect due to its 1300 km baseline. Indeed, DUNE 
 shows better sensitivity than JD+KD in constraining the NUNM parameters, which are influenced by matter effects, namely, $\alpha_{33}$, $|\alpha_{31}|$, and $|\alpha_{32}|$.
We observe that JD+KD will provide a stringent constraint on $\alpha_{22}$ as compared to DUNE because it has less systematic uncertainties in the disappearance channel. 
Also, due to the larger statistics in the appearance channel, JD+KD will give significantly better limit on the NUNM parameter $|\alpha_{21}|$ in comparison to DUNE.
Lastly, $\alpha_{11}$ is expected to be constrained almost in the same way by these two experiments.
We show how the limits on the NUNM parameters get improved in case of DUNE, if the total exposure is increased from 336 kt-MW-yrs to 480 kt-MW-yrs (corresponding to an increase in the total run-time  from 7 years to 10 years), as proposed in the recent TDR~\cite{DUNE:2021cuw}.
 We also estimate how much the sensitivities can be improved by adding the prospective data from DUNE and JD+KD. Finally, we compare our results with the constraints that can be achieved using the full exposure of currently running experiments T2K and NO$\nu$A.
 
 Due to the so-called zero-distance effects which are induced by the non-unitary neutrino mixing in neutrino oscillation probabilities,
the prospective data from near detectors in both DUNE and JD+KD experiments could be in principle used to bound the three NUNM parameters $|\alpha_{21}|$, $\alpha_{11}$ and $\alpha_{22}$. However, the zero-distance effect should also be taken into account if the near detectors data will be used to measure the initial neutrino flux for both experiments. This would lead to a substantial deterioration of the  limits that the far detectors could set on $\alpha_{11}$ and $\alpha_{22}$, as summarized in Table~\ref{Table:cons_w_norm}.
Moreover, in DUNE, the expected limits on $|\alpha_{32}|$ and $\alpha_{33}$ get improved by ${\cal O}(20)\%$ when we also add the $\nu_{\tau}$ appearance sample in our analysis. 


\blankpage 
\chapter{Constraints on long-range interactions using astrophyisical neutrino flavor composition in IceCube}
\label{C6} 
\section{Introduction}

High-energy astrophysical neutrinos provide unique opportunities to test fundamental physics. These neutrinos produced in the extremely high-energy environment inside the galactic sources travel Gpc-scale distances before they reach Earth. Because of their high energy and large propagation distances, minute physics effects, ordinarily undetectable, may accumulate en route to Earth and become observable at the detector. In this work, we explore the possible flavor-dependent long-range interactions of astrophysical neutrinos with the ordinary matter particles in the universe. Theoretically, these interaction can be introduced by gauging 
global lepton-number symmetries of the Standard Model, generated by $L_\alpha-L_\beta$, where $L_\alpha$ is $\alpha$ flavor lepton number. These models induce a new neutrino-matter interaction mediated by a new gauge boson $Z'_{\alpha\beta}$. In the ultralight mass limit of the new mediator, the resulting interaction can be long-range in nature; the interaction range depends on the mediator mass. 

Due to the long-range property of the interaction, matter particles across the universe could source the long-range potential. As a result, vast numbers of electrons, protons, and neutrons in the Earth, the Moon, the Sun, the Milky Way, and in
the distant universe can possibly influence the propagation of the astrophysical neutrinos.
Since the interaction is flavor-dependent, {\it i.e.}, interactions that affect different neutrino flavors differently, neutrino flavor transition probabilities may also get modified. In this work, we use the flavor ratio of the diffused astrophysical neutrino flux observed at several present and future neutrino telescopes. We estimate the flavor composition of the astrophysical neutrino at Earth in the standard case~(no LRI case), as well as in the presence of the LRI induced from the gauged $L_e-L_\mu$, $L_e-L_\tau$, and $L_\mu-L_\tau$ models, given a particular neutrino production at the source. We contrast the calculated flavor composition in the presence of LRI against the flavor composition estimates from the IceCube and other neutrino telescopes and estimate the limits on the mediator mass and the coupling of LRI in each case. For IceCube, we use estimated flavor composition data for eight years of runtime. Also, we use flavor composition projections for future experiments like IceCube-Gen2, Baikal-GVD, KM3NeT, P-ONE, and TAMBO for estimated runtimes till the year 2040. We compare our limits with the other existing limits.

This \sout{paper} \blue{chapter} is organized as follows: section~\ref{sec:models} introduces long-range interactions and
their effect on neutrino oscillations. We discuss how the LRI is generated in the case of gauged $L_\alpha-L_\beta$ symmetry, mediated by $Z'_{\alpha\beta}$ and through the $Z-Z'_{\alpha\beta}$ mixing. In section~\ref{sec:LRI_potential}, we estimate the long-range potential for the three underlying symmetries sourced by the ordinary matter particles in the universe. Section~\ref{sec:osc_LRI} computes the flavor transition probabilities of the astrophysical neutrinos in the presence of LRI potential. The interesting features of oscillation probabilities are explained in terms of the evolution of the modified oscillation parameters in the presence of LRI. The possible production mechanism of the high-energy astrophysical neutrinos and diffused neutrino flux is discussed in section~\ref{sec:Astro_nu_det}. This section also illustrates the detection of astrophysical neutrinos in the IceCube detector. Section~\ref{sec:flav_comp_earth} estimates the flavor composition of the astrophysical neutrinos after reaching the Earth using the computed oscillation probabilities. A detailed discussion of our statistical analysis is discussed in section~\ref{sec:stat_analysis}. We demonstrate our results in section~\ref{sec:results_lri}. In section~\ref{sec:future_improvements}, we discuss the assumptions and the possible improvements in our analysis. Finally, we summarize in section~\ref{sec:conclusion_lri}.

\section{Anomaly free $U(1)'$ models with $L_\alpha-L_\beta$ symmetry}
\label{sec:models}

The Standard Model (SM), in addition to the gauge group $SU(3)_{\rm C}\times SU(2)_{\rm L}\times U(1)_{\rm Y}$, contains global  $U(1)$ symmetries associated to the baryon number and the three lepton numbers, $L_e$, $L_\mu$, and $L_\tau$.  While they cannot be gauged individually without introducing anomalies, certain combinations of them can be; for an extensive list, see ref.~\cite{Coloma:2020gfv}. In this chapter,
we focus on three well-motivated, anomaly-free symmetries that conserve gauge lepton number differences~\cite{Foot:1990mn, He:1990pn, He:1991qd}: $L_e-L_\mu$, $L_e-L_\tau$, and $L_\mu-L_\tau$.  Each one introduces a new neutral gauge vector boson, $Z_{e \mu}^\prime$, $Z_{e \tau}^\prime$, and $Z_{\mu\tau}^\prime$, that mediates new neutrino interactions~\cite{He:1990pn,He:1991qd,Foot:1994vd} with electrons and neutrons, as we show below.  
Note that the Higgs sector in these models differentiates between different lepton flavors~\cite{Foot:2005uc}; however, here we focus only on the gauge interactions through the new boson.  
The $L_{\alpha}-L_{\beta}$ gauge symmetries and their extensions have been explored in numerous scenarios, including as possible solutions of the Hubble tension~\cite{Araki:2021xdk}, of the electron and muon $(g-2)$ anomalies~\cite{Chen:2020jvl, Bodas:2021fsy, Panda:2022kbn, Borah:2021khc}, considering the new boson to be dark matter~\cite{Baek:2015fea, Asai:2020qlp, Alonso-Alvarez:2023tii} and as radiation from compact binary systems~\cite{KumarPoddar:2019ceq}, leptogenesis~\cite{Asai:2020qax}, their influence on muon-beam dumps at the TeV-scale~\cite{Cesarotti:2022ttv}, possible production of dark photons at the MUonE experiment~\cite{GrillidiCortona:2022kbq}, and explaining the electron and positron excess in the cosmic-ray flux~\cite{Duan:2017qwj,He:2009ra}.  However, these studies consider mediators that are heavier than the ones we consider here, and couplings (see below) that are stronger.

In the $U(1)'_{L_\alpha-L_\beta}$ models, the effective neutrino-matter interaction receives three contributions, mediated by the SM $Z$ boson, by the new $Z^\prime_{\alpha \beta}$ boson, and via the mixing between $Z^\prime_{\alpha \beta}$ and $Z$, \ie,
\begin{equation}
\label{eq:Lagrangian_1} 
\mathcal{L}_{\text{eff}} = \mathcal{L}_{\text{Z}} +\mathcal{L}_{Z'}+\mathcal{L}_{ZZ'} \;.
\end{equation}
The first term in eq.~\ref{eq:Lagrangian_1}, $\mathcal{L}_{\text{SM}}$, is the SM contribution, \ie,
\begin{equation}
\label{eq:lag_sm}
\mathcal{L}_{\text{SM}}
=
\frac{e}{\sin \theta_W \cos \theta_W}Z_\mu \left[-\frac{1}{2}\bar{l}_{\alpha}\gamma^\mu P_L l_{\alpha}+\frac{1}{2}\bar{\nu}_{\alpha}\gamma^\mu P_L \nu_{\alpha}+\frac{1}{2}\bar{u}\gamma^\mu P_L u-\frac{1}{2}\bar{d}\gamma^\mu P_L d\right] \;,
\end{equation}
where $e$ is the unit electric charge, $\theta_W$ is the Weinberg angle, $l_\alpha$ and $\nu_\alpha$ are, respectively, the charged lepton and neutrino of flavor $\alpha = e$, $\mu$, or $\tau$, $u$ and $d$ are the up and down quarks, and $P_L$ is the left-handed projection operator.  

The second term in eq.~\ref{eq:Lagrangian_1}, $\mathcal{L}_{Z'}$, describes neutrino-matter interactions via the new mediator~\cite{He:1990pn,He:1991qd,Heeck:2010pg,Coloma:2020gfv}, \ie,
\begin{equation}
\label{eq:lag_zprime}
\mathcal{L}_{Z'}
=
g_{\alpha \beta}' Z'_{\sigma}(\bar{l}_{\alpha}\gamma^{\sigma}l_{\alpha}-\bar{l}_{\beta}\gamma^{\sigma}l_{\beta}+ \bar{\nu}_{\alpha}\gamma^{\sigma}P_{L}\nu_{\alpha}-\bar{\nu}_{\beta}\gamma^{\sigma}P_{L}\nu_{\beta}) \;,
\end{equation}
where $g^{\prime}_{\alpha\beta}$ ($\alpha,\beta$ = $e,\mu,\tau$, $\alpha \neq \beta$) is the adimensional coupling constant associated to the $L_\alpha-L_\beta$ gauge symmetry.  Because naturally occurring muons and tauons are scarce, we consider this contribution only under $L_e-L_\mu$ and $L_e-L_\tau$, and the corresponding interaction to be sourced only by abundant electrons.

The third term in eq.~\ref{eq:Lagrangian_1}, $\mathcal{L}_{\text{mix}}$, contains terms that mix $Z^\prime_{\alpha\beta}$ with $Z$~\cite{Babu:1997st,Heeck:2010pg,Joshipura:2019qxz}, \ie,
$\mathcal{L}_{ZZ^\prime} = -\frac{1}{2} \sin \chi \hat{Z}'_{\mu \nu}\hat{B}^{\mu \nu}+\delta \hat{M}^2 \hat{Z}'_{\mu}\hat{Z}^{\mu}$, where $\hat{Z}'_{\mu \nu}$ and $\hat{B}^{\mu \nu}$ are the field strength tensors for $U(1)^{\prime}$ and $U(1)_Y$, respectively, $\hat{Z}$ and $\hat{Z}^\prime$ are the gauge eigenstates corresponding to the neutral massive gauge bosons of the SM and the new $U(1)^{\prime}$ gauge symmetry, $\chi$ is the kinetic mixing angle, and $\delta \hat{M}^2$ is the squared-mass difference between $\hat{Z}$ and $\hat{Z}^\prime$. Diagonalizing this Lagrangian redefines the fields in terms of physical states: the photon and two massive bosons, $Z$ and $Z'$, that are related to $\hat{Z}$ and $\hat{Z'}$ through a new mixing angle $\xi$~\cite{Babu:1997st}, \ie, $\mathcal{L}_{ZZ^\prime} \supset (\xi-\sin\theta_W\chi) Z'_{\mu}Z^{\mu}$.  This introduces a four-fermion neutrino-matter interaction term via $Z$--$Z_{\alpha\beta}^\prime$ mixing, \ie,
\begin{equation}
\label{equ:lag_mix}
\mathcal{L}_{\rm mix}
=
-g_{\alpha\beta}^\prime
(\xi-\sin\theta_W\chi)\frac{e}{\sin\theta_W \cos\theta_W}
J'_\rho J_3^\rho \;,
\end{equation}
where $J^\prime_\rho = \bar{\nu}_\alpha \gamma_\rho P_L\nu_\alpha-\bar{\nu}_\beta \gamma_\rho P_L\nu_\beta$ and $J_3^\rho = -\frac{1}{2}\bar{e}\gamma^\rho P_L e+\frac{1}{2}\bar{u}\gamma^\rho P_L u-\frac{1}{2}\bar{d}\gamma^\rho P_L d$, and the term $(\xi-\sin\theta_W\chi)$ accounts for $Z$--$Z^\prime$ mixing.
The time component of $J_3^\rho$, relevant for static matter, is
\begin{equation}
J_3^0
=
-\frac{1}{4}\left[\bar{e}\gamma^0(1-\gamma^5)e-\bar{u}\gamma^0(1-\gamma^5)u+\bar{d}\gamma^0(1-\gamma^5)d\right] \;,
\end{equation}
which, for an unpolarized medium, becomes
\begin{equation}
J_3^0
=
-\frac{1}{4}(n_e-n_u+n_d)
=
-\frac{1}{4}(n_e-n_p+n_n) \;,
\end{equation}
where $n_f$ ($f = e, u, d)$ is the number density of electrons, up quarks, and down quarks.  For electrically neutral medium, $n_p = n_e$, so, $J_3^0 = -n_n/4$. So the neutrino-matter interaction induced from the $Z-Z'_{\alpha\beta}$ in sourced by the ambient neutrons.
To illustrate the effect of mixing, below we fix the value of the mixing strength to $(\xi-\sin \theta_W \chi) = 5 \times 10^{-24}$~\cite{Heeck:2010pg}, the upper limit for a new interaction whose range is of the order of the distance between the Earth and the Sun~\cite{Schlamminger:2007ht,Adelberger:2009zz,Heeck:2010pg}.  (The upper limit on the mixing strength for an interaction range of the order of the size of the Earth is slightly weaker and allows for larger mixing~\cite{Schlamminger:2007ht,Adelberger:2009zz,Heeck:2010pg}.)  
In our analysis, for the $L_e-L_\beta$~($\beta = \mu,\tau$) symmetry, we consider the leading contribution to the neutrino-matter interaction induced from the Lagrangian terms in eq.~\ref{eq:lag_zprime}. Whereas, for $L_\mu-L_\tau$, neutrino-matter interaction occurs only the through the $Z-Z'_{\alpha\beta}$ mixing, as shown by eq.~\ref{equ:lag_mix}.

\section{Long-range interaction with light mediators}
\label{sec:LRI_potential}
In this work, we are interested in the ultra-light mass limit of new mediator $Z'_{\alpha\beta}$, for which the respective interactions are long-range in nature.
A neutrino separated by a distance $d$ from a collection of $N_e$ electrons, under $L_e-L_\mu$ and $L_e-L_\tau$, or of $N_n$ neutrons, under $L_\mu-L_\tau$, experiences a Yukawa potential of
\begin{equation}
	\label{eq:pot_general}
	V_{\rm \alpha \beta}
	=
	\mathcal{G}_{\alpha \beta}\frac{e^{-m_{\alpha \beta}^{\prime} d}}{4 \pi d}
	\times
	\left\{
	\begin{array}{lll}
		N_e & , & {\rm for}~\alpha, \beta = e, \mu ~{\rm or}~ e, \tau \\
		N_n & , & {\rm for}~\alpha, \beta = \mu, \tau \\
	\end{array}
	\right. \;,
\end{equation}
where $m_{\alpha \beta}^{\prime}$ is the mass of the $Z'_{\alpha \beta}$ boson and the coupling strength is
\begin{equation}
	\label{eq:Gab}
	\mathcal{G}_{\alpha \beta}
	=
	\left\{
	\begin{array}{lll}
		g^{\prime 2}_{e \mu} & , & ~{\rm for}~\alpha, \beta = e, \mu \\
		g^{\prime 2}_{e \tau} & , & ~{\rm for}~\alpha, \beta = e, \tau \\
		g^{\prime}_{\mu \tau} (\xi-\sin \theta_W \chi) \frac{e}{4 \sin \theta_W \cos \theta_W} & , & ~{\rm for}~\alpha, \beta = \mu, \tau\,. \\
	\end{array}
	\right. 
\end{equation}
The range of the interaction is $\sim$$1/m_{\alpha\beta}^\prime$; beyond this distance from the source of the potential, it is suppressed due to the mediator mass.  Depending on the interaction range, the potential receives contributions from nearby bodies --- if the mediator is heavy --- or also from distant bodies --- if the mediator is light.  We consider masses of $10^{-35}$--$10^{-10}$~eV, for which the interaction range is km--Gpc, which encompasses the Earth ($\oplus$), Moon 
($\leftmoon$), Sun ($\astrosun$), Milky Way (MW) and the cosmological matter distribution (cos).  Thus, the potential experienced by neutrinos is the sum of the contributions sourced by each of them, \ie,
\begin{eqnarray}
	\label{eq:pot_total}
	V_{\alpha \beta} = V_{\alpha \beta}^\oplus + V_{\alpha \beta}^{\leftmoon} + V_{\alpha \beta}^{\astrosun} + V_{\alpha \beta}^{\rm MW} + \langle V_{\alpha \beta}^{\rm cos} \rangle \;,
\end{eqnarray}
and the relative contribution of each term depends on the value of $m_{\alpha\beta}^\prime$.  

We do not compute the changing potential along the underground trajectories of the neutrinos inside the Earth or inside the Sun; see ref.~\cite{Coloma:2020gfv} for such treatment. Instead, like ref.~\cite{Bustamante:2018mzu}, we compute the average potential experienced by the neutrinos at their point of detection in IceCube.  This approximation is especially valid for mediators lighter than about $10^{-18}$~eV, for which the interaction range is longer than the Earth-Sun distance, and so all of the electrons and neutrons on the Earth or the Sun contribute to the potential experienced by a neutrino regardless of its position inside them.  Below $10^{-18}$~eV is also where we place limits in an unexplored range of mediator mass.

In what follows, we use eq.~\ref{eq:pot_general} as a basis to calculate the potential induced by these sources. We adopt the methods introduced in ref.~\cite{Bustamante:2018mzu} for the $L_e-L_\mu$ and $L_e-L_\tau$ cases.  Below, we revisit them, extend them to the $L_\mu-L_\tau$ case, and introduce refinements in the computation of the potential due to solar and cosmological electrons and neutrons. Throughout, we assume that matter is electrically neutral, so that the number of electrons and protons is the same, and that matter is isoscalar, so that the number of electrons and neutrons is the same, except for the Sun and for the cosmological distribution of matter.

\paragraph{The Earth.}
Astrophysical neutrinos travel inside the Earth from its surface to IceCube, located $d_{\rm IC} = 1.5$~km underground at the South Pole.  Along the way, they undergo long-range interactions with underground electrons or neutrons.  Under the $L_\alpha - L_\beta$ symmetry, the potential sourced by them is
\begin{equation}
\label{equ:pot_earth}
V_{\alpha \beta}^\oplus
=
\frac{\mathcal{G}_{\alpha\beta}}{2} 
\int_0^\pi d\theta_z \int_0^{r_{\max}(\theta_z)} 
dr ~r 
\sin \theta ~e^{-m_{\alpha \beta}^{\prime} r}
\times
\left\{
\begin{array}{lll}
\langle n_{e,\oplus} \rangle_{\theta_z} & , & {\rm for}~\alpha, \beta = e, \mu ~{\rm or}~ e, \tau \\
\langle n_{n,\oplus} \rangle_{\theta_z} & , & {\rm for}~\alpha, \beta = \mu, \tau \\
\end{array}
\right. \;,
\end{equation}
where $\theta_z$ is the zenith angle along which the neutrinos travel, measured from the South Pole, the chord length traveled along this direction is $r_{\max}(\theta_z) = (R_\oplus-d_{\rm IC}) \cos \theta_z + \left[ (R_\oplus-d_{\rm IC})^2 \cos^2 \theta_z + (2R_\oplus-d_{\rm IC}) d_{\rm IC} \right]^{1/2}$, and the radius of the Earth is $R_\oplus = 6371$~km.  The average densities of electrons and neutrons along direction $\theta_z$ are $\langle n_{e,\oplus} \rangle_{\theta_z} = \langle n_{n,\oplus} \rangle_{\theta_z}$, assuming matter is electrically neutral and isoscalar.  To compute them, we adopt the matter density profile of the Preliminary Reference Earth Model~\cite{Dziewonski:1981xy}, and we assume an electron fraction of $Y_e \equiv N_e / (N_e + N_p) = 0.5$.  The total number of electrons and neutrons in the Earth is $N_{e, \oplus} \approx N_{n, \oplus} \sim 4 \times 10^{51}$.  Because we do not track the propagation of neutrinos inside the Earth and the changing long-range potential that they experience at different points along their trajectories, our treatment is approximate; yet, it allows us explore efficiently the parameter space.  For a detailed treatment, see ref.~\cite{Coloma:2020gfv}.

\paragraph{The Moon and the Sun.}
We treat the Moon and the Sun as point sources of electrons and neutrons.  Under the $L_\alpha-L_\beta$ symmetry, the potential sourced by the Moon is
\begin{equation}
\label{eq:pot_moon}
V^{\leftmoon}_{\alpha \beta}
=
-\mathcal{G}_{\alpha\beta}
\frac{e^{-m_{\alpha \beta}^\prime d_{\leftmoon}}} { 4\pi d_{\leftmoon}}
\times
\left\{
\begin{array}{lll}
N_{e,\leftmoon} & , & {\rm for}~\alpha, \beta = e, \mu ~{\rm or}~ e, \tau \\
N_{n,\leftmoon} & , & {\rm for}~\alpha, \beta = \mu, \tau \\
\end{array}
\right. \;,
\end{equation}
where the distance between the Earth and the Moon is $d_{\leftmoon} \approx 4 \cdot 10^5$~km, and the number of electrons and neutrons in the Moon is $N_{e,\leftmoon} = N_{n,\leftmoon}\sim 5 \cdot 10^{49}$, assuming the lunar matter is electrically neutral and isoscalar.  Similarly, the potential sourced by the Sun is
\begin{equation}
V^{\astrosun}_{\alpha \beta} 
=
-\mathcal{G}_{\alpha\beta}
\frac{e^{-m_{\alpha \beta}^\prime d_{\astrosun}}} { 4\pi d_{\astrosun}}
\times
\left\{
\begin{array}{lll}
N_{e,\astrosun} & , & {\rm for}~\alpha, \beta = e, \mu ~{\rm or}~ e, \tau \\
N_{n,\astrosun} & , & {\rm for}~\alpha, \beta = \mu, \tau \\
\end{array}
\right. \;,
\end{equation}
where the distance between the Earth and the Sun is $d_{\astrosun} =1$~A.U., the number of electrons in the Sun is $N_{e,\astrosun} \sim 10^{57}$, and the number of neutrons in it is $N_{n,\astrosun} = N_{e,\astrosun}/4$.  Like for Earth, we do not compute the changing long-range matter potential inside the Moon and Sun as neutrinos propagate inside them; for the latter, see ref.~\cite{Coloma:2020gfv}.

\paragraph{The Milky Way.}
A neutrino of extragalactic origin traverses the Milky Way before reaching the Earth and may be affected by the long-range potential sourced by Galactic electrons and neutrons.  We do not track the propagation of neutrinos inside the Milky Way.  Instead, we estimate the effect of long-range interactions by computing the potential experienced by neutrinos at the location of the Earth by integrating the Galactic electron and neutron column densities across all possible neutrino trajectories that have the Earth as the endpoint.  Under the $L_\alpha - L_\beta$ symmetry, this is
\begin{equation}
\label{eq:pot_mw}
V_{\alpha \beta}^{\rm MW}
=
\frac{\mathcal{G}_{\alpha\beta}}{4\pi} 
\int_0^\infty
dr 
\int_0^\pi d\theta \int_0^{2\pi} 
d\phi 
~r 
\sin\theta 
~e^{-m_{\alpha \beta}^{\prime} r}
\times
\left\{
\begin{array}{lll}
n_{e, {\rm MW}}(r, \theta, \phi) \, ,\,  {\rm for}~\alpha, \beta = e, \mu ~{\rm or}~ e, \tau \\
n_{n, {\rm MW}}(r, \theta, \phi) \, , \, {\rm for}~\alpha, \beta = \mu, \tau \\
\end{array}
\right.\, 
\end{equation}

where the coordinate system is centered at the position of the Earth, 8.33~kpc away from the Galactic Center, and the densities of electrons and neutrons are $n_{e, {\rm MW}} = n_{n, {\rm MW}}$, assuming matter is electrically neutral and isoscalar.  In the Milky Way, $N_{e, {\rm MW}} \approx N_{n, {\rm MW}} \sim 10^{67}$ electrons and neutrons are contained in stars and cold gas --- distributed in a central bulge, a thick disc, and a thin disc --- and in hot gas --- distributed in a diffuse halo.  Following ref.~\cite{Bustamante:2018mzu}, we adopt the ``conventional model'' from ref.~\cite{McMillan:2011wd} for the matter density of the central bulge, thick disc, and thin disc, and the spherical saturated matter density from ref.~\cite{Miller:2013nza} for the diffuse halo; Fig.~A1 in ref.~\cite{Bustamante:2018mzu} shows the total matter distribution.

\paragraph{Cosmological electrons and neutrons.}
For $m_{\alpha\beta}^\prime \lesssim 10^{-25}$~eV, the dominant contribution to the long-range potential is from the large-scale cosmological distribution of electrons and neutrons.  We follow ref.~\cite{Bustamante:2018mzu} to compute the potential, including the effect of the adiabatic cosmological expansion on the densities of electrons and neutrons; we defer to it for a derivation of the potential.  While ref.~\cite{Bustamante:2018mzu} assumed that the cosmological distribution is isoscalar, we instead account for the fact that, after the recombination epoch, the neutron-to-proton ratio saturates to about $1/7$~\cite{Steigman:2007xt}.  Using this and assuming that cosmological matter is electrically neutral, the number of cosmological electrons is seven times higher than that of neutrons.  Under the $L_\alpha - L_\beta$ symmetry, the potential at redshift $z$ is
\begin{eqnarray}
\label{equ:pot_cos_z}
V_{\alpha \beta}^{\rm cos}(z)
&=&
\frac{3}{4\pi} 
\frac{\mathcal{G}_{\alpha\beta}}{m_{\alpha \beta}^{\prime~2} d_{\rm H}^3(z)} 
\left\{ 
1 - e^{-m_{\alpha \beta}^\prime d_{\rm H}(z)} [1+m_{\alpha \beta}^\prime d_{\rm H}(z)] 
\right\} 
\nonumber \\
&&
\qquad
\times
\left\{
\begin{array}{lll}
N_{e, {\rm cos}}(z) & , & {\rm for}~\alpha, \beta = e, \mu ~{\rm or}~ e, \tau \\
N_{n, {\rm cos}}(z) & , & {\rm for}~\alpha, \beta = \mu, \tau \\
\end{array}
\right. \;,
\end{eqnarray}
where $N_{e, {\rm cos}} \left( z \right) \simeq 7 M_{\rm H}(z) / (8 m_p + 7 m_e)$ is the number of electrons, $m_p$ and $m_e$ are the proton and electron mass, respectively, $N_{n, {\rm cos}}(z) = N_{e, {\rm cos}}(z)/7$ is the number of neutrons, $M_{\rm H}$ is the baryonic mass inside casual horizon (see Eq.~(16.105) in ref.~\cite{Giunti:2007ry}), and $d_{\rm H}$ is the size of the causal horizon, i.e., the radius of the largest sphere centered on the Earth within which events can be causally connected~\cite{Weinberg:2008zzc}.  We adopt a $\Lambda$CDM cosmology with Hubble constant $H_0 = 100 h$ km s$^{-1}$ Mpc$^{-1}$, where $h = 0.673$~\cite{ParticleDataGroup:2014cgo}, the vacuum energy density $\Omega_\Lambda = 0.692$ and the matter density $\Omega_{\rm M} = 0.308$~\cite{Planck:2015fie}.  Because we use the diffuse flux of high-energy astrophysical neutrinos, we account for the evolution with redshift of the number density of neutrino sources, $\rho_{\rm src}$, by averaging the  potential over $z$, \ie,
\begin{equation}
\langle V_{\alpha \beta}^{\rm cos} \rangle \propto
\int dz~ 
\rho_{\rm src}(z) 
\frac{dV_{\rm c}}{dz} 
V_{\alpha \beta}^{\rm cos}(z) \;,
\end{equation}
where $V_{\rm c}$ is the comoving volume~\cite{Hogg:1999ad} and  $\rho_{\rm src}$ follows the star formation rate~\cite{Hopkins:2006bw, Yuksel:2008cu, Kistler:2009mv}.

\begin{figure}[t!]
	\centering
	\includegraphics[width=.49\textwidth]{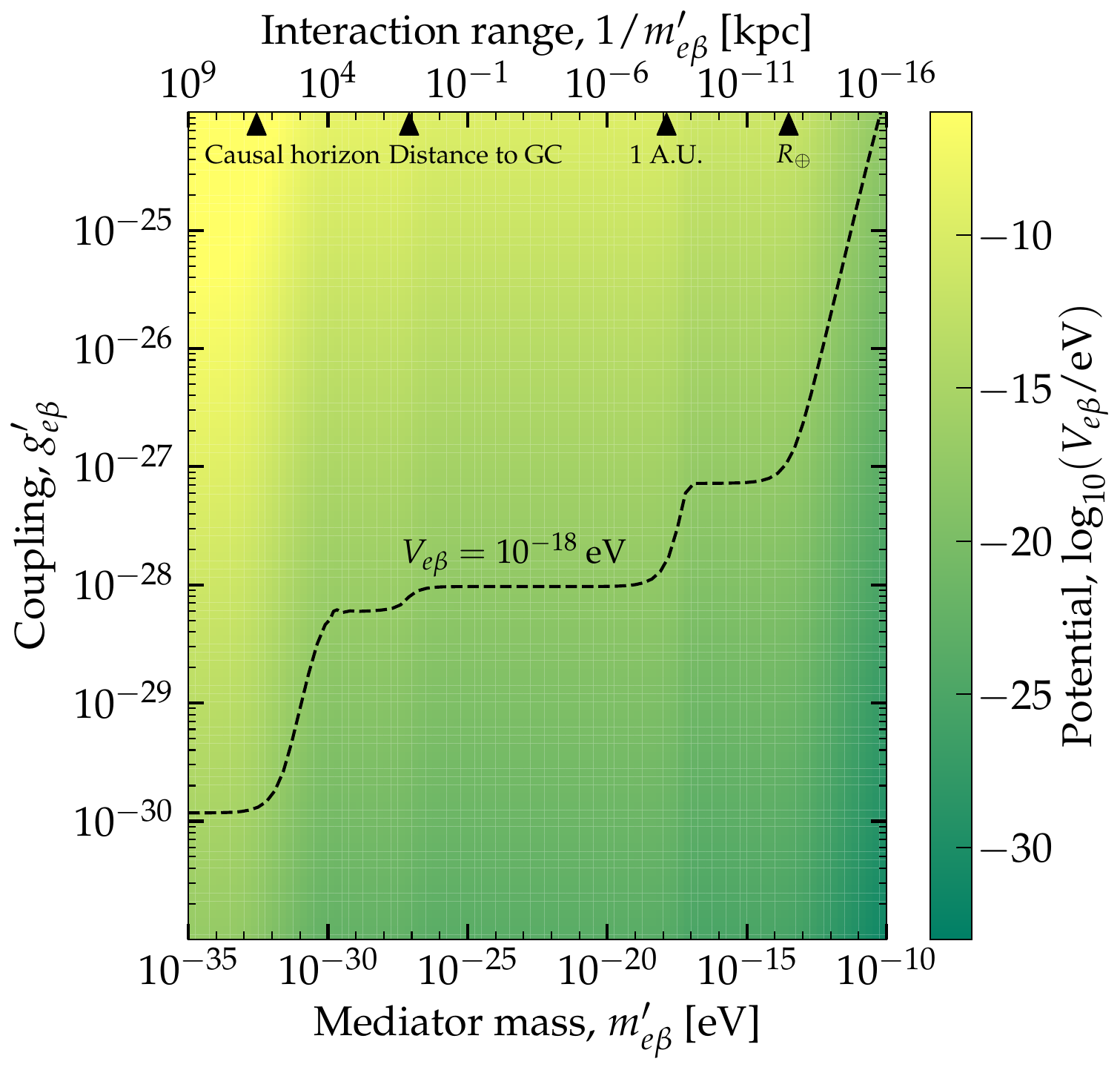}
	\includegraphics[width=.49\textwidth]{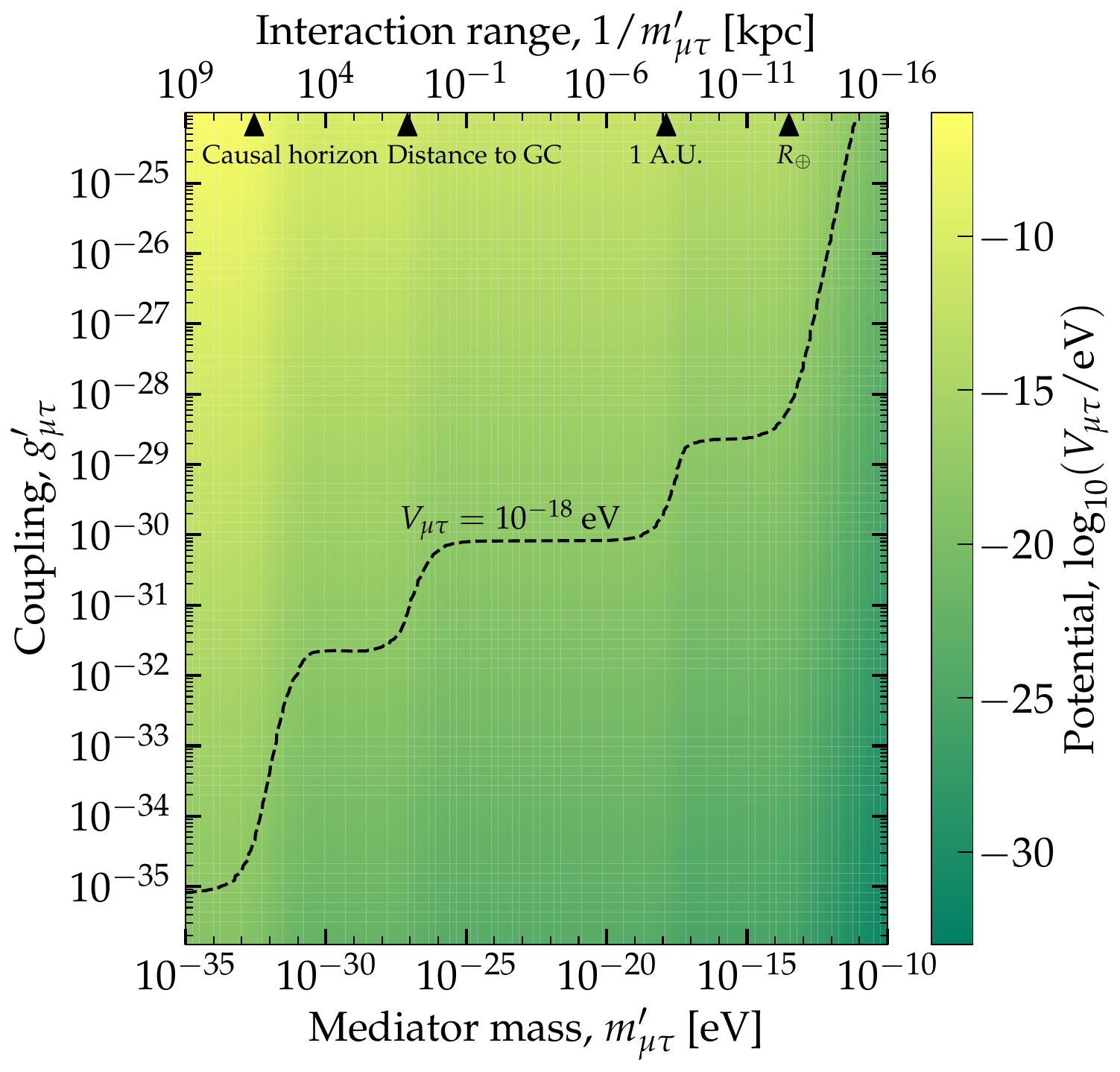}
	\mycaption{Long-range matter potential, $V_{\alpha \beta}$, experienced by a neutrino at Earth, eq.~\ref{eq:pot_total}. The potential is sourced by electrons and neutrons in the Earth, Moon, Sun, Milky Way, and the distant Universe as a function of the mass and adimensional coupling of the new $U(1)$ gauge symmetry.  \textit{Left:} For the $L_e-L_{\beta}$ symmetries ($\beta = \mu, \tau$), sourced by electrons. \textit{Right:} For the $L_{\mu}-L_{\tau}$ symmetry, sourced by neutrons. Isocontours are drawn at the illustrative values of $V_{e\beta} = V_{\mu\tau} =  10^{-18}$~eV.  (For the $L_{\mu}-L_{\tau}$ symmetry, values of the coupling run lower due to the mixing between the new mediator and the SM $Z$ boson.)  The step-like transitions in the potential signal the onset of contributions from sources of electrons and neutrons located at different distances relative to the interaction range.}
	\label{fig:potential}
\end{figure}

Figure~\ref{fig:potential} shows the total potential, eq.~\ref{eq:pot_total}, as a function of the mediator mass and coupling, for the $L_e-L_\beta$ ($\beta = \mu, \tau$) and $L_\mu-L_\tau$ symmetries.  To illustrate its behavior, we show the isocontour for the illustrative value of $V_{\alpha \beta} = 10^{-18}$~eV (which is close to the final constraint that we obtain in Section~\ref{sec:results}). As the mass shrinks, the interaction range grows.  As a result, the potential isocontour undergoes several step-like transitions, each of which represents the inclusion of the contribution of a more distant collection of electrons and neutrons into the total potential. From $m_{\alpha\beta}^\prime \sim 10^{-10}$~eV to $10^{-18}$~eV, the potential is dominated by the Earth, with a minor contribution from the Moon. The step at $m_{\alpha\beta}^\prime \sim 10^{-18}$~eV represents the inclusion of the potential sourced by the Sun, which becomes dominant. Because the Sun contains far more electrons and neutrons than the Earth and the Moon, a smaller coupling is enough to achieve the same potential.  The step at $m_{\alpha\beta}^\prime \sim 10^{-27}$~eV represents the onset of dominance of Milky-Way electrons and neutrons, which far outnumber those in the Sun, and are concentrated in the Galactic Center. The last step, at $m_{\alpha\beta}^\prime \sim 10^{-33}$~eV, represents the onset of the dominance of cosmological distribution of electrons and neutrons, which far outnumber those in the Milky Way.  

For the $L_\mu-L_\tau$ case, fig.~\ref{fig:potential} shows that smaller couplings are needed to achieve the same illustrative value of the potential because it scales $\propto g_{\mu\tau}^\prime$, rather than $\propto g_{e\beta}^{\prime~2}$, as in the $L_e-L_\mu$ and $L_e-L_\tau$ cases; see eq.~\ref{eq:Gab}.  In our work, including in fig.~\ref{fig:potential}, for the $L_\mu-L_\tau$ case we fixed the $Z_{\mu\tau}^\prime$--$Z$ mixing strength, $(\xi-\sin \theta_W \chi)$, to its maximum allowed value (see Section~\ref{sec:models}).  Decreasing or increasing the mixing strength would shift the potential isocontour in fig.~\ref{fig:potential} up or down, respectively.

\section{Neutrino flavor transition with LRI}
\label{sec:osc_LRI}
Neutrino propagation Hamiltonian in the presence of LRI induced from $L_\alpha-L_\beta$ gauge symmetry, in the flavor basis, is written as
\begin{equation}
\label{eq:hamiltonian_tot}
H
=
H_{\rm vac}
+
V_{\rm mat}
+
V_{\alpha\beta}\,,
\end{equation}
where the three terms on the right-hand side~(RHS) represent oscillations in vacuum, standard neutrino-matter interactions, and the new neutrino-matter long-range interactions, respectively.

\begin{figure}[t!]
	\centering
	\includegraphics[width=\textwidth]{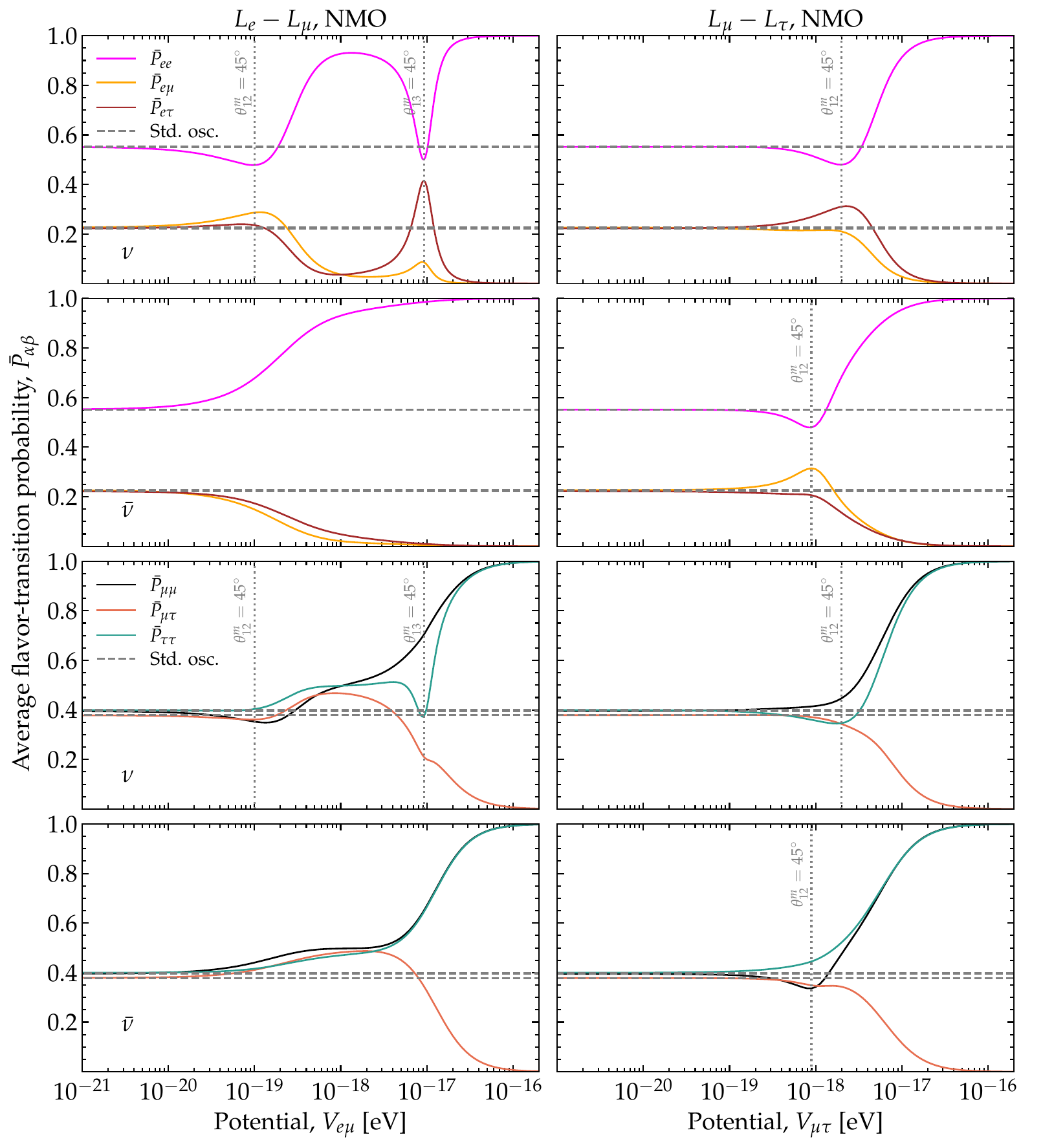}
	\mycaption{Average flavor-transition probabilities $\bar{P}(\nu_\alpha\to\nu_\beta)$ (see eq.~\ref{eq:prob_avg}, as functions of the LRI potential induced by the $U(1)$ gauge symmetries $L_e-L_\mu$ (left column) and $L_\mu-L_\tau$ (right column).  We show probabilities for both neutrinos (first and third rows) and antineutrinos (second and fourth rows), at a fixed energy of 100~TeV, assuming normal mass ordering (NMO).  
     Probabilities under standard oscillations (``Std.~osc.''), \ie, for $V_{\alpha\beta} = 0$, are shown by dashed lines for comparison. The mixing parameters are fixed at their present-day best-fit values from NuFit 5.1~\cite{Esteban:2020cvm, NuFIT} as quoted in the text. } 
	\label{fig:prob_lemmt_nmo}
\end{figure}

Hamiltonian for the oscillation in vacuum has the form,
\begin{equation}
\label{eq:H_lrf}
H_{\rm vac}
=
\frac{1}{2 E}
U
{\rm diag}(0, \Delta m^2_{21}, \Delta m^2_{31})
~U^{\dagger} \;,
\end{equation}
where $E$ is the neutrino energy, $U$ is standard three-flavor neutrino mixing matrix.
The best-fit values of the mixing parameters from the global oscillation analysis of ref.~\cite{Esteban:2020cvm, NuFIT} are: $\theta_{23} = 42.1^\circ$, $\theta_{13} = 8.62^\circ$, $\theta_{12} = 33.45^\circ$, $\delta_{\text{CP}}=230^\circ$, $\Delta m^2_{31} = 2.51\times10^{-3}$~eV$^2$, and $\Delta m^2_{21} = 7.42\times10^{-5}$~eV$^2$. We use these benchmark values to compute flavor transition probability shown in fig.~\ref{fig:prob_lemmt_nmo}. Later, in our statistical analysis, we let their values float within their present-day and future predicted allowed ranges.

Standard neutrino matter interaction term in the RHS of eq~\ref{eq:hamiltonian_tot} is given by
\begin{equation}
V_{\rm mat}
=
{\rm diag}(V_{\rm CC}, 0, 0) \;,
\end{equation}
where $V_{\rm CC} = \sqrt{2}G_{F} N_e $ is the charged-current neutrino-electron interaction potential, defined in earlier chapters. Note that in our calculations, this contribution is relevant only inside Earth, where matter densities are high. The potential above is for neutrinos; for antineutrinos, it flips sign, \ie, $V_{\rm mat} \to -V_{\rm mat}$.

Finally, the contribution from the new LRI interaction in eq.~\ref{eq:hamiltonian_tot} is
\begin{equation}
\label{equ:pot_lri_matrix}
V_{\alpha\beta}
=
\left\{
\begin{array}{ll}
{\rm diag}(V_{e\mu}, -V_{e\mu}, 0), & {\rm for}~ \alpha, \beta = e, \mu \\
{\rm diag}(V_{e\tau}, 0, -V_{e\tau}), & {\rm for}~ \alpha, \beta = e, \tau \\
{\rm diag}(0, V_{\mu\tau}, -V_{\mu\tau}), & {\rm for}~ \alpha, \beta = \mu, \tau \\   
\end{array}
\right. \;,
\end{equation}
where $V_{\alpha\beta}$ is the potential, calculated using eq.~\ref{eq:pot_total}. The potential is sourced by the electrons or neutrons in the universe, depending on the gauge symmetry. As mentioned earlier, the number of electrons/neutrons that contribute to the potential is determined by the mass of the mediator. The potential above is for neutrinos; for antineutrinos, it flips sign, \ie, $V_{\alpha\beta} \to -V_{\alpha\beta}$.

Using the Hamiltonian defined above, one can calculate the $\nu_\alpha \to \nu_\beta$ flavor-transition probability using eq.~\ref{eq:hamiltonian_tot},
\begin{equation}
\label{eq:osc_prob_1}
P(\nu_\alpha\to\nu_\beta)
=
\left\vert
\sum^3_{i=1} U^m_{\alpha i}
\exp\left[-\frac{\Delta \Tilde{m}^2_{i1}L}{2E}\right]U^{m^{\ast}}_{\beta i}
\right\vert^2 \;.
\end{equation}
In the above equation, $L$ is the distance traveled by the neutrino from its point of production to the Earth, $\Delta \tilde{m}^2_{ij} \equiv \tilde{m}_i^2 - \tilde{m}_j^2$, with $\tilde{m}^2_i/2E$ the eigenvalues of the Hamiltonian, and $U^m$ is the matrix that diagonalizes the Hamiltonian, parameterized like the PMNS matrix, but in terms of new mixing parameters $\theta_{23}^m$, $\theta_{13}^m$, $\theta_{12}^m$,  $\delta_{\rm CP}^m$. Note that the expression of the new mixing angles has similar forms to the modified mixing angles shown in Chapter~\ref{C3}, with appropriate changes in the Hamiltonian with new interaction. In Appendix~\ref{app:evol_mix_angles}, we show the evolution of these mixing angles as a function of new potential at fixed neutrino energy 100 TeV.
For high-energy neutrinos that travel cosmological-scale distances, like the ones we consider, the oscillations due to mass-square term in eq.~\ref{eq:osc_prob_1} are rapid, as $\Delta \Tilde{m}^2_{ij} L / (2E) \gg 1$. The present and upcoming neutrino telescopes, given the limited energy resolutions, are only sensitive to the average probability (see Appendix~\ref{sec:averaged_osc}),
\begin{equation}
\label{eq:prob_avg}
\bar{P}(\nu_\alpha\to\nu_\beta)
=
\sum^3_{i=1}|U^m_{\alpha i}|^2|U^m_{\beta i}|^2 \;.
\end{equation}

Using this equation, we calculate the favor transition probability and estimate the flavor composition of the later sections. In the absence of neutrino interactions, the average oscillation probability loses its dependence on neutrino energy. In contrast, in the presence of long-range interactions, it retains the energy dependence via the interplay of the vacuum and interaction potential contributions to the Hamiltonian, eq.~\ref{eq:hamiltonian_tot}.

In fig.~\ref{fig:prob_lemmt_nmo}, we show the average oscillation probabilities as a function of LRI potentials computed at a fixed energy of 100 TeV representative of high-energy astrophysical neutrinos. For explanations, we focus on the probabilities under the $L_e-L_\mu$ symmetry, but the discussions are valid for other symmetries. In the figure, we observe for low values of the potential, \ie, $V_{\alpha\beta} \ll \Delta m_{ij}^2 / (2 E)$, oscillations are standard. As the strength of the potential starts to increase and becomes comparable to the vacuum Hamiltonian~($V_{\alpha\beta} \approx 10^{-20}$~eV), deviation from the standard oscillation starts to appear. We observe some sharp features --- dips or peaks --- appear at $V_{e\mu} \approx 10^{-17}$~eV in the $\nu_e\to\nu_\beta$ ($\beta = e,\mu,\tau$) probabilities for neutrinos. This happens as the value of new mixing angle $\theta_{13}^m$ attains the maximal value, \ie $45^\circ$~(see appendix~\ref{app:evol_mix_angles}). Also, at $V_{e\mu} \approx 10^{-19}$~eV, less prominent features reflects $\theta_{12}^m = 45^\circ$; they are less prominent because in vacuum value $\theta_{12}$ is close to $45^\circ$. Similar features appear in the $\nu_\mu \to \nu_\tau$ and $\nu_\tau \to \nu_\tau$ probabilities, for analogous reasons.  For large values of the potential, \ie, $V_{\alpha\beta} \gg 10^{-17}$~eV, the term $V_{\alpha\beta}$ dominates the Hamiltonian and, because it is diagonal, flavor transition is suppressed. For $L_e-L_\tau$, shown in Appendix~\ref{app:results_le_ltau}, the flavor-transition probabilities are similar to $L_e-L_\mu$.  Figure~\ref{fig:prob_lemmt_nmo} shows that, for $L_\mu-L_\tau$, no resonance due to $\theta_{13}^m$ occurs neither for neutrinos nor antineutrinos because $\theta_{13}^m$ never reaches $45^\circ$, but there is a small dip or jump in the probabilities around $V_{\mu\tau}\approx 10^{-18}$~eV, due to $\theta_{12}^m$ resonance. For antineutrinos, the sharp features above do not appear because $\theta_{13}^m$ and $\theta_{12}^m$ never become resonant under normal neutrino ordering. However, we checked they do appear under inverted mass ordering, as $\Delta m_{31}^2$ becomes negative.

\section{Astropgysical neutrino flux and its detection at IceCube}
\label{sec:Astro_nu_det}

\subsection{Astrophysical neutrinos}
Astrophysical neutrinos are one of the essential components of multi-messenger astronomy.
First observed by IceCube in 2013~\cite{Aartsen:2013bka, Aartsen:2013jdh}, with an energy few tens of TeV to few PeV, makes it a neutrino source with the highest energies detected so far. Although exact information about their origin remains unknown, several possible sources of high-energy astrophysical neutrinos, namely, AGN, gamma-ray bursts, supernovae, black holes, pulses, and numerous others, make up their diffuse flux. Our study considers a diffuse neutrino flux with energy in the TeV to PeV range. 

Inside the sources, protons and other charged particles get accelerated to very high energy, possibly in the magnetized environment. Accelerated protons interact with the ambient matter and radiation to produce pions and Kaons  which further decay to produce neutrinos. There are three well-motivated production channels of astrophysical neutrinos which lead to three distinct flavor compositions --- the fraction of $\nu_e$ $\nu_\mu$, and $\nu_\tau$ at the source ($f_{e,S},f_{\mu,S},f_{\tau,S})$. First, the full pion decay channel results in $(1/3,2/3,0)$ flavor ratio at the source, where the neutrino production takes place through the decay of charged pions to neutrinos and muons ($\pi^{+}\rightarrow\mu^{+}+\nu_{\mu}$) which further decay into neutrinos ($\mu^{+}\rightarrow\overline{\nu}_{\mu}+e^{+}+\nu_{e}$). Second is the muon-damped channel, which leads to the $(0:1:0)$ ratio at the source. In this case, the strong magnetic field at the source induces the cooling of intermediate muons via the synchrotron radiation. As a result, only the muon flavor neutrinos from the pion decay come out of the source. The onset of the synchrotron losses results in the transition of flavor ratio at source from $(1/3,2/3,0)$ to $(0,1,0)$~\cite{Kashti:2005qa,Lipari:2007su,Hummer:2010vx}, observing which can give us information about the strength of the magnetic field at the source~\cite{Winter:2013cla,Bustamante:2020bxp}.
The third channel is the neutron decay channel, which produces a $(1,0,0)$ ratio at the source. This scenario corresponds to the production of $\overline{\nu}_{e}$ through the beta decay of neutrons. The neutron decay channel is the least likely out of the three production scenarios since the resultant neutrinos are of relatively less energy than those produced via the other two channels. In this work, we analyze the flavor composition corresponding to the pion decay channel at the source.

Astrophysical neutrino fluxes are expected to follow a power law spectrum: $\phi({E_\nu})\propto E^{-\gamma}$. In the case of $pp$ collision, produced neutrino flux follows the power law of the parent protons, with the value of $\gamma$ loosely expected to be in the interval [2,3]. In the $p\gamma$ scenario, the shape of the power law depends on those of the parent protons and photons, and it is bump-like on account of the photon spectrum peaking at a characteristic energy. Since the diffused neutrino flux includes all the neutrino sources, the feature of a particular neutrino source spectrum softened.
So far, IceCube analyses prefer an unbroken power-law spectrum with the value of $\gamma$ dependent on the data set~\cite{IceCube:2020wum, IceCube:2021uhz}. However, with more data in the coming decades, an alternate spectral shape may be used to describe the neutrino flux~\cite{Fiorillo:2022rft,Liu:2023flr}.

\subsection{Detection of Astrophysical neutrinos at IceCube}
\label{subsec:nu_detection}

A remarkable amount of diffused neutrino flux with energy in the hundreds of TeV to possibly up to EeV range passes through the Earth. Their detection at the various neutrino telescopes is crucial for understanding cosmic accelerators and neutrino propagation through the universe.
At present, IceCube, the largest neutrino telescope~\cite{IceCube:2016zyt, Ahlers:2018fkn}, with 1~km$^3$ of deep Antarctic ice instrumented with photomultipliers, plays a significant role in this direction.
In it, incoming high-energy neutrinos interact with nucleons in the ice, predominantly via deep inelastic scattering~\cite{IceCube:2017roe, Bustamante:2017xuy, Aartsen:2018vez, IceCube:2020rnc}, either neutral-current (NC), \ie, $\nu_l + N \to \nu_l + X$, where $l = e, \mu, \tau$ and $X$ are final-state hadrons, or charged-current (CC), \ie, $\nu_l + N \to l + X$. Produced final-state charge particles shower and emit Cherenkov radiation while propagating through the ice. Photomultipliers placed inside the detector volume detect the radiation. Using the number of photons detected and from their spatial and temporal distributions, IceCube reconstructs the energy, direction, and flavor of the interacting neutrino~\cite{IceCube:2013dkx, IceCube:2016zyt}.

The number of neutrino events detected at IceCube is classified into three categories: track, cascade, and double cascade. Track-like events are the muon neutrino events produced by CC interaction of $\nu_\mu$, where the final-state hadrons initiate a shower around the interaction vertex, and the final-state muon propagates through the detector volume, leaving a track of Cherenkov light. These tracks can be several kilometers in length and easily identifiable.   Tracks are also made in 17\% of $\nu_\tau$ CC interactions where the final-state tauon decays into a muon~\cite{ParticleDataGroup:2022pth}. Cascades are made by CC interactions of $\nu_e$ and $\nu_\tau$, where both the final-state lepton and hadrons shower around the interaction vertex, and by NC interactions of neutrinos of all flavors, where only the final-state hadrons shower. Double cascades are made by CC interactions of $\nu_\tau$ in which the final-state hadrons make a first shower, centered around the interaction vertex. The final-state tauon is energetic enough to decay some distance away, generating a second shower~\cite{Learned:1994wg, Athar:2000rx, IceCube:2020fpi}.

The number of neutrino events observed at IceCube are broadly categorized into {\it starting events} and {\it through-going muons}. {\it Starting events} are those where the neutrino interacts within the detector volume. Starting events with energy above 60~TeV are known as High-Energy Starting Events (HESE)~\cite{IceCube:2020wum}. These events are subjected to a self-veto~\cite{Schonert:2008is, Gaisser:2014bja} that reduces the contamination of atmospheric neutrinos and muons and, as a result, they have the highest content of neutrinos of astrophysical origin. However, its detection rate is relatively low, of approximately ten events per year~\cite{IceCube:2020wum}. {\it Through-going tracks} are those where $\nu_\mu$ (and $\nu_\tau$) interact outside the detector and produce muon tracks that cross part of it~\cite{IceCube:2021uhz}. Their detection rate is orders of magnitude higher than that of HESE. Still, because they have lower energies, they are dominated by atmospheric neutrinos and muons made in cosmic-ray interactions in the atmosphere. Further, for through-going tracks, because the location of the neutrino interaction is unknown, the neutrino energy can only be inferred uncertainly~\cite{IceCube:2013dkx}. In our analysis below, we exploit the combined capabilities of HESE and through-going events for flavor measurements, motivated by ref.~\cite{IceCube:2015gsk}.

The three categories of the neutrino events discussed above can be made by more than one neutrino flavor.
As a result, a straightforward event-by-event correspondence between detected event type and neutrino flavor is typically unfeasible, except for double cascades, which are only made by $\nu_\tau$.  In addition, showers made by the Glashow resonance~\cite{Glashow:1960zz}, recently discovered by IceCube~\cite{IceCube:2021rpz}, are triggered exclusively by the interaction of 6.3-PeV $\bar{\nu}_e$ with electrons~\cite{Bhattacharya:2011qu, Bhattacharya:2012fh, Biehl:2016psj, Huang:2019hgs}.  Still, from the relative number of detected events of different types, it is possible to infer the flavor composition of the neutrino flux statistically, even if, because of the above limitations, the measurement uncertainties are significant.  

The flavor composition can be inferred either using only HESE showers, tracks, and double cascades or by combining them with through-going muons.  IceCube has reported the flavor composition in several analyses: using the first 3 years of HESE data~\cite{IceCube:2015rro}, a combination of 4 years of contained events plus 2 years of through-going tracks~\cite{IceCube:2015gsk},
5 years of contained events starting at lower energies~\cite{Aartsen:2018vez}, and, most recently, 7.5 years of HESE~data, including a dedicated search for double cascades~\cite{IceCube:2020fpi}.  Independent analyses have reported complementary results; see, e.g., refs~\cite{Mena:2014sja, Palomares-Ruiz:2015mka, Vincent:2016nut, Brdar:2018tce, Palladino:2019pid}.  The precision of these analyses is limited by the relative scarcity of HESE events.  Presently, the most precise measurements come from the combined analysis of ref.~\cite{IceCube:2015gsk}, where a large number of through-going tracks pins down the flavor content of $\nu_\mu$.  Such a combined analysis has not been updated since 2015~\cite{IceCube:2015gsk} despite the availability of larger event samples due to its complexity.  
Our analysis below rests on the capabilities of IceCube and upcoming neutrino telescopes to infer the flavor composition in such a way, following the same spirit as ref.~\cite{Song:2020nfh, Schumacher:2021hhm, Fiorillo:2022rft}. It is based on realistic estimates of what such analysis could look like with present and future data.

\section{Neutrino flavor composition at Earth}
\label{sec:flav_comp_earth}

The flavor composition of the astrophysical neutrinos modifies en route to the Earth due to neutrino oscillation. For a given flavor composition of the source, the flavor composition of the astrophysical neutrinos after reaching the Earth is written as
\begin{equation}
\label{eq:flavatearth}
f_{\alpha,\oplus}
=
\sum_{\beta=e,\mu,\tau} \bar{P}_{\beta\alpha} f_{\beta, {\rm S}} \;,
\end{equation}
where $f_{\alpha,\oplus}$ is the neutrino flavor composition $(f_{e,\oplus},f_{\mu,\oplus},f_{\tau,\oplus})$ after reaching the Earth,and $f_{\beta, {\rm S}}$ is the flavor composition~$(f_{e,{\rm S}},f_{\mu,{\rm S}},f_{\tau,{\rm S}})$ at the source. Using the averaged oscillation probabilities calculated in section~\ref{sec:osc_LRI}, we estimate the flavor composition of astrophysical neutrinos after reaching the Earth. For the flavor ratio at the source, we use the three benchmark scenarios discussed in section~\ref{sec:Astro_nu_det}, namely, $(1/3,2/3,0)$, $(0,1,0)$, and $(1,0,0)$.
\begin{figure}[t!]
	\centering
	\includegraphics[width=.49\textwidth]{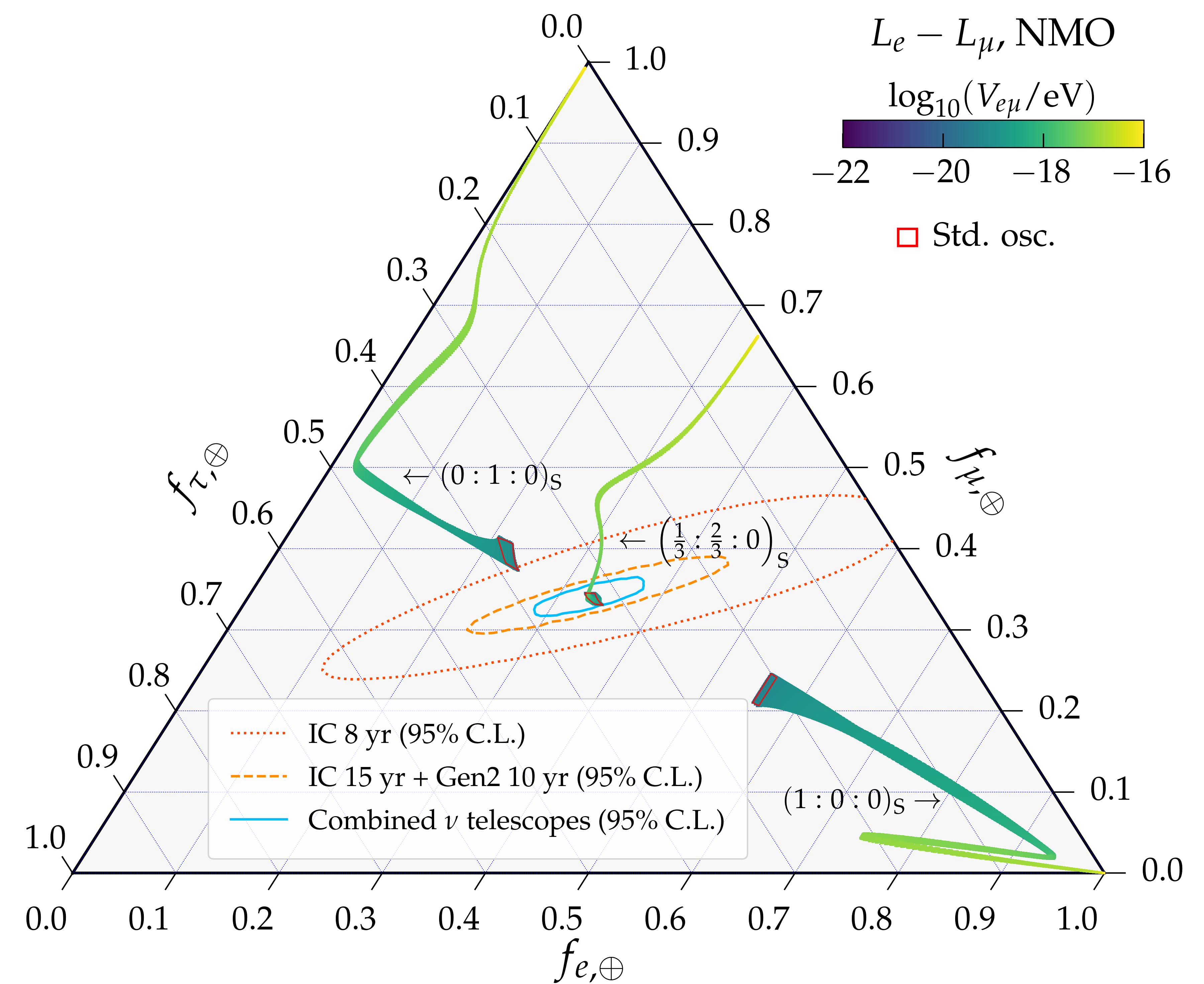}
	\includegraphics[width=.49\textwidth]{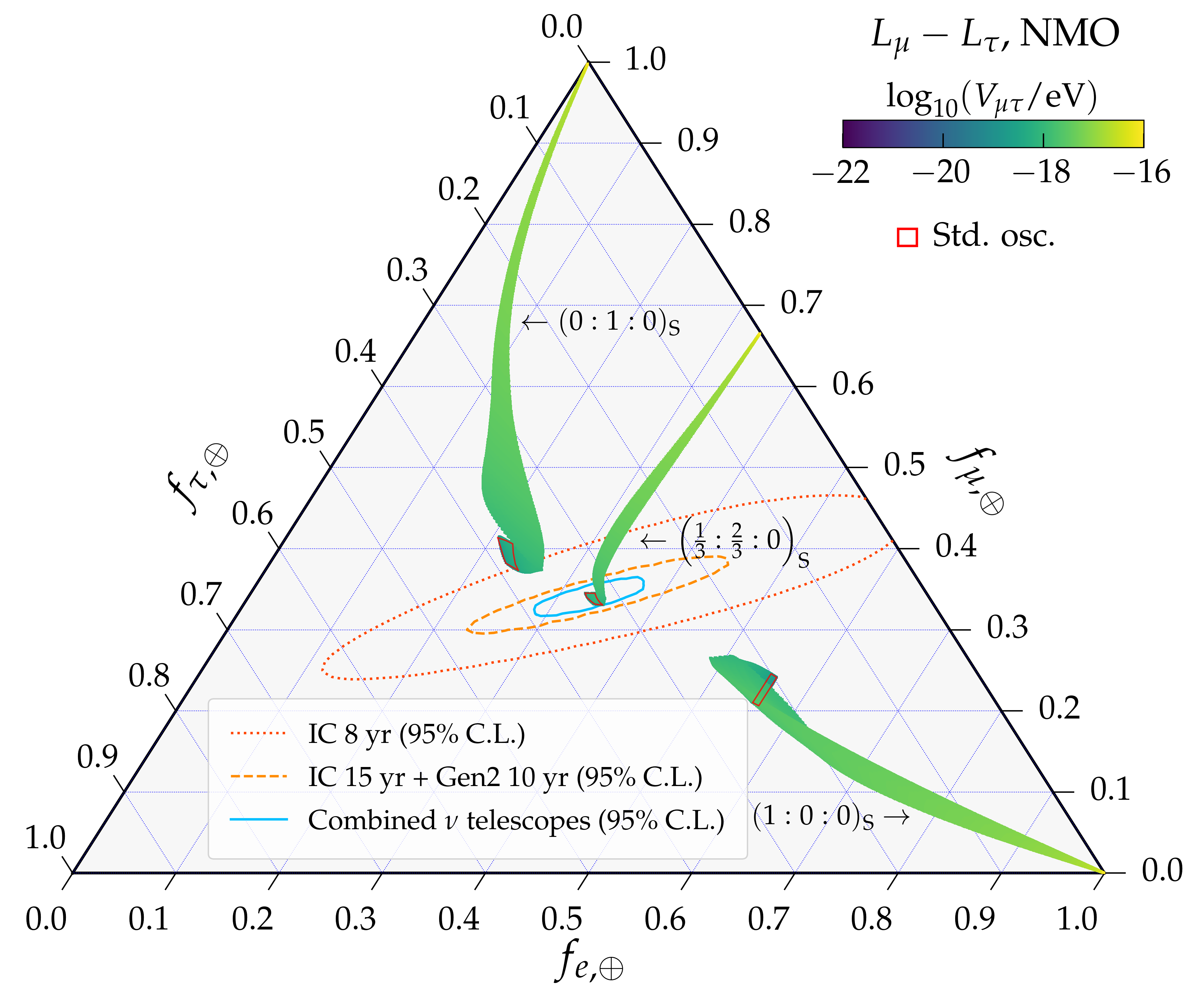}
	\mycaption{\label{fig:flav_ratio}Flavor composition of high-energy astrophysical neutrinos at Earth, $f_{\alpha, \oplus}$, as a function of the long-range matter potential $V_{e\mu}$, under $L_e-L_\mu$ (left), or $V_{\mu\tau}$, under $L_\mu-L_\tau$ (right). We show results for three benchmark choices of the flavor composition at the sources, $(f_{e, {\rm S}}, f_{\mu, {\rm S}}, f_{\tau, {\rm S}}) = (1/3, 2/3, 0)$, the canonical expectation, $(0,1,0)$, and $(1,0,0)$, compared to the expectation from standard oscillations (``Std.~osc.''). The flavor composition is averaged between neutrinos and antineutrinos, assuming they exist in equal proportion in the flux. For this plot only, we fix the neutrino energy to 100~TeV, assume normal neutrino mass ordering, and vary the standard mixing parameters within their $1\sigma$ allowed ranges; our statistical methods systematically vary these choices. Flavor-composition predictions are compared against three estimates of the flavor sensitivity of neutrino telescopes using 8~years of IceCube data (``IC 8~yr''), 15~years of IceCube data plus 10~years of IceCube-Gen2 (``IC 15~yr + Gen2 10~yr''), and the combined exposure of all neutrino telescopes available by 2040 (``Combined $\nu$ telescopes''). For $L_e-L_\tau$, results are similar as for $L_e-L_\mu$ as shown in fig.~\ref{fig:plots_for_et}.}
\end{figure}

In Figure~\ref{fig:flav_ratio}, we show the allowed flavor composition of the astrophysical neutrinos at Earth as a function of the long-range potential in the case of $L_e-L_\mu$ and $L_\mu-L_\tau$ symmetry assuming NMO. Results under the $L_e-L_\tau$ symmetry are similar to those under $L_e-L_\mu$~(see fig.~\ref{fig:plots_for_et}). For the IMO case, the corresponding flavor composition at Earth is discussed in Appendix~\ref{lrf_results_imo}. The three axes denote the flavor fraction of $\nu_e$, $\nu_\mu$, and $\nu_\tau$, respectively. 
Red contours  
show the allowed regions of flavor composition at Earth under standard oscillations~\cite{Bustamante:2015waa, Song:2020nfh}, \ie, when the potential $V_{\alpha \beta} = 0$ for the three scenarios of initial flavor composition. While computing flavor composition at Earth, we vary the values of the standard mixing parameters within their present-day $1\sigma$ allowed ranges~\cite{Esteban:2020cvm, NuFIT}. In the presence of long-range interactions, the flavor-transition probabilities are modified, and so is the flavor composition at Earth.  
In this case, the flavor composition becomes energy-dependent via the probability, in contrast to the flavor composition computed under standard oscillations. In this plot, we fix the neutrino energy at 100 TeV. Also, we average over the flavor composition of neutrinos and antineutrinos.
The three bands show the modification in the flavor composition as the strength of the long-range potential changes for the three benchmark flavor compositions at the source.
For small values of the potential, \ie, $V_{\alpha\beta} \lesssim \Delta m^2_{21}/(2E) \sim 10^{-20}$~eV, the flavor composition is close to the expectation from standard oscillations. As the potential grows, the behavior of the flavor composition traces that of the flavor-transition probabilities (Figs.~\ref{fig:prob_lemmt_nmo}): the wiggles in $f_{\alpha, \oplus}$ reflect the resonant features in the probabilities. For large values of the potential, \ie, $V_{\alpha\beta} \gtrsim 10^{-16}$~eV, the new matter potential becomes the dominant contribution to the Hamiltonian, eq.~\ref{eq:hamiltonian_tot}. In that case, because the contribution of the new potential is flavor-diagonal, flavor transitions are suppressed (Section~\ref{sec:osc_LRI}). Accordingly, fig.~\ref{fig:flav_ratio} shows that the flavor composition at Earth is the same as at the sources, \ie, $f_{\alpha, \oplus} \approx f_{\alpha, {\rm S}}$.

\section{Statistical analysis}
\label{sec:stat_analysis}

\subsection{Analysis choices}
\label{analysis_choice}
\paragraph{Flavor composition at the sources independent of energy.} 
Different production channels of the astrophysical neutrinos are expected to have different energies~\cite{Mucke:1999yb, Kelner:2006tc, Hummer:2010vx, Morejon:2019pfu}. Consequently, flavor composition at the source may vary with energy~\cite{Kashti:2005qa, Kachelriess:2006fi, Lipari:2007su}, contingent on key source properties like the magnetic field intensity~\cite{Winter:2013cla, Bustamante:2020bxp}. However, following the same practice as IceCube flavor measurements ~\cite{IceCube:2015rro, IceCube:2015gsk, Aartsen:2018vez, IceCube:2020fpi}, we consider flavor composition at the source to be independent of energy. Our choice is justified as a limited number of HESE data will be further diluted while allocating them to multiple energy bins. As a result, any energy dependence in flavor composition on Earth is expected to appear solely from neutrino-matter interactions. Reference~\cite{Mehta:2011qb} explores the interplay between energy dependence that stems from neutrino production and from new physics.

\paragraph{Averaging $f_{\alpha, \oplus}$ between $\nu$ and $\bar{\nu}$.} As mentioned in section~\ref{sec:flav_comp_earth}, we average over neutrino and antineutrinos to estimate the flavor composition at the Earth, assuming they are present in equal proportions in the flux that reaches Earth. We use this assumption further for the analysis. We use this assumption because in high-energy neutrino telescopes, events triggered by neutrinos and antineutrinos are so far indistinguishable from one another. However, in reality, proportion of $\nu$ and $\bar{\nu}$ in the are practically unknown (see, however, ref~\cite{Bustamante:2020niz, IceCube:2021rpz}), but the above assumption aligns with theoretical expectations, especially from multi-pion neutrino production at high energies; see, e.g., ref.~\cite{Hummer:2010vx}. As a result of averaging, the resonant features in the oscillation probabilities, which are more prominent for neutrinos than for antineutrinos (fig.~\ref{fig:prob_lemmt_nmo}), appear softened in the flavor composition at Earth that we use.

\paragraph{Estimates of present and future flavor sensitivity.} In our analysis, we use estimates of the present-day flavor sensitivity of IceCube and of future detectors, as shown in fig.~\ref{fig:flav_ratio}.  Present-day sensitivity at IceCube is based on estimated combined analyses of HESE events plus through-going tracks (see section~\ref{subsec:nu_detection}), motivated by ref.~\cite{IceCube:2015gsk}, which offer the greatest sensitivity.  For the future measurements, we adopt the flavor measurement projections for IceCube-Gen2 from ref.~\cite{IceCube-Gen2:2020qha}, which were generated assuming a plausible neutrino spectrum $\propto E^{-2.5}$.  Because these projections were only generated assuming a flavor composition at the sources of $(1/3,2/3,0)_{\rm S}$, later, we only derive constraints on long-range neutrino interactions for that one benchmark scenario.  As in ref.~\cite{Song:2020nfh}, we isolate the contributions of IceCube and IceCube-Gen2 to the flavor sensitivity from ref.~\cite{IceCube-Gen2:2020qha}.  This allows us to generate results based on estimates of the present-day flavor sensitivity for the year 2020, using 8 years of IceCube, and forecasts for the year 2040, using 15 years of IceCube plus 10 years of IceCube-Gen2. For the future neutrino telescopes,
Baikal-GVD~\cite{Avrorin:2019vfc}, KM3NeT~\cite{Adrian-Martinez:2016fdl}, P-ONE~\cite{P-ONE:2020ljt}, and TAMBO~\cite{Romero-Wolf:2020pzh}.  Following ref.~\cite{Song:2020nfh}, we assume that future detectors will have the same detection efficiency of HESE and through-going tracks as IceCube, though with different rates, and rescale the IceCube flavor sensitivity to their respective expected exposures.  This is admittedly a simplification, made necessary by the current absence of details on the capabilities of future detectors.

\subsection{Analysis details}
\label{statistical_analysis}

In this work, our aim is to estimate the limits on the coupling and mediator mass of the long-range interaction that IceCube and other future neutrino telescopes would place.
To do this, we contrast our predictions of the flavor composition obtained in the presence of long-range interactions (section~\ref{sec:flav_comp_earth}) with the present-day flavor composition estimates from IceCube and projected data from the other next-generation telescopes as discussed in subsection~\ref{analysis_choice}.
First, we compute bounds on the long-range matter potential, $V_{\alpha\beta}$; then, we translate them into bounds on the mediator mass and coupling.
\begin{table}[t!]
	\centering
	\resizebox{\linewidth}{!}{%
		\begin{tabular}{|c c c|}
			\hline 
			Observation epoch & Neutrino telescopes & Neutrino mixing parameters \\ 
			\hline
			2020 (estimated) & IC 8 yr & NuFit~5.1 (2021) \\ 
			2040 (projected) & IC 15 yr + IC-Gen2 10 yr & NuFit~5.1 + JUNO + DUNE + HK \\ 
			2040 (projected) & Combined $\nu$ telescopes     & NuFit~5.1 + JUNO + DUNE + HK\\ 
			\hline 
	\end{tabular}}
	\mycaption{Observation epochs of our analysis: years 2020 and 2040. For each epoch, we show the neutrino telescopes that  measure the flavor composition of high-energy astrophysical neutrinos and the oscillation experiments that constrain the value of the neutrino mixing parameters.  We assume that upcoming neutrinos have flavor-measuring capabilities similar to those of IceCube.  In 2040, the combined telescopes include IceCube, IceCube-Gen2~\cite{IceCube-Gen2:2020qha}, Baikal-GVD~\cite{Avrorin:2019vfc}, KM3NeT~\cite{Adrian-Martinez:2016fdl}, P-ONE~\cite{P-ONE:2020ljt}, and TAMBO~\cite{Romero-Wolf:2020pzh}.  See section~\ref{subsec:nu_detection} for details.  For our 2040 projections of the measurement of mixing parameters, we assume that their real values are the present-day best-fit values from the NuFit~5.1 global oscillation fit~\cite{Esteban:2020cvm, NuFIT}.  See section~\ref{sec:osc_LRI} for details.}
	\label{tab:proj}
\end{table}

We adopt the Bayesian statistical methods in our analisys, introduced in ref.~\cite{Song:2020nfh}. First, we compute the energy-averaged flavor composition at Earth, $\langle \pmb{f}_\oplus \rangle \equiv ( \langle f_{e, \oplus} \rangle,\allowbreak \langle f_{\mu, \oplus} \rangle,\allowbreak \langle f_{\tau, \oplus} \rangle )$, using the test values of the long-range potential, $V_{\alpha\beta}$, and of the mixing parameters, $\pmb{\vartheta} \equiv (s_{12}^2, s_{23}^2, s_{13}^2, \dcp)$, with $s_{ij} \equiv \sin \theta_{ij}$. As mentioned earlier, we consider the flavor composition at source $f_{\alpha, \rm S}$ correspond to the pion decay scenario, \ie, $(1/3,2/3,0)$.  
 Then, we assess the compatibility of these predictions in the presence of long-range potential with measurements of the flavor composition in neutrino telescopes through three factors: $\pi(V_{\alpha\beta})$, the prior associated to the value of $V_{\alpha\beta}$; $\pi(\pmb{\vartheta})$, the prior associated to the value of $\pmb{\vartheta}$; and $\mathcal{L}(\langle \pmb{f}_{\oplus} \rangle)$, the likelihood of having measured the flavor composition $\langle \pmb{f}_{\oplus} \rangle$.  We compute the posterior probability density of $V_{\alpha \beta}$ in terms of these three factors, marginalized over all possible values of the mixing parameters, \ie,
\begin{equation}
\label{eq:posterior}
\mathcal{P}\left(V_{\alpha \beta}\right)
=
\int d\pmb{\vartheta}
\mathcal{L}
\left(\langle\pmb{f}_\oplus\left(V_{\alpha \beta}, \pmb{\vartheta}\right)\rangle\right)
\pi(\pmb{\vartheta})
\pi\left(V_{\alpha \beta}\right) \;.
\end{equation}

In Table~\ref{tab:proj}, we summarize the two epochs for which we perform the above analysis: the present, represented by the year 2020, and the future, denoted by the year 2040.  Each epoch has an associated precision on mixing parameters and flavor measurements associated with it, encoded in the prior $\pi(\pmb{\vartheta})$ and the likelihood $\mathcal{L}\left(\pmb{f}_\oplus \right)$.  We discuss these in detail below.

\paragraph{Prior on the long-range potential, $\pi(V_{\alpha\beta})$.} We use a uniform prior on the long-range potential $V_{\alpha\beta}$ in the interval $10^{-24}$--$10^{-16}$~eV in order to avoid any unnecessary bias. The interval is chosen in such a way that it covers the full range of the long-range-interaction effects, \ie, below this potential range, there are no effects of $V_{\alpha\beta}$ in the neutrino flavor transition, and above this potential highly dominates, suppressing the flavor transition.

\paragraph{Prior on the mixing parameters, $\pi(\pmb{\vartheta})$.}  For the prior on the mixing parameters, we assume them to be a normal distribution centered at its present-day best-fit value from the NuFit~5.1~\cite{Esteban:2020cvm, NuFIT} global oscillation fit.  
The distributions for $s_{12}^2$ and $s_{13}^2$ are uncorrelated with other parameters, while those of $s_{23}^2$ and $\delta_{\rm CP}$ are correlated.  For the 2020 estimates, we compute the joint likelihood, $-2 \ln\mathcal{L}_{2020} \equiv \chi^2(s_{12}^2) + \chi^2(s_{13}^2) + \chi^2(s_{23}^2, \dcp)$, where the $\chi^2$ distributions of the oscillation parameters are taken from NuFit~5.1. 
For our 2040 projections, we combine the present likelihood with future measurements of $\theta_{12}$ by JUNO, as computed in ref.~\cite{Song:2020nfh}, and of $\theta_{23}$ and $\dcp$ by DUNE~\cite{Abi:2020wmh} and HK~\cite{Abe:2018uyc}, which we generate ourselves.  We do so in dedicated simulations using {\sc GLoBES}~\cite{Huber:2007ji}, and adopting the detector descriptions for DUNE~\cite{DUNE:2021cuw,DUNE:2016ymp}, using the same configuration discussed in table~\ref{tab:exp_details}.
Note that for $\theta_{23}$ and $\dcp$, our projections differ from those of ref.~\cite{Song:2020nfh} because we generate them using the best-fit values from NuFit~5.1~\cite{Esteban:2020cvm, NuFIT} instead of NuFit~5.0, as a result of which the value of $\theta_{23}$ has shifted from the upper octant to the lower octant~\cite{FernandezMenendez:2021jfk}.)  For $s_{13}^2$, we keep the width of its prior fixed to its present-day size, which is dominated by measurements in Daya Bay~\cite{DayaBay:2016ssb}, as they are not expected to be improved upon significantly by 2040.  In summary, for 2040, we use the likelihood $-2\ln \mathcal{L}_{2040} \equiv -2\ln \mathcal{L}_{2020} + \sum_\mathcal{E} \chi_{\mathcal{E}}^2$, where the contribution of DUNE, HK, and JUNO is each $\chi_{\mathcal{E}}^2 =\sum_{i,j}(\vartheta_i-\bar \vartheta_i)\Sigma^{-1}_{\mathcal{E},ij}(\vartheta_j-\bar \vartheta_j)$, $\Sigma_{ij}$ is the covariance matrix for the mixing parameters $\vartheta_i$, $\vartheta_j$, and $\bar{\vartheta}_i$, $\bar{\vartheta}_j$ are their assumed real values. 

\paragraph{Likelihood of flavor measurement, $\mathcal{L}\left(\langle \pmb{f}_\oplus \rangle \right)$.}  The likelihood of a flavor ratio at Earth $\langle\pmb{f}_\oplus\rangle$ (computed in presence of long-range potential)
represents the precision with which neutrino telescopes can measure the particular flavor composition. We use the same flavor likelihood as ref.~\cite{Song:2020nfh} for both 2020 and 2040 epochs.
By construction, they are centered on the canonical flavor composition at Earth, near $(1/3,1/3,1/3)_\oplus$, that is expected from neutrino production via the full pion decay chain. In other words, we assume this composition as the true value of the measured flavor composition in order to estimate the bounds of long-range interactions. 
For our 2020 estimates, we adopt an estimate of the IceCube flavor sensitivity that would be obtained using 8 years of HESE events and through-going tracks~\cite{IceCube-Gen2:2020qha}, assuming a flux with spectral index $\gamma = 2.5$. For our 2040 projections, we adopt either the expected flavor sensitivity from 15 years of IceCube plus 10 years of IceCube-Gen2 measurements, extracted from ref.~\cite{IceCube-Gen2:2020qha}, or the combined sensitivity from that plus the expected 2040 sensitivity of Baikal-GVD~\cite{Avrorin:2019vfc}, KM3NeT~\cite{Adrian-Martinez:2016fdl}, P-ONE~\cite{P-ONE:2020ljt}, and TAMBO~\cite{Romero-Wolf:2020pzh}, extracted from ref.~\cite{Song:2020nfh}, which we defer to for details. As an illustration, fig.~\ref{fig:flav_ratio} shows 95\% C.L.~contours for the 2020 and 2040 flavor-measurement likelihoods.

Note that our analysis in this work represents an improvement over the one used in the previous study of long-range interactions using the flavor composition, ref.~\cite{Bustamante:2018mzu}.  There, the sensitivity to long-range interactions was derived from a straightforward comparison of the predicted flavor composition vs.~the flavor sensitivity of IceCube, and the method was not amenable to being generalized to compute sensitivity beyond the $1\sigma$ statistical significance.  Here, in contrast, the Bayesian approach allows us to derive limits at higher statistical significance and to account for the prior on mixing parameters properly.

\section{Results}
\label{sec:results_lri}
\begin{figure}[t!]
	\centering
	\includegraphics[width=.49\textwidth]{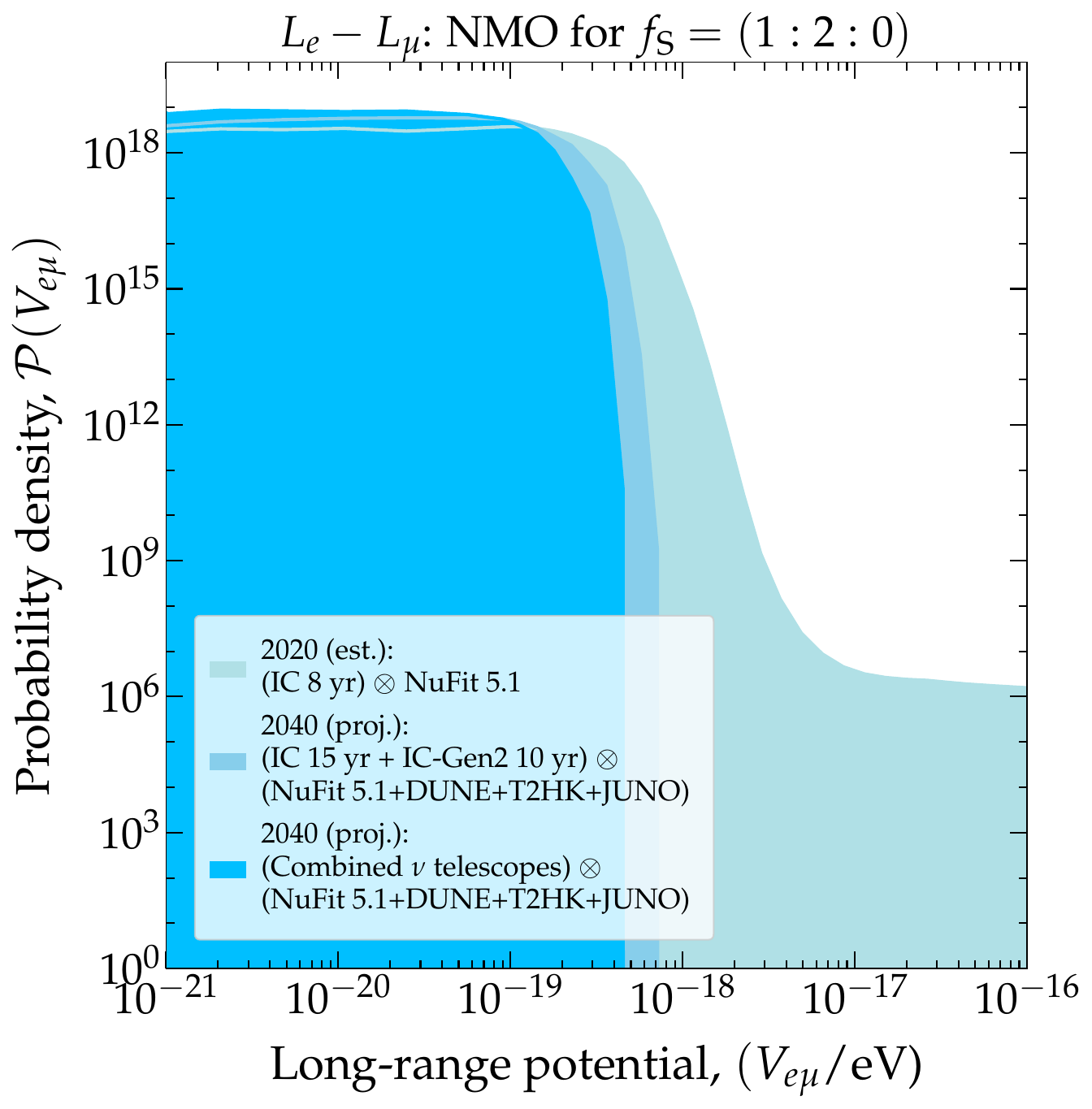}
	\includegraphics[width=.49\textwidth]{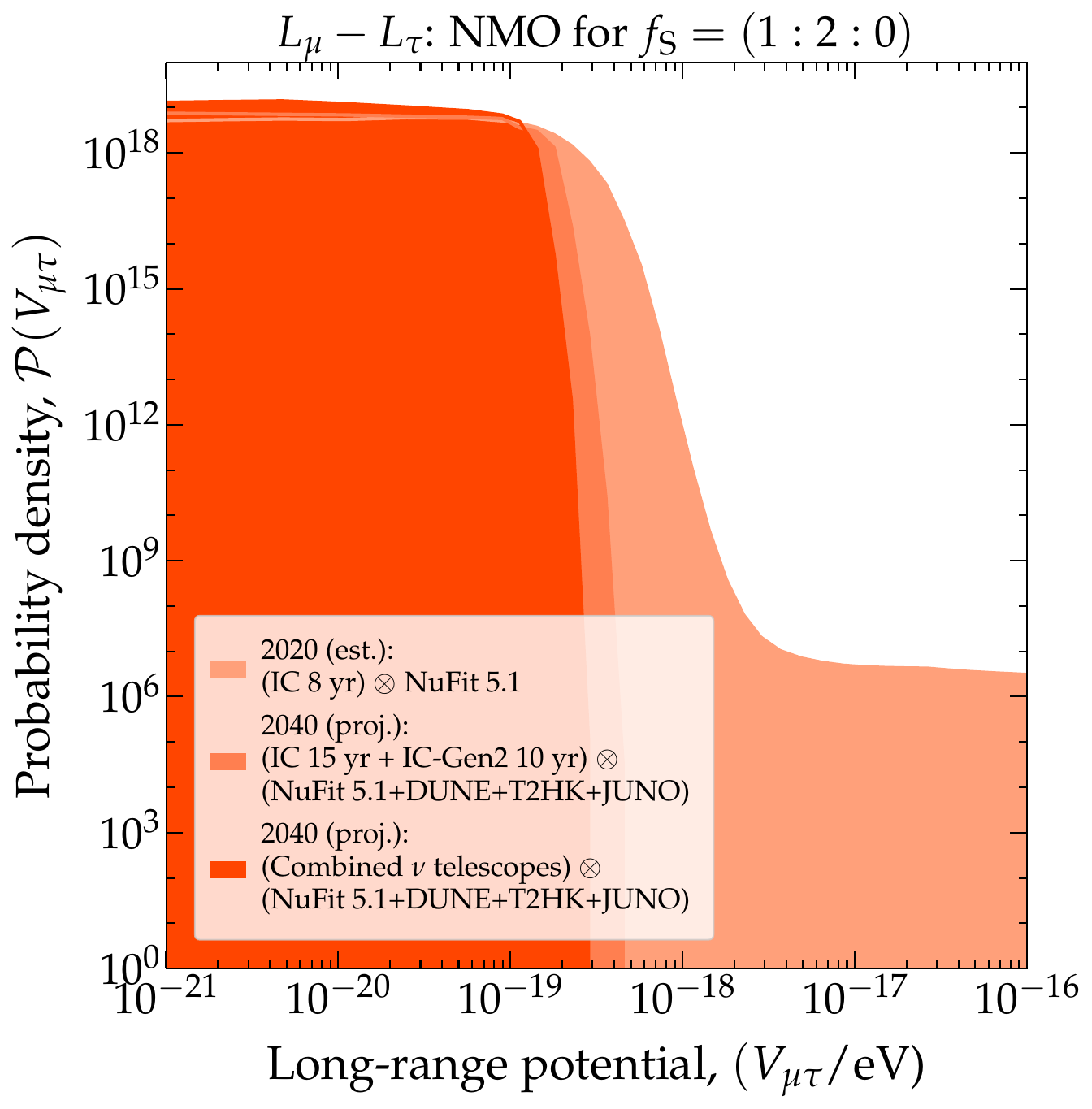}
\mycaption{Posterior probability density of the long-range matter potentials $V_{e \mu}$, under the $L_e-L_\mu$ symmetry ({left}) and $V_{\mu \tau}$, under the $L_\mu-L_\tau$ symmetry ({right}).  We assume a measurement of the flavor composition at Earth centered around the canonical expectation of about $(1/3:1/3:1/3)_{\oplus}$, corresponding to $f_{\rm S} = (1/3:2/3:0)$ at the sources; see fig.~\ref{fig:flav_ratio}.  See fig.~\ref{fig:potential_limits} for the limits on the potential derived from the posterior.  Results for $V_{e\tau}$, under the $L_e-L_\tau$ symmetry, are similar to $V_{e\mu}$, shown in fig.~\ref{fig:plots_for_et}.  We assume that all neutrino telescopes available by 2040 have detection efficiencies similar to that of IceCube.  Results here are obtained assuming normal neutrino mass ordering (NMO).} 
	\label{fig:posterior}
\end{figure}

Using the prescription discussed in the previous section, we estimate the posterior probability of the long-range potential $V_{\alpha\beta}$ with the flavor ratio measurements corresponding to the two epochs, 2020 and 2040.
In fig.~\ref{fig:posterior}, we show the resulting posteriors of $V_{e\mu}$, under the $L_e-L_\mu$ symmetry in the left panel, and of $V_{\mu\tau}$, under the $L_\mu-L_\tau$ symmetry in the right panel assuming NMO.  Results for $V_{e\tau}$, under the $L_e-L_\tau$ symmetry, are similar to $V_{e\mu}$, which is shown in Appendix~\ref{app:results_le_ltau} (see fig.~\ref{fig:plots_for_et}).  We observe the posteriors are maximum at a potential of roughly $10^{-19}$~eV.  Because the posteriors plateau at lower values, we can place upper limits on the values of the potentials.  Values above roughly $10^{-18}$~eV are noticeably less favored, and they become more so when moving from 2020 to 2040 due to improvements in the precision of the mixing parameters and flavor composition, \ie, to narrower $\pi(\pmb{\vartheta})$ and $\mathcal{L}\left(\pmb{f}_\oplus\right)$, respectively. The posteriors corresponds to the three symmetries assuming IMO are shown in Appendix~\ref{lrf_results_imo} (see fig.~\ref{fig:posterior_IMO}).

\begin{figure}[t!]
	\centering
	\includegraphics[width=\textwidth]{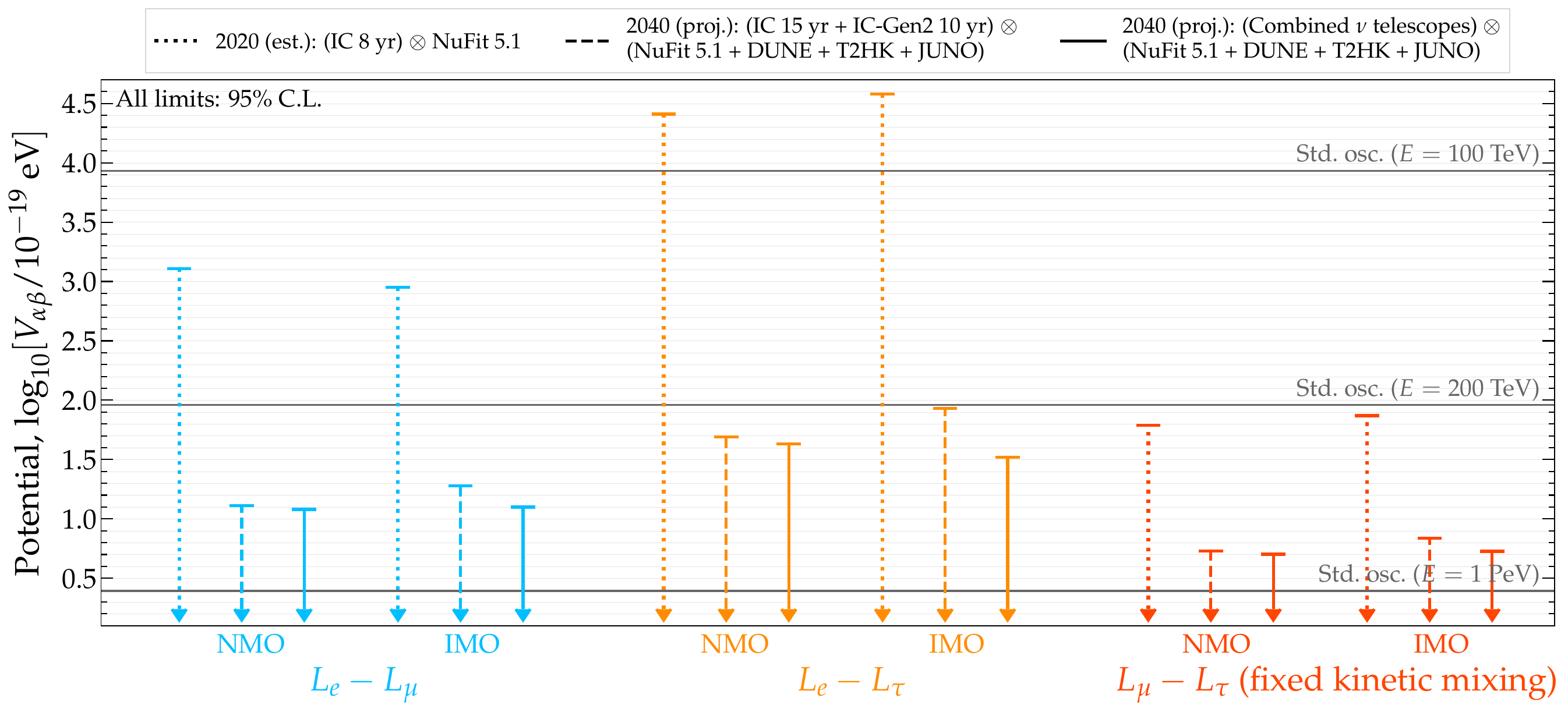}
	\mycaption{Upper limits (95\% C.L.) on the long-range potential induced by the $L_e-L_\mu$, $L_e-L_{\tau}$, and $L_\mu-L_\tau$ symmetries, estimated for 2020 and projected for 2040.  For comparison, we show the standard-oscillation contribution (``Std.~osc.'') to the Hamiltonian, $\Delta m_{21}^2/(2E)$, evaluated at the present-day best-fit value of $\Delta m_{21}^2$~\cite{Esteban:2020cvm, NuFIT} and at different values of the neutrino energy, $E$. We assume that all neutrino telescopes available by 2040 have detection efficiencies similar to that of IceCube (Section~\ref{analysis_choice}). Figure~\ref{fig:g_vs_m_120} shows these limits translated into limits on the coupling, $g_{\alpha\beta}^\prime$.  See Section~\ref{sec:stat_analysis} for details.}
	\label{fig:potential_limits}
\end{figure}

Figure~\ref{fig:potential_limits} (also Table~\ref{tab:bounds}) shows the resulting upper limits on the long-range potentials. A limit has value of $V_{\alpha\beta}^\star$ such that the integral of the posterior, $\int_0^{V_{\alpha\beta}^\star} \mathcal{P}(V_{\alpha\beta}) dV_{\alpha\beta}$, equals a desired credible level (C.L.).  We show limits at the 95\%~C.L in fig.~\ref{fig:potential_limits}.
Differences in the limits obtained under different symmetries are moderate because, under the assumption of $(1/3,2/3,0)_{\rm S}$, their predicted flavor compositions at Earth are only moderately different; see figs.~\ref{fig:flav_ratio}. Also, we observe a significant improvement in the limits between 2020 and 2040, almost by a factor of 2.5--3. However, combining the project data from future telescopes, we do not see any significant improvement in the limits. However, we expect a similar improvement in the limit for both cases in the 2040 epochs, as combining data from IceCube-Gen2 and other neutrino telescopes will significantly increase detection rates. 
Yet, a close inspection of the variation of the flavor composition at Earth with the long-range potential in fig.~\ref{fig:flav_ratio} reveals the subtle reason. For the flavor ratio at the sources, $(1/3, 2/3, 0)_{\rm S}$, which we have adopted to obtain the limits on the long-range potential, the flavor composition at Earth approaches the center of the flavor triangle~$(1,1,1)$ --- where the likelihood of flavor measurement is largest. The value of the potential where the band enters the contour corresponds to the IceCube 8-year data at 95\% C.L., should be significantly different from the potential value where it enters the contour corresponding to the IceCube 15 yr data with Gen2 10 yr data, which leads to significant improvement in the limit between these two cases. However, when we go from the IceCube+Gen2 case to the combined neutrino telescope case, we do not observe the same, leading to a marginal improvement.
 
\begin{figure}[t!]
	\centering
	\includegraphics[width=.49\textwidth]{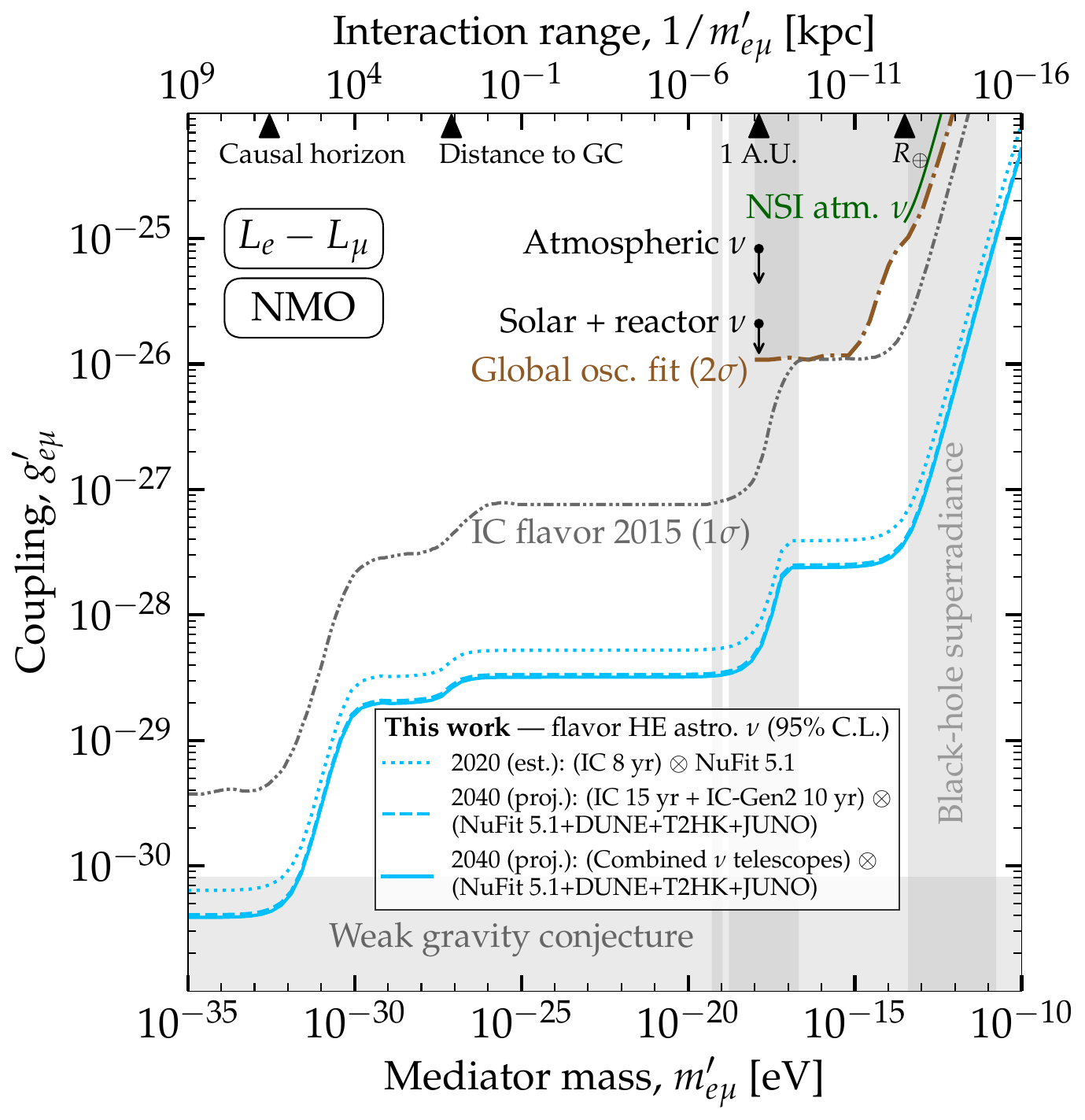}
	\includegraphics[width=.49\textwidth]{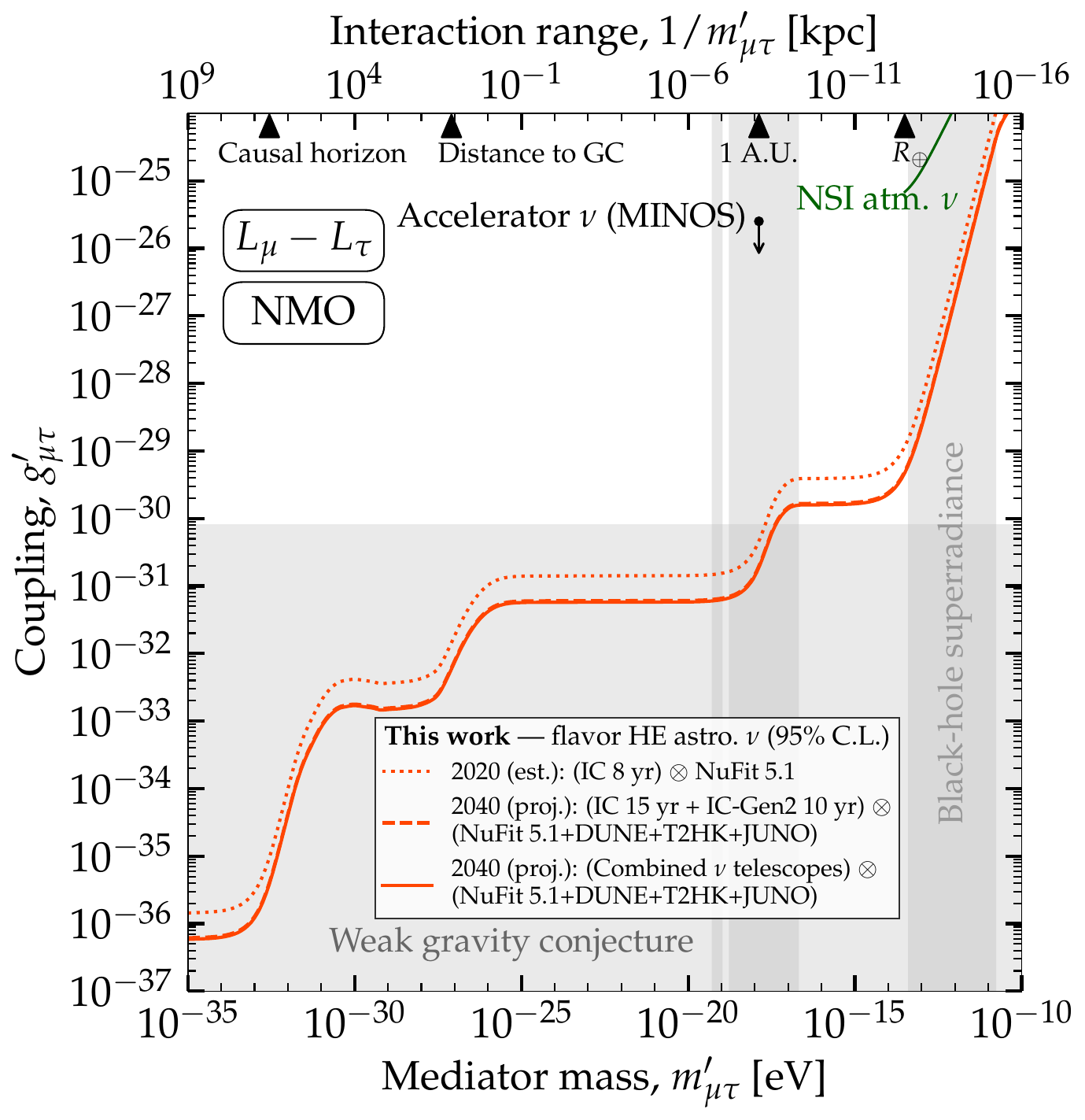}
	\mycaption{
		Estimated present-day (2020) and projected (2040) upper limits (95\%~C.L.) on the coupling strength, $g_{\alpha\beta}^\prime$, of the new boson, $Z_{\alpha\beta}^\prime$, with mass $m_{\alpha\beta}^\prime$, that mediates flavor-dependent long-range neutrino interactions.  
		{\it Left:} Limits on neutrino-electron interactions under the $L_e-L_\mu$ gauge symmetry.  {\it Right:} Limits on neutrino-neutron interactions under the $L_\mu-L_\tau$ gauge symmetry; assuming a $Z_{\mu\tau}^\prime$--$Z$ mixing strength of $(\xi - \sin\theta_W\chi) = 5 \times 10^{-24}$~\cite{Heeck:2010pg}.  Existing limits are from a recent global oscillation fit~\cite{Coloma:2020gfv}, atmospheric neutrinos~\cite{Joshipura:2003jh}, solar and reactor neutrinos~\cite{Bandyopadhyay:2006uh}, and non-standard interactions (NSI)~\cite{Super-Kamiokande:2011dam, Ohlsson:2012kf, Gonzalez-Garcia:2013usa}.  For comparison, we show the proof-of-principle sensitivity ($1\sigma$) based on 2015 IceCube flavor-composition measurements~\cite{IceCube:2015gsk} from ref.~\cite{Bustamante:2018mzu}.  Indirect limits~\cite{Wise:2018rnb} are from 
		black-hole superradiance (90\% C.L.)~\cite{Baryakhtar:2017ngi} and the weak gravity conjecture~\cite{Arkani-Hamed:2006emk}, assuming a lightest neutrino mass of $0.01$~eV.
		existing limits from accelerator neutrinos in MINOS (95\% C.L.) are from ref.~\cite{Heeck:2010pg}.   See section~\ref{sec:stat_analysis} for details.}
	\label{fig:g_vs_m_120}
\end{figure}

Using the upper limit on the potential derived from the posterior probability analysis, in fig.~\ref{fig:g_vs_m_120}, we show limits on $g_{\alpha\beta}^\prime$, as a function of the mediator mass, $m_{\alpha\beta}^\prime$ for the NMO case. 
In the figure, each isocontour in the $g_{\alpha \beta}^\prime$--$\,m_{\alpha \beta}^\prime$ plane corresponds to the upper limit on the potential with a particular set of flavor measurement epochs and mixing parameters.
We use eq.~\ref{eq:pot_total} to translate the limits on $V_{\alpha\beta}$ in fig.~\ref{fig:potential_limits} into the limits in fig.~\ref{fig:g_vs_m_120}.  Thus, the step-like transitions in the figure have the same origin as those of the potential itself (fig.~\ref{fig:potential}): the limits on $g_{\alpha\beta}^\prime$ are stronger for smaller mediator masses, where the potential is due to a larger number of electrons or neutrons. 
\begin{table}[h!]
	\centering
	\begin{tabular}{ | c | *{7}{>{\centering\arraybackslash}p{1.0cm} |}}
		\hline
		\multirow{3}{*}{Observation epoch} &
		\multicolumn{6}{c|}{Upper limit (95\%~C.L.) on potential [$10^{-19}$~eV]} \\
		&
		\multicolumn{2}{c|}{$V_{e\mu}$} &
		\multicolumn{2}{c|}{$V_{e\tau}$} &
		\multicolumn{2}{c|}{$V_{\mu\tau}$} \\
		& NMO & IMO & NMO & IMO & NMO & IMO \\
		\hline
		2020 (est.): IC 8 yr 
		& 3.11 & 2.95 & 4.41 & 4.58 & 1.79 & 1.87 \\
		2040 (proj.): IC 15 yr + Gen2 10 yr 
		& 1.11 & 1.28 & 1.69 & 1.93 & 0.731 & 0.837 \\
		2040 (proj.): Combined $\nu$ telescopes
		& 1.08 & 1.10 & 1.63 & 1.52 & 0.702 & 0.727 \\
		\hline
	\end{tabular}
	\mycaption{Estimated present-day (2020) and projected (2040) upper limits (95\%~C.L.) on the long-range matter potentials $V_{e\mu}$, $V_{e\tau}$, and $V_{\mu\tau}$.  Results are for normal (NMO) and inverted neutrino mass ordering (IMO).  See Section~\ref{sec:results} for details.}
	\label{tab:bounds}
\end{table}

\begin{figure}[t!]
	\centering
	\includegraphics[width=.7\textwidth]{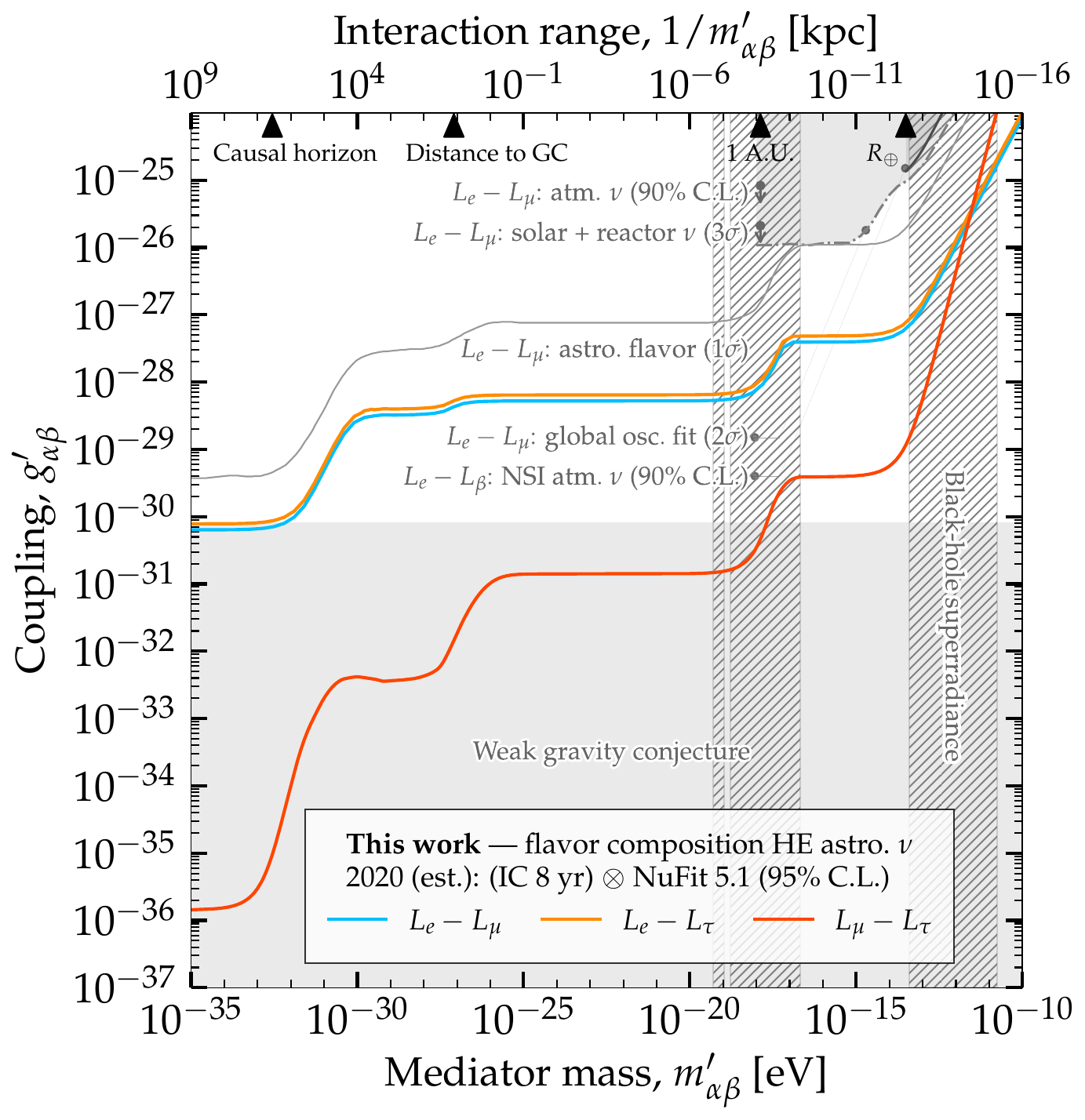}
	\mycaption{\label{fig:limits_3models}Estimated present-day upper limits on the coupling strength, $g_{\alpha\beta}^\prime$, of the new boson, $Z_{\alpha\beta}^\prime$, with mass $m_{\alpha\beta}^\prime$, that mediates flavor-dependent long-range neutrino interactions.  All the results are based on estimates of the measurement of flavor composition of high-energy astrophysical neutrinos in IceCube using 8~years of data~\cite{IceCube-Gen2:2020qha} (through-going muons plus HESE, see section~\ref{sec:stat_analysis}) and on present-day uncertainties in the neutrino mixing parameters~\cite{Esteban:2020cvm, NuFIT}, assuming normal neutrino mass ordering.  Limits are on neutrino-electron interactions, under the $L_e-L_\mu$ and $L_e-L_\tau$ symmetries, and on neutrino-neutron interactions, under the $L_\mu-L_\tau$ symmetry.    Figure~\ref{fig:g_vs_m_120} shows projections for the year 2040;  
		See section~\ref{sec:results} for details.}  
\end{figure}
 
From the figure, we observe that for $L_e-L_\mu$ (and also for $L_e-L_\tau$, which has similar results, is shown in fig.~\ref{fig:plots_for_et}), limits obtained for the epoch 2020 is already better, by about one order of magnitude, than the proof-of-principle sensitivity based on 2015 IceCube flavor measurements from ref.~\cite{Bustamante:2018mzu}. 
However, a one-to-one comparison is not possible due to differences in their statistical significance and methods. 
For the $L_\mu-L_\tau$ case, we estimate and forecast limits based on the flavor composition for the first time. Note that the limits for $g_{\mu\tau}^\prime$ shown in fig.~\ref{fig:g_vs_m_120} were obtained by fixing the mixing strength to $(\xi-\sin \theta_W \chi)=5 \times 10^{-24}$ (as mentioned earlier in section~\ref{sec:models}).  Because $V_{\mu \tau} \propto g_{\mu\tau}^\prime (\xi-\sin \theta_W \chi)$ (see eqs.~(\ref{eq:pot_general}) and (\ref{eq:Gab})), given an upper limit on $V_{\mu \tau}$, a smaller value of the mixing strength, which is plausible, would entail a weaker limit on $g_{\mu\tau}^\prime$. (Alternatively, the limits for $L_\mu-L_\tau$ in fig.~\ref{fig:g_vs_m_120} can be interpreted as being on the product of the coupling times the mixing strength, without any assumption on the value of the latter.) To summarize our result for the present day, using estimated IceCube data for the 2020 epoch and the current uncertainties on the mixing parameters, in fig.~\ref{fig:limits_3models}, we show limits on the coupling $g'_{\alpha\beta}$ as a function of the mediator mass $m_{\alpha\beta}$, for the symmetry in the same panel.
Figure~\ref{fig:g_vs_m_120} and also fig~\ref{fig:limits_3models} show that for $L_e-L_\mu$ and $L_e-L_\tau$, our limits, both estimated for 2020 and forecast for 2040, improve on existing ones obtained from atmospheric neutrinos~\cite{Joshipura:2003jh}, solar and reactor neutrinos~\cite{Bandyopadhyay:2006uh}, reinterpretations~\cite{Wise:2018rnb} of bounds on the coefficients of non-standard neutrino interactions (NSI)~\cite{Super-Kamiokande:2011dam,Ohlsson:2012kf,Gonzalez-Garcia:2013usa,Coloma:2020gfv}, and the recent global oscillation fit from ref.~\cite{Coloma:2020gfv}. For $L_\mu-L_\tau$, light mediators in the mass range that we consider are largely unconstrained, except towards high masses, from accelerator neutrinos~\cite{Heeck:2010pg}
and NSI. For the IMO case, limits are shown in Fig.~\ref{fig:g_vs_m_120_IMO}.

In summary, our findings indicate that the current estimated flavor sensitivity of IceCube, coupled with the uncertainty in the mixing parameters, holds promise for substantially enhancing the constraints on flavor-dependent long-range neutrino interactions.

\section{Future improvements}
\label{sec:future_improvements}

In this work, we estimate the present-day sensitivity of IceCube and the projected sensitivity of the other next-generation telescopes to probe possible long-range interactions arising from gauged $L_e-L_\mu$, $L_e-L_\tau$, and $L_\mu-L_\tau$ models. While doing the analysis, we have incorporated several assumptions for the simplified analysis. We list them below with a discussion on the possible improvements:
\begin{description}
	\item[Alternative flavor composition choices.] 
	As mentioned in section~\ref{statistical_analysis}, in our analysis, we limited ourselves to the case with neutrino production via the full pion decay production, $(1/3,2/3,0)_{\rm S}$, which leads to the flavor composition at Earth as the canonical expectation of $(1/3,1/3,1/3)_\oplus$.
	We take this assumption due to the unavailability of the projected flavor sensitivity of IceCube and other next-generation neutrino telescopes to alternative choices of flavor composition (Section~\ref{analysis_choice}). However, our approach is general and applicable to those alternatives. 
	 
	 \item[Flavor-measurement capabilities of upcoming detectors.]  
	 We discuss in section~\ref{analysis_choice} that for upcoming neutrino telescopes, namely, Baikal-GVD, IceCube-Gen2, KM3NeT, P-ONE, and TAMBO, we assume that their capabilities to measure the flavor composition will be similar to those of IceCube. In a more realistic scenario, they should be different. However, this is a necessary assumption given the absence of their realistic capabilities at the time of writing. Further, we have not considered the possibility of proposed improvements in flavor-tagging techniques, like those provided by muon and neutron echoes~\cite{Li:2016kra}. As the specific flavor-measurement capabilities of different upcoming experiments become better defined, future versions of our analysis will be able to incorporate them.
	 
	  \item[Using the energy dependence of the flavor composition.]
	In our analysis, we assume the measured flavor compositions are constant over an energy interval and compare the same with the energy-averaged flavor composition at Earth (Section~\ref{analysis_choice}). However, long-range interactions modify the flavor composition in an energy-dependent way, so there may be additional probing power to be reaped from measuring the flavor composition in multiple energy bins. With an insufficient event rate at present, it is beyond the scope of the analysis using present data, 
	but IceCube-Gen2 may be able to~\cite{IceCube-Gen2:2020qha,Liu:2023flr} do the same. Further, doing this could allow us to distinguish between the variation of the flavor composition with energy that stems from long-range interactions from the one that stems from the neutrino production mechanism changing with energy, e.g., from full pion decay chain at low energies to muon-damped at high energies~\cite{Kashti:2005qa, Kachelriess:2006fi, Mehta:2011qb, Winter:2013cla, Bustamante:2015waa, Bustamante:2020bxp}.
	  
	  \item[Astrophysical uncertainties.] 
	  We assume a power law spectrum of the high-energy astrophysical neutrinos~$\propto E^{-\gamma}$, ith fixed $\gamma = 2.5$, compatible with IceCube analyses that combine HESE events and through-going tracks~\cite{IceCube:2015gsk}. 
	  However, IceCube data prefers a single power law spectrum; the value of $\gamma$ is only uncertainly known and varies somewhat depending on the data set used to measure it~\cite{IceCube:2020wum, IceCube:2021uhz}.  Further, the IceCube observations admit a description with differently shaped energy spectra, e.g., refs.~\cite{Palladino:2018evm, Capanema:2020rjj, IceCube:2020wum, Ambrosone:2020evo, IceCube:2021uhz, Fiorillo:2022rft}.  Finally, because the diffuse flux of neutrinos is the addition of contributions from sources located at different distances, it depends on how the number density of sources evolves with redshift, which we have neglected in our analysis.  Incorporating uncertainties in the shape of the neutrino spectrum and the redshift distribution of sources would likely weaken the bounds that we report, though a full analysis is beyond the scope of this paper.
	  
	  \item[Computing neutrino propagation.]
	  We compute the long-range matter potential, assuming neutrino location at the IceCube, rather than propagating neutrinos from their astrophysical sources to Earth and computing the changing potential at every point along their trajectory. We take this assumption for the simplified analysis, which allows us first to limit the long-range potential and then to translate those into limits on $g_{\alpha\beta}^\prime$. Our method overestimates the influence of faraway electrons and neutrons, but mainly for limits at large mediator masses. Reference~\cite{Coloma:2020gfv} is an example of how to compute long-range interactions during propagation, though only for neutrino propagation inside the Earth and the Sun.
	  
	  \item[Screening due to relic neutrinos.]
	 We do not consider the screening effect that may arise from the relic neutrinos. If the background of relic neutrinos contains $\nu_e$ and $\bar{\nu}_e$ in equal proportions, it may partially screen the long-range potential due to cosmological electrons~\cite{Dolgov:1995hc, Blinnikov:1995kp, Joshipura:2003jh}. The Debye length~\cite{Joshipura:2003jh}, \ie, the distance at which the screening becomes important, is a factor-of-10 shorter than the long-range interaction range, so screening would affect limits at mediator masses below about $10^{-30}$~eV.
 \end{description}

\section{Summary}
\label{sec:conclusion_lri}
A new neutrino-matter interaction would be compelling evidence for the physics beyond the standard model. If they exist, they must be very weak in order to get undetected so far from the present-day experiments. At higher energy, accessible
via recently detected astrophysical neutrinos, they may be significant and even prominent with a possible signature at neutrino telescopes like IceCube and several others.

With this motivation, in this work, we explore the sensitivity of IceCube, its successor IceCube-Gen2 and several other future neutrino telescopes to probe new flavor-dependent interactions between neutrinos
and electrons and neutrinos and neutrons arising from gauged lepton number symmetries $L_e-L_\mu$, $L_e-L_\tau$, and $L_\mu-L_\tau$. This interaction is mediated by a new neutral gauge boson $Z'_{\alpha\beta}$ $(\alpha,\beta = e,\mu,\tau)$. For the $L_e-L_\beta$ $(\beta=\mu,\tau)$ case, we consider the interaction mediated solely by $Z'$, whereas for $L_\mu-L_\tau$ case interaction would appear through $Z-Z'$ mixing.
In the ultralight mass limit of the mediator, the interaction will be long-range, spanning from km to Gpc scale, depending on the mass of it mass. As a result, neutrinos at particular locations may be affected by the aggregated matter
potential sourced by large collections of nearby and faraway matter. At a large value of the potential, flavor transition probabilities of the astrophysical neutrino may have a significant impact, even suppressed at a very high value of the potential.

We compute the oscillation probabilities of astrophysical neutrinos in the presence of long-range potential induced by the three lepton number symmetries. Using this, estimate the flavor composition at Earth, using three benchmark scenarios of neutrino production at the source. While estimating the flavor composition in the presence of the long-range potential, we use the present best-fit value of mixing parameters with their associated uncertainties. Furthermore, We combine the estimated flavor-measuring capabilities of IceCube with present-day uncertainties on the standard neutrino mixing parameters and compare them with the computed flavor composition in the presence of long-range potential in order to derive expected constraints on the coupling and mediator mass of the new interaction. We repeat our analysis to improve the limits using projected flavor composition measurement from the combination of IceCube and other upcoming neutrino telescopes for the year 2040, along with the improved uncertainties in the mixing parameters in the future.
Our work improves and extends ref.~\cite{Bustamante:2018mzu}, chiefly by adopting superior statistical methods and by considering the $L_\mu-L_\tau$ symmetry. Our analysis choices are conservative and realistic (Section~\ref{analysis_choice}). Because we rely on estimates of the flavor sensitivity, our goal is to showcase the capacity of neutrino telescopes rather than provide high-precision results and to motivate further work. 

Presently, there is large uncertainty in the measurement of the flavor composition at IceCube and moderate-to-large uncertainties on the neutrino mixing parameters. At first glance, this should reduce sensitivity to long-range interactions. Surprisingly, we find that using the estimated current flavor sensitivity of IceCube and present mixing parameter uncertainties, high-energy astrophysical neutrinos could tightly constrain long-range interactions, surpassing existing limits (see fig.~\ref{fig:limits_3models}).

In the coming few decades, several upcoming neutrino oscillation experiment like, DUNE~\cite{Abi:2020wmh}, Hyper-Kamiokande~\cite{Abe:2018uyc}, and JUNO~\cite{An:2015jdp}, as well as high-energy neutrino telescopes, IceCube-Gen2~\cite{IceCube-Gen2:2020qha}, Baikal-GVD \cite{Avrorin:2019vfc}, KM3NeT~\cite{Adrian-Martinez:2016fdl}, P-ONE~\cite{P-ONE:2020ljt}, and TAMBO~\cite{Romero-Wolf:2020pzh}, will start collecting data, improving the precision on the oscillation parameters as well flavor composition measurement. Our projected results state that by 2040, nominally, we anticipate modest improvements from repeating our analysis unchanged but possible substantial gains by upgrading it to leverage higher event rates and potential advancements in flavor composition measurement (Section~\ref{sec:future_improvements}).

The discovery of a new fundamental interaction would signify a remarkable leap forward. Our research indicates that high-energy astrophysical neutrinos possess the capability, even with the present-day data, to investigate this with greater precision than existing constraints.



\blankpage 
\chapter{Summary and concluding remarks}
\label{C7} 
Among all the fundamental particles, neutrino is the most mysterious one. The story of neutrino began in 1932 when 
Prof. Wolfgang Pauli hypothesized the particle in order to account for the ``{\it missing energy}" in $\beta$-decay. The existence of neutrinos was later confirmed by Frederick Reines and Clyde Cowan in 1956 using the inverse $\beta$-decay process, in which electron antineutrino interacts with a proton, producing a neutron and a positron. Since then, several groundbreaking discoveries have enriched the field of neutrino physics. In the SM,  neutrinos are the massless neutral lepton paired with their charged counterparts. They interact with other particles through weak interaction only.

The observation of neutrino oscillation, establishing the fact that neutrinos have non-degenerate masses, makes the SM framework incomplete and compels researchers to think beyond the Standard Model~(BSM).
This phenomenon, initially put forward by Bruno Pontecorvo~\cite{Bilenky:2013wna}, was later confirmed using the data from atmospheric and solar neutrino experiments and led to the resolution of so-called ``{\it Atmospheric neutrino anomaly}" and ``{\it Solar neutrino deficit}," problems. In chapter~\ref{C2}, we discuss in detail the formalism of neutrino oscillation as well as the present status of the three-flavor oscillation parameters.
 At present, pioneering experiments involving atmospheric, solar, accelerator, and reactor neutrinos have improved our understanding of the neutrino oscillation phenomenon considerably within a three-flavor framework. Precise measurements of the six oscillation parameters in this framework have been achieved through comprehensive analyses of global oscillation data.
However, there exist several unknowns and uncertainties, --- the major three are --- the correct ordering of neutrino mass, in which octant the mixing and $\theta_{23}$ belongs if it turns out to be non-maximal, and the precise value of the CP phase. With the high-precision data from the currently running oscillation experiments and several upcoming experiments, these issues are expected to be resolved with unprecedented precision in the coming decades. 

Given the current precision in neutrino oscillation parameters and because of a few existing anomalies in the neutrino oscillation data, it offers an excellent opportunity to explore the sub-leading BSM physics in neutrino oscillation experiments. While the neutrino oscillation phenomenon itself is a compelling evidence for BSM physics, several other interesting BSM scenarios can be explored through neutrino oscillation phenomena. Chapter~\ref{C3} discusses some of these ineteresting BSM scenarios.
Theoretically, a plethora of BSM models have been proposed to account for the tiny neutrino masses and large neutrino mixings. Many of these models allow additional neutrino states, which could significantly influence oscillations of three light active neutrinos, thereby leaving observable signatures in oscillation experiments.
Moreover, certain BSM models introduce non-standard interactions between neutrinos and ambient matter particles, potentially affecting neutrino oscillation. Also, in the experimental front, anomalies observed in short-baseline~(SBL) and reactor antineutrino experiments may hint towards active-sterile oscillation. Additionally, current discrepancies between the data from the long-baseline experiments such as T2K and NO$\nu$A could possibly be due to some new physics. Upcoming long-baseline~(LBL) neutrino oscillation experiments such as the Deep Underground neutrino experiment~(DUNE), Tokai to Hyper-Kamiokande (T2HK), The European Spallation Source neutrino Super-Beam~(ESS$\nu$B), and the atmospheric neutrino data collected by the DUNE far detector,
 Hyper-KamioKande, KM3NeT-ORCA, IceCube Deepcore, and Upgrade will play a crucial role in exploring various new physics scenarios through mass-induced neutrino flavor transition.

In this direction, we explore the sub-leading impact of a few interesting BSM physics scenarios in three-flavor neutrino oscillation and the potential of the present and upcoming neutrino oscillation experiments to constrain various new physics parameters. In chapter~\ref{C4},
we investigate the role of matter effect in the evolution 
of neutrino oscillation parameters in the presence of neutral-current non-standard interaction~(NC-NSI). We perform an approximate analytical diagonalization of the neutrino Hamiltonian in the presence of NC-NSI parameters to 
derive simple 
analytical expressions showing the evolution
of mass-mixing parameters in matter. We examine how oscillation parameters get modified in the presence of NC-NSI as compared to the standard scenario.
We observe while the standard charge-current interaction does not have any impact on the evolution of $\theta_{23}$,
the NSI 
parameters in the (2,3) block, namely $\varepsilon_{\mu\tau}$
and $(\gamma - \beta) \equiv (\varepsilon_{\tau\tau}
- \varepsilon_{\mu\mu})$, significantly alter its value. 
Although all the NSI parameters influence 
the evolution of $\theta_{13}$, $\varepsilon_{e\mu}$ 
and $\varepsilon_{e\tau}$ show a stronger impact in the mult-GeV neutrino energy range.
 Interestingly, when the modified $\theta_{13}$ approaches the maximal mixing value~({\it i.e.}, $45^\circ$), the impact of these two NSI parameters vanishes.
The solar mixing angle, $\theta_{12}$, rapidly converges towards $90^\circ$ with increasing energy in both standard interaction~(SI) and NC-NSI scenarios.
The change in $\Delta m^2_{21,m}$ is quite significant
as compared to $\Delta m^2_{31,m}$ both in SI and 
SI+NSI frameworks. Flipping the sign of the NC-NSI parameters alters 
the way in which the mass-mixing parameters evolve with energy. 
Using the derived approximate analytical expressions for the modified oscillation parameters, we address several important features observed in neutrino oscillation. This includes the interesting degeneracies between $\theta_{23}$ and the new BSM parameters from the (2,3) block of the NC-NSI matrix, estimating the energy corresponding to the $\theta_{13}$-resonance --- the energy at which modified mixing angle $\theta^m_{13}$ attains the maximal value, {\it i.e.} $45^\circ$ in the presence of various NC-NSI parameters. Moreover, we observe a significant shift in the regions of oscillation maxima in the oscillograms for the $\nu_\mu\to\nu_e$ appearance channel towards smaller baselines in the presence of $\varepsilon_{e\mu}$ or $\varepsilon_{e\tau}$ with their positive strength of around $\simeq 0.2$. These regions of oscillation maxima disappear when the values of the NSI parameters are negative. Using the expressions of the modified oscillation parameters, we estimate the baselines and energies that correspond to the oscillation maxima, which help us to explain these observations analytically. We also derive a simple approximate expression of the $\nu_\mu\rightarrow\nu_\mu$ oscillation probabilities,  facilitating the interpretation of observable features in the oscillograms under both standard and NSI scenarios. This comprehensive analytical analysis offers valuable insights into the complexities of neutrino oscillation phenomena and underscores the potential impact of NSI on experimental observations.

In chapter~\ref{C5}, we study the impact of possible non-unitary neutrino
mixing (NUNM) in the context of upcoming LBL experiments 
DUNE and T2HKK having one detector in Japan (JD) and a second 
detector in Korea (KD). This chapter focuses on assessing the sensitivities of these future facilities to impose direct and model-independent constraints on various NUNM parameters.
We explore potential correlations between oscillation parameters, $\theta_{23}$, $\delta_{\mathrm{CP}}$ and various NUNM parameters.
Our comprehensive numerical analyses, using far detector data and supported by simple analytical expressions of oscillation probabilities in matter, unveil crucial insights.
Notably, we find  that the T2HKK set-up has better sensitivities for $|\alpha_{21}|$ 
and $\alpha_{22}$ as compared to DUNE, owing to its larger statistics
in the $\nu_\mu\to\nu_e$ appearance channel and less systematic uncertainties in the $\nu_\mu\to\nu_\mu$
disappearance channel, respectively. Conversely, DUNE can be more effective in constraining $|\alpha_{31}|$, $|\alpha_{32}|$, and $\alpha_{33}$, due to its 
large matter effect and wide-band neutrino energy spectra.
For $\alpha_{11}$, both DUNE and T2HKK give similar constraints.
We also show how much the bounds on the NUNM parameters can be improved 
by combining the prospective data from DUNE and T2HKK set-ups. We also estimate the limits on $\alpha_{11}$, $|\alpha_{21}|$, and $\alpha_{22}$ using the proposed near detector setups in both the experiments, leveraging the zero-distance effect in $\nu_\mu\to\nu_e$ and $\nu_\mu\to\nu_\mu$ oscillation channels. We find that the sensitivity of the far detector on these three parameters deteriorates once the normalization factor in the zero-distance effect is taken into consideration.
Lastly, we highlight the significant impact of $\nu_\tau$ appearance event samples in DUNE, improving the constraints on $|\alpha_{32}|$ and $\alpha_{33}$ by approximately $13\%$ and $60\%$, respectively.

In chapter~\ref{C6}, we explore the physics reach of the high-energy astrophysical neutrino experiments IceCube and its successor IceCube-Gen2 along with other next-generation neutrino telescopes to probe the long-range interactions~(LRI) of neutrinos. 
We focus on interactions arising from anomaly-free, gauged, abelian lepton-number symmetries, such as $L_e-L_\mu$, $L_e-L_\tau$, and $L_\mu-L_\tau$. These interactions introduce a novel matter potential sourced by electrons and neutrons, potentially influencing the flavor transition of astrophysical neutrinos. 
Given our assumption of ultra-light mediators, with masses lighter than $10^{-10}$ eV, these interactions are long-range in nature, spanning from kilometers to gigaparsecs, allowing vast numbers of electrons and neutrons in big celestial bodies and the cosmological matter distributions to contribute to this new potential. We analyze the flavor composition of astrophysical neutrinos at Earth~($f_e:f_\mu:f_\tau)_{\oplus}$ in the presence of the flavor-dependent long-range interactions. Notably, we observe significant deviations from the standard mixing scenario at Earth~($\simeq 1:1:1$) when the potential strength approaches the vacuum oscillation strength~($V_{\alpha\beta}\approx\Delta m^2_{ij}/2E$). 
When the long-range potential strength dominates over the vacuum oscillation scale, flavor transitions are suppressed, maintaining the flavor ratio at Earth same as the initial ratio at the source.
We revisit, refine, and improve the constraints on these interactions that can be placed via the flavor composition of the diffuse flux of high-energy astrophysical neutrinos at TeV--PeV energies.
Leveraging both present-day and future sensitivities of high-energy neutrino telescopes, as well as oscillation experiments, we estimate the potential constraints on the coupling strength of these interactions. 
We find that, already today, the IceCube neutrino telescope demonstrates the potential to constrain flavor-dependent long-range interactions significantly better than existing constraints, motivating further analysis. We also estimate
the improvement in the sensitivity due to the next-generation neutrino telescopes such as IceCube-Gen2, Baikal-GVD, KM3NeT, P-ONE, 
and TAMBO. Our analysis reveals that incorporating projected flavor composition data from IceCube-Gen2, with a runtime of 15 years, yields substantial improvements in LRI constraints, while the inclusion of data from other next-generation telescopes leads to marginal improvement in the constraints.

This thesis explores the possibility of probing various BSM scenarios through neutrino oscillation experiments. Specifically, we study in detail the three BSM scenarios: neutrino non-standard interaction~(NSI), non-unitarity of neutrino mixing matrix that may arise due to the presence of the heavy neutrino states, and long-range neutrino interaction mediated by ultralight gauge boson. The present oscillation data does not show any concrete evidence of these three scenarios or, in general, any BSM physics. However, the high-precision data from ongoing and upcoming oscillation experiments will offer a unique opportunity to test these scenarios with high statistics and improved precision on the standard three-flavor oscillation parameters. 
We evaluate the physics reach of next-generation neutrino oscillation experiments, with their outstanding precision, to probe these BSM scenarios if they exist in Nature. However, an inherent challenge in these BSM studies in the context of neutrino oscillation experiments is that any potential BSM signal could be attributed to multiple BSM scenarios. Therefore, it is crucial to study the degeneracies among various BSM effects and develop methods to differentiate among them. A deeper analytical understanding of oscillation probabilities in the presence of new physics is essential to address this issue. 
Also, neutrino oscillation experiments involving various neutrino sources have different advantages over others. Accelerator or long-baseline neutrino experiments have a better handle over the incoming neutrino flux; atmospheric neutrino experiments have large matter effects; astrophysical neutrino experiments operate at larger energy distance scales. Therefore, new physics effects may show up at different energy and length scales depending on the models. Combined analyses of the data from various neutrino experiments may play an important role along this direction.
With numerous next-generation neutrino experiments set to begin data collection and a growing research community, the future of neutrino physics looks promising.


\begin{appendices}
\chapter{Appendices}
\label{C8} 
\renewcommand\thesection{A}
\section{Expressions for the oscillation probabilities in various oscillation channel in the NUNM scenario}
\label{appendix:osc_prob_nunm}

In chapter~\ref{C5}, we mainly discuss $\mue$ and $\mumu$ oscillation probability in the NUNM scenario.
For the sake of completeness,  we give here the perturbative expressions of  the probabilities $P_{ee}$, $P_{e\tau}$ and $P_{\mu\tau}$, using the same approximations discussed in section~\ref{sec:osc_prob_NUNM}. We start with the $\nu_e\rightarrow\nu_e$ disappearance channel: 
\begin{eqnarray}
P_{e e } &=& 1 + 4 \alpha_{11} +6 \alpha_{11}^2-\left(\frac{4 r }{\Delta_{31}}\right)\Delta_n \left[|\alpha_{21}| \cos(\delta-\phi_{21})+ \alpha_{31} \cos(\delta-\phi_{31})\right]\sin^2\Delta_{31} +\nonumber \\
&& -\left(\frac{2 r }{\Delta_{31}}\right) \sin\Delta_{31}\left[(\Delta_{31}+2\Delta_e)\sin\Delta_{31}-2\Delta_{31}\Delta_e \cos\Delta_{31}\right] \,.
\label{eq:e-disapp}
\end{eqnarray}
The $\nu_e\rightarrow\nu_\tau$ transition is governed by the following expression: 
\begin{eqnarray}
P_{e \tau} &=&  \left(\frac{r^2}{\Delta_{31}}\right) \sin \Delta_{13} \left[(\Delta_{31}+2 \Delta_e) \sin \Delta_{31}-2 \Delta_{31} \Delta_e \cos\Delta_{31}\right]\nonumber \\
&& \left(\frac{|\alpha_{31}|^2}{\Delta_{31}}\right)\left[\Delta_{31}-\Delta_n (1-\cos2\Delta_{31})\right]+
\nonumber \\
&& \left(\frac{2 |\alpha_{21}|  r }{\Delta_{31}}\right)\Delta_n\sin\Delta_{31} \left[\sin\Delta_{31} \cos (\delta_{\mathrm{CP}}-\phi_{21})-\Delta_{31} \cos
(\delta_{\mathrm{CP}}-\Delta_{31}-\phi_{21})\right]+
\\
&&\left(\frac{|\alpha_{31}| r }{\Delta_{31}}\right)\left[2 \Delta_{31} \sin\Delta_{31} (\Delta_e+\Delta_n) \cos (\delta_{\mathrm{CP}}-\Delta_{13}-\phi_{31})+\right. \nonumber \\
&&\left.\sin (\delta_{\mathrm{CP}}-\Delta_{31}-\phi_{31}) (\sin\Delta_{31} (2 \Delta_{31}+2\Delta_e-\Delta_n)-2 \Delta_{31} \Delta_e \cos \Delta_{31})+\right. \nonumber \\ && \nonumber \left. \Delta_n \sin\Delta_{31} \sin (\delta
+\Delta_{31}-\phi_{31})\right]+\nonumber\\
&&
\left(\frac{|\alpha_{21}| |\alpha_{31}|}{\Delta_{31}}\right) \Delta_n \left[-2 \Delta_{31} \sin (\phi_{21}-\phi_{31})+\cos (2 \Delta_{31}-\phi_{21}+\phi_{31})-\cos (\phi_{21}-\phi_{31})\right]\,.\nonumber
\end{eqnarray}
Eventually, the $\nu_{\tau}$ appearance from a $\nu_{\mu}$ beam is regulated by the following probability expression:
\begin{align}
\label{eq:P_mutau}
P_{\mu \tau} &= \sin^2\Delta_{31}\left(1+2 \alpha_{22}+2 \alpha_{33}-4 a^2+ \alpha_{22}^2+ \alpha_{33}^2 + 4 \alpha_{22} \alpha_{33}\right)  +\nonumber \\
&|\alpha_{32}| \sin2\Delta_{13} \left[2 \Delta_n   \cos \phi _{32}-\sin\phi_{32}\right] +\nonumber \\
& \left(\frac{r^2}{\Delta_{31}}\right) \sin (\Delta_{31}) \left[2 \Delta_{31} \Delta_e\cos\Delta_{31}-\sin\Delta_{31} (\Delta_{31}+2 \Delta_e )\right] + \nonumber \\
& \left(\frac{2 }{\Delta_{31}}\right)\left(|\alpha_{21}|^2+|\alpha_{31}|^2\right)\Delta_{n} \sin\Delta_{31} \left[\sin\Delta_{31}-\Delta_{31} \cos\Delta_{31}\right] + \nonumber \\
&\left(\frac{|\alpha_{32}|^2}{\Delta_{31}}\right) \left[\Delta_n  \sin2 \phi_{32} (\sin2 \Delta_{31}-2 \Delta_{31} \cos2 \Delta_{31})+\Delta_{31} \cos ^2\Delta_{13}\right] +  \nonumber \\
&\left(\frac{2 |\alpha_{21}| r} {\Delta_{31}}\right)\sin \Delta_{31} \left[\sin\Delta_{31} \cos (\delta_{\mathrm{CP}}-\phi_{21}) (\Delta_{31}+\Delta_e -\Delta_n )-\right.\nonumber \\
& \qquad \qquad \left.\Delta_{31} \Delta_e   \cos (\delta_{\mathrm{CP}}+\Delta_{31}-\phi_{21})+ \Delta_{31} \Delta_n \cos (\delta_{\mathrm{CP}}-\Delta_{31}-\phi _{21})\right] + \nonumber \\
&\left(\frac{2 |\alpha_{31}| r}{\Delta_{31}}\right) \left[\sin \Delta_{31}(\sin\Delta_{31} \cos(\delta_{\mathrm{CP}}-\phi_{31}) (\Delta_{31}+\Delta_e -\Delta_n )-
\right.\nonumber \\ & \qquad\qquad  \left.\Delta_{31} \Delta_e  \cos (\delta_{\mathrm{CP}}-\Delta_{31}-\phi_{31})+  \Delta_{31} \Delta_n   \cos (\delta_{\mathrm{CP}}+\Delta_{31}-\phi_{31})\right] + \\
 & \left(\frac{8 a}{\Delta_{31}}\right)\left(\alpha_{22}-\alpha_{33}\right) \Delta_n \sin\Delta_{31} (\Delta_{31} \cos\Delta_{31}-\sin \Delta_{31}) + 4 a |\alpha_{32}| \sin^2\Delta_{31} \cos\phi_{32} +\nonumber \\
& \left(\frac{2 |\alpha_{21}| |\alpha_{31}|}{\Delta_{31}}\right) \sin\Delta_{31} \left[2 \Delta_{31} (\Delta_e+\Delta_n) \cos (\Delta_{31}-\phi_{21}+\phi_{31})+2 \Delta_n 
\sin\Delta_{31} \cos (\phi_{21}-\phi_{31})-\right.\nonumber \\
&  \left.\Delta_{31} \sin (\Delta_{31}-\phi_{21}+\phi_{31})\right] + \nonumber \\ 
&4 \alpha_{22} |\alpha_{32}| \sin\Delta_{31} \left(\frac{\Delta_n   \cos\phi_{32} (\sin\Delta_{31}+2 \Delta_{31} \cos\Delta_{31})}{\Delta_{31}}-\cos \Delta_{31} \sin\phi_{32}\right) + \nonumber \\
& -\left(\frac{2 |\alpha_{32}| \alpha_{33}}{\Delta_{31}}\right) \sin\Delta_{31} \left[2 \Delta_{n} \cos\phi_{32} (\sin\Delta_{31}-3 \Delta_{31} \cos \Delta_{31})+\Delta_{31} \cos\Delta_{31}\sin\phi _{32}\right]\,.
\end{align}
The zero-distance effects are as follows:
\begin{eqnarray}
\label{eq:zero-distance}
P_{e e}^{L = 0} &\sim& 1 + 4 \alpha_{11} + 6 \alpha_{11}^2 \,, \nonumber \\
P_{e\tau}^{L = 0} &\sim&  |\alpha_{31}|^2\,,\\
P_{\mu\tau}^{L = 0} &\sim&  |\alpha_{32}|^2\,.\nonumber 
\end{eqnarray}
It is easy to check that, when all $\alpha$ parameters are set to zero, we recover the unitary relation $P_{\mu e}+P_{\mu \mu}+P_{\mu \tau}=1$.

\renewcommand\thesection{B}
\section{Evolution of the modified mixing angles with long-range potential}
\label{app:evol_mix_angles}
\setcounter{figure}{0} 
\setcounter{section}{0}
\setcounter{equation}{0}
\renewcommand\thefigure{B\arabic{figure}}
\renewcommand\theHfigure{B\arabic{figure}}

\begin{figure}[h!]
	\centering
	\includegraphics[width=\textwidth]{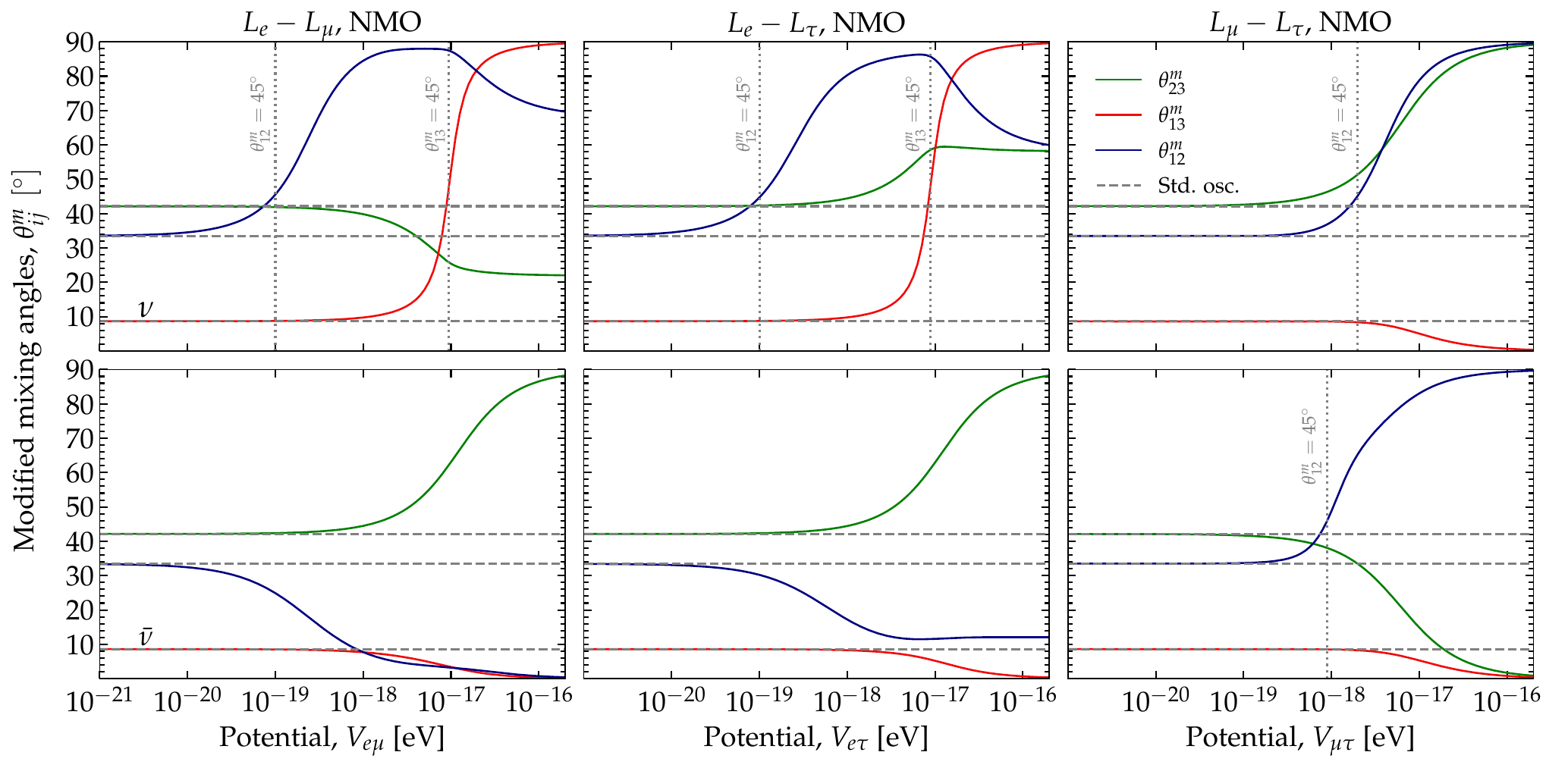}
	\mycaption{Comparison between the DUNE sensitivities on $|\alpha_{32}|$ (left panel) and $\alpha_{33}$ (right panel) when $\nu_\tau$ appearance channel is included in the analysis (red lines) and the case where no $\tau$ events are analyzed (blue lines). True values of the standard oscillation parameters are taken from table~\ref{table:vac}. All results have been obtained marginalizing over $\delta_{\mathrm{CP}}$ in the range $[-180^{\circ}, 180^{\circ}]$ and $\theta_{23}$ in the range $[40^{\circ}, 50^{\circ}]$. For $|\alpha_{32}|$ (left panel), we also marginalize over $\phi_{32}$ in the range $[-180^{\circ}, 180^{\circ}]$. }
	\label{fig:taub}
\end{figure}

Figure~\ref{fig:taub} shows the evolution of the modified mixing angles with the long-range potential.  In the main text, the average oscillation probability, eq.~\ref{eq:prob_avg}, depends on the unitarity matrix $\mathbf{U}^m$ that diagonalizes the Hamiltonian, eq.~\ref{eq:H_lrf}, which includes contribution from the long-range matter potential.  The matrix $\mathbf{U}^m$ is parametrized as the PMNS matrix, but in terms of three new mixing angles, $\theta_{12}^m$, $\theta_{23}^m$ and $\theta_{13}^m$, and one CP-violating phase, $\dcp^m$, that depend on the long-range potential.  In chapter~\ref{C4} showed analytically that $\delta^m$ remains unchanged from its vacuum value, $\dcp$.
\renewcommand\thesection{C}
\section{Average oscillation probability of the astrophysical neutrinos}
\label{sec:averaged_osc}
\renewcommand\thefigure{C\arabic{figure}}
\renewcommand\theHfigure{C\arabic{figure}}
\renewcommand\thetable{C\arabic{table}}
\renewcommand\theHtable{C\arabic{table}}
\setcounter{figure}{0} 
\setcounter{equation}{0}
\setcounter{table}{0}

The general expression for the neutrino oscillation probability $\nu_\alpha\to\nu_\beta$ is given by
\begin{align}
P(\nu_\alpha\to\nu_\beta) &= \left|\sum_k U_{\alpha k} {\rm Exp}\left[-i\frac{\Delta m^2_{k1}L}{2E}\right]U^\ast_{\beta k}\right|^2\\
& = \sum_{k,j}U^\ast_{\alpha k}U_{\beta k}U^\ast_{\alpha j}U_{\beta j}{\rm Exp}\left[-i\phi_{kj}\right]\,,
\end{align}
where, we redefine the phase $\phi_{k1} = \frac{\Delta m^2_{k1}L}{2E}$ simplifying the above equation, we get
\begin{align}
P(\nu_\alpha\to\nu_\beta) = \underbrace{\sum_{k}|U_{\alpha k}|^2|U_{\beta k}|^2}_{\rm T1} + \underbrace{\sum_{k\neq j} U^\ast_{\alpha k}U_{\beta k}U^\ast_{\alpha j}U_{\beta j} {\rm Exp}\left[-i\phi_{kj}\right]}_{\rm T2}\,.
\end{align}
Simplifying the second term, we get

\begin{align}
{\rm T2} &=  \sum_{k>j} {\rm Re}(U^\ast_{\alpha k}U_{\beta k}U_{\alpha j}U^\ast_{\beta j})\left({\rm Exp[-i\phi_{kj}]}+{\rm Exp[i\phi_{kj}]}\right)\\\nonumber
& + i\sum_{k>j} {\rm Im}(U^\ast_{\alpha k}U_{\beta k}U_{\alpha j}U^\ast_{\beta j})\left({\rm Exp[-i\phi_{kj}]}-{\rm Exp[i\phi_{kj}]}\right)\\\nonumber
& = 2\sum_{k>j} {\rm Re}(U^\ast_{\alpha k}U_{\beta k}U_{\alpha j}U^\ast_{\beta j})\cos\phi_{kj} - 2 \sum_{k>j} {\rm Im}(U^\ast_{\alpha k}U_{\beta k}U_{\alpha j}U^\ast_{\beta j})\sin\phi_{kj}\,,
\end{align}
where we have used the relation
\begin{align}
{\rm Re}(U^\ast_{\alpha k}U_{\beta k}U_{\alpha j}U^\ast_{\beta j}) = {\rm Re}(U^\ast_{\alpha j}U_{\beta j}U_{\alpha k}U^\ast_{\beta k})\,, \\
{\rm Im}(U^\ast_{\alpha k}U_{\beta k}U_{\alpha j}U^\ast_{\beta j}) = -{\rm Im}(U^\ast_{\alpha j}U_{\beta j}U_{\alpha k}U^\ast_{\beta k})\,.
\end{align}
The expression of the flavor transition probability now becomes
\begin{align}
P(\nu_\alpha\to\nu_\beta) &= \sum_k |U_{\alpha k}|^2|U_{\beta k}|^2\\
&+2\sum_{k>j} {\rm Re}(U^\ast_{\alpha k}U_{\beta k}U_{\alpha j}U^\ast_{\beta j})\cos\phi_{kj} - 2 \sum_{k>j} {\rm Im}(U^\ast_{\alpha k}U_{\beta k}U_{\alpha j}U^\ast_{\beta j})\sin\phi_{kj}\,.
\end{align}
Now, for astrophysical neutrinos reaching the Earth from various distant sources, phase $\phi_{kj} = \frac{\Delta m^2_{kj}L}{2E}>>1$. In this limit, the contribution from the phases are averaged out, {\it i.e.} $\braket{\sin\phi_{kj}}=\bra{\cos\phi_{kj}} = 0$.

After averaging out the phase, the expression of the oscillation probability becomes
\begin{align}
\bar{P}(\nu_\alpha\to\nu_\beta) = \sum_k |U_{\alpha k}|^2 |U_{\beta k}|^2\,.
\end{align} 

\renewcommand\thesection{D}
\section{Additional results under the $L_e-L_\tau$ symmetry}
\label{app:results_le_ltau}
\renewcommand\thefigure{D\arabic{figure}}
\renewcommand\theHfigure{D\arabic{figure}}
\renewcommand\thetable{D\arabic{table}}
\renewcommand\theHtable{D\arabic{table}}
\setcounter{figure}{0} 
\setcounter{equation}{0}
\setcounter{table}{0}

In chapter~\ref{C6}, we show mainly results for the $L_e-L_\mu$ and $L_\mu-L_\tau$ symmetries. Below we show results for the $L_e-L_\tau$ symmetry, under the normal mass ordering.

\begin{figure}[t!]
	\centering
	\includegraphics[width=\textwidth]{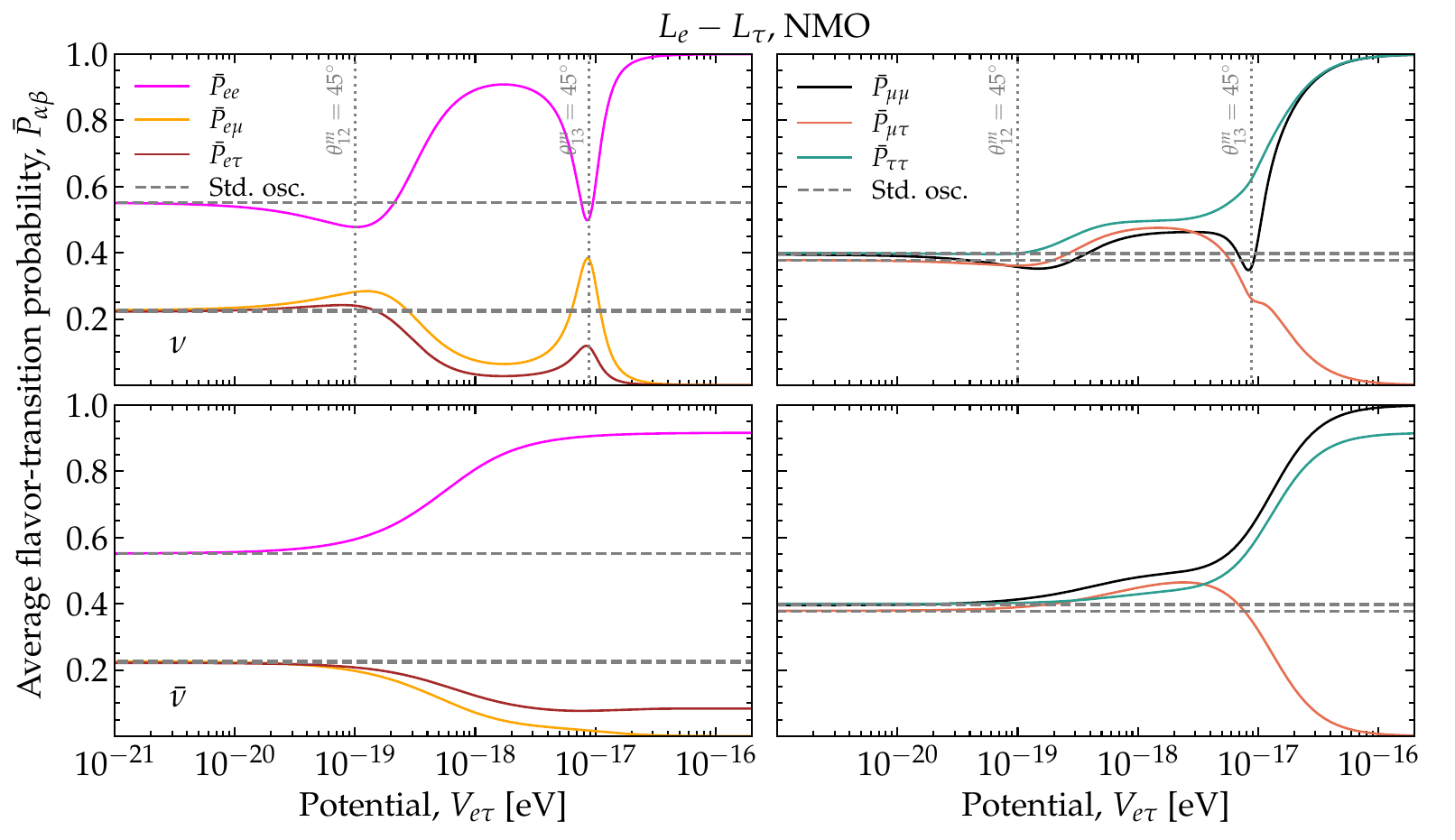}
	\mycaption{Average flavor-transition probabilities, eq.~\ref{eq:prob_avg}, as functions of the new matter potential induced by the $U(1)$ gauge symmetry $L_e-L_\tau$.  Same as fig.~\ref{fig:prob_lemmt_nmo}, but for the $L_e-L_\tau$ symmetry.  See Section~\ref{sec:osc_LRI} and Appendix~\ref{app:results_le_ltau} for details.
		\label{fig:prob_let_nmo}}
\end{figure}

Figure~\ref{fig:prob_let_nmo} shows the oscillation probabilities as functions of the long-range matter potential, $V_{e\tau}$.  Their features are similar to those for the $L_e-L_\mu$ symmetry in fig.~\ref{fig:prob_lemmt_nmo}, except swapping $\bar{P}_{e\mu}$ and $\bar{P}_{e\tau}$.

\begin{figure}[t!]
	\centering
	\includegraphics[scale=0.3]{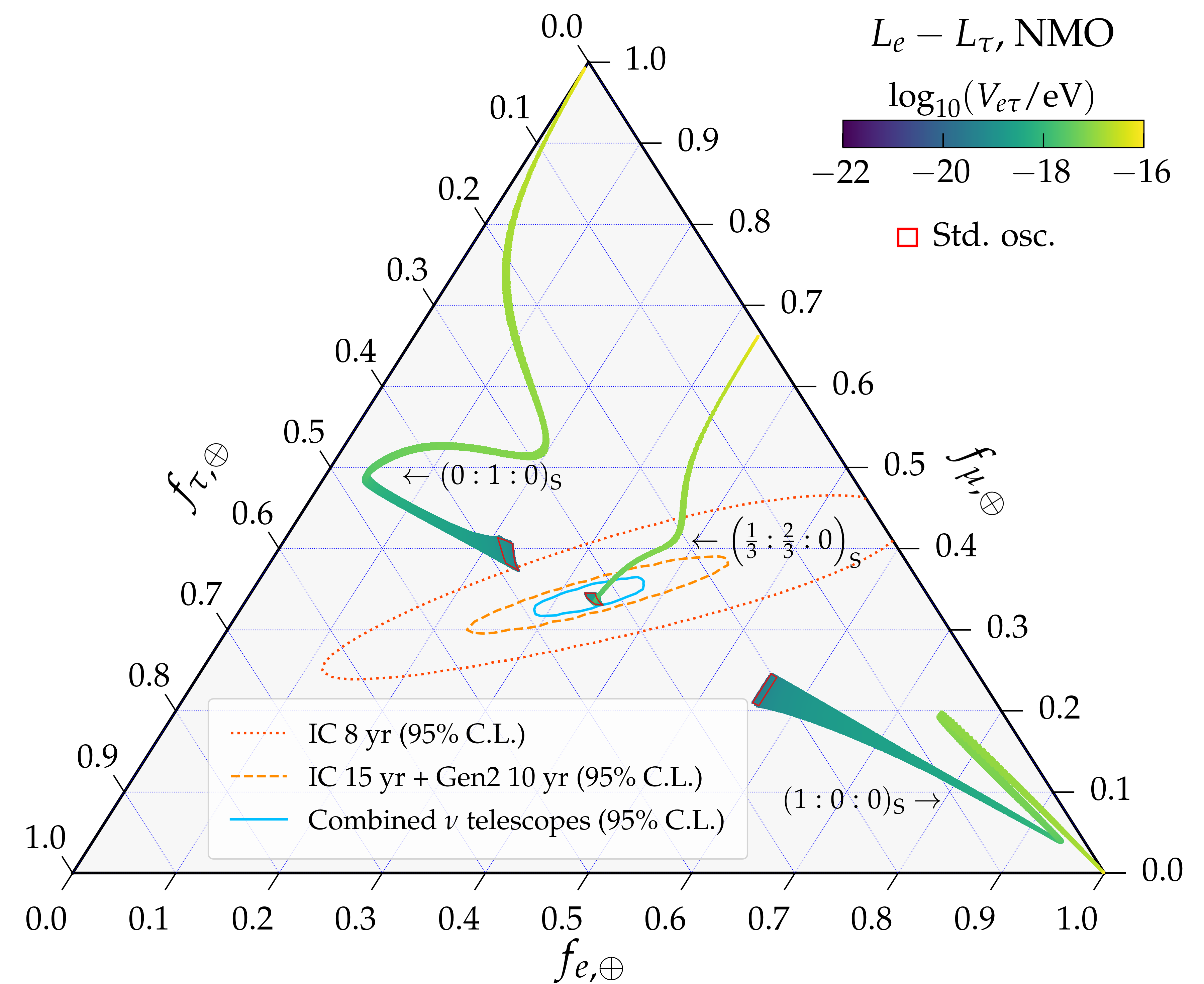}
	\mycaption{Flavor composition of high-energy astrophysical neutrinos at Earth, $f_{\alpha, \oplus}$, as a function of the long-range matter potential $V_{e\tau}$ under $L_e-L_\tau$.  Same as fig.~\ref{fig:flav_ratio}, but for the $L_e-L_\tau$ symmetry.  See section~\ref{sec:flav_comp_earth} and Appendix~\ref{app:results_le_ltau} for details.}
	\label{fig:flav_ratio_let}
\end{figure}

Figure~\ref{fig:flav_ratio_let} shows the corresponding neutrino flavor composition at Earth.  Their features track those of the oscillation probabilities in fig~\ref{fig:prob_let_nmo}.  

\begin{figure}[b!]
	\centering
	\includegraphics[width=.49\textwidth]{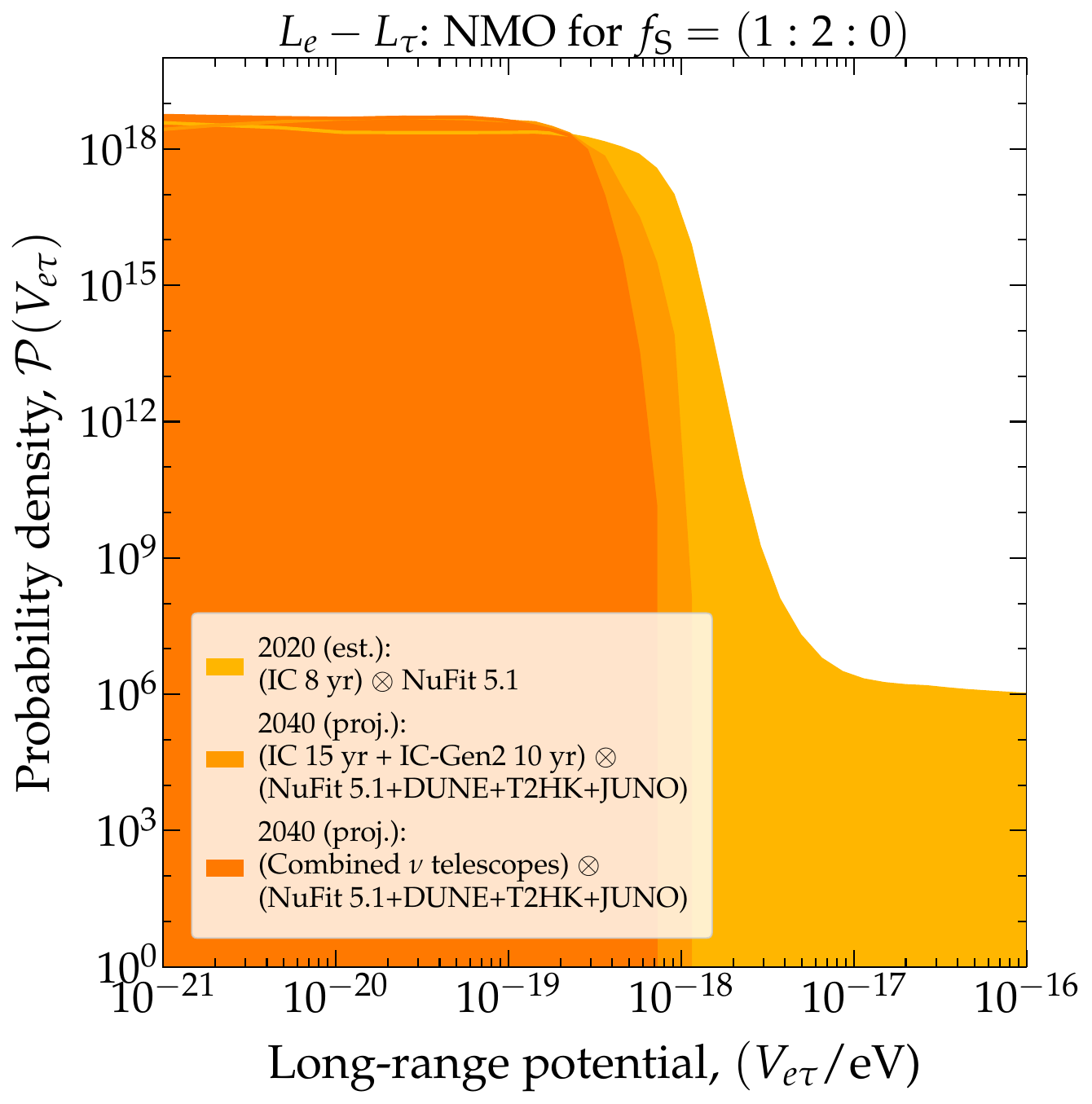}
	\includegraphics[width=.49\textwidth]{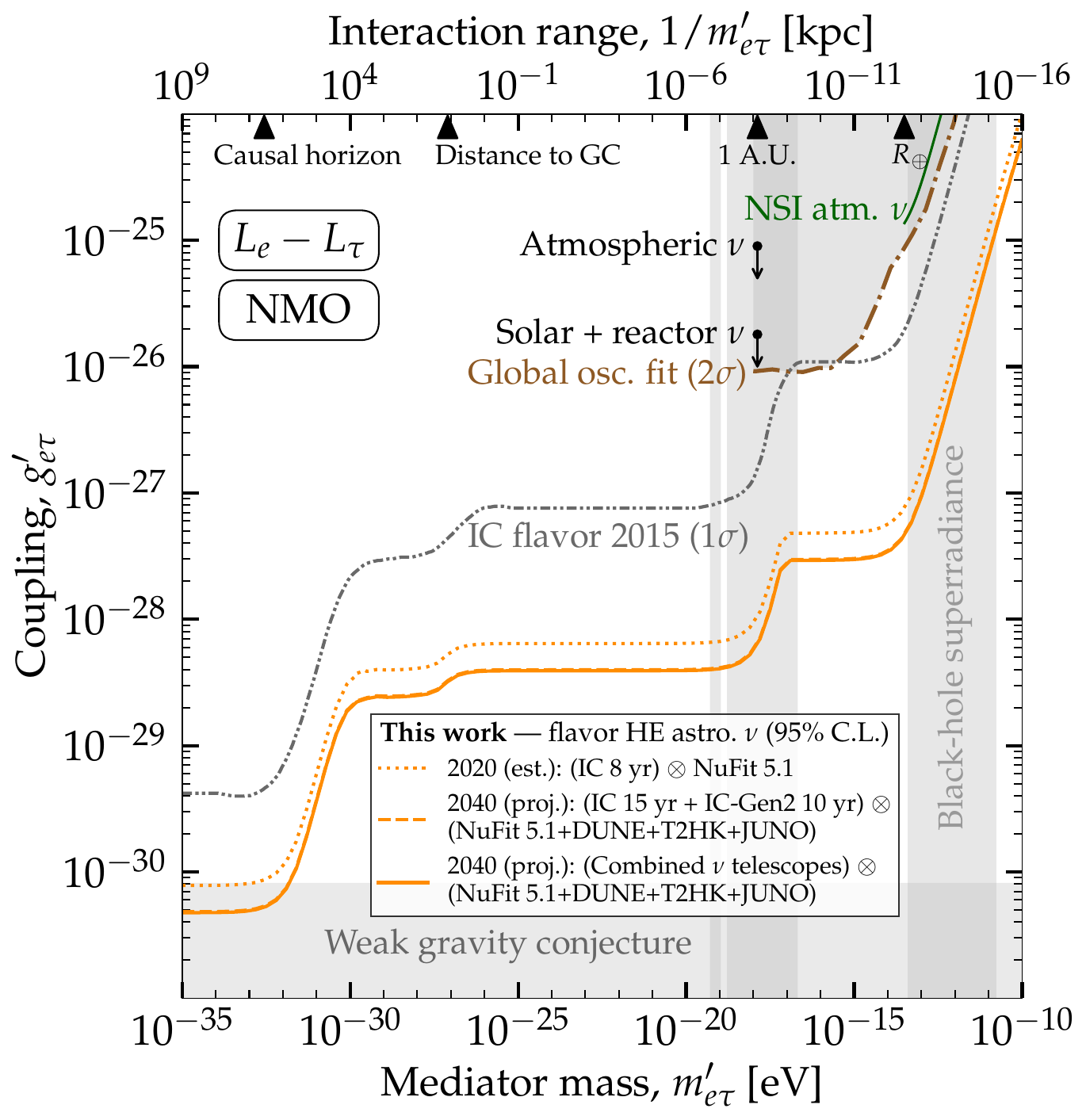}
	\mycaption{Constraints on the long-range matter potential $V_{e\tau}$, under the $L_e-L_\tau$ symmetry.  {\it Left:}  Posterior probability density of $V_{e\tau}$.  Same as fig.~\ref{fig:posterior}, but for $V_{e\tau}$.  {\it Right:}  Estimated present-day (2020) and projected (2040) upper limits (95\%~C.L.) on the coupling strength, $g_{e\tau}^\prime$, of the new boson, $Z_{e\tau}^\prime$, with mass $m_{e\tau}^\prime$.  Same as fig.~\ref{fig:g_vs_m_120}, but for $V_{e\tau}$.95\% C.L. limits on $V_{e\tau}$ is listed in table~\ref{tab:bounds}.
		 See Section~\ref{sec:results} and Appendix~\ref{app:results_le_ltau} for details.}
	\label{fig:plots_for_et}
\end{figure}

Figure~\ref{fig:plots_for_et} shows the posterior on $V_{e\tau}$ and limits on the coupling, $g_{e\tau}^\prime$, obtained following the procedure described in Section~\ref{sec:stat_analysis}. The limits are similar to those for $V_{e\mu}$ under the $L_e-L_\mu$ symmetry.

\renewcommand\thesection{E}
\section{Results assuming inverted neutrino mass ordering}
\label{lrf_results_imo}
\renewcommand\thefigure{E\arabic{figure}}
\renewcommand\theHfigure{E\arabic{figure}}
\renewcommand\thetable{E\arabic{table}}
\renewcommand\theHtable{E\arabic{table}}
\setcounter{figure}{0} 
\setcounter{equation}{0}
\setcounter{table}{0}
In the main text, we show mainly results obtained assuming normal neutrino mass ordering.  Below, we show results assuming inverted ordering.  There are differences in the oscillation probabilities: under inverted ordering, resonant features are prominent for antineutrinos rather than for neutrinos.  However, because we average the flavor composition at Earth between neutrinos and antineutrinos, the limits on the long-range matter potentials that we obtain (fig.~\ref{fig:potential_limits}) are largely insensitive to the mass ordering.

\begin{figure}[t!]
	\centering
	\includegraphics[width=\textwidth]{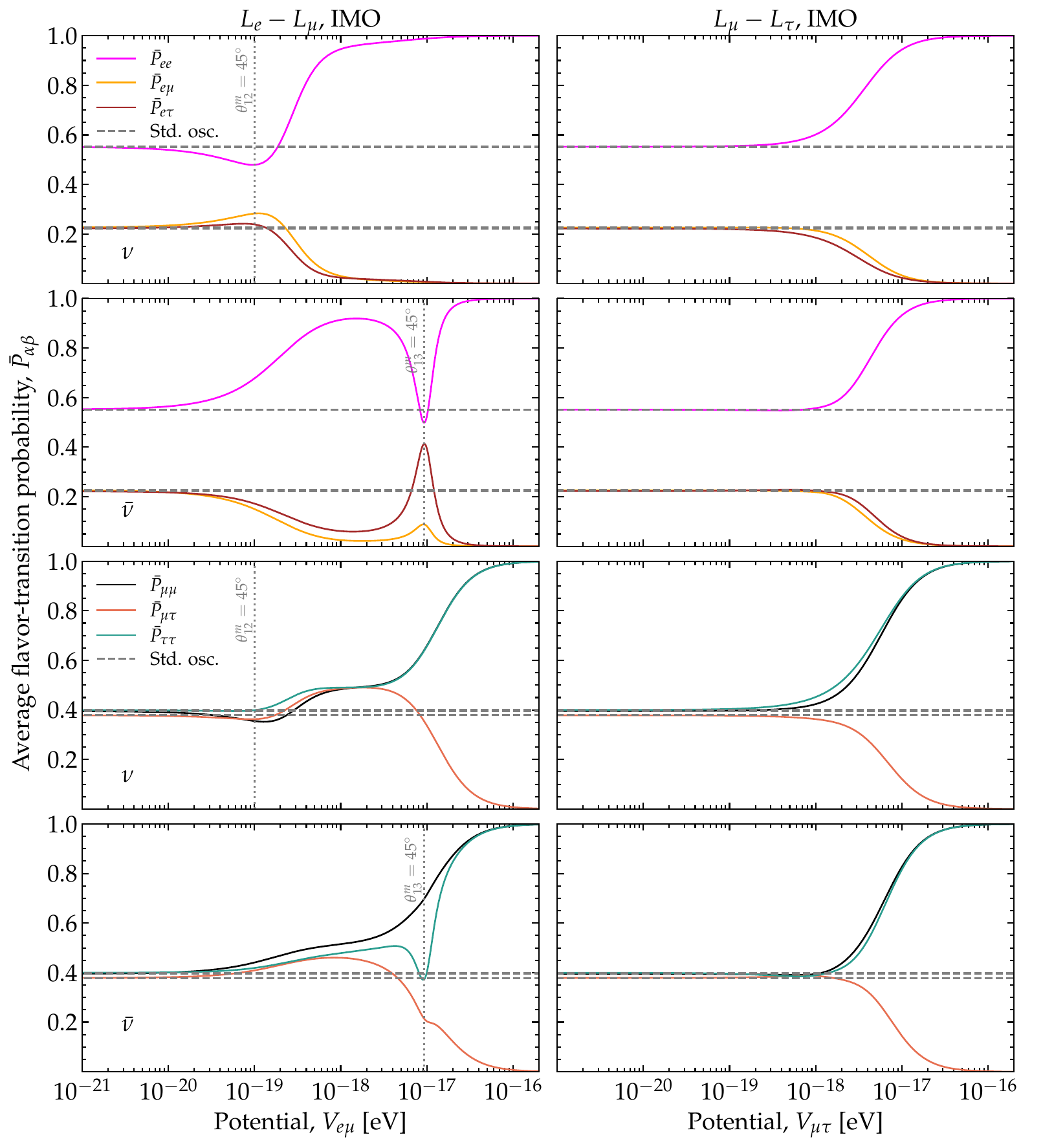}
	\mycaption{Average flavor-transition probabilities, eq.~\ref{eq:prob_avg}, as functions of the new matter potential induced by the $U(1)$ gauge symmetries $L_e-L_\mu$ (left column) and $L_\mu-L_\tau$ (right column).  Same as Fig.~\ref{fig:prob_lemmt_nmo}, but assuming inverted neutrino mass ordering (IMO).  See fig.~\ref{fig:prob_let_imo} for the probabilities under $L_e-L_\tau$.  See section~\ref{sec:osc_LRI} for details.}
	\label{fig:prob_lemmt_imo}
\end{figure}

\begin{figure}[t!]
	\centering
	\includegraphics[width=\textwidth]{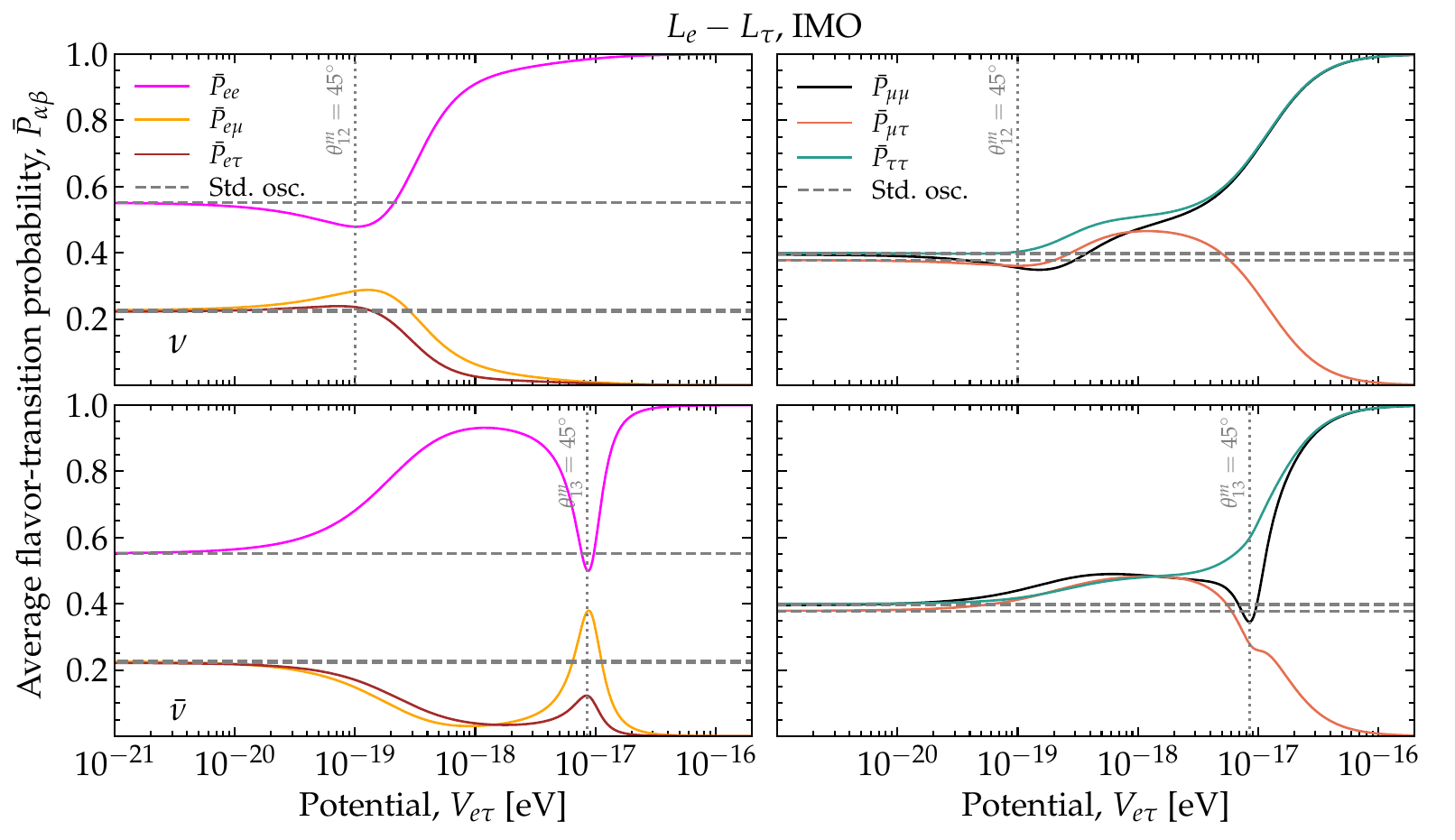}
	\mycaption{Average flavor-transition probabilities, eq.~\ref{eq:prob_avg}, as functions of the new matter potential induced by the $U(1)$ gauge symmetry $L_e-L_\tau$.  Same as fig.~\ref{fig:prob_let_nmo}, but for inverted mass ordering (IMO).  See fig.~\ref{fig:prob_lemmt_imo} for the probabilities under the other two symmetries.  See Section~\ref{sec:osc_LRI} for details.}
	\label{fig:prob_let_imo}
\end{figure}

Figures~\ref{fig:prob_lemmt_imo} and~\ref{fig:prob_let_imo} show the oscillation probabilities as functions of the long-range matter potentials.  

\begin{figure}[t!]
	\centering
	\includegraphics[width=.49\textwidth]{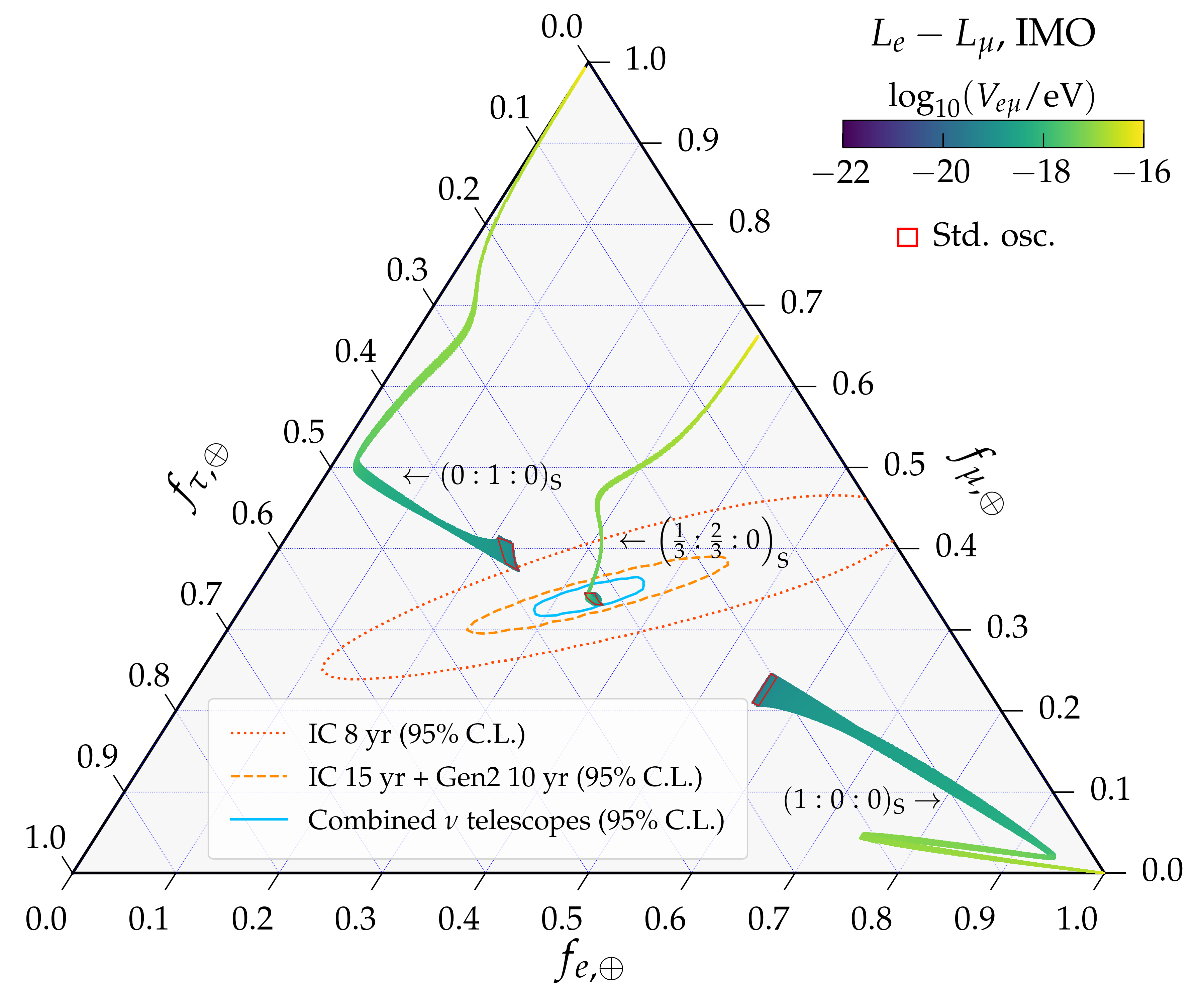}
	\includegraphics[width=.49\textwidth]{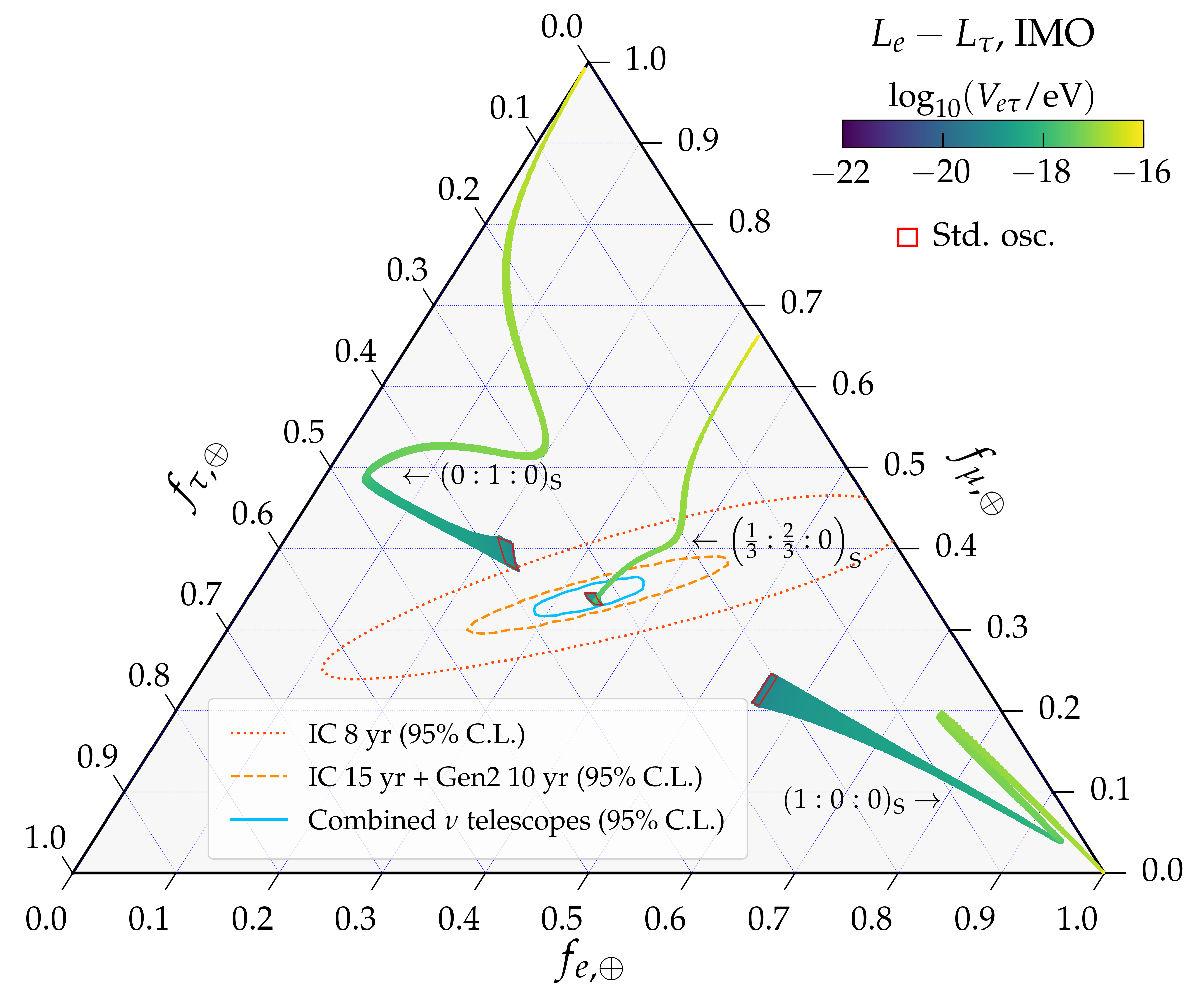}    
	\includegraphics[width=.49\textwidth]{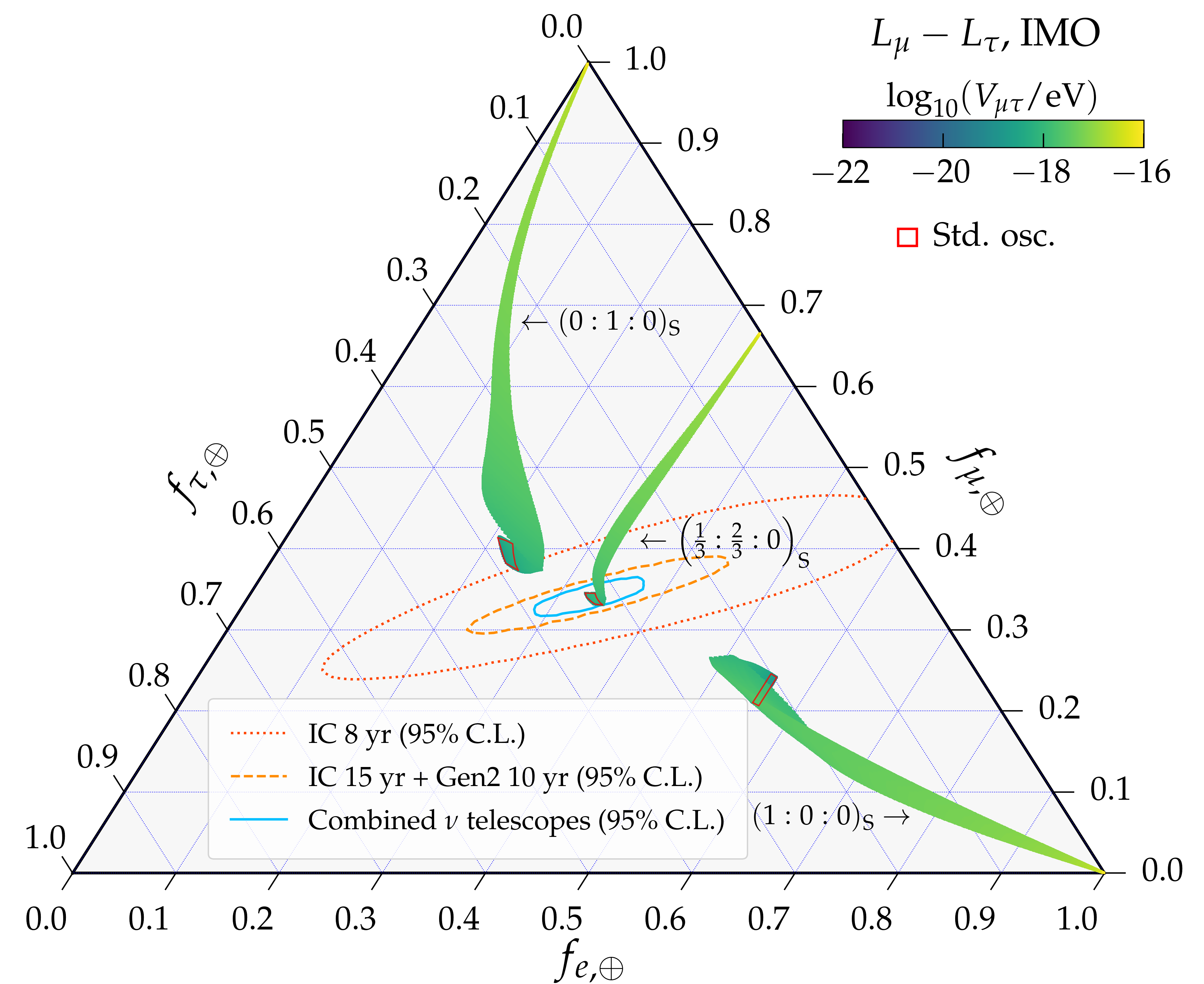}
	\mycaption{Flavor composition of high-energy astrophysical neutrinos at Earth, $f_{\alpha, \oplus}$, as a function of the long-range matter potential $V_{e\mu}$, under $L_e-L_\mu$ (top left), $V_{e\tau}$, under $L_e-L_\tau$ (top right), and $V_{\mu\tau}$, under $L_\mu-L_\tau$ (bottom).  Same as Figs.~\ref{fig:flav_ratio} and \ref{fig:flav_ratio_let}, but for inverted mass ordering (IMO).  See Section~\ref{fig:flav_ratio} for details.}
	\label{fig:flavor_IMO}
\end{figure}

Figure~\ref{fig:flavor_IMO} shows the corresponding neutrino flavor composition at Earth.

\begin{figure}[t!]
	\centering
	\includegraphics[width=.49\textwidth]{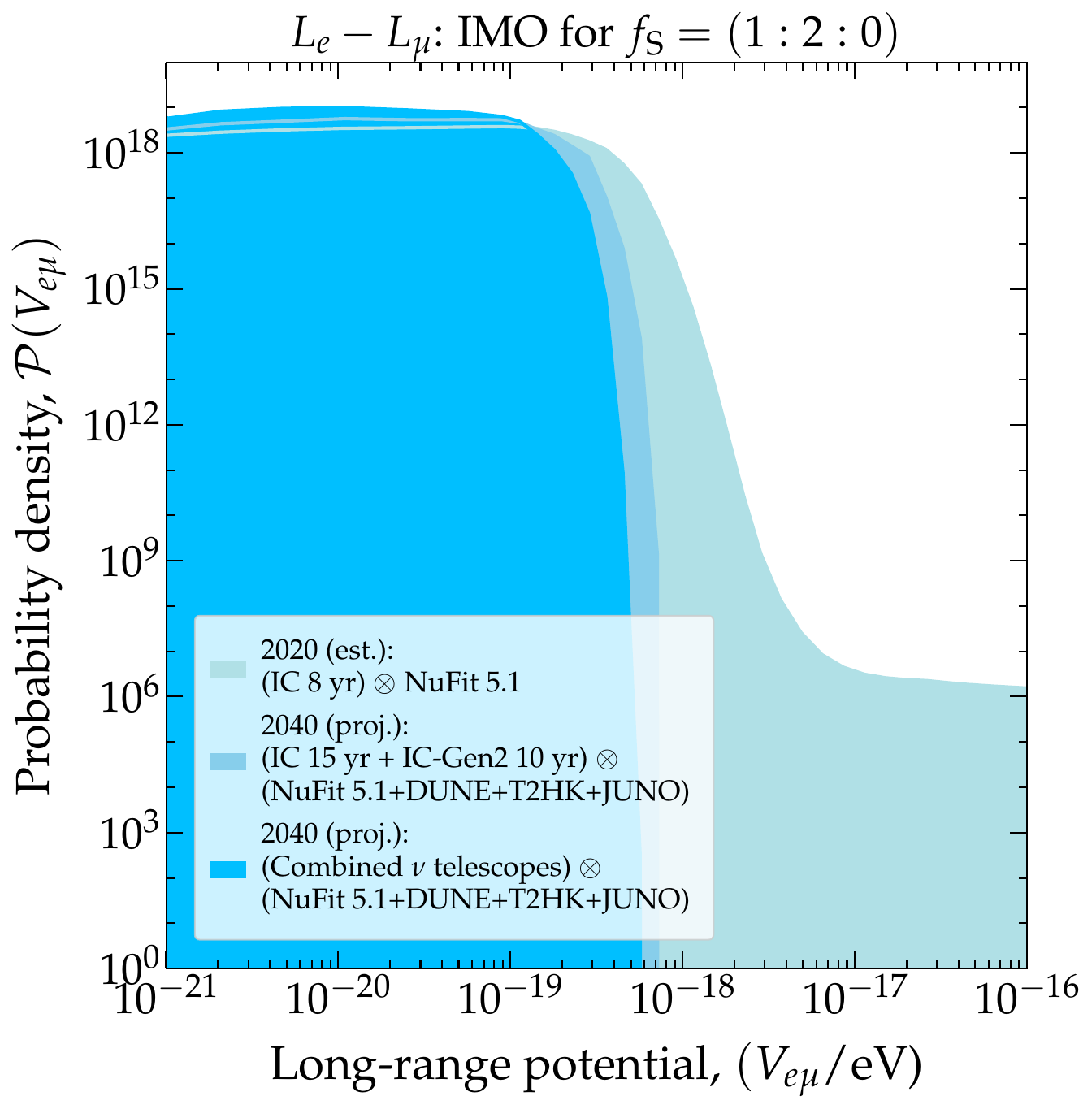}
	\includegraphics[width=.49\textwidth]{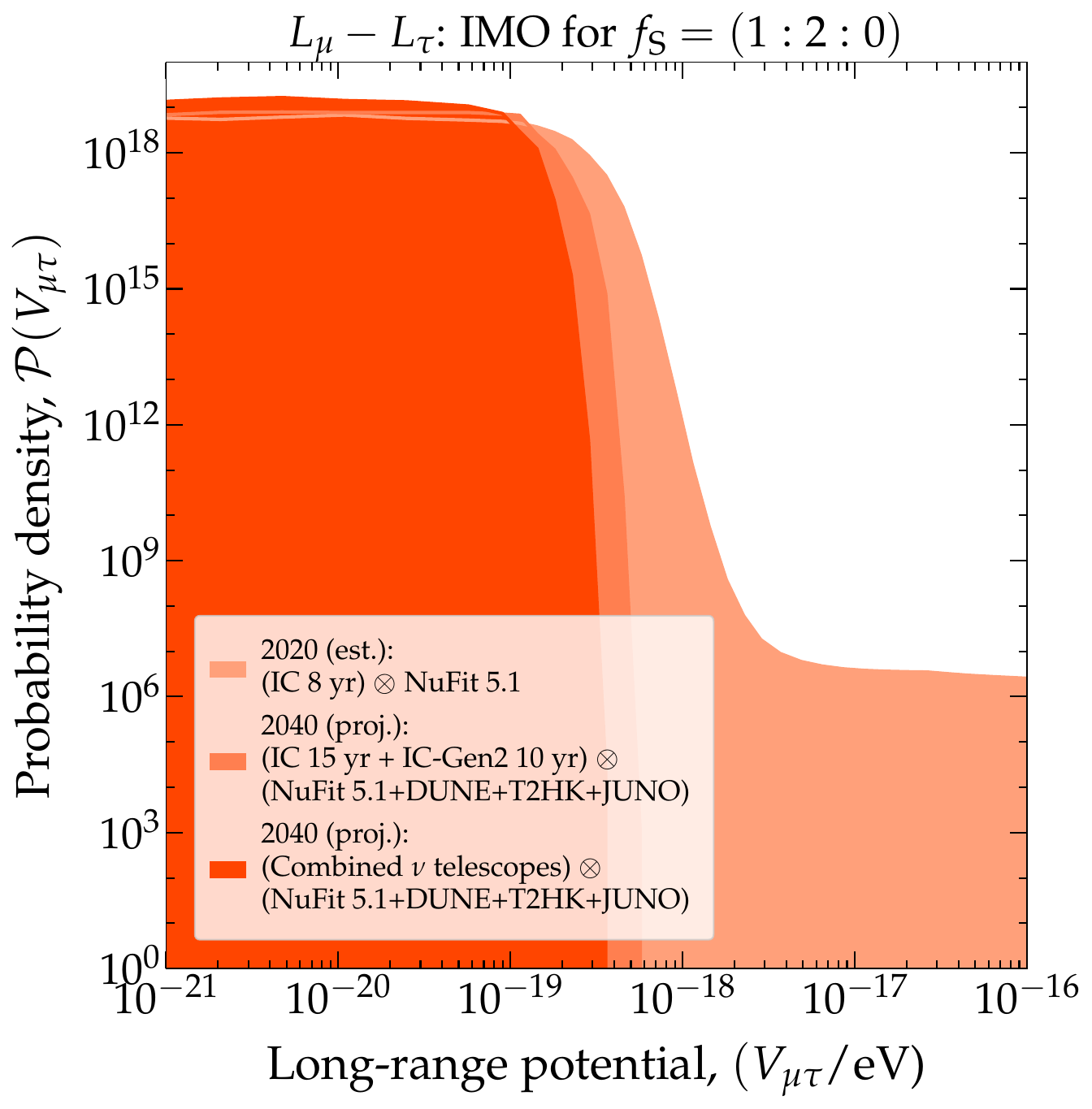}    
	\includegraphics[width=.49\textwidth]{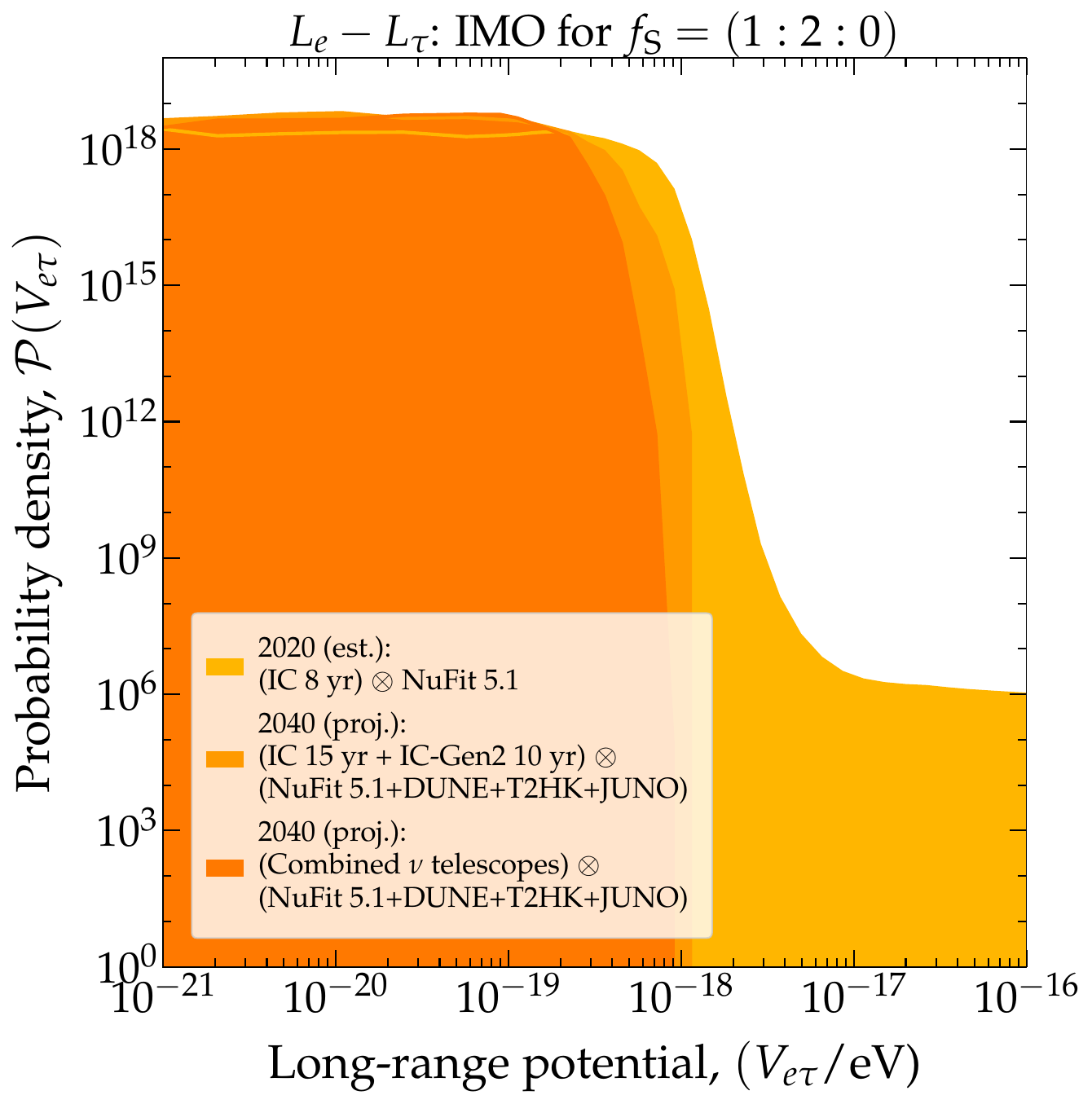}
	\mycaption{Posterior probability density of the long-range matter potentials $V_{e \mu}$, under $L_e-L_\mu$ (top left), $V_{e\tau}$, under $L_e-L_\tau$ (top right), and $V_{\mu \tau}$, under $L_\mu-L_\tau$ (bottom). Same as Figs.~\ref{fig:posterior} and \ref{fig:plots_for_et}, but for inverted neutrino mass ordering (IMO).  See Section~\ref{sec:stat_analysis} for details.}
	\label{fig:posterior_IMO}
\end{figure}

Figure~\ref{fig:posterior_IMO} shows the posterior probability distribution of the long-range potentials, obtained following the procedure described in Section~\ref{sec:stat_analysis}.

\begin{figure}[t!]
	\centering
	\includegraphics[width=.49\textwidth]{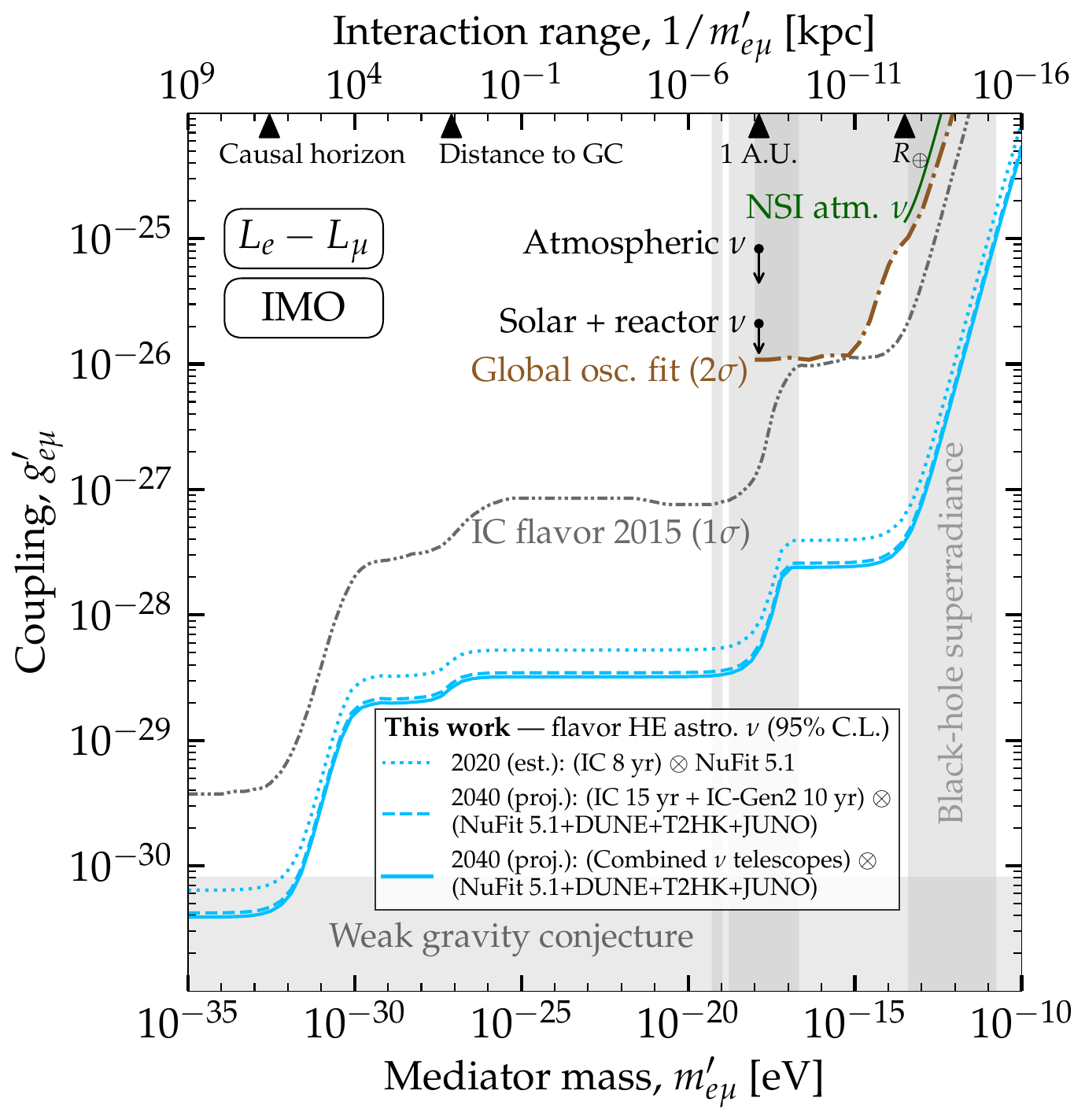}
	\includegraphics[width=.49\textwidth]{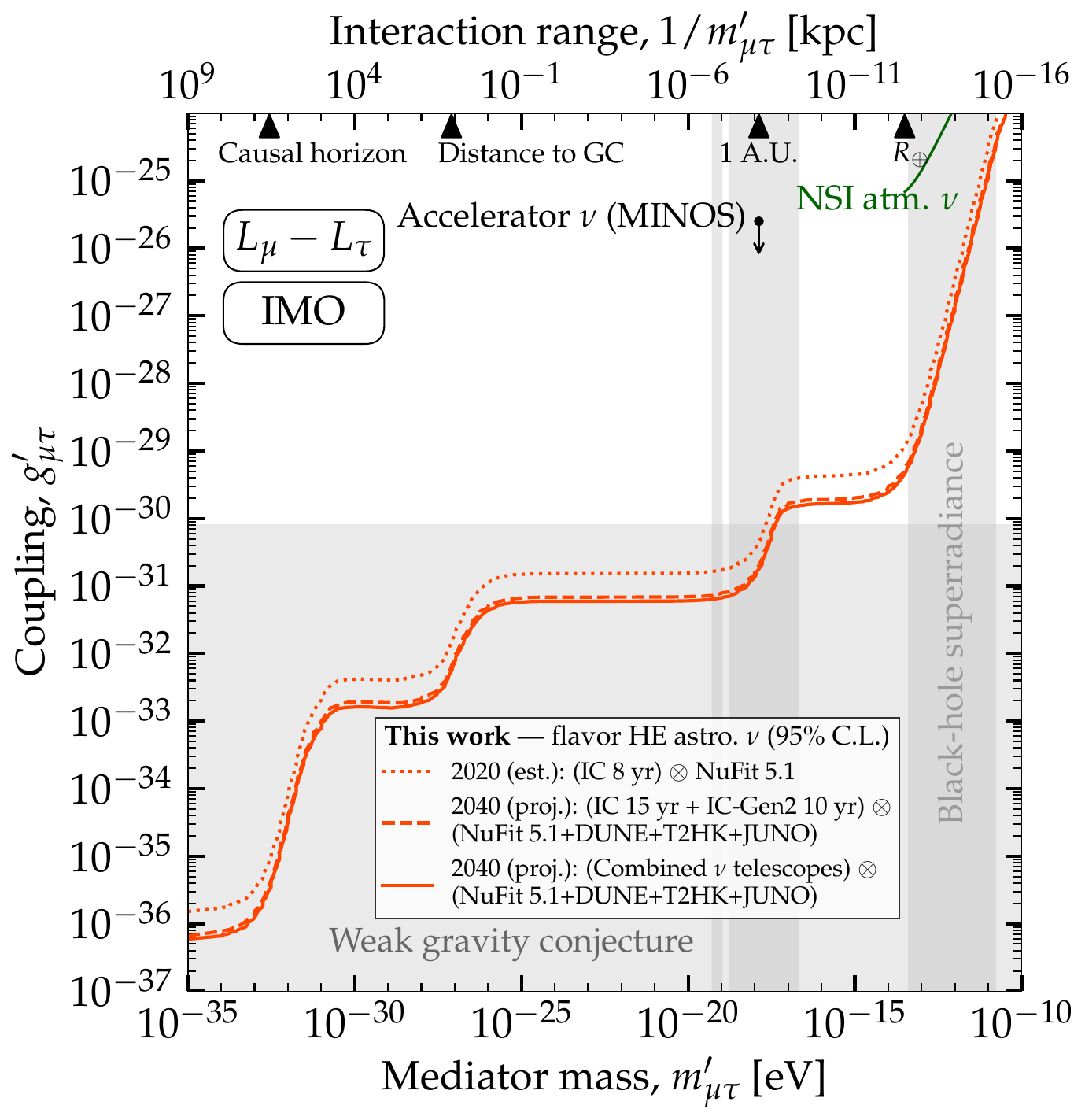}    
	\includegraphics[width=.49\textwidth]{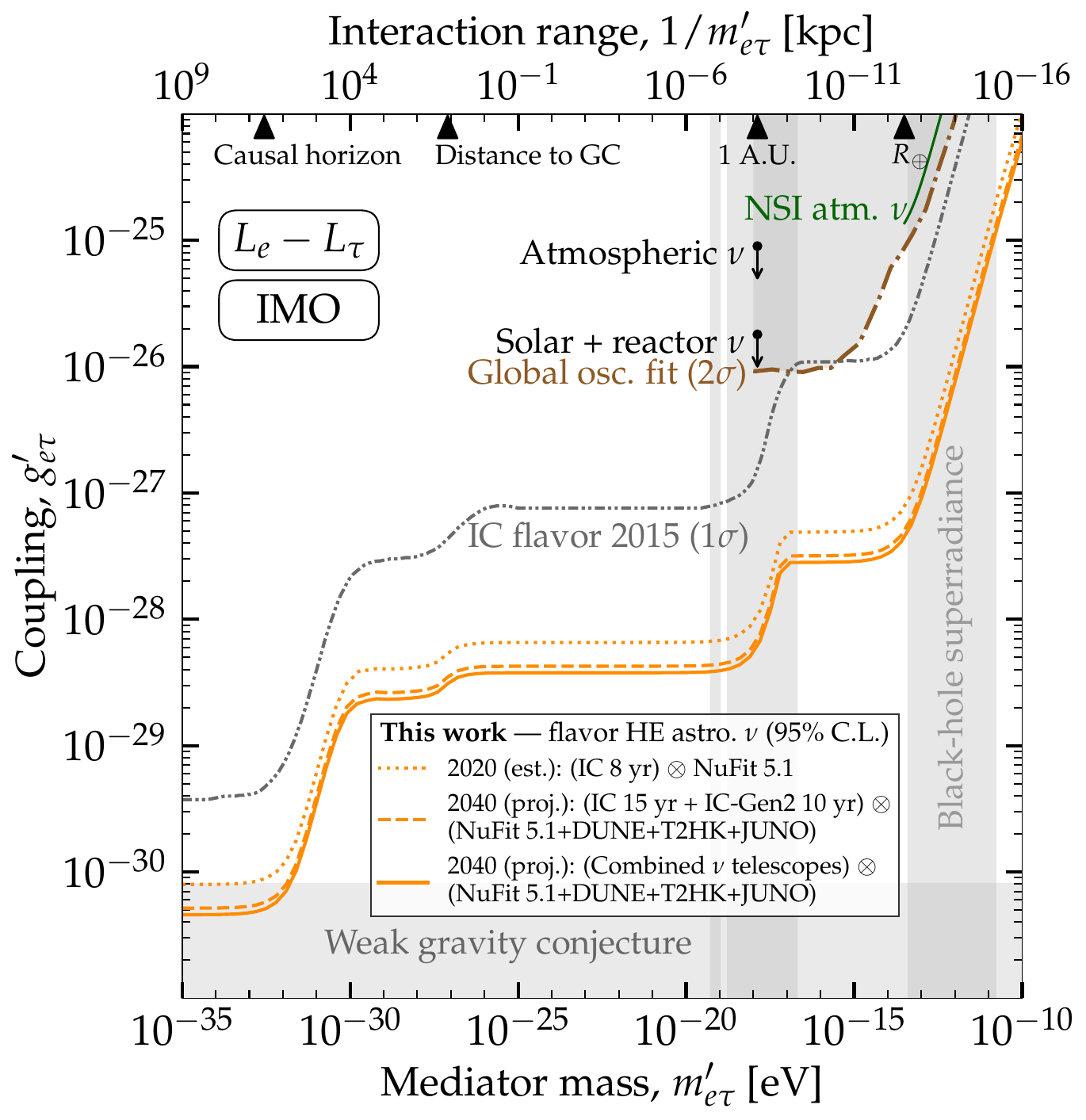}
	\mycaption{
		Estimated present-day (2020) and projected (2040) upper limits (95\%~C.L.) on the coupling strength, $g_{\alpha\beta}^\prime$, of the new boson, $Z_{\alpha\beta}^\prime$, with mass $m_{\alpha\beta}^\prime$, that mediates flavor-dependent long-range neutrino interactions.  Same as Figs.~\ref{fig:g_vs_m_120} and \ref{fig:plots_for_et}, but for inverted neutrino mass ordering (IMO).  See Section~\ref{sec:results} for details.}
	\label{fig:g_vs_m_120_IMO}
\end{figure}

Figure~\ref{fig:g_vs_m_120_IMO} shows the corresponding upper limits on the couplings, $g_{\alpha\beta}^\prime$.

\end{appendices}
\addcontentsline{toc}{chapter}{REFERENCES}
\bibliographystyle{hcdas}
\bibliography{thesis}

\begin{thebibliography}{363}%
\makeatletter
\providecommand \@ifxundefined [1]{%
 \@ifx{#1\undefined}
}%
\providecommand \@ifnum [1]{%
 \ifnum #1\expandafter \@firstoftwo
 \else \expandafter \@secondoftwo
 \fi
}%
\providecommand \@ifx [1]{%
 \ifx #1\expandafter \@firstoftwo
 \else \expandafter \@secondoftwo
 \fi
}%
\providecommand \natexlab [1]{#1}%
\providecommand \enquote  [1]{``#1''}%
\providecommand \bibnamefont  [1]{#1}%
\providecommand \bibfnamefont [1]{#1}%
\providecommand \citenamefont [1]{#1}%
\providecommand \href@noop [0]{\@secondoftwo}%
\providecommand \href [0]{\begingroup \@sanitize@url \@href}%
\providecommand \@href[1]{\@@startlink{#1}\@@href}%
\providecommand \@@href[1]{\endgroup#1\@@endlink}%
\providecommand \@sanitize@url [0]{\catcode `\\12\catcode `\$12\catcode
  `\&12\catcode `\#12\catcode `\^12\catcode `\_12\catcode `\%12\relax}%
\providecommand \@@startlink[1]{}%
\providecommand \@@endlink[0]{}%
\providecommand \url  [0]{\begingroup\@sanitize@url \@url }%
\providecommand \@url [1]{\endgroup\@href {#1}{\urlprefix }}%
\providecommand \urlprefix  [0]{URL }%
\providecommand \Eprint [0]{\href }%
\providecommand \doibase [0]{http://dx.doi.org/}%
\providecommand \selectlanguage [0]{\@gobble}%
\providecommand \bibinfo  [0]{\@secondoftwo}%
\providecommand \bibfield  [0]{\@secondoftwo}%
\providecommand \translation [1]{[#1]}%
\providecommand \BibitemOpen [0]{}%
\providecommand \bibitemStop [0]{}%
\providecommand \bibitemNoStop [0]{.\EOS\space}%
\providecommand \EOS [0]{\spacefactor3000\relax}%
\providecommand \BibitemShut  [1]{\csname bibitem#1\endcsname}%
\let\auto@bib@innerbib\@empty
\bibitem [{\citenamefont {{Fermi}}(1934)}]{1934ZPhy...88..161F}%
  \BibitemOpen
  \bibfield  {author} {\bibinfo {author} {\bibfnamefont {E.}~\bibnamefont
  {{Fermi}}},\ }\bibfield  {title} {\emph {\enquote {\bibinfo {title} {{Versuch
  einer Theorie der {\ensuremath{\beta}}-Strahlen. I}},}\ }}\href {\doibase
  10.1007/BF01351864} {\bibfield  {journal} {\bibinfo  {journal} {Zeitschrift
  fur Physik}\ }\textbf {\bibinfo {volume} {88}},\ \bibinfo {pages} {161}
  (\bibinfo {year} {1934})}\BibitemShut {NoStop}%
\bibitem [{\citenamefont {Danby}\ \emph {et~al.}(1962)\citenamefont {Danby},
  \citenamefont {Gaillard}, \citenamefont {Goulianos}, \citenamefont
  {Lederman}, \citenamefont {Mistry}, \citenamefont {Schwartz},\ and\
  \citenamefont {Steinberger}}]{Danby:1962nd}%
  \BibitemOpen
  \bibfield  {author} {\bibinfo {author} {\bibfnamefont {G.}~\bibnamefont
  {Danby}}, \bibinfo {author} {\bibfnamefont {J.~M.}\ \bibnamefont {Gaillard}},
  \bibinfo {author} {\bibfnamefont {K.~A.}\ \bibnamefont {Goulianos}}, \bibinfo
  {author} {\bibfnamefont {L.~M.}\ \bibnamefont {Lederman}}, \bibinfo {author}
  {\bibfnamefont {N.~B.}\ \bibnamefont {Mistry}}, \bibinfo {author}
  {\bibfnamefont {M.}~\bibnamefont {Schwartz}}, \ and\ \bibinfo {author}
  {\bibfnamefont {J.}~\bibnamefont {Steinberger}},\ }\bibfield  {title} {\emph
  {\enquote {\bibinfo {title} {{Observation of High-Energy Neutrino Reactions
  and the Existence of Two Kinds of Neutrinos}},}\ }}\href {\doibase
  10.1103/PhysRevLett.9.36} {\bibfield  {journal} {\bibinfo  {journal} {Phys.
  Rev. Lett.}\ }\textbf {\bibinfo {volume} {9}},\ \bibinfo {pages} {36}
  (\bibinfo {year} {1962})}\BibitemShut {NoStop}%
\bibitem [{\citenamefont {Kodama}\ \emph {et~al.}(2001)\citenamefont {Kodama}
  \emph {et~al.}}]{DONUT:2000fbd}%
  \BibitemOpen
  \bibfield  {author} {\bibinfo {author} {\bibfnamefont {K.}~\bibnamefont
  {Kodama}} \emph {et~al.} (\bibinfo {collaboration} {DONUT}),\ }\bibfield
  {title} {\emph {\enquote {\bibinfo {title} {{Observation of tau neutrino
  interactions}},}\ }}\href {\doibase 10.1016/S0370-2693(01)00307-0} {\bibfield
   {journal} {\bibinfo  {journal} {Phys. Lett. B}\ }\textbf {\bibinfo {volume}
  {504}},\ \bibinfo {pages} {218} (\bibinfo {year} {2001})},\ \Eprint
  {http://arxiv.org/abs/hep-ex/0012035}{arXiv:hep-ex/0012035}\BibitemShut
  {NoStop}%
\bibitem [{\citenamefont {Fukuda}\ \emph
  {et~al.}(1998{\natexlab{a}})\citenamefont {Fukuda} \emph
  {et~al.}}]{Super-Kamiokande:1998kpq}%
  \BibitemOpen
  \bibfield  {author} {\bibinfo {author} {\bibfnamefont {Y.}~\bibnamefont
  {Fukuda}} \emph {et~al.} (\bibinfo {collaboration} {Super-Kamiokande}),\
  }\bibfield  {title} {\emph {\enquote {\bibinfo {title} {{Evidence for
  oscillation of atmospheric neutrinos}},}\ }}\href {\doibase
  10.1103/PhysRevLett.81.1562} {\bibfield  {journal} {\bibinfo  {journal}
  {Phys. Rev. Lett.}\ }\textbf {\bibinfo {volume} {81}},\ \bibinfo {pages}
  {1562} (\bibinfo {year} {1998}{\natexlab{a}})},\ \Eprint
  {http://arxiv.org/abs/hep-ex/9807003}{arXiv:hep-ex/9807003}\BibitemShut
  {NoStop}%
\bibitem [{\citenamefont {Fukuda}\ \emph
  {et~al.}(1998{\natexlab{b}})\citenamefont {Fukuda} \emph
  {et~al.}}]{Super-Kamiokande:1998qwk}%
  \BibitemOpen
  \bibfield  {author} {\bibinfo {author} {\bibfnamefont {Y.}~\bibnamefont
  {Fukuda}} \emph {et~al.} (\bibinfo {collaboration} {Super-Kamiokande}),\
  }\bibfield  {title} {\emph {\enquote {\bibinfo {title} {{Measurements of the
  solar neutrino flux from Super-Kamiokande's first 300 days}},}\ }}\href
  {\doibase 10.1103/PhysRevLett.81.1158} {\bibfield  {journal} {\bibinfo
  {journal} {Phys. Rev. Lett.}\ }\textbf {\bibinfo {volume} {81}},\ \bibinfo
  {pages} {1158} (\bibinfo {year} {1998}{\natexlab{b}})},\ \bibinfo {note}
  {[Erratum: Phys.Rev.Lett. 81, 4279 (1998)]},\ \Eprint
  {http://arxiv.org/abs/hep-ex/9805021}{arXiv:hep-ex/9805021}\BibitemShut
  {NoStop}%
\bibitem [{\citenamefont {Cleveland}\ \emph {et~al.}(1998)\citenamefont
  {Cleveland}, \citenamefont {Daily}, \citenamefont {Davis}, \citenamefont
  {Distel}, \citenamefont {Lande}, \citenamefont {Lee}, \citenamefont
  {Wildenhain},\ and\ \citenamefont {Ullman}}]{Cleveland:1998nv}%
  \BibitemOpen
  \bibfield  {author} {\bibinfo {author} {\bibfnamefont {B.~T.}\ \bibnamefont
  {Cleveland}}, \bibinfo {author} {\bibfnamefont {T.}~\bibnamefont {Daily}},
  \bibinfo {author} {\bibfnamefont {R.}~\bibnamefont {Davis}, \bibfnamefont
  {Jr.}}, \bibinfo {author} {\bibfnamefont {J.~R.}\ \bibnamefont {Distel}},
  \bibinfo {author} {\bibfnamefont {K.}~\bibnamefont {Lande}}, \bibinfo
  {author} {\bibfnamefont {C.~K.}\ \bibnamefont {Lee}}, \bibinfo {author}
  {\bibfnamefont {P.~S.}\ \bibnamefont {Wildenhain}}, \ and\ \bibinfo {author}
  {\bibfnamefont {J.}~\bibnamefont {Ullman}},\ }\bibfield  {title} {\emph
  {\enquote {\bibinfo {title} {{Measurement of the solar electron neutrino flux
  with the Homestake chlorine detector}},}\ }}\href {\doibase 10.1086/305343}
  {\bibfield  {journal} {\bibinfo  {journal} {Astrophys. J.}\ }\textbf
  {\bibinfo {volume} {496}},\ \bibinfo {pages} {505} (\bibinfo {year}
  {1998})}\BibitemShut {NoStop}%
\bibitem [{\citenamefont {Hampel}\ \emph {et~al.}(1999)\citenamefont {Hampel}
  \emph {et~al.}}]{GALLEX:1998kcz}%
  \BibitemOpen
  \bibfield  {author} {\bibinfo {author} {\bibfnamefont {W.}~\bibnamefont
  {Hampel}} \emph {et~al.} (\bibinfo {collaboration} {GALLEX}),\ }\bibfield
  {title} {\emph {\enquote {\bibinfo {title} {{GALLEX solar neutrino
  observations: Results for GALLEX IV}},}\ }}\href {\doibase
  10.1016/S0370-2693(98)01579-2} {\bibfield  {journal} {\bibinfo  {journal}
  {Phys. Lett. B}\ }\textbf {\bibinfo {volume} {447}},\ \bibinfo {pages} {127}
  (\bibinfo {year} {1999})}\BibitemShut {NoStop}%
\bibitem [{\citenamefont {Anchordoqui}\ \emph {et~al.}(2014)\citenamefont
  {Anchordoqui} \emph {et~al.}}]{Anchordoqui:2013dnh}%
  \BibitemOpen
  \bibfield  {author} {\bibinfo {author} {\bibfnamefont {L.~A.}\ \bibnamefont
  {Anchordoqui}} \emph {et~al.},\ }\bibfield  {title} {\emph {\enquote
  {\bibinfo {title} {{Cosmic Neutrino Pevatrons: A Brand New Pathway to
  Astronomy, Astrophysics, and Particle Physics}},}\ }}\href {\doibase
  10.1016/j.jheap.2014.01.001} {\bibfield  {journal} {\bibinfo  {journal}
  {JHEAp}\ }\textbf {\bibinfo {volume} {1-2}},\ \bibinfo {pages} {1} (\bibinfo
  {year} {2014})},\ \Eprint {http://arxiv.org/abs/1312.6587}{arXiv:1312.6587
  [astro-ph.HE]}\BibitemShut {NoStop}%
\bibitem [{\citenamefont {Wu}\ \emph {et~al.}(1957)\citenamefont {Wu},
  \citenamefont {Ambler}, \citenamefont {Hayward}, \citenamefont {Hoppes},\
  and\ \citenamefont {Hudson}}]{PhysRev.105.1413}%
  \BibitemOpen
  \bibfield  {author} {\bibinfo {author} {\bibfnamefont {C.~S.}\ \bibnamefont
  {Wu}}, \bibinfo {author} {\bibfnamefont {E.}~\bibnamefont {Ambler}}, \bibinfo
  {author} {\bibfnamefont {R.~W.}\ \bibnamefont {Hayward}}, \bibinfo {author}
  {\bibfnamefont {D.~D.}\ \bibnamefont {Hoppes}}, \ and\ \bibinfo {author}
  {\bibfnamefont {R.~P.}\ \bibnamefont {Hudson}},\ }\bibfield  {title} {\emph
  {\enquote {\bibinfo {title} {Experimental Test of Parity Conservation in Beta
  Decay},}\ }}\href {\doibase 10.1103/PhysRev.105.1413} {\bibfield  {journal}
  {\bibinfo  {journal} {Phys. Rev.}\ }\textbf {\bibinfo {volume} {105}},\
  \bibinfo {pages} {1413} (\bibinfo {year} {1957})}\BibitemShut {NoStop}%
\bibitem [{\citenamefont {Friedman}\ and\ \citenamefont
  {Telegdi}(1957)}]{PhysRev.105.1681.2}%
  \BibitemOpen
  \bibfield  {author} {\bibinfo {author} {\bibfnamefont {J.~I.}\ \bibnamefont
  {Friedman}}\ and\ \bibinfo {author} {\bibfnamefont {V.~L.}\ \bibnamefont
  {Telegdi}},\ }\bibfield  {title} {\emph {\enquote {\bibinfo {title} {Nuclear
  Emulsion Evidence for Parity Nonconservation in the Decay Chain
  ${\ensuremath{\pi}}^{+}\ensuremath{-}{\ensuremath{\mu}}^{+}\ensuremath{-}{e}^{+}$},}\
  }}\href {\doibase 10.1103/PhysRev.105.1681.2} {\bibfield  {journal} {\bibinfo
   {journal} {Phys. Rev.}\ }\textbf {\bibinfo {volume} {105}},\ \bibinfo
  {pages} {1681} (\bibinfo {year} {1957})}\BibitemShut {NoStop}%
\bibitem [{\citenamefont {Agarwalla}\ \emph
  {et~al.}(2014{\natexlab{a}})\citenamefont {Agarwalla}, \citenamefont {Kao},\
  and\ \citenamefont {Takeuchi}}]{Agarwalla:2013tza}%
  \BibitemOpen
  \bibfield  {author} {\bibinfo {author} {\bibfnamefont {S.~K.}\ \bibnamefont
  {Agarwalla}}, \bibinfo {author} {\bibfnamefont {Y.}~\bibnamefont {Kao}}, \
  and\ \bibinfo {author} {\bibfnamefont {T.}~\bibnamefont {Takeuchi}},\
  }\bibfield  {title} {\emph {\enquote {\bibinfo {title} {{Analytical
  approximation of the neutrino oscillation matter effects at large
  $\theta_{13}$}},}\ }}\href {\doibase 10.1007/JHEP04(2014)047} {\bibfield
  {journal} {\bibinfo  {journal} {JHEP}\ }\textbf {\bibinfo {volume} {04}},\
  \bibinfo {pages} {047} (\bibinfo {year} {2014}{\natexlab{a}})},\ \Eprint
  {http://arxiv.org/abs/1302.6773}{arXiv:1302.6773 [hep-ph]}\BibitemShut
  {NoStop}%
\bibitem [{\citenamefont {Jarlskog}(1985)}]{PhysRevLett.55.1039}%
  \BibitemOpen
  \bibfield  {author} {\bibinfo {author} {\bibfnamefont {C.}~\bibnamefont
  {Jarlskog}},\ }\bibfield  {title} {\emph {\enquote {\bibinfo {title}
  {Commutator of the Quark Mass Matrices in the Standard Electroweak Model and
  a Measure of Maximal $\mathrm{CP}$ Nonconservation},}\ }}\href {\doibase
  10.1103/PhysRevLett.55.1039} {\bibfield  {journal} {\bibinfo  {journal}
  {Phys. Rev. Lett.}\ }\textbf {\bibinfo {volume} {55}},\ \bibinfo {pages}
  {1039} (\bibinfo {year} {1985})}\BibitemShut {NoStop}%
\bibitem [{\citenamefont {Dziewonski}\ and\ \citenamefont
  {Anderson}(1981{\natexlab{a}})}]{DZIEWONSKI1981297}%
  \BibitemOpen
  \bibfield  {author} {\bibinfo {author} {\bibfnamefont {A.~M.}\ \bibnamefont
  {Dziewonski}}\ and\ \bibinfo {author} {\bibfnamefont {D.~L.}\ \bibnamefont
  {Anderson}},\ }\bibfield  {title} {\emph {\enquote {\bibinfo {title}
  {Preliminary reference Earth model},}\ }}\href {\doibase
  https://doi.org/10.1016/0031-9201(81)90046-7} {\bibfield  {journal} {\bibinfo
   {journal} {Physics of the Earth and Planetary Interiors}\ }\textbf {\bibinfo
  {volume} {25}},\ \bibinfo {pages} {297} (\bibinfo {year}
  {1981}{\natexlab{a}})}\BibitemShut {NoStop}%
\bibitem [{\citenamefont {Gonzalez-Garcia}\ \emph
  {et~al.}(2001{\natexlab{a}})\citenamefont {Gonzalez-Garcia}, \citenamefont
  {Grossman}, \citenamefont {Gusso},\ and\ \citenamefont
  {Nir}}]{Gonzalez-Garcia:2001snt}%
  \BibitemOpen
  \bibfield  {author} {\bibinfo {author} {\bibfnamefont {M.~C.}\ \bibnamefont
  {Gonzalez-Garcia}}, \bibinfo {author} {\bibfnamefont {Y.}~\bibnamefont
  {Grossman}}, \bibinfo {author} {\bibfnamefont {A.}~\bibnamefont {Gusso}}, \
  and\ \bibinfo {author} {\bibfnamefont {Y.}~\bibnamefont {Nir}},\ }\bibfield
  {title} {\emph {\enquote {\bibinfo {title} {{New CP violation in neutrino
  oscillations}},}\ }}\href {\doibase 10.1103/PhysRevD.64.096006} {\bibfield
  {journal} {\bibinfo  {journal} {Phys. Rev. D}\ }\textbf {\bibinfo {volume}
  {64}},\ \bibinfo {pages} {096006} (\bibinfo {year} {2001}{\natexlab{a}})},\
  \Eprint
  {http://arxiv.org/abs/hep-ph/0105159}{arXiv:hep-ph/0105159}\BibitemShut
  {NoStop}%
\bibitem [{\citenamefont {Abdurashitov}\ \emph {et~al.}(1999)\citenamefont
  {Abdurashitov} \emph {et~al.}}]{SAGE:1999nng}%
  \BibitemOpen
  \bibfield  {author} {\bibinfo {author} {\bibfnamefont {J.~N.}\ \bibnamefont
  {Abdurashitov}} \emph {et~al.} (\bibinfo {collaboration} {SAGE}),\ }\bibfield
   {title} {\emph {\enquote {\bibinfo {title} {{Measurement of the solar
  neutrino capture rate with gallium metal}},}\ }}\href {\doibase
  10.1103/PhysRevC.60.055801} {\bibfield  {journal} {\bibinfo  {journal} {Phys.
  Rev. C}\ }\textbf {\bibinfo {volume} {60}},\ \bibinfo {pages} {055801}
  (\bibinfo {year} {1999})},\ \Eprint
  {http://arxiv.org/abs/astro-ph/9907113}{arXiv:astro-ph/9907113}\BibitemShut
  {NoStop}%
\bibitem [{\citenamefont {Anselmann}\ \emph {et~al.}(1992)\citenamefont
  {Anselmann} \emph {et~al.}}]{GALLEX:1992gcp}%
  \BibitemOpen
  \bibfield  {author} {\bibinfo {author} {\bibfnamefont {P.}~\bibnamefont
  {Anselmann}} \emph {et~al.} (\bibinfo {collaboration} {GALLEX}),\ }\bibfield
  {title} {\emph {\enquote {\bibinfo {title} {{Solar neutrinos observed by
  GALLEX at Gran Sasso.}}}\ }}\href {\doibase 10.1016/0370-2693(92)91521-A}
  {\bibfield  {journal} {\bibinfo  {journal} {Phys. Lett. B}\ }\textbf
  {\bibinfo {volume} {285}},\ \bibinfo {pages} {376} (\bibinfo {year}
  {1992})}\BibitemShut {NoStop}%
\bibitem [{\citenamefont {Abdurashitov}\ \emph {et~al.}(2009)\citenamefont
  {Abdurashitov} \emph {et~al.}}]{SAGE:2009eeu}%
  \BibitemOpen
  \bibfield  {author} {\bibinfo {author} {\bibfnamefont {J.~N.}\ \bibnamefont
  {Abdurashitov}} \emph {et~al.} (\bibinfo {collaboration} {SAGE}),\ }\bibfield
   {title} {\emph {\enquote {\bibinfo {title} {{Measurement of the solar
  neutrino capture rate with gallium metal. III: Results for the 2002--2007
  data-taking period}},}\ }}\href {\doibase 10.1103/PhysRevC.80.015807}
  {\bibfield  {journal} {\bibinfo  {journal} {Phys. Rev. C}\ }\textbf {\bibinfo
  {volume} {80}},\ \bibinfo {pages} {015807} (\bibinfo {year} {2009})},\
  \Eprint {http://arxiv.org/abs/0901.2200}{arXiv:0901.2200
  [nucl-ex]}\BibitemShut {NoStop}%
\bibitem [{\citenamefont {Fukuda}\ \emph {et~al.}(2003)\citenamefont {Fukuda}
  \emph {et~al.}}]{Super-Kamiokande:2002weg}%
  \BibitemOpen
  \bibfield  {author} {\bibinfo {author} {\bibfnamefont {Y.}~\bibnamefont
  {Fukuda}} \emph {et~al.} (\bibinfo {collaboration} {Super-Kamiokande}),\
  }\bibfield  {title} {\emph {\enquote {\bibinfo {title} {{The Super-Kamiokande
  detector}},}\ }}\href {\doibase 10.1016/S0168-9002(03)00425-X} {\bibfield
  {journal} {\bibinfo  {journal} {Nucl. Instrum. Meth. A}\ }\textbf {\bibinfo
  {volume} {501}},\ \bibinfo {pages} {418} (\bibinfo {year}
  {2003})}\BibitemShut {NoStop}%
\bibitem [{\citenamefont {Hosaka}\ \emph {et~al.}(2006)\citenamefont {Hosaka}
  \emph {et~al.}}]{Super-Kamiokande:2005wtt}%
  \BibitemOpen
  \bibfield  {author} {\bibinfo {author} {\bibfnamefont {J.}~\bibnamefont
  {Hosaka}} \emph {et~al.} (\bibinfo {collaboration} {Super-Kamiokande}),\
  }\bibfield  {title} {\emph {\enquote {\bibinfo {title} {{Solar neutrino
  measurements in super-Kamiokande-I}},}\ }}\href {\doibase
  10.1103/PhysRevD.73.112001} {\bibfield  {journal} {\bibinfo  {journal} {Phys.
  Rev. D}\ }\textbf {\bibinfo {volume} {73}},\ \bibinfo {pages} {112001}
  (\bibinfo {year} {2006})},\ \Eprint
  {http://arxiv.org/abs/hep-ex/0508053}{arXiv:hep-ex/0508053}\BibitemShut
  {NoStop}%
\bibitem [{\citenamefont {Cowan}\ \emph {et~al.}(1953)\citenamefont {Cowan},
  \citenamefont {Reines}, \citenamefont {Harrison}, \citenamefont {Anderson},\
  and\ \citenamefont {Hayes}}]{PhysRev.90.493}%
  \BibitemOpen
  \bibfield  {author} {\bibinfo {author} {\bibfnamefont {C.~L.}\ \bibnamefont
  {Cowan}}, \bibinfo {author} {\bibfnamefont {F.}~\bibnamefont {Reines}},
  \bibinfo {author} {\bibfnamefont {F.~B.}\ \bibnamefont {Harrison}}, \bibinfo
  {author} {\bibfnamefont {E.~C.}\ \bibnamefont {Anderson}}, \ and\ \bibinfo
  {author} {\bibfnamefont {F.~N.}\ \bibnamefont {Hayes}},\ }\bibfield  {title}
  {\emph {\enquote {\bibinfo {title} {Large Liquid Scintillation Detectors},}\
  }}\href {\doibase 10.1103/PhysRev.90.493} {\bibfield  {journal} {\bibinfo
  {journal} {Phys. Rev.}\ }\textbf {\bibinfo {volume} {90}},\ \bibinfo {pages}
  {493} (\bibinfo {year} {1953})}\BibitemShut {NoStop}%
\bibitem [{\citenamefont {Reines}\ and\ \citenamefont
  {Cowan}(1953)}]{PhysRev.92.830}%
  \BibitemOpen
  \bibfield  {author} {\bibinfo {author} {\bibfnamefont {F.}~\bibnamefont
  {Reines}}\ and\ \bibinfo {author} {\bibfnamefont {C.~L.}\ \bibnamefont
  {Cowan}},\ }\bibfield  {title} {\emph {\enquote {\bibinfo {title} {Detection
  of the Free Neutrino},}\ }}\href {\doibase 10.1103/PhysRev.92.830} {\bibfield
   {journal} {\bibinfo  {journal} {Phys. Rev.}\ }\textbf {\bibinfo {volume}
  {92}},\ \bibinfo {pages} {830} (\bibinfo {year} {1953})}\BibitemShut
  {NoStop}%
\bibitem [{\citenamefont {Reines}\ \emph {et~al.}(1980)\citenamefont {Reines},
  \citenamefont {Sobel},\ and\ \citenamefont {Pasierb}}]{PhysRevLett.45.1307}%
  \BibitemOpen
  \bibfield  {author} {\bibinfo {author} {\bibfnamefont {F.}~\bibnamefont
  {Reines}}, \bibinfo {author} {\bibfnamefont {H.~W.}\ \bibnamefont {Sobel}}, \
  and\ \bibinfo {author} {\bibfnamefont {E.}~\bibnamefont {Pasierb}},\
  }\bibfield  {title} {\emph {\enquote {\bibinfo {title} {Evidence for Neutrino
  Instability},}\ }}\href {\doibase 10.1103/PhysRevLett.45.1307} {\bibfield
  {journal} {\bibinfo  {journal} {Phys. Rev. Lett.}\ }\textbf {\bibinfo
  {volume} {45}},\ \bibinfo {pages} {1307} (\bibinfo {year}
  {1980})}\BibitemShut {NoStop}%
\bibitem [{\citenamefont {Boehm}\ \emph {et~al.}(2000)\citenamefont {Boehm},
  \citenamefont {Busenitz}, \citenamefont {Cook}, \citenamefont {Gratta},
  \citenamefont {Henrikson}, \citenamefont {Kornis}, \citenamefont {Lawrence},
  \citenamefont {Lee}, \citenamefont {McKinny}, \citenamefont {Miller},
  \citenamefont {Novikov}, \citenamefont {Piepke}, \citenamefont {Ritchie},
  \citenamefont {Tracy}, \citenamefont {Vogel}, \citenamefont {Wang},\ and\
  \citenamefont {Wolf}}]{PhysRevLett.84.3764}%
  \BibitemOpen
  \bibfield  {author} {\bibinfo {author} {\bibfnamefont {F.}~\bibnamefont
  {Boehm}}, \bibinfo {author} {\bibfnamefont {J.}~\bibnamefont {Busenitz}},
  \bibinfo {author} {\bibfnamefont {B.}~\bibnamefont {Cook}}, \bibinfo {author}
  {\bibfnamefont {G.}~\bibnamefont {Gratta}}, \bibinfo {author} {\bibfnamefont
  {H.}~\bibnamefont {Henrikson}}, \bibinfo {author} {\bibfnamefont
  {J.}~\bibnamefont {Kornis}}, \bibinfo {author} {\bibfnamefont
  {D.}~\bibnamefont {Lawrence}}, \bibinfo {author} {\bibfnamefont {K.~B.}\
  \bibnamefont {Lee}}, \bibinfo {author} {\bibfnamefont {K.}~\bibnamefont
  {McKinny}}, \bibinfo {author} {\bibfnamefont {L.}~\bibnamefont {Miller}},
  \bibinfo {author} {\bibfnamefont {V.}~\bibnamefont {Novikov}}, \bibinfo
  {author} {\bibfnamefont {A.}~\bibnamefont {Piepke}}, \bibinfo {author}
  {\bibfnamefont {B.}~\bibnamefont {Ritchie}}, \bibinfo {author} {\bibfnamefont
  {D.}~\bibnamefont {Tracy}}, \bibinfo {author} {\bibfnamefont
  {P.}~\bibnamefont {Vogel}}, \bibinfo {author} {\bibfnamefont {Y.-F.}\
  \bibnamefont {Wang}}, \ and\ \bibinfo {author} {\bibfnamefont
  {J.}~\bibnamefont {Wolf}},\ }\bibfield  {title} {\emph {\enquote {\bibinfo
  {title} {Search for Neutrino Oscillations at the Palo Verde Nuclear
  Reactors},}\ }}\href {\doibase 10.1103/PhysRevLett.84.3764} {\bibfield
  {journal} {\bibinfo  {journal} {Phys. Rev. Lett.}\ }\textbf {\bibinfo
  {volume} {84}},\ \bibinfo {pages} {3764} (\bibinfo {year}
  {2000})}\BibitemShut {NoStop}%
\bibitem [{\citenamefont {Eguchi}\ \emph {et~al.}(2003)\citenamefont {Eguchi}
  \emph {et~al.}}]{KamLAND:2002uet}%
  \BibitemOpen
  \bibfield  {author} {\bibinfo {author} {\bibfnamefont {K.}~\bibnamefont
  {Eguchi}} \emph {et~al.} (\bibinfo {collaboration} {KamLAND}),\ }\bibfield
  {title} {\emph {\enquote {\bibinfo {title} {{First results from KamLAND:
  Evidence for reactor anti-neutrino disappearance}},}\ }}\href {\doibase
  10.1103/PhysRevLett.90.021802} {\bibfield  {journal} {\bibinfo  {journal}
  {Phys. Rev. Lett.}\ }\textbf {\bibinfo {volume} {90}},\ \bibinfo {pages}
  {021802} (\bibinfo {year} {2003})},\ \Eprint
  {http://arxiv.org/abs/hep-ex/0212021}{arXiv:hep-ex/0212021}\BibitemShut
  {NoStop}%
\bibitem [{\citenamefont {Apollonio}\ \emph {et~al.}(1999)\citenamefont
  {Apollonio} \emph {et~al.}}]{CHOOZ:1999hei}%
  \BibitemOpen
  \bibfield  {author} {\bibinfo {author} {\bibfnamefont {M.}~\bibnamefont
  {Apollonio}} \emph {et~al.} (\bibinfo {collaboration} {CHOOZ}),\ }\bibfield
  {title} {\emph {\enquote {\bibinfo {title} {{Limits on neutrino oscillations
  from the CHOOZ experiment}},}\ }}\href {\doibase
  10.1016/S0370-2693(99)01072-2} {\bibfield  {journal} {\bibinfo  {journal}
  {Phys. Lett. B}\ }\textbf {\bibinfo {volume} {466}},\ \bibinfo {pages} {415}
  (\bibinfo {year} {1999})},\ \Eprint
  {http://arxiv.org/abs/hep-ex/9907037}{arXiv:hep-ex/9907037}\BibitemShut
  {NoStop}%
\bibitem [{\citenamefont {An}\ \emph {et~al.}(2012)\citenamefont {An} \emph
  {et~al.}}]{DayaBay:2012fng}%
  \BibitemOpen
  \bibfield  {author} {\bibinfo {author} {\bibfnamefont {F.~P.}\ \bibnamefont
  {An}} \emph {et~al.} (\bibinfo {collaboration} {Daya Bay}),\ }\bibfield
  {title} {\emph {\enquote {\bibinfo {title} {{Observation of
  electron-antineutrino disappearance at Daya Bay}},}\ }}\href {\doibase
  10.1103/PhysRevLett.108.171803} {\bibfield  {journal} {\bibinfo  {journal}
  {Phys. Rev. Lett.}\ }\textbf {\bibinfo {volume} {108}},\ \bibinfo {pages}
  {171803} (\bibinfo {year} {2012})},\ \Eprint
  {http://arxiv.org/abs/1203.1669}{arXiv:1203.1669 [hep-ex]}\BibitemShut
  {NoStop}%
\bibitem [{\citenamefont {Abe}\ \emph {et~al.}(2012)\citenamefont {Abe} \emph
  {et~al.}}]{DoubleChooz:2011ymz}%
  \BibitemOpen
  \bibfield  {author} {\bibinfo {author} {\bibfnamefont {Y.}~\bibnamefont
  {Abe}} \emph {et~al.} (\bibinfo {collaboration} {Double Chooz}),\ }\bibfield
  {title} {\emph {\enquote {\bibinfo {title} {{Indication of Reactor
  $\bar{\nu}_e$ Disappearance in the Double Chooz Experiment}},}\ }}\href
  {\doibase 10.1103/PhysRevLett.108.131801} {\bibfield  {journal} {\bibinfo
  {journal} {Phys. Rev. Lett.}\ }\textbf {\bibinfo {volume} {108}},\ \bibinfo
  {pages} {131801} (\bibinfo {year} {2012})},\ \Eprint
  {http://arxiv.org/abs/1112.6353}{arXiv:1112.6353 [hep-ex]}\BibitemShut
  {NoStop}%
\bibitem [{\citenamefont {Ahn}\ \emph {et~al.}(2012)\citenamefont {Ahn} \emph
  {et~al.}}]{RENO:2012mkc}%
  \BibitemOpen
  \bibfield  {author} {\bibinfo {author} {\bibfnamefont {J.~K.}\ \bibnamefont
  {Ahn}} \emph {et~al.} (\bibinfo {collaboration} {RENO}),\ }\bibfield  {title}
  {\emph {\enquote {\bibinfo {title} {{Observation of Reactor Electron
  Antineutrino Disappearance in the RENO Experiment}},}\ }}\href {\doibase
  10.1103/PhysRevLett.108.191802} {\bibfield  {journal} {\bibinfo  {journal}
  {Phys. Rev. Lett.}\ }\textbf {\bibinfo {volume} {108}},\ \bibinfo {pages}
  {191802} (\bibinfo {year} {2012})},\ \Eprint
  {http://arxiv.org/abs/1204.0626}{arXiv:1204.0626 [hep-ex]}\BibitemShut
  {NoStop}%
\bibitem [{\citenamefont {Abe}\ \emph {et~al.}(2008)\citenamefont {Abe} \emph
  {et~al.}}]{KamLAND:2008dgz}%
  \BibitemOpen
  \bibfield  {author} {\bibinfo {author} {\bibfnamefont {S.}~\bibnamefont
  {Abe}} \emph {et~al.} (\bibinfo {collaboration} {KamLAND}),\ }\bibfield
  {title} {\emph {\enquote {\bibinfo {title} {{Precision Measurement of
  Neutrino Oscillation Parameters with KamLAND}},}\ }}\href {\doibase
  10.1103/PhysRevLett.100.221803} {\bibfield  {journal} {\bibinfo  {journal}
  {Phys. Rev. Lett.}\ }\textbf {\bibinfo {volume} {100}},\ \bibinfo {pages}
  {221803} (\bibinfo {year} {2008})},\ \Eprint
  {http://arxiv.org/abs/0801.4589}{arXiv:0801.4589 [hep-ex]}\BibitemShut
  {NoStop}%
\bibitem [{\citenamefont {Abe}\ \emph {et~al.}(2011{\natexlab{a}})\citenamefont
  {Abe} \emph {et~al.}}]{T2K:2011ypd}%
  \BibitemOpen
  \bibfield  {author} {\bibinfo {author} {\bibfnamefont {K.}~\bibnamefont
  {Abe}} \emph {et~al.} (\bibinfo {collaboration} {T2K}),\ }\bibfield  {title}
  {\emph {\enquote {\bibinfo {title} {{Indication of Electron Neutrino
  Appearance from an Accelerator-produced Off-axis Muon Neutrino Beam}},}\
  }}\href {\doibase 10.1103/PhysRevLett.107.041801} {\bibfield  {journal}
  {\bibinfo  {journal} {Phys. Rev. Lett.}\ }\textbf {\bibinfo {volume} {107}},\
  \bibinfo {pages} {041801} (\bibinfo {year} {2011}{\natexlab{a}})},\ \Eprint
  {http://arxiv.org/abs/1106.2822}{arXiv:1106.2822 [hep-ex]}\BibitemShut
  {NoStop}%
\bibitem [{\citenamefont {Adamson}\ \emph {et~al.}(2011)\citenamefont {Adamson}
  \emph {et~al.}}]{MINOS:2011amj}%
  \BibitemOpen
  \bibfield  {author} {\bibinfo {author} {\bibfnamefont {P.}~\bibnamefont
  {Adamson}} \emph {et~al.} (\bibinfo {collaboration} {MINOS}),\ }\bibfield
  {title} {\emph {\enquote {\bibinfo {title} {{Improved search for
  muon-neutrino to electron-neutrino oscillations in MINOS}},}\ }}\href
  {\doibase 10.1103/PhysRevLett.107.181802} {\bibfield  {journal} {\bibinfo
  {journal} {Phys. Rev. Lett.}\ }\textbf {\bibinfo {volume} {107}},\ \bibinfo
  {pages} {181802} (\bibinfo {year} {2011})},\ \Eprint
  {http://arxiv.org/abs/1108.0015}{arXiv:1108.0015 [hep-ex]}\BibitemShut
  {NoStop}%
\bibitem [{\citenamefont {Tanabashi}\ \emph {et~al.}(2018)\citenamefont
  {Tanabashi} \emph {et~al.}}]{ParticleDataGroup:2018ovx}%
  \BibitemOpen
  \bibfield  {author} {\bibinfo {author} {\bibfnamefont {M.}~\bibnamefont
  {Tanabashi}} \emph {et~al.} (\bibinfo {collaboration} {Particle Data
  Group}),\ }\bibfield  {title} {\emph {\enquote {\bibinfo {title} {{Review of
  Particle Physics}},}\ }}\href {\doibase 10.1103/PhysRevD.98.030001}
  {\bibfield  {journal} {\bibinfo  {journal} {Phys. Rev. D}\ }\textbf {\bibinfo
  {volume} {98}},\ \bibinfo {pages} {030001} (\bibinfo {year}
  {2018})}\BibitemShut {NoStop}%
\bibitem [{\citenamefont {Abbasi}\ \emph {et~al.}(2012)\citenamefont {Abbasi}
  \emph {et~al.}}]{IceCube:2011ucd}%
  \BibitemOpen
  \bibfield  {author} {\bibinfo {author} {\bibfnamefont {R.}~\bibnamefont
  {Abbasi}} \emph {et~al.} (\bibinfo {collaboration} {IceCube}),\ }\bibfield
  {title} {\emph {\enquote {\bibinfo {title} {{The Design and Performance of
  IceCube DeepCore}},}\ }}\href {\doibase 10.1016/j.astropartphys.2012.01.004}
  {\bibfield  {journal} {\bibinfo  {journal} {Astropart. Phys.}\ }\textbf
  {\bibinfo {volume} {35}},\ \bibinfo {pages} {615} (\bibinfo {year} {2012})},\
  \Eprint {http://arxiv.org/abs/1109.6096}{arXiv:1109.6096
  [astro-ph.IM]}\BibitemShut {NoStop}%
\bibitem [{\citenamefont {Abbasi}\ \emph {et~al.}(2023)\citenamefont {Abbasi}
  \emph {et~al.}}]{IceCubeCollaboration:2023wtb}%
  \BibitemOpen
  \bibfield  {author} {\bibinfo {author} {\bibfnamefont {R.}~\bibnamefont
  {Abbasi}} \emph {et~al.} (\bibinfo {collaboration} {(IceCube Collaboration)*,
  IceCube}),\ }\bibfield  {title} {\emph {\enquote {\bibinfo {title}
  {{Measurement of atmospheric neutrino mixing with improved IceCube DeepCore
  calibration and data processing}},}\ }}\href {\doibase
  10.1103/PhysRevD.108.012014} {\bibfield  {journal} {\bibinfo  {journal}
  {Phys. Rev. D}\ }\textbf {\bibinfo {volume} {108}},\ \bibinfo {pages}
  {012014} (\bibinfo {year} {2023})},\ \Eprint
  {http://arxiv.org/abs/2304.12236}{arXiv:2304.12236 [hep-ex]}\BibitemShut
  {NoStop}%
\bibitem [{\citenamefont {Katz}(2013)}]{Katz:2013svu}%
  \BibitemOpen
  \bibfield  {author} {\bibinfo {author} {\bibfnamefont {U.~F.}\ \bibnamefont
  {Katz}} (\bibinfo {collaboration} {KM3NeT}),\ }\bibfield  {title} {\emph
  {\enquote {\bibinfo {title} {{The ORCA Option for KM3NeT}},}\ }}\href@noop {}
  {\  (\bibinfo {year} {2013})},\ \Eprint
  {http://arxiv.org/abs/1402.1022}{arXiv:1402.1022 [astro-ph.IM]}\BibitemShut
  {NoStop}%
\bibitem [{\citenamefont {Adrian-Martinez}\ \emph {et~al.}(2016)\citenamefont
  {Adrian-Martinez} \emph {et~al.}}]{KM3Net:2016zxf}%
  \BibitemOpen
  \bibfield  {author} {\bibinfo {author} {\bibfnamefont {S.}~\bibnamefont
  {Adrian-Martinez}} \emph {et~al.} (\bibinfo {collaboration} {KM3Net}),\
  }\bibfield  {title} {\emph {\enquote {\bibinfo {title} {{Letter of intent for
  KM3NeT 2.0}},}\ }}\href {\doibase 10.1088/0954-3899/43/8/084001} {\bibfield
  {journal} {\bibinfo  {journal} {J. Phys. G}\ }\textbf {\bibinfo {volume}
  {43}},\ \bibinfo {pages} {084001} (\bibinfo {year} {2016})},\ \Eprint
  {http://arxiv.org/abs/1601.07459}{arXiv:1601.07459 [astro-ph.IM]}\BibitemShut
  {NoStop}%
\bibitem [{\citenamefont {Schumann}(2023)}]{psf2023008035}%
  \BibitemOpen
  \bibfield  {author} {\bibinfo {author} {\bibfnamefont {J.}~\bibnamefont
  {Schumann}},\ }\bibfield  {title} {\emph {\enquote {\bibinfo {title}
  {Neutrino Oscillation Measurements with KM3NeT/ORCA},}\ }}\href {\doibase
  10.3390/psf2023008035} {\bibfield  {journal} {\bibinfo  {journal} {Physical
  Sciences Forum}\ }\textbf {\bibinfo {volume} {8}} (\bibinfo {year} {2023}),\
  10.3390/psf2023008035}\BibitemShut {NoStop}%
\bibitem [{\citenamefont {Athar}\ \emph {et~al.}(2006)\citenamefont {Athar}
  \emph {et~al.}}]{INO:2006vde}%
  \BibitemOpen
  \bibfield  {author} {\bibinfo {author} {\bibfnamefont {M.~S.}\ \bibnamefont
  {Athar}} \emph {et~al.} (\bibinfo {collaboration} {INO})\ }(\bibinfo {year}
  {2006})\BibitemShut {NoStop}%
\bibitem [{\citenamefont {Khindri}\ \emph {et~al.}(2022)\citenamefont {Khindri}
  \emph {et~al.}}]{Khindri:2022elz}%
  \BibitemOpen
  \bibfield  {author} {\bibinfo {author} {\bibfnamefont {H.}~\bibnamefont
  {Khindri}} \emph {et~al.},\ }\bibfield  {title} {\emph {\enquote {\bibinfo
  {title} {{Magnetic field measurements on the mini-ICAL detector using Hall
  probes}},}\ }}\href {\doibase 10.1088/1748-0221/17/10/T10006} {\bibfield
  {journal} {\bibinfo  {journal} {JINST}\ }\textbf {\bibinfo {volume} {17}},\
  \bibinfo {pages} {T10006} (\bibinfo {year} {2022})},\ \Eprint
  {http://arxiv.org/abs/2206.15082}{arXiv:2206.15082
  [physics.ins-det]}\BibitemShut {NoStop}%
\bibitem [{\citenamefont {Ahmed}\ \emph {et~al.}(2017)\citenamefont {Ahmed}
  \emph {et~al.}}]{ICAL:2015stm}%
  \BibitemOpen
  \bibfield  {author} {\bibinfo {author} {\bibfnamefont {S.}~\bibnamefont
  {Ahmed}} \emph {et~al.} (\bibinfo {collaboration} {ICAL}),\ }\bibfield
  {title} {\emph {\enquote {\bibinfo {title} {{Physics Potential of the ICAL
  detector at the India-based Neutrino Observatory (INO)}},}\ }}\href {\doibase
  10.1007/s12043-017-1373-4} {\bibfield  {journal} {\bibinfo  {journal}
  {Pramana}\ }\textbf {\bibinfo {volume} {88}},\ \bibinfo {pages} {79}
  (\bibinfo {year} {2017})},\ \Eprint
  {http://arxiv.org/abs/1505.07380}{arXiv:1505.07380
  [physics.ins-det]}\BibitemShut {NoStop}%
\bibitem [{\citenamefont {Abe}\ \emph {et~al.}(2011{\natexlab{b}})\citenamefont
  {Abe} \emph {et~al.}}]{T2K:2011qtm}%
  \BibitemOpen
  \bibfield  {author} {\bibinfo {author} {\bibfnamefont {K.}~\bibnamefont
  {Abe}} \emph {et~al.} (\bibinfo {collaboration} {T2K}),\ }\bibfield  {title}
  {\emph {\enquote {\bibinfo {title} {{The T2K Experiment}},}\ }}\href
  {\doibase 10.1016/j.nima.2011.06.067} {\bibfield  {journal} {\bibinfo
  {journal} {Nucl. Instrum. Meth. A}\ }\textbf {\bibinfo {volume} {659}},\
  \bibinfo {pages} {106} (\bibinfo {year} {2011}{\natexlab{b}})},\ \Eprint
  {http://arxiv.org/abs/1106.1238}{arXiv:1106.1238
  [physics.ins-det]}\BibitemShut {NoStop}%
\bibitem [{\citenamefont {Abe}\ \emph {et~al.}(2023{\natexlab{a}})\citenamefont
  {Abe} \emph {et~al.}}]{T2K:2023mcm}%
  \BibitemOpen
  \bibfield  {author} {\bibinfo {author} {\bibfnamefont {K.}~\bibnamefont
  {Abe}} \emph {et~al.} (\bibinfo {collaboration} {T2K}),\ }\bibfield  {title}
  {\emph {\enquote {\bibinfo {title} {{Updated T2K measurements of muon
  neutrino and antineutrino disappearance using 3.6\texttimes{}1021 protons on
  target}},}\ }}\href {\doibase 10.1103/PhysRevD.108.072011} {\bibfield
  {journal} {\bibinfo  {journal} {Phys. Rev. D}\ }\textbf {\bibinfo {volume}
  {108}},\ \bibinfo {pages} {072011} (\bibinfo {year} {2023}{\natexlab{a}})},\
  \Eprint {http://arxiv.org/abs/2305.09916}{arXiv:2305.09916
  [hep-ex]}\BibitemShut {NoStop}%
\bibitem [{\citenamefont {Ayres}\ \emph {et~al.}(2004)\citenamefont {Ayres}
  \emph {et~al.}}]{NOvA:2004blv}%
  \BibitemOpen
  \bibfield  {author} {\bibinfo {author} {\bibfnamefont {D.~S.}\ \bibnamefont
  {Ayres}} \emph {et~al.} (\bibinfo {collaboration} {NOvA}),\ }\bibfield
  {title} {\emph {\enquote {\bibinfo {title} {{NOvA: Proposal to Build a 30
  Kiloton Off-Axis Detector to Study $\nu_{\mu} \to \nu_e$ Oscillations in the
  NuMI Beamline}},}\ }}\href@noop {} {\  (\bibinfo {year} {2004})},\ \Eprint
  {http://arxiv.org/abs/hep-ex/0503053}{arXiv:hep-ex/0503053}\BibitemShut
  {NoStop}%
\bibitem [{\citenamefont {Acero}\ \emph {et~al.}(2019)\citenamefont {Acero}
  \emph {et~al.}}]{NOvA:2019cyt}%
  \BibitemOpen
  \bibfield  {author} {\bibinfo {author} {\bibfnamefont {M.~A.}\ \bibnamefont
  {Acero}} \emph {et~al.} (\bibinfo {collaboration} {NOvA}),\ }\bibfield
  {title} {\emph {\enquote {\bibinfo {title} {{First Measurement of Neutrino
  Oscillation Parameters using Neutrinos and Antineutrinos by NOvA}},}\ }}\href
  {\doibase 10.1103/PhysRevLett.123.151803} {\bibfield  {journal} {\bibinfo
  {journal} {Phys. Rev. Lett.}\ }\textbf {\bibinfo {volume} {123}},\ \bibinfo
  {pages} {151803} (\bibinfo {year} {2019})},\ \Eprint
  {http://arxiv.org/abs/1906.04907}{arXiv:1906.04907 [hep-ex]}\BibitemShut
  {NoStop}%
\bibitem [{\citenamefont {Adamson}\ \emph {et~al.}(2016)\citenamefont {Adamson}
  \emph {et~al.}}]{Adamson:2015dkw}%
  \BibitemOpen
  \bibfield  {author} {\bibinfo {author} {\bibfnamefont {P.}~\bibnamefont
  {Adamson}} \emph {et~al.},\ }\bibfield  {title} {\emph {\enquote {\bibinfo
  {title} {{The NuMI Neutrino Beam}},}\ }}\href {\doibase
  10.1016/j.nima.2015.08.063} {\bibfield  {journal} {\bibinfo  {journal} {Nucl.
  Instrum. Meth. A}\ }\textbf {\bibinfo {volume} {806}},\ \bibinfo {pages}
  {279} (\bibinfo {year} {2016})},\ \Eprint
  {http://arxiv.org/abs/1507.06690}{arXiv:1507.06690
  [physics.acc-ph]}\BibitemShut {NoStop}%
\bibitem [{\citenamefont {Acero}\ \emph {et~al.}(2022)\citenamefont {Acero}
  \emph {et~al.}}]{NOvA:2021nfi}%
  \BibitemOpen
  \bibfield  {author} {\bibinfo {author} {\bibfnamefont {M.~A.}\ \bibnamefont
  {Acero}} \emph {et~al.} (\bibinfo {collaboration} {NOvA}),\ }\bibfield
  {title} {\emph {\enquote {\bibinfo {title} {{Improved measurement of neutrino
  oscillation parameters by the NOvA experiment}},}\ }}\href {\doibase
  10.1103/PhysRevD.106.032004} {\bibfield  {journal} {\bibinfo  {journal}
  {Phys. Rev. D}\ }\textbf {\bibinfo {volume} {106}},\ \bibinfo {pages}
  {032004} (\bibinfo {year} {2022})},\ \Eprint
  {http://arxiv.org/abs/2108.08219}{arXiv:2108.08219 [hep-ex]}\BibitemShut
  {NoStop}%
\bibitem [{\citenamefont {Abi}\ \emph {et~al.}(2020{\natexlab{a}})\citenamefont
  {Abi} \emph {et~al.}}]{DUNE:2020lwj}%
  \BibitemOpen
  \bibfield  {author} {\bibinfo {author} {\bibfnamefont {B.}~\bibnamefont
  {Abi}} \emph {et~al.} (\bibinfo {collaboration} {DUNE}),\ }\bibfield  {title}
  {\emph {\enquote {\bibinfo {title} {{Deep Underground Neutrino Experiment
  (DUNE), Far Detector Technical Design Report, Volume I Introduction to
  DUNE}},}\ }}\href {\doibase 10.1088/1748-0221/15/08/T08008} {\bibfield
  {journal} {\bibinfo  {journal} {JINST}\ }\textbf {\bibinfo {volume} {15}},\
  \bibinfo {pages} {T08008} (\bibinfo {year} {2020}{\natexlab{a}})},\ \Eprint
  {http://arxiv.org/abs/2002.02967}{arXiv:2002.02967
  [physics.ins-det]}\BibitemShut {NoStop}%
\bibitem [{\citenamefont {Abed~Abud}\ \emph {et~al.}(2021)\citenamefont
  {Abed~Abud} \emph {et~al.}}]{DUNE:2021tad}%
  \BibitemOpen
  \bibfield  {author} {\bibinfo {author} {\bibfnamefont {A.}~\bibnamefont
  {Abed~Abud}} \emph {et~al.} (\bibinfo {collaboration} {DUNE}),\ }\bibfield
  {title} {\emph {\enquote {\bibinfo {title} {{Deep Underground Neutrino
  Experiment (DUNE) Near Detector Conceptual Design Report}},}\ }}\href@noop {}
  {\  (\bibinfo {year} {2021})},\ \Eprint
  {http://arxiv.org/abs/2103.13910}{arXiv:2103.13910
  [physics.ins-det]}\BibitemShut {NoStop}%
\bibitem [{\citenamefont {Abi}\ \emph {et~al.}(2020{\natexlab{b}})\citenamefont
  {Abi} \emph {et~al.}}]{DUNE:2020jqi}%
  \BibitemOpen
  \bibfield  {author} {\bibinfo {author} {\bibfnamefont {B.}~\bibnamefont
  {Abi}} \emph {et~al.} (\bibinfo {collaboration} {DUNE}),\ }\bibfield  {title}
  {\emph {\enquote {\bibinfo {title} {{Long-baseline neutrino oscillation
  physics potential of the DUNE experiment}},}\ }}\href {\doibase
  10.1140/epjc/s10052-020-08456-z} {\bibfield  {journal} {\bibinfo  {journal}
  {Eur. Phys. J. C}\ }\textbf {\bibinfo {volume} {80}},\ \bibinfo {pages} {978}
  (\bibinfo {year} {2020}{\natexlab{b}})},\ \Eprint
  {http://arxiv.org/abs/2006.16043}{arXiv:2006.16043 [hep-ex]}\BibitemShut
  {NoStop}%
\bibitem [{\citenamefont {Abe}\ \emph {et~al.}(2014)\citenamefont {Abe} \emph
  {et~al.}}]{Hyper-KamiokandeWorkingGroup:2014czz}%
  \BibitemOpen
  \bibfield  {author} {\bibinfo {author} {\bibfnamefont {K.}~\bibnamefont
  {Abe}} \emph {et~al.} (\bibinfo {collaboration} {Hyper-Kamiokande Working
  Group})\ }(\bibinfo {year} {2014})\ \Eprint
  {http://arxiv.org/abs/1412.4673}{arXiv:1412.4673
  [physics.ins-det]}\BibitemShut {NoStop}%
\bibitem [{\citenamefont {Abe}\ \emph {et~al.}(2018{\natexlab{a}})\citenamefont
  {Abe} \emph {et~al.}}]{Hyper-Kamiokande:2018ofw}%
  \BibitemOpen
  \bibfield  {author} {\bibinfo {author} {\bibfnamefont {K.}~\bibnamefont
  {Abe}} \emph {et~al.} (\bibinfo {collaboration} {Hyper-Kamiokande}),\
  }\bibfield  {title} {\emph {\enquote {\bibinfo {title} {{Hyper-Kamiokande
  Design Report}},}\ }}\href@noop {} {\  (\bibinfo {year}
  {2018}{\natexlab{a}})},\ \Eprint
  {http://arxiv.org/abs/1805.04163}{arXiv:1805.04163
  [physics.ins-det]}\BibitemShut {NoStop}%
\bibitem [{\citenamefont {Abe}\ \emph {et~al.}(2015{\natexlab{a}})\citenamefont
  {Abe} \emph {et~al.}}]{Hyper-KamiokandeProto-:2015xww}%
  \BibitemOpen
  \bibfield  {author} {\bibinfo {author} {\bibfnamefont {K.}~\bibnamefont
  {Abe}} \emph {et~al.} (\bibinfo {collaboration} {Hyper-Kamiokande Proto-}),\
  }\bibfield  {title} {\emph {\enquote {\bibinfo {title} {{Physics potential of
  a long-baseline neutrino oscillation experiment using a J-PARC neutrino beam
  and Hyper-Kamiokande}},}\ }}\href {\doibase 10.1093/ptep/ptv061} {\bibfield
  {journal} {\bibinfo  {journal} {PTEP}\ }\textbf {\bibinfo {volume} {2015}},\
  \bibinfo {pages} {053C02} (\bibinfo {year} {2015}{\natexlab{a}})},\ \Eprint
  {http://arxiv.org/abs/1502.05199}{arXiv:1502.05199 [hep-ex]}\BibitemShut
  {NoStop}%
\bibitem [{\citenamefont {Abbasi}\ \emph {et~al.}(2021)\citenamefont {Abbasi}
  \emph {et~al.}}]{IceCube:2020wum}%
  \BibitemOpen
  \bibfield  {author} {\bibinfo {author} {\bibfnamefont {R.}~\bibnamefont
  {Abbasi}} \emph {et~al.} (\bibinfo {collaboration} {IceCube}),\ }\bibfield
  {title} {\emph {\enquote {\bibinfo {title} {{The IceCube high-energy starting
  event sample: Description and flux characterization with 7.5 years of
  data}},}\ }}\href {\doibase 10.1103/PhysRevD.104.022002} {\bibfield
  {journal} {\bibinfo  {journal} {Phys. Rev. D}\ }\textbf {\bibinfo {volume}
  {104}},\ \bibinfo {pages} {022002} (\bibinfo {year} {2021})},\ \Eprint
  {http://arxiv.org/abs/2011.03545}{arXiv:2011.03545 [astro-ph.HE]}\BibitemShut
  {NoStop}%
\bibitem [{\citenamefont {Aartsen}\ \emph
  {et~al.}(2018{\natexlab{a}})\citenamefont {Aartsen} \emph
  {et~al.}}]{IceCube:2018dnn}%
  \BibitemOpen
  \bibfield  {author} {\bibinfo {author} {\bibfnamefont {M.~G.}\ \bibnamefont
  {Aartsen}} \emph {et~al.} (\bibinfo {collaboration} {IceCube, Fermi-LAT,
  MAGIC, AGILE, ASAS-SN, HAWC, H.E.S.S., INTEGRAL, Kanata, Kiso, Kapteyn,
  Liverpool Telescope, Subaru, Swift NuSTAR, VERITAS, VLA/17B-403}),\
  }\bibfield  {title} {\emph {\enquote {\bibinfo {title} {{Multimessenger
  observations of a flaring blazar coincident with high-energy neutrino
  IceCube-170922A}},}\ }}\href {\doibase 10.1126/science.aat1378} {\bibfield
  {journal} {\bibinfo  {journal} {Science}\ }\textbf {\bibinfo {volume}
  {361}},\ \bibinfo {pages} {eaat1378} (\bibinfo {year}
  {2018}{\natexlab{a}})},\ \Eprint
  {http://arxiv.org/abs/1807.08816}{arXiv:1807.08816 [astro-ph.HE]}\BibitemShut
  {NoStop}%
\bibitem [{\citenamefont {Aartsen}\ \emph
  {et~al.}(2018{\natexlab{b}})\citenamefont {Aartsen} \emph
  {et~al.}}]{IceCube:2018cha}%
  \BibitemOpen
  \bibfield  {author} {\bibinfo {author} {\bibfnamefont {M.~G.}\ \bibnamefont
  {Aartsen}} \emph {et~al.} (\bibinfo {collaboration} {IceCube}),\ }\bibfield
  {title} {\emph {\enquote {\bibinfo {title} {{Neutrino emission from the
  direction of the blazar TXS 0506+056 prior to the IceCube-170922A alert}},}\
  }}\href {\doibase 10.1126/science.aat2890} {\bibfield  {journal} {\bibinfo
  {journal} {Science}\ }\textbf {\bibinfo {volume} {361}},\ \bibinfo {pages}
  {147} (\bibinfo {year} {2018}{\natexlab{b}})},\ \Eprint
  {http://arxiv.org/abs/1807.08794}{arXiv:1807.08794 [astro-ph.HE]}\BibitemShut
  {NoStop}%
\bibitem [{\citenamefont {Abbasi}\ \emph
  {et~al.}(2022{\natexlab{a}})\citenamefont {Abbasi} \emph
  {et~al.}}]{IceCube:2022der}%
  \BibitemOpen
  \bibfield  {author} {\bibinfo {author} {\bibfnamefont {R.}~\bibnamefont
  {Abbasi}} \emph {et~al.} (\bibinfo {collaboration} {IceCube}),\ }\bibfield
  {title} {\emph {\enquote {\bibinfo {title} {{Evidence for neutrino emission
  from the nearby active galaxy NGC 1068}},}\ }}\href {\doibase
  10.1126/science.abg3395} {\bibfield  {journal} {\bibinfo  {journal}
  {Science}\ }\textbf {\bibinfo {volume} {378}},\ \bibinfo {pages} {538}
  (\bibinfo {year} {2022}{\natexlab{a}})},\ \Eprint
  {http://arxiv.org/abs/2211.09972}{arXiv:2211.09972 [astro-ph.HE]}\BibitemShut
  {NoStop}%
\bibitem [{\citenamefont {D.}\ \emph {et~al.}(2021)\citenamefont {D.} \emph
  {et~al.}}]{Baikal-GVD:2020xgh}%
  \BibitemOpen
  \bibfield  {author} {\bibinfo {author} {\bibfnamefont {A.~A.}\ \bibnamefont
  {D.}} \emph {et~al.} (\bibinfo {collaboration} {Baikal-GVD}),\ }\bibfield
  {title} {\emph {\enquote {\bibinfo {title} {{Baikal-GVD: status and first
  results}},}\ }}\href {\doibase 10.22323/1.390.0606} {\bibfield  {journal}
  {\bibinfo  {journal} {PoS}\ }\textbf {\bibinfo {volume} {ICHEP2020}},\
  \bibinfo {pages} {606} (\bibinfo {year} {2021})},\ \Eprint
  {http://arxiv.org/abs/2012.03373}{arXiv:2012.03373 [astro-ph.HE]}\BibitemShut
  {NoStop}%
\bibitem [{\citenamefont {Avrorin}\ \emph {et~al.}(2021)\citenamefont {Avrorin}
  \emph {et~al.}}]{Baikal-GVD:2020irv}%
  \BibitemOpen
  \bibfield  {author} {\bibinfo {author} {\bibfnamefont {A.~D.}\ \bibnamefont
  {Avrorin}} \emph {et~al.} (\bibinfo {collaboration} {Baikal-GVD}),\
  }\bibfield  {title} {\emph {\enquote {\bibinfo {title} {{High-Energy Neutrino
  Astronomy and the Baikal-GVD Neutrino Telescope}},}\ }}\href {\doibase
  10.1134/S1063778821040062} {\bibfield  {journal} {\bibinfo  {journal} {Phys.
  Atom. Nucl.}\ }\textbf {\bibinfo {volume} {84}},\ \bibinfo {pages} {513}
  (\bibinfo {year} {2021})},\ \Eprint
  {http://arxiv.org/abs/2011.09209}{arXiv:2011.09209 [astro-ph.HE]}\BibitemShut
  {NoStop}%
\bibitem [{\citenamefont {Aiello}\ \emph {et~al.}(2024)\citenamefont {Aiello}
  \emph {et~al.}}]{Aiello:2024jbp}%
  \BibitemOpen
  \bibfield  {author} {\bibinfo {author} {\bibfnamefont {S.}~\bibnamefont
  {Aiello}} \emph {et~al.},\ }\bibfield  {title} {\emph {\enquote {\bibinfo
  {title} {{Astronomy potential of KM3NeT/ARCA}},}\ }}\href@noop {} {\
  (\bibinfo {year} {2024})},\ \Eprint
  {http://arxiv.org/abs/2402.08363}{arXiv:2402.08363 [astro-ph.HE]}\BibitemShut
  {NoStop}%
\bibitem [{\citenamefont {Agostini}\ \emph {et~al.}(2020)\citenamefont
  {Agostini} \emph {et~al.}}]{P-ONE:2020ljt}%
  \BibitemOpen
  \bibfield  {author} {\bibinfo {author} {\bibfnamefont {M.}~\bibnamefont
  {Agostini}} \emph {et~al.} (\bibinfo {collaboration} {P-ONE}),\ }\bibfield
  {title} {\emph {\enquote {\bibinfo {title} {{The Pacific Ocean Neutrino
  Experiment}},}\ }}\href {\doibase 10.1038/s41550-020-1182-4} {\bibfield
  {journal} {\bibinfo  {journal} {Nature Astron.}\ }\textbf {\bibinfo {volume}
  {4}},\ \bibinfo {pages} {913} (\bibinfo {year} {2020})},\ \Eprint
  {http://arxiv.org/abs/2005.09493}{arXiv:2005.09493 [astro-ph.HE]}\BibitemShut
  {NoStop}%
\bibitem [{\citenamefont {de~Salas}\ \emph {et~al.}(2021)\citenamefont
  {de~Salas}, \citenamefont {Forero}, \citenamefont {Gariazzo}, \citenamefont
  {Mart\'\i{}nez-Mirav\'e}, \citenamefont {Mena}, \citenamefont {Ternes},
  \citenamefont {T\'ortola},\ and\ \citenamefont {Valle}}]{deSalas:2020pgw}%
  \BibitemOpen
  \bibfield  {author} {\bibinfo {author} {\bibfnamefont {P.~F.}\ \bibnamefont
  {de~Salas}}, \bibinfo {author} {\bibfnamefont {D.~V.}\ \bibnamefont
  {Forero}}, \bibinfo {author} {\bibfnamefont {S.}~\bibnamefont {Gariazzo}},
  \bibinfo {author} {\bibfnamefont {P.}~\bibnamefont {Mart\'\i{}nez-Mirav\'e}},
  \bibinfo {author} {\bibfnamefont {O.}~\bibnamefont {Mena}}, \bibinfo {author}
  {\bibfnamefont {C.~A.}\ \bibnamefont {Ternes}}, \bibinfo {author}
  {\bibfnamefont {M.}~\bibnamefont {T\'ortola}}, \ and\ \bibinfo {author}
  {\bibfnamefont {J.~W.~F.}\ \bibnamefont {Valle}},\ }\bibfield  {title} {\emph
  {\enquote {\bibinfo {title} {{2020 global reassessment of the neutrino
  oscillation picture}},}\ }}\href {\doibase 10.1007/JHEP02(2021)071}
  {\bibfield  {journal} {\bibinfo  {journal} {JHEP}\ }\textbf {\bibinfo
  {volume} {02}},\ \bibinfo {pages} {071} (\bibinfo {year} {2021})},\ \Eprint
  {http://arxiv.org/abs/2006.11237}{arXiv:2006.11237 [hep-ph]}\BibitemShut
  {NoStop}%
\bibitem [{\citenamefont {Abusleme}\ \emph {et~al.}(2022)\citenamefont
  {Abusleme} \emph {et~al.}}]{JUNO:2022mxj}%
  \BibitemOpen
  \bibfield  {author} {\bibinfo {author} {\bibfnamefont {A.}~\bibnamefont
  {Abusleme}} \emph {et~al.} (\bibinfo {collaboration} {JUNO}),\ }\bibfield
  {title} {\emph {\enquote {\bibinfo {title} {{Sub-percent precision
  measurement of neutrino oscillation parameters with JUNO}},}\ }}\href
  {\doibase 10.1088/1674-1137/ac8bc9} {\bibfield  {journal} {\bibinfo
  {journal} {Chin. Phys. C}\ }\textbf {\bibinfo {volume} {46}},\ \bibinfo
  {pages} {123001} (\bibinfo {year} {2022})},\ \Eprint
  {http://arxiv.org/abs/2204.13249}{arXiv:2204.13249 [hep-ex]}\BibitemShut
  {NoStop}%
\bibitem [{\citenamefont {Kaether}\ \emph {et~al.}(2010)\citenamefont
  {Kaether}, \citenamefont {Hampel}, \citenamefont {Heusser}, \citenamefont
  {Kiko},\ and\ \citenamefont {Kirsten}}]{Kaether:2010ag}%
  \BibitemOpen
  \bibfield  {author} {\bibinfo {author} {\bibfnamefont {F.}~\bibnamefont
  {Kaether}}, \bibinfo {author} {\bibfnamefont {W.}~\bibnamefont {Hampel}},
  \bibinfo {author} {\bibfnamefont {G.}~\bibnamefont {Heusser}}, \bibinfo
  {author} {\bibfnamefont {J.}~\bibnamefont {Kiko}}, \ and\ \bibinfo {author}
  {\bibfnamefont {T.}~\bibnamefont {Kirsten}},\ }\bibfield  {title} {\emph
  {\enquote {\bibinfo {title} {{Reanalysis of the GALLEX solar neutrino flux
  and source experiments}},}\ }}\href {\doibase 10.1016/j.physletb.2010.01.030}
  {\bibfield  {journal} {\bibinfo  {journal} {Phys. Lett. B}\ }\textbf
  {\bibinfo {volume} {685}},\ \bibinfo {pages} {47} (\bibinfo {year} {2010})},\
  \Eprint {http://arxiv.org/abs/1001.2731}{arXiv:1001.2731
  [hep-ex]}\BibitemShut {NoStop}%
\bibitem [{\citenamefont {Bellini}\ \emph {et~al.}(2014)\citenamefont {Bellini}
  \emph {et~al.}}]{Borexino:2013zhu}%
  \BibitemOpen
  \bibfield  {author} {\bibinfo {author} {\bibfnamefont {G.}~\bibnamefont
  {Bellini}} \emph {et~al.} (\bibinfo {collaboration} {Borexino}),\ }\bibfield
  {title} {\emph {\enquote {\bibinfo {title} {{Final results of Borexino
  Phase-I on low energy solar neutrino spectroscopy}},}\ }}\href {\doibase
  10.1103/PhysRevD.89.112007} {\bibfield  {journal} {\bibinfo  {journal} {Phys.
  Rev. D}\ }\textbf {\bibinfo {volume} {89}},\ \bibinfo {pages} {112007}
  (\bibinfo {year} {2014})},\ \Eprint
  {http://arxiv.org/abs/1308.0443}{arXiv:1308.0443 [hep-ex]}\BibitemShut
  {NoStop}%
\bibitem [{\citenamefont {Abe}\ \emph {et~al.}(2011{\natexlab{c}})\citenamefont
  {Abe} \emph {et~al.}}]{Super-Kamiokande:2010tar}%
  \BibitemOpen
  \bibfield  {author} {\bibinfo {author} {\bibfnamefont {K.}~\bibnamefont
  {Abe}} \emph {et~al.} (\bibinfo {collaboration} {Super-Kamiokande}),\
  }\bibfield  {title} {\emph {\enquote {\bibinfo {title} {{Solar neutrino
  results in Super-Kamiokande-III}},}\ }}\href {\doibase
  10.1103/PhysRevD.83.052010} {\bibfield  {journal} {\bibinfo  {journal} {Phys.
  Rev. D}\ }\textbf {\bibinfo {volume} {83}},\ \bibinfo {pages} {052010}
  (\bibinfo {year} {2011}{\natexlab{c}})},\ \Eprint
  {http://arxiv.org/abs/1010.0118}{arXiv:1010.0118 [hep-ex]}\BibitemShut
  {NoStop}%
\bibitem [{\citenamefont {Gando}\ \emph {et~al.}(2011)\citenamefont {Gando}
  \emph {et~al.}}]{KamLAND:2010fvi}%
  \BibitemOpen
  \bibfield  {author} {\bibinfo {author} {\bibfnamefont {A.}~\bibnamefont
  {Gando}} \emph {et~al.} (\bibinfo {collaboration} {KamLAND}),\ }\bibfield
  {title} {\emph {\enquote {\bibinfo {title} {{Constraints on $\theta_{13}$
  from A Three-Flavor Oscillation Analysis of Reactor Antineutrinos at
  KamLAND}},}\ }}\href {\doibase 10.1103/PhysRevD.83.052002} {\bibfield
  {journal} {\bibinfo  {journal} {Phys. Rev. D}\ }\textbf {\bibinfo {volume}
  {83}},\ \bibinfo {pages} {052002} (\bibinfo {year} {2011})},\ \Eprint
  {http://arxiv.org/abs/1009.4771}{arXiv:1009.4771 [hep-ex]}\BibitemShut
  {NoStop}%
\bibitem [{\citenamefont {Abe}\ \emph {et~al.}(2023{\natexlab{b}})\citenamefont
  {Abe} \emph {et~al.}}]{T2K:2023smv}%
  \BibitemOpen
  \bibfield  {author} {\bibinfo {author} {\bibfnamefont {K.}~\bibnamefont
  {Abe}} \emph {et~al.} (\bibinfo {collaboration} {T2K}),\ }\bibfield  {title}
  {\emph {\enquote {\bibinfo {title} {{Measurements of neutrino oscillation
  parameters from the T2K experiment using $3.6\times 10^{21}$ protons on
  target}},}\ }}\href {\doibase 10.1140/epjc/s10052-023-11819-x} {\bibfield
  {journal} {\bibinfo  {journal} {Eur. Phys. J. C}\ }\textbf {\bibinfo {volume}
  {83}},\ \bibinfo {pages} {782} (\bibinfo {year} {2023}{\natexlab{b}})},\
  \Eprint {http://arxiv.org/abs/2303.03222}{arXiv:2303.03222
  [hep-ex]}\BibitemShut {NoStop}%
\bibitem [{\citenamefont {Fernandez}(2021)}]{pablo_fernandez_2021_5779075}%
  \BibitemOpen
  \bibfield  {author} {\bibinfo {author} {\bibfnamefont {P.}~\bibnamefont
  {Fernandez}} (\bibinfo {collaboration} {Super-Kamiokande}),\ }\href {\doibase
  10.5281/zenodo.5779075} {\enquote {\bibinfo {title} {{Atmospheric neutrino
  oscillations with Super- Kamiokande and prospects for SuperK-Gd}},}\ }
  (\bibinfo {year} {2021}),\ \bibinfo {note} {" XIX International Workshop on
  Neutrino Telescopes"}\BibitemShut {NoStop}%
\bibitem [{\citenamefont {Adamson}\ \emph {et~al.}(2020)\citenamefont {Adamson}
  \emph {et~al.}}]{MINOS:2020llm}%
  \BibitemOpen
  \bibfield  {author} {\bibinfo {author} {\bibfnamefont {P.}~\bibnamefont
  {Adamson}} \emph {et~al.} (\bibinfo {collaboration} {MINOS+}),\ }\bibfield
  {title} {\emph {\enquote {\bibinfo {title} {{Precision Constraints for
  Three-Flavor Neutrino Oscillations from the Full MINOS+ and MINOS
  Dataset}},}\ }}\href {\doibase 10.1103/PhysRevLett.125.131802} {\bibfield
  {journal} {\bibinfo  {journal} {Phys. Rev. Lett.}\ }\textbf {\bibinfo
  {volume} {125}},\ \bibinfo {pages} {131802} (\bibinfo {year} {2020})},\
  \Eprint {http://arxiv.org/abs/2006.15208}{arXiv:2006.15208
  [hep-ex]}\BibitemShut {NoStop}%
\bibitem [{\citenamefont {Abbasi}\ \emph {et~al.}(2024)\citenamefont {Abbasi}
  \emph {et~al.}}]{IceCube:2024xjj}%
  \BibitemOpen
  \bibfield  {author} {\bibinfo {author} {\bibfnamefont {R.}~\bibnamefont
  {Abbasi}} \emph {et~al.} (\bibinfo {collaboration} {IceCube}),\ }\bibfield
  {title} {\emph {\enquote {\bibinfo {title} {{Measurement of atmospheric
  neutrino oscillation parameters using convolutional neural networks with 9.3
  years of data in IceCube DeepCore}},}\ }}\href@noop {} {\  (\bibinfo {year}
  {2024})},\ \Eprint {http://arxiv.org/abs/2405.02163}{arXiv:2405.02163
  [hep-ex]}\BibitemShut {NoStop}%
\bibitem [{\citenamefont {Adey}\ \emph {et~al.}(2018)\citenamefont {Adey} \emph
  {et~al.}}]{DayaBay:2018yms}%
  \BibitemOpen
  \bibfield  {author} {\bibinfo {author} {\bibfnamefont {D.}~\bibnamefont
  {Adey}} \emph {et~al.} (\bibinfo {collaboration} {Daya Bay}),\ }\bibfield
  {title} {\emph {\enquote {\bibinfo {title} {{Measurement of the Electron
  Antineutrino Oscillation with 1958 Days of Operation at Daya Bay}},}\ }}\href
  {\doibase 10.1103/PhysRevLett.121.241805} {\bibfield  {journal} {\bibinfo
  {journal} {Phys. Rev. Lett.}\ }\textbf {\bibinfo {volume} {121}},\ \bibinfo
  {pages} {241805} (\bibinfo {year} {2018})},\ \Eprint
  {http://arxiv.org/abs/1809.02261}{arXiv:1809.02261 [hep-ex]}\BibitemShut
  {NoStop}%
\bibitem [{\citenamefont {Yoon}\ \emph {et~al.}(2021)\citenamefont {Yoon} \emph
  {et~al.}}]{RENO:2020dxd}%
  \BibitemOpen
  \bibfield  {author} {\bibinfo {author} {\bibfnamefont {S.~G.}\ \bibnamefont
  {Yoon}} \emph {et~al.} (\bibinfo {collaboration} {RENO}),\ }\bibfield
  {title} {\emph {\enquote {\bibinfo {title} {{Measurement of reactor
  antineutrino flux and spectrum at RENO}},}\ }}\href {\doibase
  10.1103/PhysRevD.104.L111301} {\bibfield  {journal} {\bibinfo  {journal}
  {Phys. Rev. D}\ }\textbf {\bibinfo {volume} {104}},\ \bibinfo {pages}
  {L111301} (\bibinfo {year} {2021})},\ \Eprint
  {http://arxiv.org/abs/2010.14989}{arXiv:2010.14989 [hep-ex]}\BibitemShut
  {NoStop}%
\bibitem [{\citenamefont {Himmel}()}]{Himmel:2020}%
  \BibitemOpen
  \bibfield  {author} {\bibinfo {author} {\bibfnamefont {A.}~\bibnamefont
  {Himmel}},\ }\bibinfo {note} {talk given at the Neutrino 2020 meeting on
  July, 2nd,
  2020,\url{https://indico.fnal.gov/event/43209/contributions/187840/attachments/130740/159597/NOvA-Oscilations-NEUTRINO2020.pdf}}\BibitemShut
  {NoStop}%
\bibitem [{\citenamefont {Dunne}()}]{Dunne:2020}%
  \BibitemOpen
  \bibfield  {author} {\bibinfo {author} {\bibfnamefont {P.}~\bibnamefont
  {Dunne}},\ }\bibinfo {note} {talk given at the Neutrino 2020 meeting on July,
  2nd,
  2020,\url{https://indico.fnal.gov/event/43209/contributions/187830/attachments/129636/159603/T2K_Neutrino2020.pdf}}\BibitemShut
  {NoStop}%
\bibitem [{\citenamefont {Abe}\ \emph {et~al.}(2021)\citenamefont {Abe} \emph
  {et~al.}}]{T2K:2021xwb}%
  \BibitemOpen
  \bibfield  {author} {\bibinfo {author} {\bibfnamefont {K.}~\bibnamefont
  {Abe}} \emph {et~al.} (\bibinfo {collaboration} {T2K}),\ }\bibfield  {title}
  {\emph {\enquote {\bibinfo {title} {{Improved constraints on neutrino mixing
  from the T2K experiment with $\mathbf{3.13\times10^{21}}$ protons on
  target}},}\ }}\href {\doibase 10.1103/PhysRevD.103.112008} {\bibfield
  {journal} {\bibinfo  {journal} {Phys. Rev. D}\ }\textbf {\bibinfo {volume}
  {103}},\ \bibinfo {pages} {112008} (\bibinfo {year} {2021})},\ \Eprint
  {http://arxiv.org/abs/2101.03779}{arXiv:2101.03779 [hep-ex]}\BibitemShut
  {NoStop}%
\bibitem [{NuF()}]{NuFIT}%
  \BibitemOpen
  \href {http://www.nu-fit.org/} {}\bibinfo {note} {NuFIT 5.0 (2020),
  http://www.nu-fit.org/}\BibitemShut {NoStop}%
\bibitem [{\citenamefont {Esteban}\ \emph {et~al.}(2020)\citenamefont
  {Esteban}, \citenamefont {Gonzalez-Garcia}, \citenamefont {Maltoni},
  \citenamefont {Schwetz},\ and\ \citenamefont {Zhou}}]{Esteban:2020cvm}%
  \BibitemOpen
  \bibfield  {author} {\bibinfo {author} {\bibfnamefont {I.}~\bibnamefont
  {Esteban}}, \bibinfo {author} {\bibfnamefont {M.~C.}\ \bibnamefont
  {Gonzalez-Garcia}}, \bibinfo {author} {\bibfnamefont {M.}~\bibnamefont
  {Maltoni}}, \bibinfo {author} {\bibfnamefont {T.}~\bibnamefont {Schwetz}}, \
  and\ \bibinfo {author} {\bibfnamefont {A.}~\bibnamefont {Zhou}},\ }\bibfield
  {title} {\emph {\enquote {\bibinfo {title} {{The fate of hints: updated
  global analysis of three-flavor neutrino oscillations}},}\ }}\href {\doibase
  10.1007/JHEP09(2020)178} {\bibfield  {journal} {\bibinfo  {journal} {JHEP}\
  }\textbf {\bibinfo {volume} {09}},\ \bibinfo {pages} {178} (\bibinfo {year}
  {2020})},\ \Eprint {http://arxiv.org/abs/2007.14792}{arXiv:2007.14792
  [hep-ph]}\BibitemShut {NoStop}%
\bibitem [{\citenamefont {Capozzi}\ \emph {et~al.}(2021)\citenamefont
  {Capozzi}, \citenamefont {Di~Valentino}, \citenamefont {Lisi}, \citenamefont
  {Marrone}, \citenamefont {Melchiorri},\ and\ \citenamefont
  {Palazzo}}]{Capozzi:2021fjo}%
  \BibitemOpen
  \bibfield  {author} {\bibinfo {author} {\bibfnamefont {F.}~\bibnamefont
  {Capozzi}}, \bibinfo {author} {\bibfnamefont {E.}~\bibnamefont
  {Di~Valentino}}, \bibinfo {author} {\bibfnamefont {E.}~\bibnamefont {Lisi}},
  \bibinfo {author} {\bibfnamefont {A.}~\bibnamefont {Marrone}}, \bibinfo
  {author} {\bibfnamefont {A.}~\bibnamefont {Melchiorri}}, \ and\ \bibinfo
  {author} {\bibfnamefont {A.}~\bibnamefont {Palazzo}},\ }\bibfield  {title}
  {\emph {\enquote {\bibinfo {title} {{Unfinished fabric of the three neutrino
  paradigm}},}\ }}\href {\doibase 10.1103/PhysRevD.104.083031} {\bibfield
  {journal} {\bibinfo  {journal} {Phys. Rev. D}\ }\textbf {\bibinfo {volume}
  {104}},\ \bibinfo {pages} {083031} (\bibinfo {year} {2021})},\ \Eprint
  {http://arxiv.org/abs/2107.00532}{arXiv:2107.00532 [hep-ph]}\BibitemShut
  {NoStop}%
\bibitem [{\citenamefont {Mohapatra}\ and\ \citenamefont
  {Smirnov}(2006)}]{Mohapatra:2006gs}%
  \BibitemOpen
  \bibfield  {author} {\bibinfo {author} {\bibfnamefont {R.~N.}\ \bibnamefont
  {Mohapatra}}\ and\ \bibinfo {author} {\bibfnamefont {A.~Y.}\ \bibnamefont
  {Smirnov}},\ }\bibfield  {title} {\emph {\enquote {\bibinfo {title}
  {{Neutrino Mass and New Physics}},}\ }}\href {\doibase
  10.1146/annurev.nucl.56.080805.140534} {\bibfield  {journal} {\bibinfo
  {journal} {Ann. Rev. Nucl. Part. Sci.}\ }\textbf {\bibinfo {volume} {56}},\
  \bibinfo {pages} {569} (\bibinfo {year} {2006})},\ \Eprint
  {http://arxiv.org/abs/hep-ph/0603118}{arXiv:hep-ph/0603118}\BibitemShut
  {NoStop}%
\bibitem [{\citenamefont {Albright}\ and\ \citenamefont
  {Chen}(2006)}]{Albright:2006cw}%
  \BibitemOpen
  \bibfield  {author} {\bibinfo {author} {\bibfnamefont {C.~H.}\ \bibnamefont
  {Albright}}\ and\ \bibinfo {author} {\bibfnamefont {M.-C.}\ \bibnamefont
  {Chen}},\ }\bibfield  {title} {\emph {\enquote {\bibinfo {title} {{Model
  Predictions for Neutrino Oscillation Parameters}},}\ }}\href {\doibase
  10.1103/PhysRevD.74.113006} {\bibfield  {journal} {\bibinfo  {journal} {Phys.
  Rev. D}\ }\textbf {\bibinfo {volume} {74}},\ \bibinfo {pages} {113006}
  (\bibinfo {year} {2006})},\ \Eprint
  {http://arxiv.org/abs/hep-ph/0608137}{arXiv:hep-ph/0608137}\BibitemShut
  {NoStop}%
\bibitem [{\citenamefont {Albright}\ \emph {et~al.}(2010)\citenamefont
  {Albright}, \citenamefont {Dueck},\ and\ \citenamefont
  {Rodejohann}}]{Albright:2010ap}%
  \BibitemOpen
  \bibfield  {author} {\bibinfo {author} {\bibfnamefont {C.~H.}\ \bibnamefont
  {Albright}}, \bibinfo {author} {\bibfnamefont {A.}~\bibnamefont {Dueck}}, \
  and\ \bibinfo {author} {\bibfnamefont {W.}~\bibnamefont {Rodejohann}},\
  }\bibfield  {title} {\emph {\enquote {\bibinfo {title} {{Possible
  Alternatives to Tri-bimaximal Mixing}},}\ }}\href {\doibase
  10.1140/epjc/s10052-010-1492-2} {\bibfield  {journal} {\bibinfo  {journal}
  {Eur. Phys. J. C}\ }\textbf {\bibinfo {volume} {70}},\ \bibinfo {pages}
  {1099} (\bibinfo {year} {2010})},\ \Eprint
  {http://arxiv.org/abs/1004.2798}{arXiv:1004.2798 [hep-ph]}\BibitemShut
  {NoStop}%
\bibitem [{\citenamefont {King}\ and\ \citenamefont
  {Luhn}(2013)}]{King:2013eh}%
  \BibitemOpen
  \bibfield  {author} {\bibinfo {author} {\bibfnamefont {S.~F.}\ \bibnamefont
  {King}}\ and\ \bibinfo {author} {\bibfnamefont {C.}~\bibnamefont {Luhn}},\
  }\bibfield  {title} {\emph {\enquote {\bibinfo {title} {{Neutrino Mass and
  Mixing with Discrete Symmetry}},}\ }}\href {\doibase
  10.1088/0034-4885/76/5/056201} {\bibfield  {journal} {\bibinfo  {journal}
  {Rept. Prog. Phys.}\ }\textbf {\bibinfo {volume} {76}},\ \bibinfo {pages}
  {056201} (\bibinfo {year} {2013})},\ \Eprint
  {http://arxiv.org/abs/1301.1340}{arXiv:1301.1340 [hep-ph]}\BibitemShut
  {NoStop}%
\bibitem [{\citenamefont {Antusch}\ \emph {et~al.}(2004)\citenamefont
  {Antusch}, \citenamefont {Huber}, \citenamefont {Kersten}, \citenamefont
  {Schwetz},\ and\ \citenamefont {Winter}}]{Antusch:2004yx}%
  \BibitemOpen
  \bibfield  {author} {\bibinfo {author} {\bibfnamefont {S.}~\bibnamefont
  {Antusch}}, \bibinfo {author} {\bibfnamefont {P.}~\bibnamefont {Huber}},
  \bibinfo {author} {\bibfnamefont {J.}~\bibnamefont {Kersten}}, \bibinfo
  {author} {\bibfnamefont {T.}~\bibnamefont {Schwetz}}, \ and\ \bibinfo
  {author} {\bibfnamefont {W.}~\bibnamefont {Winter}},\ }\bibfield  {title}
  {\emph {\enquote {\bibinfo {title} {{Is there maximal mixing in the lepton
  sector?}}}\ }}\href {\doibase 10.1103/PhysRevD.70.097302} {\bibfield
  {journal} {\bibinfo  {journal} {Phys. Rev. D}\ }\textbf {\bibinfo {volume}
  {70}},\ \bibinfo {pages} {097302} (\bibinfo {year} {2004})},\ \Eprint
  {http://arxiv.org/abs/hep-ph/0404268}{arXiv:hep-ph/0404268}\BibitemShut
  {NoStop}%
\bibitem [{\citenamefont {Hagiwara}\ and\ \citenamefont
  {Okamura}(2008)}]{Hagiwara:2006nn}%
  \BibitemOpen
  \bibfield  {author} {\bibinfo {author} {\bibfnamefont {K.}~\bibnamefont
  {Hagiwara}}\ and\ \bibinfo {author} {\bibfnamefont {N.}~\bibnamefont
  {Okamura}},\ }\bibfield  {title} {\emph {\enquote {\bibinfo {title} {{Solving
  the degeneracy of the lepton-flavor mixing angle theta(ATM) by the T2KK two
  detector neutrino oscillation experiment}},}\ }}\href {\doibase
  10.1088/1126-6708/2008/01/022} {\bibfield  {journal} {\bibinfo  {journal}
  {JHEP}\ }\textbf {\bibinfo {volume} {01}},\ \bibinfo {pages} {022} (\bibinfo
  {year} {2008})},\ \Eprint
  {http://arxiv.org/abs/hep-ph/0611058}{arXiv:hep-ph/0611058}\BibitemShut
  {NoStop}%
\bibitem [{\citenamefont {Chatterjee}\ \emph {et~al.}(2013)\citenamefont
  {Chatterjee}, \citenamefont {Ghoshal}, \citenamefont {Goswami},\ and\
  \citenamefont {Raut}}]{Chatterjee:2013qus}%
  \BibitemOpen
  \bibfield  {author} {\bibinfo {author} {\bibfnamefont {A.}~\bibnamefont
  {Chatterjee}}, \bibinfo {author} {\bibfnamefont {P.}~\bibnamefont {Ghoshal}},
  \bibinfo {author} {\bibfnamefont {S.}~\bibnamefont {Goswami}}, \ and\
  \bibinfo {author} {\bibfnamefont {S.~K.}\ \bibnamefont {Raut}},\ }\bibfield
  {title} {\emph {\enquote {\bibinfo {title} {{Octant sensitivity for large
  theta(13) in atmospheric and long baseline neutrino experiments}},}\ }}\href
  {\doibase 10.1007/JHEP06(2013)010} {\bibfield  {journal} {\bibinfo  {journal}
  {JHEP}\ }\textbf {\bibinfo {volume} {06}},\ \bibinfo {pages} {010} (\bibinfo
  {year} {2013})},\ \Eprint {http://arxiv.org/abs/1302.1370}{arXiv:1302.1370
  [hep-ph]}\BibitemShut {NoStop}%
\bibitem [{\citenamefont {Agarwalla}\ \emph
  {et~al.}(2014{\natexlab{b}})\citenamefont {Agarwalla}, \citenamefont
  {Prakash},\ and\ \citenamefont {Uma~Sankar}}]{Agarwalla:2013hma}%
  \BibitemOpen
  \bibfield  {author} {\bibinfo {author} {\bibfnamefont {S.~K.}\ \bibnamefont
  {Agarwalla}}, \bibinfo {author} {\bibfnamefont {S.}~\bibnamefont {Prakash}},
  \ and\ \bibinfo {author} {\bibfnamefont {S.}~\bibnamefont {Uma~Sankar}},\
  }\bibfield  {title} {\emph {\enquote {\bibinfo {title} {{Exploring the three
  flavor effects with future superbeams using liquid argon detectors}},}\
  }}\href {\doibase 10.1007/JHEP03(2014)087} {\bibfield  {journal} {\bibinfo
  {journal} {JHEP}\ }\textbf {\bibinfo {volume} {03}},\ \bibinfo {pages} {087}
  (\bibinfo {year} {2014}{\natexlab{b}})},\ \Eprint
  {http://arxiv.org/abs/1304.3251}{arXiv:1304.3251 [hep-ph]}\BibitemShut
  {NoStop}%
\bibitem [{\citenamefont {Agarwalla}\ \emph {et~al.}(2022)\citenamefont
  {Agarwalla}, \citenamefont {Kundu}, \citenamefont {Prakash},\ and\
  \citenamefont {Singh}}]{Agarwalla:2021bzs}%
  \BibitemOpen
  \bibfield  {author} {\bibinfo {author} {\bibfnamefont {S.~K.}\ \bibnamefont
  {Agarwalla}}, \bibinfo {author} {\bibfnamefont {R.}~\bibnamefont {Kundu}},
  \bibinfo {author} {\bibfnamefont {S.}~\bibnamefont {Prakash}}, \ and\
  \bibinfo {author} {\bibfnamefont {M.}~\bibnamefont {Singh}},\ }\bibfield
  {title} {\emph {\enquote {\bibinfo {title} {{A close look on 2-3 mixing angle
  with DUNE in light of current neutrino oscillation data}},}\ }}\href
  {\doibase 10.1007/JHEP03(2022)206} {\bibfield  {journal} {\bibinfo  {journal}
  {JHEP}\ }\textbf {\bibinfo {volume} {03}},\ \bibinfo {pages} {206} (\bibinfo
  {year} {2022})},\ \Eprint {http://arxiv.org/abs/2111.11748}{arXiv:2111.11748
  [hep-ph]}\BibitemShut {NoStop}%
\bibitem [{\citenamefont {Prado~Rodriguez}(2023)}]{psf2023008007}%
  \BibitemOpen
  \bibfield  {author} {\bibinfo {author} {\bibfnamefont {M.}~\bibnamefont
  {Prado~Rodriguez}},\ }\bibfield  {title} {\emph {\enquote {\bibinfo {title}
  {Neutrino Mass Ordering with IceCube DeepCore},}\ }}\href {\doibase
  10.3390/psf2023008007} {\bibfield  {journal} {\bibinfo  {journal} {Physical
  Sciences Forum}\ }\textbf {\bibinfo {volume} {8}} (\bibinfo {year} {2023}),\
  10.3390/psf2023008007}\BibitemShut {NoStop}%
\bibitem [{\citenamefont {Aiello}\ \emph {et~al.}(2022)\citenamefont {Aiello}
  \emph {et~al.}}]{KM3NeT:2021ozk}%
  \BibitemOpen
  \bibfield  {author} {\bibinfo {author} {\bibfnamefont {S.}~\bibnamefont
  {Aiello}} \emph {et~al.} (\bibinfo {collaboration} {KM3NeT}),\ }\bibfield
  {title} {\emph {\enquote {\bibinfo {title} {{Determining the neutrino mass
  ordering and oscillation parameters with KM3NeT/ORCA}},}\ }}\href {\doibase
  10.1140/epjc/s10052-021-09893-0} {\bibfield  {journal} {\bibinfo  {journal}
  {Eur. Phys. J. C}\ }\textbf {\bibinfo {volume} {82}},\ \bibinfo {pages} {26}
  (\bibinfo {year} {2022})},\ \Eprint
  {http://arxiv.org/abs/2103.09885}{arXiv:2103.09885 [hep-ex]}\BibitemShut
  {NoStop}%
\bibitem [{\citenamefont {Cao}(2023)}]{Cao:2023hwq}%
  \BibitemOpen
  \bibfield  {author} {\bibinfo {author} {\bibfnamefont {S.}~\bibnamefont
  {Cao}}\ }(\bibinfo {year} {2023})\ \Eprint
  {http://arxiv.org/abs/2310.09855}{arXiv:2310.09855 [hep-ex]}\BibitemShut
  {NoStop}%
\bibitem [{\citenamefont {Gariazzo}\ \emph {et~al.}(2022)\citenamefont
  {Gariazzo} \emph {et~al.}}]{Gariazzo:2022ahe}%
  \BibitemOpen
  \bibfield  {author} {\bibinfo {author} {\bibfnamefont {S.}~\bibnamefont
  {Gariazzo}} \emph {et~al.},\ }\bibfield  {title} {\emph {\enquote {\bibinfo
  {title} {{Neutrino mass and mass ordering: no conclusive evidence for normal
  ordering}},}\ }}\href {\doibase 10.1088/1475-7516/2022/10/010} {\bibfield
  {journal} {\bibinfo  {journal} {JCAP}\ }\textbf {\bibinfo {volume} {10}},\
  \bibinfo {pages} {010} (\bibinfo {year} {2022})},\ \Eprint
  {http://arxiv.org/abs/2205.02195}{arXiv:2205.02195 [hep-ph]}\BibitemShut
  {NoStop}%
\bibitem [{\citenamefont {Heinz}(2024)}]{Heinz:2024dyt}%
  \BibitemOpen
  \bibfield  {author} {\bibinfo {author} {\bibfnamefont {T.}~\bibnamefont
  {Heinz}} (\bibinfo {collaboration} {JUNO}),\ }\bibfield  {title} {\emph
  {\enquote {\bibinfo {title} {{JUNO\textquoteright{}s Sensitivity to the
  Neutrino Mass Ordering}},}\ }}\href {\doibase 10.22323/1.441.0264} {\bibfield
   {journal} {\bibinfo  {journal} {PoS}\ }\textbf {\bibinfo {volume}
  {TAUP2023}},\ \bibinfo {pages} {264} (\bibinfo {year} {2024})}\BibitemShut
  {NoStop}%
\bibitem [{\citenamefont {Abe}\ \emph {et~al.}(2018{\natexlab{b}})\citenamefont
  {Abe} \emph {et~al.}}]{Hyper-Kamiokande:2016srs}%
  \BibitemOpen
  \bibfield  {author} {\bibinfo {author} {\bibfnamefont {K.}~\bibnamefont
  {Abe}} \emph {et~al.} (\bibinfo {collaboration} {Hyper-Kamiokande}),\
  }\bibfield  {title} {\emph {\enquote {\bibinfo {title} {{Physics potentials
  with the second Hyper-Kamiokande detector in Korea}},}\ }}\href {\doibase
  10.1093/ptep/pty044} {\bibfield  {journal} {\bibinfo  {journal} {PTEP}\
  }\textbf {\bibinfo {volume} {2018}},\ \bibinfo {pages} {063C01} (\bibinfo
  {year} {2018}{\natexlab{b}})},\ \Eprint
  {http://arxiv.org/abs/1611.06118}{arXiv:1611.06118 [hep-ex]}\BibitemShut
  {NoStop}%
\bibitem [{\citenamefont {Panda}\ \emph {et~al.}(2023)\citenamefont {Panda},
  \citenamefont {Ghosh},\ and\ \citenamefont {Mohanta}}]{Panda:2023rxa}%
  \BibitemOpen
  \bibfield  {author} {\bibinfo {author} {\bibfnamefont {P.}~\bibnamefont
  {Panda}}, \bibinfo {author} {\bibfnamefont {M.}~\bibnamefont {Ghosh}}, \ and\
  \bibinfo {author} {\bibfnamefont {R.}~\bibnamefont {Mohanta}},\ }\bibfield
  {title} {\emph {\enquote {\bibinfo {title} {{Determination of neutrino mass
  ordering from Supernova neutrinos with T2HK and DUNE}},}\ }}\href {\doibase
  10.1088/1475-7516/2023/10/033} {\bibfield  {journal} {\bibinfo  {journal}
  {JCAP}\ }\textbf {\bibinfo {volume} {10}},\ \bibinfo {pages} {033} (\bibinfo
  {year} {2023})},\ \Eprint {http://arxiv.org/abs/2304.13303}{arXiv:2304.13303
  [hep-ph]}\BibitemShut {NoStop}%
\bibitem [{\citenamefont {Choubey}\ \emph
  {et~al.}(2018{\natexlab{a}})\citenamefont {Choubey}, \citenamefont {Dutta},\
  and\ \citenamefont {Pramanik}}]{Choubey:2018cfz}%
  \BibitemOpen
  \bibfield  {author} {\bibinfo {author} {\bibfnamefont {S.}~\bibnamefont
  {Choubey}}, \bibinfo {author} {\bibfnamefont {D.}~\bibnamefont {Dutta}}, \
  and\ \bibinfo {author} {\bibfnamefont {D.}~\bibnamefont {Pramanik}},\
  }\bibfield  {title} {\emph {\enquote {\bibinfo {title} {{Invisible neutrino
  decay in the light of NOvA and T2K data}},}\ }}\href {\doibase
  10.1007/JHEP08(2018)141} {\bibfield  {journal} {\bibinfo  {journal} {JHEP}\
  }\textbf {\bibinfo {volume} {08}},\ \bibinfo {pages} {141} (\bibinfo {year}
  {2018}{\natexlab{a}})},\ \Eprint
  {http://arxiv.org/abs/1805.01848}{arXiv:1805.01848 [hep-ph]}\BibitemShut
  {NoStop}%
\bibitem [{\citenamefont {Chatterjee}\ and\ \citenamefont
  {Palazzo}(2021)}]{Chatterjee:2020kkm}%
  \BibitemOpen
  \bibfield  {author} {\bibinfo {author} {\bibfnamefont {S.~S.}\ \bibnamefont
  {Chatterjee}}\ and\ \bibinfo {author} {\bibfnamefont {A.}~\bibnamefont
  {Palazzo}},\ }\bibfield  {title} {\emph {\enquote {\bibinfo {title}
  {{Nonstandard Neutrino Interactions as a Solution to the $NO\nu A$ and T2K
  Discrepancy}},}\ }}\href {\doibase 10.1103/PhysRevLett.126.051802} {\bibfield
   {journal} {\bibinfo  {journal} {Phys. Rev. Lett.}\ }\textbf {\bibinfo
  {volume} {126}},\ \bibinfo {pages} {051802} (\bibinfo {year} {2021})},\
  \Eprint {http://arxiv.org/abs/2008.04161}{arXiv:2008.04161
  [hep-ph]}\BibitemShut {NoStop}%
\bibitem [{\citenamefont {de~Gouv\^ea}\ \emph {et~al.}(2022)\citenamefont
  {de~Gouv\^ea}, \citenamefont {Jusino~S\'anchez},\ and\ \citenamefont
  {Kelly}}]{deGouvea:2022kma}%
  \BibitemOpen
  \bibfield  {author} {\bibinfo {author} {\bibfnamefont {A.}~\bibnamefont
  {de~Gouv\^ea}}, \bibinfo {author} {\bibfnamefont {G.}~\bibnamefont
  {Jusino~S\'anchez}}, \ and\ \bibinfo {author} {\bibfnamefont {K.~J.}\
  \bibnamefont {Kelly}},\ }\bibfield  {title} {\emph {\enquote {\bibinfo
  {title} {{Very light sterile neutrinos at NOvA and T2K}},}\ }}\href {\doibase
  10.1103/PhysRevD.106.055025} {\bibfield  {journal} {\bibinfo  {journal}
  {Phys. Rev. D}\ }\textbf {\bibinfo {volume} {106}},\ \bibinfo {pages}
  {055025} (\bibinfo {year} {2022})},\ \Eprint
  {http://arxiv.org/abs/2204.09130}{arXiv:2204.09130 [hep-ph]}\BibitemShut
  {NoStop}%
\bibitem [{\citenamefont {Konwar}\ \emph {et~al.}(2024)\citenamefont {Konwar},
  \citenamefont {Vardani},\ and\ \citenamefont {Yadav}}]{Konwar:2024nwc}%
  \BibitemOpen
  \bibfield  {author} {\bibinfo {author} {\bibfnamefont {L.}~\bibnamefont
  {Konwar}}, \bibinfo {author} {\bibfnamefont {J.}~\bibnamefont {Vardani}}, \
  and\ \bibinfo {author} {\bibfnamefont {B.}~\bibnamefont {Yadav}},\ }\bibfield
   {title} {\emph {\enquote {\bibinfo {title} {{Violation of LGtI inequalities
  in the light of NO$\nu$A and T2K anomaly}},}\ }}\href@noop {} {\  (\bibinfo
  {year} {2024})},\ \Eprint {http://arxiv.org/abs/2401.02886}{arXiv:2401.02886
  [hep-ph]}\BibitemShut {NoStop}%
\bibitem [{\citenamefont {Gariazzo}\ \emph {et~al.}(2016)\citenamefont
  {Gariazzo}, \citenamefont {Giunti}, \citenamefont {Laveder}, \citenamefont
  {Li},\ and\ \citenamefont {Zavanin}}]{Gariazzo:2015rra}%
  \BibitemOpen
  \bibfield  {author} {\bibinfo {author} {\bibfnamefont {S.}~\bibnamefont
  {Gariazzo}}, \bibinfo {author} {\bibfnamefont {C.}~\bibnamefont {Giunti}},
  \bibinfo {author} {\bibfnamefont {M.}~\bibnamefont {Laveder}}, \bibinfo
  {author} {\bibfnamefont {Y.~F.}\ \bibnamefont {Li}}, \ and\ \bibinfo {author}
  {\bibfnamefont {E.~M.}\ \bibnamefont {Zavanin}},\ }\bibfield  {title} {\emph
  {\enquote {\bibinfo {title} {{Light sterile neutrinos}},}\ }}\href {\doibase
  10.1088/0954-3899/43/3/033001} {\bibfield  {journal} {\bibinfo  {journal} {J.
  Phys. G}\ }\textbf {\bibinfo {volume} {43}},\ \bibinfo {pages} {033001}
  (\bibinfo {year} {2016})},\ \Eprint
  {http://arxiv.org/abs/1507.08204}{arXiv:1507.08204 [hep-ph]}\BibitemShut
  {NoStop}%
\bibitem [{\citenamefont {Giunti}\ and\ \citenamefont
  {Lasserre}(2019)}]{Giunti:2019aiy}%
  \BibitemOpen
  \bibfield  {author} {\bibinfo {author} {\bibfnamefont {C.}~\bibnamefont
  {Giunti}}\ and\ \bibinfo {author} {\bibfnamefont {T.}~\bibnamefont
  {Lasserre}},\ }\bibfield  {title} {\emph {\enquote {\bibinfo {title}
  {{eV-scale Sterile Neutrinos}},}\ }}\href {\doibase
  10.1146/annurev-nucl-101918-023755} {\bibfield  {journal} {\bibinfo
  {journal} {Ann. Rev. Nucl. Part. Sci.}\ }\textbf {\bibinfo {volume} {69}},\
  \bibinfo {pages} {163} (\bibinfo {year} {2019})},\ \Eprint
  {http://arxiv.org/abs/1901.08330}{arXiv:1901.08330 [hep-ph]}\BibitemShut
  {NoStop}%
\bibitem [{\citenamefont {Diaz}\ \emph {et~al.}(2020)\citenamefont {Diaz},
  \citenamefont {Arg\"uelles}, \citenamefont {Collin}, \citenamefont {Conrad},\
  and\ \citenamefont {Shaevitz}}]{Diaz:2019fwt}%
  \BibitemOpen
  \bibfield  {author} {\bibinfo {author} {\bibfnamefont {A.}~\bibnamefont
  {Diaz}}, \bibinfo {author} {\bibfnamefont {C.~A.}\ \bibnamefont
  {Arg\"uelles}}, \bibinfo {author} {\bibfnamefont {G.~H.}\ \bibnamefont
  {Collin}}, \bibinfo {author} {\bibfnamefont {J.~M.}\ \bibnamefont {Conrad}},
  \ and\ \bibinfo {author} {\bibfnamefont {M.~H.}\ \bibnamefont {Shaevitz}},\
  }\bibfield  {title} {\emph {\enquote {\bibinfo {title} {{Where Are We With
  Light Sterile Neutrinos?}}}\ }}\href {\doibase 10.1016/j.physrep.2020.08.005}
  {\bibfield  {journal} {\bibinfo  {journal} {Phys. Rept.}\ }\textbf {\bibinfo
  {volume} {884}},\ \bibinfo {pages} {1} (\bibinfo {year} {2020})},\ \Eprint
  {http://arxiv.org/abs/1906.00045}{arXiv:1906.00045 [hep-ex]}\BibitemShut
  {NoStop}%
\bibitem [{\citenamefont {B\"oser}\ \emph {et~al.}(2020)\citenamefont
  {B\"oser}, \citenamefont {Buck}, \citenamefont {Giunti}, \citenamefont
  {Lesgourgues}, \citenamefont {Ludhova}, \citenamefont {Mertens},
  \citenamefont {Schukraft},\ and\ \citenamefont {Wurm}}]{Boser:2019rta}%
  \BibitemOpen
  \bibfield  {author} {\bibinfo {author} {\bibfnamefont {S.}~\bibnamefont
  {B\"oser}}, \bibinfo {author} {\bibfnamefont {C.}~\bibnamefont {Buck}},
  \bibinfo {author} {\bibfnamefont {C.}~\bibnamefont {Giunti}}, \bibinfo
  {author} {\bibfnamefont {J.}~\bibnamefont {Lesgourgues}}, \bibinfo {author}
  {\bibfnamefont {L.}~\bibnamefont {Ludhova}}, \bibinfo {author} {\bibfnamefont
  {S.}~\bibnamefont {Mertens}}, \bibinfo {author} {\bibfnamefont
  {A.}~\bibnamefont {Schukraft}}, \ and\ \bibinfo {author} {\bibfnamefont
  {M.}~\bibnamefont {Wurm}},\ }\bibfield  {title} {\emph {\enquote {\bibinfo
  {title} {{Status of Light Sterile Neutrino Searches}},}\ }}\href {\doibase
  10.1016/j.ppnp.2019.103736} {\bibfield  {journal} {\bibinfo  {journal} {Prog.
  Part. Nucl. Phys.}\ }\textbf {\bibinfo {volume} {111}},\ \bibinfo {pages}
  {103736} (\bibinfo {year} {2020})},\ \Eprint
  {http://arxiv.org/abs/1906.01739}{arXiv:1906.01739 [hep-ex]}\BibitemShut
  {NoStop}%
\bibitem [{\citenamefont {Aguilar}\ \emph
  {et~al.}(2001{\natexlab{a}})\citenamefont {Aguilar} \emph
  {et~al.}}]{LSND:2001aii}%
  \BibitemOpen
  \bibfield  {author} {\bibinfo {author} {\bibfnamefont {A.}~\bibnamefont
  {Aguilar}} \emph {et~al.} (\bibinfo {collaboration} {LSND}),\ }\bibfield
  {title} {\emph {\enquote {\bibinfo {title} {{Evidence for neutrino
  oscillations from the observation of $\bar{\nu}_e$ appearance in a
  $\bar{\nu}_\mu$ beam}},}\ }}\href {\doibase 10.1103/PhysRevD.64.112007}
  {\bibfield  {journal} {\bibinfo  {journal} {Phys. Rev. D}\ }\textbf {\bibinfo
  {volume} {64}},\ \bibinfo {pages} {112007} (\bibinfo {year}
  {2001}{\natexlab{a}})},\ \Eprint
  {http://arxiv.org/abs/hep-ex/0104049}{arXiv:hep-ex/0104049}\BibitemShut
  {NoStop}%
\bibitem [{\citenamefont {Aguilar-Arevalo}\ \emph {et~al.}(2009)\citenamefont
  {Aguilar-Arevalo} \emph {et~al.}}]{MiniBooNE:2008yuf}%
  \BibitemOpen
  \bibfield  {author} {\bibinfo {author} {\bibfnamefont {A.~A.}\ \bibnamefont
  {Aguilar-Arevalo}} \emph {et~al.} (\bibinfo {collaboration} {MiniBooNE}),\
  }\bibfield  {title} {\emph {\enquote {\bibinfo {title} {{Unexplained Excess
  of Electron-Like Events From a 1-GeV Neutrino Beam}},}\ }}\href {\doibase
  10.1103/PhysRevLett.102.101802} {\bibfield  {journal} {\bibinfo  {journal}
  {Phys. Rev. Lett.}\ }\textbf {\bibinfo {volume} {102}},\ \bibinfo {pages}
  {101802} (\bibinfo {year} {2009})},\ \Eprint
  {http://arxiv.org/abs/0812.2243}{arXiv:0812.2243 [hep-ex]}\BibitemShut
  {NoStop}%
\bibitem [{\citenamefont {Aguilar}\ \emph
  {et~al.}(2001{\natexlab{b}})\citenamefont {Aguilar}, \citenamefont
  {Auerbach}, \citenamefont {Burman}, \citenamefont {Caldwell}, \citenamefont
  {Church}, \citenamefont {Cochran}, \citenamefont {Donahue}, \citenamefont
  {Fazely}, \citenamefont {Garvey}, \citenamefont {Gunasingha}, \citenamefont
  {Imlay}, \citenamefont {Louis}, \citenamefont {Majkic}, \citenamefont
  {Malik}, \citenamefont {Metcalf}, \citenamefont {Mills}, \citenamefont
  {Sandberg}, \citenamefont {Smith}, \citenamefont {Stancu}, \citenamefont
  {Sung}, \citenamefont {Tayloe}, \citenamefont {VanDalen}, \citenamefont
  {Vernon}, \citenamefont {Wadia}, \citenamefont {White},\ and\ \citenamefont
  {Yellin}}]{PhysRevD.64.112007}%
  \BibitemOpen
  \bibfield  {author} {\bibinfo {author} {\bibfnamefont {A.}~\bibnamefont
  {Aguilar}}, \bibinfo {author} {\bibfnamefont {L.~B.}\ \bibnamefont
  {Auerbach}}, \bibinfo {author} {\bibfnamefont {R.~L.}\ \bibnamefont
  {Burman}}, \bibinfo {author} {\bibfnamefont {D.~O.}\ \bibnamefont
  {Caldwell}}, \bibinfo {author} {\bibfnamefont {E.~D.}\ \bibnamefont
  {Church}}, \bibinfo {author} {\bibfnamefont {A.~K.}\ \bibnamefont {Cochran}},
  \bibinfo {author} {\bibfnamefont {J.~B.}\ \bibnamefont {Donahue}}, \bibinfo
  {author} {\bibfnamefont {A.}~\bibnamefont {Fazely}}, \bibinfo {author}
  {\bibfnamefont {G.~T.}\ \bibnamefont {Garvey}}, \bibinfo {author}
  {\bibfnamefont {R.~M.}\ \bibnamefont {Gunasingha}}, \bibinfo {author}
  {\bibfnamefont {R.}~\bibnamefont {Imlay}}, \bibinfo {author} {\bibfnamefont
  {W.~C.}\ \bibnamefont {Louis}}, \bibinfo {author} {\bibfnamefont
  {R.}~\bibnamefont {Majkic}}, \bibinfo {author} {\bibfnamefont
  {A.}~\bibnamefont {Malik}}, \bibinfo {author} {\bibfnamefont
  {W.}~\bibnamefont {Metcalf}}, \bibinfo {author} {\bibfnamefont {G.~B.}\
  \bibnamefont {Mills}}, \bibinfo {author} {\bibfnamefont {V.}~\bibnamefont
  {Sandberg}}, \bibinfo {author} {\bibfnamefont {D.}~\bibnamefont {Smith}},
  \bibinfo {author} {\bibfnamefont {I.}~\bibnamefont {Stancu}}, \bibinfo
  {author} {\bibfnamefont {M.}~\bibnamefont {Sung}}, \bibinfo {author}
  {\bibfnamefont {R.}~\bibnamefont {Tayloe}}, \bibinfo {author} {\bibfnamefont
  {G.~J.}\ \bibnamefont {VanDalen}}, \bibinfo {author} {\bibfnamefont
  {W.}~\bibnamefont {Vernon}}, \bibinfo {author} {\bibfnamefont
  {N.}~\bibnamefont {Wadia}}, \bibinfo {author} {\bibfnamefont {D.~H.}\
  \bibnamefont {White}}, \ and\ \bibinfo {author} {\bibfnamefont
  {S.}~\bibnamefont {Yellin}} (\bibinfo {collaboration} {LSND Collaboration}),\
  }\bibfield  {title} {\emph {\enquote {\bibinfo {title} {Evidence for neutrino
  oscillations from the observation of ${\overline{\ensuremath{\nu}}}_{e}$
  appearance in a ${\overline{\ensuremath{\nu}}}_{\ensuremath{\mu}}$ beam},}\
  }}\href {\doibase 10.1103/PhysRevD.64.112007} {\bibfield  {journal} {\bibinfo
   {journal} {Phys. Rev. D}\ }\textbf {\bibinfo {volume} {64}},\ \bibinfo
  {pages} {112007} (\bibinfo {year} {2001}{\natexlab{b}})}\BibitemShut
  {NoStop}%
\bibitem [{\citenamefont {Aguilar-Arevalo}\ \emph {et~al.}(2021)\citenamefont
  {Aguilar-Arevalo} \emph {et~al.}}]{MiniBooNE:2020pnu}%
  \BibitemOpen
  \bibfield  {author} {\bibinfo {author} {\bibfnamefont {A.~A.}\ \bibnamefont
  {Aguilar-Arevalo}} \emph {et~al.} (\bibinfo {collaboration} {MiniBooNE}),\
  }\bibfield  {title} {\emph {\enquote {\bibinfo {title} {{Updated MiniBooNE
  neutrino oscillation results with increased data and new background
  studies}},}\ }}\href {\doibase 10.1103/PhysRevD.103.052002} {\bibfield
  {journal} {\bibinfo  {journal} {Phys. Rev. D}\ }\textbf {\bibinfo {volume}
  {103}},\ \bibinfo {pages} {052002} (\bibinfo {year} {2021})},\ \Eprint
  {http://arxiv.org/abs/2006.16883}{arXiv:2006.16883 [hep-ex]}\BibitemShut
  {NoStop}%
\bibitem [{\citenamefont {Aguilar-Arevalo}\ \emph {et~al.}(2013)\citenamefont
  {Aguilar-Arevalo} \emph {et~al.}}]{MiniBooNE:2013uba}%
  \BibitemOpen
  \bibfield  {author} {\bibinfo {author} {\bibfnamefont {A.~A.}\ \bibnamefont
  {Aguilar-Arevalo}} \emph {et~al.} (\bibinfo {collaboration} {MiniBooNE}),\
  }\bibfield  {title} {\emph {\enquote {\bibinfo {title} {{Improved Search for
  $\bar \nu_\mu \rightarrow \bar \nu_e$ Oscillations in the MiniBooNE
  Experiment}},}\ }}\href {\doibase 10.1103/PhysRevLett.110.161801} {\bibfield
  {journal} {\bibinfo  {journal} {Phys. Rev. Lett.}\ }\textbf {\bibinfo
  {volume} {110}},\ \bibinfo {pages} {161801} (\bibinfo {year} {2013})},\
  \Eprint {http://arxiv.org/abs/1303.2588}{arXiv:1303.2588
  [hep-ex]}\BibitemShut {NoStop}%
\bibitem [{\citenamefont {Aguilar-Arevalo}\ \emph {et~al.}(2018)\citenamefont
  {Aguilar-Arevalo} \emph {et~al.}}]{MiniBooNE:2018esg}%
  \BibitemOpen
  \bibfield  {author} {\bibinfo {author} {\bibfnamefont {A.~A.}\ \bibnamefont
  {Aguilar-Arevalo}} \emph {et~al.} (\bibinfo {collaboration} {MiniBooNE}),\
  }\bibfield  {title} {\emph {\enquote {\bibinfo {title} {{Significant Excess
  of ElectronLike Events in the MiniBooNE Short-Baseline Neutrino
  Experiment}},}\ }}\href {\doibase 10.1103/PhysRevLett.121.221801} {\bibfield
  {journal} {\bibinfo  {journal} {Phys. Rev. Lett.}\ }\textbf {\bibinfo
  {volume} {121}},\ \bibinfo {pages} {221801} (\bibinfo {year} {2018})},\
  \Eprint {http://arxiv.org/abs/1805.12028}{arXiv:1805.12028
  [hep-ex]}\BibitemShut {NoStop}%
\bibitem [{\citenamefont {Giunti}\ \emph
  {et~al.}(2022{\natexlab{a}})\citenamefont {Giunti}, \citenamefont {Li},
  \citenamefont {Ternes}, \citenamefont {Tyagi},\ and\ \citenamefont
  {Xin}}]{Giunti:2022btk}%
  \BibitemOpen
  \bibfield  {author} {\bibinfo {author} {\bibfnamefont {C.}~\bibnamefont
  {Giunti}}, \bibinfo {author} {\bibfnamefont {Y.~F.}\ \bibnamefont {Li}},
  \bibinfo {author} {\bibfnamefont {C.~A.}\ \bibnamefont {Ternes}}, \bibinfo
  {author} {\bibfnamefont {O.}~\bibnamefont {Tyagi}}, \ and\ \bibinfo {author}
  {\bibfnamefont {Z.}~\bibnamefont {Xin}},\ }\bibfield  {title} {\emph
  {\enquote {\bibinfo {title} {{Gallium Anomaly: critical view from the global
  picture of \ensuremath{\nu}$_{e}$ and $ {\overline{\nu}}_e $
  disappearance}},}\ }}\href {\doibase 10.1007/JHEP10(2022)164} {\bibfield
  {journal} {\bibinfo  {journal} {JHEP}\ }\textbf {\bibinfo {volume} {10}},\
  \bibinfo {pages} {164} (\bibinfo {year} {2022}{\natexlab{a}})},\ \Eprint
  {http://arxiv.org/abs/2209.00916}{arXiv:2209.00916 [hep-ph]}\BibitemShut
  {NoStop}%
\bibitem [{\citenamefont {Mention}\ \emph {et~al.}(2011)\citenamefont
  {Mention}, \citenamefont {Fechner}, \citenamefont {Lasserre}, \citenamefont
  {Mueller}, \citenamefont {Lhuillier}, \citenamefont {Cribier},\ and\
  \citenamefont {Letourneau}}]{Mention:2011rk}%
  \BibitemOpen
  \bibfield  {author} {\bibinfo {author} {\bibfnamefont {G.}~\bibnamefont
  {Mention}}, \bibinfo {author} {\bibfnamefont {M.}~\bibnamefont {Fechner}},
  \bibinfo {author} {\bibfnamefont {T.}~\bibnamefont {Lasserre}}, \bibinfo
  {author} {\bibfnamefont {T.~A.}\ \bibnamefont {Mueller}}, \bibinfo {author}
  {\bibfnamefont {D.}~\bibnamefont {Lhuillier}}, \bibinfo {author}
  {\bibfnamefont {M.}~\bibnamefont {Cribier}}, \ and\ \bibinfo {author}
  {\bibfnamefont {A.}~\bibnamefont {Letourneau}},\ }\bibfield  {title} {\emph
  {\enquote {\bibinfo {title} {{The Reactor Antineutrino Anomaly}},}\ }}\href
  {\doibase 10.1103/PhysRevD.83.073006} {\bibfield  {journal} {\bibinfo
  {journal} {Phys. Rev. D}\ }\textbf {\bibinfo {volume} {83}},\ \bibinfo
  {pages} {073006} (\bibinfo {year} {2011})},\ \Eprint
  {http://arxiv.org/abs/1101.2755}{arXiv:1101.2755 [hep-ex]}\BibitemShut
  {NoStop}%
\bibitem [{\citenamefont {Giunti}\ and\ \citenamefont
  {Laveder}(2007)}]{Giunti:2006bj}%
  \BibitemOpen
  \bibfield  {author} {\bibinfo {author} {\bibfnamefont {C.}~\bibnamefont
  {Giunti}}\ and\ \bibinfo {author} {\bibfnamefont {M.}~\bibnamefont
  {Laveder}},\ }\bibfield  {title} {\emph {\enquote {\bibinfo {title}
  {{Short-Baseline Active-Sterile Neutrino Oscillations?}}}\ }}\href {\doibase
  10.1142/S0217732307025455} {\bibfield  {journal} {\bibinfo  {journal} {Mod.
  Phys. Lett. A}\ }\textbf {\bibinfo {volume} {22}},\ \bibinfo {pages} {2499}
  (\bibinfo {year} {2007})},\ \Eprint
  {http://arxiv.org/abs/hep-ph/0610352}{arXiv:hep-ph/0610352}\BibitemShut
  {NoStop}%
\bibitem [{\citenamefont {Giunti}\ and\ \citenamefont
  {Laveder}(2011)}]{Giunti:2010zu}%
  \BibitemOpen
  \bibfield  {author} {\bibinfo {author} {\bibfnamefont {C.}~\bibnamefont
  {Giunti}}\ and\ \bibinfo {author} {\bibfnamefont {M.}~\bibnamefont
  {Laveder}},\ }\bibfield  {title} {\emph {\enquote {\bibinfo {title}
  {{Statistical Significance of the Gallium Anomaly}},}\ }}\href {\doibase
  10.1103/PhysRevC.83.065504} {\bibfield  {journal} {\bibinfo  {journal} {Phys.
  Rev. C}\ }\textbf {\bibinfo {volume} {83}},\ \bibinfo {pages} {065504}
  (\bibinfo {year} {2011})},\ \Eprint
  {http://arxiv.org/abs/1006.3244}{arXiv:1006.3244 [hep-ph]}\BibitemShut
  {NoStop}%
\bibitem [{\citenamefont {Giunti}\ \emph {et~al.}(2012)\citenamefont {Giunti},
  \citenamefont {Laveder}, \citenamefont {Li}, \citenamefont {Liu},\ and\
  \citenamefont {Long}}]{Giunti:2012tn}%
  \BibitemOpen
  \bibfield  {author} {\bibinfo {author} {\bibfnamefont {C.}~\bibnamefont
  {Giunti}}, \bibinfo {author} {\bibfnamefont {M.}~\bibnamefont {Laveder}},
  \bibinfo {author} {\bibfnamefont {Y.~F.}\ \bibnamefont {Li}}, \bibinfo
  {author} {\bibfnamefont {Q.~Y.}\ \bibnamefont {Liu}}, \ and\ \bibinfo
  {author} {\bibfnamefont {H.~W.}\ \bibnamefont {Long}},\ }\bibfield  {title}
  {\emph {\enquote {\bibinfo {title} {{Update of Short-Baseline Electron
  Neutrino and Antineutrino Disappearance}},}\ }}\href {\doibase
  10.1103/PhysRevD.86.113014} {\bibfield  {journal} {\bibinfo  {journal} {Phys.
  Rev. D}\ }\textbf {\bibinfo {volume} {86}},\ \bibinfo {pages} {113014}
  (\bibinfo {year} {2012})},\ \Eprint
  {http://arxiv.org/abs/1210.5715}{arXiv:1210.5715 [hep-ph]}\BibitemShut
  {NoStop}%
\bibitem [{\citenamefont {Giunti}\ \emph
  {et~al.}(2022{\natexlab{b}})\citenamefont {Giunti}, \citenamefont {Li},
  \citenamefont {Ternes},\ and\ \citenamefont {Xin}}]{Giunti:2021kab}%
  \BibitemOpen
  \bibfield  {author} {\bibinfo {author} {\bibfnamefont {C.}~\bibnamefont
  {Giunti}}, \bibinfo {author} {\bibfnamefont {Y.~F.}\ \bibnamefont {Li}},
  \bibinfo {author} {\bibfnamefont {C.~A.}\ \bibnamefont {Ternes}}, \ and\
  \bibinfo {author} {\bibfnamefont {Z.}~\bibnamefont {Xin}},\ }\bibfield
  {title} {\emph {\enquote {\bibinfo {title} {{Reactor antineutrino anomaly in
  light of recent flux model refinements}},}\ }}\href {\doibase
  10.1016/j.physletb.2022.137054} {\bibfield  {journal} {\bibinfo  {journal}
  {Phys. Lett. B}\ }\textbf {\bibinfo {volume} {829}},\ \bibinfo {pages}
  {137054} (\bibinfo {year} {2022}{\natexlab{b}})},\ \Eprint
  {http://arxiv.org/abs/2110.06820}{arXiv:2110.06820 [hep-ph]}\BibitemShut
  {NoStop}%
\bibitem [{\citenamefont {Elliott}\ \emph {et~al.}(2024)\citenamefont
  {Elliott}, \citenamefont {Gavrin},\ and\ \citenamefont
  {Haxton}}]{Elliott:2023cvh}%
  \BibitemOpen
  \bibfield  {author} {\bibinfo {author} {\bibfnamefont {S.~R.}\ \bibnamefont
  {Elliott}}, \bibinfo {author} {\bibfnamefont {V.}~\bibnamefont {Gavrin}}, \
  and\ \bibinfo {author} {\bibfnamefont {W.}~\bibnamefont {Haxton}},\
  }\bibfield  {title} {\emph {\enquote {\bibinfo {title} {{The gallium
  anomaly}},}\ }}\href {\doibase 10.1016/j.ppnp.2023.104082} {\bibfield
  {journal} {\bibinfo  {journal} {Prog. Part. Nucl. Phys.}\ }\textbf {\bibinfo
  {volume} {134}},\ \bibinfo {pages} {104082} (\bibinfo {year} {2024})},\
  \Eprint {http://arxiv.org/abs/2306.03299}{arXiv:2306.03299
  [nucl-ex]}\BibitemShut {NoStop}%
\bibitem [{\citenamefont {Barinov}\ \emph {et~al.}(2022)\citenamefont {Barinov}
  \emph {et~al.}}]{Barinov:2022wfh}%
  \BibitemOpen
  \bibfield  {author} {\bibinfo {author} {\bibfnamefont {V.~V.}\ \bibnamefont
  {Barinov}} \emph {et~al.},\ }\bibfield  {title} {\emph {\enquote {\bibinfo
  {title} {{Search for electron-neutrino transitions to sterile states in the
  BEST experiment}},}\ }}\href {\doibase 10.1103/PhysRevC.105.065502}
  {\bibfield  {journal} {\bibinfo  {journal} {Phys. Rev. C}\ }\textbf {\bibinfo
  {volume} {105}},\ \bibinfo {pages} {065502} (\bibinfo {year} {2022})},\
  \Eprint {http://arxiv.org/abs/2201.07364}{arXiv:2201.07364
  [nucl-ex]}\BibitemShut {NoStop}%
\bibitem [{\citenamefont {Abratenko}\ \emph {et~al.}(2023)\citenamefont
  {Abratenko} \emph {et~al.}}]{MicroBooNE:2022sdp}%
  \BibitemOpen
  \bibfield  {author} {\bibinfo {author} {\bibfnamefont {P.}~\bibnamefont
  {Abratenko}} \emph {et~al.} (\bibinfo {collaboration} {MicroBooNE}),\
  }\bibfield  {title} {\emph {\enquote {\bibinfo {title} {{First Constraints on
  Light Sterile Neutrino Oscillations from Combined Appearance and
  Disappearance Searches with the MicroBooNE Detector}},}\ }}\href {\doibase
  10.1103/PhysRevLett.130.011801} {\bibfield  {journal} {\bibinfo  {journal}
  {Phys. Rev. Lett.}\ }\textbf {\bibinfo {volume} {130}},\ \bibinfo {pages}
  {011801} (\bibinfo {year} {2023})},\ \Eprint
  {http://arxiv.org/abs/2210.10216}{arXiv:2210.10216 [hep-ex]}\BibitemShut
  {NoStop}%
\bibitem [{\citenamefont {Denton}(2022)}]{Denton:2021czb}%
  \BibitemOpen
  \bibfield  {author} {\bibinfo {author} {\bibfnamefont {P.~B.}\ \bibnamefont
  {Denton}},\ }\bibfield  {title} {\emph {\enquote {\bibinfo {title} {{Sterile
  Neutrino Search with MicroBooNE\textquoteright{}s Electron Neutrino
  Disappearance Data}},}\ }}\href {\doibase 10.1103/PhysRevLett.129.061801}
  {\bibfield  {journal} {\bibinfo  {journal} {Phys. Rev. Lett.}\ }\textbf
  {\bibinfo {volume} {129}},\ \bibinfo {pages} {061801} (\bibinfo {year}
  {2022})},\ \Eprint {http://arxiv.org/abs/2111.05793}{arXiv:2111.05793
  [hep-ph]}\BibitemShut {NoStop}%
\bibitem [{\citenamefont {Arg\"uelles}\ \emph {et~al.}(2022)\citenamefont
  {Arg\"uelles}, \citenamefont {Esteban}, \citenamefont {Hostert},
  \citenamefont {Kelly}, \citenamefont {Kopp}, \citenamefont {Machado},
  \citenamefont {Martinez-Soler},\ and\ \citenamefont
  {Perez-Gonzalez}}]{Arguelles:2021meu}%
  \BibitemOpen
  \bibfield  {author} {\bibinfo {author} {\bibfnamefont {C.~A.}\ \bibnamefont
  {Arg\"uelles}}, \bibinfo {author} {\bibfnamefont {I.}~\bibnamefont
  {Esteban}}, \bibinfo {author} {\bibfnamefont {M.}~\bibnamefont {Hostert}},
  \bibinfo {author} {\bibfnamefont {K.~J.}\ \bibnamefont {Kelly}}, \bibinfo
  {author} {\bibfnamefont {J.}~\bibnamefont {Kopp}}, \bibinfo {author}
  {\bibfnamefont {P.~A.~N.}\ \bibnamefont {Machado}}, \bibinfo {author}
  {\bibfnamefont {I.}~\bibnamefont {Martinez-Soler}}, \ and\ \bibinfo {author}
  {\bibfnamefont {Y.~F.}\ \bibnamefont {Perez-Gonzalez}},\ }\bibfield  {title}
  {\emph {\enquote {\bibinfo {title} {{MicroBooNE and the \ensuremath{\nu}e
  Interpretation of the MiniBooNE Low-Energy Excess}},}\ }}\href {\doibase
  10.1103/PhysRevLett.128.241802} {\bibfield  {journal} {\bibinfo  {journal}
  {Phys. Rev. Lett.}\ }\textbf {\bibinfo {volume} {128}},\ \bibinfo {pages}
  {241802} (\bibinfo {year} {2022})},\ \Eprint
  {http://arxiv.org/abs/2111.10359}{arXiv:2111.10359 [hep-ph]}\BibitemShut
  {NoStop}%
\bibitem [{\citenamefont {Gavrin}\ \emph {et~al.}(2017)\citenamefont {Gavrin},
  \citenamefont {Cleveland}, \citenamefont {Gorbachev}, \citenamefont
  {Ibragimova}, \citenamefont {Kalikhov}, \citenamefont {Kozlova},
  \citenamefont {Malyshkin}, \citenamefont {Mirmov}, \citenamefont {Shikhin},\
  and\ \citenamefont {Veretenkin}}]{Gavrin_2017}%
  \BibitemOpen
  \bibfield  {author} {\bibinfo {author} {\bibfnamefont {V.~N.}\ \bibnamefont
  {Gavrin}}, \bibinfo {author} {\bibfnamefont {B.~T.}\ \bibnamefont
  {Cleveland}}, \bibinfo {author} {\bibfnamefont {V.~V.}\ \bibnamefont
  {Gorbachev}}, \bibinfo {author} {\bibfnamefont {T.~V.}\ \bibnamefont
  {Ibragimova}}, \bibinfo {author} {\bibfnamefont {A.~V.}\ \bibnamefont
  {Kalikhov}}, \bibinfo {author} {\bibfnamefont {Y.~P.}\ \bibnamefont
  {Kozlova}}, \bibinfo {author} {\bibfnamefont {Y.~A.}\ \bibnamefont
  {Malyshkin}}, \bibinfo {author} {\bibfnamefont {I.~N.}\ \bibnamefont
  {Mirmov}}, \bibinfo {author} {\bibfnamefont {A.~A.}\ \bibnamefont {Shikhin}},
  \ and\ \bibinfo {author} {\bibfnamefont {E.~P.}\ \bibnamefont {Veretenkin}},\
  }\bibfield  {title} {\emph {\enquote {\bibinfo {title} {Search for sterile
  neutrinos on the Gallium Germanium Neutrino Telescope with artificial
  neutrino sources in the BEST experiment},}\ }}\href {\doibase
  10.1088/1742-6596/798/1/012113} {\bibfield  {journal} {\bibinfo  {journal}
  {Journal of Physics: Conference Series}\ }\textbf {\bibinfo {volume} {798}},\
  \bibinfo {pages} {012113} (\bibinfo {year} {2017})}\BibitemShut {NoStop}%
\bibitem [{\citenamefont {Acciarri}\ \emph {et~al.}(2015)\citenamefont
  {Acciarri} \emph {et~al.}}]{MicroBooNE:2015bmn}%
  \BibitemOpen
  \bibfield  {author} {\bibinfo {author} {\bibfnamefont {R.}~\bibnamefont
  {Acciarri}} \emph {et~al.} (\bibinfo {collaboration} {MicroBooNE, LAr1-ND,
  ICARUS-WA104}),\ }\bibfield  {title} {\emph {\enquote {\bibinfo {title} {{A
  Proposal for a Three Detector Short-Baseline Neutrino Oscillation Program in
  the Fermilab Booster Neutrino Beam}},}\ }}\href@noop {} {\  (\bibinfo {year}
  {2015})},\ \Eprint {http://arxiv.org/abs/1503.01520}{arXiv:1503.01520
  [physics.ins-det]}\BibitemShut {NoStop}%
\bibitem [{\citenamefont {Harada}\ \emph {et~al.}(2013)\citenamefont {Harada}
  \emph {et~al.}}]{JSNS2:2013jdh}%
  \BibitemOpen
  \bibfield  {author} {\bibinfo {author} {\bibfnamefont {M.}~\bibnamefont
  {Harada}} \emph {et~al.} (\bibinfo {collaboration} {JSNS2}),\ }\bibfield
  {title} {\emph {\enquote {\bibinfo {title} {{Proposal: A Search for Sterile
  Neutrino at J-PARC Materials and Life Science Experimental Facility}},}\
  }}\href@noop {} {\  (\bibinfo {year} {2013})},\ \Eprint
  {http://arxiv.org/abs/1310.1437}{arXiv:1310.1437
  [physics.ins-det]}\BibitemShut {NoStop}%
\bibitem [{\citenamefont {Alonso}\ \emph {et~al.}(2022)\citenamefont {Alonso}
  \emph {et~al.}}]{Alonso:2022mup}%
  \BibitemOpen
  \bibfield  {author} {\bibinfo {author} {\bibfnamefont {J.~R.}\ \bibnamefont
  {Alonso}} \emph {et~al.},\ }\bibfield  {title} {\emph {\enquote {\bibinfo
  {title} {{IsoDAR@Yemilab: A report on the technology, capabilities, and
  deployment}},}\ }}\href {\doibase 10.1088/1748-0221/17/09/P09042} {\bibfield
  {journal} {\bibinfo  {journal} {JINST}\ }\textbf {\bibinfo {volume} {17}},\
  \bibinfo {pages} {P09042} (\bibinfo {year} {2022})},\ \Eprint
  {http://arxiv.org/abs/2201.10040}{arXiv:2201.10040
  [physics.ins-det]}\BibitemShut {NoStop}%
\bibitem [{\citenamefont {Andriamirado}\ \emph {et~al.}(2022)\citenamefont
  {Andriamirado} \emph {et~al.}}]{PROSPECT:2021jey}%
  \BibitemOpen
  \bibfield  {author} {\bibinfo {author} {\bibfnamefont {M.}~\bibnamefont
  {Andriamirado}} \emph {et~al.} (\bibinfo {collaboration} {PROSPECT}),\
  }\bibfield  {title} {\emph {\enquote {\bibinfo {title} {{PROSPECT-II physics
  opportunities}},}\ }}\href {\doibase 10.1088/1361-6471/ac48a4} {\bibfield
  {journal} {\bibinfo  {journal} {J. Phys. G}\ }\textbf {\bibinfo {volume}
  {49}},\ \bibinfo {pages} {070501} (\bibinfo {year} {2022})},\ \Eprint
  {http://arxiv.org/abs/2107.03934}{arXiv:2107.03934 [hep-ex]}\BibitemShut
  {NoStop}%
\bibitem [{\citenamefont {Langacker}\ and\ \citenamefont
  {Sankar}(1989)}]{PhysRevD.40.1569}%
  \BibitemOpen
  \bibfield  {author} {\bibinfo {author} {\bibfnamefont {P.}~\bibnamefont
  {Langacker}}\ and\ \bibinfo {author} {\bibfnamefont {S.~U.}\ \bibnamefont
  {Sankar}},\ }\bibfield  {title} {\emph {\enquote {\bibinfo {title} {Bounds on
  the mass of ${W}_{R}$ and the ${W}_{L}\ensuremath{-}{W}_{R}$ mixing angle
  $\ensuremath{\zeta}$ in general
  $\mathrm{SU}{(2)}_{L}\ifmmode\times\else\texttimes\fi{}\mathrm{SU}{(2)}_{R}\ifmmode\times\else\texttimes\fi{}\mathrm{U}(1)$
  models},}\ }}\href {\doibase 10.1103/PhysRevD.40.1569} {\bibfield  {journal}
  {\bibinfo  {journal} {Phys. Rev. D}\ }\textbf {\bibinfo {volume} {40}},\
  \bibinfo {pages} {1569} (\bibinfo {year} {1989})}\BibitemShut {NoStop}%
\bibitem [{\citenamefont {Davidson}\ \emph {et~al.}(1994)\citenamefont
  {Davidson}, \citenamefont {Bailey},\ and\ \citenamefont
  {Campbell}}]{Davidson:1993qk}%
  \BibitemOpen
  \bibfield  {author} {\bibinfo {author} {\bibfnamefont {S.}~\bibnamefont
  {Davidson}}, \bibinfo {author} {\bibfnamefont {D.~C.}\ \bibnamefont
  {Bailey}}, \ and\ \bibinfo {author} {\bibfnamefont {B.~A.}\ \bibnamefont
  {Campbell}},\ }\bibfield  {title} {\emph {\enquote {\bibinfo {title} {{Model
  independent constraints on leptoquarks from rare processes}},}\ }}\href
  {\doibase 10.1007/BF01552629} {\bibfield  {journal} {\bibinfo  {journal} {Z.
  Phys. C}\ }\textbf {\bibinfo {volume} {61}},\ \bibinfo {pages} {613}
  (\bibinfo {year} {1994})},\ \Eprint
  {http://arxiv.org/abs/hep-ph/9309310}{arXiv:hep-ph/9309310}\BibitemShut
  {NoStop}%
\bibitem [{\citenamefont {Gavela}\ \emph {et~al.}(2009)\citenamefont {Gavela},
  \citenamefont {Hernandez}, \citenamefont {Ota},\ and\ \citenamefont
  {Winter}}]{Gavela:2008ra}%
  \BibitemOpen
  \bibfield  {author} {\bibinfo {author} {\bibfnamefont {M.~B.}\ \bibnamefont
  {Gavela}}, \bibinfo {author} {\bibfnamefont {D.}~\bibnamefont {Hernandez}},
  \bibinfo {author} {\bibfnamefont {T.}~\bibnamefont {Ota}}, \ and\ \bibinfo
  {author} {\bibfnamefont {W.}~\bibnamefont {Winter}},\ }\bibfield  {title}
  {\emph {\enquote {\bibinfo {title} {{Large gauge invariant non-standard
  neutrino interactions}},}\ }}\href {\doibase 10.1103/PhysRevD.79.013007}
  {\bibfield  {journal} {\bibinfo  {journal} {Phys. Rev. D}\ }\textbf {\bibinfo
  {volume} {79}},\ \bibinfo {pages} {013007} (\bibinfo {year} {2009})},\
  \Eprint {http://arxiv.org/abs/0809.3451}{arXiv:0809.3451
  [hep-ph]}\BibitemShut {NoStop}%
\bibitem [{\citenamefont {Dor\v{s}ner}\ \emph {et~al.}(2016)\citenamefont
  {Dor\v{s}ner}, \citenamefont {Fajfer}, \citenamefont {Greljo}, \citenamefont
  {Kamenik},\ and\ \citenamefont {Ko\v{s}nik}}]{Dorsner:2016wpm}%
  \BibitemOpen
  \bibfield  {author} {\bibinfo {author} {\bibfnamefont {I.}~\bibnamefont
  {Dor\v{s}ner}}, \bibinfo {author} {\bibfnamefont {S.}~\bibnamefont {Fajfer}},
  \bibinfo {author} {\bibfnamefont {A.}~\bibnamefont {Greljo}}, \bibinfo
  {author} {\bibfnamefont {J.~F.}\ \bibnamefont {Kamenik}}, \ and\ \bibinfo
  {author} {\bibfnamefont {N.}~\bibnamefont {Ko\v{s}nik}},\ }\bibfield  {title}
  {\emph {\enquote {\bibinfo {title} {{Physics of leptoquarks in precision
  experiments and at particle colliders}},}\ }}\href {\doibase
  10.1016/j.physrep.2016.06.001} {\bibfield  {journal} {\bibinfo  {journal}
  {Phys. Rept.}\ }\textbf {\bibinfo {volume} {641}},\ \bibinfo {pages} {1}
  (\bibinfo {year} {2016})},\ \Eprint
  {http://arxiv.org/abs/1603.04993}{arXiv:1603.04993 [hep-ph]}\BibitemShut
  {NoStop}%
\bibitem [{\citenamefont {Babu}\ \emph {et~al.}(2020)\citenamefont {Babu},
  \citenamefont {Dev}, \citenamefont {Jana},\ and\ \citenamefont
  {Thapa}}]{Babu:2019mfe}%
  \BibitemOpen
  \bibfield  {author} {\bibinfo {author} {\bibfnamefont {K.~S.}\ \bibnamefont
  {Babu}}, \bibinfo {author} {\bibfnamefont {P.~S.~B.}\ \bibnamefont {Dev}},
  \bibinfo {author} {\bibfnamefont {S.}~\bibnamefont {Jana}}, \ and\ \bibinfo
  {author} {\bibfnamefont {A.}~\bibnamefont {Thapa}},\ }\bibfield  {title}
  {\emph {\enquote {\bibinfo {title} {{Non-Standard Interactions in Radiative
  Neutrino Mass Models}},}\ }}\href {\doibase 10.1007/JHEP03(2020)006}
  {\bibfield  {journal} {\bibinfo  {journal} {JHEP}\ }\textbf {\bibinfo
  {volume} {03}},\ \bibinfo {pages} {006} (\bibinfo {year} {2020})},\ \Eprint
  {http://arxiv.org/abs/1907.09498}{arXiv:1907.09498 [hep-ph]}\BibitemShut
  {NoStop}%
\bibitem [{\citenamefont {Wolfenstein}(1978)}]{PhysRevD.17.2369}%
  \BibitemOpen
  \bibfield  {author} {\bibinfo {author} {\bibfnamefont {L.}~\bibnamefont
  {Wolfenstein}},\ }\bibfield  {title} {\emph {\enquote {\bibinfo {title}
  {Neutrino oscillations in matter},}\ }}\href {\doibase
  10.1103/PhysRevD.17.2369} {\bibfield  {journal} {\bibinfo  {journal} {Phys.
  Rev. D}\ }\textbf {\bibinfo {volume} {17}},\ \bibinfo {pages} {2369}
  (\bibinfo {year} {1978})}\BibitemShut {NoStop}%
\bibitem [{\citenamefont {Ohlsson}\ and\ \citenamefont
  {Zhang}(2009)}]{Ohlsson:2008gx}%
  \BibitemOpen
  \bibfield  {author} {\bibinfo {author} {\bibfnamefont {T.}~\bibnamefont
  {Ohlsson}}\ and\ \bibinfo {author} {\bibfnamefont {H.}~\bibnamefont
  {Zhang}},\ }\bibfield  {title} {\emph {\enquote {\bibinfo {title}
  {{Non-Standard Interaction Effects at Reactor Neutrino Experiments}},}\
  }}\href {\doibase 10.1016/j.physletb.2008.12.005} {\bibfield  {journal}
  {\bibinfo  {journal} {Phys. Lett. B}\ }\textbf {\bibinfo {volume} {671}},\
  \bibinfo {pages} {99} (\bibinfo {year} {2009})},\ \Eprint
  {http://arxiv.org/abs/0809.4835}{arXiv:0809.4835 [hep-ph]}\BibitemShut
  {NoStop}%
\bibitem [{\citenamefont {Meloni}\ \emph {et~al.}(2010)\citenamefont {Meloni},
  \citenamefont {Ohlsson}, \citenamefont {Winter},\ and\ \citenamefont
  {Zhang}}]{Meloni:2009cg}%
  \BibitemOpen
  \bibfield  {author} {\bibinfo {author} {\bibfnamefont {D.}~\bibnamefont
  {Meloni}}, \bibinfo {author} {\bibfnamefont {T.}~\bibnamefont {Ohlsson}},
  \bibinfo {author} {\bibfnamefont {W.}~\bibnamefont {Winter}}, \ and\ \bibinfo
  {author} {\bibfnamefont {H.}~\bibnamefont {Zhang}},\ }\bibfield  {title}
  {\emph {\enquote {\bibinfo {title} {{Non-standard interactions versus
  non-unitary lepton flavor mixing at a neutrino factory}},}\ }}\href {\doibase
  10.1007/JHEP04(2010)041} {\bibfield  {journal} {\bibinfo  {journal} {JHEP}\
  }\textbf {\bibinfo {volume} {04}},\ \bibinfo {pages} {041} (\bibinfo {year}
  {2010})},\ \Eprint {http://arxiv.org/abs/0912.2735}{arXiv:0912.2735
  [hep-ph]}\BibitemShut {NoStop}%
\bibitem [{\citenamefont {Giarnetti}\ and\ \citenamefont
  {Meloni}(2021)}]{Giarnetti:2020bmf}%
  \BibitemOpen
  \bibfield  {author} {\bibinfo {author} {\bibfnamefont {A.}~\bibnamefont
  {Giarnetti}}\ and\ \bibinfo {author} {\bibfnamefont {D.}~\bibnamefont
  {Meloni}},\ }\bibfield  {title} {\emph {\enquote {\bibinfo {title} {{Probing
  source and detector nonstandard interaction parameters at the DUNE near
  detector}},}\ }}\href {\doibase 10.1103/PhysRevD.104.015027} {\bibfield
  {journal} {\bibinfo  {journal} {Phys. Rev. D}\ }\textbf {\bibinfo {volume}
  {104}},\ \bibinfo {pages} {015027} (\bibinfo {year} {2021})},\ \Eprint
  {http://arxiv.org/abs/2005.10272}{arXiv:2005.10272 [hep-ph]}\BibitemShut
  {NoStop}%
\bibitem [{\citenamefont {Ohlsson}(2013)}]{Ohlsson:2012kf}%
  \BibitemOpen
  \bibfield  {author} {\bibinfo {author} {\bibfnamefont {T.}~\bibnamefont
  {Ohlsson}},\ }\bibfield  {title} {\emph {\enquote {\bibinfo {title} {{Status
  of non-standard neutrino interactions}},}\ }}\href {\doibase
  10.1088/0034-4885/76/4/044201} {\bibfield  {journal} {\bibinfo  {journal}
  {Rept. Prog. Phys.}\ }\textbf {\bibinfo {volume} {76}},\ \bibinfo {pages}
  {044201} (\bibinfo {year} {2013})},\ \Eprint
  {http://arxiv.org/abs/1209.2710}{arXiv:1209.2710 [hep-ph]}\BibitemShut
  {NoStop}%
\bibitem [{\citenamefont {Kopp}\ \emph {et~al.}(2008)\citenamefont {Kopp},
  \citenamefont {Lindner}, \citenamefont {Ota},\ and\ \citenamefont
  {Sato}}]{Kopp:2007ne}%
  \BibitemOpen
  \bibfield  {author} {\bibinfo {author} {\bibfnamefont {J.}~\bibnamefont
  {Kopp}}, \bibinfo {author} {\bibfnamefont {M.}~\bibnamefont {Lindner}},
  \bibinfo {author} {\bibfnamefont {T.}~\bibnamefont {Ota}}, \ and\ \bibinfo
  {author} {\bibfnamefont {J.}~\bibnamefont {Sato}},\ }\bibfield  {title}
  {\emph {\enquote {\bibinfo {title} {{Non-standard neutrino interactions in
  reactor and superbeam experiments}},}\ }}\href {\doibase
  10.1103/PhysRevD.77.013007} {\bibfield  {journal} {\bibinfo  {journal} {Phys.
  Rev. D}\ }\textbf {\bibinfo {volume} {77}},\ \bibinfo {pages} {013007}
  (\bibinfo {year} {2008})},\ \Eprint
  {http://arxiv.org/abs/0708.0152}{arXiv:0708.0152 [hep-ph]}\BibitemShut
  {NoStop}%
\bibitem [{\citenamefont {Abdurashitov}\ \emph {et~al.}(2006)\citenamefont
  {Abdurashitov} \emph {et~al.}}]{Abdurashitov:2005tb}%
  \BibitemOpen
  \bibfield  {author} {\bibinfo {author} {\bibfnamefont {J.~N.}\ \bibnamefont
  {Abdurashitov}} \emph {et~al.},\ }\bibfield  {title} {\emph {\enquote
  {\bibinfo {title} {{Measurement of the response of a Ga solar neutrino
  experiment to neutrinos from an Ar-37 source}},}\ }}\href {\doibase
  10.1103/PhysRevC.73.045805} {\bibfield  {journal} {\bibinfo  {journal} {Phys.
  Rev. C}\ }\textbf {\bibinfo {volume} {73}},\ \bibinfo {pages} {045805}
  (\bibinfo {year} {2006})},\ \Eprint
  {http://arxiv.org/abs/nucl-ex/0512041}{arXiv:nucl-ex/0512041}\BibitemShut
  {NoStop}%
\bibitem [{\citenamefont {Abazajian}\ \emph {et~al.}(2012)\citenamefont
  {Abazajian} \emph {et~al.}}]{Abazajian:2012ys}%
  \BibitemOpen
  \bibfield  {author} {\bibinfo {author} {\bibfnamefont {K.~N.}\ \bibnamefont
  {Abazajian}} \emph {et~al.},\ }\bibfield  {title} {\emph {\enquote {\bibinfo
  {title} {{Light Sterile Neutrinos: A White Paper}},}\ }}\href@noop {} {\
  (\bibinfo {year} {2012})},\ \Eprint
  {http://arxiv.org/abs/1204.5379}{arXiv:1204.5379 [hep-ph]}\BibitemShut
  {NoStop}%
\bibitem [{\citenamefont {Palazzo}(2013)}]{Palazzo:2013me}%
  \BibitemOpen
  \bibfield  {author} {\bibinfo {author} {\bibfnamefont {A.}~\bibnamefont
  {Palazzo}},\ }\bibfield  {title} {\emph {\enquote {\bibinfo {title}
  {{Phenomenology of light sterile neutrinos: a brief review}},}\ }}\href
  {\doibase 10.1142/S0217732313300048} {\bibfield  {journal} {\bibinfo
  {journal} {Mod. Phys. Lett. A}\ }\textbf {\bibinfo {volume} {28}},\ \bibinfo
  {pages} {1330004} (\bibinfo {year} {2013})},\ \Eprint
  {http://arxiv.org/abs/1302.1102}{arXiv:1302.1102 [hep-ph]}\BibitemShut
  {NoStop}%
\bibitem [{\citenamefont {Capozzi}\ \emph {et~al.}(2017)\citenamefont
  {Capozzi}, \citenamefont {Giunti}, \citenamefont {Laveder},\ and\
  \citenamefont {Palazzo}}]{Capozzi:2016vac}%
  \BibitemOpen
  \bibfield  {author} {\bibinfo {author} {\bibfnamefont {F.}~\bibnamefont
  {Capozzi}}, \bibinfo {author} {\bibfnamefont {C.}~\bibnamefont {Giunti}},
  \bibinfo {author} {\bibfnamefont {M.}~\bibnamefont {Laveder}}, \ and\
  \bibinfo {author} {\bibfnamefont {A.}~\bibnamefont {Palazzo}},\ }\bibfield
  {title} {\emph {\enquote {\bibinfo {title} {{Joint short- and long-baseline
  constraints on light sterile neutrinos}},}\ }}\href {\doibase
  10.1103/PhysRevD.95.033006} {\bibfield  {journal} {\bibinfo  {journal} {Phys.
  Rev. D}\ }\textbf {\bibinfo {volume} {95}},\ \bibinfo {pages} {033006}
  (\bibinfo {year} {2017})},\ \Eprint
  {http://arxiv.org/abs/1612.07764}{arXiv:1612.07764 [hep-ph]}\BibitemShut
  {NoStop}%
\bibitem [{\citenamefont {Boyarsky}\ \emph {et~al.}(2014)\citenamefont
  {Boyarsky}, \citenamefont {Ruchayskiy}, \citenamefont {Iakubovskyi},\ and\
  \citenamefont {Franse}}]{Boyarsky:2014jta}%
  \BibitemOpen
  \bibfield  {author} {\bibinfo {author} {\bibfnamefont {A.}~\bibnamefont
  {Boyarsky}}, \bibinfo {author} {\bibfnamefont {O.}~\bibnamefont
  {Ruchayskiy}}, \bibinfo {author} {\bibfnamefont {D.}~\bibnamefont
  {Iakubovskyi}}, \ and\ \bibinfo {author} {\bibfnamefont {J.}~\bibnamefont
  {Franse}},\ }\bibfield  {title} {\emph {\enquote {\bibinfo {title}
  {{Unidentified Line in X-Ray Spectra of the Andromeda Galaxy and Perseus
  Galaxy Cluster}},}\ }}\href {\doibase 10.1103/PhysRevLett.113.251301}
  {\bibfield  {journal} {\bibinfo  {journal} {Phys. Rev. Lett.}\ }\textbf
  {\bibinfo {volume} {113}},\ \bibinfo {pages} {251301} (\bibinfo {year}
  {2014})},\ \Eprint {http://arxiv.org/abs/1402.4119}{arXiv:1402.4119
  [astro-ph.CO]}\BibitemShut {NoStop}%
\bibitem [{\citenamefont {Bulbul}\ \emph {et~al.}(2014)\citenamefont {Bulbul},
  \citenamefont {Markevitch}, \citenamefont {Foster}, \citenamefont {Smith},
  \citenamefont {Loewenstein},\ and\ \citenamefont {Randall}}]{Bulbul:2014sua}%
  \BibitemOpen
  \bibfield  {author} {\bibinfo {author} {\bibfnamefont {E.}~\bibnamefont
  {Bulbul}}, \bibinfo {author} {\bibfnamefont {M.}~\bibnamefont {Markevitch}},
  \bibinfo {author} {\bibfnamefont {A.}~\bibnamefont {Foster}}, \bibinfo
  {author} {\bibfnamefont {R.~K.}\ \bibnamefont {Smith}}, \bibinfo {author}
  {\bibfnamefont {M.}~\bibnamefont {Loewenstein}}, \ and\ \bibinfo {author}
  {\bibfnamefont {S.~W.}\ \bibnamefont {Randall}},\ }\bibfield  {title} {\emph
  {\enquote {\bibinfo {title} {{Detection of An Unidentified Emission Line in
  the Stacked X-ray spectrum of Galaxy Clusters}},}\ }}\href {\doibase
  10.1088/0004-637X/789/1/13} {\bibfield  {journal} {\bibinfo  {journal}
  {Astrophys. J.}\ }\textbf {\bibinfo {volume} {789}},\ \bibinfo {pages} {13}
  (\bibinfo {year} {2014})},\ \Eprint
  {http://arxiv.org/abs/1402.2301}{arXiv:1402.2301 [astro-ph.CO]}\BibitemShut
  {NoStop}%
\bibitem [{\citenamefont {Abazajian}(2014)}]{Abazajian:2014gza}%
  \BibitemOpen
  \bibfield  {author} {\bibinfo {author} {\bibfnamefont {K.~N.}\ \bibnamefont
  {Abazajian}},\ }\bibfield  {title} {\emph {\enquote {\bibinfo {title}
  {{Resonantly Produced 7 keV Sterile Neutrino Dark Matter Models and the
  Properties of Milky Way Satellites}},}\ }}\href {\doibase
  10.1103/PhysRevLett.112.161303} {\bibfield  {journal} {\bibinfo  {journal}
  {Phys. Rev. Lett.}\ }\textbf {\bibinfo {volume} {112}},\ \bibinfo {pages}
  {161303} (\bibinfo {year} {2014})},\ \Eprint
  {http://arxiv.org/abs/1403.0954}{arXiv:1403.0954 [astro-ph.CO]}\BibitemShut
  {NoStop}%
\bibitem [{\citenamefont {Ng}\ \emph {et~al.}(2015)\citenamefont {Ng},
  \citenamefont {Horiuchi}, \citenamefont {Gaskins}, \citenamefont {Smith},\
  and\ \citenamefont {Preece}}]{Ng:2015gfa}%
  \BibitemOpen
  \bibfield  {author} {\bibinfo {author} {\bibfnamefont {K.~C.~Y.}\
  \bibnamefont {Ng}}, \bibinfo {author} {\bibfnamefont {S.}~\bibnamefont
  {Horiuchi}}, \bibinfo {author} {\bibfnamefont {J.~M.}\ \bibnamefont
  {Gaskins}}, \bibinfo {author} {\bibfnamefont {M.}~\bibnamefont {Smith}}, \
  and\ \bibinfo {author} {\bibfnamefont {R.}~\bibnamefont {Preece}},\
  }\bibfield  {title} {\emph {\enquote {\bibinfo {title} {{Improved Limits on
  Sterile Neutrino Dark Matter using Full-Sky Fermi Gamma-Ray Burst Monitor
  Data}},}\ }}\href {\doibase 10.1103/PhysRevD.92.043503} {\bibfield  {journal}
  {\bibinfo  {journal} {Phys. Rev. D}\ }\textbf {\bibinfo {volume} {92}},\
  \bibinfo {pages} {043503} (\bibinfo {year} {2015})},\ \Eprint
  {http://arxiv.org/abs/1504.04027}{arXiv:1504.04027 [astro-ph.CO]}\BibitemShut
  {NoStop}%
\bibitem [{\citenamefont {Schneider}(2016)}]{Schneider:2016uqi}%
  \BibitemOpen
  \bibfield  {author} {\bibinfo {author} {\bibfnamefont {A.}~\bibnamefont
  {Schneider}},\ }\bibfield  {title} {\emph {\enquote {\bibinfo {title}
  {{Astrophysical constraints on resonantly produced sterile neutrino dark
  matter}},}\ }}\href {\doibase 10.1088/1475-7516/2016/04/059} {\bibfield
  {journal} {\bibinfo  {journal} {JCAP}\ }\textbf {\bibinfo {volume} {04}},\
  \bibinfo {pages} {059} (\bibinfo {year} {2016})},\ \Eprint
  {http://arxiv.org/abs/1601.07553}{arXiv:1601.07553 [astro-ph.CO]}\BibitemShut
  {NoStop}%
\bibitem [{\citenamefont {Klop}\ and\ \citenamefont
  {Palazzo}(2015)}]{Klop:2014ima}%
  \BibitemOpen
  \bibfield  {author} {\bibinfo {author} {\bibfnamefont {N.}~\bibnamefont
  {Klop}}\ and\ \bibinfo {author} {\bibfnamefont {A.}~\bibnamefont {Palazzo}},\
  }\bibfield  {title} {\emph {\enquote {\bibinfo {title} {{Imprints of CP
  violation induced by sterile neutrinos in T2K data}},}\ }}\href {\doibase
  10.1103/PhysRevD.91.073017} {\bibfield  {journal} {\bibinfo  {journal} {Phys.
  Rev. D}\ }\textbf {\bibinfo {volume} {91}},\ \bibinfo {pages} {073017}
  (\bibinfo {year} {2015})},\ \Eprint
  {http://arxiv.org/abs/1412.7524}{arXiv:1412.7524 [hep-ph]}\BibitemShut
  {NoStop}%
\bibitem [{\citenamefont {Berryman}\ \emph
  {et~al.}(2015{\natexlab{a}})\citenamefont {Berryman}, \citenamefont
  {de~Gouv\^ea}, \citenamefont {Kelly},\ and\ \citenamefont
  {Kobach}}]{Berryman:2015nua}%
  \BibitemOpen
  \bibfield  {author} {\bibinfo {author} {\bibfnamefont {J.~M.}\ \bibnamefont
  {Berryman}}, \bibinfo {author} {\bibfnamefont {A.}~\bibnamefont
  {de~Gouv\^ea}}, \bibinfo {author} {\bibfnamefont {K.~J.}\ \bibnamefont
  {Kelly}}, \ and\ \bibinfo {author} {\bibfnamefont {A.}~\bibnamefont
  {Kobach}},\ }\bibfield  {title} {\emph {\enquote {\bibinfo {title} {{Sterile
  neutrino at the Deep Underground Neutrino Experiment}},}\ }}\href {\doibase
  10.1103/PhysRevD.92.073012} {\bibfield  {journal} {\bibinfo  {journal} {Phys.
  Rev. D}\ }\textbf {\bibinfo {volume} {92}},\ \bibinfo {pages} {073012}
  (\bibinfo {year} {2015}{\natexlab{a}})},\ \Eprint
  {http://arxiv.org/abs/1507.03986}{arXiv:1507.03986 [hep-ph]}\BibitemShut
  {NoStop}%
\bibitem [{\citenamefont {Gandhi}\ \emph {et~al.}(2015)\citenamefont {Gandhi},
  \citenamefont {Kayser}, \citenamefont {Masud},\ and\ \citenamefont
  {Prakash}}]{Gandhi:2015xza}%
  \BibitemOpen
  \bibfield  {author} {\bibinfo {author} {\bibfnamefont {R.}~\bibnamefont
  {Gandhi}}, \bibinfo {author} {\bibfnamefont {B.}~\bibnamefont {Kayser}},
  \bibinfo {author} {\bibfnamefont {M.}~\bibnamefont {Masud}}, \ and\ \bibinfo
  {author} {\bibfnamefont {S.}~\bibnamefont {Prakash}},\ }\bibfield  {title}
  {\emph {\enquote {\bibinfo {title} {{The impact of sterile neutrinos on CP
  measurements at long baselines}},}\ }}\href {\doibase
  10.1007/JHEP11(2015)039} {\bibfield  {journal} {\bibinfo  {journal} {JHEP}\
  }\textbf {\bibinfo {volume} {11}},\ \bibinfo {pages} {039} (\bibinfo {year}
  {2015})},\ \Eprint {http://arxiv.org/abs/1508.06275}{arXiv:1508.06275
  [hep-ph]}\BibitemShut {NoStop}%
\bibitem [{\citenamefont {Agarwalla}\ \emph {et~al.}(2016)\citenamefont
  {Agarwalla}, \citenamefont {Chatterjee}, \citenamefont {Dasgupta},\ and\
  \citenamefont {Palazzo}}]{Agarwalla:2016mrc}%
  \BibitemOpen
  \bibfield  {author} {\bibinfo {author} {\bibfnamefont {S.~K.}\ \bibnamefont
  {Agarwalla}}, \bibinfo {author} {\bibfnamefont {S.~S.}\ \bibnamefont
  {Chatterjee}}, \bibinfo {author} {\bibfnamefont {A.}~\bibnamefont
  {Dasgupta}}, \ and\ \bibinfo {author} {\bibfnamefont {A.}~\bibnamefont
  {Palazzo}},\ }\bibfield  {title} {\emph {\enquote {\bibinfo {title}
  {{Discovery Potential of T2K and NOvA in the Presence of a Light Sterile
  Neutrino}},}\ }}\href {\doibase 10.1007/JHEP02(2016)111} {\bibfield
  {journal} {\bibinfo  {journal} {JHEP}\ }\textbf {\bibinfo {volume} {02}},\
  \bibinfo {pages} {111} (\bibinfo {year} {2016})},\ \Eprint
  {http://arxiv.org/abs/1601.05995}{arXiv:1601.05995 [hep-ph]}\BibitemShut
  {NoStop}%
\bibitem [{\citenamefont {Choubey}\ \emph {et~al.}(2017)\citenamefont
  {Choubey}, \citenamefont {Dutta},\ and\ \citenamefont
  {Pramanik}}]{Choubey:2017cba}%
  \BibitemOpen
  \bibfield  {author} {\bibinfo {author} {\bibfnamefont {S.}~\bibnamefont
  {Choubey}}, \bibinfo {author} {\bibfnamefont {D.}~\bibnamefont {Dutta}}, \
  and\ \bibinfo {author} {\bibfnamefont {D.}~\bibnamefont {Pramanik}},\
  }\bibfield  {title} {\emph {\enquote {\bibinfo {title} {{Imprints of a light
  Sterile Neutrino at DUNE, T2HK and T2HKK}},}\ }}\href {\doibase
  10.1103/PhysRevD.96.056026} {\bibfield  {journal} {\bibinfo  {journal} {Phys.
  Rev. D}\ }\textbf {\bibinfo {volume} {96}},\ \bibinfo {pages} {056026}
  (\bibinfo {year} {2017})},\ \Eprint
  {http://arxiv.org/abs/1704.07269}{arXiv:1704.07269 [hep-ph]}\BibitemShut
  {NoStop}%
\bibitem [{\citenamefont {Agarwalla}\ \emph {et~al.}(2018)\citenamefont
  {Agarwalla}, \citenamefont {Chatterjee},\ and\ \citenamefont
  {Palazzo}}]{Agarwalla:2018nlx}%
  \BibitemOpen
  \bibfield  {author} {\bibinfo {author} {\bibfnamefont {S.~K.}\ \bibnamefont
  {Agarwalla}}, \bibinfo {author} {\bibfnamefont {S.~S.}\ \bibnamefont
  {Chatterjee}}, \ and\ \bibinfo {author} {\bibfnamefont {A.}~\bibnamefont
  {Palazzo}},\ }\bibfield  {title} {\emph {\enquote {\bibinfo {title}
  {{Signatures of a Light Sterile Neutrino in T2HK}},}\ }}\href {\doibase
  10.1007/JHEP04(2018)091} {\bibfield  {journal} {\bibinfo  {journal} {JHEP}\
  }\textbf {\bibinfo {volume} {04}},\ \bibinfo {pages} {091} (\bibinfo {year}
  {2018})},\ \Eprint {http://arxiv.org/abs/1801.04855}{arXiv:1801.04855
  [hep-ph]}\BibitemShut {NoStop}%
\bibitem [{\citenamefont {Miranda}\ \emph {et~al.}(2020)\citenamefont
  {Miranda}, \citenamefont {Papoulias}, \citenamefont {Sanders}, \citenamefont
  {T\'ortola},\ and\ \citenamefont {Valle}}]{Miranda:2020syh}%
  \BibitemOpen
  \bibfield  {author} {\bibinfo {author} {\bibfnamefont {O.~G.}\ \bibnamefont
  {Miranda}}, \bibinfo {author} {\bibfnamefont {D.~K.}\ \bibnamefont
  {Papoulias}}, \bibinfo {author} {\bibfnamefont {O.}~\bibnamefont {Sanders}},
  \bibinfo {author} {\bibfnamefont {M.}~\bibnamefont {T\'ortola}}, \ and\
  \bibinfo {author} {\bibfnamefont {J.~W.~F.}\ \bibnamefont {Valle}},\
  }\bibfield  {title} {\emph {\enquote {\bibinfo {title} {{Future CEvNS
  experiments as probes of lepton unitarity and light-sterile neutrinos}},}\
  }}\href {\doibase 10.1103/PhysRevD.102.113014} {\bibfield  {journal}
  {\bibinfo  {journal} {Phys. Rev. D}\ }\textbf {\bibinfo {volume} {102}},\
  \bibinfo {pages} {113014} (\bibinfo {year} {2020})},\ \Eprint
  {http://arxiv.org/abs/2008.02759}{arXiv:2008.02759 [hep-ph]}\BibitemShut
  {NoStop}%
\bibitem [{\citenamefont {Fiza}\ \emph {et~al.}(2021)\citenamefont {Fiza},
  \citenamefont {Masud},\ and\ \citenamefont {Mitra}}]{Fiza:2021gvq}%
  \BibitemOpen
  \bibfield  {author} {\bibinfo {author} {\bibfnamefont {N.}~\bibnamefont
  {Fiza}}, \bibinfo {author} {\bibfnamefont {M.}~\bibnamefont {Masud}}, \ and\
  \bibinfo {author} {\bibfnamefont {M.}~\bibnamefont {Mitra}},\ }\bibfield
  {title} {\emph {\enquote {\bibinfo {title} {{Exploring the new physics phases
  in 3+1 scenario in neutrino oscillation experiments}},}\ }}\href {\doibase
  10.1007/JHEP09(2021)162} {\bibfield  {journal} {\bibinfo  {journal} {JHEP}\
  }\textbf {\bibinfo {volume} {09}},\ \bibinfo {pages} {162} (\bibinfo {year}
  {2021})},\ \Eprint {http://arxiv.org/abs/2102.05063}{arXiv:2102.05063
  [hep-ph]}\BibitemShut {NoStop}%
\bibitem [{\citenamefont {Escrihuela}\ \emph {et~al.}(2015)\citenamefont
  {Escrihuela}, \citenamefont {Forero}, \citenamefont {Miranda}, \citenamefont
  {Tortola},\ and\ \citenamefont {Valle}}]{Escrihuela:2015wra}%
  \BibitemOpen
  \bibfield  {author} {\bibinfo {author} {\bibfnamefont {F.~J.}\ \bibnamefont
  {Escrihuela}}, \bibinfo {author} {\bibfnamefont {D.~V.}\ \bibnamefont
  {Forero}}, \bibinfo {author} {\bibfnamefont {O.~G.}\ \bibnamefont {Miranda}},
  \bibinfo {author} {\bibfnamefont {M.}~\bibnamefont {Tortola}}, \ and\
  \bibinfo {author} {\bibfnamefont {J.~W.~F.}\ \bibnamefont {Valle}},\
  }\bibfield  {title} {\emph {\enquote {\bibinfo {title} {{On the description
  of nonunitary neutrino mixing}},}\ }}\href {\doibase
  10.1103/PhysRevD.92.053009} {\bibfield  {journal} {\bibinfo  {journal} {Phys.
  Rev. D}\ }\textbf {\bibinfo {volume} {92}},\ \bibinfo {pages} {053009}
  (\bibinfo {year} {2015})},\ \bibinfo {note} {[Erratum: Phys.Rev.D 93, 119905
  (2016)]},\ \Eprint {http://arxiv.org/abs/1503.08879}{arXiv:1503.08879
  [hep-ph]}\BibitemShut {NoStop}%
\bibitem [{\citenamefont {Miranda}\ \emph {et~al.}(2016)\citenamefont
  {Miranda}, \citenamefont {Tortola},\ and\ \citenamefont
  {Valle}}]{Miranda:2016wdr}%
  \BibitemOpen
  \bibfield  {author} {\bibinfo {author} {\bibfnamefont {O.~G.}\ \bibnamefont
  {Miranda}}, \bibinfo {author} {\bibfnamefont {M.}~\bibnamefont {Tortola}}, \
  and\ \bibinfo {author} {\bibfnamefont {J.~W.~F.}\ \bibnamefont {Valle}},\
  }\bibfield  {title} {\emph {\enquote {\bibinfo {title} {{New ambiguity in
  probing CP violation in neutrino oscillations}},}\ }}\href {\doibase
  10.1103/PhysRevLett.117.061804} {\bibfield  {journal} {\bibinfo  {journal}
  {Phys. Rev. Lett.}\ }\textbf {\bibinfo {volume} {117}},\ \bibinfo {pages}
  {061804} (\bibinfo {year} {2016})},\ \Eprint
  {http://arxiv.org/abs/1604.05690}{arXiv:1604.05690 [hep-ph]}\BibitemShut
  {NoStop}%
\bibitem [{\citenamefont {Ge}\ \emph {et~al.}(2017)\citenamefont {Ge},
  \citenamefont {Pasquini}, \citenamefont {Tortola},\ and\ \citenamefont
  {Valle}}]{Ge:2016xya}%
  \BibitemOpen
  \bibfield  {author} {\bibinfo {author} {\bibfnamefont {S.-F.}\ \bibnamefont
  {Ge}}, \bibinfo {author} {\bibfnamefont {P.}~\bibnamefont {Pasquini}},
  \bibinfo {author} {\bibfnamefont {M.}~\bibnamefont {Tortola}}, \ and\
  \bibinfo {author} {\bibfnamefont {J.~W.~F.}\ \bibnamefont {Valle}},\
  }\bibfield  {title} {\emph {\enquote {\bibinfo {title} {{Measuring the
  leptonic CP phase in neutrino oscillations with nonunitary mixing}},}\
  }}\href {\doibase 10.1103/PhysRevD.95.033005} {\bibfield  {journal} {\bibinfo
   {journal} {Phys. Rev. D}\ }\textbf {\bibinfo {volume} {95}},\ \bibinfo
  {pages} {033005} (\bibinfo {year} {2017})},\ \Eprint
  {http://arxiv.org/abs/1605.01670}{arXiv:1605.01670 [hep-ph]}\BibitemShut
  {NoStop}%
\bibitem [{\citenamefont {Blennow}\ \emph {et~al.}(2017)\citenamefont
  {Blennow}, \citenamefont {Coloma}, \citenamefont {Fernandez-Martinez},
  \citenamefont {Hernandez-Garcia},\ and\ \citenamefont
  {Lopez-Pavon}}]{Blennow:2016jkn}%
  \BibitemOpen
  \bibfield  {author} {\bibinfo {author} {\bibfnamefont {M.}~\bibnamefont
  {Blennow}}, \bibinfo {author} {\bibfnamefont {P.}~\bibnamefont {Coloma}},
  \bibinfo {author} {\bibfnamefont {E.}~\bibnamefont {Fernandez-Martinez}},
  \bibinfo {author} {\bibfnamefont {J.}~\bibnamefont {Hernandez-Garcia}}, \
  and\ \bibinfo {author} {\bibfnamefont {J.}~\bibnamefont {Lopez-Pavon}},\
  }\bibfield  {title} {\emph {\enquote {\bibinfo {title} {{Non-Unitarity,
  sterile neutrinos, and Non-Standard neutrino Interactions}},}\ }}\href
  {\doibase 10.1007/JHEP04(2017)153} {\bibfield  {journal} {\bibinfo  {journal}
  {JHEP}\ }\textbf {\bibinfo {volume} {04}},\ \bibinfo {pages} {153} (\bibinfo
  {year} {2017})},\ \Eprint {http://arxiv.org/abs/1609.08637}{arXiv:1609.08637
  [hep-ph]}\BibitemShut {NoStop}%
\bibitem [{\citenamefont {Escrihuela}\ \emph {et~al.}(2017)\citenamefont
  {Escrihuela}, \citenamefont {Forero}, \citenamefont {Miranda}, \citenamefont
  {T\'ortola},\ and\ \citenamefont {Valle}}]{Escrihuela:2016ube}%
  \BibitemOpen
  \bibfield  {author} {\bibinfo {author} {\bibfnamefont {F.~J.}\ \bibnamefont
  {Escrihuela}}, \bibinfo {author} {\bibfnamefont {D.~V.}\ \bibnamefont
  {Forero}}, \bibinfo {author} {\bibfnamefont {O.~G.}\ \bibnamefont {Miranda}},
  \bibinfo {author} {\bibfnamefont {M.}~\bibnamefont {T\'ortola}}, \ and\
  \bibinfo {author} {\bibfnamefont {J.~W.~F.}\ \bibnamefont {Valle}},\
  }\bibfield  {title} {\emph {\enquote {\bibinfo {title} {{Probing CP violation
  with non-unitary mixing in long-baseline neutrino oscillation experiments:
  DUNE as a case study}},}\ }}\href {\doibase 10.1088/1367-2630/aa79ec}
  {\bibfield  {journal} {\bibinfo  {journal} {New J. Phys.}\ }\textbf {\bibinfo
  {volume} {19}},\ \bibinfo {pages} {093005} (\bibinfo {year} {2017})},\
  \Eprint {http://arxiv.org/abs/1612.07377}{arXiv:1612.07377
  [hep-ph]}\BibitemShut {NoStop}%
\bibitem [{\citenamefont {Fong}\ \emph {et~al.}(2019)\citenamefont {Fong},
  \citenamefont {Minakata},\ and\ \citenamefont {Nunokawa}}]{Fong:2017gke}%
  \BibitemOpen
  \bibfield  {author} {\bibinfo {author} {\bibfnamefont {C.~S.}\ \bibnamefont
  {Fong}}, \bibinfo {author} {\bibfnamefont {H.}~\bibnamefont {Minakata}}, \
  and\ \bibinfo {author} {\bibfnamefont {H.}~\bibnamefont {Nunokawa}},\
  }\bibfield  {title} {\emph {\enquote {\bibinfo {title} {{Non-unitary
  evolution of neutrinos in matter and the leptonic unitarity test}},}\ }}\href
  {\doibase 10.1007/JHEP02(2019)015} {\bibfield  {journal} {\bibinfo  {journal}
  {JHEP}\ }\textbf {\bibinfo {volume} {02}},\ \bibinfo {pages} {015} (\bibinfo
  {year} {2019})},\ \Eprint {http://arxiv.org/abs/1712.02798}{arXiv:1712.02798
  [hep-ph]}\BibitemShut {NoStop}%
\bibitem [{\citenamefont {De~Gouv\^ea}\ \emph {et~al.}(2019)\citenamefont
  {De~Gouv\^ea}, \citenamefont {Kelly}, \citenamefont {Stenico},\ and\
  \citenamefont {Pasquini}}]{DeGouvea:2019kea}%
  \BibitemOpen
  \bibfield  {author} {\bibinfo {author} {\bibfnamefont {A.}~\bibnamefont
  {De~Gouv\^ea}}, \bibinfo {author} {\bibfnamefont {K.~J.}\ \bibnamefont
  {Kelly}}, \bibinfo {author} {\bibfnamefont {G.~V.}\ \bibnamefont {Stenico}},
  \ and\ \bibinfo {author} {\bibfnamefont {P.}~\bibnamefont {Pasquini}},\
  }\bibfield  {title} {\emph {\enquote {\bibinfo {title} {{Physics with Beam
  Tau-Neutrino Appearance at DUNE}},}\ }}\href {\doibase
  10.1103/PhysRevD.100.016004} {\bibfield  {journal} {\bibinfo  {journal}
  {Phys. Rev. D}\ }\textbf {\bibinfo {volume} {100}},\ \bibinfo {pages}
  {016004} (\bibinfo {year} {2019})},\ \Eprint
  {http://arxiv.org/abs/1904.07265}{arXiv:1904.07265 [hep-ph]}\BibitemShut
  {NoStop}%
\bibitem [{\citenamefont {Miranda}\ \emph {et~al.}(2021)\citenamefont
  {Miranda}, \citenamefont {Pasquini}, \citenamefont {Rahaman},\ and\
  \citenamefont {Razzaque}}]{Miranda:2019ynh}%
  \BibitemOpen
  \bibfield  {author} {\bibinfo {author} {\bibfnamefont {L.~S.}\ \bibnamefont
  {Miranda}}, \bibinfo {author} {\bibfnamefont {P.}~\bibnamefont {Pasquini}},
  \bibinfo {author} {\bibfnamefont {U.}~\bibnamefont {Rahaman}}, \ and\
  \bibinfo {author} {\bibfnamefont {S.}~\bibnamefont {Razzaque}},\ }\bibfield
  {title} {\emph {\enquote {\bibinfo {title} {{Searching for non-unitary
  neutrino oscillations in the present T2K and NO$\nu $A data}},}\ }}\href
  {\doibase 10.1140/epjc/s10052-021-09227-0} {\bibfield  {journal} {\bibinfo
  {journal} {Eur. Phys. J. C}\ }\textbf {\bibinfo {volume} {81}},\ \bibinfo
  {pages} {444} (\bibinfo {year} {2021})},\ \Eprint
  {http://arxiv.org/abs/1911.09398}{arXiv:1911.09398 [hep-ph]}\BibitemShut
  {NoStop}%
\bibitem [{\citenamefont {Martinez-Soler}\ and\ \citenamefont
  {Minakata}(2020)}]{Martinez-Soler:2019noy}%
  \BibitemOpen
  \bibfield  {author} {\bibinfo {author} {\bibfnamefont {I.}~\bibnamefont
  {Martinez-Soler}}\ and\ \bibinfo {author} {\bibfnamefont {H.}~\bibnamefont
  {Minakata}},\ }\bibfield  {title} {\emph {\enquote {\bibinfo {title}
  {{Physics of parameter correlations around the solar-scale enhancement in
  neutrino theory with unitarity violation}},}\ }}\href {\doibase
  10.1093/ptep/ptaa112} {\bibfield  {journal} {\bibinfo  {journal} {PTEP}\
  }\textbf {\bibinfo {volume} {2020}},\ \bibinfo {pages} {113B01} (\bibinfo
  {year} {2020})},\ \Eprint {http://arxiv.org/abs/1908.04855}{arXiv:1908.04855
  [hep-ph]}\BibitemShut {NoStop}%
\bibitem [{\citenamefont {Ellis}\ \emph
  {et~al.}(2020{\natexlab{a}})\citenamefont {Ellis}, \citenamefont {Kelly},\
  and\ \citenamefont {Li}}]{Ellis:2020hus}%
  \BibitemOpen
  \bibfield  {author} {\bibinfo {author} {\bibfnamefont {S.~A.~R.}\
  \bibnamefont {Ellis}}, \bibinfo {author} {\bibfnamefont {K.~J.}\ \bibnamefont
  {Kelly}}, \ and\ \bibinfo {author} {\bibfnamefont {S.~W.}\ \bibnamefont
  {Li}},\ }\bibfield  {title} {\emph {\enquote {\bibinfo {title} {{Current and
  Future Neutrino Oscillation Constraints on Leptonic Unitarity}},}\ }}\href
  {\doibase 10.1007/JHEP12(2020)068} {\bibfield  {journal} {\bibinfo  {journal}
  {JHEP}\ }\textbf {\bibinfo {volume} {12}},\ \bibinfo {pages} {068} (\bibinfo
  {year} {2020}{\natexlab{a}})},\ \Eprint
  {http://arxiv.org/abs/2008.01088}{arXiv:2008.01088 [hep-ph]}\BibitemShut
  {NoStop}%
\bibitem [{\citenamefont {Hu}\ \emph {et~al.}(2021)\citenamefont {Hu},
  \citenamefont {Ling}, \citenamefont {Tang},\ and\ \citenamefont
  {Wang}}]{Hu:2020oba}%
  \BibitemOpen
  \bibfield  {author} {\bibinfo {author} {\bibfnamefont {Z.}~\bibnamefont
  {Hu}}, \bibinfo {author} {\bibfnamefont {J.}~\bibnamefont {Ling}}, \bibinfo
  {author} {\bibfnamefont {J.}~\bibnamefont {Tang}}, \ and\ \bibinfo {author}
  {\bibfnamefont {T.}~\bibnamefont {Wang}},\ }\bibfield  {title} {\emph
  {\enquote {\bibinfo {title} {{Global oscillation data analysis on the $3\nu$
  mixing without unitarity}},}\ }}\href {\doibase 10.1007/JHEP01(2021)124}
  {\bibfield  {journal} {\bibinfo  {journal} {JHEP}\ }\textbf {\bibinfo
  {volume} {01}},\ \bibinfo {pages} {124} (\bibinfo {year} {2021})},\ \Eprint
  {http://arxiv.org/abs/2008.09730}{arXiv:2008.09730 [hep-ph]}\BibitemShut
  {NoStop}%
\bibitem [{\citenamefont {Forero}\ \emph {et~al.}(2021)\citenamefont {Forero},
  \citenamefont {Giunti}, \citenamefont {Ternes},\ and\ \citenamefont
  {Tortola}}]{Forero:2021azc}%
  \BibitemOpen
  \bibfield  {author} {\bibinfo {author} {\bibfnamefont {D.~V.}\ \bibnamefont
  {Forero}}, \bibinfo {author} {\bibfnamefont {C.}~\bibnamefont {Giunti}},
  \bibinfo {author} {\bibfnamefont {C.~A.}\ \bibnamefont {Ternes}}, \ and\
  \bibinfo {author} {\bibfnamefont {M.}~\bibnamefont {Tortola}},\ }\bibfield
  {title} {\emph {\enquote {\bibinfo {title} {{Nonunitary neutrino mixing in
  short and long-baseline experiments}},}\ }}\href {\doibase
  10.1103/PhysRevD.104.075030} {\bibfield  {journal} {\bibinfo  {journal}
  {Phys. Rev. D}\ }\textbf {\bibinfo {volume} {104}},\ \bibinfo {pages}
  {075030} (\bibinfo {year} {2021})},\ \Eprint
  {http://arxiv.org/abs/2103.01998}{arXiv:2103.01998 [hep-ph]}\BibitemShut
  {NoStop}%
\bibitem [{\citenamefont {Coloma}\ \emph
  {et~al.}(2021{\natexlab{a}})\citenamefont {Coloma}, \citenamefont
  {L\'opez-Pav\'on}, \citenamefont {Rosauro-Alcaraz},\ and\ \citenamefont
  {Urrea}}]{Coloma:2021uhq}%
  \BibitemOpen
  \bibfield  {author} {\bibinfo {author} {\bibfnamefont {P.}~\bibnamefont
  {Coloma}}, \bibinfo {author} {\bibfnamefont {J.}~\bibnamefont
  {L\'opez-Pav\'on}}, \bibinfo {author} {\bibfnamefont {S.}~\bibnamefont
  {Rosauro-Alcaraz}}, \ and\ \bibinfo {author} {\bibfnamefont {S.}~\bibnamefont
  {Urrea}},\ }\bibfield  {title} {\emph {\enquote {\bibinfo {title} {{New
  physics from oscillations at the DUNE near detector, and the role of
  systematic uncertainties}},}\ }}\href {\doibase 10.1007/JHEP08(2021)065} {\
  (\bibinfo {year} {2021}{\natexlab{a}}),\ 10.1007/JHEP08(2021)065},\ \Eprint
  {http://arxiv.org/abs/2105.11466}{arXiv:2105.11466 [hep-ph]}\BibitemShut
  {NoStop}%
\bibitem [{\citenamefont {Chatterjee}\ \emph {et~al.}(2022)\citenamefont
  {Chatterjee}, \citenamefont {Miranda}, \citenamefont {T\'ortola},\ and\
  \citenamefont {Valle}}]{Chatterjee:2021xyu}%
  \BibitemOpen
  \bibfield  {author} {\bibinfo {author} {\bibfnamefont {S.~S.}\ \bibnamefont
  {Chatterjee}}, \bibinfo {author} {\bibfnamefont {O.~G.}\ \bibnamefont
  {Miranda}}, \bibinfo {author} {\bibfnamefont {M.}~\bibnamefont {T\'ortola}},
  \ and\ \bibinfo {author} {\bibfnamefont {J.~W.~F.}\ \bibnamefont {Valle}},\
  }\bibfield  {title} {\emph {\enquote {\bibinfo {title} {{Nonunitarity of the
  lepton mixing matrix at the European Spallation Source}},}\ }}\href {\doibase
  10.1103/PhysRevD.106.075016} {\bibfield  {journal} {\bibinfo  {journal}
  {Phys. Rev. D}\ }\textbf {\bibinfo {volume} {106}},\ \bibinfo {pages}
  {075016} (\bibinfo {year} {2022})},\ \Eprint
  {http://arxiv.org/abs/2111.08673}{arXiv:2111.08673 [hep-ph]}\BibitemShut
  {NoStop}%
\bibitem [{\citenamefont {Okubo}(1962)}]{Okubo:1962zzc}%
  \BibitemOpen
  \bibfield  {author} {\bibinfo {author} {\bibfnamefont {S.}~\bibnamefont
  {Okubo}},\ }\bibfield  {title} {\emph {\enquote {\bibinfo {title} {{Note on
  Unitary Symmetry in Strong Interaction. II Excited States of Baryons}},}\
  }}\href {\doibase 10.1143/PTP.28.24} {\bibfield  {journal} {\bibinfo
  {journal} {Prog. Theor. Phys.}\ }\textbf {\bibinfo {volume} {28}},\ \bibinfo
  {pages} {24} (\bibinfo {year} {1962})}\BibitemShut {NoStop}%
\bibitem [{\citenamefont {Fernandez-Martinez}\ \emph
  {et~al.}(2016)\citenamefont {Fernandez-Martinez}, \citenamefont
  {Hernandez-Garcia},\ and\ \citenamefont
  {Lopez-Pavon}}]{Fernandez-Martinez:2016lgt}%
  \BibitemOpen
  \bibfield  {author} {\bibinfo {author} {\bibfnamefont {E.}~\bibnamefont
  {Fernandez-Martinez}}, \bibinfo {author} {\bibfnamefont {J.}~\bibnamefont
  {Hernandez-Garcia}}, \ and\ \bibinfo {author} {\bibfnamefont
  {J.}~\bibnamefont {Lopez-Pavon}},\ }\bibfield  {title} {\emph {\enquote
  {\bibinfo {title} {{Global constraints on heavy neutrino mixing}},}\ }}\href
  {\doibase 10.1007/JHEP08(2016)033} {\bibfield  {journal} {\bibinfo  {journal}
  {JHEP}\ }\textbf {\bibinfo {volume} {08}},\ \bibinfo {pages} {033} (\bibinfo
  {year} {2016})},\ \Eprint {http://arxiv.org/abs/1605.08774}{arXiv:1605.08774
  [hep-ph]}\BibitemShut {NoStop}%
\bibitem [{\citenamefont {Blennow}\ \emph {et~al.}(2023)\citenamefont
  {Blennow}, \citenamefont {Fern\'andez-Mart\'\i{}nez}, \citenamefont
  {Hern\'andez-Garc\'\i{}a}, \citenamefont {L\'opez-Pav\'on}, \citenamefont
  {Marcano},\ and\ \citenamefont {Naredo-Tuero}}]{Blennow:2023mqx}%
  \BibitemOpen
  \bibfield  {author} {\bibinfo {author} {\bibfnamefont {M.}~\bibnamefont
  {Blennow}}, \bibinfo {author} {\bibfnamefont {E.}~\bibnamefont
  {Fern\'andez-Mart\'\i{}nez}}, \bibinfo {author} {\bibfnamefont
  {J.}~\bibnamefont {Hern\'andez-Garc\'\i{}a}}, \bibinfo {author}
  {\bibfnamefont {J.}~\bibnamefont {L\'opez-Pav\'on}}, \bibinfo {author}
  {\bibfnamefont {X.}~\bibnamefont {Marcano}}, \ and\ \bibinfo {author}
  {\bibfnamefont {D.}~\bibnamefont {Naredo-Tuero}},\ }\bibfield  {title} {\emph
  {\enquote {\bibinfo {title} {{Bounds on lepton non-unitarity and heavy
  neutrino mixing}},}\ }}\href {\doibase 10.1007/JHEP08(2023)030} {\bibfield
  {journal} {\bibinfo  {journal} {JHEP}\ }\textbf {\bibinfo {volume} {08}},\
  \bibinfo {pages} {030} (\bibinfo {year} {2023})},\ \Eprint
  {http://arxiv.org/abs/2306.01040}{arXiv:2306.01040 [hep-ph]}\BibitemShut
  {NoStop}%
\bibitem [{\citenamefont {Schwingenheuer}\ \emph {et~al.}(1995)\citenamefont
  {Schwingenheuer}, \citenamefont {Briere}, \citenamefont {Barker},
  \citenamefont {Cheu}, \citenamefont {Gibbons}, \citenamefont {Harris},
  \citenamefont {Makoff}, \citenamefont {McFarland}, \citenamefont {Roodman},
  \citenamefont {Wah}, \citenamefont {Winstein}, \citenamefont {Winston},
  \citenamefont {Swallow}, \citenamefont {Bock}, \citenamefont {Coleman},
  \citenamefont {Crisler}, \citenamefont {Enagonio}, \citenamefont {Ford},
  \citenamefont {Hsiung}, \citenamefont {Jensen}, \citenamefont {Ramberg},
  \citenamefont {Tschirhart}, \citenamefont {Yamanaka}, \citenamefont
  {Collins}, \citenamefont {Gollin}, \citenamefont {Gu}, \citenamefont {Haas},
  \citenamefont {Hogan}, \citenamefont {Kim}, \citenamefont {Matthews},
  \citenamefont {Myung}, \citenamefont {Schnetzer}, \citenamefont {Somalwar},
  \citenamefont {Thomson},\ and\ \citenamefont {Zou}}]{PhysRevLett.74.4376}%
  \BibitemOpen
  \bibfield  {author} {\bibinfo {author} {\bibfnamefont {B.}~\bibnamefont
  {Schwingenheuer}}, \bibinfo {author} {\bibfnamefont {R.~A.}\ \bibnamefont
  {Briere}}, \bibinfo {author} {\bibfnamefont {A.~R.}\ \bibnamefont {Barker}},
  \bibinfo {author} {\bibfnamefont {E.}~\bibnamefont {Cheu}}, \bibinfo {author}
  {\bibfnamefont {L.~K.}\ \bibnamefont {Gibbons}}, \bibinfo {author}
  {\bibfnamefont {D.~A.}\ \bibnamefont {Harris}}, \bibinfo {author}
  {\bibfnamefont {G.}~\bibnamefont {Makoff}}, \bibinfo {author} {\bibfnamefont
  {K.~S.}\ \bibnamefont {McFarland}}, \bibinfo {author} {\bibfnamefont
  {A.}~\bibnamefont {Roodman}}, \bibinfo {author} {\bibfnamefont {Y.~W.}\
  \bibnamefont {Wah}}, \bibinfo {author} {\bibfnamefont {B.}~\bibnamefont
  {Winstein}}, \bibinfo {author} {\bibfnamefont {R.}~\bibnamefont {Winston}},
  \bibinfo {author} {\bibfnamefont {E.~C.}\ \bibnamefont {Swallow}}, \bibinfo
  {author} {\bibfnamefont {G.~J.}\ \bibnamefont {Bock}}, \bibinfo {author}
  {\bibfnamefont {R.}~\bibnamefont {Coleman}}, \bibinfo {author} {\bibfnamefont
  {M.}~\bibnamefont {Crisler}}, \bibinfo {author} {\bibfnamefont
  {J.}~\bibnamefont {Enagonio}}, \bibinfo {author} {\bibfnamefont
  {R.}~\bibnamefont {Ford}}, \bibinfo {author} {\bibfnamefont {Y.~B.}\
  \bibnamefont {Hsiung}}, \bibinfo {author} {\bibfnamefont {D.~A.}\
  \bibnamefont {Jensen}}, \bibinfo {author} {\bibfnamefont {E.}~\bibnamefont
  {Ramberg}}, \bibinfo {author} {\bibfnamefont {R.}~\bibnamefont {Tschirhart}},
  \bibinfo {author} {\bibfnamefont {T.}~\bibnamefont {Yamanaka}}, \bibinfo
  {author} {\bibfnamefont {E.~M.}\ \bibnamefont {Collins}}, \bibinfo {author}
  {\bibfnamefont {G.~D.}\ \bibnamefont {Gollin}}, \bibinfo {author}
  {\bibfnamefont {P.}~\bibnamefont {Gu}}, \bibinfo {author} {\bibfnamefont
  {P.}~\bibnamefont {Haas}}, \bibinfo {author} {\bibfnamefont {W.~P.}\
  \bibnamefont {Hogan}}, \bibinfo {author} {\bibfnamefont {S.~K.}\ \bibnamefont
  {Kim}}, \bibinfo {author} {\bibfnamefont {J.~N.}\ \bibnamefont {Matthews}},
  \bibinfo {author} {\bibfnamefont {S.~S.}\ \bibnamefont {Myung}}, \bibinfo
  {author} {\bibfnamefont {S.}~\bibnamefont {Schnetzer}}, \bibinfo {author}
  {\bibfnamefont {S.~V.}\ \bibnamefont {Somalwar}}, \bibinfo {author}
  {\bibfnamefont {G.~B.}\ \bibnamefont {Thomson}}, \ and\ \bibinfo {author}
  {\bibfnamefont {Y.}~\bibnamefont {Zou}},\ }\bibfield  {title} {\emph
  {\enquote {\bibinfo {title} {$\mathit{CPT}$ Tests in the Neutral Kaon
  System},}\ }}\href {\doibase 10.1103/PhysRevLett.74.4376} {\bibfield
  {journal} {\bibinfo  {journal} {Phys. Rev. Lett.}\ }\textbf {\bibinfo
  {volume} {74}},\ \bibinfo {pages} {4376} (\bibinfo {year}
  {1995})}\BibitemShut {NoStop}%
\bibitem [{\citenamefont {Barenboim}\ and\ \citenamefont
  {Lykken}(2009)}]{Barenboim:2009ts}%
  \BibitemOpen
  \bibfield  {author} {\bibinfo {author} {\bibfnamefont {G.}~\bibnamefont
  {Barenboim}}\ and\ \bibinfo {author} {\bibfnamefont {J.~D.}\ \bibnamefont
  {Lykken}},\ }\bibfield  {title} {\emph {\enquote {\bibinfo {title} {{MINOS
  and CPT-violating neutrinos}},}\ }}\href {\doibase
  10.1103/PhysRevD.80.113008} {\bibfield  {journal} {\bibinfo  {journal} {Phys.
  Rev. D}\ }\textbf {\bibinfo {volume} {80}},\ \bibinfo {pages} {113008}
  (\bibinfo {year} {2009})},\ \Eprint
  {http://arxiv.org/abs/0908.2993}{arXiv:0908.2993 [hep-ph]}\BibitemShut
  {NoStop}%
\bibitem [{\citenamefont {Barenboim}\ \emph {et~al.}(2018)\citenamefont
  {Barenboim}, \citenamefont {Ternes},\ and\ \citenamefont
  {T\'ortola}}]{Barenboim:2017ewj}%
  \BibitemOpen
  \bibfield  {author} {\bibinfo {author} {\bibfnamefont {G.}~\bibnamefont
  {Barenboim}}, \bibinfo {author} {\bibfnamefont {C.~A.}\ \bibnamefont
  {Ternes}}, \ and\ \bibinfo {author} {\bibfnamefont {M.}~\bibnamefont
  {T\'ortola}},\ }\bibfield  {title} {\emph {\enquote {\bibinfo {title}
  {{Neutrinos, DUNE and the world best bound on CPT invariance}},}\ }}\href
  {\doibase 10.1016/j.physletb.2018.03.060} {\bibfield  {journal} {\bibinfo
  {journal} {Phys. Lett. B}\ }\textbf {\bibinfo {volume} {780}},\ \bibinfo
  {pages} {631} (\bibinfo {year} {2018})},\ \Eprint
  {http://arxiv.org/abs/1712.01714}{arXiv:1712.01714 [hep-ph]}\BibitemShut
  {NoStop}%
\bibitem [{\citenamefont {Kaur}(2020)}]{Kaur:2020ggv}%
  \BibitemOpen
  \bibfield  {author} {\bibinfo {author} {\bibfnamefont {D.}~\bibnamefont
  {Kaur}},\ }\bibfield  {title} {\emph {\enquote {\bibinfo {title}
  {{Model-independent test for $CPT$ violation using long-baseline and
  atmospheric neutrino experiments}},}\ }}\href {\doibase
  10.1103/PhysRevD.101.055017} {\bibfield  {journal} {\bibinfo  {journal}
  {Phys. Rev. D}\ }\textbf {\bibinfo {volume} {101}},\ \bibinfo {pages}
  {055017} (\bibinfo {year} {2020})},\ \Eprint
  {http://arxiv.org/abs/2004.00349}{arXiv:2004.00349 [hep-ex]}\BibitemShut
  {NoStop}%
\bibitem [{\citenamefont {Ternes}(2021)}]{Ternes_2021}%
  \BibitemOpen
  \bibfield  {author} {\bibinfo {author} {\bibfnamefont {C.~A.}\ \bibnamefont
  {Ternes}},\ }\bibfield  {title} {\emph {\enquote {\bibinfo {title} {CPT
  violation in neutrino oscillations},}\ }}\href {\doibase
  10.1088/1742-6596/2156/1/012108} {\bibfield  {journal} {\bibinfo  {journal}
  {Journal of Physics: Conference Series}\ }\textbf {\bibinfo {volume}
  {2156}},\ \bibinfo {pages} {012108} (\bibinfo {year} {2021})}\BibitemShut
  {NoStop}%
\bibitem [{\citenamefont {Capolupo}\ \emph {et~al.}(2019)\citenamefont
  {Capolupo}, \citenamefont {Giampaolo},\ and\ \citenamefont
  {Lambiase}}]{CAPOLUPO2019298}%
  \BibitemOpen
  \bibfield  {author} {\bibinfo {author} {\bibfnamefont {A.}~\bibnamefont
  {Capolupo}}, \bibinfo {author} {\bibfnamefont {S.}~\bibnamefont {Giampaolo}},
  \ and\ \bibinfo {author} {\bibfnamefont {G.}~\bibnamefont {Lambiase}},\
  }\bibfield  {title} {\emph {\enquote {\bibinfo {title} {Decoherence in
  neutrino oscillations, neutrino nature and CPT violation},}\ }}\href
  {\doibase https://doi.org/10.1016/j.physletb.2019.03.062} {\bibfield
  {journal} {\bibinfo  {journal} {Physics Letters B}\ }\textbf {\bibinfo
  {volume} {792}},\ \bibinfo {pages} {298} (\bibinfo {year}
  {2019})}\BibitemShut {NoStop}%
\bibitem [{\citenamefont {Sahoo}\ \emph {et~al.}(2023)\citenamefont {Sahoo},
  \citenamefont {Kumar}, \citenamefont {Agarwalla},\ and\ \citenamefont
  {Dighe}}]{Sahoo:2022nbu}%
  \BibitemOpen
  \bibfield  {author} {\bibinfo {author} {\bibfnamefont {S.}~\bibnamefont
  {Sahoo}}, \bibinfo {author} {\bibfnamefont {A.}~\bibnamefont {Kumar}},
  \bibinfo {author} {\bibfnamefont {S.~K.}\ \bibnamefont {Agarwalla}}, \ and\
  \bibinfo {author} {\bibfnamefont {A.}~\bibnamefont {Dighe}},\ }\bibfield
  {title} {\emph {\enquote {\bibinfo {title} {{Discriminating between Lorentz
  violation and non-standard interactions using core-passing atmospheric
  neutrinos at INO-ICAL}},}\ }}\href {\doibase 10.1016/j.physletb.2023.137949}
  {\bibfield  {journal} {\bibinfo  {journal} {Phys. Lett. B}\ }\textbf
  {\bibinfo {volume} {841}},\ \bibinfo {pages} {137949} (\bibinfo {year}
  {2023})},\ \Eprint {http://arxiv.org/abs/2205.05134}{arXiv:2205.05134
  [hep-ph]}\BibitemShut {NoStop}%
\bibitem [{\citenamefont {Majhi}\ \emph {et~al.}(2023)\citenamefont {Majhi},
  \citenamefont {Singha}, \citenamefont {Ghosh},\ and\ \citenamefont
  {Mohanta}}]{Majhi:2022fed}%
  \BibitemOpen
  \bibfield  {author} {\bibinfo {author} {\bibfnamefont {R.}~\bibnamefont
  {Majhi}}, \bibinfo {author} {\bibfnamefont {D.~K.}\ \bibnamefont {Singha}},
  \bibinfo {author} {\bibfnamefont {M.}~\bibnamefont {Ghosh}}, \ and\ \bibinfo
  {author} {\bibfnamefont {R.}~\bibnamefont {Mohanta}},\ }\bibfield  {title}
  {\emph {\enquote {\bibinfo {title} {{Distinguishing nonstandard interaction
  and Lorentz invariance violation at the Protvino to super-ORCA
  experiment}},}\ }}\href {\doibase 10.1103/PhysRevD.107.075036} {\bibfield
  {journal} {\bibinfo  {journal} {Phys. Rev. D}\ }\textbf {\bibinfo {volume}
  {107}},\ \bibinfo {pages} {075036} (\bibinfo {year} {2023})},\ \Eprint
  {http://arxiv.org/abs/2212.07244}{arXiv:2212.07244 [hep-ph]}\BibitemShut
  {NoStop}%
\bibitem [{\citenamefont {Abe}\ \emph {et~al.}(2015{\natexlab{b}})\citenamefont
  {Abe} \emph {et~al.}}]{Super-Kamiokande:2014exs}%
  \BibitemOpen
  \bibfield  {author} {\bibinfo {author} {\bibfnamefont {K.}~\bibnamefont
  {Abe}} \emph {et~al.} (\bibinfo {collaboration} {Super-Kamiokande}),\
  }\bibfield  {title} {\emph {\enquote {\bibinfo {title} {{Test of Lorentz
  invariance with atmospheric neutrinos}},}\ }}\href {\doibase
  10.1103/PhysRevD.91.052003} {\bibfield  {journal} {\bibinfo  {journal} {Phys.
  Rev. D}\ }\textbf {\bibinfo {volume} {91}},\ \bibinfo {pages} {052003}
  (\bibinfo {year} {2015}{\natexlab{b}})},\ \Eprint
  {http://arxiv.org/abs/1410.4267}{arXiv:1410.4267 [hep-ex]}\BibitemShut
  {NoStop}%
\bibitem [{\citenamefont {Aartsen}\ \emph
  {et~al.}(2018{\natexlab{c}})\citenamefont {Aartsen} \emph
  {et~al.}}]{IceCube:2017qyp}%
  \BibitemOpen
  \bibfield  {author} {\bibinfo {author} {\bibfnamefont {M.~G.}\ \bibnamefont
  {Aartsen}} \emph {et~al.} (\bibinfo {collaboration} {IceCube}),\ }\bibfield
  {title} {\emph {\enquote {\bibinfo {title} {{Neutrino Interferometry for
  High-Precision Tests of Lorentz Symmetry with IceCube}},}\ }}\href {\doibase
  10.1038/s41567-018-0172-2} {\bibfield  {journal} {\bibinfo  {journal} {Nature
  Phys.}\ }\textbf {\bibinfo {volume} {14}},\ \bibinfo {pages} {961} (\bibinfo
  {year} {2018}{\natexlab{c}})},\ \Eprint
  {http://arxiv.org/abs/1709.03434}{arXiv:1709.03434 [hep-ex]}\BibitemShut
  {NoStop}%
\bibitem [{\citenamefont {Agarwalla}\ \emph {et~al.}(2023)\citenamefont
  {Agarwalla}, \citenamefont {Das}, \citenamefont {Sahoo},\ and\ \citenamefont
  {Swain}}]{Agarwalla:2023wft}%
  \BibitemOpen
  \bibfield  {author} {\bibinfo {author} {\bibfnamefont {S.~K.}\ \bibnamefont
  {Agarwalla}}, \bibinfo {author} {\bibfnamefont {S.}~\bibnamefont {Das}},
  \bibinfo {author} {\bibfnamefont {S.}~\bibnamefont {Sahoo}}, \ and\ \bibinfo
  {author} {\bibfnamefont {P.}~\bibnamefont {Swain}},\ }\bibfield  {title}
  {\emph {\enquote {\bibinfo {title} {{Constraining Lorentz invariance
  violation with next-generation long-baseline experiments}},}\ }}\href
  {\doibase 10.1007/JHEP07(2023)216} {\bibfield  {journal} {\bibinfo  {journal}
  {JHEP}\ }\textbf {\bibinfo {volume} {07}},\ \bibinfo {pages} {216} (\bibinfo
  {year} {2023})},\ \Eprint {http://arxiv.org/abs/2302.12005}{arXiv:2302.12005
  [hep-ph]}\BibitemShut {NoStop}%
\bibitem [{\citenamefont {Acker}\ \emph
  {et~al.}(1992{\natexlab{a}})\citenamefont {Acker}, \citenamefont
  {Joshipura},\ and\ \citenamefont {Pakvasa}}]{Acker:1992eh}%
  \BibitemOpen
  \bibfield  {author} {\bibinfo {author} {\bibfnamefont {A.}~\bibnamefont
  {Acker}}, \bibinfo {author} {\bibfnamefont {A.}~\bibnamefont {Joshipura}}, \
  and\ \bibinfo {author} {\bibfnamefont {S.}~\bibnamefont {Pakvasa}},\
  }\bibfield  {title} {\emph {\enquote {\bibinfo {title} {{A Neutrino decay
  model, solar anti-neutrinos and atmospheric neutrinos}},}\ }}\href {\doibase
  10.1016/0370-2693(92)91520-J} {\bibfield  {journal} {\bibinfo  {journal}
  {Phys. Lett. B}\ }\textbf {\bibinfo {volume} {285}},\ \bibinfo {pages} {371}
  (\bibinfo {year} {1992}{\natexlab{a}})}\BibitemShut {NoStop}%
\bibitem [{\citenamefont {Schechter}\ and\ \citenamefont
  {Valle}(1982)}]{PhysRevD.25.774}%
  \BibitemOpen
  \bibfield  {author} {\bibinfo {author} {\bibfnamefont {J.}~\bibnamefont
  {Schechter}}\ and\ \bibinfo {author} {\bibfnamefont {J.~W.~F.}\ \bibnamefont
  {Valle}},\ }\bibfield  {title} {\emph {\enquote {\bibinfo {title} {Neutrino
  decay and spontaneous violation of lepton number},}\ }}\href {\doibase
  10.1103/PhysRevD.25.774} {\bibfield  {journal} {\bibinfo  {journal} {Phys.
  Rev. D}\ }\textbf {\bibinfo {volume} {25}},\ \bibinfo {pages} {774} (\bibinfo
  {year} {1982})}\BibitemShut {NoStop}%
\bibitem [{\citenamefont {Nussinov}(1987)}]{NUSSINOV1987171}%
  \BibitemOpen
  \bibfield  {author} {\bibinfo {author} {\bibfnamefont {S.}~\bibnamefont
  {Nussinov}},\ }\bibfield  {title} {\emph {\enquote {\bibinfo {title} {Some
  comments on decaying neutrinos and the triplet majoron model},}\ }}\href
  {\doibase https://doi.org/10.1016/0370-2693(87)91548-6} {\bibfield  {journal}
  {\bibinfo  {journal} {Physics Letters B}\ }\textbf {\bibinfo {volume}
  {185}},\ \bibinfo {pages} {171} (\bibinfo {year} {1987})}\BibitemShut
  {NoStop}%
\bibitem [{\citenamefont {Kim}\ and\ \citenamefont {Lam}(1990)}]{Kim:1990km}%
  \BibitemOpen
  \bibfield  {author} {\bibinfo {author} {\bibfnamefont {C.~W.}\ \bibnamefont
  {Kim}}\ and\ \bibinfo {author} {\bibfnamefont {W.~P.}\ \bibnamefont {Lam}},\
  }\bibfield  {title} {\emph {\enquote {\bibinfo {title} {{Some remarks on
  neutrino decay via a Nambu-Goldstone boson}},}\ }}\href {\doibase
  10.1142/S0217732390000354} {\bibfield  {journal} {\bibinfo  {journal} {Mod.
  Phys. Lett. A}\ }\textbf {\bibinfo {volume} {5}},\ \bibinfo {pages} {297}
  (\bibinfo {year} {1990})}\BibitemShut {NoStop}%
\bibitem [{\citenamefont {Biller}\ \emph {et~al.}(1998)\citenamefont {Biller}
  \emph {et~al.}}]{Biller:1998nc}%
  \BibitemOpen
  \bibfield  {author} {\bibinfo {author} {\bibfnamefont {S.~D.}\ \bibnamefont
  {Biller}} \emph {et~al.},\ }\bibfield  {title} {\emph {\enquote {\bibinfo
  {title} {{New limits to the IR background: Bounds on radiative neutrino decay
  and on VMO contributions to the dark matter problem}},}\ }}\href {\doibase
  10.1103/PhysRevLett.80.2992} {\bibfield  {journal} {\bibinfo  {journal}
  {Phys. Rev. Lett.}\ }\textbf {\bibinfo {volume} {80}},\ \bibinfo {pages}
  {2992} (\bibinfo {year} {1998})},\ \Eprint
  {http://arxiv.org/abs/astro-ph/9802234}{arXiv:astro-ph/9802234}\BibitemShut
  {NoStop}%
\bibitem [{\citenamefont {Gonzalez-Garcia}\ and\ \citenamefont
  {Maltoni}(2008)}]{Gonzalez-Garcia:2008mgl}%
  \BibitemOpen
  \bibfield  {author} {\bibinfo {author} {\bibfnamefont {M.~C.}\ \bibnamefont
  {Gonzalez-Garcia}}\ and\ \bibinfo {author} {\bibfnamefont {M.}~\bibnamefont
  {Maltoni}},\ }\bibfield  {title} {\emph {\enquote {\bibinfo {title} {{Status
  of Oscillation plus Decay of Atmospheric and Long-Baseline Neutrinos}},}\
  }}\href {\doibase 10.1016/j.physletb.2008.04.041} {\bibfield  {journal}
  {\bibinfo  {journal} {Phys. Lett. B}\ }\textbf {\bibinfo {volume} {663}},\
  \bibinfo {pages} {405} (\bibinfo {year} {2008})},\ \Eprint
  {http://arxiv.org/abs/0802.3699}{arXiv:0802.3699 [hep-ph]}\BibitemShut
  {NoStop}%
\bibitem [{\citenamefont {Berryman}\ \emph
  {et~al.}(2015{\natexlab{b}})\citenamefont {Berryman}, \citenamefont
  {de~Gouvea},\ and\ \citenamefont {Hernandez}}]{Berryman:2014qha}%
  \BibitemOpen
  \bibfield  {author} {\bibinfo {author} {\bibfnamefont {J.~M.}\ \bibnamefont
  {Berryman}}, \bibinfo {author} {\bibfnamefont {A.}~\bibnamefont {de~Gouvea}},
  \ and\ \bibinfo {author} {\bibfnamefont {D.}~\bibnamefont {Hernandez}},\
  }\bibfield  {title} {\emph {\enquote {\bibinfo {title} {{Solar Neutrinos and
  the Decaying Neutrino Hypothesis}},}\ }}\href {\doibase
  10.1103/PhysRevD.92.073003} {\bibfield  {journal} {\bibinfo  {journal} {Phys.
  Rev. D}\ }\textbf {\bibinfo {volume} {92}},\ \bibinfo {pages} {073003}
  (\bibinfo {year} {2015}{\natexlab{b}})},\ \Eprint
  {http://arxiv.org/abs/1411.0308}{arXiv:1411.0308 [hep-ph]}\BibitemShut
  {NoStop}%
\bibitem [{\citenamefont {Picoreti}\ \emph {et~al.}(2016)\citenamefont
  {Picoreti}, \citenamefont {Guzzo}, \citenamefont {de~Holanda},\ and\
  \citenamefont {Peres}}]{Picoreti:2015ika}%
  \BibitemOpen
  \bibfield  {author} {\bibinfo {author} {\bibfnamefont {R.}~\bibnamefont
  {Picoreti}}, \bibinfo {author} {\bibfnamefont {M.~M.}\ \bibnamefont {Guzzo}},
  \bibinfo {author} {\bibfnamefont {P.~C.}\ \bibnamefont {de~Holanda}}, \ and\
  \bibinfo {author} {\bibfnamefont {O.~L.~G.}\ \bibnamefont {Peres}},\
  }\bibfield  {title} {\emph {\enquote {\bibinfo {title} {{Neutrino Decay and
  Solar Neutrino Seasonal Effect}},}\ }}\href {\doibase
  10.1016/j.physletb.2016.08.007} {\bibfield  {journal} {\bibinfo  {journal}
  {Phys. Lett. B}\ }\textbf {\bibinfo {volume} {761}},\ \bibinfo {pages} {70}
  (\bibinfo {year} {2016})},\ \Eprint
  {http://arxiv.org/abs/1506.08158}{arXiv:1506.08158 [hep-ph]}\BibitemShut
  {NoStop}%
\bibitem [{\citenamefont {Frieman}\ \emph {et~al.}(1988)\citenamefont
  {Frieman}, \citenamefont {Haber},\ and\ \citenamefont
  {Freese}}]{FRIEMAN1988115}%
  \BibitemOpen
  \bibfield  {author} {\bibinfo {author} {\bibfnamefont {J.~A.}\ \bibnamefont
  {Frieman}}, \bibinfo {author} {\bibfnamefont {H.~E.}\ \bibnamefont {Haber}},
  \ and\ \bibinfo {author} {\bibfnamefont {K.}~\bibnamefont {Freese}},\
  }\bibfield  {title} {\emph {\enquote {\bibinfo {title} {Neutrino mixing,
  decays and supernova 1987A},}\ }}\href {\doibase
  https://doi.org/10.1016/0370-2693(88)91120-3} {\bibfield  {journal} {\bibinfo
   {journal} {Physics Letters B}\ }\textbf {\bibinfo {volume} {200}},\ \bibinfo
  {pages} {115} (\bibinfo {year} {1988})}\BibitemShut {NoStop}%
\bibitem [{\citenamefont {Hirata}\ \emph {et~al.}(1987)\citenamefont {Hirata},
  \citenamefont {Kajita}, \citenamefont {Koshiba}, \citenamefont {Nakahata},
  \citenamefont {Oyama}, \citenamefont {Sato}, \citenamefont {Suzuki},
  \citenamefont {Takita}, \citenamefont {Totsuka}, \citenamefont {Kifune},
  \citenamefont {Suda}, \citenamefont {Takahashi}, \citenamefont {Tanimori},
  \citenamefont {Miyano}, \citenamefont {Yamada}, \citenamefont {Beier},
  \citenamefont {Feldscher}, \citenamefont {Kim}, \citenamefont {Mann},
  \citenamefont {Newcomer}, \citenamefont {Van}, \citenamefont {Zhang},\ and\
  \citenamefont {Cortez}}]{PhysRevLett.58.1490}%
  \BibitemOpen
  \bibfield  {author} {\bibinfo {author} {\bibfnamefont {K.}~\bibnamefont
  {Hirata}}, \bibinfo {author} {\bibfnamefont {T.}~\bibnamefont {Kajita}},
  \bibinfo {author} {\bibfnamefont {M.}~\bibnamefont {Koshiba}}, \bibinfo
  {author} {\bibfnamefont {M.}~\bibnamefont {Nakahata}}, \bibinfo {author}
  {\bibfnamefont {Y.}~\bibnamefont {Oyama}}, \bibinfo {author} {\bibfnamefont
  {N.}~\bibnamefont {Sato}}, \bibinfo {author} {\bibfnamefont {A.}~\bibnamefont
  {Suzuki}}, \bibinfo {author} {\bibfnamefont {M.}~\bibnamefont {Takita}},
  \bibinfo {author} {\bibfnamefont {Y.}~\bibnamefont {Totsuka}}, \bibinfo
  {author} {\bibfnamefont {T.}~\bibnamefont {Kifune}}, \bibinfo {author}
  {\bibfnamefont {T.}~\bibnamefont {Suda}}, \bibinfo {author} {\bibfnamefont
  {K.}~\bibnamefont {Takahashi}}, \bibinfo {author} {\bibfnamefont
  {T.}~\bibnamefont {Tanimori}}, \bibinfo {author} {\bibfnamefont
  {K.}~\bibnamefont {Miyano}}, \bibinfo {author} {\bibfnamefont
  {M.}~\bibnamefont {Yamada}}, \bibinfo {author} {\bibfnamefont {E.~W.}\
  \bibnamefont {Beier}}, \bibinfo {author} {\bibfnamefont {L.~R.}\ \bibnamefont
  {Feldscher}}, \bibinfo {author} {\bibfnamefont {S.~B.}\ \bibnamefont {Kim}},
  \bibinfo {author} {\bibfnamefont {A.~K.}\ \bibnamefont {Mann}}, \bibinfo
  {author} {\bibfnamefont {F.~M.}\ \bibnamefont {Newcomer}}, \bibinfo {author}
  {\bibfnamefont {R.}~\bibnamefont {Van}}, \bibinfo {author} {\bibfnamefont
  {W.}~\bibnamefont {Zhang}}, \ and\ \bibinfo {author} {\bibfnamefont {B.~G.}\
  \bibnamefont {Cortez}},\ }\bibfield  {title} {\emph {\enquote {\bibinfo
  {title} {Observation of a neutrino burst from the supernova SN1987A},}\
  }}\href {\doibase 10.1103/PhysRevLett.58.1490} {\bibfield  {journal}
  {\bibinfo  {journal} {Phys. Rev. Lett.}\ }\textbf {\bibinfo {volume} {58}},\
  \bibinfo {pages} {1490} (\bibinfo {year} {1987})}\BibitemShut {NoStop}%
\bibitem [{\citenamefont {Bionta}\ \emph {et~al.}(1987)\citenamefont {Bionta},
  \citenamefont {Blewitt}, \citenamefont {Bratton}, \citenamefont {Casper},
  \citenamefont {Ciocio}, \citenamefont {Claus}, \citenamefont {Cortez},
  \citenamefont {Crouch}, \citenamefont {Dye}, \citenamefont {Errede},
  \citenamefont {Foster}, \citenamefont {Gajewski}, \citenamefont {Ganezer},
  \citenamefont {Goldhaber}, \citenamefont {Haines}, \citenamefont {Jones},
  \citenamefont {Kielczewska}, \citenamefont {Kropp}, \citenamefont {Learned},
  \citenamefont {LoSecco}, \citenamefont {Matthews}, \citenamefont {Miller},
  \citenamefont {Mudan}, \citenamefont {Park}, \citenamefont {Price},
  \citenamefont {Reines}, \citenamefont {Schultz}, \citenamefont {Seidel},
  \citenamefont {Shumard}, \citenamefont {Sinclair}, \citenamefont {Sobel},
  \citenamefont {Stone}, \citenamefont {Sulak}, \citenamefont {Svoboda},
  \citenamefont {Thornton}, \citenamefont {van~der Velde},\ and\ \citenamefont
  {Wuest}}]{PhysRevLett.58.1494}%
  \BibitemOpen
  \bibfield  {author} {\bibinfo {author} {\bibfnamefont {R.~M.}\ \bibnamefont
  {Bionta}}, \bibinfo {author} {\bibfnamefont {G.}~\bibnamefont {Blewitt}},
  \bibinfo {author} {\bibfnamefont {C.~B.}\ \bibnamefont {Bratton}}, \bibinfo
  {author} {\bibfnamefont {D.}~\bibnamefont {Casper}}, \bibinfo {author}
  {\bibfnamefont {A.}~\bibnamefont {Ciocio}}, \bibinfo {author} {\bibfnamefont
  {R.}~\bibnamefont {Claus}}, \bibinfo {author} {\bibfnamefont
  {B.}~\bibnamefont {Cortez}}, \bibinfo {author} {\bibfnamefont
  {M.}~\bibnamefont {Crouch}}, \bibinfo {author} {\bibfnamefont {S.~T.}\
  \bibnamefont {Dye}}, \bibinfo {author} {\bibfnamefont {S.}~\bibnamefont
  {Errede}}, \bibinfo {author} {\bibfnamefont {G.~W.}\ \bibnamefont {Foster}},
  \bibinfo {author} {\bibfnamefont {W.}~\bibnamefont {Gajewski}}, \bibinfo
  {author} {\bibfnamefont {K.~S.}\ \bibnamefont {Ganezer}}, \bibinfo {author}
  {\bibfnamefont {M.}~\bibnamefont {Goldhaber}}, \bibinfo {author}
  {\bibfnamefont {T.~J.}\ \bibnamefont {Haines}}, \bibinfo {author}
  {\bibfnamefont {T.~W.}\ \bibnamefont {Jones}}, \bibinfo {author}
  {\bibfnamefont {D.}~\bibnamefont {Kielczewska}}, \bibinfo {author}
  {\bibfnamefont {W.~R.}\ \bibnamefont {Kropp}}, \bibinfo {author}
  {\bibfnamefont {J.~G.}\ \bibnamefont {Learned}}, \bibinfo {author}
  {\bibfnamefont {J.~M.}\ \bibnamefont {LoSecco}}, \bibinfo {author}
  {\bibfnamefont {J.}~\bibnamefont {Matthews}}, \bibinfo {author}
  {\bibfnamefont {R.}~\bibnamefont {Miller}}, \bibinfo {author} {\bibfnamefont
  {M.~S.}\ \bibnamefont {Mudan}}, \bibinfo {author} {\bibfnamefont {H.~S.}\
  \bibnamefont {Park}}, \bibinfo {author} {\bibfnamefont {L.~R.}\ \bibnamefont
  {Price}}, \bibinfo {author} {\bibfnamefont {F.}~\bibnamefont {Reines}},
  \bibinfo {author} {\bibfnamefont {J.}~\bibnamefont {Schultz}}, \bibinfo
  {author} {\bibfnamefont {S.}~\bibnamefont {Seidel}}, \bibinfo {author}
  {\bibfnamefont {E.}~\bibnamefont {Shumard}}, \bibinfo {author} {\bibfnamefont
  {D.}~\bibnamefont {Sinclair}}, \bibinfo {author} {\bibfnamefont {H.~W.}\
  \bibnamefont {Sobel}}, \bibinfo {author} {\bibfnamefont {J.~L.}\ \bibnamefont
  {Stone}}, \bibinfo {author} {\bibfnamefont {L.~R.}\ \bibnamefont {Sulak}},
  \bibinfo {author} {\bibfnamefont {R.}~\bibnamefont {Svoboda}}, \bibinfo
  {author} {\bibfnamefont {G.}~\bibnamefont {Thornton}}, \bibinfo {author}
  {\bibfnamefont {J.~C.}\ \bibnamefont {van~der Velde}}, \ and\ \bibinfo
  {author} {\bibfnamefont {C.}~\bibnamefont {Wuest}},\ }\bibfield  {title}
  {\emph {\enquote {\bibinfo {title} {Observation of a neutrino burst in
  coincidence with supernova 1987A in the Large Magellanic Cloud},}\ }}\href
  {\doibase 10.1103/PhysRevLett.58.1494} {\bibfield  {journal} {\bibinfo
  {journal} {Phys. Rev. Lett.}\ }\textbf {\bibinfo {volume} {58}},\ \bibinfo
  {pages} {1494} (\bibinfo {year} {1987})}\BibitemShut {NoStop}%
\bibitem [{\citenamefont {Chikashige}\ \emph {et~al.}(1981)\citenamefont
  {Chikashige}, \citenamefont {Mohapatra},\ and\ \citenamefont
  {Peccei}}]{CHIKASHIGE1981265}%
  \BibitemOpen
  \bibfield  {author} {\bibinfo {author} {\bibfnamefont {Y.}~\bibnamefont
  {Chikashige}}, \bibinfo {author} {\bibfnamefont {R.}~\bibnamefont
  {Mohapatra}}, \ and\ \bibinfo {author} {\bibfnamefont {R.}~\bibnamefont
  {Peccei}},\ }\bibfield  {title} {\emph {\enquote {\bibinfo {title} {Are there
  real goldstone bosons associated with broken lepton number?}}\ }}\href
  {\doibase https://doi.org/10.1016/0370-2693(81)90011-3} {\bibfield  {journal}
  {\bibinfo  {journal} {Physics Letters B}\ }\textbf {\bibinfo {volume} {98}},\
  \bibinfo {pages} {265} (\bibinfo {year} {1981})}\BibitemShut {NoStop}%
\bibitem [{\citenamefont {Gelmini}\ and\ \citenamefont
  {Roncadelli}(1981)}]{GELMINI1981411}%
  \BibitemOpen
  \bibfield  {author} {\bibinfo {author} {\bibfnamefont {G.}~\bibnamefont
  {Gelmini}}\ and\ \bibinfo {author} {\bibfnamefont {M.}~\bibnamefont
  {Roncadelli}},\ }\bibfield  {title} {\emph {\enquote {\bibinfo {title}
  {Left-handed neutrino mass scale and spontaneously broken lepton number},}\
  }}\href {\doibase https://doi.org/10.1016/0370-2693(81)90559-1} {\bibfield
  {journal} {\bibinfo  {journal} {Physics Letters B}\ }\textbf {\bibinfo
  {volume} {99}},\ \bibinfo {pages} {411} (\bibinfo {year} {1981})}\BibitemShut
  {NoStop}%
\bibitem [{\citenamefont {Acker}\ \emph
  {et~al.}(1992{\natexlab{b}})\citenamefont {Acker}, \citenamefont {Pakvasa},\
  and\ \citenamefont {Pantaleone}}]{PhysRevD.45.R1}%
  \BibitemOpen
  \bibfield  {author} {\bibinfo {author} {\bibfnamefont {A.}~\bibnamefont
  {Acker}}, \bibinfo {author} {\bibfnamefont {S.}~\bibnamefont {Pakvasa}}, \
  and\ \bibinfo {author} {\bibfnamefont {J.}~\bibnamefont {Pantaleone}},\
  }\bibfield  {title} {\emph {\enquote {\bibinfo {title} {Decaying Dirac
  neutrinos},}\ }}\href {\doibase 10.1103/PhysRevD.45.R1} {\bibfield  {journal}
  {\bibinfo  {journal} {Phys. Rev. D}\ }\textbf {\bibinfo {volume} {45}},\
  \bibinfo {pages} {R1} (\bibinfo {year} {1992}{\natexlab{b}})}\BibitemShut
  {NoStop}%
\bibitem [{\citenamefont {Coloma}\ and\ \citenamefont
  {Peres}(2017)}]{Coloma:2017zpg}%
  \BibitemOpen
  \bibfield  {author} {\bibinfo {author} {\bibfnamefont {P.}~\bibnamefont
  {Coloma}}\ and\ \bibinfo {author} {\bibfnamefont {O.~L.~G.}\ \bibnamefont
  {Peres}},\ }\bibfield  {title} {\emph {\enquote {\bibinfo {title} {{Visible
  neutrino decay at DUNE}},}\ }}\href@noop {} {\  (\bibinfo {year} {2017})},\
  \Eprint {http://arxiv.org/abs/1705.03599}{arXiv:1705.03599
  [hep-ph]}\BibitemShut {NoStop}%
\bibitem [{\citenamefont {Petcov}\ and\ \citenamefont
  {Toshev}(1987)}]{Petcov:1986qg}%
  \BibitemOpen
  \bibfield  {author} {\bibinfo {author} {\bibfnamefont {S.~T.}\ \bibnamefont
  {Petcov}}\ and\ \bibinfo {author} {\bibfnamefont {S.}~\bibnamefont
  {Toshev}},\ }\bibfield  {title} {\emph {\enquote {\bibinfo {title} {{Three
  Neutrino Oscillations in Matter: Analytical Results in the Adiabatic
  Approximation}},}\ }}\href {\doibase 10.1016/0370-2693(87)90083-9} {\bibfield
   {journal} {\bibinfo  {journal} {Phys. Lett. B}\ }\textbf {\bibinfo {volume}
  {187}},\ \bibinfo {pages} {120} (\bibinfo {year} {1987})}\BibitemShut
  {NoStop}%
\bibitem [{\citenamefont {Kim}\ and\ \citenamefont {Sze}(1987)}]{Kim:1986vg}%
  \BibitemOpen
  \bibfield  {author} {\bibinfo {author} {\bibfnamefont {C.~W.}\ \bibnamefont
  {Kim}}\ and\ \bibinfo {author} {\bibfnamefont {W.~K.}\ \bibnamefont {Sze}},\
  }\bibfield  {title} {\emph {\enquote {\bibinfo {title} {{Adiabatic Resonant
  Oscillations of Solar Neutrinos in Three Generations}},}\ }}\href {\doibase
  10.1103/PhysRevD.35.1404} {\bibfield  {journal} {\bibinfo  {journal} {Phys.
  Rev. D}\ }\textbf {\bibinfo {volume} {35}},\ \bibinfo {pages} {1404}
  (\bibinfo {year} {1987})}\BibitemShut {NoStop}%
\bibitem [{\citenamefont {Arafune}\ and\ \citenamefont
  {Sato}(1997)}]{Arafune:1996bt}%
  \BibitemOpen
  \bibfield  {author} {\bibinfo {author} {\bibfnamefont {J.}~\bibnamefont
  {Arafune}}\ and\ \bibinfo {author} {\bibfnamefont {J.}~\bibnamefont {Sato}},\
  }\bibfield  {title} {\emph {\enquote {\bibinfo {title} {{CP and T violation
  test in neutrino oscillation}},}\ }}\href {\doibase 10.1103/PhysRevD.55.1653}
  {\bibfield  {journal} {\bibinfo  {journal} {Phys. Rev. D}\ }\textbf {\bibinfo
  {volume} {55}},\ \bibinfo {pages} {1653} (\bibinfo {year} {1997})},\ \Eprint
  {http://arxiv.org/abs/hep-ph/9607437}{arXiv:hep-ph/9607437}\BibitemShut
  {NoStop}%
\bibitem [{\citenamefont {Arafune}\ \emph {et~al.}(1997)\citenamefont
  {Arafune}, \citenamefont {Koike},\ and\ \citenamefont
  {Sato}}]{Arafune:1997hd}%
  \BibitemOpen
  \bibfield  {author} {\bibinfo {author} {\bibfnamefont {J.}~\bibnamefont
  {Arafune}}, \bibinfo {author} {\bibfnamefont {M.}~\bibnamefont {Koike}}, \
  and\ \bibinfo {author} {\bibfnamefont {J.}~\bibnamefont {Sato}},\ }\bibfield
  {title} {\emph {\enquote {\bibinfo {title} {{CP violation and matter effect
  in long baseline neutrino oscillation experiments}},}\ }}\href {\doibase
  10.1103/PhysRevD.60.119905} {\bibfield  {journal} {\bibinfo  {journal} {Phys.
  Rev. D}\ }\textbf {\bibinfo {volume} {56}},\ \bibinfo {pages} {3093}
  (\bibinfo {year} {1997})},\ \bibinfo {note} {[Erratum: Phys.Rev.D 60, 119905
  (1999)]},\ \Eprint
  {http://arxiv.org/abs/hep-ph/9703351}{arXiv:hep-ph/9703351}\BibitemShut
  {NoStop}%
\bibitem [{\citenamefont {Ohlsson}\ and\ \citenamefont
  {Snellman}(2000)}]{Ohlsson:1999xb}%
  \BibitemOpen
  \bibfield  {author} {\bibinfo {author} {\bibfnamefont {T.}~\bibnamefont
  {Ohlsson}}\ and\ \bibinfo {author} {\bibfnamefont {H.}~\bibnamefont
  {Snellman}},\ }\bibfield  {title} {\emph {\enquote {\bibinfo {title} {{Three
  flavor neutrino oscillations in matter}},}\ }}\href {\doibase
  10.1063/1.533270} {\bibfield  {journal} {\bibinfo  {journal} {J. Math.
  Phys.}\ }\textbf {\bibinfo {volume} {41}},\ \bibinfo {pages} {2768} (\bibinfo
  {year} {2000})},\ \bibinfo {note} {[Erratum: J.Math.Phys. 42, 2345 (2001)]},\
  \Eprint
  {http://arxiv.org/abs/hep-ph/9910546}{arXiv:hep-ph/9910546}\BibitemShut
  {NoStop}%
\bibitem [{\citenamefont {Freund}(2001)}]{Freund:2001pn}%
  \BibitemOpen
  \bibfield  {author} {\bibinfo {author} {\bibfnamefont {M.}~\bibnamefont
  {Freund}},\ }\bibfield  {title} {\emph {\enquote {\bibinfo {title} {{Analytic
  approximations for three neutrino oscillation parameters and probabilities in
  matter}},}\ }}\href {\doibase 10.1103/PhysRevD.64.053003} {\bibfield
  {journal} {\bibinfo  {journal} {Phys. Rev. D}\ }\textbf {\bibinfo {volume}
  {64}},\ \bibinfo {pages} {053003} (\bibinfo {year} {2001})},\ \Eprint
  {http://arxiv.org/abs/hep-ph/0103300}{arXiv:hep-ph/0103300}\BibitemShut
  {NoStop}%
\bibitem [{\citenamefont {Cervera}\ \emph {et~al.}(2000)\citenamefont
  {Cervera}, \citenamefont {Donini}, \citenamefont {Gavela}, \citenamefont
  {Gomez~Cadenas}, \citenamefont {Hernandez}, \citenamefont {Mena},\ and\
  \citenamefont {Rigolin}}]{Cervera:2000kp}%
  \BibitemOpen
  \bibfield  {author} {\bibinfo {author} {\bibfnamefont {A.}~\bibnamefont
  {Cervera}}, \bibinfo {author} {\bibfnamefont {A.}~\bibnamefont {Donini}},
  \bibinfo {author} {\bibfnamefont {M.}~\bibnamefont {Gavela}}, \bibinfo
  {author} {\bibfnamefont {J.}~\bibnamefont {Gomez~Cadenas}}, \bibinfo {author}
  {\bibfnamefont {P.}~\bibnamefont {Hernandez}}, \bibinfo {author}
  {\bibfnamefont {O.}~\bibnamefont {Mena}}, \ and\ \bibinfo {author}
  {\bibfnamefont {S.}~\bibnamefont {Rigolin}},\ }\bibfield  {title} {\emph
  {\enquote {\bibinfo {title} {{Golden measurements at a neutrino factory}},}\
  }}\href {\doibase 10.1016/S0550-3213(00)00221-2} {\bibfield  {journal}
  {\bibinfo  {journal} {Nucl. Phys. B}\ }\textbf {\bibinfo {volume} {579}},\
  \bibinfo {pages} {17} (\bibinfo {year} {2000})},\ \bibinfo {note} {[Erratum:
  Nucl.Phys.B 593, 731--732 (2001)]},\ \Eprint
  {http://arxiv.org/abs/hep-ph/0002108}{arXiv:hep-ph/0002108}\BibitemShut
  {NoStop}%
\bibitem [{\citenamefont {Akhmedov}\ \emph {et~al.}(2004)\citenamefont
  {Akhmedov}, \citenamefont {Johansson}, \citenamefont {Lindner}, \citenamefont
  {Ohlsson},\ and\ \citenamefont {Schwetz}}]{Akhmedov:2004ny}%
  \BibitemOpen
  \bibfield  {author} {\bibinfo {author} {\bibfnamefont {E.~K.}\ \bibnamefont
  {Akhmedov}}, \bibinfo {author} {\bibfnamefont {R.}~\bibnamefont {Johansson}},
  \bibinfo {author} {\bibfnamefont {M.}~\bibnamefont {Lindner}}, \bibinfo
  {author} {\bibfnamefont {T.}~\bibnamefont {Ohlsson}}, \ and\ \bibinfo
  {author} {\bibfnamefont {T.}~\bibnamefont {Schwetz}},\ }\bibfield  {title}
  {\emph {\enquote {\bibinfo {title} {{Series expansions for three flavor
  neutrino oscillation probabilities in matter}},}\ }}\href {\doibase
  10.1088/1126-6708/2004/04/078} {\bibfield  {journal} {\bibinfo  {journal}
  {JHEP}\ }\textbf {\bibinfo {volume} {04}},\ \bibinfo {pages} {078} (\bibinfo
  {year} {2004})},\ \Eprint
  {http://arxiv.org/abs/hep-ph/0402175}{arXiv:hep-ph/0402175}\BibitemShut
  {NoStop}%
\bibitem [{\citenamefont {Asano}\ and\ \citenamefont
  {Minakata}(2011)}]{Asano:2011nj}%
  \BibitemOpen
  \bibfield  {author} {\bibinfo {author} {\bibfnamefont {K.}~\bibnamefont
  {Asano}}\ and\ \bibinfo {author} {\bibfnamefont {H.}~\bibnamefont
  {Minakata}},\ }\bibfield  {title} {\emph {\enquote {\bibinfo {title}
  {{Large-Theta(13) Perturbation Theory of Neutrino Oscillation for
  Long-Baseline Experiments}},}\ }}\href {\doibase 10.1007/JHEP06(2011)022}
  {\bibfield  {journal} {\bibinfo  {journal} {JHEP}\ }\textbf {\bibinfo
  {volume} {06}},\ \bibinfo {pages} {022} (\bibinfo {year} {2011})},\ \Eprint
  {http://arxiv.org/abs/1103.4387}{arXiv:1103.4387 [hep-ph]}\BibitemShut
  {NoStop}%
\bibitem [{\citenamefont {Minakata}\ and\ \citenamefont
  {Parke}(2016)}]{Minakata:2015gra}%
  \BibitemOpen
  \bibfield  {author} {\bibinfo {author} {\bibfnamefont {H.}~\bibnamefont
  {Minakata}}\ and\ \bibinfo {author} {\bibfnamefont {S.~J.}\ \bibnamefont
  {Parke}},\ }\bibfield  {title} {\emph {\enquote {\bibinfo {title} {{Simple
  and Compact Expressions for Neutrino Oscillation Probabilities in Matter}},}\
  }}\href {\doibase 10.1007/JHEP01(2016)180} {\bibfield  {journal} {\bibinfo
  {journal} {JHEP}\ }\textbf {\bibinfo {volume} {01}},\ \bibinfo {pages} {180}
  (\bibinfo {year} {2016})},\ \Eprint
  {http://arxiv.org/abs/1505.01826}{arXiv:1505.01826 [hep-ph]}\BibitemShut
  {NoStop}%
\bibitem [{\citenamefont {Denton}\ \emph {et~al.}(2016)\citenamefont {Denton},
  \citenamefont {Minakata},\ and\ \citenamefont {Parke}}]{Denton:2016wmg}%
  \BibitemOpen
  \bibfield  {author} {\bibinfo {author} {\bibfnamefont {P.~B.}\ \bibnamefont
  {Denton}}, \bibinfo {author} {\bibfnamefont {H.}~\bibnamefont {Minakata}}, \
  and\ \bibinfo {author} {\bibfnamefont {S.~J.}\ \bibnamefont {Parke}},\
  }\bibfield  {title} {\emph {\enquote {\bibinfo {title} {{Compact Perturbative
  Expressions For Neutrino Oscillations in Matter}},}\ }}\href {\doibase
  10.1007/JHEP06(2016)051} {\bibfield  {journal} {\bibinfo  {journal} {JHEP}\
  }\textbf {\bibinfo {volume} {06}},\ \bibinfo {pages} {051} (\bibinfo {year}
  {2016})},\ \Eprint {http://arxiv.org/abs/1604.08167}{arXiv:1604.08167
  [hep-ph]}\BibitemShut {NoStop}%
\bibitem [{\citenamefont {Gonzalez-Garcia}\ \emph
  {et~al.}(2001{\natexlab{b}})\citenamefont {Gonzalez-Garcia}, \citenamefont
  {Grossman}, \citenamefont {Gusso},\ and\ \citenamefont
  {Nir}}]{GonzalezGarcia:2001mp}%
  \BibitemOpen
  \bibfield  {author} {\bibinfo {author} {\bibfnamefont {M.}~\bibnamefont
  {Gonzalez-Garcia}}, \bibinfo {author} {\bibfnamefont {Y.}~\bibnamefont
  {Grossman}}, \bibinfo {author} {\bibfnamefont {A.}~\bibnamefont {Gusso}}, \
  and\ \bibinfo {author} {\bibfnamefont {Y.}~\bibnamefont {Nir}},\ }\bibfield
  {title} {\emph {\enquote {\bibinfo {title} {{New CP violation in neutrino
  oscillations}},}\ }}\href {\doibase 10.1103/PhysRevD.64.096006} {\bibfield
  {journal} {\bibinfo  {journal} {Phys. Rev. D}\ }\textbf {\bibinfo {volume}
  {64}},\ \bibinfo {pages} {096006} (\bibinfo {year} {2001}{\natexlab{b}})},\
  \Eprint
  {http://arxiv.org/abs/hep-ph/0105159}{arXiv:hep-ph/0105159}\BibitemShut
  {NoStop}%
\bibitem [{\citenamefont {Ota}\ \emph {et~al.}(2002)\citenamefont {Ota},
  \citenamefont {Sato},\ and\ \citenamefont {Yamashita}}]{Ota:2001pw}%
  \BibitemOpen
  \bibfield  {author} {\bibinfo {author} {\bibfnamefont {T.}~\bibnamefont
  {Ota}}, \bibinfo {author} {\bibfnamefont {J.}~\bibnamefont {Sato}}, \ and\
  \bibinfo {author} {\bibfnamefont {N.-a.}\ \bibnamefont {Yamashita}},\
  }\bibfield  {title} {\emph {\enquote {\bibinfo {title} {{Oscillation enhanced
  search for new interaction with neutrinos}},}\ }}\href {\doibase
  10.1103/PhysRevD.65.093015} {\bibfield  {journal} {\bibinfo  {journal} {Phys.
  Rev. D}\ }\textbf {\bibinfo {volume} {65}},\ \bibinfo {pages} {093015}
  (\bibinfo {year} {2002})},\ \Eprint
  {http://arxiv.org/abs/hep-ph/0112329}{arXiv:hep-ph/0112329}\BibitemShut
  {NoStop}%
\bibitem [{\citenamefont {Yasuda}(2007)}]{Yasuda:2007jp}%
  \BibitemOpen
  \bibfield  {author} {\bibinfo {author} {\bibfnamefont {O.}~\bibnamefont
  {Yasuda}},\ }\bibfield  {title} {\emph {\enquote {\bibinfo {title} {{On the
  exact formula for neutrino oscillation probability by Kimura, Takamura and
  Yokomakura}},}\ }}\href@noop {} {\  (\bibinfo {year} {2007})},\ \Eprint
  {http://arxiv.org/abs/0704.1531}{arXiv:0704.1531 [hep-ph]}\BibitemShut
  {NoStop}%
\bibitem [{\citenamefont {Ribeiro}\ \emph {et~al.}(2007)\citenamefont
  {Ribeiro}, \citenamefont {Minakata}, \citenamefont {Nunokawa}, \citenamefont
  {Uchinami},\ and\ \citenamefont {Zukanovich-Funchal}}]{Ribeiro:2007ud}%
  \BibitemOpen
  \bibfield  {author} {\bibinfo {author} {\bibfnamefont {N.~C.}\ \bibnamefont
  {Ribeiro}}, \bibinfo {author} {\bibfnamefont {H.}~\bibnamefont {Minakata}},
  \bibinfo {author} {\bibfnamefont {H.}~\bibnamefont {Nunokawa}}, \bibinfo
  {author} {\bibfnamefont {S.}~\bibnamefont {Uchinami}}, \ and\ \bibinfo
  {author} {\bibfnamefont {R.}~\bibnamefont {Zukanovich-Funchal}},\ }\bibfield
  {title} {\emph {\enquote {\bibinfo {title} {{Probing Non-Standard Neutrino
  Interactions with Neutrino Factories}},}\ }}\href {\doibase
  10.1088/1126-6708/2007/12/002} {\bibfield  {journal} {\bibinfo  {journal}
  {JHEP}\ }\textbf {\bibinfo {volume} {12}},\ \bibinfo {pages} {002} (\bibinfo
  {year} {2007})},\ \Eprint {http://arxiv.org/abs/0709.1980}{arXiv:0709.1980
  [hep-ph]}\BibitemShut {NoStop}%
\bibitem [{\citenamefont {Blennow}\ and\ \citenamefont
  {Ohlsson}(2008)}]{Blennow:2008eb}%
  \BibitemOpen
  \bibfield  {author} {\bibinfo {author} {\bibfnamefont {M.}~\bibnamefont
  {Blennow}}\ and\ \bibinfo {author} {\bibfnamefont {T.}~\bibnamefont
  {Ohlsson}},\ }\bibfield  {title} {\emph {\enquote {\bibinfo {title}
  {{Approximative two-flavor framework for neutrino oscillations with
  non-standard interactions}},}\ }}\href {\doibase 10.1103/PhysRevD.78.093002}
  {\bibfield  {journal} {\bibinfo  {journal} {Phys. Rev. D}\ }\textbf {\bibinfo
  {volume} {78}},\ \bibinfo {pages} {093002} (\bibinfo {year} {2008})},\
  \Eprint {http://arxiv.org/abs/0805.2301}{arXiv:0805.2301
  [hep-ph]}\BibitemShut {NoStop}%
\bibitem [{\citenamefont {Kikuchi}\ \emph {et~al.}(2009)\citenamefont
  {Kikuchi}, \citenamefont {Minakata},\ and\ \citenamefont
  {Uchinami}}]{Kikuchi:2008vq}%
  \BibitemOpen
  \bibfield  {author} {\bibinfo {author} {\bibfnamefont {T.}~\bibnamefont
  {Kikuchi}}, \bibinfo {author} {\bibfnamefont {H.}~\bibnamefont {Minakata}}, \
  and\ \bibinfo {author} {\bibfnamefont {S.}~\bibnamefont {Uchinami}},\
  }\bibfield  {title} {\emph {\enquote {\bibinfo {title} {{Perturbation Theory
  of Neutrino Oscillation with Nonstandard Neutrino Interactions}},}\ }}\href
  {\doibase 10.1088/1126-6708/2009/03/114} {\bibfield  {journal} {\bibinfo
  {journal} {JHEP}\ }\textbf {\bibinfo {volume} {03}},\ \bibinfo {pages} {114}
  (\bibinfo {year} {2009})},\ \Eprint
  {http://arxiv.org/abs/0809.3312}{arXiv:0809.3312 [hep-ph]}\BibitemShut
  {NoStop}%
\bibitem [{\citenamefont {Meloni}\ \emph {et~al.}(2009)\citenamefont {Meloni},
  \citenamefont {Ohlsson},\ and\ \citenamefont {Zhang}}]{Meloni:2009ia}%
  \BibitemOpen
  \bibfield  {author} {\bibinfo {author} {\bibfnamefont {D.}~\bibnamefont
  {Meloni}}, \bibinfo {author} {\bibfnamefont {T.}~\bibnamefont {Ohlsson}}, \
  and\ \bibinfo {author} {\bibfnamefont {H.}~\bibnamefont {Zhang}},\ }\bibfield
   {title} {\emph {\enquote {\bibinfo {title} {{Exact and Approximate Formulas
  for Neutrino Mixing and Oscillations with Non-Standard Interactions}},}\
  }}\href {\doibase 10.1088/1126-6708/2009/04/033} {\bibfield  {journal}
  {\bibinfo  {journal} {JHEP}\ }\textbf {\bibinfo {volume} {04}},\ \bibinfo
  {pages} {033} (\bibinfo {year} {2009})},\ \Eprint
  {http://arxiv.org/abs/0901.1784}{arXiv:0901.1784 [hep-ph]}\BibitemShut
  {NoStop}%
\bibitem [{\citenamefont {Agarwalla}\ \emph {et~al.}(2015)\citenamefont
  {Agarwalla}, \citenamefont {Kao}, \citenamefont {Saha},\ and\ \citenamefont
  {Takeuchi}}]{Agarwalla:2015cta}%
  \BibitemOpen
  \bibfield  {author} {\bibinfo {author} {\bibfnamefont {S.~K.}\ \bibnamefont
  {Agarwalla}}, \bibinfo {author} {\bibfnamefont {Y.}~\bibnamefont {Kao}},
  \bibinfo {author} {\bibfnamefont {D.}~\bibnamefont {Saha}}, \ and\ \bibinfo
  {author} {\bibfnamefont {T.}~\bibnamefont {Takeuchi}},\ }\bibfield  {title}
  {\emph {\enquote {\bibinfo {title} {{Running of Oscillation Parameters in
  Matter with Flavor-Diagonal Non-Standard Interactions of the Neutrino}},}\
  }}\href {\doibase 10.1007/JHEP11(2015)035} {\bibfield  {journal} {\bibinfo
  {journal} {JHEP}\ }\textbf {\bibinfo {volume} {11}},\ \bibinfo {pages} {035}
  (\bibinfo {year} {2015})},\ \Eprint
  {http://arxiv.org/abs/1506.08464}{arXiv:1506.08464 [hep-ph]}\BibitemShut
  {NoStop}%
\bibitem [{\citenamefont {Pontecorvo}(1958)}]{Pontecorvo:1957qd}%
  \BibitemOpen
  \bibfield  {author} {\bibinfo {author} {\bibfnamefont {B.}~\bibnamefont
  {Pontecorvo}},\ }\bibfield  {title} {\emph {\enquote {\bibinfo {title}
  {{Inverse beta processes and nonconservation of lepton charge}},}\
  }}\href@noop {} {\bibfield  {journal} {\bibinfo  {journal} {Sov. Phys. JETP}\
  }\textbf {\bibinfo {volume} {7}},\ \bibinfo {pages} {172} (\bibinfo {year}
  {1958})}\BibitemShut {NoStop}%
\bibitem [{\citenamefont {Maki}\ \emph {et~al.}(1962)\citenamefont {Maki},
  \citenamefont {Nakagawa},\ and\ \citenamefont {Sakata}}]{Maki:1962mu}%
  \BibitemOpen
  \bibfield  {author} {\bibinfo {author} {\bibfnamefont {Z.}~\bibnamefont
  {Maki}}, \bibinfo {author} {\bibfnamefont {M.}~\bibnamefont {Nakagawa}}, \
  and\ \bibinfo {author} {\bibfnamefont {S.}~\bibnamefont {Sakata}},\
  }\bibfield  {title} {\emph {\enquote {\bibinfo {title} {{Remarks on the
  unified model of elementary particles}},}\ }}\href {\doibase
  10.1143/PTP.28.870} {\bibfield  {journal} {\bibinfo  {journal}
  {Prog.Theor.Phys.}\ }\textbf {\bibinfo {volume} {28}},\ \bibinfo {pages}
  {870} (\bibinfo {year} {1962})}\BibitemShut {NoStop}%
\bibitem [{\citenamefont {Pontecorvo}(1968)}]{Pontecorvo:1967fh}%
  \BibitemOpen
  \bibfield  {author} {\bibinfo {author} {\bibfnamefont {B.}~\bibnamefont
  {Pontecorvo}},\ }\bibfield  {title} {\emph {\enquote {\bibinfo {title}
  {{Neutrino Experiments and the Problem of Conservation of Leptonic
  Charge}},}\ }}\href@noop {} {\bibfield  {journal} {\bibinfo  {journal}
  {Sov.Phys.JETP}\ }\textbf {\bibinfo {volume} {26}},\ \bibinfo {pages} {984}
  (\bibinfo {year} {1968})}\BibitemShut {NoStop}%
\bibitem [{\citenamefont {Huber}\ \emph {et~al.}(2005)\citenamefont {Huber},
  \citenamefont {Lindner},\ and\ \citenamefont {Winter}}]{Huber:2004ka}%
  \BibitemOpen
  \bibfield  {author} {\bibinfo {author} {\bibfnamefont {P.}~\bibnamefont
  {Huber}}, \bibinfo {author} {\bibfnamefont {M.}~\bibnamefont {Lindner}}, \
  and\ \bibinfo {author} {\bibfnamefont {W.}~\bibnamefont {Winter}},\
  }\bibfield  {title} {\emph {\enquote {\bibinfo {title} {{Simulation of
  long-baseline neutrino oscillation experiments with GLoBES (General Long
  Baseline Experiment Simulator)}},}\ }}\href {\doibase
  10.1016/j.cpc.2005.01.003} {\bibfield  {journal} {\bibinfo  {journal}
  {Comput. Phys. Commun.}\ }\textbf {\bibinfo {volume} {167}},\ \bibinfo
  {pages} {195} (\bibinfo {year} {2005})},\ \Eprint
  {http://arxiv.org/abs/hep-ph/0407333}{arXiv:hep-ph/0407333
  [hep-ph]}\BibitemShut {NoStop}%
\bibitem [{\citenamefont {Huber}\ \emph {et~al.}(2007)\citenamefont {Huber},
  \citenamefont {Kopp}, \citenamefont {Lindner}, \citenamefont {Rolinec},\ and\
  \citenamefont {Winter}}]{Huber:2007ji}%
  \BibitemOpen
  \bibfield  {author} {\bibinfo {author} {\bibfnamefont {P.}~\bibnamefont
  {Huber}}, \bibinfo {author} {\bibfnamefont {J.}~\bibnamefont {Kopp}},
  \bibinfo {author} {\bibfnamefont {M.}~\bibnamefont {Lindner}}, \bibinfo
  {author} {\bibfnamefont {M.}~\bibnamefont {Rolinec}}, \ and\ \bibinfo
  {author} {\bibfnamefont {W.}~\bibnamefont {Winter}},\ }\bibfield  {title}
  {\emph {\enquote {\bibinfo {title} {{New features in the simulation of
  neutrino oscillation experiments with GLoBES 3.0: General Long Baseline
  Experiment Simulator}},}\ }}\href {\doibase 10.1016/j.cpc.2007.05.004}
  {\bibfield  {journal} {\bibinfo  {journal} {Comput. Phys. Commun.}\ }\textbf
  {\bibinfo {volume} {177}},\ \bibinfo {pages} {432} (\bibinfo {year}
  {2007})},\ \Eprint {http://arxiv.org/abs/hep-ph/0701187}{arXiv:hep-ph/0701187
  [hep-ph]}\BibitemShut {NoStop}%
\bibitem [{\citenamefont {Marrone}(2021)}]{Marrone:2021}%
  \BibitemOpen
  \bibfield  {author} {\bibinfo {author} {\bibfnamefont {A.}~\bibnamefont
  {Marrone}},\ }\href@noop {} {\enquote {\bibinfo {title} {{Phenomenology of
  Three Neutrino Oscillations}},}\ } (\bibinfo {year} {2021}),\ \bibinfo {note}
  {talk given at the XIX International Workshop on Neutrino Telescopes, 18th to
  26th February, 2021, Padova, Italy, {\tt
  https://agenda.infn.it/event/24250/overview}}\BibitemShut {NoStop}%
\bibitem [{\citenamefont {Dziewonski}\ and\ \citenamefont
  {Anderson}(1981{\natexlab{b}})}]{Dziewonski:1981xy}%
  \BibitemOpen
  \bibfield  {author} {\bibinfo {author} {\bibfnamefont {A.~M.}\ \bibnamefont
  {Dziewonski}}\ and\ \bibinfo {author} {\bibfnamefont {D.~L.}\ \bibnamefont
  {Anderson}},\ }\bibfield  {title} {\emph {\enquote {\bibinfo {title}
  {{Preliminary reference earth model}},}\ }}\href {\doibase
  10.1016/0031-9201(81)90046-7} {\bibfield  {journal} {\bibinfo  {journal}
  {Phys. Earth Planet. Interiors}\ }\textbf {\bibinfo {volume} {25}},\ \bibinfo
  {pages} {297} (\bibinfo {year} {1981}{\natexlab{b}})}\BibitemShut {NoStop}%
\bibitem [{\citenamefont {Banuls}\ \emph {et~al.}(2001)\citenamefont {Banuls},
  \citenamefont {Barenboim},\ and\ \citenamefont {Bernabeu}}]{Banuls:2001zn}%
  \BibitemOpen
  \bibfield  {author} {\bibinfo {author} {\bibfnamefont {M.~C.}\ \bibnamefont
  {Banuls}}, \bibinfo {author} {\bibfnamefont {G.}~\bibnamefont {Barenboim}}, \
  and\ \bibinfo {author} {\bibfnamefont {J.}~\bibnamefont {Bernabeu}},\
  }\bibfield  {title} {\emph {\enquote {\bibinfo {title} {{Medium effects for
  terrestrial and atmospheric neutrino oscillations}},}\ }}\href {\doibase
  10.1016/S0370-2693(01)00723-7} {\bibfield  {journal} {\bibinfo  {journal}
  {Phys. Lett. B}\ }\textbf {\bibinfo {volume} {513}},\ \bibinfo {pages} {391}
  (\bibinfo {year} {2001})},\ \Eprint
  {http://arxiv.org/abs/hep-ph/0102184}{arXiv:hep-ph/0102184}\BibitemShut
  {NoStop}%
\bibitem [{\citenamefont {Gandhi}\ \emph {et~al.}(2005)\citenamefont {Gandhi},
  \citenamefont {Ghoshal}, \citenamefont {Goswami}, \citenamefont {Mehta},\
  and\ \citenamefont {Sankar}}]{Gandhi:2004md}%
  \BibitemOpen
  \bibfield  {author} {\bibinfo {author} {\bibfnamefont {R.}~\bibnamefont
  {Gandhi}}, \bibinfo {author} {\bibfnamefont {P.}~\bibnamefont {Ghoshal}},
  \bibinfo {author} {\bibfnamefont {S.}~\bibnamefont {Goswami}}, \bibinfo
  {author} {\bibfnamefont {P.}~\bibnamefont {Mehta}}, \ and\ \bibinfo {author}
  {\bibfnamefont {S.}~\bibnamefont {Sankar}},\ }\bibfield  {title} {\emph
  {\enquote {\bibinfo {title} {{Large matter effects in $\nu_{\mu}$
  $\rightarrow$ $\nu_{\tau}$ oscillations}},}\ }}\href {\doibase
  10.1103/PhysRevLett.94.051801} {\bibfield  {journal} {\bibinfo  {journal}
  {Phys. Rev. Lett.}\ }\textbf {\bibinfo {volume} {94}},\ \bibinfo {pages}
  {051801} (\bibinfo {year} {2005})},\ \Eprint
  {http://arxiv.org/abs/hep-ph/0408361}{arXiv:hep-ph/0408361}\BibitemShut
  {NoStop}%
\bibitem [{\citenamefont {Gandhi}\ \emph {et~al.}(2006)\citenamefont {Gandhi},
  \citenamefont {Ghoshal}, \citenamefont {Goswami}, \citenamefont {Mehta},\
  and\ \citenamefont {Sankar}}]{Gandhi:2004bj}%
  \BibitemOpen
  \bibfield  {author} {\bibinfo {author} {\bibfnamefont {R.}~\bibnamefont
  {Gandhi}}, \bibinfo {author} {\bibfnamefont {P.}~\bibnamefont {Ghoshal}},
  \bibinfo {author} {\bibfnamefont {S.}~\bibnamefont {Goswami}}, \bibinfo
  {author} {\bibfnamefont {P.}~\bibnamefont {Mehta}}, \ and\ \bibinfo {author}
  {\bibfnamefont {S.~U.}\ \bibnamefont {Sankar}},\ }\bibfield  {title} {\emph
  {\enquote {\bibinfo {title} {{Earth matter effects at very long baselines and
  the neutrino mass hierarchy}},}\ }}\href {\doibase
  10.1103/PhysRevD.73.053001} {\bibfield  {journal} {\bibinfo  {journal} {Phys.
  Rev. D}\ }\textbf {\bibinfo {volume} {73}},\ \bibinfo {pages} {053001}
  (\bibinfo {year} {2006})},\ \Eprint
  {http://arxiv.org/abs/hep-ph/0411252}{arXiv:hep-ph/0411252}\BibitemShut
  {NoStop}%
\bibitem [{\citenamefont {Mocioiu}\ and\ \citenamefont
  {Wright}(2015)}]{Mocioiu:2014gua}%
  \BibitemOpen
  \bibfield  {author} {\bibinfo {author} {\bibfnamefont {I.}~\bibnamefont
  {Mocioiu}}\ and\ \bibinfo {author} {\bibfnamefont {W.}~\bibnamefont
  {Wright}},\ }\bibfield  {title} {\emph {\enquote {\bibinfo {title}
  {{Non-standard neutrino interactions in the $\mu$\textendash{}$\tau$
  sector}},}\ }}\href {\doibase 10.1016/j.nuclphysb.2015.02.016} {\bibfield
  {journal} {\bibinfo  {journal} {Nucl. Phys. B}\ }\textbf {\bibinfo {volume}
  {893}},\ \bibinfo {pages} {376} (\bibinfo {year} {2015})},\ \Eprint
  {http://arxiv.org/abs/1410.6193}{arXiv:1410.6193 [hep-ph]}\BibitemShut
  {NoStop}%
\bibitem [{\citenamefont {Li}\ and\ \citenamefont {Luo}(2016)}]{Li:2015oal}%
  \BibitemOpen
  \bibfield  {author} {\bibinfo {author} {\bibfnamefont {Y.-F.}\ \bibnamefont
  {Li}}\ and\ \bibinfo {author} {\bibfnamefont {S.}~\bibnamefont {Luo}},\
  }\bibfield  {title} {\emph {\enquote {\bibinfo {title} {{Neutrino Oscillation
  Probabilities in Matter with Direct and Indirect Unitarity Violation in the
  Lepton Mixing Matrix}},}\ }}\href {\doibase 10.1103/PhysRevD.93.033008}
  {\bibfield  {journal} {\bibinfo  {journal} {Phys. Rev. D}\ }\textbf {\bibinfo
  {volume} {93}},\ \bibinfo {pages} {033008} (\bibinfo {year} {2016})},\
  \Eprint {http://arxiv.org/abs/1508.00052}{arXiv:1508.00052
  [hep-ph]}\BibitemShut {NoStop}%
\bibitem [{\citenamefont {Parke}\ and\ \citenamefont
  {Ross-Lonergan}(2016)}]{Parke:2015goa}%
  \BibitemOpen
  \bibfield  {author} {\bibinfo {author} {\bibfnamefont {S.}~\bibnamefont
  {Parke}}\ and\ \bibinfo {author} {\bibfnamefont {M.}~\bibnamefont
  {Ross-Lonergan}},\ }\bibfield  {title} {\emph {\enquote {\bibinfo {title}
  {{Unitarity and the three flavor neutrino mixing matrix}},}\ }}\href
  {\doibase 10.1103/PhysRevD.93.113009} {\bibfield  {journal} {\bibinfo
  {journal} {Phys. Rev. D}\ }\textbf {\bibinfo {volume} {93}},\ \bibinfo
  {pages} {113009} (\bibinfo {year} {2016})},\ \Eprint
  {http://arxiv.org/abs/1508.05095}{arXiv:1508.05095 [hep-ph]}\BibitemShut
  {NoStop}%
\bibitem [{\citenamefont {Dutta}\ \emph {et~al.}(2017)\citenamefont {Dutta},
  \citenamefont {Ghoshal},\ and\ \citenamefont {Roy}}]{Dutta:2016czj}%
  \BibitemOpen
  \bibfield  {author} {\bibinfo {author} {\bibfnamefont {D.}~\bibnamefont
  {Dutta}}, \bibinfo {author} {\bibfnamefont {P.}~\bibnamefont {Ghoshal}}, \
  and\ \bibinfo {author} {\bibfnamefont {S.}~\bibnamefont {Roy}},\ }\bibfield
  {title} {\emph {\enquote {\bibinfo {title} {{Effect of Non Unitarity on
  Neutrino Mass Hierarchy determination at DUNE, NO$\nu$A and T2K}},}\ }}\href
  {\doibase 10.1016/j.nuclphysb.2017.04.018} {\bibfield  {journal} {\bibinfo
  {journal} {Nucl. Phys. B}\ }\textbf {\bibinfo {volume} {920}},\ \bibinfo
  {pages} {385} (\bibinfo {year} {2017})},\ \Eprint
  {http://arxiv.org/abs/1609.07094}{arXiv:1609.07094 [hep-ph]}\BibitemShut
  {NoStop}%
\bibitem [{\citenamefont {P\"as}\ and\ \citenamefont
  {Sicking}(2017)}]{Pas:2016qbg}%
  \BibitemOpen
  \bibfield  {author} {\bibinfo {author} {\bibfnamefont {H.}~\bibnamefont
  {P\"as}}\ and\ \bibinfo {author} {\bibfnamefont {P.}~\bibnamefont
  {Sicking}},\ }\bibfield  {title} {\emph {\enquote {\bibinfo {title}
  {{Discriminating sterile neutrinos and unitarity violation with CP
  invariants}},}\ }}\href {\doibase 10.1103/PhysRevD.95.075004} {\bibfield
  {journal} {\bibinfo  {journal} {Phys. Rev. D}\ }\textbf {\bibinfo {volume}
  {95}},\ \bibinfo {pages} {075004} (\bibinfo {year} {2017})},\ \Eprint
  {http://arxiv.org/abs/1611.08450}{arXiv:1611.08450 [hep-ph]}\BibitemShut
  {NoStop}%
\bibitem [{\citenamefont {Antusch}\ \emph {et~al.}(2006)\citenamefont
  {Antusch}, \citenamefont {Biggio}, \citenamefont {Fernandez-Martinez},
  \citenamefont {Gavela},\ and\ \citenamefont {Lopez-Pavon}}]{Antusch:2006vwa}%
  \BibitemOpen
  \bibfield  {author} {\bibinfo {author} {\bibfnamefont {S.}~\bibnamefont
  {Antusch}}, \bibinfo {author} {\bibfnamefont {C.}~\bibnamefont {Biggio}},
  \bibinfo {author} {\bibfnamefont {E.}~\bibnamefont {Fernandez-Martinez}},
  \bibinfo {author} {\bibfnamefont {M.~B.}\ \bibnamefont {Gavela}}, \ and\
  \bibinfo {author} {\bibfnamefont {J.}~\bibnamefont {Lopez-Pavon}},\
  }\bibfield  {title} {\emph {\enquote {\bibinfo {title} {{Unitarity of the
  Leptonic Mixing Matrix}},}\ }}\href {\doibase 10.1088/1126-6708/2006/10/084}
  {\bibfield  {journal} {\bibinfo  {journal} {JHEP}\ }\textbf {\bibinfo
  {volume} {10}},\ \bibinfo {pages} {084} (\bibinfo {year} {2006})},\ \Eprint
  {http://arxiv.org/abs/hep-ph/0607020}{arXiv:hep-ph/0607020}\BibitemShut
  {NoStop}%
\bibitem [{\citenamefont {Abe}\ \emph {et~al.}(2017)\citenamefont {Abe},
  \citenamefont {Asano}, \citenamefont {Haba},\ and\ \citenamefont
  {Yamada}}]{Abe:2017jit}%
  \BibitemOpen
  \bibfield  {author} {\bibinfo {author} {\bibfnamefont {Y.}~\bibnamefont
  {Abe}}, \bibinfo {author} {\bibfnamefont {Y.}~\bibnamefont {Asano}}, \bibinfo
  {author} {\bibfnamefont {N.}~\bibnamefont {Haba}}, \ and\ \bibinfo {author}
  {\bibfnamefont {T.}~\bibnamefont {Yamada}},\ }\bibfield  {title} {\emph
  {\enquote {\bibinfo {title} {{Heavy neutrino mixing in the T2HK, the T2HKK
  and an extension of the T2HK with a detector at Oki Islands}},}\ }}\href
  {\doibase 10.1140/epjc/s10052-017-5294-7} {\bibfield  {journal} {\bibinfo
  {journal} {Eur. Phys. J. C}\ }\textbf {\bibinfo {volume} {77}},\ \bibinfo
  {pages} {851} (\bibinfo {year} {2017})},\ \Eprint
  {http://arxiv.org/abs/1705.03818}{arXiv:1705.03818 [hep-ph]}\BibitemShut
  {NoStop}%
\bibitem [{\citenamefont {Dutta}\ and\ \citenamefont
  {Roy}(2021)}]{Dutta:2019hmb}%
  \BibitemOpen
  \bibfield  {author} {\bibinfo {author} {\bibfnamefont {D.}~\bibnamefont
  {Dutta}}\ and\ \bibinfo {author} {\bibfnamefont {S.}~\bibnamefont {Roy}},\
  }\bibfield  {title} {\emph {\enquote {\bibinfo {title} {{Non-Unitarity at
  DUNE and T2HK with Charged and Neutral Current Measurements}},}\ }}\href
  {\doibase 10.1088/1361-6471/abdc03} {\bibfield  {journal} {\bibinfo
  {journal} {J. Phys. G}\ }\textbf {\bibinfo {volume} {48}},\ \bibinfo {pages}
  {045004} (\bibinfo {year} {2021})},\ \Eprint
  {http://arxiv.org/abs/1901.11298}{arXiv:1901.11298 [hep-ph]}\BibitemShut
  {NoStop}%
\bibitem [{\citenamefont {Xing}(2013)}]{Xing:2012kh}%
  \BibitemOpen
  \bibfield  {author} {\bibinfo {author} {\bibfnamefont {Z.-z.}\ \bibnamefont
  {Xing}},\ }\bibfield  {title} {\emph {\enquote {\bibinfo {title} {{Towards
  testing the unitarity of the 3X3 lepton flavor mixing matrix in a precision
  reactor antineutrino oscillation experiment}},}\ }}\href {\doibase
  10.1016/j.physletb.2012.12.062} {\bibfield  {journal} {\bibinfo  {journal}
  {Phys. Lett. B}\ }\textbf {\bibinfo {volume} {718}},\ \bibinfo {pages} {1447}
  (\bibinfo {year} {2013})},\ \Eprint
  {http://arxiv.org/abs/1210.1523}{arXiv:1210.1523 [hep-ph]}\BibitemShut
  {NoStop}%
\bibitem [{\citenamefont {Flieger}\ \emph {et~al.}(2020)\citenamefont
  {Flieger}, \citenamefont {Gluza},\ and\ \citenamefont
  {Porwit}}]{Flieger:2019eor}%
  \BibitemOpen
  \bibfield  {author} {\bibinfo {author} {\bibfnamefont {W.}~\bibnamefont
  {Flieger}}, \bibinfo {author} {\bibfnamefont {J.}~\bibnamefont {Gluza}}, \
  and\ \bibinfo {author} {\bibfnamefont {K.}~\bibnamefont {Porwit}},\
  }\bibfield  {title} {\emph {\enquote {\bibinfo {title} {{New limits on
  neutrino non-unitary mixings based on prescribed singular values}},}\ }}\href
  {\doibase 10.1007/JHEP03(2020)169} {\bibfield  {journal} {\bibinfo  {journal}
  {JHEP}\ }\textbf {\bibinfo {volume} {03}},\ \bibinfo {pages} {169} (\bibinfo
  {year} {2020})},\ \Eprint {http://arxiv.org/abs/1910.01233}{arXiv:1910.01233
  [hep-ph]}\BibitemShut {NoStop}%
\bibitem [{\citenamefont {Bielas}\ \emph {et~al.}(2018)\citenamefont {Bielas},
  \citenamefont {Flieger}, \citenamefont {Gluza},\ and\ \citenamefont
  {Gluza}}]{Bielas:2017lok}%
  \BibitemOpen
  \bibfield  {author} {\bibinfo {author} {\bibfnamefont {K.}~\bibnamefont
  {Bielas}}, \bibinfo {author} {\bibfnamefont {W.}~\bibnamefont {Flieger}},
  \bibinfo {author} {\bibfnamefont {J.}~\bibnamefont {Gluza}}, \ and\ \bibinfo
  {author} {\bibfnamefont {M.}~\bibnamefont {Gluza}},\ }\bibfield  {title}
  {\emph {\enquote {\bibinfo {title} {{Neutrino mixing, interval matrices and
  singular values}},}\ }}\href {\doibase 10.1103/PhysRevD.98.053001} {\bibfield
   {journal} {\bibinfo  {journal} {Phys. Rev. D}\ }\textbf {\bibinfo {volume}
  {98}},\ \bibinfo {pages} {053001} (\bibinfo {year} {2018})},\ \Eprint
  {http://arxiv.org/abs/1708.09196}{arXiv:1708.09196 [hep-ph]}\BibitemShut
  {NoStop}%
\bibitem [{\citenamefont {Ellis}\ \emph
  {et~al.}(2020{\natexlab{b}})\citenamefont {Ellis}, \citenamefont {Kelly},\
  and\ \citenamefont {Li}}]{Ellis:2020ehi}%
  \BibitemOpen
  \bibfield  {author} {\bibinfo {author} {\bibfnamefont {S.~A.~R.}\
  \bibnamefont {Ellis}}, \bibinfo {author} {\bibfnamefont {K.~J.}\ \bibnamefont
  {Kelly}}, \ and\ \bibinfo {author} {\bibfnamefont {S.~W.}\ \bibnamefont
  {Li}},\ }\bibfield  {title} {\emph {\enquote {\bibinfo {title} {{Leptonic
  Unitarity Triangles}},}\ }}\href {\doibase 10.1103/PhysRevD.102.115027}
  {\bibfield  {journal} {\bibinfo  {journal} {Phys. Rev. D}\ }\textbf {\bibinfo
  {volume} {102}},\ \bibinfo {pages} {115027} (\bibinfo {year}
  {2020}{\natexlab{b}})},\ \Eprint
  {http://arxiv.org/abs/2004.13719}{arXiv:2004.13719 [hep-ph]}\BibitemShut
  {NoStop}%
\bibitem [{\citenamefont {King}(2008)}]{King:2007pr}%
  \BibitemOpen
  \bibfield  {author} {\bibinfo {author} {\bibfnamefont {S.~F.}\ \bibnamefont
  {King}},\ }\bibfield  {title} {\emph {\enquote {\bibinfo {title}
  {{Parametrizing the lepton mixing matrix in terms of deviations from
  tri-bimaximal mixing}},}\ }}\href {\doibase 10.1016/j.physletb.2007.10.078}
  {\bibfield  {journal} {\bibinfo  {journal} {Phys. Lett. B}\ }\textbf
  {\bibinfo {volume} {659}},\ \bibinfo {pages} {244} (\bibinfo {year}
  {2008})},\ \Eprint {http://arxiv.org/abs/0710.0530}{arXiv:0710.0530
  [hep-ph]}\BibitemShut {NoStop}%
\bibitem [{\citenamefont {Pakvasa}\ \emph {et~al.}(2008)\citenamefont
  {Pakvasa}, \citenamefont {Rodejohann},\ and\ \citenamefont
  {Weiler}}]{Pakvasa:2007zj}%
  \BibitemOpen
  \bibfield  {author} {\bibinfo {author} {\bibfnamefont {S.}~\bibnamefont
  {Pakvasa}}, \bibinfo {author} {\bibfnamefont {W.}~\bibnamefont {Rodejohann}},
  \ and\ \bibinfo {author} {\bibfnamefont {T.~J.}\ \bibnamefont {Weiler}},\
  }\bibfield  {title} {\emph {\enquote {\bibinfo {title} {{Unitary
  parametrization of perturbations to tribimaximal neutrino mixing}},}\ }}\href
  {\doibase 10.1103/PhysRevLett.100.111801} {\bibfield  {journal} {\bibinfo
  {journal} {Phys. Rev. Lett.}\ }\textbf {\bibinfo {volume} {100}},\ \bibinfo
  {pages} {111801} (\bibinfo {year} {2008})},\ \Eprint
  {http://arxiv.org/abs/0711.0052}{arXiv:0711.0052 [hep-ph]}\BibitemShut
  {NoStop}%
\bibitem [{\citenamefont {Blennow}\ and\ \citenamefont
  {Fernandez-Martinez}(2010)}]{Blennow:2009pk}%
  \BibitemOpen
  \bibfield  {author} {\bibinfo {author} {\bibfnamefont {M.}~\bibnamefont
  {Blennow}}\ and\ \bibinfo {author} {\bibfnamefont {E.}~\bibnamefont
  {Fernandez-Martinez}},\ }\bibfield  {title} {\emph {\enquote {\bibinfo
  {title} {{Neutrino oscillation parameter sampling with MonteCUBES}},}\
  }}\href {\doibase 10.1016/j.cpc.2009.09.014} {\bibfield  {journal} {\bibinfo
  {journal} {Comput. Phys. Commun.}\ }\textbf {\bibinfo {volume} {181}},\
  \bibinfo {pages} {227} (\bibinfo {year} {2010})},\ \Eprint
  {http://arxiv.org/abs/0903.3985}{arXiv:0903.3985 [hep-ph]}\BibitemShut
  {NoStop}%
\bibitem [{\citenamefont {Abi}\ \emph {et~al.}(2021)\citenamefont {Abi} \emph
  {et~al.}}]{DUNE:2021cuw}%
  \BibitemOpen
  \bibfield  {author} {\bibinfo {author} {\bibfnamefont {B.}~\bibnamefont
  {Abi}} \emph {et~al.} (\bibinfo {collaboration} {DUNE}),\ }\bibfield  {title}
  {\emph {\enquote {\bibinfo {title} {{Experiment Simulation Configurations
  Approximating DUNE TDR}},}\ }}\href@noop {} {\  (\bibinfo {year} {2021})},\
  \Eprint {http://arxiv.org/abs/2103.04797}{arXiv:2103.04797
  [hep-ex]}\BibitemShut {NoStop}%
\bibitem [{\citenamefont {Kelly}(2017)}]{Kelly:2017kch}%
  \BibitemOpen
  \bibfield  {author} {\bibinfo {author} {\bibfnamefont {K.~J.}\ \bibnamefont
  {Kelly}},\ }\bibfield  {title} {\emph {\enquote {\bibinfo {title} {{Searches
  for new physics at the Hyper-Kamiokande experiment}},}\ }}\href {\doibase
  10.1103/PhysRevD.95.115009} {\bibfield  {journal} {\bibinfo  {journal} {Phys.
  Rev. D}\ }\textbf {\bibinfo {volume} {95}},\ \bibinfo {pages} {115009}
  (\bibinfo {year} {2017})},\ \Eprint
  {http://arxiv.org/abs/1703.00448}{arXiv:1703.00448 [hep-ph]}\BibitemShut
  {NoStop}%
\bibitem [{\citenamefont {Choubey}\ \emph
  {et~al.}(2018{\natexlab{b}})\citenamefont {Choubey}, \citenamefont {Dutta},\
  and\ \citenamefont {Pramanik}}]{Choubey:2017ppj}%
  \BibitemOpen
  \bibfield  {author} {\bibinfo {author} {\bibfnamefont {S.}~\bibnamefont
  {Choubey}}, \bibinfo {author} {\bibfnamefont {D.}~\bibnamefont {Dutta}}, \
  and\ \bibinfo {author} {\bibfnamefont {D.}~\bibnamefont {Pramanik}},\
  }\bibfield  {title} {\emph {\enquote {\bibinfo {title} {{Measuring the
  Sterile Neutrino CP Phase at DUNE and T2HK}},}\ }}\href {\doibase
  10.1140/epjc/s10052-018-5816-y} {\bibfield  {journal} {\bibinfo  {journal}
  {Eur. Phys. J. C}\ }\textbf {\bibinfo {volume} {78}},\ \bibinfo {pages} {339}
  (\bibinfo {year} {2018}{\natexlab{b}})},\ \Eprint
  {http://arxiv.org/abs/1711.07464}{arXiv:1711.07464 [hep-ph]}\BibitemShut
  {NoStop}%
\bibitem [{\citenamefont {Abe}\ \emph {et~al.}(2011{\natexlab{d}})\citenamefont
  {Abe} \emph {et~al.}}]{Abe:2011ts}%
  \BibitemOpen
  \bibfield  {author} {\bibinfo {author} {\bibfnamefont {K.}~\bibnamefont
  {Abe}} \emph {et~al.},\ }\bibfield  {title} {\emph {\enquote {\bibinfo
  {title} {{Letter of Intent: The Hyper-Kamiokande Experiment --- Detector
  Design and Physics Potential ---}},}\ }}\href@noop {} {\  (\bibinfo {year}
  {2011}{\natexlab{d}})},\ \Eprint
  {http://arxiv.org/abs/1109.3262}{arXiv:1109.3262 [hep-ex]}\BibitemShut
  {NoStop}%
\bibitem [{\citenamefont {Seo}(2019)}]{Seo:2019dpr}%
  \BibitemOpen
  \bibfield  {author} {\bibinfo {author} {\bibfnamefont {S.-H.}\ \bibnamefont
  {Seo}},\ }\bibfield  {title} {\emph {\enquote {\bibinfo {title} {{Neutrino
  Telescope at Yemilab, Korea}},}\ }}\href@noop {} {\  (\bibinfo {year}
  {2019})},\ \Eprint {http://arxiv.org/abs/1903.05368}{arXiv:1903.05368
  [physics.ins-det]}\BibitemShut {NoStop}%
\bibitem [{\citenamefont {Abe}\ \emph {et~al.}(2019)\citenamefont {Abe} \emph
  {et~al.}}]{T2K:2019bbb}%
  \BibitemOpen
  \bibfield  {author} {\bibinfo {author} {\bibfnamefont {K.}~\bibnamefont
  {Abe}} \emph {et~al.} (\bibinfo {collaboration} {T2K}),\ }\bibfield  {title}
  {\emph {\enquote {\bibinfo {title} {{T2K ND280 Upgrade - Technical Design
  Report}},}\ }}\href@noop {} {\  (\bibinfo {year} {2019})},\ \Eprint
  {http://arxiv.org/abs/1901.03750}{arXiv:1901.03750
  [physics.ins-det]}\BibitemShut {NoStop}%
\bibitem [{\citenamefont {Drakopoulou}(2018)}]{Drakopoulou:2017qdu}%
  \BibitemOpen
  \bibfield  {author} {\bibinfo {author} {\bibfnamefont {E.}~\bibnamefont
  {Drakopoulou}} (\bibinfo {collaboration} {J-PARC E61}),\ }\bibfield  {title}
  {\emph {\enquote {\bibinfo {title} {{An intermediate water Cherenkov detector
  for the Hyper-Kamiokande experiment: overview and status}},}\ }}\href
  {\doibase 10.22323/1.301.1021} {\bibfield  {journal} {\bibinfo  {journal}
  {PoS}\ }\textbf {\bibinfo {volume} {ICRC2017}},\ \bibinfo {pages} {1021}
  (\bibinfo {year} {2018})}\BibitemShut {NoStop}%
\bibitem [{\citenamefont {Wilson}(2020)}]{Wilson:2020trq}%
  \BibitemOpen
  \bibfield  {author} {\bibinfo {author} {\bibfnamefont {J.~R.}\ \bibnamefont
  {Wilson}},\ }\bibfield  {title} {\emph {\enquote {\bibinfo {title} {{The
  Hyper-K Near Detector Programme}},}\ }}\href {\doibase
  10.1088/1742-6596/1342/1/012053} {\bibfield  {journal} {\bibinfo  {journal}
  {J. Phys. Conf. Ser.}\ }\textbf {\bibinfo {volume} {1342}},\ \bibinfo {pages}
  {012053} (\bibinfo {year} {2020})}\BibitemShut {NoStop}%
\bibitem [{\citenamefont {Blennow}\ \emph {et~al.}(2014)\citenamefont
  {Blennow}, \citenamefont {Coloma}, \citenamefont {Huber},\ and\ \citenamefont
  {Schwetz}}]{Blennow:2013oma}%
  \BibitemOpen
  \bibfield  {author} {\bibinfo {author} {\bibfnamefont {M.}~\bibnamefont
  {Blennow}}, \bibinfo {author} {\bibfnamefont {P.}~\bibnamefont {Coloma}},
  \bibinfo {author} {\bibfnamefont {P.}~\bibnamefont {Huber}}, \ and\ \bibinfo
  {author} {\bibfnamefont {T.}~\bibnamefont {Schwetz}},\ }\bibfield  {title}
  {\emph {\enquote {\bibinfo {title} {{Quantifying the sensitivity of
  oscillation experiments to the neutrino mass ordering}},}\ }}\href {\doibase
  10.1007/JHEP03(2014)028} {\bibfield  {journal} {\bibinfo  {journal} {JHEP}\
  }\textbf {\bibinfo {volume} {03}},\ \bibinfo {pages} {028} (\bibinfo {year}
  {2014})},\ \Eprint {http://arxiv.org/abs/1311.1822}{arXiv:1311.1822
  [hep-ph]}\BibitemShut {NoStop}%
\bibitem [{\citenamefont {Huber}\ \emph {et~al.}(2002)\citenamefont {Huber},
  \citenamefont {Lindner},\ and\ \citenamefont {Winter}}]{Huber:2002mx}%
  \BibitemOpen
  \bibfield  {author} {\bibinfo {author} {\bibfnamefont {P.}~\bibnamefont
  {Huber}}, \bibinfo {author} {\bibfnamefont {M.}~\bibnamefont {Lindner}}, \
  and\ \bibinfo {author} {\bibfnamefont {W.}~\bibnamefont {Winter}},\
  }\bibfield  {title} {\emph {\enquote {\bibinfo {title} {{Superbeams versus
  neutrino factories}},}\ }}\href {\doibase 10.1016/S0550-3213(02)00825-8}
  {\bibfield  {journal} {\bibinfo  {journal} {Nucl. Phys. B}\ }\textbf
  {\bibinfo {volume} {645}},\ \bibinfo {pages} {3} (\bibinfo {year} {2002})},\
  \Eprint
  {http://arxiv.org/abs/hep-ph/0204352}{arXiv:hep-ph/0204352}\BibitemShut
  {NoStop}%
\bibitem [{\citenamefont {Fogli}\ \emph {et~al.}(2002)\citenamefont {Fogli},
  \citenamefont {Lisi}, \citenamefont {Marrone}, \citenamefont {Montanino},\
  and\ \citenamefont {Palazzo}}]{Fogli:2002pt}%
  \BibitemOpen
  \bibfield  {author} {\bibinfo {author} {\bibfnamefont {G.~L.}\ \bibnamefont
  {Fogli}}, \bibinfo {author} {\bibfnamefont {E.}~\bibnamefont {Lisi}},
  \bibinfo {author} {\bibfnamefont {A.}~\bibnamefont {Marrone}}, \bibinfo
  {author} {\bibfnamefont {D.}~\bibnamefont {Montanino}}, \ and\ \bibinfo
  {author} {\bibfnamefont {A.}~\bibnamefont {Palazzo}},\ }\bibfield  {title}
  {\emph {\enquote {\bibinfo {title} {{Getting the most from the statistical
  analysis of solar neutrino oscillations}},}\ }}\href {\doibase
  10.1103/PhysRevD.66.053010} {\bibfield  {journal} {\bibinfo  {journal} {Phys.
  Rev. D}\ }\textbf {\bibinfo {volume} {66}},\ \bibinfo {pages} {053010}
  (\bibinfo {year} {2002})},\ \Eprint
  {http://arxiv.org/abs/hep-ph/0206162}{arXiv:hep-ph/0206162}\BibitemShut
  {NoStop}%
\bibitem [{\citenamefont {Gonzalez-Garcia}\ and\ \citenamefont
  {Maltoni}(2004)}]{Gonzalez-Garcia:2004pka}%
  \BibitemOpen
  \bibfield  {author} {\bibinfo {author} {\bibfnamefont {M.~C.}\ \bibnamefont
  {Gonzalez-Garcia}}\ and\ \bibinfo {author} {\bibfnamefont {M.}~\bibnamefont
  {Maltoni}},\ }\bibfield  {title} {\emph {\enquote {\bibinfo {title}
  {{Atmospheric neutrino oscillations and new physics}},}\ }}\href {\doibase
  10.1103/PhysRevD.70.033010} {\bibfield  {journal} {\bibinfo  {journal} {Phys.
  Rev. D}\ }\textbf {\bibinfo {volume} {70}},\ \bibinfo {pages} {033010}
  (\bibinfo {year} {2004})},\ \Eprint
  {http://arxiv.org/abs/hep-ph/0404085}{arXiv:hep-ph/0404085}\BibitemShut
  {NoStop}%
\bibitem [{\citenamefont {Aliaga}\ \emph {et~al.}(2016)\citenamefont {Aliaga}
  \emph {et~al.}}]{MINERvA:2016iqn}%
  \BibitemOpen
  \bibfield  {author} {\bibinfo {author} {\bibfnamefont {L.}~\bibnamefont
  {Aliaga}} \emph {et~al.} (\bibinfo {collaboration} {MINERvA}),\ }\bibfield
  {title} {\emph {\enquote {\bibinfo {title} {{Neutrino Flux Predictions for
  the NuMI Beam}},}\ }}\href {\doibase 10.1103/PhysRevD.94.092005} {\bibfield
  {journal} {\bibinfo  {journal} {Phys. Rev. D}\ }\textbf {\bibinfo {volume}
  {94}},\ \bibinfo {pages} {092005} (\bibinfo {year} {2016})},\ \bibinfo {note}
  {[Addendum: Phys.Rev.D 95, 039903 (2017)]},\ \Eprint
  {http://arxiv.org/abs/1607.00704}{arXiv:1607.00704 [hep-ex]}\BibitemShut
  {NoStop}%
\bibitem [{\citenamefont {Miranda}\ \emph {et~al.}(2018)\citenamefont
  {Miranda}, \citenamefont {Pasquini}, \citenamefont {T\'ortola},\ and\
  \citenamefont {Valle}}]{Miranda:2018yym}%
  \BibitemOpen
  \bibfield  {author} {\bibinfo {author} {\bibfnamefont {O.~G.}\ \bibnamefont
  {Miranda}}, \bibinfo {author} {\bibfnamefont {P.}~\bibnamefont {Pasquini}},
  \bibinfo {author} {\bibfnamefont {M.}~\bibnamefont {T\'ortola}}, \ and\
  \bibinfo {author} {\bibfnamefont {J.~W.~F.}\ \bibnamefont {Valle}},\
  }\bibfield  {title} {\emph {\enquote {\bibinfo {title} {{Exploring the
  Potential of Short-Baseline Physics at Fermilab}},}\ }}\href {\doibase
  10.1103/PhysRevD.97.095026} {\bibfield  {journal} {\bibinfo  {journal} {Phys.
  Rev. D}\ }\textbf {\bibinfo {volume} {97}},\ \bibinfo {pages} {095026}
  (\bibinfo {year} {2018})},\ \Eprint
  {http://arxiv.org/abs/1802.02133}{arXiv:1802.02133 [hep-ph]}\BibitemShut
  {NoStop}%
\bibitem [{\citenamefont {Agafonova}\ \emph {et~al.}(2010)\citenamefont
  {Agafonova} \emph {et~al.}}]{OPERA:2010pne}%
  \BibitemOpen
  \bibfield  {author} {\bibinfo {author} {\bibfnamefont {N.}~\bibnamefont
  {Agafonova}} \emph {et~al.} (\bibinfo {collaboration} {OPERA}),\ }\bibfield
  {title} {\emph {\enquote {\bibinfo {title} {{Observation of a first
  $\nu_\tau$ candidate in the OPERA experiment in the CNGS beam}},}\ }}\href
  {\doibase 10.1016/j.physletb.2010.06.022} {\bibfield  {journal} {\bibinfo
  {journal} {Phys. Lett. B}\ }\textbf {\bibinfo {volume} {691}},\ \bibinfo
  {pages} {138} (\bibinfo {year} {2010})},\ \Eprint
  {http://arxiv.org/abs/1006.1623}{arXiv:1006.1623 [hep-ex]}\BibitemShut
  {NoStop}%
\bibitem [{\citenamefont {Machado}\ \emph {et~al.}(2020)\citenamefont
  {Machado}, \citenamefont {Schulz},\ and\ \citenamefont
  {Turner}}]{Machado:2020yxl}%
  \BibitemOpen
  \bibfield  {author} {\bibinfo {author} {\bibfnamefont {P.}~\bibnamefont
  {Machado}}, \bibinfo {author} {\bibfnamefont {H.}~\bibnamefont {Schulz}}, \
  and\ \bibinfo {author} {\bibfnamefont {J.}~\bibnamefont {Turner}},\
  }\bibfield  {title} {\emph {\enquote {\bibinfo {title} {{Tau neutrinos at
  DUNE: New strategies, new opportunities}},}\ }}\href {\doibase
  10.1103/PhysRevD.102.053010} {\bibfield  {journal} {\bibinfo  {journal}
  {Phys. Rev. D}\ }\textbf {\bibinfo {volume} {102}},\ \bibinfo {pages}
  {053010} (\bibinfo {year} {2020})},\ \Eprint
  {http://arxiv.org/abs/2007.00015}{arXiv:2007.00015 [hep-ph]}\BibitemShut
  {NoStop}%
\bibitem [{\citenamefont {Ghoshal}\ \emph {et~al.}(2019)\citenamefont
  {Ghoshal}, \citenamefont {Giarnetti},\ and\ \citenamefont
  {Meloni}}]{Ghoshal:2019pab}%
  \BibitemOpen
  \bibfield  {author} {\bibinfo {author} {\bibfnamefont {A.}~\bibnamefont
  {Ghoshal}}, \bibinfo {author} {\bibfnamefont {A.}~\bibnamefont {Giarnetti}},
  \ and\ \bibinfo {author} {\bibfnamefont {D.}~\bibnamefont {Meloni}},\
  }\bibfield  {title} {\emph {\enquote {\bibinfo {title} {{On the role of the
  $\nu_{\tau}$ appearance in DUNE in constraining standard neutrino physics and
  beyond}},}\ }}\href {\doibase 10.1007/JHEP12(2019)126} {\bibfield  {journal}
  {\bibinfo  {journal} {JHEP}\ }\textbf {\bibinfo {volume} {12}},\ \bibinfo
  {pages} {126} (\bibinfo {year} {2019})},\ \Eprint
  {http://arxiv.org/abs/1906.06212}{arXiv:1906.06212 [hep-ph]}\BibitemShut
  {NoStop}%
\bibitem [{\citenamefont {Martinez-Soler}\ and\ \citenamefont
  {Minakata}(2021)}]{Martinez-Soler:2021sir}%
  \BibitemOpen
  \bibfield  {author} {\bibinfo {author} {\bibfnamefont {I.}~\bibnamefont
  {Martinez-Soler}}\ and\ \bibinfo {author} {\bibfnamefont {H.}~\bibnamefont
  {Minakata}},\ }\bibfield  {title} {\emph {\enquote {\bibinfo {title}
  {{Measuring tau neutrino appearance probability via unitarity}},}\
  }}\href@noop {} {\  (\bibinfo {year} {2021})},\ \Eprint
  {http://arxiv.org/abs/2109.06933}{arXiv:2109.06933 [hep-ph]}\BibitemShut
  {NoStop}%
\bibitem [{\citenamefont {Denton}\ and\ \citenamefont
  {Gehrlein}(2021)}]{Denton:2021mso}%
  \BibitemOpen
  \bibfield  {author} {\bibinfo {author} {\bibfnamefont {P.~B.}\ \bibnamefont
  {Denton}}\ and\ \bibinfo {author} {\bibfnamefont {J.}~\bibnamefont
  {Gehrlein}},\ }\bibfield  {title} {\emph {\enquote {\bibinfo {title} {{New
  tau neutrino oscillation and scattering constraints on unitarity
  violation}},}\ }}\href@noop {} {\  (\bibinfo {year} {2021})},\ \Eprint
  {http://arxiv.org/abs/2109.14575}{arXiv:2109.14575 [hep-ph]}\BibitemShut
  {NoStop}%
\bibitem [{\citenamefont {Denton}(2021)}]{Denton:2021rsa}%
  \BibitemOpen
  \bibfield  {author} {\bibinfo {author} {\bibfnamefont {P.~B.}\ \bibnamefont
  {Denton}},\ }\bibfield  {title} {\emph {\enquote {\bibinfo {title} {{Tau
  Neutrino Identification in Atmospheric Neutrino Oscillations Without Particle
  Identification or Unitarity}},}\ }}\href@noop {} {\  (\bibinfo {year}
  {2021})},\ \Eprint {http://arxiv.org/abs/2109.14576}{arXiv:2109.14576
  [hep-ph]}\BibitemShut {NoStop}%
\bibitem [{\citenamefont {Coloma}\ \emph
  {et~al.}(2021{\natexlab{b}})\citenamefont {Coloma}, \citenamefont
  {Gonzalez-Garcia},\ and\ \citenamefont {Maltoni}}]{Coloma:2020gfv}%
  \BibitemOpen
  \bibfield  {author} {\bibinfo {author} {\bibfnamefont {P.}~\bibnamefont
  {Coloma}}, \bibinfo {author} {\bibfnamefont {M.~C.}\ \bibnamefont
  {Gonzalez-Garcia}}, \ and\ \bibinfo {author} {\bibfnamefont {M.}~\bibnamefont
  {Maltoni}},\ }\bibfield  {title} {\emph {\enquote {\bibinfo {title}
  {{Neutrino oscillation constraints on U(1)' models: from non-standard
  interactions to long-range forces}},}\ }}\href {\doibase
  10.1007/JHEP01(2021)114} {\bibfield  {journal} {\bibinfo  {journal} {JHEP}\
  }\textbf {\bibinfo {volume} {01}},\ \bibinfo {pages} {114} (\bibinfo {year}
  {2021}{\natexlab{b}})},\ \Eprint
  {http://arxiv.org/abs/2009.14220}{arXiv:2009.14220 [hep-ph]}\BibitemShut
  {NoStop}%
\bibitem [{\citenamefont {Foot}(1991)}]{Foot:1990mn}%
  \BibitemOpen
  \bibfield  {author} {\bibinfo {author} {\bibfnamefont {R.}~\bibnamefont
  {Foot}},\ }\bibfield  {title} {\emph {\enquote {\bibinfo {title} {{New
  Physics From Electric Charge Quantization?}}}\ }}\href {\doibase
  10.1142/S0217732391000543} {\bibfield  {journal} {\bibinfo  {journal} {Mod.
  Phys. Lett. A}\ }\textbf {\bibinfo {volume} {6}},\ \bibinfo {pages} {527}
  (\bibinfo {year} {1991})}\BibitemShut {NoStop}%
\bibitem [{\citenamefont {He}\ \emph {et~al.}(1991{\natexlab{a}})\citenamefont
  {He}, \citenamefont {Joshi}, \citenamefont {Lew},\ and\ \citenamefont
  {Volkas}}]{He:1990pn}%
  \BibitemOpen
  \bibfield  {author} {\bibinfo {author} {\bibfnamefont {X.~G.}\ \bibnamefont
  {He}}, \bibinfo {author} {\bibfnamefont {G.~C.}\ \bibnamefont {Joshi}},
  \bibinfo {author} {\bibfnamefont {H.}~\bibnamefont {Lew}}, \ and\ \bibinfo
  {author} {\bibfnamefont {R.~R.}\ \bibnamefont {Volkas}},\ }\bibfield  {title}
  {\emph {\enquote {\bibinfo {title} {{New Z-prime phenomenology}},}\ }}\href
  {\doibase 10.1103/PhysRevD.43.R22} {\bibfield  {journal} {\bibinfo  {journal}
  {Phys. Rev. D}\ }\textbf {\bibinfo {volume} {43}},\ \bibinfo {pages} {22}
  (\bibinfo {year} {1991}{\natexlab{a}})}\BibitemShut {NoStop}%
\bibitem [{\citenamefont {He}\ \emph {et~al.}(1991{\natexlab{b}})\citenamefont
  {He}, \citenamefont {Joshi}, \citenamefont {Lew},\ and\ \citenamefont
  {Volkas}}]{He:1991qd}%
  \BibitemOpen
  \bibfield  {author} {\bibinfo {author} {\bibfnamefont {X.-G.}\ \bibnamefont
  {He}}, \bibinfo {author} {\bibfnamefont {G.~C.}\ \bibnamefont {Joshi}},
  \bibinfo {author} {\bibfnamefont {H.}~\bibnamefont {Lew}}, \ and\ \bibinfo
  {author} {\bibfnamefont {R.~R.}\ \bibnamefont {Volkas}},\ }\bibfield  {title}
  {\emph {\enquote {\bibinfo {title} {{Simplest Z-prime model}},}\ }}\href
  {\doibase 10.1103/PhysRevD.44.2118} {\bibfield  {journal} {\bibinfo
  {journal} {Phys. Rev. D}\ }\textbf {\bibinfo {volume} {44}},\ \bibinfo
  {pages} {2118} (\bibinfo {year} {1991}{\natexlab{b}})}\BibitemShut {NoStop}%
\bibitem [{\citenamefont {Foot}\ \emph {et~al.}(1994)\citenamefont {Foot},
  \citenamefont {He}, \citenamefont {Lew},\ and\ \citenamefont
  {Volkas}}]{Foot:1994vd}%
  \BibitemOpen
  \bibfield  {author} {\bibinfo {author} {\bibfnamefont {R.}~\bibnamefont
  {Foot}}, \bibinfo {author} {\bibfnamefont {X.~G.}\ \bibnamefont {He}},
  \bibinfo {author} {\bibfnamefont {H.}~\bibnamefont {Lew}}, \ and\ \bibinfo
  {author} {\bibfnamefont {R.~R.}\ \bibnamefont {Volkas}},\ }\bibfield  {title}
  {\emph {\enquote {\bibinfo {title} {{Model for a light Z-prime boson}},}\
  }}\href {\doibase 10.1103/PhysRevD.50.4571} {\bibfield  {journal} {\bibinfo
  {journal} {Phys. Rev. D}\ }\textbf {\bibinfo {volume} {50}},\ \bibinfo
  {pages} {4571} (\bibinfo {year} {1994})},\ \Eprint
  {http://arxiv.org/abs/hep-ph/9401250}{arXiv:hep-ph/9401250}\BibitemShut
  {NoStop}%
\bibitem [{\citenamefont {Foot}(2005)}]{Foot:2005uc}%
  \BibitemOpen
  \bibfield  {author} {\bibinfo {author} {\bibfnamefont {R.}~\bibnamefont
  {Foot}},\ }\bibfield  {title} {\emph {\enquote {\bibinfo {title} {{Avoiding
  the gauge hierarchy problem with see-sawed neutrino masses}},}\ }}\href
  {\doibase 10.1142/S0217732305019134} {\bibfield  {journal} {\bibinfo
  {journal} {Mod. Phys. Lett. A}\ }\textbf {\bibinfo {volume} {20}},\ \bibinfo
  {pages} {3035} (\bibinfo {year} {2005})},\ \Eprint
  {http://arxiv.org/abs/hep-ph/0505154}{arXiv:hep-ph/0505154}\BibitemShut
  {NoStop}%
\bibitem [{\citenamefont {Araki}\ \emph {et~al.}(2021)\citenamefont {Araki},
  \citenamefont {Asai}, \citenamefont {Honda}, \citenamefont {Kasuya},
  \citenamefont {Sato}, \citenamefont {Shimomura},\ and\ \citenamefont
  {Yang}}]{Araki:2021xdk}%
  \BibitemOpen
  \bibfield  {author} {\bibinfo {author} {\bibfnamefont {T.}~\bibnamefont
  {Araki}}, \bibinfo {author} {\bibfnamefont {K.}~\bibnamefont {Asai}},
  \bibinfo {author} {\bibfnamefont {K.}~\bibnamefont {Honda}}, \bibinfo
  {author} {\bibfnamefont {R.}~\bibnamefont {Kasuya}}, \bibinfo {author}
  {\bibfnamefont {J.}~\bibnamefont {Sato}}, \bibinfo {author} {\bibfnamefont
  {T.}~\bibnamefont {Shimomura}}, \ and\ \bibinfo {author} {\bibfnamefont
  {M.~J.~S.}\ \bibnamefont {Yang}},\ }\bibfield  {title} {\emph {\enquote
  {\bibinfo {title} {{Resolving the Hubble tension in a U(1)$_{L_\mu-L_\tau}$
  model with the Majoron}},}\ }}\href {\doibase 10.1093/ptep/ptab108}
  {\bibfield  {journal} {\bibinfo  {journal} {PTEP}\ }\textbf {\bibinfo
  {volume} {2021}},\ \bibinfo {pages} {103B05} (\bibinfo {year} {2021})},\
  \Eprint {http://arxiv.org/abs/2103.07167}{arXiv:2103.07167
  [hep-ph]}\BibitemShut {NoStop}%
\bibitem [{\citenamefont {Chen}\ and\ \citenamefont
  {Nomura}(2021)}]{Chen:2020jvl}%
  \BibitemOpen
  \bibfield  {author} {\bibinfo {author} {\bibfnamefont {C.-H.}\ \bibnamefont
  {Chen}}\ and\ \bibinfo {author} {\bibfnamefont {T.}~\bibnamefont {Nomura}},\
  }\bibfield  {title} {\emph {\enquote {\bibinfo {title} {{Electron and muon
  $g-2$, radiative neutrino mass, and $\ell' \to \ell \gamma$ in a
  $U(1)_{e-\mu}$ model}},}\ }}\href {\doibase 10.1016/j.nuclphysb.2021.115314}
  {\bibfield  {journal} {\bibinfo  {journal} {Nucl. Phys. B}\ }\textbf
  {\bibinfo {volume} {964}},\ \bibinfo {pages} {115314} (\bibinfo {year}
  {2021})},\ \Eprint {http://arxiv.org/abs/2003.07638}{arXiv:2003.07638
  [hep-ph]}\BibitemShut {NoStop}%
\bibitem [{\citenamefont {Bodas}\ \emph {et~al.}(2021)\citenamefont {Bodas},
  \citenamefont {Coy},\ and\ \citenamefont {King}}]{Bodas:2021fsy}%
  \BibitemOpen
  \bibfield  {author} {\bibinfo {author} {\bibfnamefont {A.}~\bibnamefont
  {Bodas}}, \bibinfo {author} {\bibfnamefont {R.}~\bibnamefont {Coy}}, \ and\
  \bibinfo {author} {\bibfnamefont {S.~J.~D.}\ \bibnamefont {King}},\
  }\bibfield  {title} {\emph {\enquote {\bibinfo {title} {{Solving the electron
  and muon $g-2$ anomalies in $Z'$ models}},}\ }}\href {\doibase
  10.1140/epjc/s10052-021-09850-x} {\bibfield  {journal} {\bibinfo  {journal}
  {Eur. Phys. J. C}\ }\textbf {\bibinfo {volume} {81}},\ \bibinfo {pages}
  {1065} (\bibinfo {year} {2021})},\ \Eprint
  {http://arxiv.org/abs/2102.07781}{arXiv:2102.07781 [hep-ph]}\BibitemShut
  {NoStop}%
\bibitem [{\citenamefont {Panda}\ \emph {et~al.}(2022)\citenamefont {Panda},
  \citenamefont {Mishra}, \citenamefont {Behera},\ and\ \citenamefont
  {Mohanta}}]{Panda:2022kbn}%
  \BibitemOpen
  \bibfield  {author} {\bibinfo {author} {\bibfnamefont {P.}~\bibnamefont
  {Panda}}, \bibinfo {author} {\bibfnamefont {P.}~\bibnamefont {Mishra}},
  \bibinfo {author} {\bibfnamefont {M.~K.}\ \bibnamefont {Behera}}, \ and\
  \bibinfo {author} {\bibfnamefont {R.}~\bibnamefont {Mohanta}},\ }\bibfield
  {title} {\emph {\enquote {\bibinfo {title} {{Neutrino phenomenology, muon and
  electron (g-2) under $U(1)$ gauged symmetries in an extended inverse seesaw
  model}},}\ }}\href@noop {} {\  (\bibinfo {year} {2022})},\ \Eprint
  {http://arxiv.org/abs/2203.14536}{arXiv:2203.14536 [hep-ph]}\BibitemShut
  {NoStop}%
\bibitem [{\citenamefont {Borah}\ \emph {et~al.}(2022)\citenamefont {Borah},
  \citenamefont {Dutta}, \citenamefont {Mahapatra},\ and\ \citenamefont
  {Sahu}}]{Borah:2021khc}%
  \BibitemOpen
  \bibfield  {author} {\bibinfo {author} {\bibfnamefont {D.}~\bibnamefont
  {Borah}}, \bibinfo {author} {\bibfnamefont {M.}~\bibnamefont {Dutta}},
  \bibinfo {author} {\bibfnamefont {S.}~\bibnamefont {Mahapatra}}, \ and\
  \bibinfo {author} {\bibfnamefont {N.}~\bibnamefont {Sahu}},\ }\bibfield
  {title} {\emph {\enquote {\bibinfo {title} {{Lepton anomalous magnetic moment
  with singlet-doublet fermion dark matter in a scotogenic
  $U(1)_{L_\ensuremath{\mu}-L_\ensuremath{\tau}}$ model}},}\ }}\href {\doibase
  10.1103/PhysRevD.105.015029} {\bibfield  {journal} {\bibinfo  {journal}
  {Phys. Rev. D}\ }\textbf {\bibinfo {volume} {105}},\ \bibinfo {pages}
  {015029} (\bibinfo {year} {2022})},\ \Eprint
  {http://arxiv.org/abs/2109.02699}{arXiv:2109.02699 [hep-ph]}\BibitemShut
  {NoStop}%
\bibitem [{\citenamefont {Baek}(2016)}]{Baek:2015fea}%
  \BibitemOpen
  \bibfield  {author} {\bibinfo {author} {\bibfnamefont {S.}~\bibnamefont
  {Baek}},\ }\bibfield  {title} {\emph {\enquote {\bibinfo {title} {{Dark
  matter and muon $(g-2)$ in local $U(1)_{L_\mu-L_\tau}$-extended Ma Model}},}\
  }}\href {\doibase 10.1016/j.physletb.2016.02.062} {\bibfield  {journal}
  {\bibinfo  {journal} {Phys. Lett. B}\ }\textbf {\bibinfo {volume} {756}},\
  \bibinfo {pages} {1} (\bibinfo {year} {2016})},\ \Eprint
  {http://arxiv.org/abs/1510.02168}{arXiv:1510.02168 [hep-ph]}\BibitemShut
  {NoStop}%
\bibitem [{\citenamefont {Asai}\ \emph {et~al.}(2021)\citenamefont {Asai},
  \citenamefont {Okawa},\ and\ \citenamefont {Tsumura}}]{Asai:2020qlp}%
  \BibitemOpen
  \bibfield  {author} {\bibinfo {author} {\bibfnamefont {K.}~\bibnamefont
  {Asai}}, \bibinfo {author} {\bibfnamefont {S.}~\bibnamefont {Okawa}}, \ and\
  \bibinfo {author} {\bibfnamefont {K.}~\bibnamefont {Tsumura}},\ }\bibfield
  {title} {\emph {\enquote {\bibinfo {title} {{Search for $
  \mathrm{U}{(1)}_{L_{\mu }-{L}_{\tau }} $ charged dark matter with neutrino
  telescope}},}\ }}\href {\doibase 10.1007/JHEP03(2021)047} {\bibfield
  {journal} {\bibinfo  {journal} {JHEP}\ }\textbf {\bibinfo {volume} {03}},\
  \bibinfo {pages} {047} (\bibinfo {year} {2021})},\ \Eprint
  {http://arxiv.org/abs/2011.03165}{arXiv:2011.03165 [hep-ph]}\BibitemShut
  {NoStop}%
\bibitem [{\citenamefont {Alonso-\'Alvarez}\ \emph {et~al.}(2023)\citenamefont
  {Alonso-\'Alvarez}, \citenamefont {Bleau},\ and\ \citenamefont
  {Cline}}]{Alonso-Alvarez:2023tii}%
  \BibitemOpen
  \bibfield  {author} {\bibinfo {author} {\bibfnamefont {G.}~\bibnamefont
  {Alonso-\'Alvarez}}, \bibinfo {author} {\bibfnamefont {K.}~\bibnamefont
  {Bleau}}, \ and\ \bibinfo {author} {\bibfnamefont {J.~M.}\ \bibnamefont
  {Cline}},\ }\bibfield  {title} {\emph {\enquote {\bibinfo {title}
  {{Distortion of neutrino oscillations by dark photon dark matter}},}\ }}\href
  {\doibase 10.1103/PhysRevD.107.055045} {\bibfield  {journal} {\bibinfo
  {journal} {Phys. Rev. D}\ }\textbf {\bibinfo {volume} {107}},\ \bibinfo
  {pages} {055045} (\bibinfo {year} {2023})},\ \Eprint
  {http://arxiv.org/abs/2301.04152}{arXiv:2301.04152 [hep-ph]}\BibitemShut
  {NoStop}%
\bibitem [{\citenamefont {Kumar~Poddar}\ \emph {et~al.}(2019)\citenamefont
  {Kumar~Poddar}, \citenamefont {Mohanty},\ and\ \citenamefont
  {Jana}}]{KumarPoddar:2019ceq}%
  \BibitemOpen
  \bibfield  {author} {\bibinfo {author} {\bibfnamefont {T.}~\bibnamefont
  {Kumar~Poddar}}, \bibinfo {author} {\bibfnamefont {S.}~\bibnamefont
  {Mohanty}}, \ and\ \bibinfo {author} {\bibfnamefont {S.}~\bibnamefont
  {Jana}},\ }\bibfield  {title} {\emph {\enquote {\bibinfo {title} {{Vector
  gauge boson radiation from compact binary systems in a gauged $L_\mu-L_\tau$
  scenario}},}\ }}\href {\doibase 10.1103/PhysRevD.100.123023} {\bibfield
  {journal} {\bibinfo  {journal} {Phys. Rev. D}\ }\textbf {\bibinfo {volume}
  {100}},\ \bibinfo {pages} {123023} (\bibinfo {year} {2019})},\ \Eprint
  {http://arxiv.org/abs/1908.09732}{arXiv:1908.09732 [hep-ph]}\BibitemShut
  {NoStop}%
\bibitem [{\citenamefont {Asai}\ \emph {et~al.}(2020)\citenamefont {Asai},
  \citenamefont {Hamaguchi}, \citenamefont {Nagata},\ and\ \citenamefont
  {Tseng}}]{Asai:2020qax}%
  \BibitemOpen
  \bibfield  {author} {\bibinfo {author} {\bibfnamefont {K.}~\bibnamefont
  {Asai}}, \bibinfo {author} {\bibfnamefont {K.}~\bibnamefont {Hamaguchi}},
  \bibinfo {author} {\bibfnamefont {N.}~\bibnamefont {Nagata}}, \ and\ \bibinfo
  {author} {\bibfnamefont {S.-Y.}\ \bibnamefont {Tseng}},\ }\bibfield  {title}
  {\emph {\enquote {\bibinfo {title} {{Leptogenesis in the minimal gauged
  U(1)$_{L_\mu-L_\tau}$ model and the sign of the cosmological baryon
  asymmetry}},}\ }}\href {\doibase 10.1088/1475-7516/2020/11/013} {\bibfield
  {journal} {\bibinfo  {journal} {JCAP}\ }\textbf {\bibinfo {volume} {11}},\
  \bibinfo {pages} {013} (\bibinfo {year} {2020})},\ \Eprint
  {http://arxiv.org/abs/2005.01039}{arXiv:2005.01039 [hep-ph]}\BibitemShut
  {NoStop}%
\bibitem [{\citenamefont {Cesarotti}\ \emph {et~al.}(2023)\citenamefont
  {Cesarotti}, \citenamefont {Homiller}, \citenamefont {Mishra},\ and\
  \citenamefont {Reece}}]{Cesarotti:2022ttv}%
  \BibitemOpen
  \bibfield  {author} {\bibinfo {author} {\bibfnamefont {C.}~\bibnamefont
  {Cesarotti}}, \bibinfo {author} {\bibfnamefont {S.}~\bibnamefont {Homiller}},
  \bibinfo {author} {\bibfnamefont {R.~K.}\ \bibnamefont {Mishra}}, \ and\
  \bibinfo {author} {\bibfnamefont {M.}~\bibnamefont {Reece}},\ }\bibfield
  {title} {\emph {\enquote {\bibinfo {title} {{Probing New Gauge Forces with a
  High-Energy Muon Beam Dump}},}\ }}\href {\doibase
  10.1103/PhysRevLett.130.071803} {\bibfield  {journal} {\bibinfo  {journal}
  {Phys. Rev. Lett.}\ }\textbf {\bibinfo {volume} {130}},\ \bibinfo {pages}
  {071803} (\bibinfo {year} {2023})},\ \Eprint
  {http://arxiv.org/abs/2202.12302}{arXiv:2202.12302 [hep-ph]}\BibitemShut
  {NoStop}%
\bibitem [{\citenamefont {Grilli~di Cortona}\ and\ \citenamefont
  {Nardi}(2022)}]{GrillidiCortona:2022kbq}%
  \BibitemOpen
  \bibfield  {author} {\bibinfo {author} {\bibfnamefont {G.}~\bibnamefont
  {Grilli~di Cortona}}\ and\ \bibinfo {author} {\bibfnamefont {E.}~\bibnamefont
  {Nardi}},\ }\bibfield  {title} {\emph {\enquote {\bibinfo {title} {{Probing
  light mediators at the MUonE experiment}},}\ }}\href {\doibase
  10.1103/PhysRevD.105.L111701} {\bibfield  {journal} {\bibinfo  {journal}
  {Phys. Rev. D}\ }\textbf {\bibinfo {volume} {105}},\ \bibinfo {pages}
  {L111701} (\bibinfo {year} {2022})},\ \Eprint
  {http://arxiv.org/abs/2204.04227}{arXiv:2204.04227 [hep-ph]}\BibitemShut
  {NoStop}%
\bibitem [{\citenamefont {Duan}\ \emph {et~al.}(2018)\citenamefont {Duan},
  \citenamefont {He}, \citenamefont {Wu},\ and\ \citenamefont
  {Yang}}]{Duan:2017qwj}%
  \BibitemOpen
  \bibfield  {author} {\bibinfo {author} {\bibfnamefont {G.~H.}\ \bibnamefont
  {Duan}}, \bibinfo {author} {\bibfnamefont {X.-G.}\ \bibnamefont {He}},
  \bibinfo {author} {\bibfnamefont {L.}~\bibnamefont {Wu}}, \ and\ \bibinfo
  {author} {\bibfnamefont {J.~M.}\ \bibnamefont {Yang}},\ }\bibfield  {title}
  {\emph {\enquote {\bibinfo {title} {{Leptophilic dark matter in gauged
  $U(1)_{L{_e}-L_{\mu }}$ model in light of DAMPE cosmic ray ${e{^+}} +
  {e{^-}}$ excess}},}\ }}\href {\doibase 10.1140/epjc/s10052-018-5805-1}
  {\bibfield  {journal} {\bibinfo  {journal} {Eur. Phys. J. C}\ }\textbf
  {\bibinfo {volume} {78}},\ \bibinfo {pages} {323} (\bibinfo {year} {2018})},\
  \Eprint {http://arxiv.org/abs/1711.11563}{arXiv:1711.11563
  [hep-ph]}\BibitemShut {NoStop}%
\bibitem [{\citenamefont {He}(2009)}]{He:2009ra}%
  \BibitemOpen
  \bibfield  {author} {\bibinfo {author} {\bibfnamefont {X.-G.}\ \bibnamefont
  {He}},\ }\bibfield  {title} {\emph {\enquote {\bibinfo {title} {{Dark Matter
  Annihilation Explanation for e+- Excesses in Cosmic Ray}},}\ }}\href
  {\doibase 10.1142/S0217732309031740} {\bibfield  {journal} {\bibinfo
  {journal} {Mod. Phys. Lett. A}\ }\textbf {\bibinfo {volume} {24}},\ \bibinfo
  {pages} {2139} (\bibinfo {year} {2009})},\ \Eprint
  {http://arxiv.org/abs/0908.2908}{arXiv:0908.2908 [hep-ph]}\BibitemShut
  {NoStop}%
\bibitem [{\citenamefont {Heeck}\ and\ \citenamefont
  {Rodejohann}(2011)}]{Heeck:2010pg}%
  \BibitemOpen
  \bibfield  {author} {\bibinfo {author} {\bibfnamefont {J.}~\bibnamefont
  {Heeck}}\ and\ \bibinfo {author} {\bibfnamefont {W.}~\bibnamefont
  {Rodejohann}},\ }\bibfield  {title} {\emph {\enquote {\bibinfo {title}
  {{Gauged $L_\mu - L_\tau$ and different Muon Neutrino and Anti-Neutrino
  Oscillations: MINOS and beyond}},}\ }}\href {\doibase
  10.1088/0954-3899/38/8/085005} {\bibfield  {journal} {\bibinfo  {journal} {J.
  Phys. G}\ }\textbf {\bibinfo {volume} {38}},\ \bibinfo {pages} {085005}
  (\bibinfo {year} {2011})},\ \Eprint
  {http://arxiv.org/abs/1007.2655}{arXiv:1007.2655 [hep-ph]}\BibitemShut
  {NoStop}%
\bibitem [{\citenamefont {Babu}\ \emph {et~al.}(1998)\citenamefont {Babu},
  \citenamefont {Kolda},\ and\ \citenamefont {March-Russell}}]{Babu:1997st}%
  \BibitemOpen
  \bibfield  {author} {\bibinfo {author} {\bibfnamefont {K.~S.}\ \bibnamefont
  {Babu}}, \bibinfo {author} {\bibfnamefont {C.~F.}\ \bibnamefont {Kolda}}, \
  and\ \bibinfo {author} {\bibfnamefont {J.}~\bibnamefont {March-Russell}},\
  }\bibfield  {title} {\emph {\enquote {\bibinfo {title} {{Implications of
  generalized $Z$--$Z^\prime$ mixing}},}\ }}\href {\doibase
  10.1103/PhysRevD.57.6788} {\bibfield  {journal} {\bibinfo  {journal} {Phys.
  Rev. D}\ }\textbf {\bibinfo {volume} {57}},\ \bibinfo {pages} {6788}
  (\bibinfo {year} {1998})},\ \Eprint
  {http://arxiv.org/abs/hep-ph/9710441}{arXiv:hep-ph/9710441}\BibitemShut
  {NoStop}%
\bibitem [{\citenamefont {Joshipura}\ \emph {et~al.}(2020)\citenamefont
  {Joshipura}, \citenamefont {Mahajan},\ and\ \citenamefont
  {Patel}}]{Joshipura:2019qxz}%
  \BibitemOpen
  \bibfield  {author} {\bibinfo {author} {\bibfnamefont {A.~S.}\ \bibnamefont
  {Joshipura}}, \bibinfo {author} {\bibfnamefont {N.}~\bibnamefont {Mahajan}},
  \ and\ \bibinfo {author} {\bibfnamefont {K.~M.}\ \bibnamefont {Patel}},\
  }\bibfield  {title} {\emph {\enquote {\bibinfo {title} {{Generalised
  $\mu$-$\tau$ symmetries and calculable gauge kinetic and mass mixing in
  $U(1)_{L_{\mu }-{L}_{\tau }}$ models}},}\ }}\href {\doibase
  10.1007/JHEP03(2020)001} {\bibfield  {journal} {\bibinfo  {journal} {JHEP}\
  }\textbf {\bibinfo {volume} {03}},\ \bibinfo {pages} {001} (\bibinfo {year}
  {2020})},\ \Eprint {http://arxiv.org/abs/1909.02331}{arXiv:1909.02331
  [hep-ph]}\BibitemShut {NoStop}%
\bibitem [{\citenamefont {Schlamminger}\ \emph {et~al.}(2008)\citenamefont
  {Schlamminger}, \citenamefont {Choi}, \citenamefont {Wagner}, \citenamefont
  {Gundlach},\ and\ \citenamefont {Adelberger}}]{Schlamminger:2007ht}%
  \BibitemOpen
  \bibfield  {author} {\bibinfo {author} {\bibfnamefont {S.}~\bibnamefont
  {Schlamminger}}, \bibinfo {author} {\bibfnamefont {K.~Y.}\ \bibnamefont
  {Choi}}, \bibinfo {author} {\bibfnamefont {T.~A.}\ \bibnamefont {Wagner}},
  \bibinfo {author} {\bibfnamefont {J.~H.}\ \bibnamefont {Gundlach}}, \ and\
  \bibinfo {author} {\bibfnamefont {E.~G.}\ \bibnamefont {Adelberger}},\
  }\bibfield  {title} {\emph {\enquote {\bibinfo {title} {{Test of the
  equivalence principle using a rotating torsion balance}},}\ }}\href {\doibase
  10.1103/PhysRevLett.100.041101} {\bibfield  {journal} {\bibinfo  {journal}
  {Phys. Rev. Lett.}\ }\textbf {\bibinfo {volume} {100}},\ \bibinfo {pages}
  {041101} (\bibinfo {year} {2008})},\ \Eprint
  {http://arxiv.org/abs/0712.0607}{arXiv:0712.0607 [gr-qc]}\BibitemShut
  {NoStop}%
\bibitem [{\citenamefont {Adelberger}\ \emph {et~al.}(2009)\citenamefont
  {Adelberger}, \citenamefont {Gundlach}, \citenamefont {Heckel}, \citenamefont
  {Hoedl},\ and\ \citenamefont {Schlamminger}}]{Adelberger:2009zz}%
  \BibitemOpen
  \bibfield  {author} {\bibinfo {author} {\bibfnamefont {E.~G.}\ \bibnamefont
  {Adelberger}}, \bibinfo {author} {\bibfnamefont {J.~H.}\ \bibnamefont
  {Gundlach}}, \bibinfo {author} {\bibfnamefont {B.~R.}\ \bibnamefont
  {Heckel}}, \bibinfo {author} {\bibfnamefont {S.}~\bibnamefont {Hoedl}}, \
  and\ \bibinfo {author} {\bibfnamefont {S.}~\bibnamefont {Schlamminger}},\
  }\bibfield  {title} {\emph {\enquote {\bibinfo {title} {{Torsion balance
  experiments: A low-energy frontier of particle physics}},}\ }}\href {\doibase
  10.1016/j.ppnp.2008.08.002} {\bibfield  {journal} {\bibinfo  {journal} {Prog.
  Part. Nucl. Phys.}\ }\textbf {\bibinfo {volume} {62}},\ \bibinfo {pages}
  {102} (\bibinfo {year} {2009})}\BibitemShut {NoStop}%
\bibitem [{\citenamefont {Bustamante}\ and\ \citenamefont
  {Agarwalla}(2019)}]{Bustamante:2018mzu}%
  \BibitemOpen
  \bibfield  {author} {\bibinfo {author} {\bibfnamefont {M.}~\bibnamefont
  {Bustamante}}\ and\ \bibinfo {author} {\bibfnamefont {S.~K.}\ \bibnamefont
  {Agarwalla}},\ }\bibfield  {title} {\emph {\enquote {\bibinfo {title}
  {{Universe's Worth of Electrons to Probe Long-Range Interactions of
  High-Energy Astrophysical Neutrinos}},}\ }}\href {\doibase
  10.1103/PhysRevLett.122.061103} {\bibfield  {journal} {\bibinfo  {journal}
  {Phys. Rev. Lett.}\ }\textbf {\bibinfo {volume} {122}},\ \bibinfo {pages}
  {061103} (\bibinfo {year} {2019})},\ \Eprint
  {http://arxiv.org/abs/1808.02042}{arXiv:1808.02042 [astro-ph.HE]}\BibitemShut
  {NoStop}%
\bibitem [{\citenamefont {McMillan}(2011)}]{McMillan:2011wd}%
  \BibitemOpen
  \bibfield  {author} {\bibinfo {author} {\bibfnamefont {P.~J.}\ \bibnamefont
  {McMillan}},\ }\bibfield  {title} {\emph {\enquote {\bibinfo {title} {{Mass
  models of the Milky Way}},}\ }}\href {\doibase
  10.1111/j.1365-2966.2011.18564.x} {\bibfield  {journal} {\bibinfo  {journal}
  {Mon. Not. Roy. Astron. Soc.}\ }\textbf {\bibinfo {volume} {414}},\ \bibinfo
  {pages} {2446} (\bibinfo {year} {2011})},\ \Eprint
  {http://arxiv.org/abs/1102.4340}{arXiv:1102.4340 [astro-ph.GA]}\BibitemShut
  {NoStop}%
\bibitem [{\citenamefont {Miller}\ and\ \citenamefont
  {Bregman}(2013)}]{Miller:2013nza}%
  \BibitemOpen
  \bibfield  {author} {\bibinfo {author} {\bibfnamefont {M.~J.}\ \bibnamefont
  {Miller}}\ and\ \bibinfo {author} {\bibfnamefont {J.~N.}\ \bibnamefont
  {Bregman}},\ }\bibfield  {title} {\emph {\enquote {\bibinfo {title} {{The
  Structure of the Milky Way's Hot Gas Halo}},}\ }}\href {\doibase
  10.1088/0004-637X/770/2/118} {\bibfield  {journal} {\bibinfo  {journal}
  {Astrophys. J.}\ }\textbf {\bibinfo {volume} {770}},\ \bibinfo {pages} {118}
  (\bibinfo {year} {2013})},\ \Eprint
  {http://arxiv.org/abs/1305.2430}{arXiv:1305.2430 [astro-ph.GA]}\BibitemShut
  {NoStop}%
\bibitem [{\citenamefont {Giunti}\ and\ \citenamefont
  {Kim}(2007)}]{Giunti:2007ry}%
  \BibitemOpen
  \bibfield  {author} {\bibinfo {author} {\bibfnamefont {C.}~\bibnamefont
  {Giunti}}\ and\ \bibinfo {author} {\bibfnamefont {C.~W.}\ \bibnamefont
  {Kim}},\ }\href@noop {} {\emph {\bibinfo {title} {{Fundamentals of Neutrino
  Physics and Astrophysics}}}}\ (\bibinfo  {publisher} {Oxford University
  Press},\ \bibinfo {year} {2007})\BibitemShut {NoStop}%
\bibitem [{\citenamefont {Weinberg}(2008)}]{Weinberg:2008zzc}%
  \BibitemOpen
  \bibfield  {author} {\bibinfo {author} {\bibfnamefont {S.}~\bibnamefont
  {Weinberg}},\ }\href@noop {} {\emph {\bibinfo {title} {{Cosmology}}}}\
  (\bibinfo  {publisher} {Oxford University Press},\ \bibinfo {year}
  {2008})\BibitemShut {NoStop}%
\bibitem [{\citenamefont {Olive}\ \emph {et~al.}(2014)\citenamefont {Olive}
  \emph {et~al.}}]{ParticleDataGroup:2014cgo}%
  \BibitemOpen
  \bibfield  {author} {\bibinfo {author} {\bibfnamefont {K.~A.}\ \bibnamefont
  {Olive}} \emph {et~al.} (\bibinfo {collaboration} {Particle Data Group}),\
  }\bibfield  {title} {\emph {\enquote {\bibinfo {title} {{Review of Particle
  Physics}},}\ }}\href {\doibase 10.1088/1674-1137/38/9/090001} {\bibfield
  {journal} {\bibinfo  {journal} {Chin. Phys. C}\ }\textbf {\bibinfo {volume}
  {38}},\ \bibinfo {pages} {090001} (\bibinfo {year} {2014})}\BibitemShut
  {NoStop}%
\bibitem [{\citenamefont {Ade}\ \emph {et~al.}(2016)\citenamefont {Ade} \emph
  {et~al.}}]{Planck:2015fie}%
  \BibitemOpen
  \bibfield  {author} {\bibinfo {author} {\bibfnamefont {P.~A.~R.}\
  \bibnamefont {Ade}} \emph {et~al.} (\bibinfo {collaboration} {Planck}),\
  }\bibfield  {title} {\emph {\enquote {\bibinfo {title} {{Planck 2015 results.
  XIII. Cosmological parameters}},}\ }}\href {\doibase
  10.1051/0004-6361/201525830} {\bibfield  {journal} {\bibinfo  {journal}
  {Astron. Astrophys.}\ }\textbf {\bibinfo {volume} {594}},\ \bibinfo {pages}
  {A13} (\bibinfo {year} {2016})},\ \Eprint
  {http://arxiv.org/abs/1502.01589}{arXiv:1502.01589 [astro-ph.CO]}\BibitemShut
  {NoStop}%
\bibitem [{\citenamefont {Hogg}(1999)}]{Hogg:1999ad}%
  \BibitemOpen
  \bibfield  {author} {\bibinfo {author} {\bibfnamefont {D.~W.}\ \bibnamefont
  {Hogg}},\ }\bibfield  {title} {\emph {\enquote {\bibinfo {title} {{Distance
  measures in cosmology}},}\ }}\href@noop {} {\  (\bibinfo {year} {1999})},\
  \Eprint
  {http://arxiv.org/abs/astro-ph/9905116}{arXiv:astro-ph/9905116}\BibitemShut
  {NoStop}%
\bibitem [{\citenamefont {Hopkins}\ and\ \citenamefont
  {Beacom}(2006)}]{Hopkins:2006bw}%
  \BibitemOpen
  \bibfield  {author} {\bibinfo {author} {\bibfnamefont {A.~M.}\ \bibnamefont
  {Hopkins}}\ and\ \bibinfo {author} {\bibfnamefont {J.~F.}\ \bibnamefont
  {Beacom}},\ }\bibfield  {title} {\emph {\enquote {\bibinfo {title} {{On the
  normalisation of the cosmic star formation history}},}\ }}\href {\doibase
  10.1086/506610} {\bibfield  {journal} {\bibinfo  {journal} {Astrophys. J.}\
  }\textbf {\bibinfo {volume} {651}},\ \bibinfo {pages} {142} (\bibinfo {year}
  {2006})},\ \Eprint
  {http://arxiv.org/abs/astro-ph/0601463}{arXiv:astro-ph/0601463}\BibitemShut
  {NoStop}%
\bibitem [{\citenamefont {Yuksel}\ \emph {et~al.}(2008)\citenamefont {Yuksel},
  \citenamefont {Kistler}, \citenamefont {Beacom},\ and\ \citenamefont
  {Hopkins}}]{Yuksel:2008cu}%
  \BibitemOpen
  \bibfield  {author} {\bibinfo {author} {\bibfnamefont {H.}~\bibnamefont
  {Yuksel}}, \bibinfo {author} {\bibfnamefont {M.~D.}\ \bibnamefont {Kistler}},
  \bibinfo {author} {\bibfnamefont {J.~F.}\ \bibnamefont {Beacom}}, \ and\
  \bibinfo {author} {\bibfnamefont {A.~M.}\ \bibnamefont {Hopkins}},\
  }\bibfield  {title} {\emph {\enquote {\bibinfo {title} {{Revealing the
  High-Redshift Star Formation Rate with Gamma-Ray Bursts}},}\ }}\href
  {\doibase 10.1086/591449} {\bibfield  {journal} {\bibinfo  {journal}
  {Astrophys. J. Lett.}\ }\textbf {\bibinfo {volume} {683}},\ \bibinfo {pages}
  {L58} (\bibinfo {year} {2008})},\ \Eprint
  {http://arxiv.org/abs/0804.4008}{arXiv:0804.4008 [astro-ph]}\BibitemShut
  {NoStop}%
\bibitem [{\citenamefont {Kistler}\ \emph {et~al.}(2009)\citenamefont
  {Kistler}, \citenamefont {Yuksel}, \citenamefont {Beacom}, \citenamefont
  {Hopkins},\ and\ \citenamefont {Wyithe}}]{Kistler:2009mv}%
  \BibitemOpen
  \bibfield  {author} {\bibinfo {author} {\bibfnamefont {M.~D.}\ \bibnamefont
  {Kistler}}, \bibinfo {author} {\bibfnamefont {H.}~\bibnamefont {Yuksel}},
  \bibinfo {author} {\bibfnamefont {J.~F.}\ \bibnamefont {Beacom}}, \bibinfo
  {author} {\bibfnamefont {A.~M.}\ \bibnamefont {Hopkins}}, \ and\ \bibinfo
  {author} {\bibfnamefont {J.~S.~B.}\ \bibnamefont {Wyithe}},\ }\bibfield
  {title} {\emph {\enquote {\bibinfo {title} {{The Star Formation Rate in the
  Reionization Era as Indicated by Gamma-ray Bursts}},}\ }}\href {\doibase
  10.1088/0004-637X/705/2/L104} {\bibfield  {journal} {\bibinfo  {journal}
  {Astrophys. J. Lett.}\ }\textbf {\bibinfo {volume} {705}},\ \bibinfo {pages}
  {L104} (\bibinfo {year} {2009})},\ \Eprint
  {http://arxiv.org/abs/0906.0590}{arXiv:0906.0590 [astro-ph.CO]}\BibitemShut
  {NoStop}%
\bibitem [{\citenamefont {Aartsen}\ \emph
  {et~al.}(2013{\natexlab{a}})\citenamefont {Aartsen} \emph
  {et~al.}}]{Aartsen:2013bka}%
  \BibitemOpen
  \bibfield  {author} {\bibinfo {author} {\bibfnamefont {M.~G.}\ \bibnamefont
  {Aartsen}} \emph {et~al.} (\bibinfo {collaboration} {IceCube}),\ }\bibfield
  {title} {\emph {\enquote {\bibinfo {title} {{First observation of PeV-energy
  neutrinos with IceCube}},}\ }}\href {\doibase 10.1103/PhysRevLett.111.021103}
  {\bibfield  {journal} {\bibinfo  {journal} {Phys. Rev. Lett.}\ }\textbf
  {\bibinfo {volume} {111}},\ \bibinfo {pages} {021103} (\bibinfo {year}
  {2013}{\natexlab{a}})},\ \Eprint
  {http://arxiv.org/abs/1304.5356}{arXiv:1304.5356 [astro-ph.HE]}\BibitemShut
  {NoStop}%
\bibitem [{\citenamefont {Aartsen}\ \emph
  {et~al.}(2013{\natexlab{b}})\citenamefont {Aartsen} \emph
  {et~al.}}]{Aartsen:2013jdh}%
  \BibitemOpen
  \bibfield  {author} {\bibinfo {author} {\bibfnamefont {M.~G.}\ \bibnamefont
  {Aartsen}} \emph {et~al.} (\bibinfo {collaboration} {IceCube}),\ }\bibfield
  {title} {\emph {\enquote {\bibinfo {title} {{Evidence for High-Energy
  Extraterrestrial Neutrinos at the IceCube Detector}},}\ }}\href {\doibase
  10.1126/science.1242856} {\bibfield  {journal} {\bibinfo  {journal}
  {Science}\ }\textbf {\bibinfo {volume} {342}},\ \bibinfo {pages} {1242856}
  (\bibinfo {year} {2013}{\natexlab{b}})},\ \Eprint
  {http://arxiv.org/abs/1311.5238}{arXiv:1311.5238 [astro-ph.HE]}\BibitemShut
  {NoStop}%
\bibitem [{\citenamefont {Kashti}\ and\ \citenamefont
  {Waxman}(2005)}]{Kashti:2005qa}%
  \BibitemOpen
  \bibfield  {author} {\bibinfo {author} {\bibfnamefont {T.}~\bibnamefont
  {Kashti}}\ and\ \bibinfo {author} {\bibfnamefont {E.}~\bibnamefont
  {Waxman}},\ }\bibfield  {title} {\emph {\enquote {\bibinfo {title}
  {{Flavoring astrophysical neutrinos: Flavor ratios depend on energy}},}\
  }}\href {\doibase 10.1103/PhysRevLett.95.181101} {\bibfield  {journal}
  {\bibinfo  {journal} {Phys. Rev. Lett.}\ }\textbf {\bibinfo {volume} {95}},\
  \bibinfo {pages} {181101} (\bibinfo {year} {2005})},\ \Eprint
  {http://arxiv.org/abs/astro-ph/0507599}{arXiv:astro-ph/0507599}\BibitemShut
  {NoStop}%
\bibitem [{\citenamefont {Lipari}\ \emph {et~al.}(2007)\citenamefont {Lipari},
  \citenamefont {Lusignoli},\ and\ \citenamefont {Meloni}}]{Lipari:2007su}%
  \BibitemOpen
  \bibfield  {author} {\bibinfo {author} {\bibfnamefont {P.}~\bibnamefont
  {Lipari}}, \bibinfo {author} {\bibfnamefont {M.}~\bibnamefont {Lusignoli}}, \
  and\ \bibinfo {author} {\bibfnamefont {D.}~\bibnamefont {Meloni}},\
  }\bibfield  {title} {\emph {\enquote {\bibinfo {title} {{Flavor Composition
  and Energy Spectrum of Astrophysical Neutrinos}},}\ }}\href {\doibase
  10.1103/PhysRevD.75.123005} {\bibfield  {journal} {\bibinfo  {journal} {Phys.
  Rev. D}\ }\textbf {\bibinfo {volume} {75}},\ \bibinfo {pages} {123005}
  (\bibinfo {year} {2007})},\ \Eprint
  {http://arxiv.org/abs/0704.0718}{arXiv:0704.0718 [astro-ph]}\BibitemShut
  {NoStop}%
\bibitem [{\citenamefont {Hummer}\ \emph {et~al.}(2010)\citenamefont {Hummer},
  \citenamefont {Ruger}, \citenamefont {Spanier},\ and\ \citenamefont
  {Winter}}]{Hummer:2010vx}%
  \BibitemOpen
  \bibfield  {author} {\bibinfo {author} {\bibfnamefont {S.}~\bibnamefont
  {Hummer}}, \bibinfo {author} {\bibfnamefont {M.}~\bibnamefont {Ruger}},
  \bibinfo {author} {\bibfnamefont {F.}~\bibnamefont {Spanier}}, \ and\
  \bibinfo {author} {\bibfnamefont {W.}~\bibnamefont {Winter}},\ }\bibfield
  {title} {\emph {\enquote {\bibinfo {title} {{Simplified models for
  photohadronic interactions in cosmic accelerators}},}\ }}\href {\doibase
  10.1088/0004-637X/721/1/630} {\bibfield  {journal} {\bibinfo  {journal}
  {Astrophys. J.}\ }\textbf {\bibinfo {volume} {721}},\ \bibinfo {pages} {630}
  (\bibinfo {year} {2010})},\ \Eprint
  {http://arxiv.org/abs/1002.1310}{arXiv:1002.1310 [astro-ph.HE]}\BibitemShut
  {NoStop}%
\bibitem [{\citenamefont {Winter}(2013)}]{Winter:2013cla}%
  \BibitemOpen
  \bibfield  {author} {\bibinfo {author} {\bibfnamefont {W.}~\bibnamefont
  {Winter}},\ }\bibfield  {title} {\emph {\enquote {\bibinfo {title}
  {{Photohadronic Origin of the TeV-PeV Neutrinos Observed in IceCube}},}\
  }}\href {\doibase 10.1103/PhysRevD.88.083007} {\bibfield  {journal} {\bibinfo
   {journal} {Phys. Rev. D}\ }\textbf {\bibinfo {volume} {88}},\ \bibinfo
  {pages} {083007} (\bibinfo {year} {2013})},\ \Eprint
  {http://arxiv.org/abs/1307.2793}{arXiv:1307.2793 [astro-ph.HE]}\BibitemShut
  {NoStop}%
\bibitem [{\citenamefont {Bustamante}\ and\ \citenamefont
  {Tamborra}(2020)}]{Bustamante:2020bxp}%
  \BibitemOpen
  \bibfield  {author} {\bibinfo {author} {\bibfnamefont {M.}~\bibnamefont
  {Bustamante}}\ and\ \bibinfo {author} {\bibfnamefont {I.}~\bibnamefont
  {Tamborra}},\ }\bibfield  {title} {\emph {\enquote {\bibinfo {title} {{Using
  high-energy neutrinos as cosmic magnetometers}},}\ }}\href {\doibase
  10.1103/PhysRevD.102.123008} {\bibfield  {journal} {\bibinfo  {journal}
  {Phys. Rev. D}\ }\textbf {\bibinfo {volume} {102}},\ \bibinfo {pages}
  {123008} (\bibinfo {year} {2020})},\ \Eprint
  {http://arxiv.org/abs/2009.01306}{arXiv:2009.01306 [astro-ph.HE]}\BibitemShut
  {NoStop}%
\bibitem [{\citenamefont {Abbasi}\ \emph
  {et~al.}(2022{\natexlab{b}})\citenamefont {Abbasi} \emph
  {et~al.}}]{IceCube:2021uhz}%
  \BibitemOpen
  \bibfield  {author} {\bibinfo {author} {\bibfnamefont {R.}~\bibnamefont
  {Abbasi}} \emph {et~al.} (\bibinfo {collaboration} {IceCube}),\ }\bibfield
  {title} {\emph {\enquote {\bibinfo {title} {{Improved Characterization of the
  Astrophysical Muon\textendash{}neutrino Flux with 9.5 Years of IceCube
  Data}},}\ }}\href {\doibase 10.3847/1538-4357/ac4d29} {\bibfield  {journal}
  {\bibinfo  {journal} {Astrophys. J.}\ }\textbf {\bibinfo {volume} {928}},\
  \bibinfo {pages} {50} (\bibinfo {year} {2022}{\natexlab{b}})},\ \Eprint
  {http://arxiv.org/abs/2111.10299}{arXiv:2111.10299 [astro-ph.HE]}\BibitemShut
  {NoStop}%
\bibitem [{\citenamefont {Fiorillo}\ and\ \citenamefont
  {Bustamante}(2023)}]{Fiorillo:2022rft}%
  \BibitemOpen
  \bibfield  {author} {\bibinfo {author} {\bibfnamefont {D.~F.~G.}\
  \bibnamefont {Fiorillo}}\ and\ \bibinfo {author} {\bibfnamefont
  {M.}~\bibnamefont {Bustamante}},\ }\bibfield  {title} {\emph {\enquote
  {\bibinfo {title} {{Bump hunting in the diffuse flux of high-energy cosmic
  neutrinos}},}\ }}\href {\doibase 10.1103/PhysRevD.107.083008} {\bibfield
  {journal} {\bibinfo  {journal} {Phys. Rev. D}\ }\textbf {\bibinfo {volume}
  {107}},\ \bibinfo {pages} {083008} (\bibinfo {year} {2023})},\ \Eprint
  {http://arxiv.org/abs/2301.00024}{arXiv:2301.00024 [astro-ph.HE]}\BibitemShut
  {NoStop}%
\bibitem [{\citenamefont {Liu}\ \emph {et~al.}(2023)\citenamefont {Liu},
  \citenamefont {Fiorillo}, \citenamefont {Arg\"uelles}, \citenamefont
  {Bustamante}, \citenamefont {Song},\ and\ \citenamefont
  {Vincent}}]{Liu:2023flr}%
  \BibitemOpen
  \bibfield  {author} {\bibinfo {author} {\bibfnamefont {Q.}~\bibnamefont
  {Liu}}, \bibinfo {author} {\bibfnamefont {D.~F.~G.}\ \bibnamefont
  {Fiorillo}}, \bibinfo {author} {\bibfnamefont {C.~A.}\ \bibnamefont
  {Arg\"uelles}}, \bibinfo {author} {\bibfnamefont {M.}~\bibnamefont
  {Bustamante}}, \bibinfo {author} {\bibfnamefont {N.}~\bibnamefont {Song}}, \
  and\ \bibinfo {author} {\bibfnamefont {A.~C.}\ \bibnamefont {Vincent}},\
  }\bibfield  {title} {\emph {\enquote {\bibinfo {title} {{Identifying
  Energy-Dependent Flavor Transitions in High-Energy Astrophysical Neutrino
  Measurements}},}\ }}\href@noop {} {\  (\bibinfo {year} {2023})},\ \Eprint
  {http://arxiv.org/abs/2312.07649}{arXiv:2312.07649 [astro-ph.HE]}\BibitemShut
  {NoStop}%
\bibitem [{\citenamefont {Aartsen}\ \emph
  {et~al.}(2017{\natexlab{a}})\citenamefont {Aartsen} \emph
  {et~al.}}]{IceCube:2016zyt}%
  \BibitemOpen
  \bibfield  {author} {\bibinfo {author} {\bibfnamefont {M.~G.}\ \bibnamefont
  {Aartsen}} \emph {et~al.} (\bibinfo {collaboration} {IceCube}),\ }\bibfield
  {title} {\emph {\enquote {\bibinfo {title} {{The IceCube Neutrino
  Observatory: Instrumentation and Online Systems}},}\ }}\href {\doibase
  10.1088/1748-0221/12/03/P03012} {\bibfield  {journal} {\bibinfo  {journal}
  {JINST}\ }\textbf {\bibinfo {volume} {12}},\ \bibinfo {pages} {P03012}
  (\bibinfo {year} {2017}{\natexlab{a}})},\ \Eprint
  {http://arxiv.org/abs/1612.05093}{arXiv:1612.05093 [astro-ph.IM]}\BibitemShut
  {NoStop}%
\bibitem [{\citenamefont {Ahlers}\ and\ \citenamefont
  {Halzen}(2018)}]{Ahlers:2018fkn}%
  \BibitemOpen
  \bibfield  {author} {\bibinfo {author} {\bibfnamefont {M.}~\bibnamefont
  {Ahlers}}\ and\ \bibinfo {author} {\bibfnamefont {F.}~\bibnamefont
  {Halzen}},\ }\bibfield  {title} {\emph {\enquote {\bibinfo {title} {{Opening
  a New Window onto the Universe with IceCube}},}\ }}\href {\doibase
  10.1016/j.ppnp.2018.05.001} {\bibfield  {journal} {\bibinfo  {journal} {Prog.
  Part. Nucl. Phys.}\ }\textbf {\bibinfo {volume} {102}},\ \bibinfo {pages}
  {73} (\bibinfo {year} {2018})},\ \Eprint
  {http://arxiv.org/abs/1805.11112}{arXiv:1805.11112 [astro-ph.HE]}\BibitemShut
  {NoStop}%
\bibitem [{\citenamefont {Aartsen}\ \emph
  {et~al.}(2017{\natexlab{b}})\citenamefont {Aartsen} \emph
  {et~al.}}]{IceCube:2017roe}%
  \BibitemOpen
  \bibfield  {author} {\bibinfo {author} {\bibfnamefont {M.~G.}\ \bibnamefont
  {Aartsen}} \emph {et~al.} (\bibinfo {collaboration} {IceCube}),\ }\bibfield
  {title} {\emph {\enquote {\bibinfo {title} {{Measurement of the multi-TeV
  neutrino cross section with IceCube using Earth absorption}},}\ }}\href
  {\doibase 10.1038/nature24459} {\bibfield  {journal} {\bibinfo  {journal}
  {Nature}\ }\textbf {\bibinfo {volume} {551}},\ \bibinfo {pages} {596}
  (\bibinfo {year} {2017}{\natexlab{b}})},\ \Eprint
  {http://arxiv.org/abs/1711.08119}{arXiv:1711.08119 [hep-ex]}\BibitemShut
  {NoStop}%
\bibitem [{\citenamefont {Bustamante}\ and\ \citenamefont
  {Connolly}(2019)}]{Bustamante:2017xuy}%
  \BibitemOpen
  \bibfield  {author} {\bibinfo {author} {\bibfnamefont {M.}~\bibnamefont
  {Bustamante}}\ and\ \bibinfo {author} {\bibfnamefont {A.}~\bibnamefont
  {Connolly}},\ }\bibfield  {title} {\emph {\enquote {\bibinfo {title}
  {{Extracting the Energy-Dependent Neutrino-Nucleon Cross Section above 10 TeV
  Using IceCube Showers}},}\ }}\href {\doibase 10.1103/PhysRevLett.122.041101}
  {\bibfield  {journal} {\bibinfo  {journal} {Phys. Rev. Lett.}\ }\textbf
  {\bibinfo {volume} {122}},\ \bibinfo {pages} {041101} (\bibinfo {year}
  {2019})},\ \Eprint {http://arxiv.org/abs/1711.11043}{arXiv:1711.11043
  [astro-ph.HE]}\BibitemShut {NoStop}%
\bibitem [{\citenamefont {Aartsen}\ \emph {et~al.}(2019)\citenamefont {Aartsen}
  \emph {et~al.}}]{Aartsen:2018vez}%
  \BibitemOpen
  \bibfield  {author} {\bibinfo {author} {\bibfnamefont {M.}~\bibnamefont
  {Aartsen}} \emph {et~al.} (\bibinfo {collaboration} {IceCube}),\ }\bibfield
  {title} {\emph {\enquote {\bibinfo {title} {{Measurements using the
  inelasticity distribution of multi-TeV neutrino interactions in IceCube}},}\
  }}\href {\doibase 10.1103/PhysRevD.99.032004} {\bibfield  {journal} {\bibinfo
   {journal} {Phys. Rev. D}\ }\textbf {\bibinfo {volume} {99}},\ \bibinfo
  {pages} {032004} (\bibinfo {year} {2019})},\ \Eprint
  {http://arxiv.org/abs/1808.07629}{arXiv:1808.07629 [hep-ex]}\BibitemShut
  {NoStop}%
\bibitem [{\citenamefont {Abbasi}\ \emph {et~al.}(2020)\citenamefont {Abbasi}
  \emph {et~al.}}]{IceCube:2020rnc}%
  \BibitemOpen
  \bibfield  {author} {\bibinfo {author} {\bibfnamefont {R.}~\bibnamefont
  {Abbasi}} \emph {et~al.} (\bibinfo {collaboration} {IceCube}),\ }\bibfield
  {title} {\emph {\enquote {\bibinfo {title} {{Measurement of the high-energy
  all-flavor neutrino-nucleon cross section with IceCube}},}\ }}\href {\doibase
  10.1103/PhysRevD.104.022001} {\bibfield  {journal} {\bibinfo  {journal}
  {Phys. Rev. D}\ }\textbf {\bibinfo {volume} {104}},\ \bibinfo {pages}
  {022001} (\bibinfo {year} {2020})},\ \Eprint
  {http://arxiv.org/abs/2011.03560}{arXiv:2011.03560 [hep-ex]}\BibitemShut
  {NoStop}%
\bibitem [{\citenamefont {Aartsen}\ \emph {et~al.}(2014)\citenamefont {Aartsen}
  \emph {et~al.}}]{IceCube:2013dkx}%
  \BibitemOpen
  \bibfield  {author} {\bibinfo {author} {\bibfnamefont {M.~G.}\ \bibnamefont
  {Aartsen}} \emph {et~al.} (\bibinfo {collaboration} {IceCube}),\ }\bibfield
  {title} {\emph {\enquote {\bibinfo {title} {{Energy Reconstruction Methods in
  the IceCube Neutrino Telescope}},}\ }}\href {\doibase
  10.1088/1748-0221/9/03/P03009} {\bibfield  {journal} {\bibinfo  {journal}
  {JINST}\ }\textbf {\bibinfo {volume} {9}},\ \bibinfo {pages} {P03009}
  (\bibinfo {year} {2014})},\ \Eprint
  {http://arxiv.org/abs/1311.4767}{arXiv:1311.4767
  [physics.ins-det]}\BibitemShut {NoStop}%
\bibitem [{\citenamefont {Workman}\ \emph {et~al.}(2022)\citenamefont {Workman}
  \emph {et~al.}}]{ParticleDataGroup:2022pth}%
  \BibitemOpen
  \bibfield  {author} {\bibinfo {author} {\bibfnamefont {R.~L.}\ \bibnamefont
  {Workman}} \emph {et~al.} (\bibinfo {collaboration} {Particle Data Group}),\
  }\bibfield  {title} {\emph {\enquote {\bibinfo {title} {{Review of Particle
  Physics}},}\ }}\href {\doibase 10.1093/ptep/ptac097} {\bibfield  {journal}
  {\bibinfo  {journal} {PTEP}\ }\textbf {\bibinfo {volume} {2022}},\ \bibinfo
  {pages} {083C01} (\bibinfo {year} {2022})}\BibitemShut {NoStop}%
\bibitem [{\citenamefont {Learned}\ and\ \citenamefont
  {Pakvasa}(1995)}]{Learned:1994wg}%
  \BibitemOpen
  \bibfield  {author} {\bibinfo {author} {\bibfnamefont {J.~G.}\ \bibnamefont
  {Learned}}\ and\ \bibinfo {author} {\bibfnamefont {S.}~\bibnamefont
  {Pakvasa}},\ }\bibfield  {title} {\emph {\enquote {\bibinfo {title}
  {{Detecting tau-neutrino oscillations at PeV energies}},}\ }}\href {\doibase
  10.1016/0927-6505(94)00043-3} {\bibfield  {journal} {\bibinfo  {journal}
  {Astropart. Phys.}\ }\textbf {\bibinfo {volume} {3}},\ \bibinfo {pages} {267}
  (\bibinfo {year} {1995})},\ \Eprint
  {http://arxiv.org/abs/hep-ph/9405296}{arXiv:hep-ph/9405296
  [hep-ph]}\BibitemShut {NoStop}%
\bibitem [{\citenamefont {Athar}\ \emph {et~al.}(2000)\citenamefont {Athar},
  \citenamefont {Parente},\ and\ \citenamefont {Zas}}]{Athar:2000rx}%
  \BibitemOpen
  \bibfield  {author} {\bibinfo {author} {\bibfnamefont {H.}~\bibnamefont
  {Athar}}, \bibinfo {author} {\bibfnamefont {G.}~\bibnamefont {Parente}}, \
  and\ \bibinfo {author} {\bibfnamefont {E.}~\bibnamefont {Zas}},\ }\bibfield
  {title} {\emph {\enquote {\bibinfo {title} {{Prospects for observations of
  high-energy cosmic tau neutrinos}},}\ }}\href {\doibase
  10.1103/PhysRevD.62.093010} {\bibfield  {journal} {\bibinfo  {journal} {Phys.
  Rev. D}\ }\textbf {\bibinfo {volume} {62}},\ \bibinfo {pages} {093010}
  (\bibinfo {year} {2000})},\ \Eprint
  {http://arxiv.org/abs/hep-ph/0006123}{arXiv:hep-ph/0006123}\BibitemShut
  {NoStop}%
\bibitem [{\citenamefont {Abbasi}\ \emph
  {et~al.}(2022{\natexlab{c}})\citenamefont {Abbasi} \emph
  {et~al.}}]{IceCube:2020fpi}%
  \BibitemOpen
  \bibfield  {author} {\bibinfo {author} {\bibfnamefont {R.}~\bibnamefont
  {Abbasi}} \emph {et~al.} (\bibinfo {collaboration} {IceCube}),\ }\bibfield
  {title} {\emph {\enquote {\bibinfo {title} {{Detection of astrophysical tau
  neutrino candidates in IceCube}},}\ }}\href {\doibase
  10.1140/epjc/s10052-022-10795-y} {\bibfield  {journal} {\bibinfo  {journal}
  {Eur. Phys. J. C}\ }\textbf {\bibinfo {volume} {82}},\ \bibinfo {pages}
  {1031} (\bibinfo {year} {2022}{\natexlab{c}})},\ \Eprint
  {http://arxiv.org/abs/2011.03561}{arXiv:2011.03561 [hep-ex]}\BibitemShut
  {NoStop}%
\bibitem [{\citenamefont {Sch{\"o}nert}\ \emph {et~al.}(2009)\citenamefont
  {Sch{\"o}nert}, \citenamefont {Gaisser}, \citenamefont {Resconi},\ and\
  \citenamefont {Schulz}}]{Schonert:2008is}%
  \BibitemOpen
  \bibfield  {author} {\bibinfo {author} {\bibfnamefont {S.}~\bibnamefont
  {Sch{\"o}nert}}, \bibinfo {author} {\bibfnamefont {T.~K.}\ \bibnamefont
  {Gaisser}}, \bibinfo {author} {\bibfnamefont {E.}~\bibnamefont {Resconi}}, \
  and\ \bibinfo {author} {\bibfnamefont {O.}~\bibnamefont {Schulz}},\
  }\bibfield  {title} {\emph {\enquote {\bibinfo {title} {{Vetoing atmospheric
  neutrinos in a high energy neutrino telescope}},}\ }}\href {\doibase
  10.1103/PhysRevD.79.043009} {\bibfield  {journal} {\bibinfo  {journal} {Phys.
  Rev. D}\ }\textbf {\bibinfo {volume} {79}},\ \bibinfo {pages} {043009}
  (\bibinfo {year} {2009})},\ \Eprint
  {http://arxiv.org/abs/0812.4308}{arXiv:0812.4308 [astro-ph]}\BibitemShut
  {NoStop}%
\bibitem [{\citenamefont {Gaisser}\ \emph {et~al.}(2014)\citenamefont
  {Gaisser}, \citenamefont {Jero}, \citenamefont {Karle},\ and\ \citenamefont
  {van Santen}}]{Gaisser:2014bja}%
  \BibitemOpen
  \bibfield  {author} {\bibinfo {author} {\bibfnamefont {T.~K.}\ \bibnamefont
  {Gaisser}}, \bibinfo {author} {\bibfnamefont {K.}~\bibnamefont {Jero}},
  \bibinfo {author} {\bibfnamefont {A.}~\bibnamefont {Karle}}, \ and\ \bibinfo
  {author} {\bibfnamefont {J.}~\bibnamefont {van Santen}},\ }\bibfield  {title}
  {\emph {\enquote {\bibinfo {title} {{Generalized self-veto probability for
  atmospheric neutrinos}},}\ }}\href {\doibase 10.1103/PhysRevD.90.023009}
  {\bibfield  {journal} {\bibinfo  {journal} {Phys. Rev. D}\ }\textbf {\bibinfo
  {volume} {90}},\ \bibinfo {pages} {023009} (\bibinfo {year} {2014})},\
  \Eprint {http://arxiv.org/abs/1405.0525}{arXiv:1405.0525
  [astro-ph.HE]}\BibitemShut {NoStop}%
\bibitem [{\citenamefont {Aartsen}\ \emph
  {et~al.}(2015{\natexlab{a}})\citenamefont {Aartsen} \emph
  {et~al.}}]{IceCube:2015gsk}%
  \BibitemOpen
  \bibfield  {author} {\bibinfo {author} {\bibfnamefont {M.~G.}\ \bibnamefont
  {Aartsen}} \emph {et~al.} (\bibinfo {collaboration} {IceCube}),\ }\bibfield
  {title} {\emph {\enquote {\bibinfo {title} {{A combined maximum-likelihood
  analysis of the high-energy astrophysical neutrino flux measured with
  IceCube}},}\ }}\href {\doibase 10.1088/0004-637X/809/1/98} {\bibfield
  {journal} {\bibinfo  {journal} {Astrophys. J.}\ }\textbf {\bibinfo {volume}
  {809}},\ \bibinfo {pages} {98} (\bibinfo {year} {2015}{\natexlab{a}})},\
  \Eprint {http://arxiv.org/abs/1507.03991}{arXiv:1507.03991
  [astro-ph.HE]}\BibitemShut {NoStop}%
\bibitem [{\citenamefont {Glashow}(1960)}]{Glashow:1960zz}%
  \BibitemOpen
  \bibfield  {author} {\bibinfo {author} {\bibfnamefont {S.~L.}\ \bibnamefont
  {Glashow}},\ }\bibfield  {title} {\emph {\enquote {\bibinfo {title}
  {{Resonant Scattering of Antineutrinos}},}\ }}\href {\doibase
  10.1103/PhysRev.118.316} {\bibfield  {journal} {\bibinfo  {journal} {Phys.
  Rev.}\ }\textbf {\bibinfo {volume} {118}},\ \bibinfo {pages} {316} (\bibinfo
  {year} {1960})}\BibitemShut {NoStop}%
\bibitem [{\citenamefont {Aartsen}\ \emph
  {et~al.}(2021{\natexlab{a}})\citenamefont {Aartsen} \emph
  {et~al.}}]{IceCube:2021rpz}%
  \BibitemOpen
  \bibfield  {author} {\bibinfo {author} {\bibfnamefont {M.~G.}\ \bibnamefont
  {Aartsen}} \emph {et~al.} (\bibinfo {collaboration} {IceCube}),\ }\bibfield
  {title} {\emph {\enquote {\bibinfo {title} {{Detection of a particle shower
  at the Glashow resonance with IceCube}},}\ }}\href {\doibase
  10.1038/s41586-021-03256-1} {\bibfield  {journal} {\bibinfo  {journal}
  {Nature}\ }\textbf {\bibinfo {volume} {591}},\ \bibinfo {pages} {220}
  (\bibinfo {year} {2021}{\natexlab{a}})},\ \bibinfo {note} {[Erratum: Nature
  592, E11 (2021)]},\ \Eprint
  {http://arxiv.org/abs/2110.15051}{arXiv:2110.15051 [hep-ex]}\BibitemShut
  {NoStop}%
\bibitem [{\citenamefont {Bhattacharya}\ \emph {et~al.}(2011)\citenamefont
  {Bhattacharya}, \citenamefont {Gandhi}, \citenamefont {Rodejohann},\ and\
  \citenamefont {Watanabe}}]{Bhattacharya:2011qu}%
  \BibitemOpen
  \bibfield  {author} {\bibinfo {author} {\bibfnamefont {A.}~\bibnamefont
  {Bhattacharya}}, \bibinfo {author} {\bibfnamefont {R.}~\bibnamefont
  {Gandhi}}, \bibinfo {author} {\bibfnamefont {W.}~\bibnamefont {Rodejohann}},
  \ and\ \bibinfo {author} {\bibfnamefont {A.}~\bibnamefont {Watanabe}},\
  }\bibfield  {title} {\emph {\enquote {\bibinfo {title} {{The Glashow
  resonance at IceCube: signatures, event rates and $pp$ vs. $p\gamma$
  interactions}},}\ }}\href {\doibase 10.1088/1475-7516/2011/10/017} {\bibfield
   {journal} {\bibinfo  {journal} {JCAP}\ }\textbf {\bibinfo {volume} {10}},\
  \bibinfo {pages} {017} (\bibinfo {year} {2011})},\ \Eprint
  {http://arxiv.org/abs/1108.3163}{arXiv:1108.3163 [astro-ph.HE]}\BibitemShut
  {NoStop}%
\bibitem [{\citenamefont {Bhattacharya}\ \emph {et~al.}(2012)\citenamefont
  {Bhattacharya}, \citenamefont {Gandhi}, \citenamefont {Rodejohann},\ and\
  \citenamefont {Watanabe}}]{Bhattacharya:2012fh}%
  \BibitemOpen
  \bibfield  {author} {\bibinfo {author} {\bibfnamefont {A.}~\bibnamefont
  {Bhattacharya}}, \bibinfo {author} {\bibfnamefont {R.}~\bibnamefont
  {Gandhi}}, \bibinfo {author} {\bibfnamefont {W.}~\bibnamefont {Rodejohann}},
  \ and\ \bibinfo {author} {\bibfnamefont {A.}~\bibnamefont {Watanabe}},\
  }\bibfield  {title} {\emph {\enquote {\bibinfo {title} {{On the
  interpretation of IceCube cascade events in terms of the Glashow
  resonance}},}\ }}\href@noop {} {\  (\bibinfo {year} {2012})},\ \Eprint
  {http://arxiv.org/abs/1209.2422}{arXiv:1209.2422 [hep-ph]}\BibitemShut
  {NoStop}%
\bibitem [{\citenamefont {Biehl}\ \emph {et~al.}(2017)\citenamefont {Biehl},
  \citenamefont {Fedynitch}, \citenamefont {Palladino}, \citenamefont
  {Weiler},\ and\ \citenamefont {Winter}}]{Biehl:2016psj}%
  \BibitemOpen
  \bibfield  {author} {\bibinfo {author} {\bibfnamefont {D.}~\bibnamefont
  {Biehl}}, \bibinfo {author} {\bibfnamefont {A.}~\bibnamefont {Fedynitch}},
  \bibinfo {author} {\bibfnamefont {A.}~\bibnamefont {Palladino}}, \bibinfo
  {author} {\bibfnamefont {T.~J.}\ \bibnamefont {Weiler}}, \ and\ \bibinfo
  {author} {\bibfnamefont {W.}~\bibnamefont {Winter}},\ }\bibfield  {title}
  {\emph {\enquote {\bibinfo {title} {{Astrophysical Neutrino Production
  Diagnostics with the Glashow Resonance}},}\ }}\href {\doibase
  10.1088/1475-7516/2017/01/033} {\bibfield  {journal} {\bibinfo  {journal}
  {JCAP}\ }\textbf {\bibinfo {volume} {01}},\ \bibinfo {pages} {033} (\bibinfo
  {year} {2017})},\ \Eprint {http://arxiv.org/abs/1611.07983}{arXiv:1611.07983
  [astro-ph.HE]}\BibitemShut {NoStop}%
\bibitem [{\citenamefont {Huang}\ and\ \citenamefont
  {Liu}(2020)}]{Huang:2019hgs}%
  \BibitemOpen
  \bibfield  {author} {\bibinfo {author} {\bibfnamefont {G.-y.}\ \bibnamefont
  {Huang}}\ and\ \bibinfo {author} {\bibfnamefont {Q.}~\bibnamefont {Liu}},\
  }\bibfield  {title} {\emph {\enquote {\bibinfo {title} {{Hunting the Glashow
  Resonance with PeV Neutrino Telescopes}},}\ }}\href {\doibase
  10.1088/1475-7516/2020/03/005} {\bibfield  {journal} {\bibinfo  {journal}
  {JCAP}\ }\textbf {\bibinfo {volume} {03}},\ \bibinfo {pages} {005} (\bibinfo
  {year} {2020})},\ \Eprint {http://arxiv.org/abs/1912.02976}{arXiv:1912.02976
  [hep-ph]}\BibitemShut {NoStop}%
\bibitem [{\citenamefont {Aartsen}\ \emph
  {et~al.}(2015{\natexlab{b}})\citenamefont {Aartsen} \emph
  {et~al.}}]{IceCube:2015rro}%
  \BibitemOpen
  \bibfield  {author} {\bibinfo {author} {\bibfnamefont {M.~G.}\ \bibnamefont
  {Aartsen}} \emph {et~al.} (\bibinfo {collaboration} {IceCube}),\ }\bibfield
  {title} {\emph {\enquote {\bibinfo {title} {{Flavor Ratio of Astrophysical
  Neutrinos above 35 TeV in IceCube}},}\ }}\href {\doibase
  10.1103/PhysRevLett.114.171102} {\bibfield  {journal} {\bibinfo  {journal}
  {Phys. Rev. Lett.}\ }\textbf {\bibinfo {volume} {114}},\ \bibinfo {pages}
  {171102} (\bibinfo {year} {2015}{\natexlab{b}})},\ \Eprint
  {http://arxiv.org/abs/1502.03376}{arXiv:1502.03376 [astro-ph.HE]}\BibitemShut
  {NoStop}%
\bibitem [{\citenamefont {Mena}\ \emph {et~al.}(2014)\citenamefont {Mena},
  \citenamefont {Palomares-Ruiz},\ and\ \citenamefont
  {Vincent}}]{Mena:2014sja}%
  \BibitemOpen
  \bibfield  {author} {\bibinfo {author} {\bibfnamefont {O.}~\bibnamefont
  {Mena}}, \bibinfo {author} {\bibfnamefont {S.}~\bibnamefont
  {Palomares-Ruiz}}, \ and\ \bibinfo {author} {\bibfnamefont {A.~C.}\
  \bibnamefont {Vincent}},\ }\bibfield  {title} {\emph {\enquote {\bibinfo
  {title} {{Flavor Composition of the High-Energy Neutrino Events in
  IceCube}},}\ }}\href {\doibase 10.1103/PhysRevLett.113.091103} {\bibfield
  {journal} {\bibinfo  {journal} {Phys. Rev. Lett.}\ }\textbf {\bibinfo
  {volume} {113}},\ \bibinfo {pages} {091103} (\bibinfo {year} {2014})},\
  \Eprint {http://arxiv.org/abs/1404.0017}{arXiv:1404.0017
  [astro-ph.HE]}\BibitemShut {NoStop}%
\bibitem [{\citenamefont {Palomares-Ruiz}\ \emph {et~al.}(2015)\citenamefont
  {Palomares-Ruiz}, \citenamefont {Vincent},\ and\ \citenamefont
  {Mena}}]{Palomares-Ruiz:2015mka}%
  \BibitemOpen
  \bibfield  {author} {\bibinfo {author} {\bibfnamefont {S.}~\bibnamefont
  {Palomares-Ruiz}}, \bibinfo {author} {\bibfnamefont {A.~C.}\ \bibnamefont
  {Vincent}}, \ and\ \bibinfo {author} {\bibfnamefont {O.}~\bibnamefont
  {Mena}},\ }\bibfield  {title} {\emph {\enquote {\bibinfo {title} {{Spectral
  analysis of the high-energy IceCube neutrinos}},}\ }}\href {\doibase
  10.1103/PhysRevD.91.103008} {\bibfield  {journal} {\bibinfo  {journal} {Phys.
  Rev. D}\ }\textbf {\bibinfo {volume} {91}},\ \bibinfo {pages} {103008}
  (\bibinfo {year} {2015})},\ \Eprint
  {http://arxiv.org/abs/1502.02649}{arXiv:1502.02649 [astro-ph.HE]}\BibitemShut
  {NoStop}%
\bibitem [{\citenamefont {Vincent}\ \emph {et~al.}(2016)\citenamefont
  {Vincent}, \citenamefont {Palomares-Ruiz},\ and\ \citenamefont
  {Mena}}]{Vincent:2016nut}%
  \BibitemOpen
  \bibfield  {author} {\bibinfo {author} {\bibfnamefont {A.~C.}\ \bibnamefont
  {Vincent}}, \bibinfo {author} {\bibfnamefont {S.}~\bibnamefont
  {Palomares-Ruiz}}, \ and\ \bibinfo {author} {\bibfnamefont {O.}~\bibnamefont
  {Mena}},\ }\bibfield  {title} {\emph {\enquote {\bibinfo {title} {{Analysis
  of the 4-year IceCube high-energy starting events}},}\ }}\href {\doibase
  10.1103/PhysRevD.94.023009} {\bibfield  {journal} {\bibinfo  {journal} {Phys.
  Rev. D}\ }\textbf {\bibinfo {volume} {94}},\ \bibinfo {pages} {023009}
  (\bibinfo {year} {2016})},\ \Eprint
  {http://arxiv.org/abs/1605.01556}{arXiv:1605.01556 [astro-ph.HE]}\BibitemShut
  {NoStop}%
\bibitem [{\citenamefont {Brdar}\ and\ \citenamefont
  {Hansen}(2019)}]{Brdar:2018tce}%
  \BibitemOpen
  \bibfield  {author} {\bibinfo {author} {\bibfnamefont {V.}~\bibnamefont
  {Brdar}}\ and\ \bibinfo {author} {\bibfnamefont {R.~S.}\ \bibnamefont
  {Hansen}},\ }\bibfield  {title} {\emph {\enquote {\bibinfo {title} {{IceCube
  Flavor Ratios with Identified Astrophysical Sources: Towards Improving New
  Physics Testability}},}\ }}\href {\doibase 10.1088/1475-7516/2019/02/023}
  {\bibfield  {journal} {\bibinfo  {journal} {JCAP}\ }\textbf {\bibinfo
  {volume} {02}},\ \bibinfo {pages} {023} (\bibinfo {year} {2019})},\ \Eprint
  {http://arxiv.org/abs/1812.05541}{arXiv:1812.05541 [hep-ph]}\BibitemShut
  {NoStop}%
\bibitem [{\citenamefont {Palladino}(2019)}]{Palladino:2019pid}%
  \BibitemOpen
  \bibfield  {author} {\bibinfo {author} {\bibfnamefont {A.}~\bibnamefont
  {Palladino}},\ }\bibfield  {title} {\emph {\enquote {\bibinfo {title} {{The
  flavor composition of astrophysical neutrinos after 8 years of IceCube: an
  indication of neutron decay scenario?}}}\ }}\href {\doibase
  10.1140/epjc/s10052-019-7018-7} {\bibfield  {journal} {\bibinfo  {journal}
  {Eur. Phys. J. C}\ }\textbf {\bibinfo {volume} {79}},\ \bibinfo {pages} {500}
  (\bibinfo {year} {2019})},\ \Eprint
  {http://arxiv.org/abs/1902.08630}{arXiv:1902.08630 [astro-ph.HE]}\BibitemShut
  {NoStop}%
\bibitem [{\citenamefont {Song}\ \emph {et~al.}(2021)\citenamefont {Song},
  \citenamefont {Li}, \citenamefont {Arg\"uelles}, \citenamefont {Bustamante},\
  and\ \citenamefont {Vincent}}]{Song:2020nfh}%
  \BibitemOpen
  \bibfield  {author} {\bibinfo {author} {\bibfnamefont {N.}~\bibnamefont
  {Song}}, \bibinfo {author} {\bibfnamefont {S.~W.}\ \bibnamefont {Li}},
  \bibinfo {author} {\bibfnamefont {C.~A.}\ \bibnamefont {Arg\"uelles}},
  \bibinfo {author} {\bibfnamefont {M.}~\bibnamefont {Bustamante}}, \ and\
  \bibinfo {author} {\bibfnamefont {A.~C.}\ \bibnamefont {Vincent}},\
  }\bibfield  {title} {\emph {\enquote {\bibinfo {title} {{The Future of
  High-Energy Astrophysical Neutrino Flavor Measurements}},}\ }}\href {\doibase
  10.1088/1475-7516/2021/04/054} {\bibfield  {journal} {\bibinfo  {journal}
  {JCAP}\ }\textbf {\bibinfo {volume} {04}},\ \bibinfo {pages} {054} (\bibinfo
  {year} {2021})},\ \Eprint {http://arxiv.org/abs/2012.12893}{arXiv:2012.12893
  [hep-ph]}\BibitemShut {NoStop}%
\bibitem [{\citenamefont {Schumacher}\ \emph {et~al.}(2021)\citenamefont
  {Schumacher}, \citenamefont {Huber}, \citenamefont {Agostini}, \citenamefont
  {Bustamante}, \citenamefont {Oikonomou},\ and\ \citenamefont
  {Resconi}}]{Schumacher:2021hhm}%
  \BibitemOpen
  \bibfield  {author} {\bibinfo {author} {\bibfnamefont {L.~J.}\ \bibnamefont
  {Schumacher}}, \bibinfo {author} {\bibfnamefont {M.}~\bibnamefont {Huber}},
  \bibinfo {author} {\bibfnamefont {M.}~\bibnamefont {Agostini}}, \bibinfo
  {author} {\bibfnamefont {M.}~\bibnamefont {Bustamante}}, \bibinfo {author}
  {\bibfnamefont {F.}~\bibnamefont {Oikonomou}}, \ and\ \bibinfo {author}
  {\bibfnamefont {E.}~\bibnamefont {Resconi}},\ }\bibfield  {title} {\emph
  {\enquote {\bibinfo {title} {{PLE$\nu$M: A global and distributed monitoring
  system of high-energy astrophysical neutrinos}},}\ }}\href {\doibase
  10.22323/1.395.1185} {\bibfield  {journal} {\bibinfo  {journal} {PoS}\
  }\textbf {\bibinfo {volume} {ICRC2021}},\ \bibinfo {pages} {1185} (\bibinfo
  {year} {2021})},\ \Eprint {http://arxiv.org/abs/2107.13534}{arXiv:2107.13534
  [astro-ph.IM]}\BibitemShut {NoStop}%
\bibitem [{\citenamefont {Bustamante}\ \emph {et~al.}(2015)\citenamefont
  {Bustamante}, \citenamefont {Beacom},\ and\ \citenamefont
  {Winter}}]{Bustamante:2015waa}%
  \BibitemOpen
  \bibfield  {author} {\bibinfo {author} {\bibfnamefont {M.}~\bibnamefont
  {Bustamante}}, \bibinfo {author} {\bibfnamefont {J.~F.}\ \bibnamefont
  {Beacom}}, \ and\ \bibinfo {author} {\bibfnamefont {W.}~\bibnamefont
  {Winter}},\ }\bibfield  {title} {\emph {\enquote {\bibinfo {title}
  {{Theoretically palatable flavor combinations of astrophysical neutrinos}},}\
  }}\href {\doibase 10.1103/PhysRevLett.115.161302} {\bibfield  {journal}
  {\bibinfo  {journal} {Phys. Rev. Lett.}\ }\textbf {\bibinfo {volume} {115}},\
  \bibinfo {pages} {161302} (\bibinfo {year} {2015})},\ \Eprint
  {http://arxiv.org/abs/1506.02645}{arXiv:1506.02645 [astro-ph.HE]}\BibitemShut
  {NoStop}%
\bibitem [{\citenamefont {M{\"u}cke}\ \emph {et~al.}(2000)\citenamefont
  {M{\"u}cke}, \citenamefont {Engel}, \citenamefont {Rachen}, \citenamefont
  {Protheroe},\ and\ \citenamefont {Stanev}}]{Mucke:1999yb}%
  \BibitemOpen
  \bibfield  {author} {\bibinfo {author} {\bibfnamefont {A.}~\bibnamefont
  {M{\"u}cke}}, \bibinfo {author} {\bibfnamefont {R.}~\bibnamefont {Engel}},
  \bibinfo {author} {\bibfnamefont {J.}~\bibnamefont {Rachen}}, \bibinfo
  {author} {\bibfnamefont {R.}~\bibnamefont {Protheroe}}, \ and\ \bibinfo
  {author} {\bibfnamefont {T.}~\bibnamefont {Stanev}},\ }\bibfield  {title}
  {\emph {\enquote {\bibinfo {title} {{SOPHIA: Monte Carlo simulations of
  photohadronic processes in astrophysics}},}\ }}\href {\doibase
  10.1016/S0010-4655(99)00446-4} {\bibfield  {journal} {\bibinfo  {journal}
  {Comput. Phys. Commun.}\ }\textbf {\bibinfo {volume} {124}},\ \bibinfo
  {pages} {290} (\bibinfo {year} {2000})},\ \Eprint
  {http://arxiv.org/abs/astro-ph/9903478}{arXiv:astro-ph/9903478}\BibitemShut
  {NoStop}%
\bibitem [{\citenamefont {Kelner}\ \emph {et~al.}(2006)\citenamefont {Kelner},
  \citenamefont {Aharonian},\ and\ \citenamefont {Bugayov}}]{Kelner:2006tc}%
  \BibitemOpen
  \bibfield  {author} {\bibinfo {author} {\bibfnamefont {S.~R.}\ \bibnamefont
  {Kelner}}, \bibinfo {author} {\bibfnamefont {F.~A.}\ \bibnamefont
  {Aharonian}}, \ and\ \bibinfo {author} {\bibfnamefont {V.~V.}\ \bibnamefont
  {Bugayov}},\ }\bibfield  {title} {\emph {\enquote {\bibinfo {title} {{Energy
  spectra of gamma-rays, electrons and neutrinos produced at proton-proton
  interactions in the very high energy regime}},}\ }}\href {\doibase
  10.1103/PhysRevD.74.034018} {\bibfield  {journal} {\bibinfo  {journal} {Phys.
  Rev. D}\ }\textbf {\bibinfo {volume} {74}},\ \bibinfo {pages} {034018}
  (\bibinfo {year} {2006})},\ \bibinfo {note} {[Erratum: Phys.~Rev.~D 79,
  039901 (2009)]},\ \Eprint
  {http://arxiv.org/abs/astro-ph/0606058}{arXiv:astro-ph/0606058}\BibitemShut
  {NoStop}%
\bibitem [{\citenamefont {Morej\'on}\ \emph {et~al.}(2019)\citenamefont
  {Morej\'on}, \citenamefont {Fedynitch}, \citenamefont {Boncioli},
  \citenamefont {Biehl},\ and\ \citenamefont {Winter}}]{Morejon:2019pfu}%
  \BibitemOpen
  \bibfield  {author} {\bibinfo {author} {\bibfnamefont {L.}~\bibnamefont
  {Morej\'on}}, \bibinfo {author} {\bibfnamefont {A.}~\bibnamefont
  {Fedynitch}}, \bibinfo {author} {\bibfnamefont {D.}~\bibnamefont {Boncioli}},
  \bibinfo {author} {\bibfnamefont {D.}~\bibnamefont {Biehl}}, \ and\ \bibinfo
  {author} {\bibfnamefont {W.}~\bibnamefont {Winter}},\ }\bibfield  {title}
  {\emph {\enquote {\bibinfo {title} {{Improved photomeson model for
  interactions of cosmic ray nuclei}},}\ }}\href {\doibase
  10.1088/1475-7516/2019/11/007} {\bibfield  {journal} {\bibinfo  {journal}
  {JCAP}\ }\textbf {\bibinfo {volume} {11}},\ \bibinfo {pages} {007} (\bibinfo
  {year} {2019})},\ \Eprint {http://arxiv.org/abs/1904.07999}{arXiv:1904.07999
  [astro-ph.HE]}\BibitemShut {NoStop}%
\bibitem [{\citenamefont {Kachelriess}\ and\ \citenamefont
  {Tom\`as}(2006)}]{Kachelriess:2006fi}%
  \BibitemOpen
  \bibfield  {author} {\bibinfo {author} {\bibfnamefont {M.}~\bibnamefont
  {Kachelriess}}\ and\ \bibinfo {author} {\bibfnamefont {R.}~\bibnamefont
  {Tom\`as}},\ }\bibfield  {title} {\emph {\enquote {\bibinfo {title} {{High
  energy neutrino yields from astrophysical sources I: Weakly magnetized
  sources}},}\ }}\href {\doibase 10.1103/PhysRevD.74.063009} {\bibfield
  {journal} {\bibinfo  {journal} {Phys. Rev. D}\ }\textbf {\bibinfo {volume}
  {74}},\ \bibinfo {pages} {063009} (\bibinfo {year} {2006})},\ \Eprint
  {http://arxiv.org/abs/astro-ph/0606406}{arXiv:astro-ph/0606406}\BibitemShut
  {NoStop}%
\bibitem [{\citenamefont {Mehta}\ and\ \citenamefont
  {Winter}(2011)}]{Mehta:2011qb}%
  \BibitemOpen
  \bibfield  {author} {\bibinfo {author} {\bibfnamefont {P.}~\bibnamefont
  {Mehta}}\ and\ \bibinfo {author} {\bibfnamefont {W.}~\bibnamefont {Winter}},\
  }\bibfield  {title} {\emph {\enquote {\bibinfo {title} {{Interplay of energy
  dependent astrophysical neutrino flavor ratios and new physics effects}},}\
  }}\href {\doibase 10.1088/1475-7516/2011/03/041} {\bibfield  {journal}
  {\bibinfo  {journal} {JCAP}\ }\textbf {\bibinfo {volume} {03}},\ \bibinfo
  {pages} {041} (\bibinfo {year} {2011})},\ \Eprint
  {http://arxiv.org/abs/1101.2673}{arXiv:1101.2673 [hep-ph]}\BibitemShut
  {NoStop}%
\bibitem [{\citenamefont {Bustamante}(2020)}]{Bustamante:2020niz}%
  \BibitemOpen
  \bibfield  {author} {\bibinfo {author} {\bibfnamefont {M.}~\bibnamefont
  {Bustamante}},\ }\bibfield  {title} {\emph {\enquote {\bibinfo {title} {{New
  limits on neutrino decay from the Glashow resonance of high-energy cosmic
  neutrinos}},}\ }}\href@noop {} {\  (\bibinfo {year} {2020})},\ \Eprint
  {http://arxiv.org/abs/2004.06844}{arXiv:2004.06844 [astro-ph.HE]}\BibitemShut
  {NoStop}%
\bibitem [{\citenamefont {Aartsen}\ \emph
  {et~al.}(2021{\natexlab{b}})\citenamefont {Aartsen} \emph
  {et~al.}}]{IceCube-Gen2:2020qha}%
  \BibitemOpen
  \bibfield  {author} {\bibinfo {author} {\bibfnamefont {M.~G.}\ \bibnamefont
  {Aartsen}} \emph {et~al.} (\bibinfo {collaboration} {IceCube-Gen2}),\
  }\bibfield  {title} {\emph {\enquote {\bibinfo {title} {{IceCube-Gen2: the
  window to the extreme Universe}},}\ }}\href {\doibase
  10.1088/1361-6471/abbd48} {\bibfield  {journal} {\bibinfo  {journal} {J.
  Phys. G}\ }\textbf {\bibinfo {volume} {48}},\ \bibinfo {pages} {060501}
  (\bibinfo {year} {2021}{\natexlab{b}})},\ \Eprint
  {http://arxiv.org/abs/2008.04323}{arXiv:2008.04323 [astro-ph.HE]}\BibitemShut
  {NoStop}%
\bibitem [{\citenamefont {Avrorin}\ \emph {et~al.}(2020)\citenamefont {Avrorin}
  \emph {et~al.}}]{Avrorin:2019vfc}%
  \BibitemOpen
  \bibfield  {author} {\bibinfo {author} {\bibfnamefont {A.}~\bibnamefont
  {Avrorin}} \emph {et~al.} (\bibinfo {collaboration} {Baikal-GVD}),\
  }\bibfield  {title} {\emph {\enquote {\bibinfo {title} {{The Baikal-GVD
  neutrino telescope: First results of multi-messenger study}},}\ }}\href
  {\doibase 10.22323/1.358.1013} {\bibfield  {journal} {\bibinfo  {journal}
  {PoS}\ }\textbf {\bibinfo {volume} {ICRC2019}},\ \bibinfo {pages} {1013}
  (\bibinfo {year} {2020})},\ \Eprint
  {http://arxiv.org/abs/1908.05450}{arXiv:1908.05450 [astro-ph.HE]}\BibitemShut
  {NoStop}%
\bibitem [{\citenamefont {Adri\'an-Mart\'inez}\ \emph
  {et~al.}(2016)\citenamefont {Adri\'an-Mart\'inez} \emph
  {et~al.}}]{Adrian-Martinez:2016fdl}%
  \BibitemOpen
  \bibfield  {author} {\bibinfo {author} {\bibfnamefont {S.}~\bibnamefont
  {Adri\'an-Mart\'inez}} \emph {et~al.} (\bibinfo {collaboration} {KM3Net}),\
  }\bibfield  {title} {\emph {\enquote {\bibinfo {title} {{Letter of intent for
  KM3NeT 2.0}},}\ }}\href {\doibase 10.1088/0954-3899/43/8/084001} {\bibfield
  {journal} {\bibinfo  {journal} {J. Phys. G}\ }\textbf {\bibinfo {volume}
  {43}},\ \bibinfo {pages} {084001} (\bibinfo {year} {2016})},\ \Eprint
  {http://arxiv.org/abs/1601.07459}{arXiv:1601.07459 [astro-ph.IM]}\BibitemShut
  {NoStop}%
\bibitem [{\citenamefont {Romero-Wolf}\ \emph {et~al.}(2020)\citenamefont
  {Romero-Wolf} \emph {et~al.}}]{Romero-Wolf:2020pzh}%
  \BibitemOpen
  \bibfield  {author} {\bibinfo {author} {\bibfnamefont {A.}~\bibnamefont
  {Romero-Wolf}} \emph {et~al.},\ }in\ \href@noop {} {\emph {\bibinfo
  {booktitle} {{Latin American Strategy Forum for Research Infrastructure}}}}\
  (\bibinfo {year} {2020})\ \Eprint
  {http://arxiv.org/abs/2002.06475}{arXiv:2002.06475 [astro-ph.IM]}\BibitemShut
  {NoStop}%
\bibitem [{\citenamefont {Abi}\ \emph {et~al.}(2020{\natexlab{c}})\citenamefont
  {Abi} \emph {et~al.}}]{Abi:2020wmh}%
  \BibitemOpen
  \bibfield  {author} {\bibinfo {author} {\bibfnamefont {B.}~\bibnamefont
  {Abi}} \emph {et~al.} (\bibinfo {collaboration} {DUNE}),\ }\bibfield  {title}
  {\emph {\enquote {\bibinfo {title} {{Deep Underground Neutrino Experiment
  (DUNE), Far Detector Technical Design Report, Volume I Introduction to
  DUNE}},}\ }}\href {\doibase 10.1088/1748-0221/15/08/T08008} {\bibfield
  {journal} {\bibinfo  {journal} {JINST}\ }\textbf {\bibinfo {volume} {15}},\
  \bibinfo {pages} {T08008} (\bibinfo {year} {2020}{\natexlab{c}})},\ \Eprint
  {http://arxiv.org/abs/2002.02967}{arXiv:2002.02967
  [physics.ins-det]}\BibitemShut {NoStop}%
\bibitem [{\citenamefont {Abe}\ \emph {et~al.}(2018{\natexlab{c}})\citenamefont
  {Abe} \emph {et~al.}}]{Abe:2018uyc}%
  \BibitemOpen
  \bibfield  {author} {\bibinfo {author} {\bibfnamefont {K.}~\bibnamefont
  {Abe}} \emph {et~al.} (\bibinfo {collaboration} {Hyper-Kamiokande}),\
  }\bibfield  {title} {\emph {\enquote {\bibinfo {title} {{Hyper-Kamiokande
  Design Report}},}\ }}\href@noop {} {\  (\bibinfo {year}
  {2018}{\natexlab{c}})},\ \Eprint
  {http://arxiv.org/abs/1805.04163}{arXiv:1805.04163
  [physics.ins-det]}\BibitemShut {NoStop}%
\bibitem [{\citenamefont {Alion}\ \emph {et~al.}(2016)\citenamefont {Alion}
  \emph {et~al.}}]{DUNE:2016ymp}%
  \BibitemOpen
  \bibfield  {author} {\bibinfo {author} {\bibfnamefont {T.}~\bibnamefont
  {Alion}} \emph {et~al.} (\bibinfo {collaboration} {DUNE}),\ }\bibfield
  {title} {\emph {\enquote {\bibinfo {title} {{Experiment Simulation
  Configurations Used in DUNE CDR}},}\ }}\href@noop {} {\  (\bibinfo {year}
  {2016})},\ \Eprint {http://arxiv.org/abs/1606.09550}{arXiv:1606.09550
  [physics.ins-det]}\BibitemShut {NoStop}%
\bibitem [{\citenamefont {An}\ \emph {et~al.}(2017)\citenamefont {An} \emph
  {et~al.}}]{DayaBay:2016ssb}%
  \BibitemOpen
  \bibfield  {author} {\bibinfo {author} {\bibfnamefont {F.~P.}\ \bibnamefont
  {An}} \emph {et~al.} (\bibinfo {collaboration} {Daya Bay}),\ }\bibfield
  {title} {\emph {\enquote {\bibinfo {title} {{Improved Measurement of the
  Reactor Antineutrino Flux and Spectrum at Daya Bay}},}\ }}\href {\doibase
  10.1088/1674-1137/41/1/013002} {\bibfield  {journal} {\bibinfo  {journal}
  {Chin. Phys. C}\ }\textbf {\bibinfo {volume} {41}},\ \bibinfo {pages}
  {013002} (\bibinfo {year} {2017})},\ \Eprint
  {http://arxiv.org/abs/1607.05378}{arXiv:1607.05378 [hep-ex]}\BibitemShut
  {NoStop}%
\bibitem [{\citenamefont {Joshipura}\ and\ \citenamefont
  {Mohanty}(2004)}]{Joshipura:2003jh}%
  \BibitemOpen
  \bibfield  {author} {\bibinfo {author} {\bibfnamefont {A.~S.}\ \bibnamefont
  {Joshipura}}\ and\ \bibinfo {author} {\bibfnamefont {S.}~\bibnamefont
  {Mohanty}},\ }\bibfield  {title} {\emph {\enquote {\bibinfo {title}
  {{Constraints on flavor dependent long range forces from atmospheric neutrino
  observations at super-Kamiokande}},}\ }}\href {\doibase
  10.1016/j.physletb.2004.01.057} {\bibfield  {journal} {\bibinfo  {journal}
  {Phys. Lett. B}\ }\textbf {\bibinfo {volume} {584}},\ \bibinfo {pages} {103}
  (\bibinfo {year} {2004})},\ \Eprint
  {http://arxiv.org/abs/hep-ph/0310210}{arXiv:hep-ph/0310210}\BibitemShut
  {NoStop}%
\bibitem [{\citenamefont {Bandyopadhyay}\ \emph {et~al.}(2007)\citenamefont
  {Bandyopadhyay}, \citenamefont {Dighe},\ and\ \citenamefont
  {Joshipura}}]{Bandyopadhyay:2006uh}%
  \BibitemOpen
  \bibfield  {author} {\bibinfo {author} {\bibfnamefont {A.}~\bibnamefont
  {Bandyopadhyay}}, \bibinfo {author} {\bibfnamefont {A.}~\bibnamefont
  {Dighe}}, \ and\ \bibinfo {author} {\bibfnamefont {A.~S.}\ \bibnamefont
  {Joshipura}},\ }\bibfield  {title} {\emph {\enquote {\bibinfo {title}
  {{Constraints on flavor-dependent long range forces from solar neutrinos and
  KamLAND}},}\ }}\href {\doibase 10.1103/PhysRevD.75.093005} {\bibfield
  {journal} {\bibinfo  {journal} {Phys. Rev. D}\ }\textbf {\bibinfo {volume}
  {75}},\ \bibinfo {pages} {093005} (\bibinfo {year} {2007})},\ \Eprint
  {http://arxiv.org/abs/hep-ph/0610263}{arXiv:hep-ph/0610263}\BibitemShut
  {NoStop}%
\bibitem [{\citenamefont {Mitsuka}\ \emph {et~al.}(2011)\citenamefont {Mitsuka}
  \emph {et~al.}}]{Super-Kamiokande:2011dam}%
  \BibitemOpen
  \bibfield  {author} {\bibinfo {author} {\bibfnamefont {G.}~\bibnamefont
  {Mitsuka}} \emph {et~al.} (\bibinfo {collaboration} {Super-Kamiokande}),\
  }\bibfield  {title} {\emph {\enquote {\bibinfo {title} {{Study of
  Non-Standard Neutrino Interactions with Atmospheric Neutrino Data in
  Super-Kamiokande I and II}},}\ }}\href {\doibase 10.1103/PhysRevD.84.113008}
  {\bibfield  {journal} {\bibinfo  {journal} {Phys. Rev. D}\ }\textbf {\bibinfo
  {volume} {84}},\ \bibinfo {pages} {113008} (\bibinfo {year} {2011})},\
  \Eprint {http://arxiv.org/abs/1109.1889}{arXiv:1109.1889
  [hep-ex]}\BibitemShut {NoStop}%
\bibitem [{\citenamefont {Gonzalez-Garcia}\ and\ \citenamefont
  {Maltoni}(2013)}]{Gonzalez-Garcia:2013usa}%
  \BibitemOpen
  \bibfield  {author} {\bibinfo {author} {\bibfnamefont {M.~C.}\ \bibnamefont
  {Gonzalez-Garcia}}\ and\ \bibinfo {author} {\bibfnamefont {M.}~\bibnamefont
  {Maltoni}},\ }\bibfield  {title} {\emph {\enquote {\bibinfo {title}
  {{Determination of matter potential from global analysis of neutrino
  oscillation data}},}\ }}\href {\doibase 10.1007/JHEP09(2013)152} {\bibfield
  {journal} {\bibinfo  {journal} {JHEP}\ }\textbf {\bibinfo {volume} {09}},\
  \bibinfo {pages} {152} (\bibinfo {year} {2013})},\ \Eprint
  {http://arxiv.org/abs/1307.3092}{arXiv:1307.3092 [hep-ph]}\BibitemShut
  {NoStop}%
\bibitem [{\citenamefont {Wise}\ and\ \citenamefont
  {Zhang}(2018)}]{Wise:2018rnb}%
  \BibitemOpen
  \bibfield  {author} {\bibinfo {author} {\bibfnamefont {M.~B.}\ \bibnamefont
  {Wise}}\ and\ \bibinfo {author} {\bibfnamefont {Y.}~\bibnamefont {Zhang}},\
  }\bibfield  {title} {\emph {\enquote {\bibinfo {title} {{Lepton Flavorful
  Fifth Force and Depth-dependent Neutrino Matter Interactions}},}\ }}\href
  {\doibase 10.1007/JHEP06(2018)053} {\bibfield  {journal} {\bibinfo  {journal}
  {JHEP}\ }\textbf {\bibinfo {volume} {06}},\ \bibinfo {pages} {053} (\bibinfo
  {year} {2018})},\ \Eprint {http://arxiv.org/abs/1803.00591}{arXiv:1803.00591
  [hep-ph]}\BibitemShut {NoStop}%
\bibitem [{\citenamefont {Baryakhtar}\ \emph {et~al.}(2017)\citenamefont
  {Baryakhtar}, \citenamefont {Lasenby},\ and\ \citenamefont
  {Teo}}]{Baryakhtar:2017ngi}%
  \BibitemOpen
  \bibfield  {author} {\bibinfo {author} {\bibfnamefont {M.}~\bibnamefont
  {Baryakhtar}}, \bibinfo {author} {\bibfnamefont {R.}~\bibnamefont {Lasenby}},
  \ and\ \bibinfo {author} {\bibfnamefont {M.}~\bibnamefont {Teo}},\ }\bibfield
   {title} {\emph {\enquote {\bibinfo {title} {{Black Hole Superradiance
  Signatures of Ultralight Vectors}},}\ }}\href {\doibase
  10.1103/PhysRevD.96.035019} {\bibfield  {journal} {\bibinfo  {journal} {Phys.
  Rev. D}\ }\textbf {\bibinfo {volume} {96}},\ \bibinfo {pages} {035019}
  (\bibinfo {year} {2017})},\ \Eprint
  {http://arxiv.org/abs/1704.05081}{arXiv:1704.05081 [hep-ph]}\BibitemShut
  {NoStop}%
\bibitem [{\citenamefont {Arkani-Hamed}\ \emph {et~al.}(2007)\citenamefont
  {Arkani-Hamed}, \citenamefont {Motl}, \citenamefont {Nicolis},\ and\
  \citenamefont {Vafa}}]{Arkani-Hamed:2006emk}%
  \BibitemOpen
  \bibfield  {author} {\bibinfo {author} {\bibfnamefont {N.}~\bibnamefont
  {Arkani-Hamed}}, \bibinfo {author} {\bibfnamefont {L.}~\bibnamefont {Motl}},
  \bibinfo {author} {\bibfnamefont {A.}~\bibnamefont {Nicolis}}, \ and\
  \bibinfo {author} {\bibfnamefont {C.}~\bibnamefont {Vafa}},\ }\bibfield
  {title} {\emph {\enquote {\bibinfo {title} {{The String landscape, black
  holes and gravity as the weakest force}},}\ }}\href {\doibase
  10.1088/1126-6708/2007/06/060} {\bibfield  {journal} {\bibinfo  {journal}
  {JHEP}\ }\textbf {\bibinfo {volume} {06}},\ \bibinfo {pages} {060} (\bibinfo
  {year} {2007})},\ \Eprint
  {http://arxiv.org/abs/hep-th/0601001}{arXiv:hep-th/0601001}\BibitemShut
  {NoStop}%
\bibitem [{\citenamefont {Li}\ \emph {et~al.}(2019)\citenamefont {Li},
  \citenamefont {Bustamante},\ and\ \citenamefont {Beacom}}]{Li:2016kra}%
  \BibitemOpen
  \bibfield  {author} {\bibinfo {author} {\bibfnamefont {S.~W.}\ \bibnamefont
  {Li}}, \bibinfo {author} {\bibfnamefont {M.}~\bibnamefont {Bustamante}}, \
  and\ \bibinfo {author} {\bibfnamefont {J.~F.}\ \bibnamefont {Beacom}},\
  }\bibfield  {title} {\emph {\enquote {\bibinfo {title} {{Echo Technique to
  Distinguish Flavors of Astrophysical Neutrinos}},}\ }}\href {\doibase
  10.1103/PhysRevLett.122.151101} {\bibfield  {journal} {\bibinfo  {journal}
  {Phys. Rev. Lett.}\ }\textbf {\bibinfo {volume} {122}},\ \bibinfo {pages}
  {151101} (\bibinfo {year} {2019})},\ \Eprint
  {http://arxiv.org/abs/1606.06290}{arXiv:1606.06290 [astro-ph.HE]}\BibitemShut
  {NoStop}%
\bibitem [{\citenamefont {Palladino}\ and\ \citenamefont
  {Winter}(2018)}]{Palladino:2018evm}%
  \BibitemOpen
  \bibfield  {author} {\bibinfo {author} {\bibfnamefont {A.}~\bibnamefont
  {Palladino}}\ and\ \bibinfo {author} {\bibfnamefont {W.}~\bibnamefont
  {Winter}},\ }\bibfield  {title} {\emph {\enquote {\bibinfo {title} {{A
  multi-component model for observed astrophysical neutrinos}},}\ }}\href
  {\doibase 10.3204/PUBDB-2018-01376} {\bibfield  {journal} {\bibinfo
  {journal} {Astron. Astrophys.}\ }\textbf {\bibinfo {volume} {615}},\ \bibinfo
  {pages} {A168} (\bibinfo {year} {2018})},\ \Eprint
  {http://arxiv.org/abs/1801.07277}{arXiv:1801.07277 [astro-ph.HE]}\BibitemShut
  {NoStop}%
\bibitem [{\citenamefont {Capanema}\ \emph {et~al.}(2020)\citenamefont
  {Capanema}, \citenamefont {Esmaili},\ and\ \citenamefont
  {Murase}}]{Capanema:2020rjj}%
  \BibitemOpen
  \bibfield  {author} {\bibinfo {author} {\bibfnamefont {A.}~\bibnamefont
  {Capanema}}, \bibinfo {author} {\bibfnamefont {A.}~\bibnamefont {Esmaili}}, \
  and\ \bibinfo {author} {\bibfnamefont {K.}~\bibnamefont {Murase}},\
  }\bibfield  {title} {\emph {\enquote {\bibinfo {title} {{New constraints on
  the origin of medium-energy neutrinos observed by IceCube}},}\ }}\href
  {\doibase 10.1103/PhysRevD.101.103012} {\bibfield  {journal} {\bibinfo
  {journal} {Phys. Rev. D}\ }\textbf {\bibinfo {volume} {101}},\ \bibinfo
  {pages} {103012} (\bibinfo {year} {2020})},\ \Eprint
  {http://arxiv.org/abs/2002.07192}{arXiv:2002.07192 [hep-ph]}\BibitemShut
  {NoStop}%
\bibitem [{\citenamefont {Ambrosone}\ \emph {et~al.}(2021)\citenamefont
  {Ambrosone}, \citenamefont {Chianese}, \citenamefont {Fiorillo},
  \citenamefont {Marinelli}, \citenamefont {Miele},\ and\ \citenamefont
  {Pisanti}}]{Ambrosone:2020evo}%
  \BibitemOpen
  \bibfield  {author} {\bibinfo {author} {\bibfnamefont {A.}~\bibnamefont
  {Ambrosone}}, \bibinfo {author} {\bibfnamefont {M.}~\bibnamefont {Chianese}},
  \bibinfo {author} {\bibfnamefont {D.~F.~G.}\ \bibnamefont {Fiorillo}},
  \bibinfo {author} {\bibfnamefont {A.}~\bibnamefont {Marinelli}}, \bibinfo
  {author} {\bibfnamefont {G.}~\bibnamefont {Miele}}, \ and\ \bibinfo {author}
  {\bibfnamefont {O.}~\bibnamefont {Pisanti}},\ }\bibfield  {title} {\emph
  {\enquote {\bibinfo {title} {{Starburst galaxies strike back: a
  multi-messenger analysis with Fermi-LAT and IceCube data}},}\ }}\href
  {\doibase 10.1093/mnras/stab659} {\bibfield  {journal} {\bibinfo  {journal}
  {Mon. Not. Roy. Astron. Soc.}\ }\textbf {\bibinfo {volume} {503}},\ \bibinfo
  {pages} {4032} (\bibinfo {year} {2021})},\ \Eprint
  {http://arxiv.org/abs/2011.02483}{arXiv:2011.02483 [astro-ph.HE]}\BibitemShut
  {NoStop}%
\bibitem [{\citenamefont {Dolgov}\ and\ \citenamefont
  {Raffelt}(1995)}]{Dolgov:1995hc}%
  \BibitemOpen
  \bibfield  {author} {\bibinfo {author} {\bibfnamefont {A.~D.}\ \bibnamefont
  {Dolgov}}\ and\ \bibinfo {author} {\bibfnamefont {G.~G.}\ \bibnamefont
  {Raffelt}},\ }\bibfield  {title} {\emph {\enquote {\bibinfo {title}
  {{Screening of long range leptonic forces by cosmic background neutrinos}},}\
  }}\href {\doibase 10.1103/PhysRevD.52.2581} {\bibfield  {journal} {\bibinfo
  {journal} {Phys. Rev. D}\ }\textbf {\bibinfo {volume} {52}},\ \bibinfo
  {pages} {2581} (\bibinfo {year} {1995})},\ \Eprint
  {http://arxiv.org/abs/hep-ph/9503438}{arXiv:hep-ph/9503438}\BibitemShut
  {NoStop}%
\bibitem [{\citenamefont {Blinnikov}\ \emph {et~al.}(1996)\citenamefont
  {Blinnikov}, \citenamefont {Dolgov}, \citenamefont {Okun},\ and\
  \citenamefont {Voloshin}}]{Blinnikov:1995kp}%
  \BibitemOpen
  \bibfield  {author} {\bibinfo {author} {\bibfnamefont {S.~I.}\ \bibnamefont
  {Blinnikov}}, \bibinfo {author} {\bibfnamefont {A.~D.}\ \bibnamefont
  {Dolgov}}, \bibinfo {author} {\bibfnamefont {L.~B.}\ \bibnamefont {Okun}}, \
  and\ \bibinfo {author} {\bibfnamefont {M.~B.}\ \bibnamefont {Voloshin}},\
  }\bibfield  {title} {\emph {\enquote {\bibinfo {title} {{How strong can the
  coupling of leptonic photons be?}}}\ }}\href {\doibase
  10.1016/0550-3213(95)00579-X} {\bibfield  {journal} {\bibinfo  {journal}
  {Nucl. Phys. B}\ }\textbf {\bibinfo {volume} {458}},\ \bibinfo {pages} {52}
  (\bibinfo {year} {1996})},\ \Eprint
  {http://arxiv.org/abs/hep-ph/9505444}{arXiv:hep-ph/9505444}\BibitemShut
  {NoStop}%
\bibitem [{\citenamefont {An}\ \emph {et~al.}(2016)\citenamefont {An} \emph
  {et~al.}}]{An:2015jdp}%
  \BibitemOpen
  \bibfield  {author} {\bibinfo {author} {\bibfnamefont {F.}~\bibnamefont {An}}
  \emph {et~al.} (\bibinfo {collaboration} {JUNO}),\ }\bibfield  {title} {\emph
  {\enquote {\bibinfo {title} {{Neutrino Physics with JUNO}},}\ }}\href
  {\doibase 10.1088/0954-3899/43/3/030401} {\bibfield  {journal} {\bibinfo
  {journal} {J. Phys. G}\ }\textbf {\bibinfo {volume} {43}},\ \bibinfo {pages}
  {030401} (\bibinfo {year} {2016})},\ \Eprint
  {http://arxiv.org/abs/1507.05613}{arXiv:1507.05613
  [physics.ins-det]}\BibitemShut {NoStop}%
\bibitem [{\citenamefont {Bilenky}(2013)}]{Bilenky:2013wna}%
  \BibitemOpen
  \bibfield  {author} {\bibinfo {author} {\bibfnamefont {S.~M.}\ \bibnamefont
  {Bilenky}},\ }\bibfield  {title} {\emph {\enquote {\bibinfo {title} {{Bruno
  Pontecorvo and Neutrino Oscillations}},}\ }}\href {\doibase
  10.1155/2013/873236} {\bibfield  {journal} {\bibinfo  {journal} {Adv. High
  Energy Phys.}\ }\textbf {\bibinfo {volume} {2013}},\ \bibinfo {pages}
  {873236} (\bibinfo {year} {2013})}\BibitemShut {NoStop}%
\end{thebibliography}%
\end{document}